%% file: ms.tex
\documentclass[pdflatex,sn-mathphys-num]{sn-jnl}

\usepackage{graphicx}%
\usepackage{multirow}%
\usepackage{amsfonts}%
\usepackage{bm}
\usepackage{mathrsfs}%
\usepackage[title]{appendix}%
\usepackage{xcolor}%
\usepackage{textcomp}%
\usepackage{manyfoot}%
\usepackage{booktabs}%
\usepackage{algorithm}%
\usepackage{algorithmicx}%
\usepackage{algpseudocode}%
\usepackage{listings}%
\usepackage{array}
\usepackage{arydshln}
\usepackage{colortbl}
\usepackage{xcolor}
\usepackage{arydshln} 
\usepackage{mathdefs}
\usepackage{my_macros}
\renewcommand{\blue}[1]{\textcolor{black}{#1}} 
\renewcommand{\bblue}[1]{\textcolor{black}{#1}} 
\begin{document}

\title[]{Rapid wavefront shaping using an optical gradient acquisition}

\author[1]{\fnm{Sagi} \sur{Monin}}

\author[1]{\fnm{Marina} \sur{Alterman}}

\author*[1]{\fnm{Anat} \sur{Levin}}\email{anat.levin.g@gmail.com}

\affil*[1]{\orgdiv{Electrical and Computer Engineering}, \orgname{Technion}, \orgaddress{\city{Haifa}, \country{Israel}}}

%

%
%

\abstract{
	
Wavefront shaping systems enable deep tissue imaging by correcting scattering aberrations, but estimating optimal modulation correction is challenging, since it depends on the unknown tissue structures. Most current methods use slow coordinate descent algorithms, which sequentially scan all modulation parameters and query them independently, thus their complexity scales prohibitively with the number of parameters. We introduce a rapid wavefront shaping system, replacing coordinate descent with gradient descent optimization. To this end, our system acquires a gradient vector, which allows simultaneous update of all modulation parameters. We start with a non-invasive, guide-star-free score function to assess modulation quality and analytically derive its gradient with respect to all modulation parameters. Although the gradient depends on unknown tissue structure, we show it can be inferred from optical measurements. This enables fast, high-resolution wavefront correction with complexity independent of parameter count. We demonstrate the system’s effectiveness in correcting aberrations in a coherent confocal microscope.}


\maketitle

\input{intro}

\input{results}
\input{discussion}
\input{methods}

\boldstart{Data availability}\\
The data that support the findings of this study is available in \cite{Monin2025Code}.

\boldstart{Code availability}
The code used for acquiring and processing the data is available at \cite{Monin2025Code}.

\bibliography{biblio_anat, proposal,biblioscattering}

\boldstart{Author contribution}\\
S.M. designed and constructed the system. S.M. and M.A. performed the experiments. S.M. analyzed the data. A.L. conceived and supervised the project. All authors contributed to writing of the manuscript.\\

\boldstart{Competing interests}\\
The authors declare no competing interests.\\

\boldstart{Supplementary information}\\
Supplementary material is attached to the submission.

\end{document}


\maketitle

\input{supp_setup}
\input{score_sup}

\input{pd_interferometry}

\input{sup_res}
\newpage

\bibliography{biblio_anat, proposal,biblioscattering}

%% file: intro.tex
\section{Introduction}

Optical imaging of tissue is challenging because cells and other tissue components scatter light. As a result, as light propagates deeper inside tissue, it becomes heavily aberrated, making it challenging to resolve clear images of deeper structures. Over the decades, several techniques, such as confocal microscopy~\cite{ConfocalMicroscopyOverView2020} and optical coherence tomography (OCT)~\cite{OCTOverview2016}, have been developed to achieve deeper imaging by filtering out scattered photons and isolating ballistic photons. Despite this progress, these techniques are inherently limited to thin layers as the number of ballistic photons decays rapidly within scattering materials. To overcome this challenge, wavefront shaping was introduced. Instead of filtering out the scattered photons, wavefront shaping aims to measure the scattered light and invert the scattering process. Such corrections follow two major approaches: digital and optical.

Digital aberration correction methods attempt to illuminate the tissue using a set of wavefronts and measure the scattered light. This data is fitted with a parametric model to estimate the aberration and reconstruct the hidden target~\cite{PMID:17694065,Choi2011,Choi2015,Badoneaay7170,Kwon2023,ChoiLyers2023,zhang2024deepimaginginsidescattering,Choi2023AngWavelength,Metzler23NeuWS,YeminyKatz2021,haim2023imageguidedcomputationalholographicwavefront,Balondrade_2024,Kang2017,Najar2024, Zhu22, Baek_2023,Jeong2018,Gil2024,Yonghyeon2022}. This approach has led to impressive results, but it is ultimately limited by signal-to-noise ratio (SNR). Even with advanced gating strategies, light scattered by deeper tissue components is weak and can be lost in measurement noise.

An alternative class of techniques uses optical wavefront-shaping~\cite{Ji2017review,HampsonBooth21review,Horstmeyer15,YU2015632,Gigan22,Cui2025}. Such techniques use spatial light modulators (SLMs) to control the incoming and/or outgoing paths of the optical system. These SLMs reshape the incoming and/or outgoing wavefront in a way that is the inverse of the aberration it undergoes inside the tissue. This allows an incoming wavefront to focus to a diffraction-limited spot, and the light emitted from a single point within the tissue can be focused onto a single detector element. Unlike digital aberration corrections, the significant advantage of optical wavefront shaping is that the correction is physical; hence, all light photons emerging from a single target point can be brought into a single sensor point and measured with a significantly higher SNR.

Wavefront shaping ideas have found applications in a wide range of imaging modalities, including sound~\cite{Horstmeyer15,doi:10.1121/1.412285} and light, coherent imaging and OCT~\cite{Jang:13}, and incoherent fluorescence imaging using single-photon~\cite{Boniface:19,Dror22,DrorNatureComm24,Stern:19} and multi-photon~\cite{Katz:14,Ji2017review} excitation. This work showcases a simple aberration correction in a coherent, reflection mode confocal microscope. However, our framework also applies to other popular imaging schemes such as OCT and fluorescent microscopy. 

The practical application of wavefront shaping is hindered by
the complexity of determining the desired modulation correction. This
wavefront correction differs between different tissue samples, and vary even within different regions of the same tissue sample. 
Earlier proof-of-concept demonstrations employed a validation camera behind the tissue to provide feedback for the algorithm~\cite{Vellekoop:07,Conkey:12,PopoffPhysRevLett2010,Vellekoop2010,Yaqoob2008,chen20203PointTM}. Other approaches utilize a guide star~\cite{Horstmeyer15,Tang2012,Katz:14,Wang20142PAdaptive,Liu2018,Fiolka:12,Jang:13,Xu11,Wang2012,Kong:11,Vellekoop2012}, where scattering originates from a strong single point source within the tissue, allowing a wavefront sensor~\cite{Vellekoop2012,Liu2018} to directly measure the scattered wavefront.

In the absence of such a guide-star, determining a wavefront shaping correction is a significantly more challenging task. Most approaches define a score function to evaluate the quality of the modulation based on the captured signal, and then optimize the SLM parameters to maximize this score. However, the optimal modulation depends on the unknown tissue structure. This structure can only be probed by projecting multiple modulation patterns and imaging their scattered output.
Thus, most wavefront shaping optimization schemes employ coordinate-descent (CD) approaches, where a single  parameter of the modulation is varied at a time. Each optimization step involves selecting a new value for one parameter to improve the modulation score.
These algorithms iterate sequentially through all degrees of freedom~\cite{Vellekoop:07,Popoff2011,Conkey:12,Katz:14,chen20203PointTM,Boniface:19,DrorNatureComm24}. The main drawback of coordinate descent schemes is their time complexity, which scales with the number of free parameters in the modulation. However, for thick tissue, the modulation should ideally use all pixels on the SLM, usually in the megapixel range. 

In this research, we derive a fast approach for estimating a wavefront-shaping correction in a non-invasive, guide-star free setting. In contrast to sequential coordinate descent approaches, our algorithm is capable of simultaneously updating all SLM parameters. 
To this end, we start by analytically differentiating the wavefront-shaping score. Although the gradient depends on the unknown tissue structure, we show that we can use optical computing to measure it. By simply capturing the back scattered field, the gradient can be evaluated in closed-form. The dimensionality of the gradient vector is equivalent to the number of SLM parameters. With this gradient at hand,    we can simultaneously update all SLM parameters, and we can  transition from slow coordinate descent to fast gradient descent optimization.

Our gradient descent scheme is significantly faster than existing coordinate descent schemes and can also recover better modulations. 
Coordinate descent methods typically restrict the number of SLM modes they optimize due to computational complexity and because the measurement of high-frequency modes is noise sensitive. 
However, in thick tissue where the scattering is wide, the optimal modulation requires a large number of degrees of freedom. In contrast,  our approach can optimize a large number of modes without an increase in computational complexity. We show that this higher number of modes allows us to find significantly better modulations.

\input{fig_system_image}

%% file: fig_system_image.tex
\begin{figure}[t!]     
	\begin{center}
				{\includegraphics[width=1\textwidth]{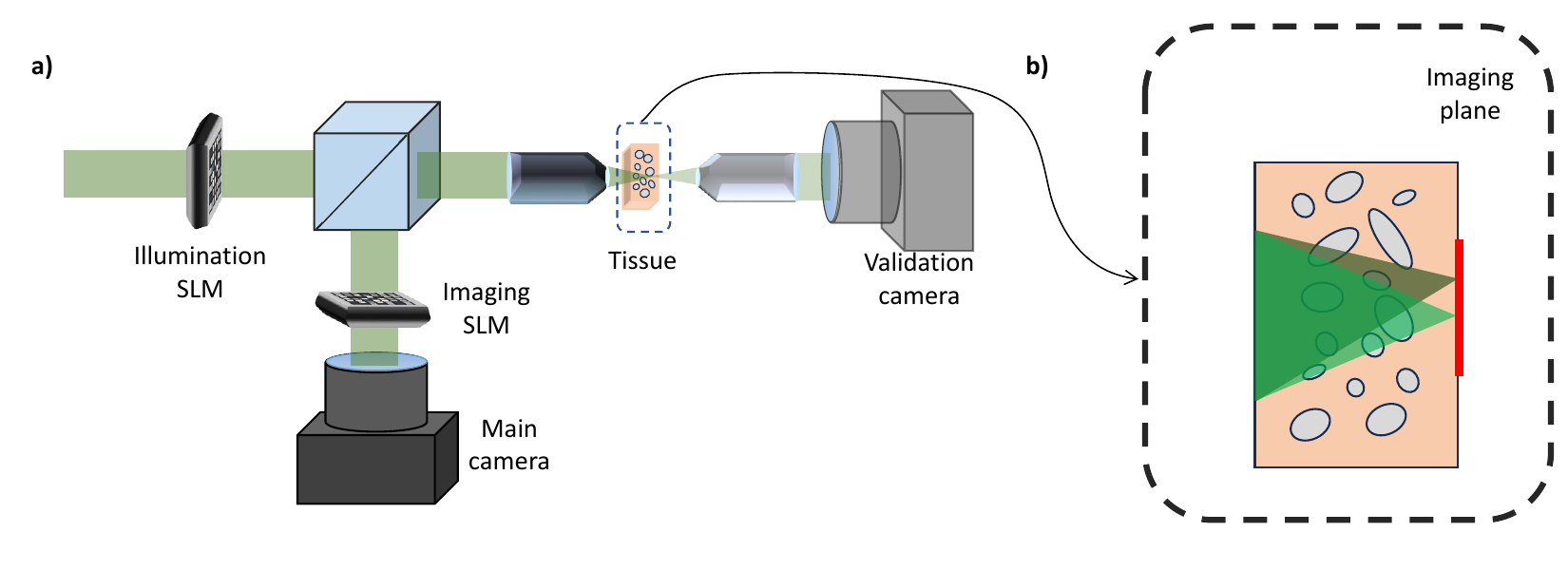}}
	\end{center}
	\caption{\textbf{System Schematic:}(a) Schematic diagram of our system consisting of two SLMs. The first SLM modulates the illuminating laser light, while the second SLM modulates the reflected light. A validation camera provides reference images and confirms focusing on the desired target. (b) Our algorithm optimizes SLM phase to focus light on multiple adjacent points, by applying simple tilt-shift to the same modulation.}
	\label{fig:setup}
\end{figure}

%% file: results.tex
\section{Results}\label{sec:results}

\subsection*{\boldstart{Imaging setup}}
 In \figref{fig:setup}, we visualize a wavefront-shaping imaging setup.  A laser beam illuminates a tissue sample via  a microscope objective. A phase SLM in the illumination arm modulates the illumination pattern.

Coherent light back-scattered from the tissue target is collected by the same objective lens and reflected by a beam-splitter. A second phase SLM in the imaging arm modulates the returning wavefront. Lastly, the modulated light is measured by the front main camera. 

Our setup incorporates a validation camera positioned behind the tissue sample. This camera serves to evaluate focusing quality and capture an undistorted reference image of the target. While earlier research demonstrations of wavefront-shaping utilized this camera to provide feedback to the algorithm, here we  develop a non-invasive technique relying solely on feedback by the main (front) camera. It is crucial to note that the validation camera serves only for reference and does not provide any input to our algorithm.

To maximize the correctable area with a single modulation, the SLM should ideally be placed conjugate to the aberration~\cite{Mertz:15}, namely to a plane in the middle of the tissue. For simplicity , in our experimental setup, the SLMs are placed in the Fourier plane. To correct aberrations across a wide field of view with a single modulation pattern, we apply  a tilt and shift operation to the Fourier pattern~\cite{SeeThroughSubmission,osnabrugge2017generalized}.Critical to our derivation is the assumption that  the two SLMs   are placed such that they are conjugate to each other.

\subsection*{\boldstart{Image formation model}}
We start by presenting a generalized image formation model for a dual SLM system, comprising an illumination SLM and an imaging SLM. Based on this model, we derive the specific model of our imaging system.

We denote the reflection matrix of the tissue as $\RM$. This matrix describes propagation of light from the illumination SLM plane through the tissue and optics, and back to the imaging SLM. Thus, for every wavefront $\binu$ placed on the illumination SLM, the resulting wavefront reaching the imaging SLM is modeled as a linear transformation by the reflection matrix:
\BE
\RM\binu.
\EE

Applying a modulation $\boutu$ to the imaging SLM, the wavefront reaching the camera is:
\BE
\bP \left( \boutu \hadprod (\RM \binu)\right),
\EE

where $\bP$ denotes a linear mapping between the imaging SLM and the camera sensor that can be calibrated and described by a matrix, as it is independent of the tissue sample. $\hadprod$ represents element-wise multiplication between two vectors.

Our goal is to use the SLM modulation to image a small area inside the volume rather than a single point. For that, we denote the modulation parameters (the phase values of the SLM) as $\PrmInVect,\PrmOutVect$. To direct this modulation to adjacent points within the scattering tissue, we apply tilt and shift to the SLM modulation~\cite{SeeThroughSubmission,osnabrugge2017generalized}. We represent the tilt-shifted modulation directed towards volume point $\ell$ as $\bu^\ell(\PrmInVect)$. This notation implies that $\PrmInVect$ are the parameters we can adjust, from which we derive multiple modulations $\bu^\ell(\PrmInVect)$ by applying fixed tilt-shift operations toward different points $\ell$.  Similarly, $\bu^\ell(\PrmOutVect)$ denotes a tilt-shifted modulation on the imaging SLM. With this tilt-shift, light emerging from different points $\ell$ in the volume arrives at the same position on the sensor, and is expressed as:
\BE
\bP \left( \bu^\ell(\PrmOutVect) \hadprod (\RM \bu^\ell(\PrmInVect))\right).
\EE
The intensity at pixel $x$ of the camera is:
\BE \label{eq:gen_sys_int}
I(x) = |(\bzeta_x\hadprod \bu^\ell(\PrmOutVect) )^T \RM \bu^\ell(\PrmInVect)|^2,
\EE
where $ \bzeta_x$ is a row of the matrix $\bP$ mapping the output of the SLM to a pixel $x$ on the camera.

In our system, $\RM$ models the coherent propagation of a monochromatic laser beam. However, this formulation can be generalized to describe other forms of propagation, such as a path-length resolved reflection matrix describing the measurements of an OCT system. Also,  $\bP$ can represent various mappings between the imaging SLM plane to the camera sensor plane, depending on an optical setup of choice. In our system, we place the SLM in the Fourier plane; hence, $\bP$ is a Fourier transform.

\subsection*{\boldstart{Score function}}
We aim to develop an optimization framework that identifies the optimal wavefront modulation for focusing light inside the scattering tissue. To achieve this, we start by defining a score $\Mtric(\PrmVect)$ which quantifies a good modulation. We will then attempt to  maximize this score to obtain a good modulation:
\BE
\hat{\PrmVect} =  \arg\max_{\PrmVect} \Mtric(\PrmVect).
\EE

To identify a good modulation in a non-invasive, guide-star free setting, we
build upon the work of~\cite{DrorNatureComm24} and recent digital correction approaches~\cite{Balondrade_2024,zhang2024deepimaginginsidescattering,Choi2023AngWavelength}, and measure the confocal energy of the corrected area. For that, we apply the same modulation $\PrmInVect=\PrmOutVect$ on both illumination and imaging SLMs, and denote it by a single parameter vector $\PrmVect$. We direct and collect light from points $\ell$ and measure the intensity at the central sensor pixel. In our system, the mapping between the SLM and the camera is a Fourier transform. Consequently, the central pixel corresponds to the DC component of the Fourier transform (i.e., $\bzeta_x $ is a uniform vector ). Following \equref{eq:gen_sys_int}, we can express the measured confocal energy as:

\BE
I(0) = |\bu^\ell(\PrmVect)^T\RM \bu^\ell(\PrmVect)|^2.
\EE

We sum the confocal intensity over an area and measure the total intensity we can collect by a confocal scanning of the modulation tilted toward different points $\ell$ in an area of interest $\Area$:
\BE\label{eq:score-confocal-area}
\Mtric(\PrmVect)\equiv \sum_{\ell\in \Area} |\bu^\ell(\PrmVect)^T\RM \bu^\ell(\PrmVect)|^2.
\EE

\input{fig_tilt_shift}

To better illustrate this score, \figref{fig:tilt_shift} visualizes the incident and reflected wavefronts, both with and without modulation, for two points within the target area. Initially, without modulation, these wavefronts exhibit minimal correlation. However, once  we find a focusing modulation, the incoming and outgoing wavefronts demonstrate higher correlation. Additionally, we observe a strong memory effect correlation among wavefronts returning from different points.

We note that in~\cite{DrorNatureComm24} incoherent fluorescence signal was collected, and the confocal energy was measured at a single pixel, without attempting to average over an area. Since we use coherent light, we notice that if we only direct light to a single spot, a high confocal score at the sensor plane does not guarantee that the light has focused into a single spot inside the tissue due to various interference effects. To address this, we propose averaging the confocal score over a small area \blue{ $\Area$}, which mitigates these interference effects. This approach aligns with recent coherent digital correction algorithms~\cite{Balondrade_2024,zhang2024deepimaginginsidescattering,Choi2023AngWavelength} that optimize over finite isoplanatic patches to maximize the diagonal elements of the reflection matrix which is equivalent to the confocal area score. \blue{The supplementary file provides a short derivation justifying the score using local memory effect correlations. Fig. 3 in the supplementary material shows a simulation comparison of an input/output modulated vs unmodulated wavefronts. } We investigate the impact of the target area size on focusing performance in our experimental section, and show that in practice one modulation can only be used over a very small area due to the limited extent of memory effect correlations.

\subsection*{\boldstart{Optical gradient acquisition}} 
The main challenge in utilizing the score defined in \equref{eq:score-confocal-area} lies in its dependence on the tissue's reflection matrix, which is inherently linked to the unknown tissue structure. Since the tissue is treated as a black box, most previous approaches have employed coordinate descent optimization strategies~\cite{Vellekoop:07,Popoff2011,Conkey:12,Katz:14,Boniface:19,DrorNatureComm24}. 
In each iteration, only one parameter is varied (often in a Hadamard basis) and adjusted to improve the measured score. For invasive setups, the score is simply the intensity measured by a validation camera at a point behind the tissue. Given that modern SLMs are in mega-pixel resolution, sequentially scanning each free parameter is an extremely time-consuming process. Consequently, researchers frequently reduce the SLM resolution and correct a significantly smaller number of modes.
Below, we derive a scheme for simultaneously measuring the gradient with respect to all SLM parameters. This approach enables us to transition from coordinate descent schemes to a gradient descent scheme that updates all SLM parameters in each iteration. 

We start by applying the chain rule and differentiate \equref{eq:score-confocal-area} with respect to the SLM wavefront   $\PrmVect$:
\BE\label{eq:deriv-vect}
\frac{\partial \Mtric(\PrmVect)}{\partial \PrmVect}=2 \sum_\ell  \underbrace{ \left(\bu^\ell(\PrmVect)^T\RM \bu^\ell(\PrmVect) \right)^*}_{(1)}\cdot\underbrace{\left(\RM \bu^\ell(\PrmVect) + (\bu^\ell(\PrmVect)^T \RM)^T \right)}_{(2)}\hadprod \underbrace{\frac{\partial \bu^\ell(\PrmVect) }{\partial \PrmVect}}_{(3)}.
\EE

This gradient is a product of three terms. The first (1) is a complex scalar. The second (2) is a complex vector whose dimensionality is equivalent to that of the modulation. The third term (3), which differentiates the wavefront with respect to the parameters $\PrmVect$, can be computed explicitly. As we explain in the supplementary, since $\bu^\ell(\PrmVect)$ simply shifts and multiplies the elements of $\PrmVect$ by complex tilt scalars, the derivative with respect to $\PrmVect$ is also a simple tilt-shift of the derivative with respect to $\bu^\ell(\PrmVect)$.

The challenge with the first two terms is that they involve the unknown reflection matrix $\RM$. Nevertheless, we can leverage the optics to measure the gradient in \equref{eq:deriv-vect} for any candidate modulation $\PrmVect$ of interest. 
We observe that the complex field $\RM \bu^\ell(\PrmVect)$ is essentially equivalent to applying a modulation at the illumination arm only, and capturing a non-modulated speckle field.
The complex field can be captured using a phase retrieval algorithm or with an interferometric imaging system, as detailed in the supplementary material. With this measurement, we can numerically compute the remaining terms of \equref{eq:deriv-vect}. 
For example, following the reciprocity principle, $\RM$ is a symmetric matrix and $(\bu^\ell(\PrmVect)^T \RM)^T$ is equal to $\RM\bu^\ell(\PrmVect)$.  Similarly, the term (1) in \equref{eq:deriv-vect} can be computed by multiplying $\RM \bu^\ell(\PrmVect)$ with the known vector $\bu^\ell(\PrmVect)$.

By measuring the gradient of the modulation score,  we can simultaneously update all entries of the SLM. This offers a substantial acceleration compared to previous approaches, which iterate through the entries and update only one degree of freedom at a time.

In our current implementation, we measured the fields $\RM \bu^\ell(\PrmVect)$ using a phase diversity scheme. This involved applying five defocus wavefronts to the imaging SLM~\cite{Dean2003,Robert1982}, capturing five speckle intensity images, and solving an optimization problem for recovering the phase~\cite{candes2015}. We measure each of the $\ell$ points and sum them. In the supplementary material, we show that the gradient can in fact be imaged directly using a point-diffraction interferometry scheme~\cite{Smartt_1975,Akondi2014PDI}, bypassing the optimization. Moreover, the summation across all points $\ell$ can be computed  within a single exposure; hence the entire gradient can be imaged with as little as three shots.

\blue{We note that the gradient formula in \equref{eq:deriv-vect} assumes $\PrmVect$ is a complex vector. In practice, most  SLMs only  modify phase. In supplementary sec. 3.1, we provide a small adaptation of \equref{eq:deriv-vect}, which  provides the derivative with respect to  phase.}

In  the supplementary file we also explain that the gradient descent approach derived above resembles fast wavefront shaping algorithms based on time reversal and the power algorithm~\cite{Dror22,doi:10.1121/1.424648,doi:10.1121/1.412285,Yang2014,Meng2012,Papadopoulos16}. In fact, it is equivalent to a power iteration of the reflection matrix if one aims to maximize confocal intensity at a single spot rather than an area.  However, as we show below, a single spot score does not result in good modulations. 
While power iterations are tied to an eigenvector calculation and may not be adjusted easily to optimize other things,  defining a general  score function and its gradient provides a more principled way to impose all sorts of desired properties on the solution.  

The optical calculation of the gradient here also resembles the derivation in~\cite{Mididoddi2025Threading}. However, their system uses transmission mode imaging and thus requires two SLMs on two sides of the scattering medium. 


\subsection*{\boldstart{Experimental results}}
In our experiments, we used \blue{three} types of targets. The first consisted of  high-reflectance chrome-coated masks (Nanofilm), patterned using an in-house lithography process to create structures with a resolution of $2\um$. We placed these behind scattering layers composed of chicken breast tissue with thickness of $130\um-240\um$ or with a number of layers of parafilm~\cite{Boniface2020}. The second target included polystyrene beads dispersed in agarose gel. \blue{ Finally, we imaged inside an onion layer.}
\input{fig_converge}
\input{fig_cmp_size}

We start by showing convergence of our algorithm. In \figref{fig:Converge}, we cover the chrome mask with $180\um$ thick chicken breast, and demonstrate the algorithm's efficiency, converging to the desired spot in as few as ten iterations, compared to the thousands required by previous coordinate descent algorithms. We show an image of one of the points $\ell$ in the optimized area $\Area$, with no modulation and when the optimized modulation is placed on both SLMs. Without modulation, we see a noisy speckle image but with the modulation, we see a sharp spot whose intensity is about $20\times$ higher. We also image the chrome mask directly using the validation camera. Without modulation, we see a wide speckle pattern, but with  the optimized modulation on the illumination arm all light propagates through the scattering layer and is focused into a single point on the chrome mask. These validation images ensure that the point we see in the main camera indeed corresponds to a point on the mask, and we are not achieving a point on the main camera by complicated interference effects inside the tissue. At the lower row, we show the progress of the phase mask during the iterations.

Next, we evaluate the effect of the area $\Area$ over which the score is optimized in \figref{fig:cmp_size}. If this area is small, we observe a sharp spot at the main camera, but the light does not actually focus into a spot inside the tissue, as we do not observe a spot at the validation camera. This indicates that the spot at the main camera results from interference of coherent light within the tissue. This problem does not occur with fluorescent wavefront shaping~\cite{DrorNatureComm24} since the incoherent emissions from nearby fluorescent points do not interfere. As we increase the size of the scanned area, requiring the same modulation to explain multiple points, such interference effects disappear, and we achieve a point in both cameras. However, as we increase the target area the intensity of the focused spot at the main camera decays because the correction required by nearby points vary spatially and one modulation cannot explain a wide area. This is because memory effect correlation does not hold for very large ranges.  
Note that our score function recovers modulations using a non-invasive, guide-star free feedback.

\input{fig_res_mask}
Next, we show that our system can retrieve an image of an area by performing a confocal scan. In \figref{fig:res-mask}, we present imaging results for a few patterns printed on the chrome mask. We display images of a single point in the area through both the main and validation cameras. We also show a confocal scan of the entire area. Without modulation, this looks like a speckle pattern; however, with the modulation, the confocal scan reveals the shape of patterns printed on the chrome mask. Note how the modulation increases light by a factor of $12-100\times$. \blue{In supplementary Fig. 7, we show the individual images captured for many points $\ell\in \Area$ in both the main and validation cameras. The result in the last row of \figref{fig:res-mask} was larger than the range of memory effect correlation. To overcome this, we imaged multiple local corrections and stitched the patches together. In supplementary Fig. 8, we show the different isoplanatic patches.}

\input{fig_beads_3D_1}
\blue{Next, we demonstrate that our algorithm can also image translucent targets with small variations in refractive indices. To do this, we disperse polystyrene beads in agarose gel. Polystyrene and agarose have refractive indices of approximately $1.598$ and $1.334$ under our $532nm$ illumination, and thus correspond to the practical range of refractive indices of tissue components. The bead diameter is $0.5\um$ and the slab thickness is $1.3mm$. The left part of \figref{fig:beads3d} illustrates the schematic of this target, where we used our algorithm to focus inside the volume and image a 3D sub-volume. We estimate that the optical depth (OD) of this sample as 2.5 by measuring the attenuation of ballistic light in the validation camera.
In \figref{fig:beads3d}, we visualize the confocal scanning result. Our target area $\Area$ was now a 3D sub-volume rather than a 2D patch, where beyond tilt-shift, we  added in $\bu^\ell(\PrmInVect)$  a quadratic phase to focus light at varying depths. In our results, we present a few $x-y$ cross-sections of  planes in the corrected sub-volume. In addition, we display maximum projection onto the three different axes, showing good agreement between our aberration corrected results and the reference. While a standard confocal scan of such beads is very noisy, with our estimated modulation we can largely increase its contrast and achieve a clear image of spots corresponding to the beads position. To obtain a reference image of the bead positions, we image beads at the further depth of the slab, closer to the validation camera, so we are able to see an aberration-free reference in the validation camera. For this reference image, we used a wide-field incoherent illumination. In supplementary Figs. 5-6 we show additional results of 2D slices inside a beads dispersion, where we use beads of various sizes.}

\input{fig_onion}

\input{fig_onion_3D_1}
\blue{Utilizing our algorithm, in \figref{fig:onion} we successfully imaged onion cells through a $130\um$ thick slice of highly scattering onion~\cite{WANG2013494}. Specifically, we visualized the intercellular boundary between two neighbouring cells. In the supplementary Fig. 9 we show a larger field of view of the onion under incoherent illumination. In addition, we show a confocal scan of an onion cell at a shallower layer in supplementary Fig. 11, where scattering does not limit image quality. The aberration-corrected result in \figref{fig:onion} matches this unaberrated structure.}

\blue{The area of the result achieved in \figref{fig:onion}  was bigger than the range of the memory effect correlation and cannot be considered as a single isoplanatic patch. Hence, we imaged this target by applying multiple local corrections and stitching a few patches together. In Fig. 10 of the supplementary material we show the different isoplanatic patches. }
\bblue{In \figref{fig:onion3d} we visualize the 3D structure of the onion, by displaying corrections at three different depth slices. We show additional slices at supplementary Fig. 12. }

In the following paragraphs, we use again the chrome mask target to evaluate various components of our algorithm.

In contrast to coordinate descent optimization whose time complexity scales with the number of modes corrected, the time complexity of our gradient descent approach  does not depend on the number of corrected modes. In practice, to reduce scan time, coordinate descent schemes optimize for a relatively small number of modes. However, for thick tissue where the scattering is wide, we require a large number of modes to obtain a good correction. In \figref{fig:cmp_resultion}, we demonstrate the impact of  resolution on the quality of the correction. As shown, a modulation with more degrees of freedom allows us to focus more energy into a single spot.

Next, we explicitly compare our gradient descent scheme with a coordinate descent approach. Following~\cite{DrorNatureComm24,PopoffPhysRevLett2010}, we express the modulation as $\PrmVect=\sum \phi_k H_k$, where $H_k$ are elements in a Hadamard basis. We scan the basis elements, adjusting the phase of one basis element at a time. Our algorithm clearly converges in a significantly fewer iterations, as all parameters can be updated simultaneously. However, we argue that the resolution of coordinate descent schemes is not only limited by acquisition time but also by SNR. For high frequency modes, varying $\phi_k$ results in a very small change to the score, potentially lower than the noise level. In \figref{fig:hadamarad}, we show that the coordinate descent score cannot improve beyond a certain point. To illustrate that noise is the limiting factor, we run another coordinate descent scheme, but this time we averaged multiple shots for each basis element to reduce noise. This oversampling approach allowed us to update the high frequency modes, leading to much better results. However, the oversampling is also $5\times$ slower than the standard coordinate descent scheme and $125\times$ slower than our gradient descent scheme. The modulation computed in \figref{fig:hadamarad} is a low-resolution one, consisting of only $256$ parameters. For higher resolution modulations, the time difference between coordinate descent and gradient descent approaches would be even larger.

\input{fig_cmp_resolution}
\input{fig_hadamard}
\bblue{Finally, in Supplementary Fig. 13 we further compare our method with the CLASS algorithm \cite{Choi2015}, a representative digital correction technique \cite{Kwon2023,ChoiLyers2023,haim2023imageguidedcomputationalholographicwavefront,Balondrade_2024}.
Digital approaches such as these measure the scattered wavefront and numerically fit it with a parametric model that describes the underlying aberration and the hidden target.
Because the measured scattered fields are typically very noisy, the reconstruction quality is ultimately limited by the signal-to-noise ratio (SNR) of the recorded data.
In contrast, performing the correction optically—by applying the estimated modulation directly on the SLM—enhances the effective SNR of subsequent measurements, leading to cleaner reconstructions and improved correction fidelity.}

%% file: fig_tilt_shift.tex
\begin{figure}[t!]     
	\begin{center}
				{\includegraphics[width=1\textwidth]{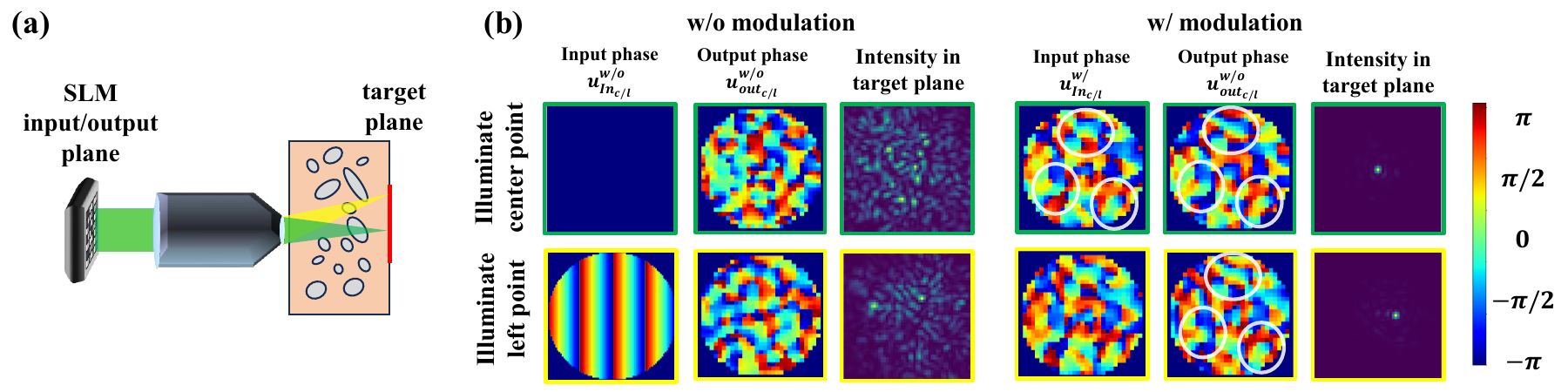}}
	\end{center}
	\caption{\textbf{Tilt-shift and time reversal effects:} (a) We show a simplified schematic of our system with two input wavefronts,  green focusing into the center of the imaging plane and yellow focusing to a neighboring point to the left. (b) Simulation results showing the incoming phase on the SLM, the conjugate of the output phase after tissue reflection (when reaching the SLM plane), and the intensity at the target plane. With light modulation, strong correlations are observed between the incoming SLM phase and the output phase, as well as between output phases for different points (white circles indicate areas of strong correlation). Without aberration correction, correlation decreases rapidly.}
	\label{fig:tilt_shift}
\end{figure}

%% file: fig_converge.tex
\begin{figure*}[t!]
	\begin{center}\begin{tabular}{@{}c@{~}c@{~}c@{~}c@{~}c@{~}c@{~}c@{~}}
			\multicolumn{1}{c}{}&
			\multicolumn{1}{c}{\hspace{-0.6cm} \scriptsize Init. }&
			\multicolumn{1}{c}{\hspace{-0.6cm} \scriptsize Iter 1 }&
			\multicolumn{1}{c}{\hspace{-0.6cm} \scriptsize Iter 5 }&
			\multicolumn{1}{c}{\hspace{-0.6cm} \scriptsize Iter 7 }&
			\multicolumn{1}{c}{\hspace{-0.6cm} \scriptsize Iter 10 }&
			\multicolumn{1}{c}{\scriptsize Cost function}\\
			
			{\raisebox{0.61cm}	{\rotatebox[origin=c]{90}{~ {\scriptsize  Main cam.} }}}&
			\includegraphics[width= 0.14\textwidth]{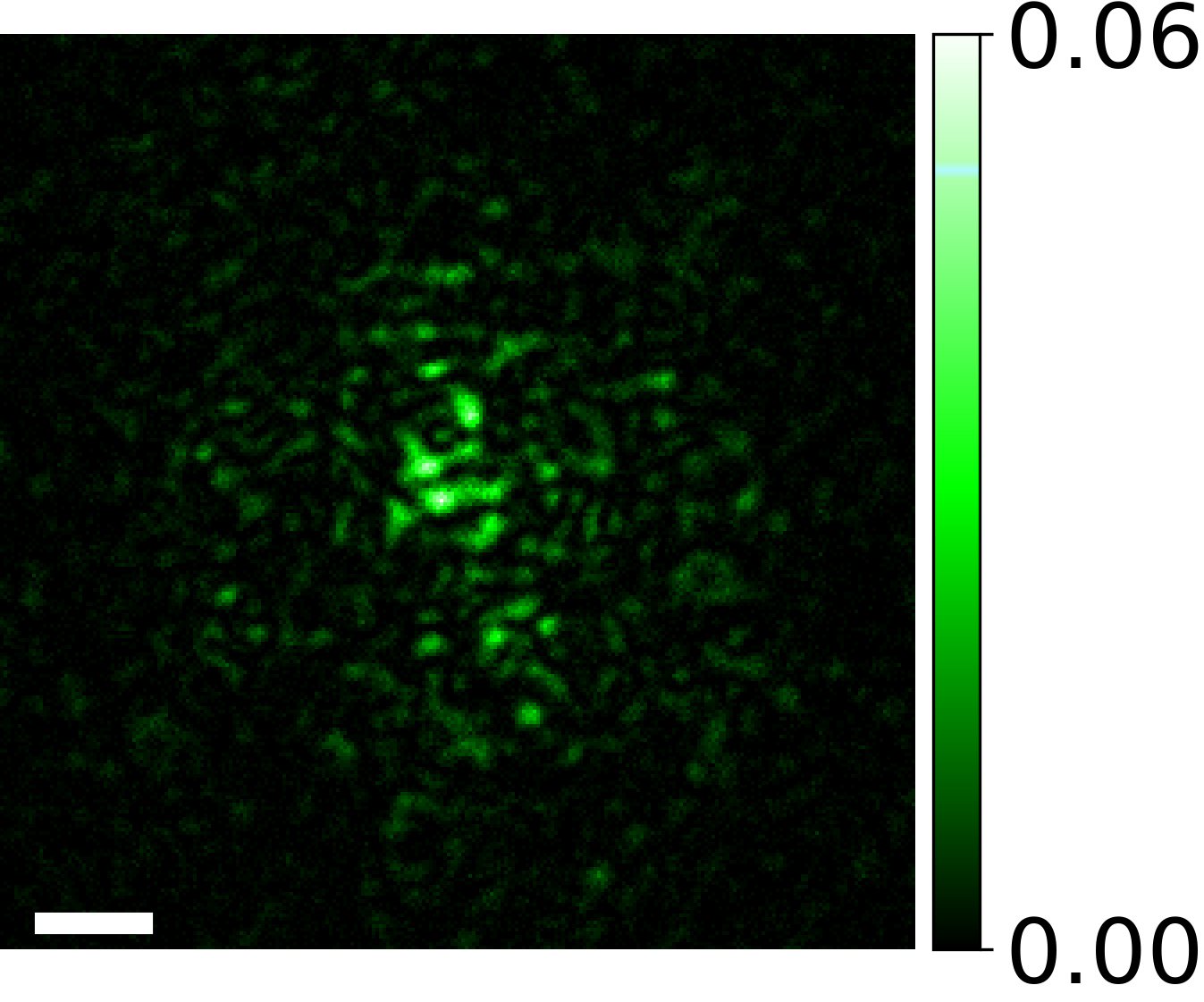}&
			\includegraphics[width= 0.14\textwidth]{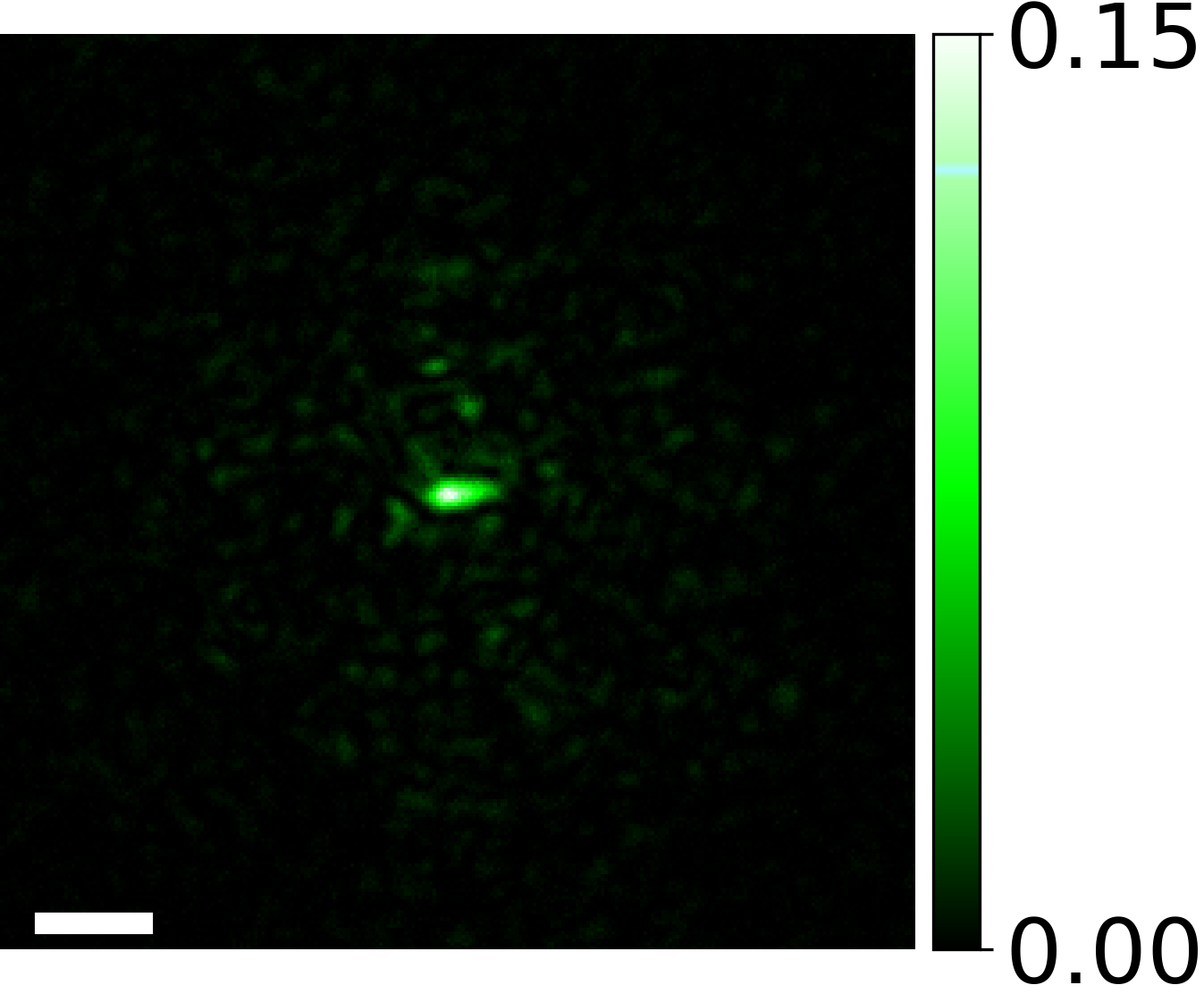}&
			\includegraphics[width= 0.14\textwidth]{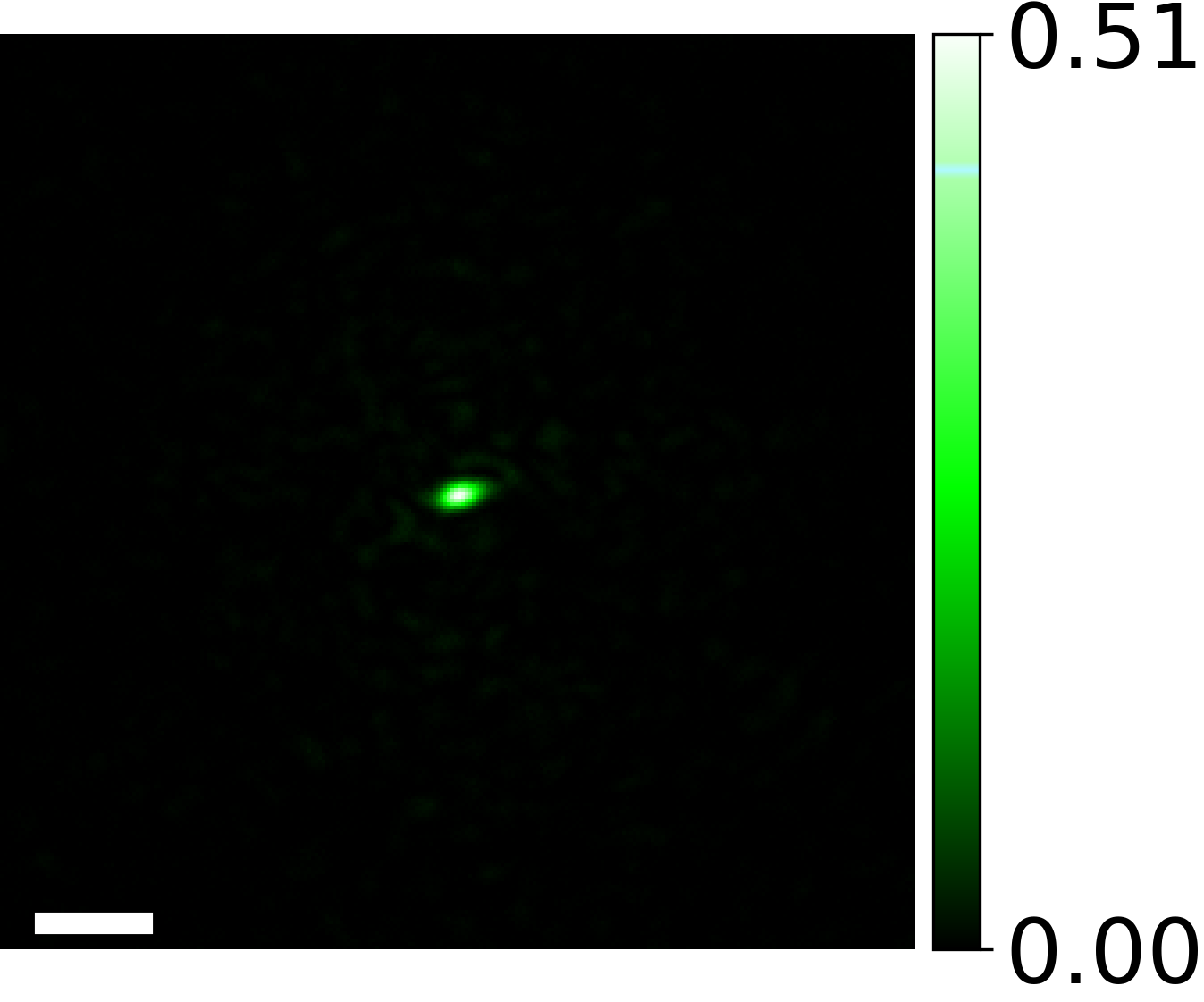}&
			\includegraphics[width= 0.14\textwidth]{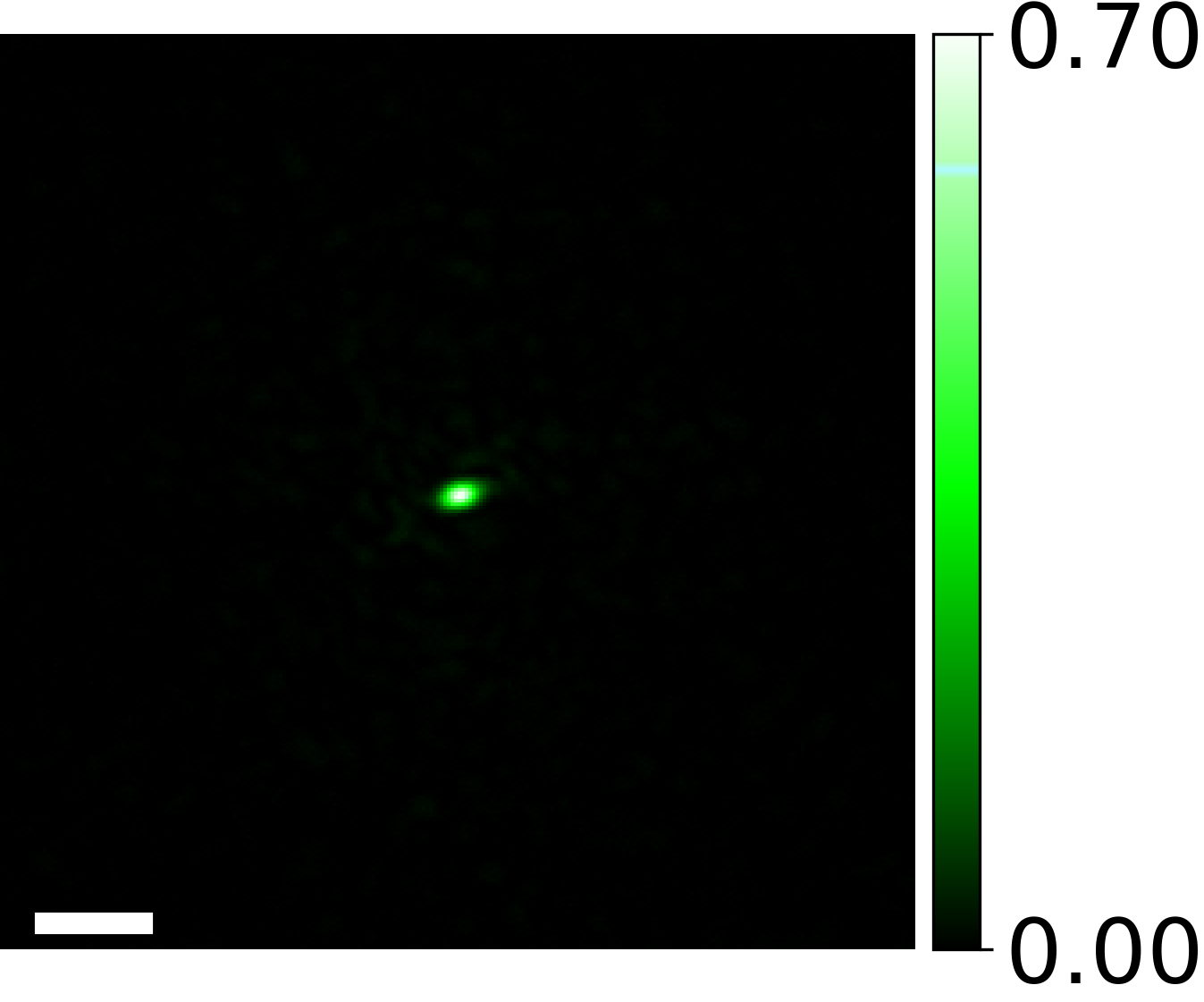}&
			\includegraphics[width= 0.14\textwidth]{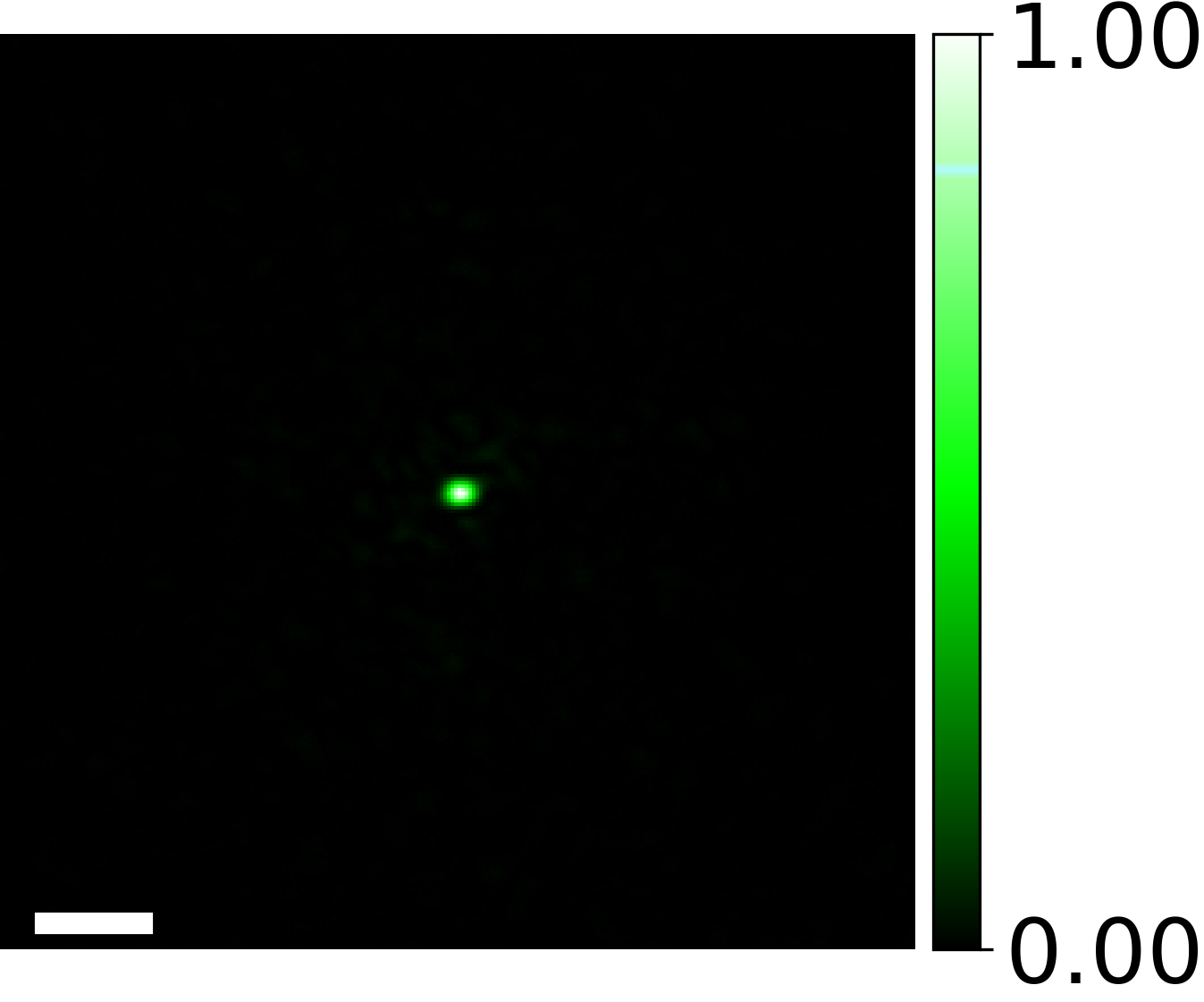}&
			\raisebox{-0.1cm}{\includegraphics[ width= 0.19\textwidth]{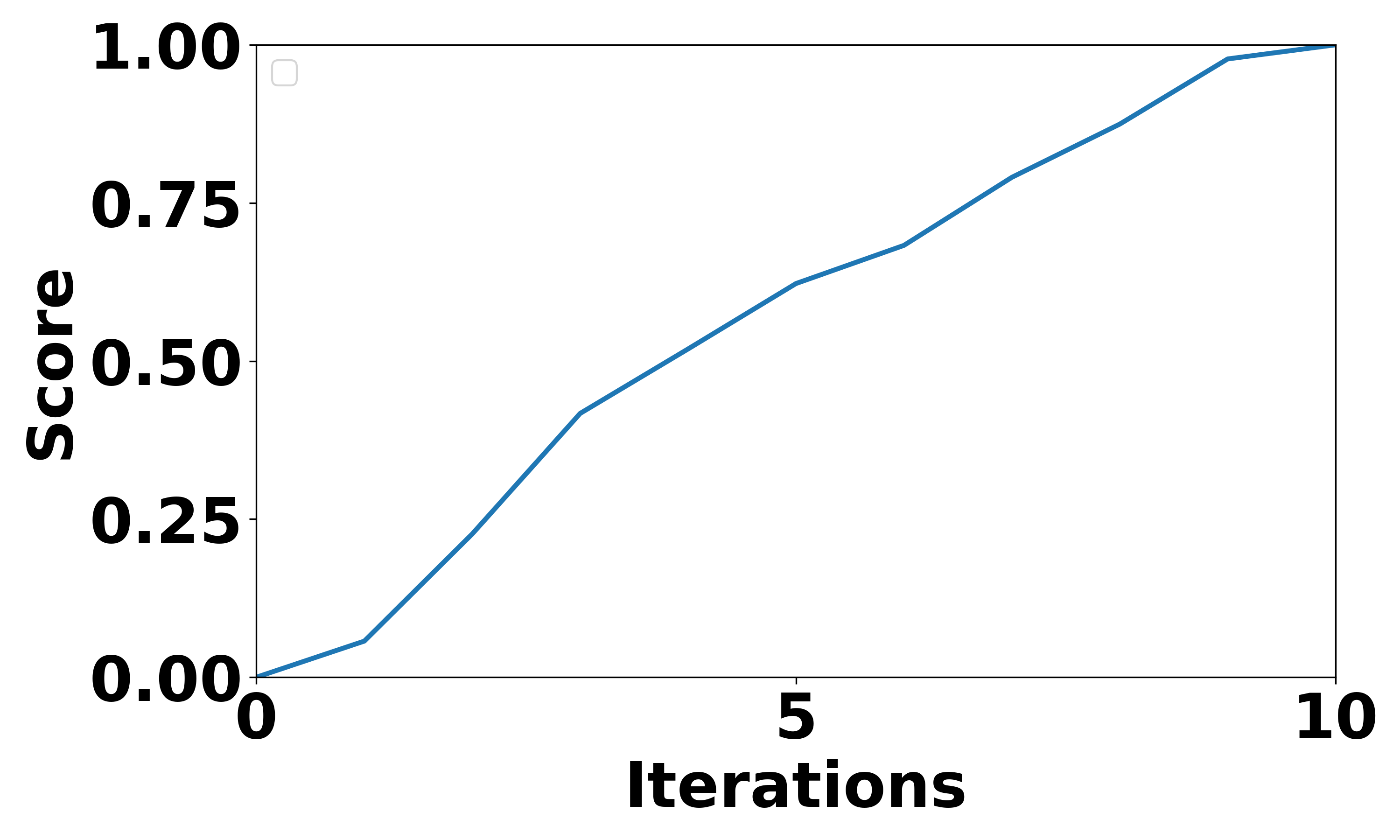}}\\

			{\raisebox{0.61cm}	{\rotatebox[origin=c]{90}{~ {\scriptsize  Valid. cam.} }}}&
			\includegraphics[width= 0.14\textwidth]{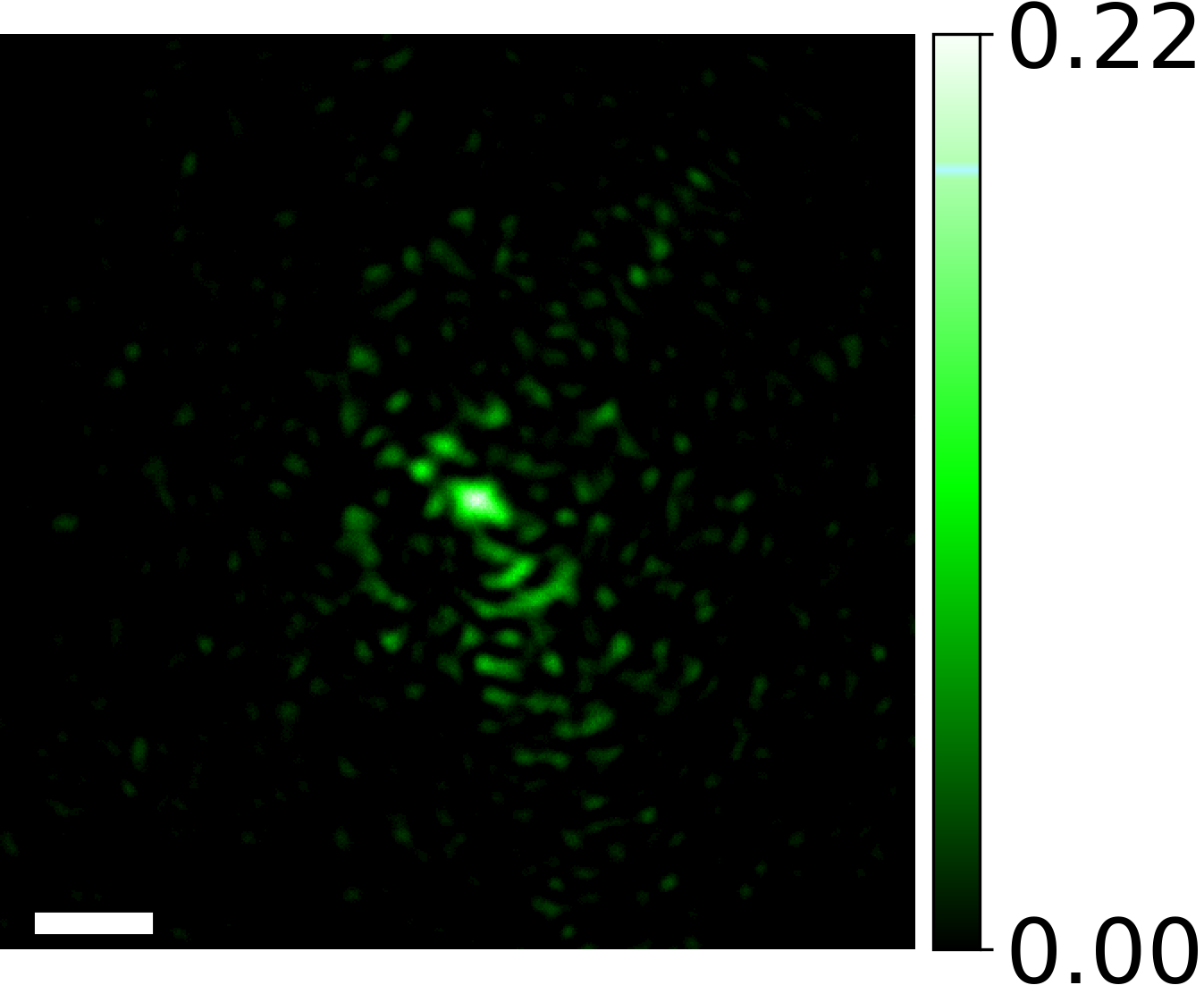}&
			\includegraphics[width= 0.14\textwidth]{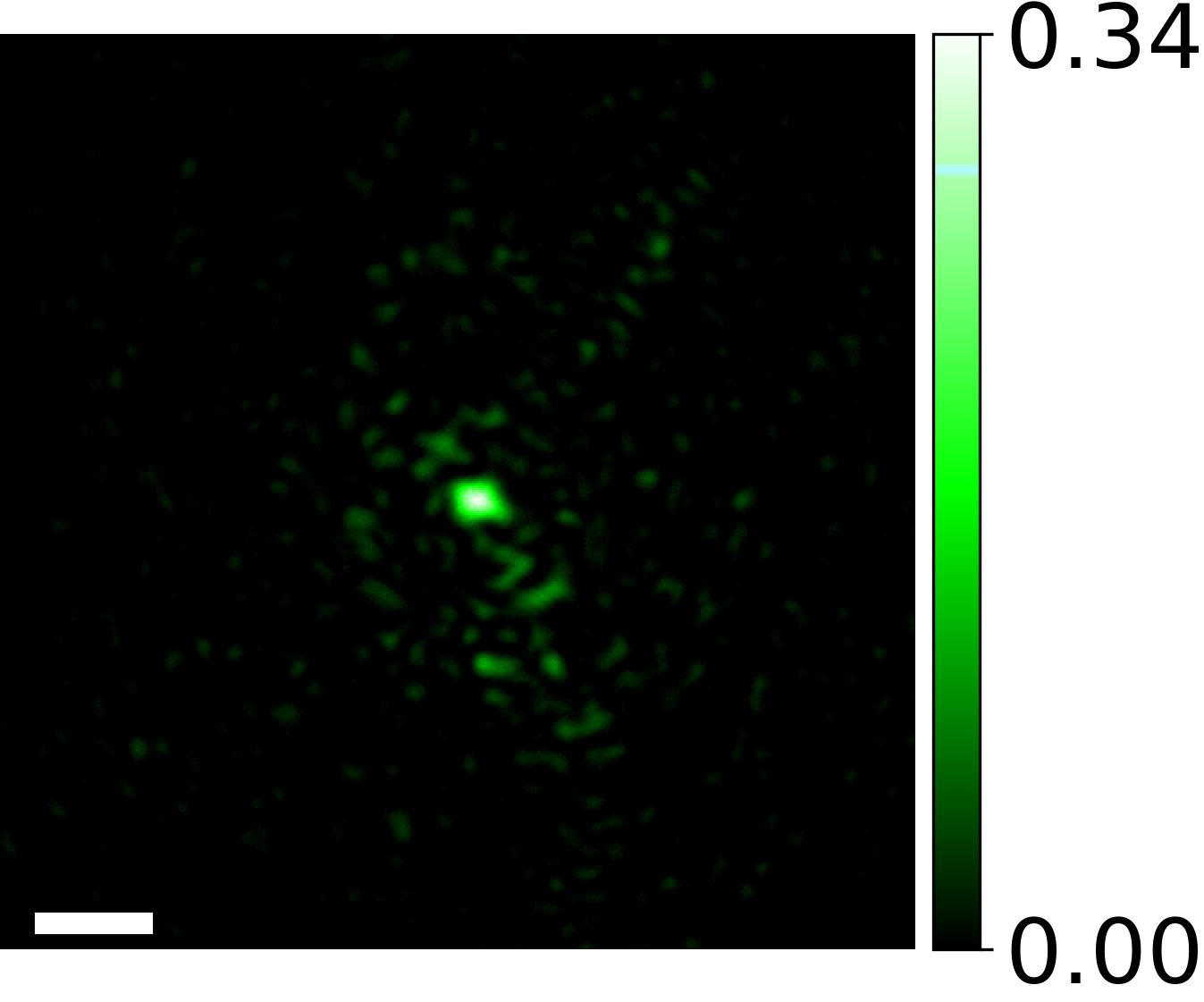}&
			\includegraphics[width= 0.14\textwidth]{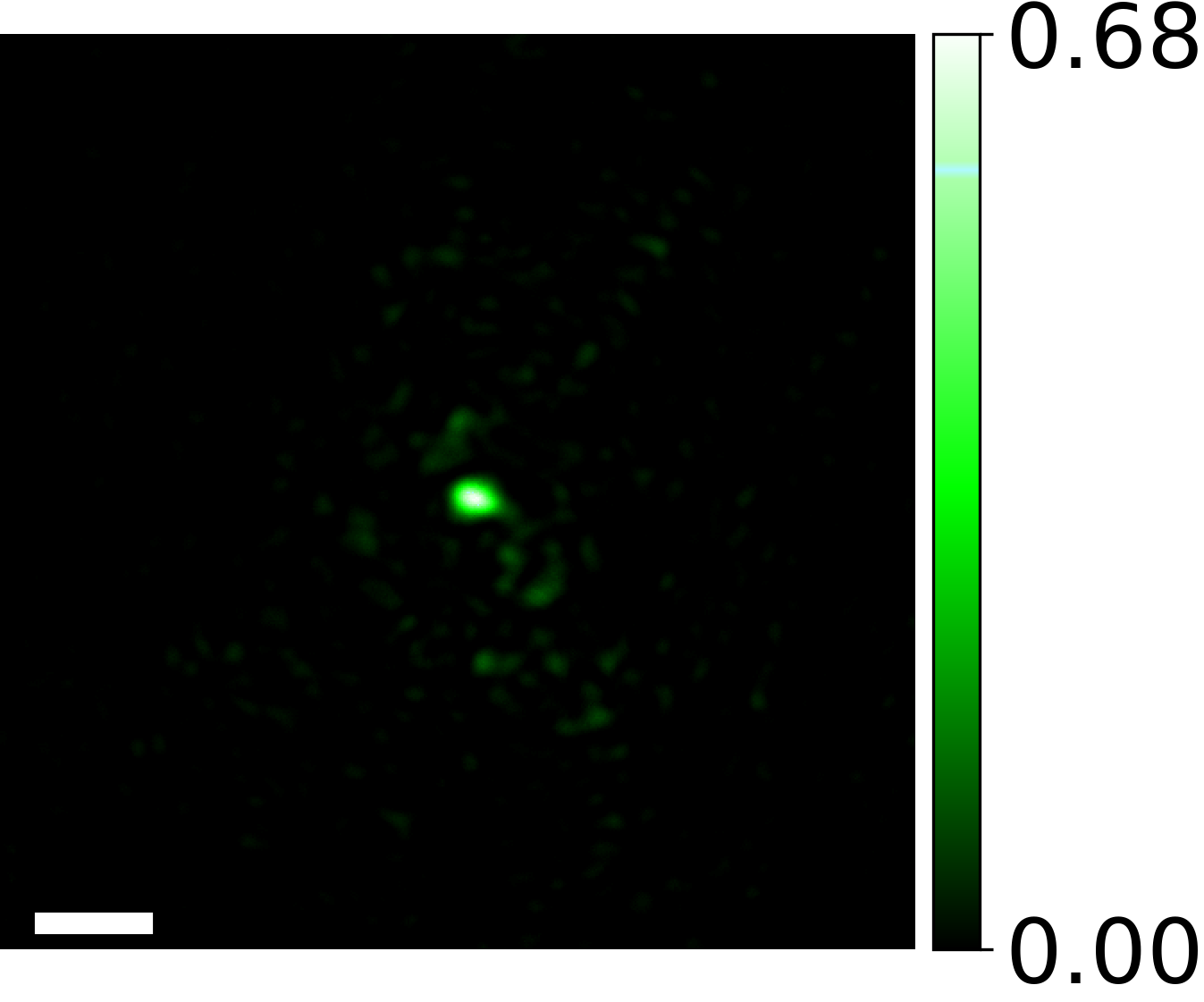}&
			\includegraphics[width= 0.14\textwidth]{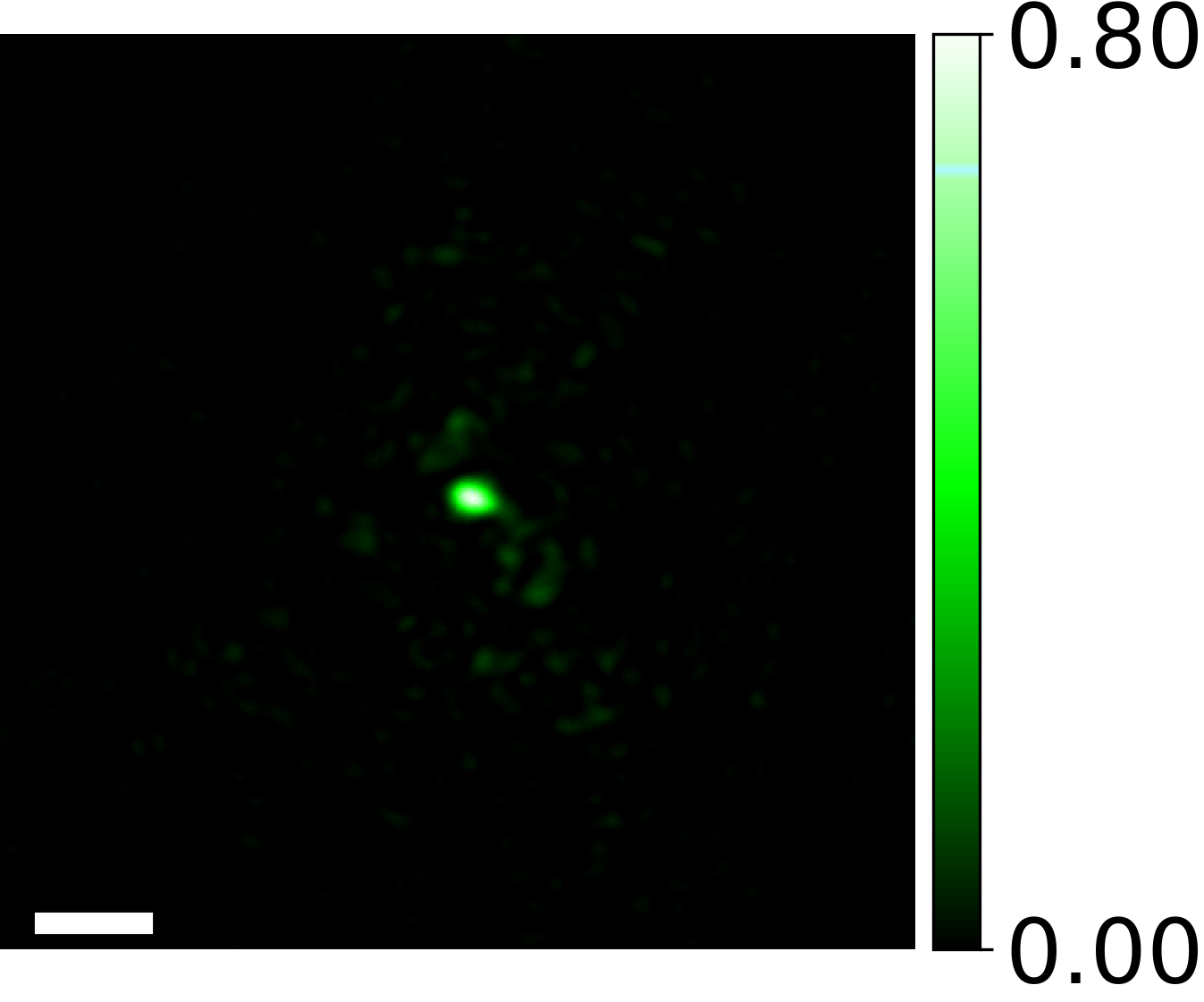}&
			\includegraphics[width= 0.14\textwidth]{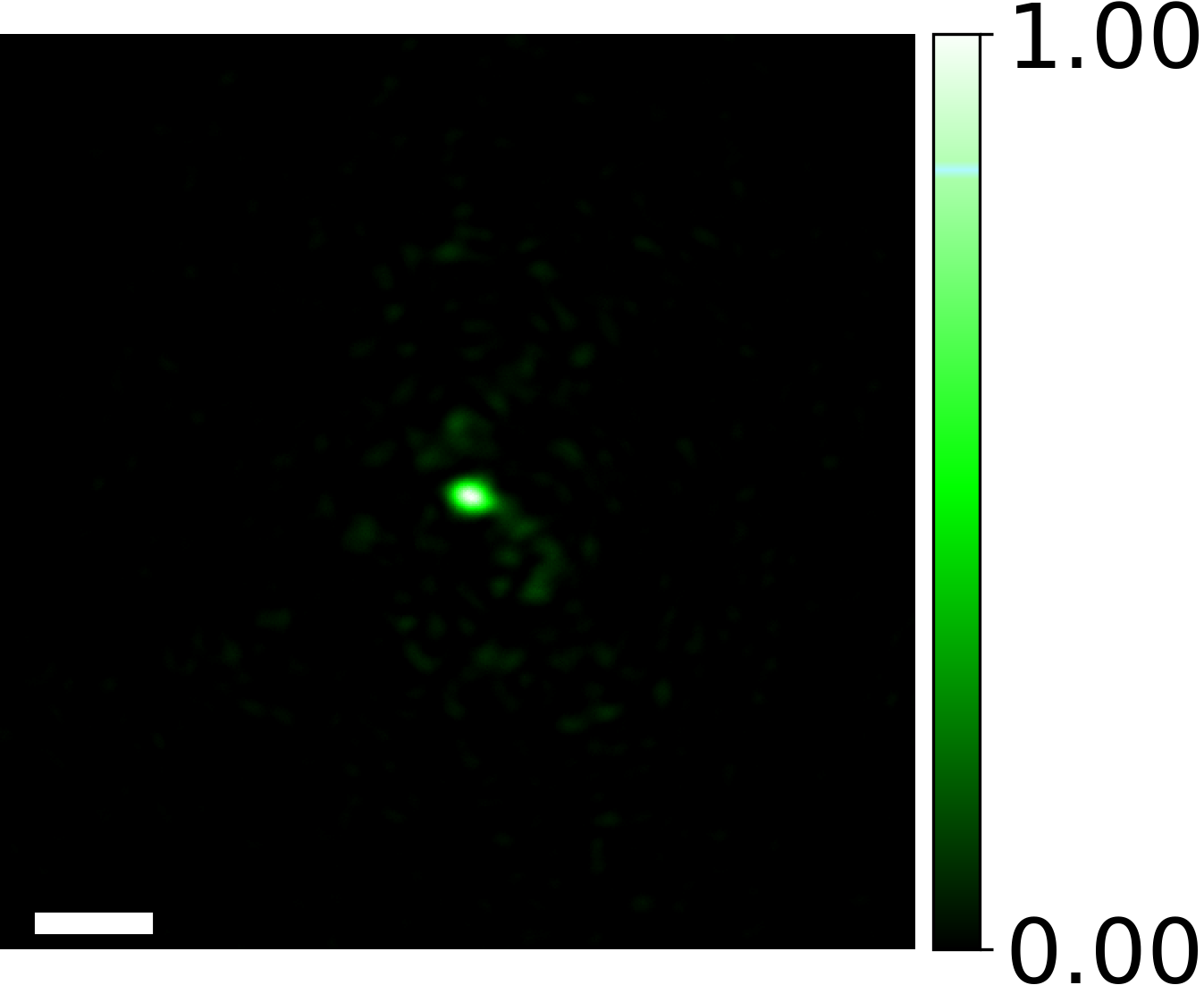}& \\		
			
			{\raisebox{0.5cm}	{\rotatebox[origin=c]{90}{~ {\scriptsize  Phase mask} }}}&
			\hspace{-0.03\textwidth}\includegraphics[width= 0.105\textwidth]{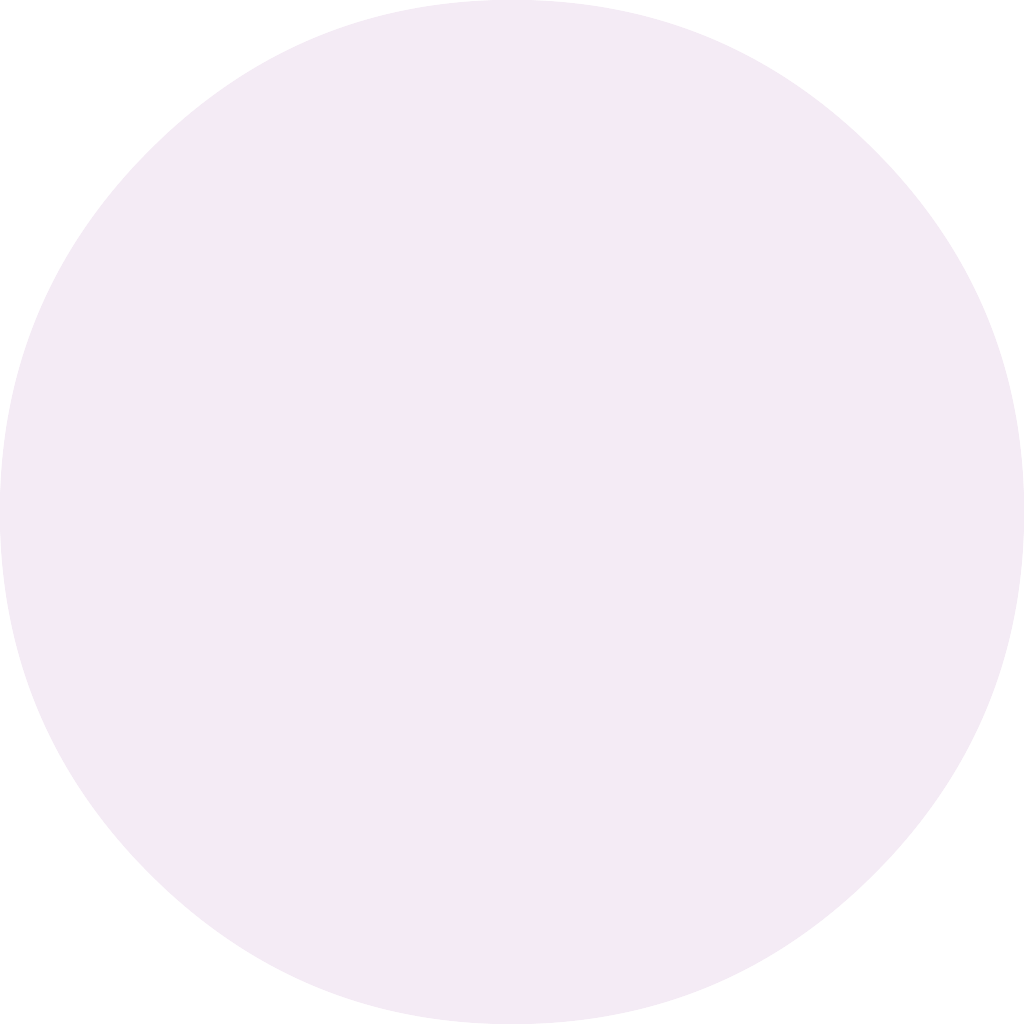}&
			\hspace{-0.03\textwidth}\includegraphics[width= 0.105\textwidth]{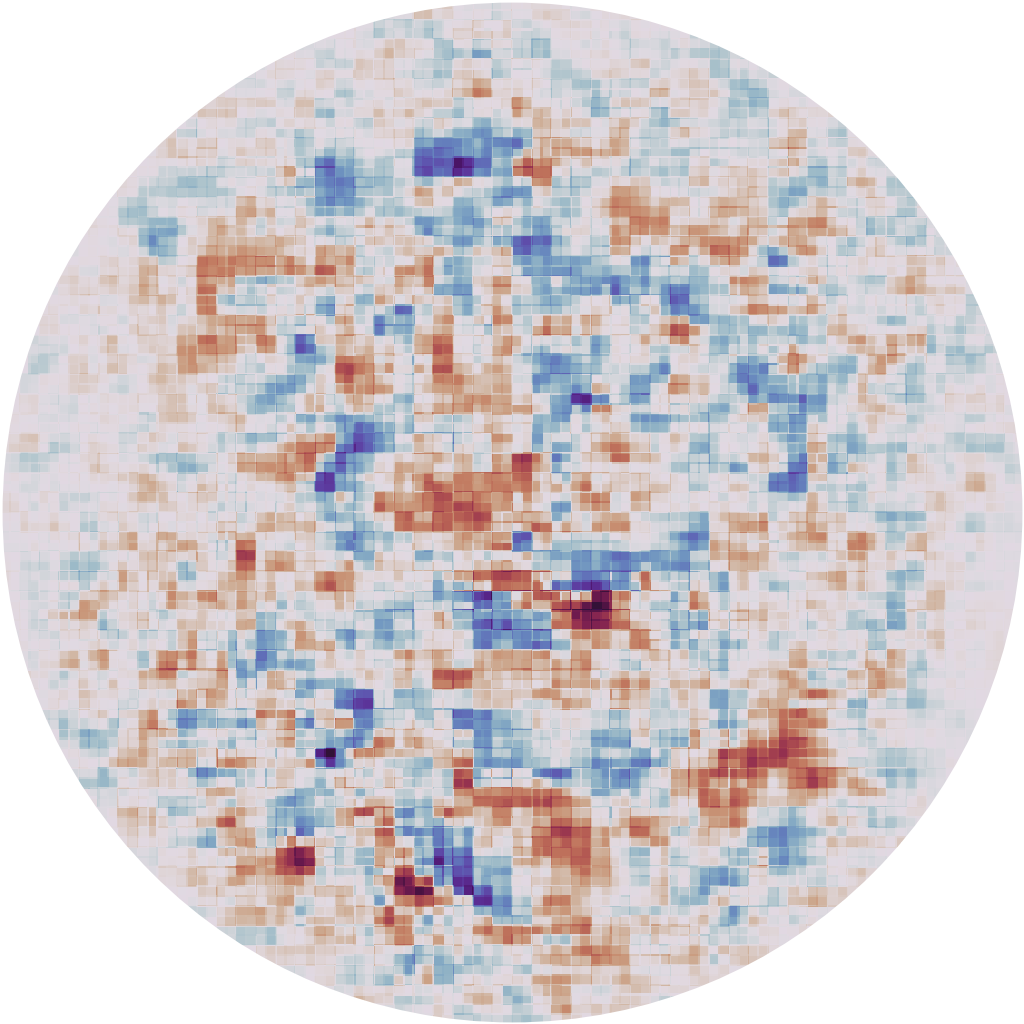}&
			\hspace{-0.03\textwidth}\includegraphics[width= 0.105\textwidth]{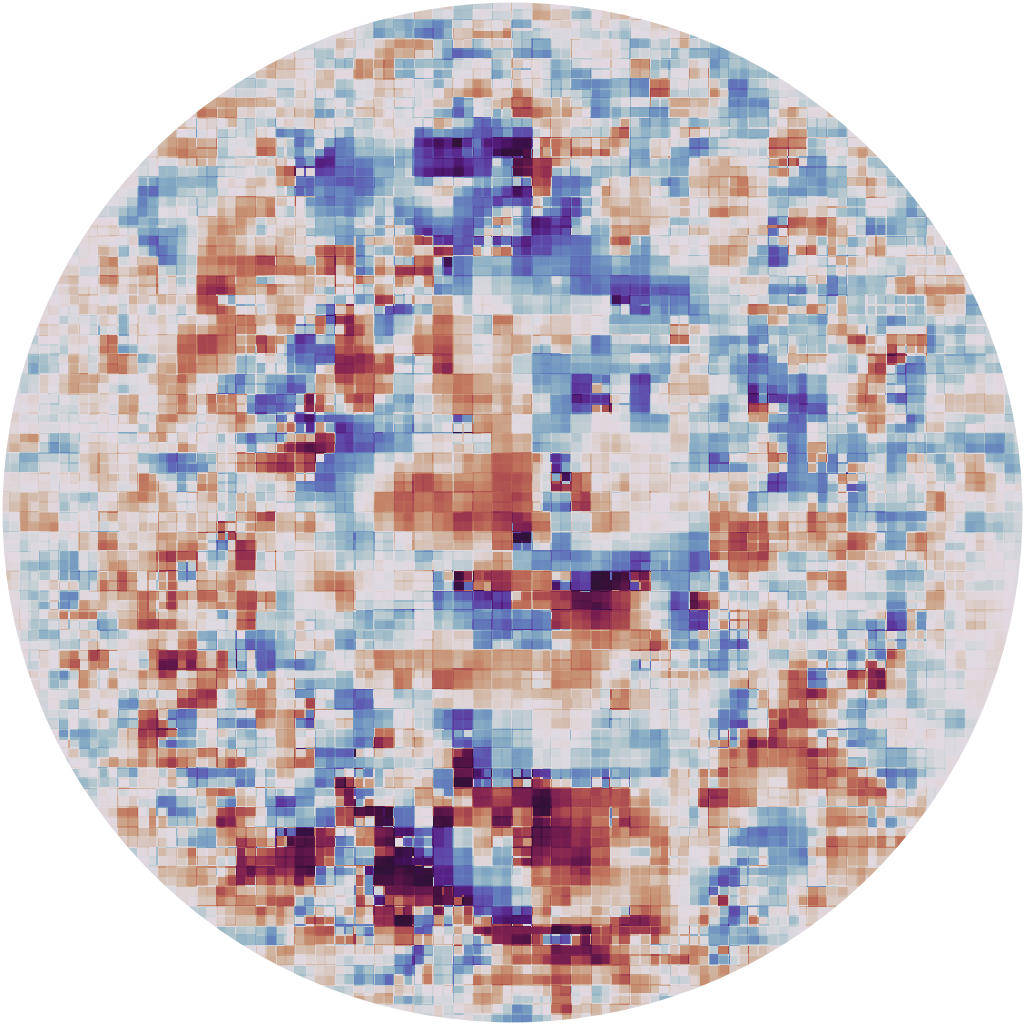}&
			\hspace{-0.03\textwidth}\includegraphics[width= 0.105\textwidth]{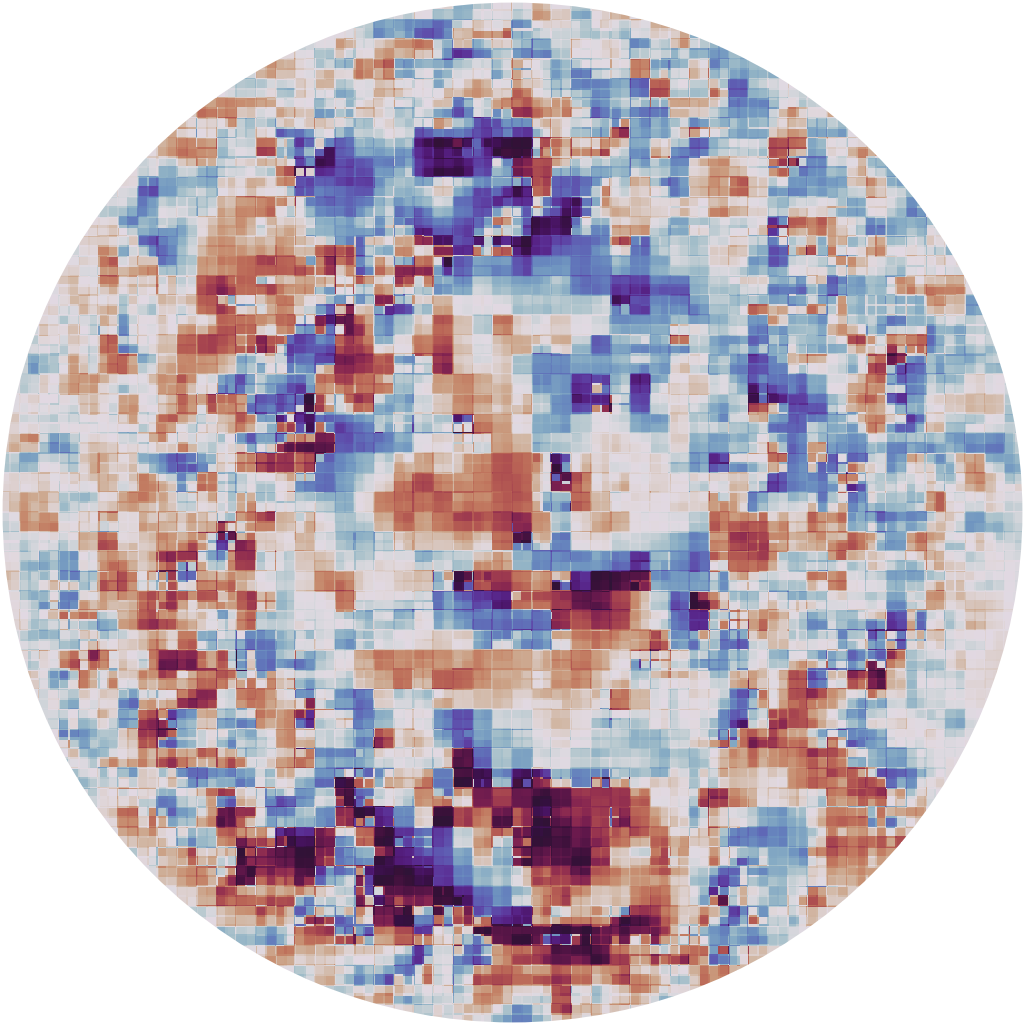}&
			\hspace{-0.03\textwidth}\includegraphics[width= 0.105\textwidth]{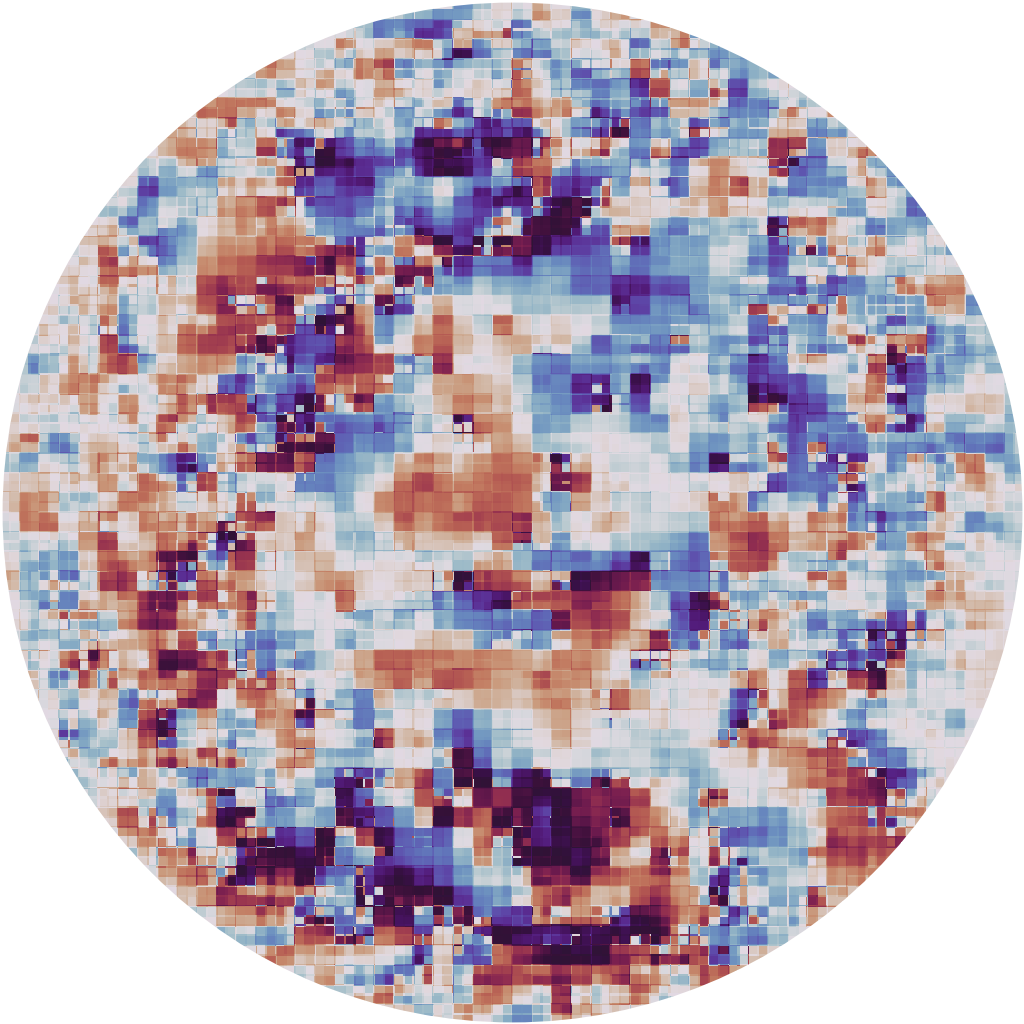}& 
				\end{tabular}
		\caption{\textbf{Algorithm convergence:} We used our algorithm to image a glass mask partially covered with chrome, placed behind a  $180\um$ thick chicken breast tissue as the scattering material. We show captured images on both the main and validation cameras before the algorithm is applied and at the end of each iteration. The first row demonstrates the algorithm's convergence over iterations on the main camera for one of the scanned points. The second row displays the convergence on the validation camera, demonstrating  our ability to also focus light on the target plane. The final row depicts the evolution of the phase mask presented on both SLMs. The last column illustrates the improvement of the cost function across iterations. Scale bar is $4\um$. }
	\label{fig:Converge}
\end{center}
\end{figure*}

%% file: fig_cmp_size.tex
\begin{figure*}[t!]
	\begin{center}\begin{tabular}{@{}c@{~}c@{~}c@{~}c@{~}c@{~}c@{~}c@{~}}
			\multicolumn{1}{c}{}&
			\multicolumn{1}{c}{\hspace{-0.65cm} \scriptsize Init.}&
			\multicolumn{1}{c}{\hspace{-0.65cm} \scriptsize $1.3{\times}1.3{\scriptstyle\um}$}&
			\multicolumn{1}{c}{\hspace{-0.65cm} \scriptsize $5.2{\times}5.2{\scriptstyle\um}$}&
			\multicolumn{1}{c}{\hspace{-0.65cm} \scriptsize $10.4{\times}10.4{\scriptstyle\um}$}&
			\multicolumn{1}{c}{\hspace{-0.65cm} \scriptsize $15.6{\times}15.6{\scriptstyle\um}$}\\
			{\raisebox{0.77cm}	{\rotatebox[origin=c]{90}{~ {\scriptsize Main cam.} }}}&
			\includegraphics[width= 0.18\textwidth]{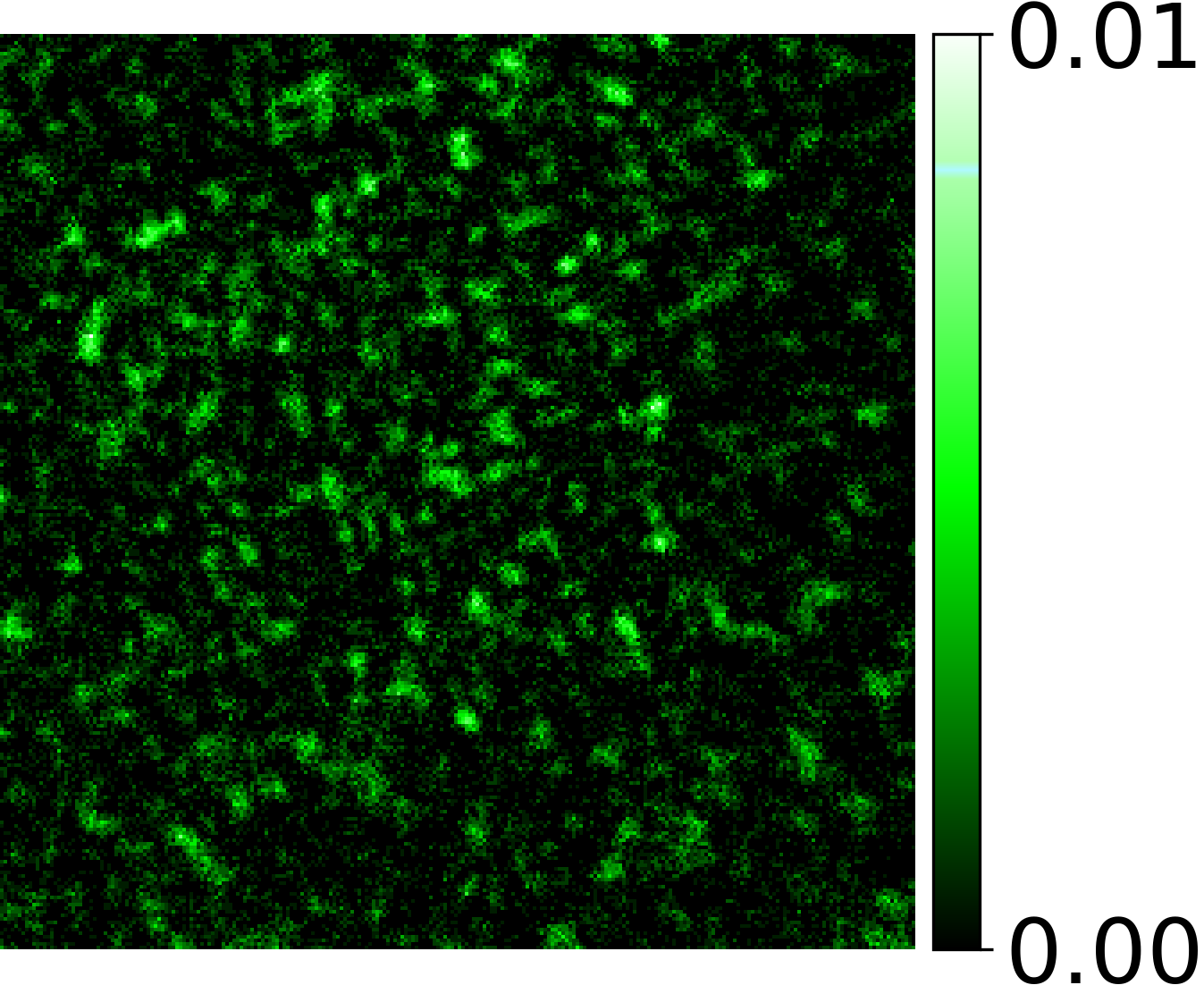}&
			\includegraphics[width= 0.18\textwidth]{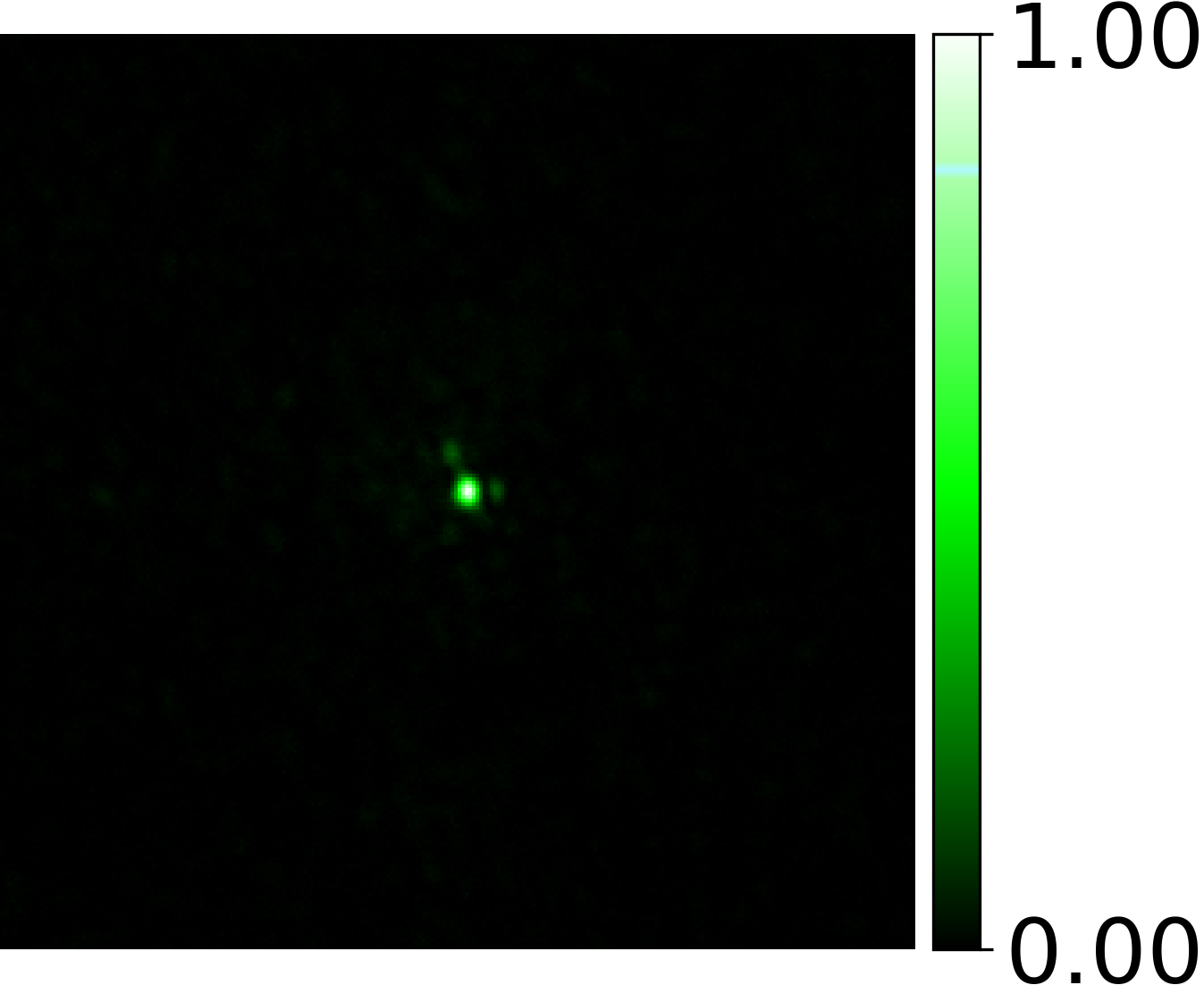}&
			\includegraphics[width= 0.18\textwidth]{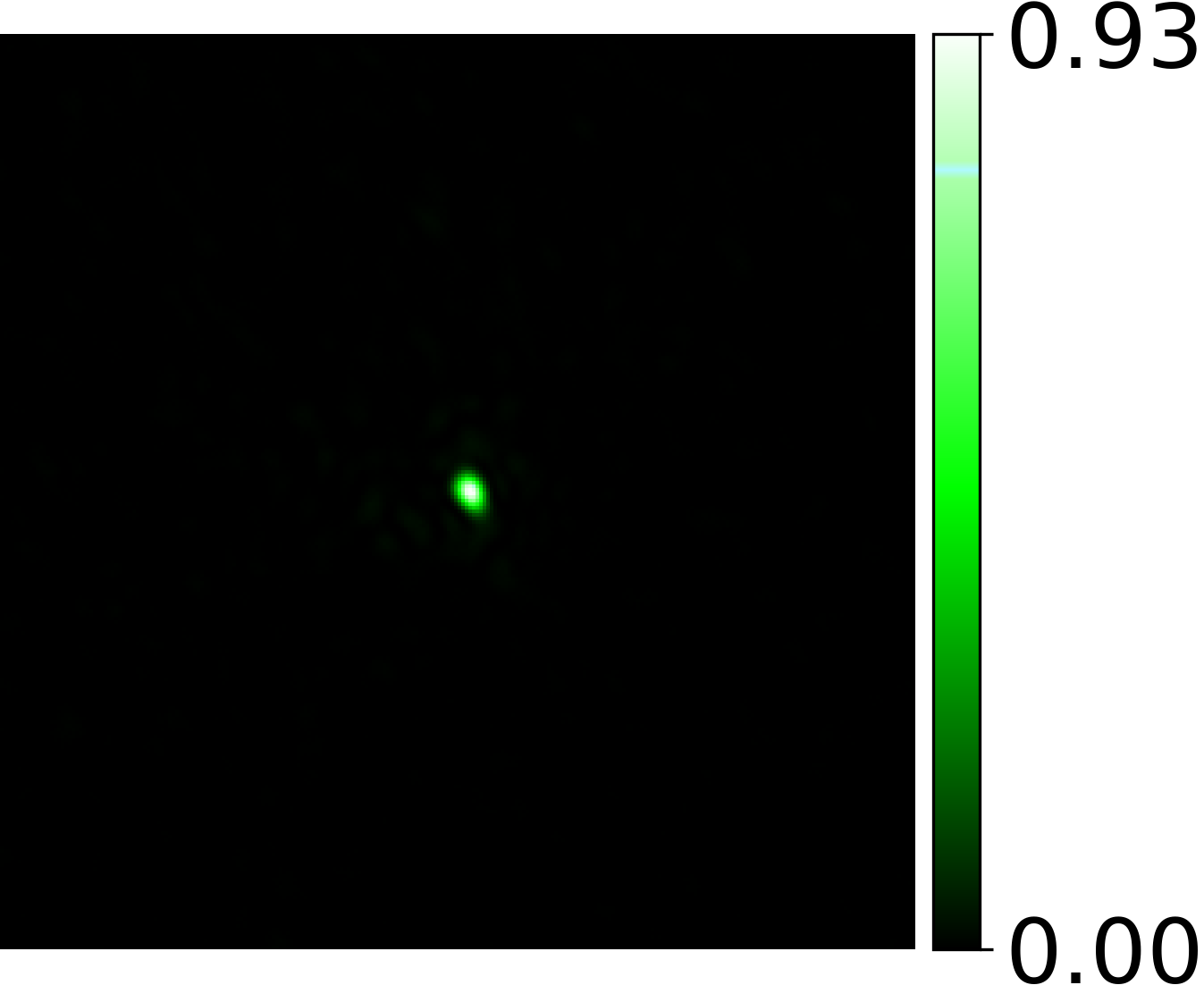}&
			\includegraphics[width= 0.18\textwidth]{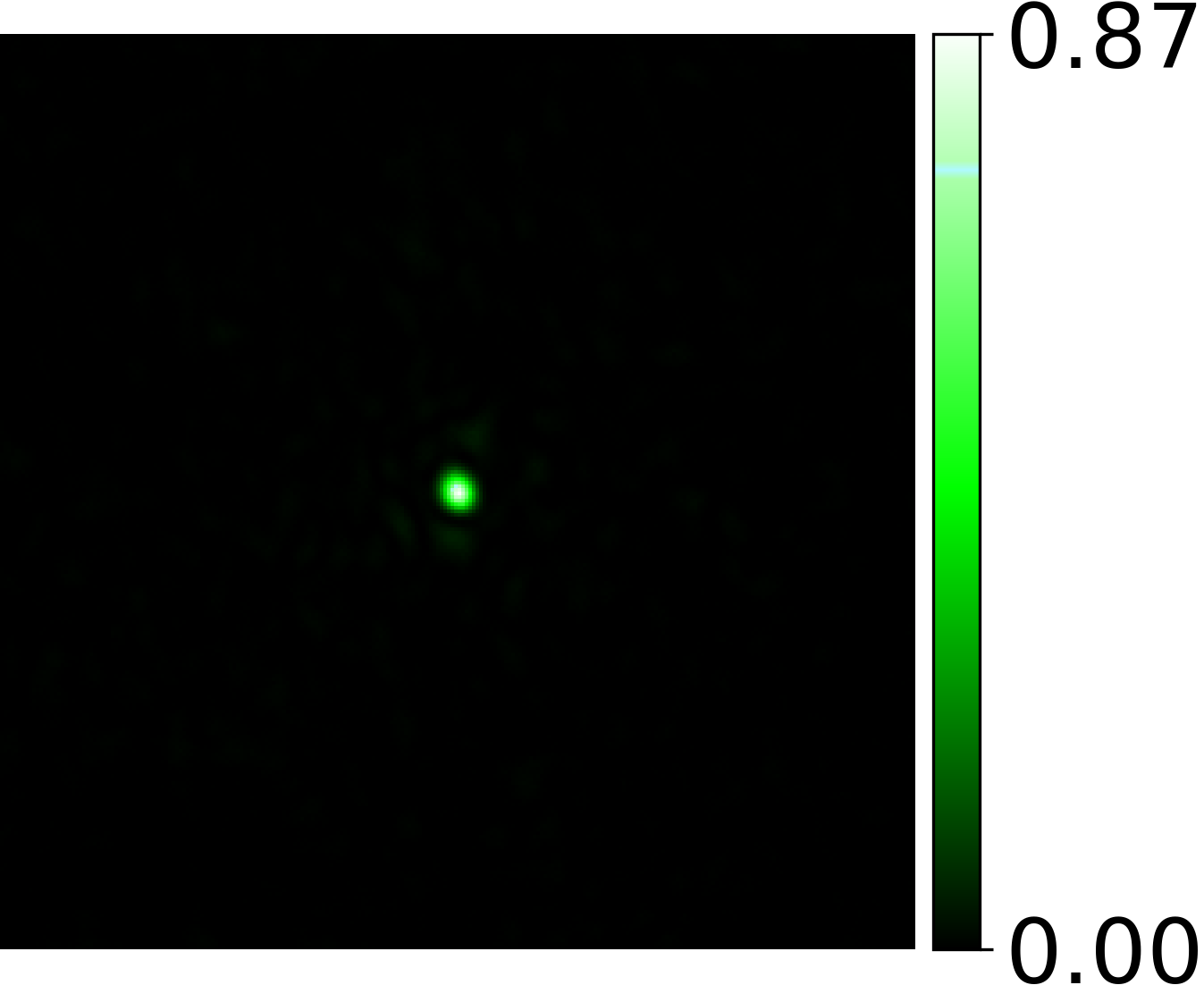}&
			\includegraphics[width= 0.18\textwidth]{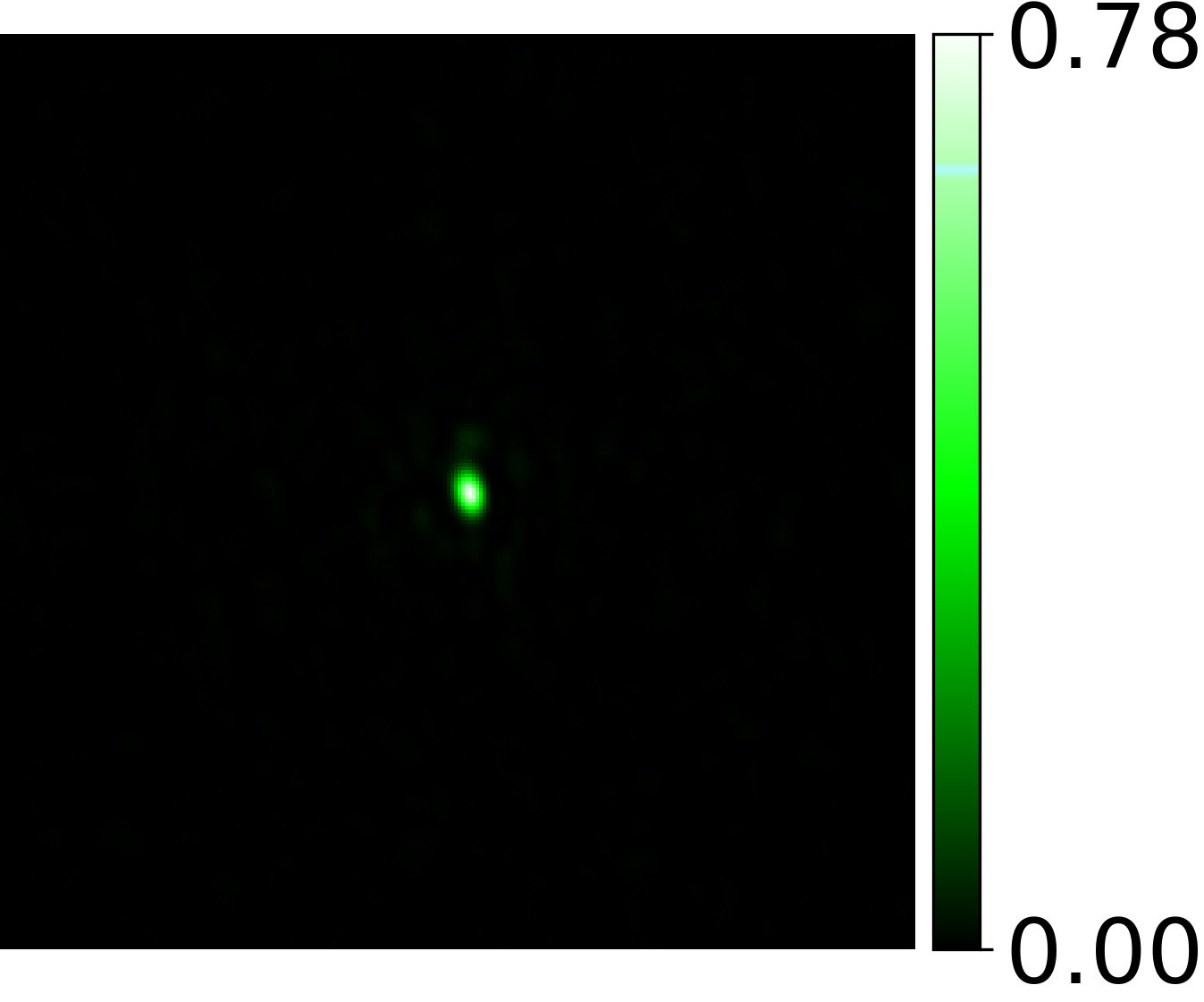}\\
			{\raisebox{0.78cm}	{\rotatebox[origin=c]{90}{~ {\scriptsize Valid. cam.} }}}&
			\includegraphics[width= 0.18\textwidth]{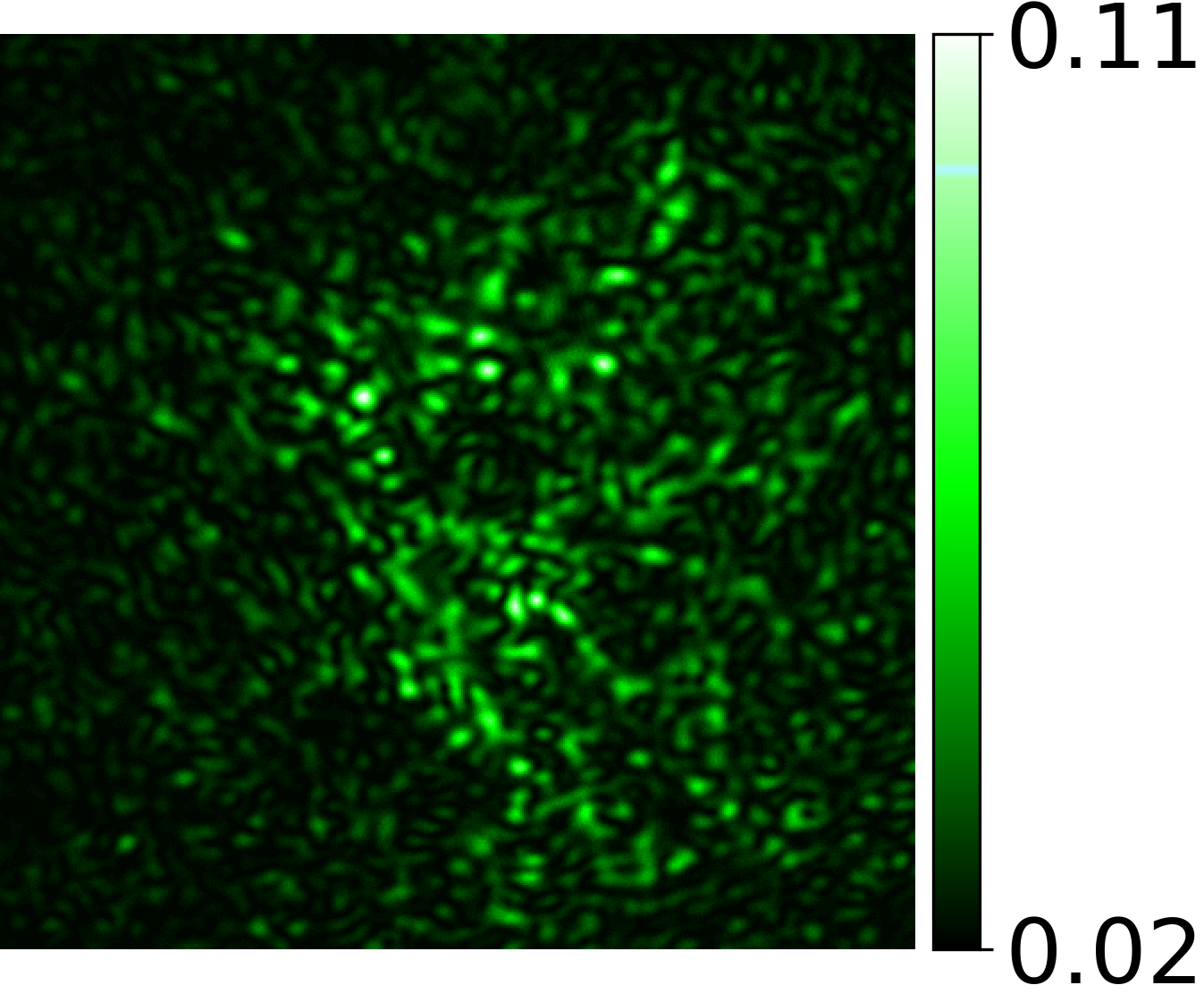}&
			\includegraphics[width= 0.18\textwidth]{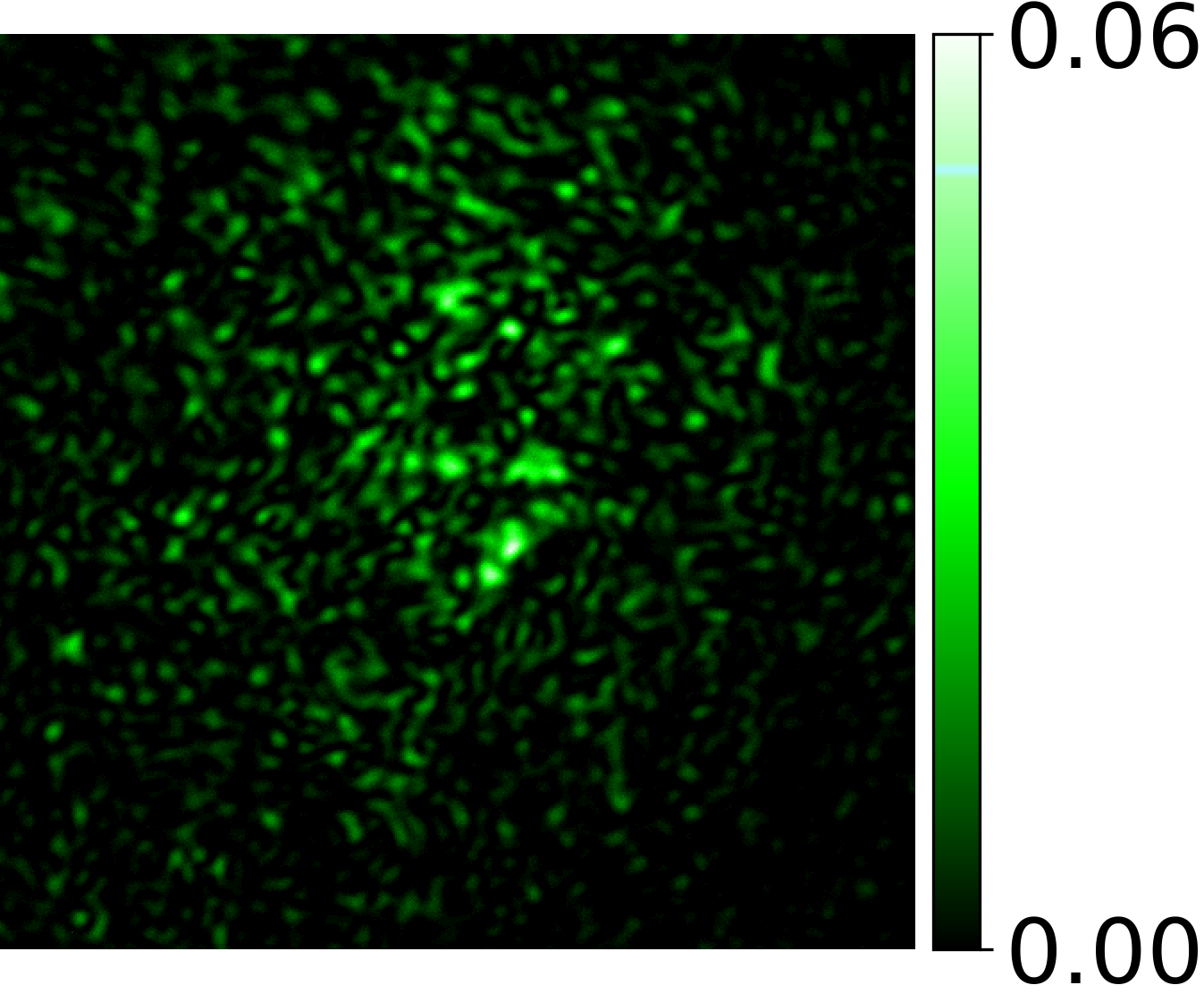}&
			\includegraphics[width= 0.18\textwidth]{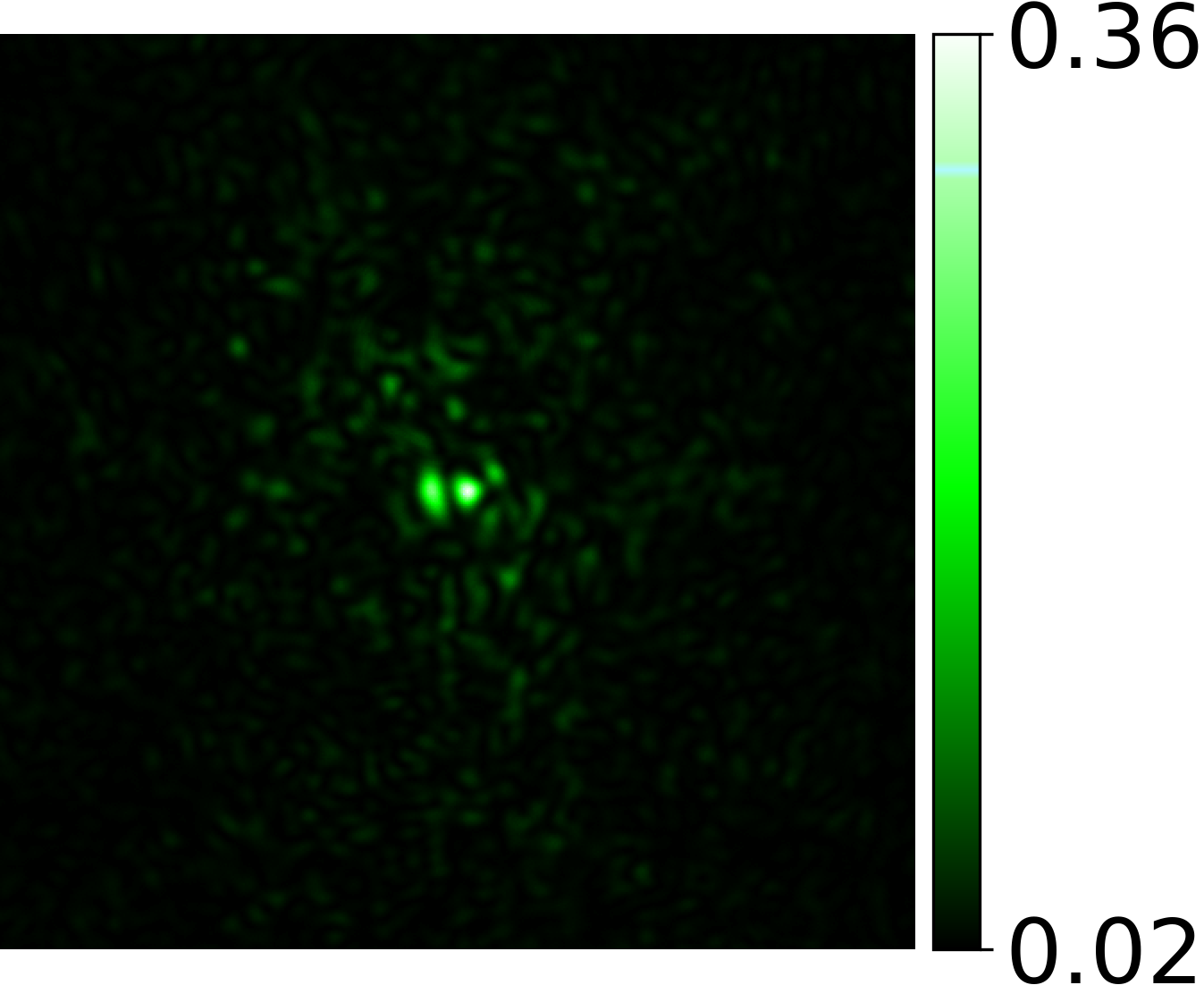}&
			\includegraphics[width= 0.18\textwidth]{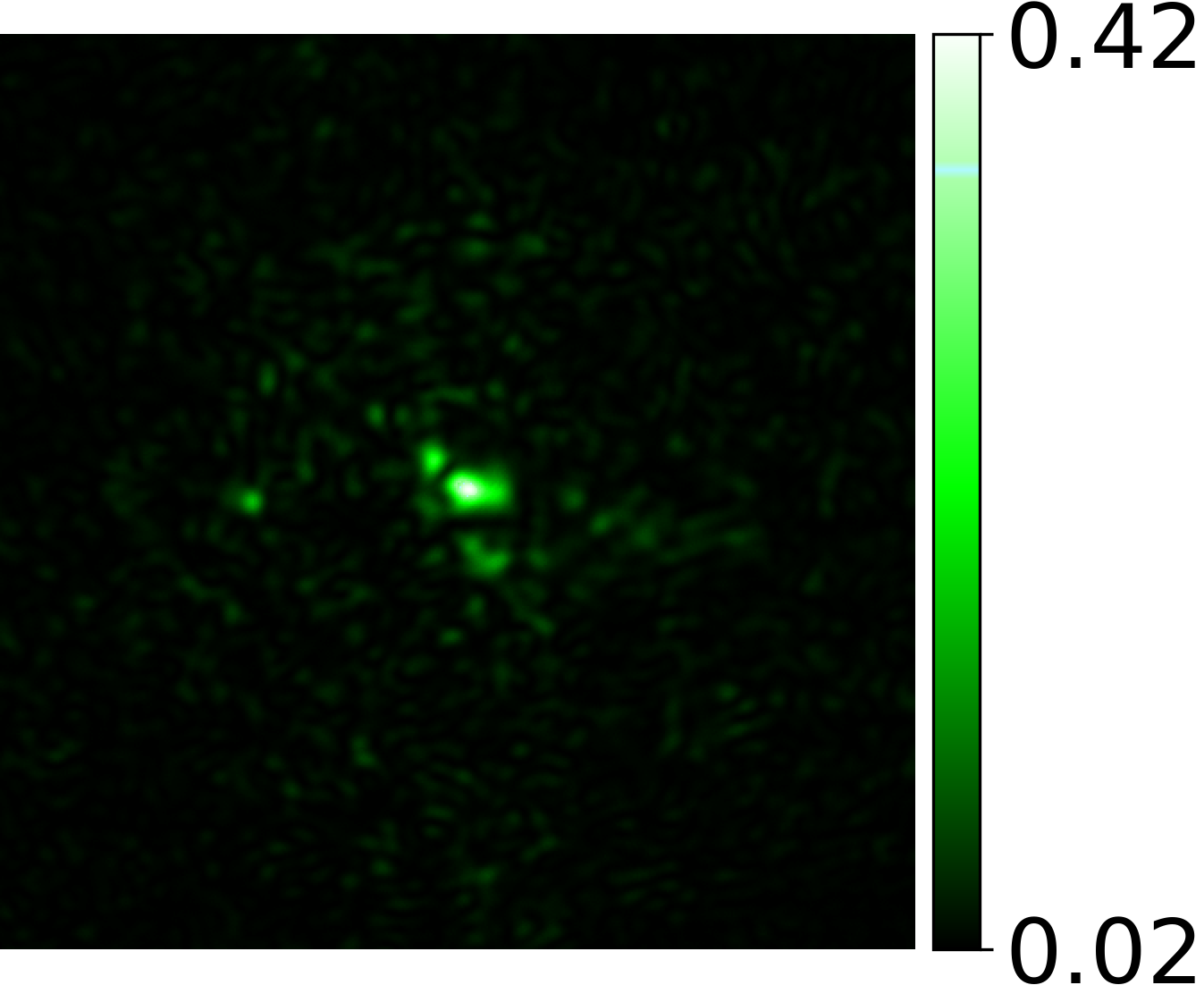}&
			\includegraphics[width= 0.18\textwidth]{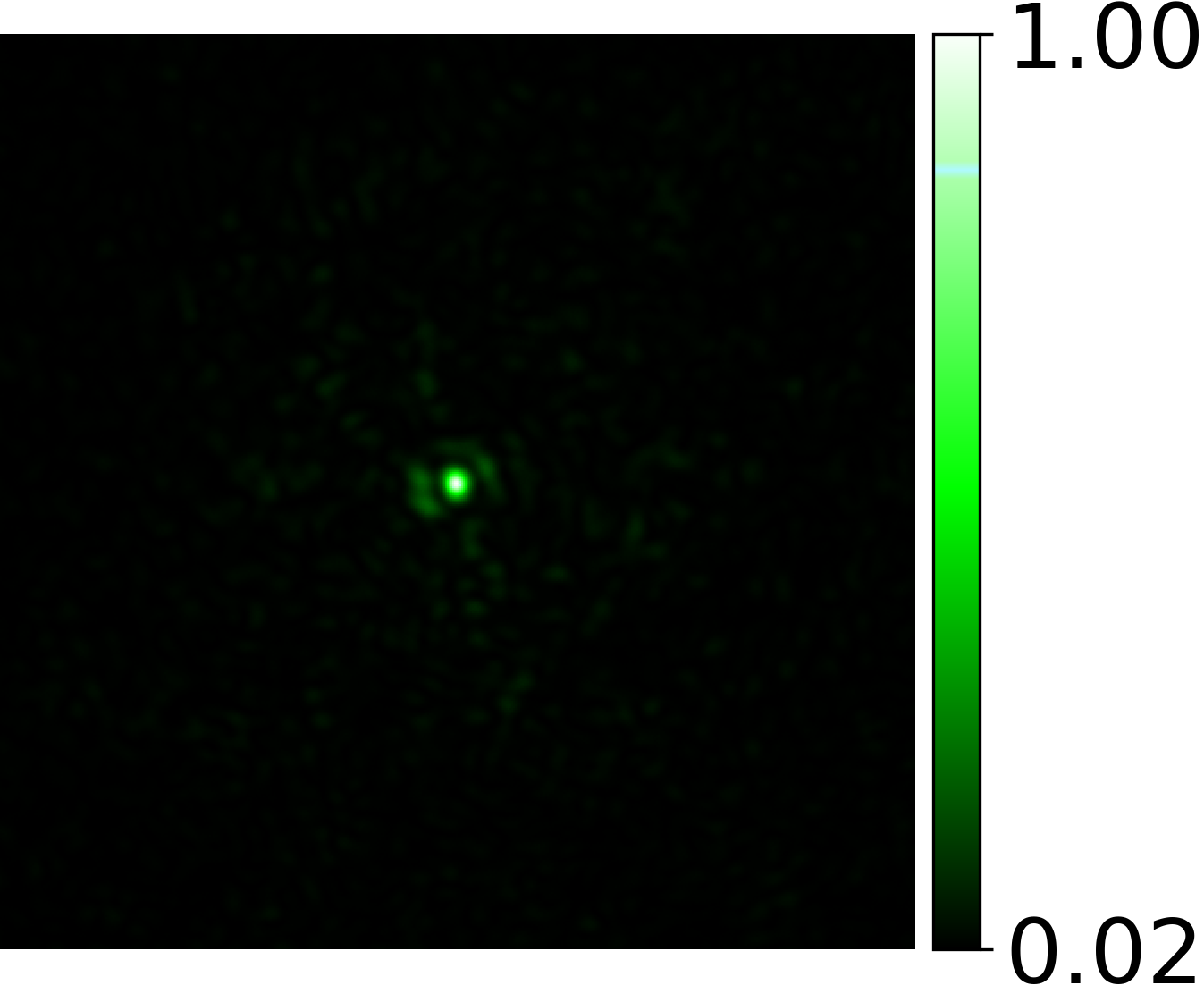}
		\end{tabular}
		\caption{\textbf{Effect of target area size:} We evaluate how focusing quality depends on the size of the target area $\Area$ over which the confocal score is computed. The experiment used a chrome-covered glass mask overlaid with $240\um$ thick chicken breast tissue. Our results show that for single-spot illumination (second column) the algorithm successfully achieved a sharp focused spot on the main camera without actually focusing light to a spot inside the tissue, as evident by the image from the validation camera. As we expanded the size of the focus area, the algorithm demonstrated progressively improved ability to focus light inside the tissue and the validation camera images are sharper. At the same time increasing the scanned area reduces the intensity of the focused spot on the main camera because a single modulation cannot correct very large areas.}
		\label{fig:cmp_size}
	\end{center}
\end{figure*}

%% file: fig_res_mask.tex
\begin{figure*}[t!]
	\begin{center}
		\begin{tabular}{@{}c@{~}c@{~}c@{~}c@{~}c@{~}c@{~}c@{~}}    	
			\multicolumn{4}{c}{\hspace{-0.6cm}\large  One point }& 
			\multicolumn{3}{c}{\hspace{-0.6cm}\large Confocal scan}\\
		  	\multicolumn{2}{c}{\hspace{-0.6cm} Main cam.} & 
		  	\multicolumn{2}{c}{\hspace{-0.6cm} Valid. cam.} & 
		  	\multicolumn{3}{c}{\hspace{-0.6cm} Main cam.} \\
			\multicolumn{1}{c}{\hspace{-0.6cm} \scriptsize w/o}&	
			\multicolumn{1}{c}{\hspace{-0.6cm} \scriptsize w/ }&	
			\multicolumn{1}{c}{\hspace{-0.6cm} \scriptsize w/o}&	
			\multicolumn{1}{c}{\hspace{-0.6cm} \scriptsize w/ }&	
			\multicolumn{1}{c}{\hspace{-0.6cm} \scriptsize w/o}&	
			\multicolumn{1}{c}{\hspace{-0.6cm} \scriptsize w/ }&	
			\multicolumn{1}{c}{\hspace{-0.6cm} \scriptsize Ground}\vspace{-0.0cm}\\
			\multicolumn{1}{c}{\hspace{-0.6cm} \scriptsize modulation}&	
			\multicolumn{1}{c}{\hspace{-0.6cm} \scriptsize modulation}&	
			\multicolumn{1}{c}{\hspace{-0.6cm} \scriptsize modulation}&	
			\multicolumn{1}{c}{\hspace{-0.6cm} \scriptsize modulation}&	
			\multicolumn{1}{c}{\hspace{-0.6cm} \scriptsize modulation}&	
			\multicolumn{1}{c}{\hspace{-0.6cm} \scriptsize modulation}&	
			\multicolumn{1}{c}{\hspace{-0.6cm} \scriptsize truth}\\
			\includegraphics[width= 0.14\textwidth]{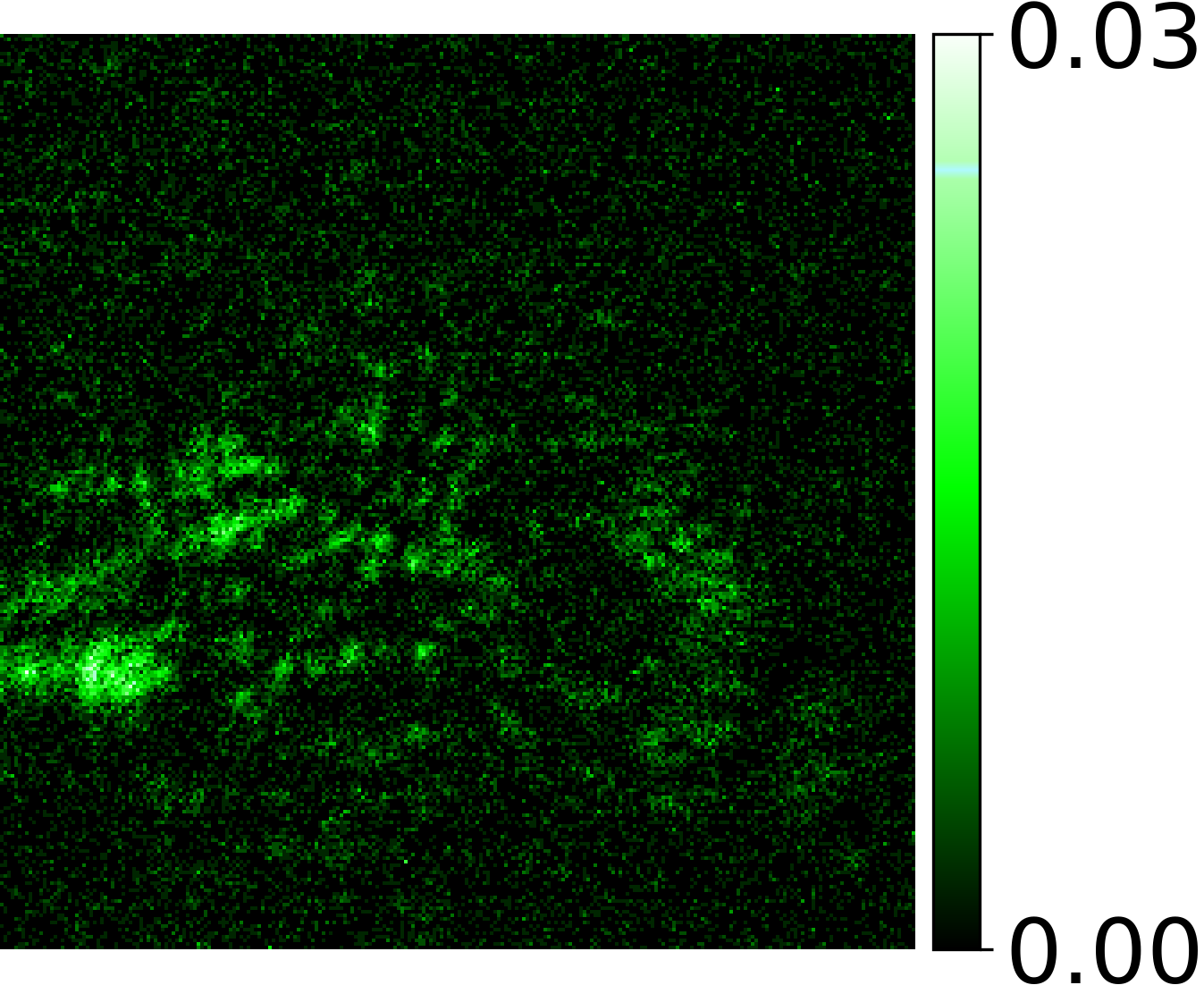}&
			\includegraphics[width= 0.14\textwidth]{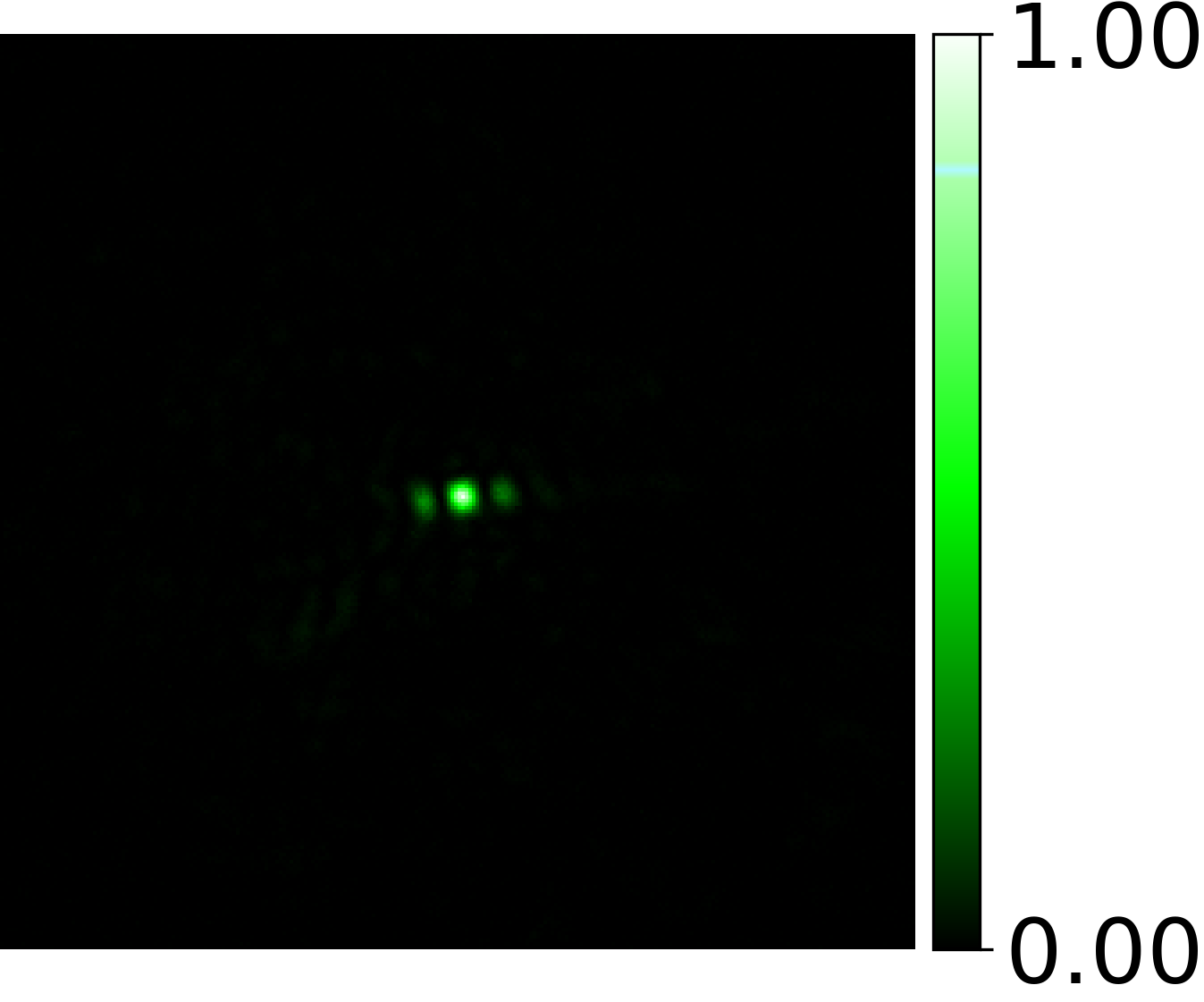}&
			\includegraphics[width= 0.14\textwidth]{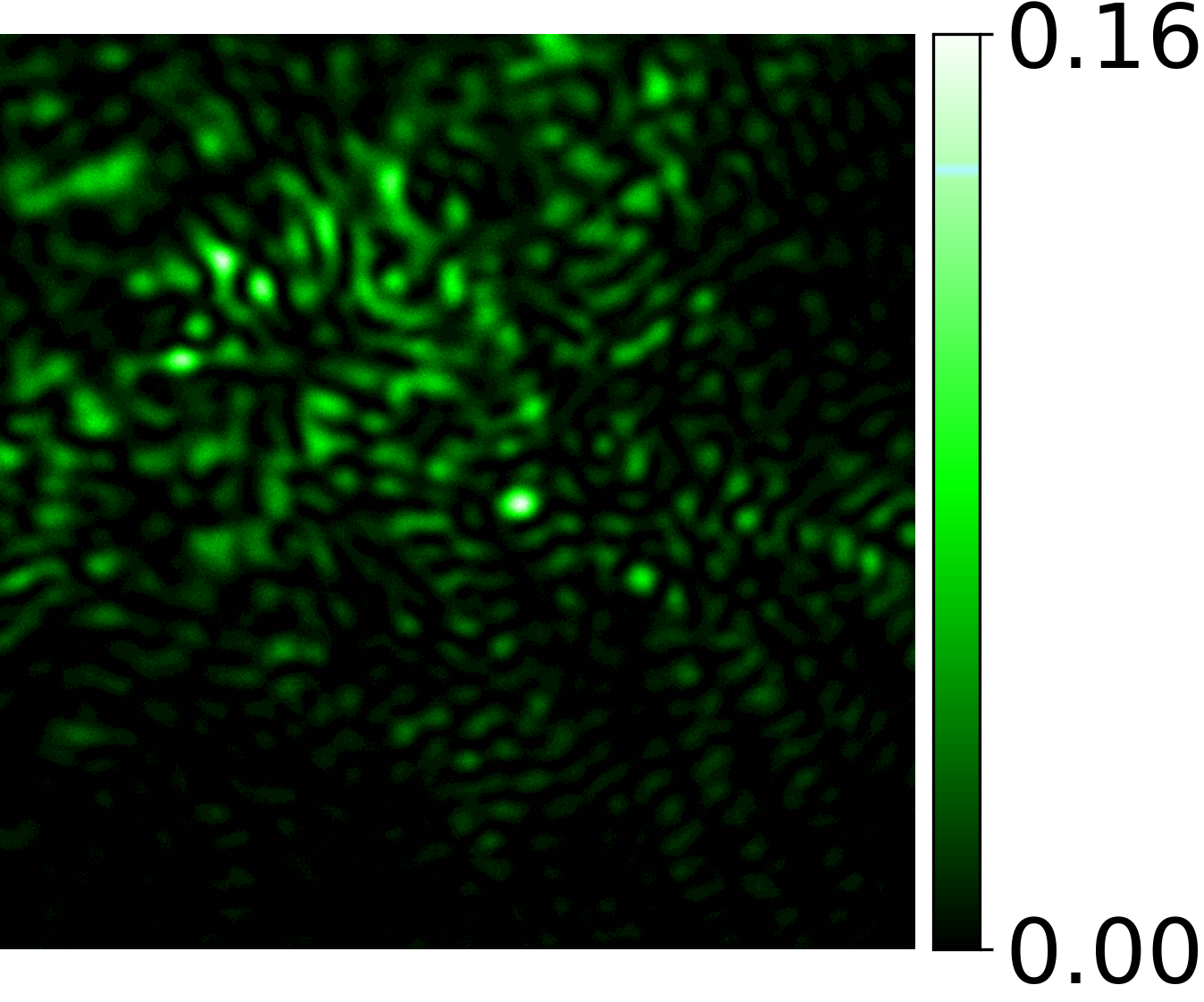}&
			\includegraphics[width= 0.14\textwidth]{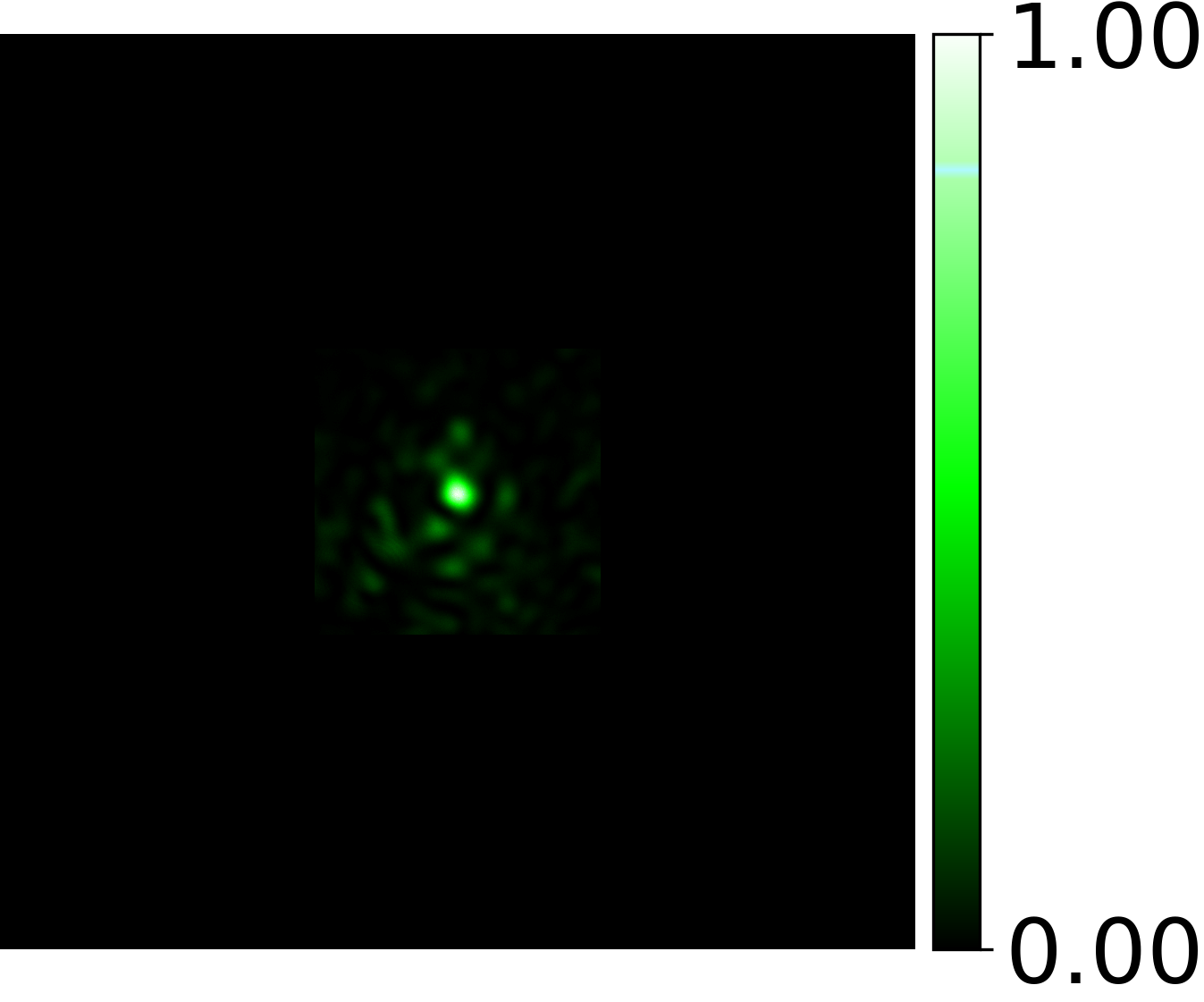}&
			\includegraphics[width= 0.14\textwidth]{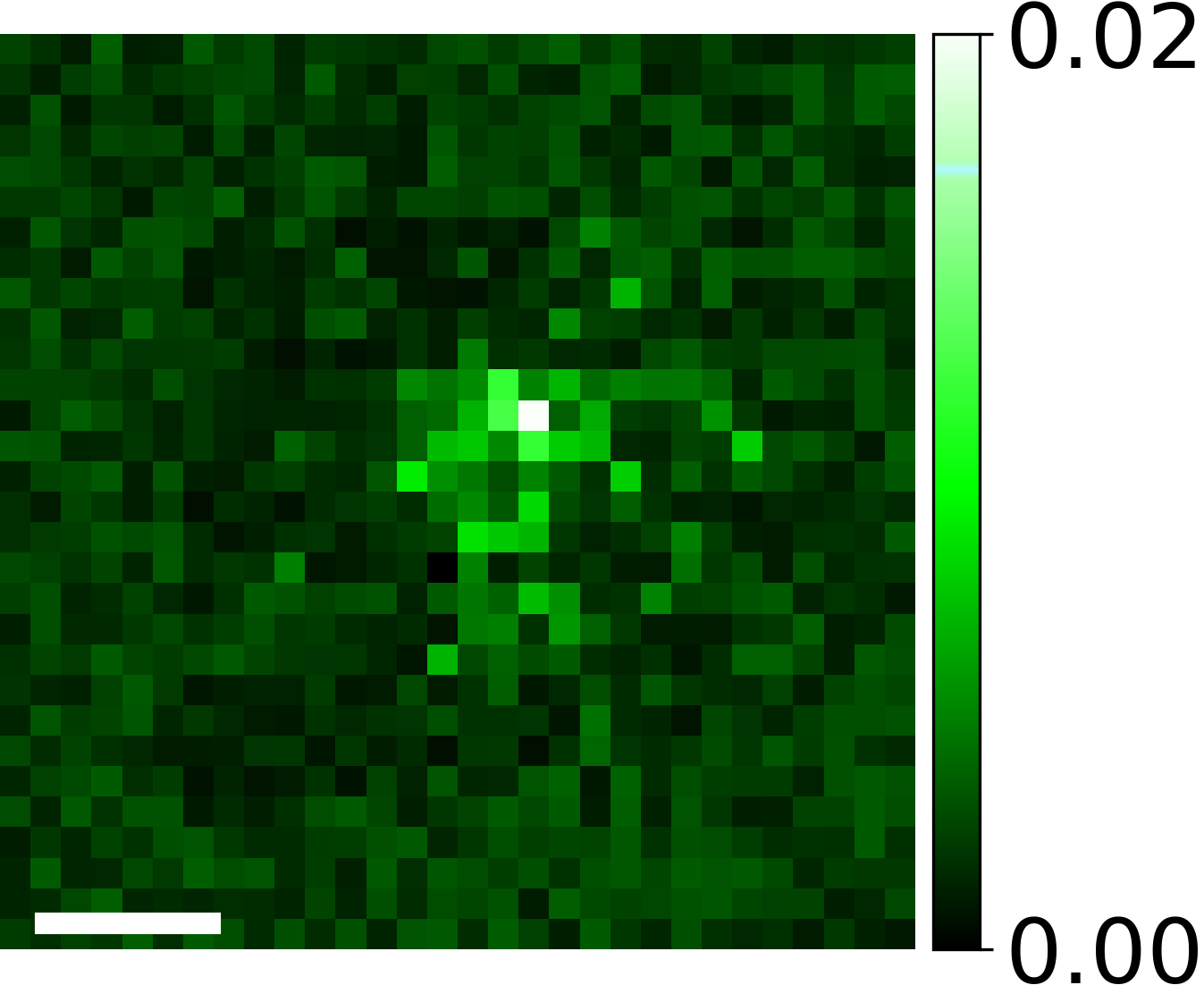}&
			\includegraphics[width= 0.14\textwidth]{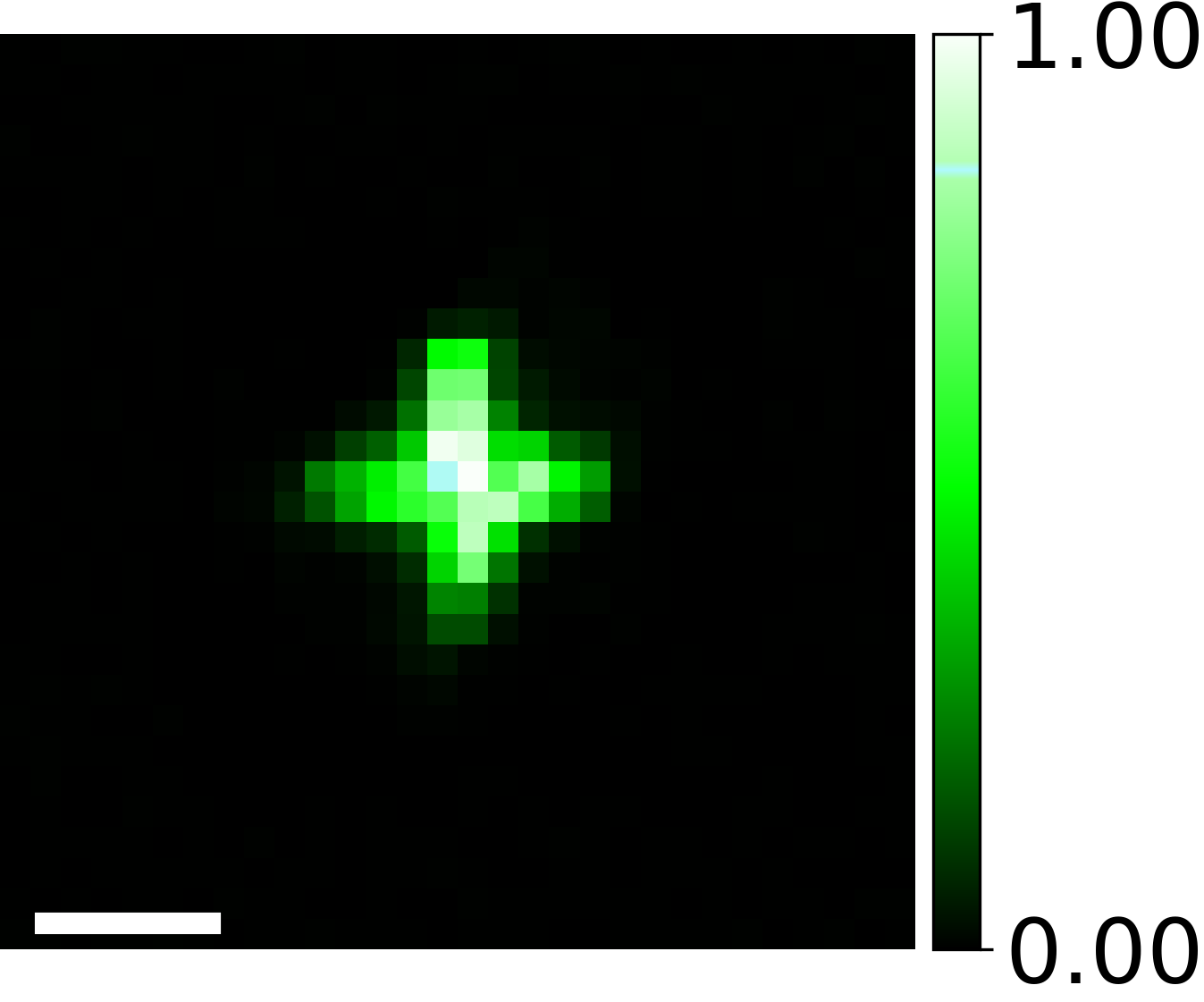}&
			\includegraphics[width= 0.14\textwidth]{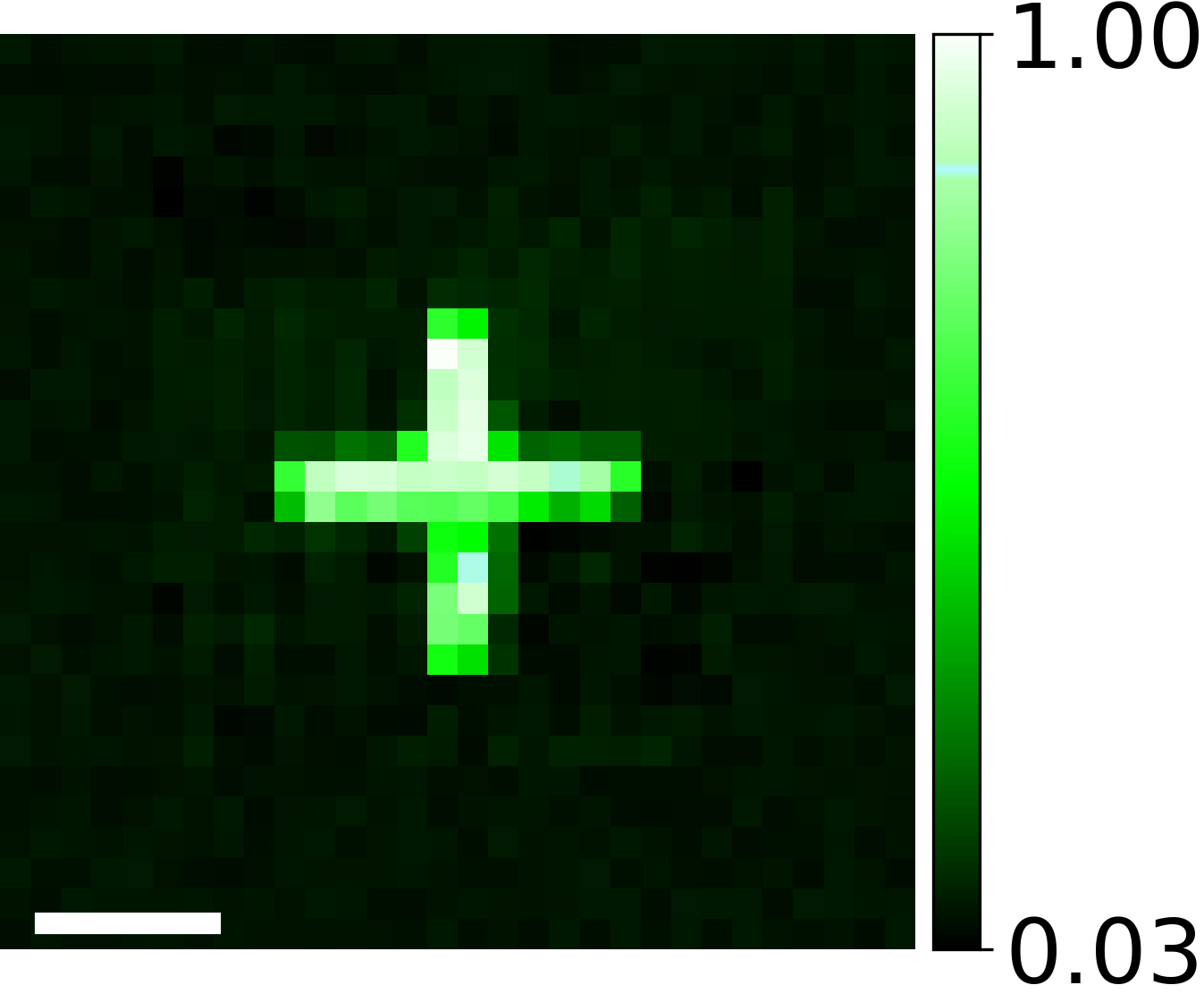}\\
			\includegraphics[width= 0.14\textwidth]{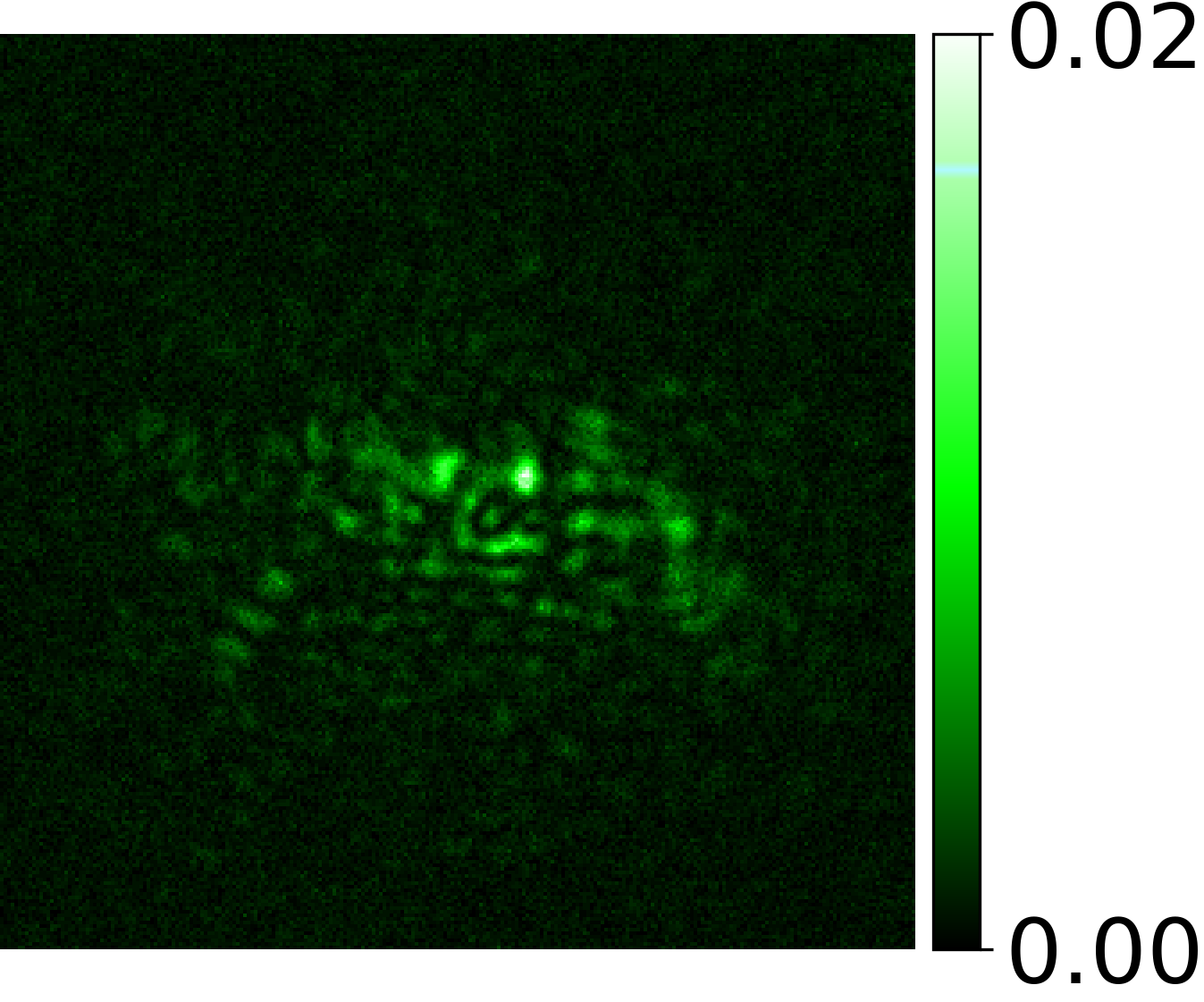}&
			\includegraphics[width= 0.14\textwidth]{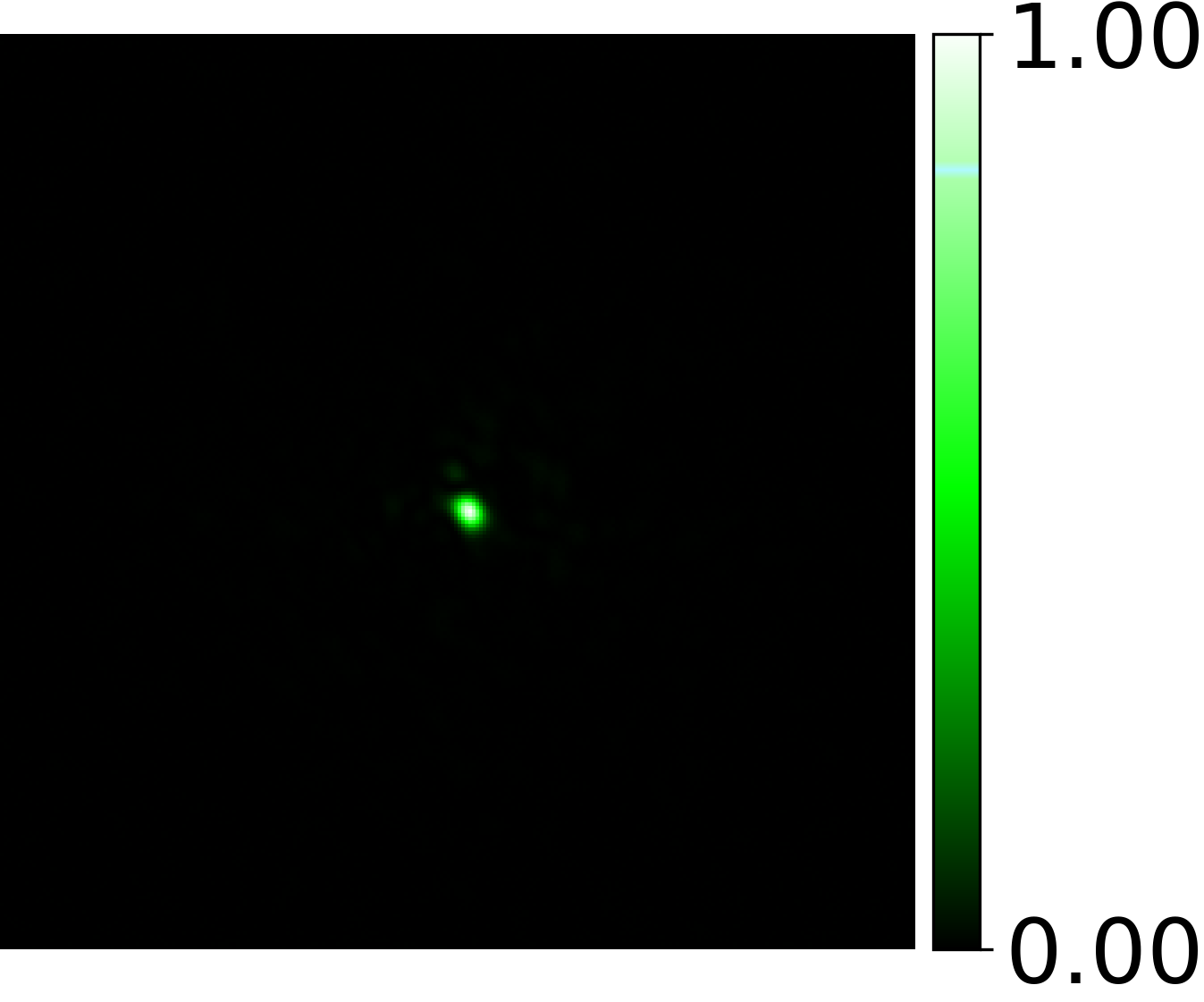}&
			\includegraphics[width= 0.14\textwidth]{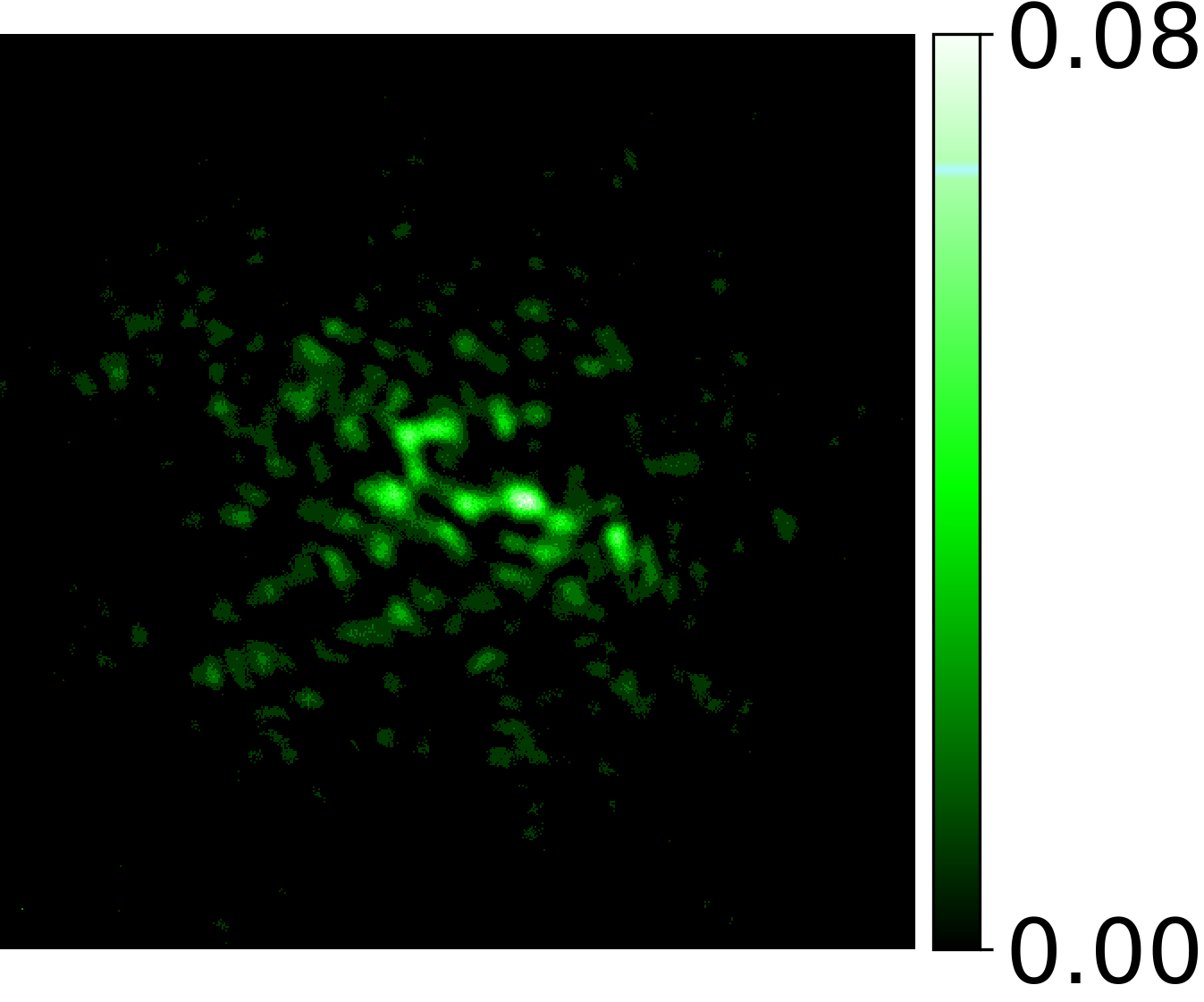}&
			\includegraphics[width= 0.14\textwidth]{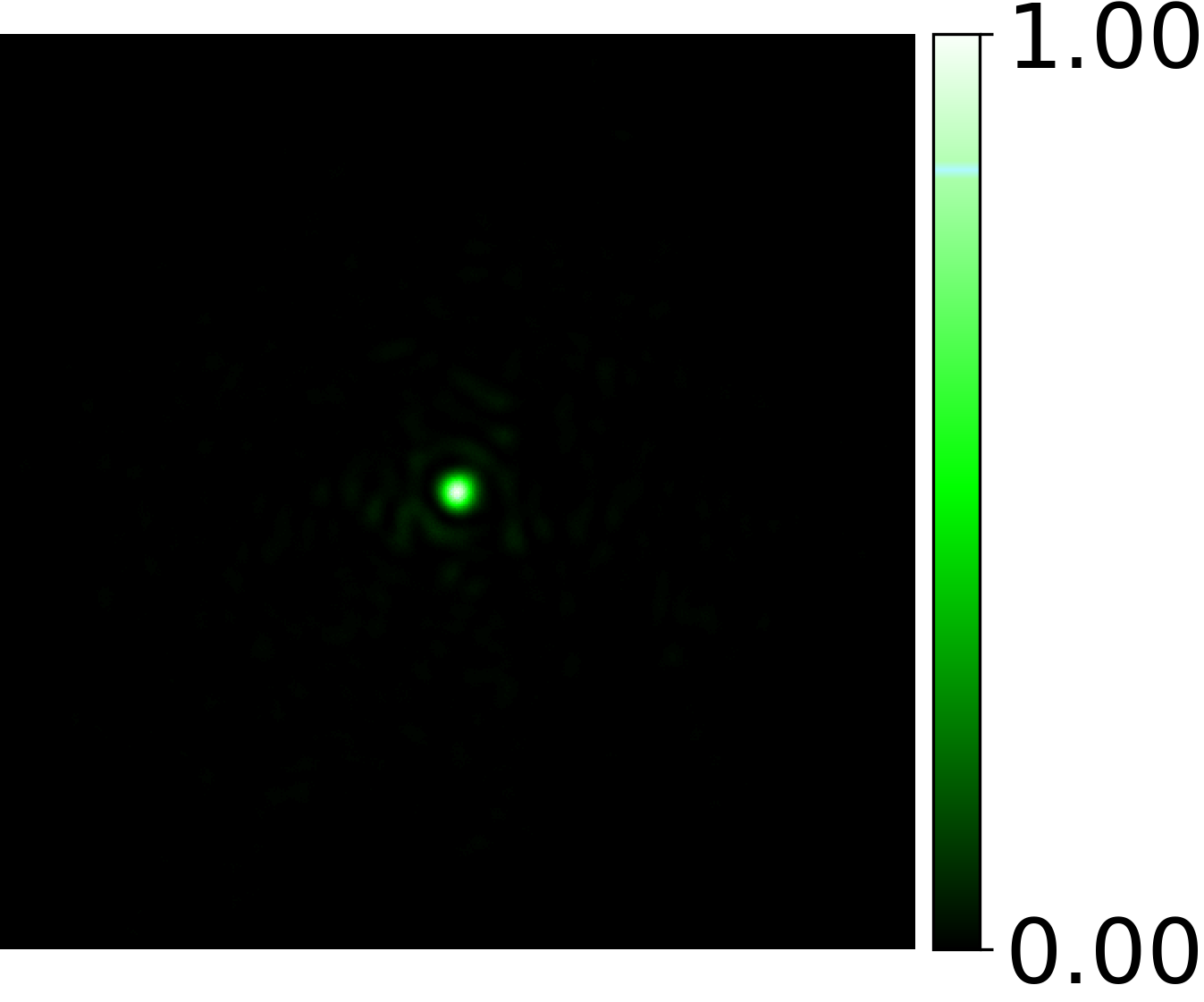}&			
			\includegraphics[width= 0.14\textwidth]{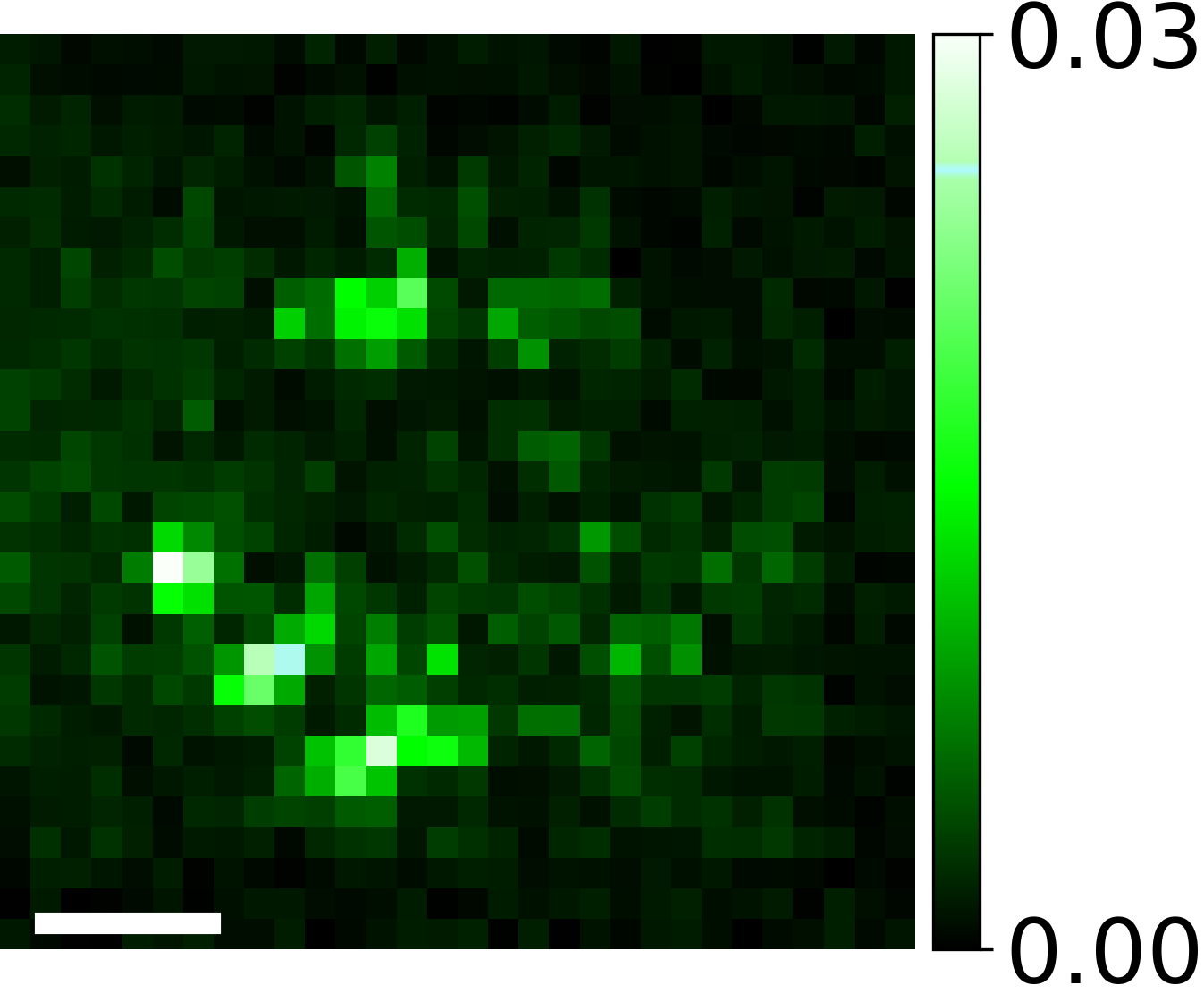}&
			\includegraphics[width= 0.14\textwidth]{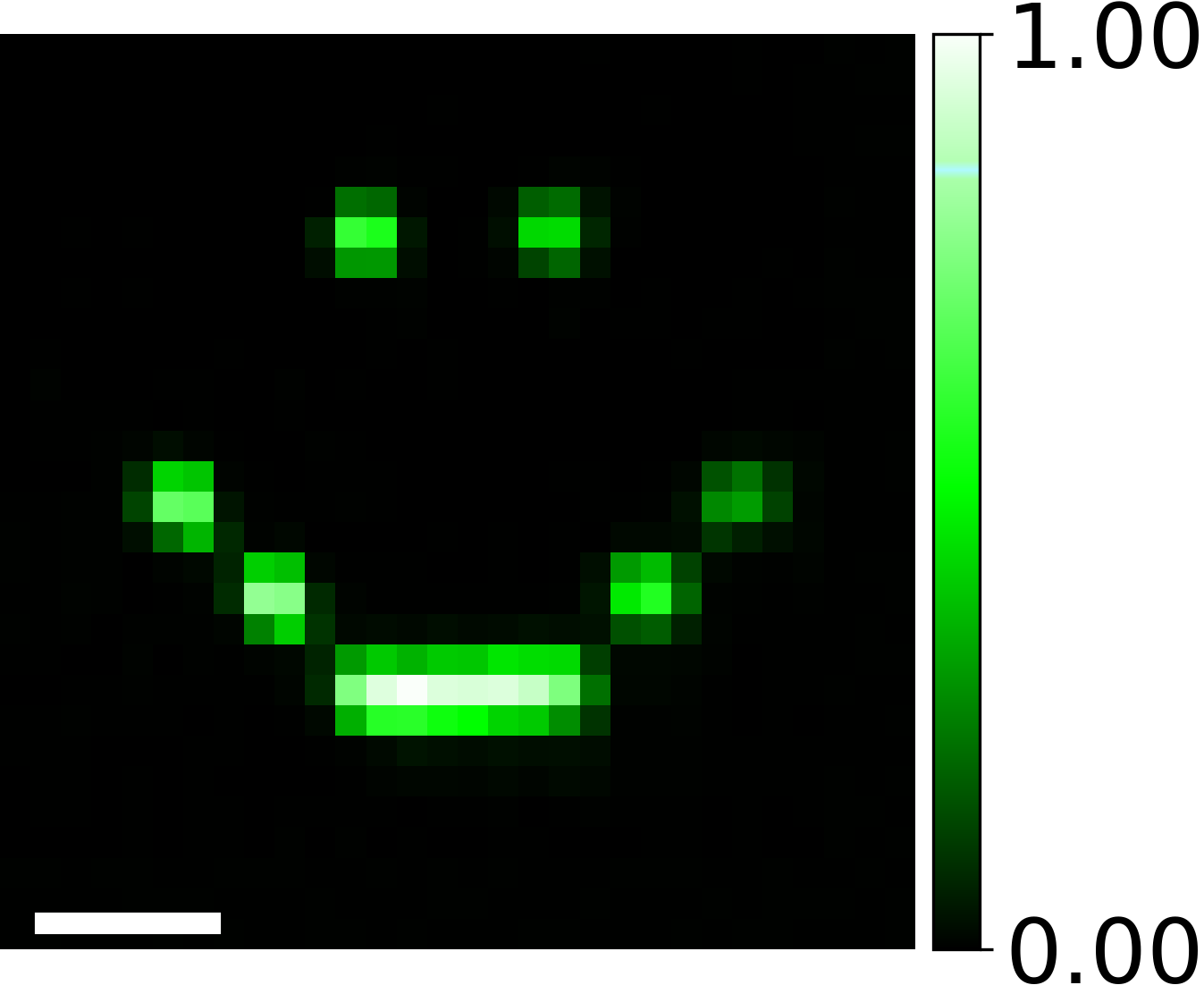}&
			\includegraphics[width= 0.14\textwidth]{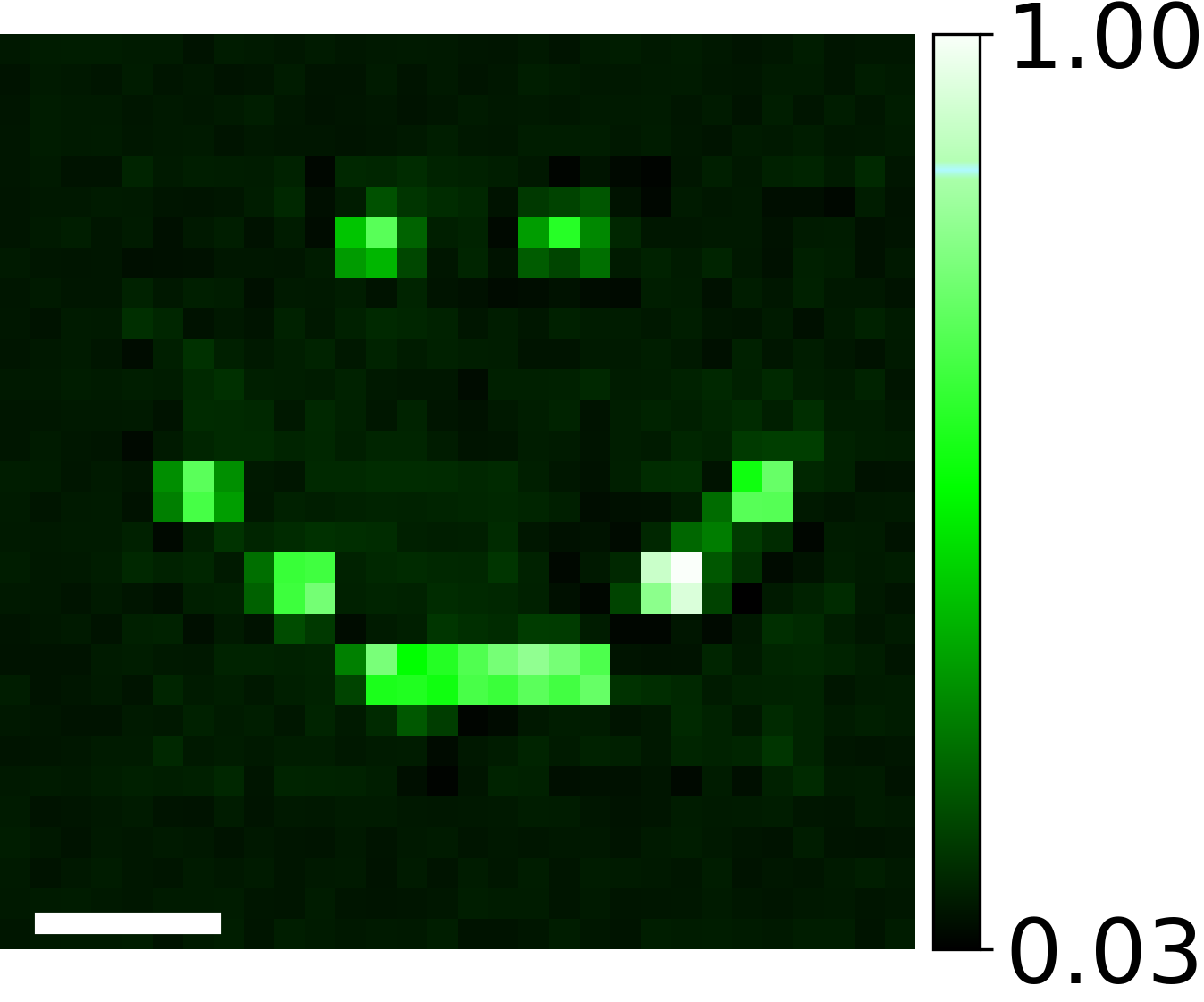}\\
			\includegraphics[width= 0.14\textwidth]{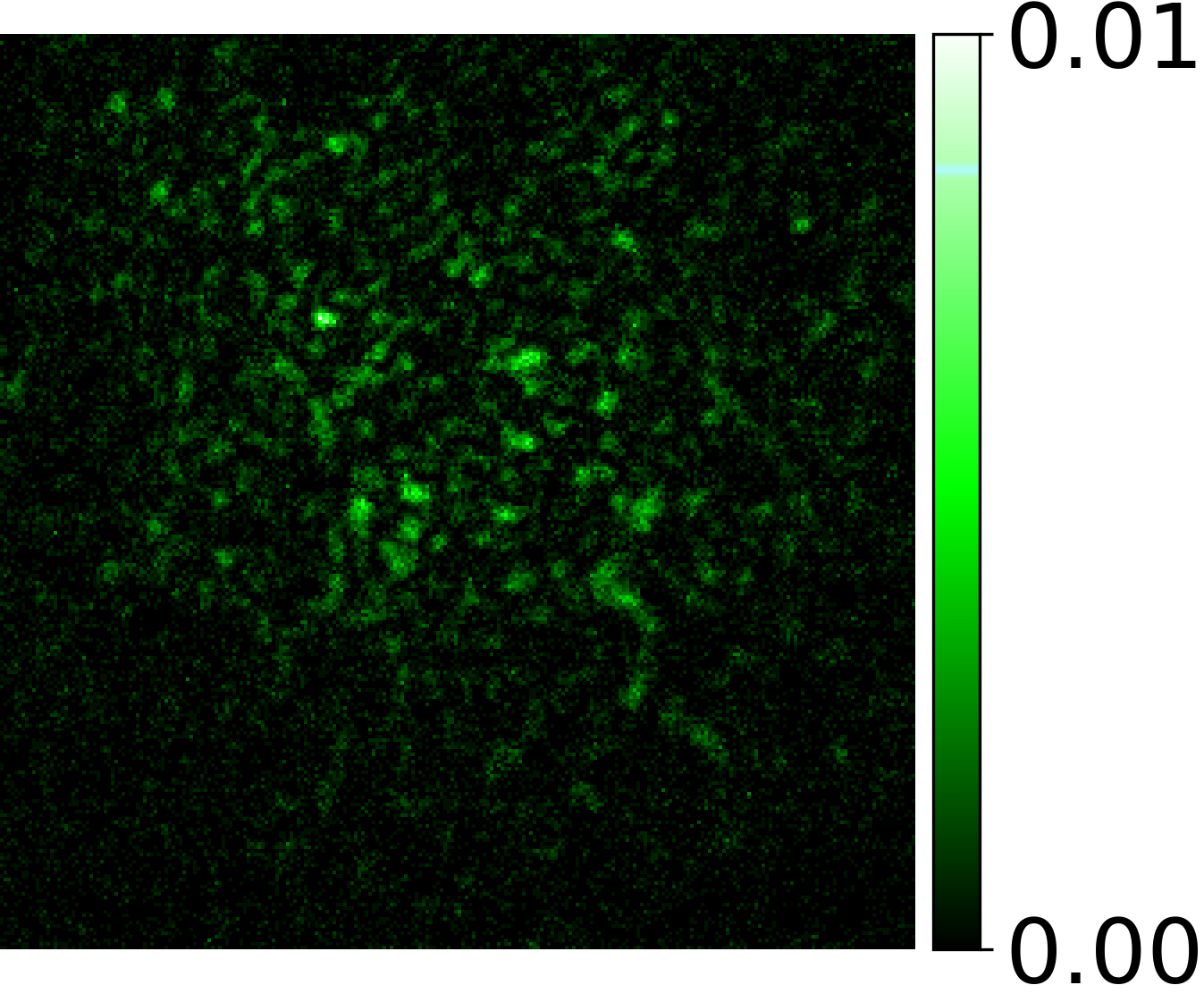}&
			\includegraphics[width= 0.14\textwidth]{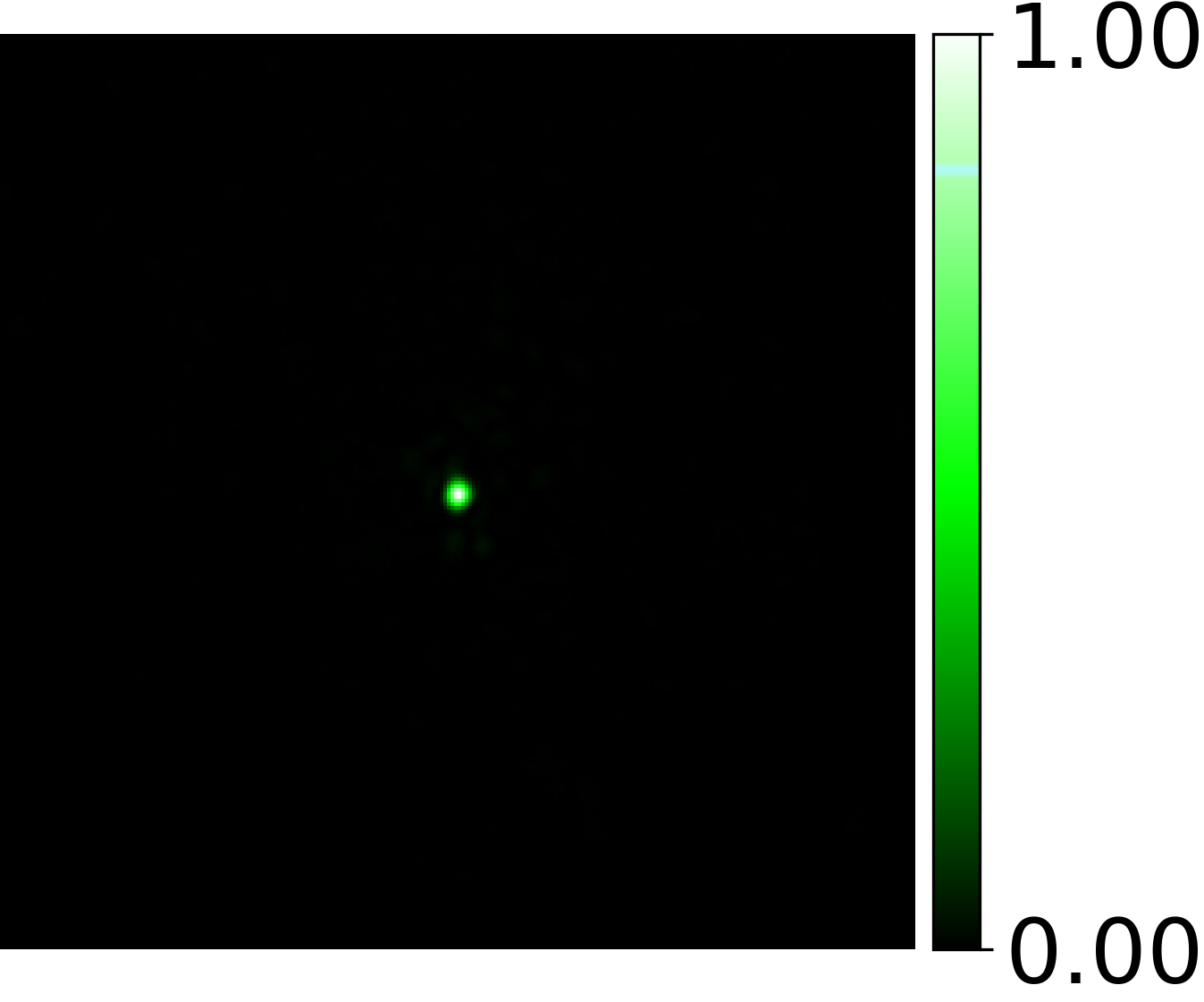}&
			\includegraphics[width= 0.14\textwidth]{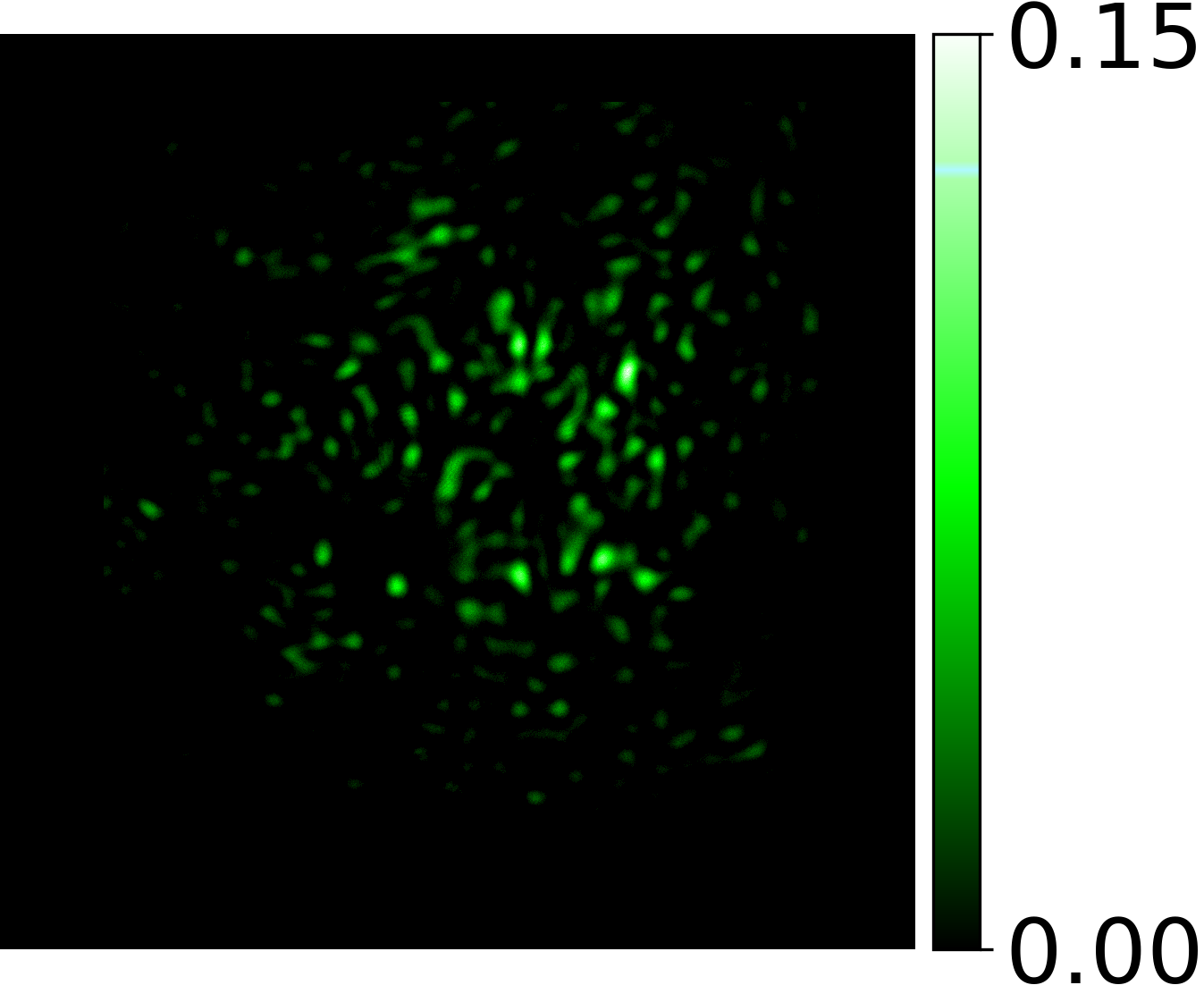}&
			\includegraphics[width= 0.14\textwidth]{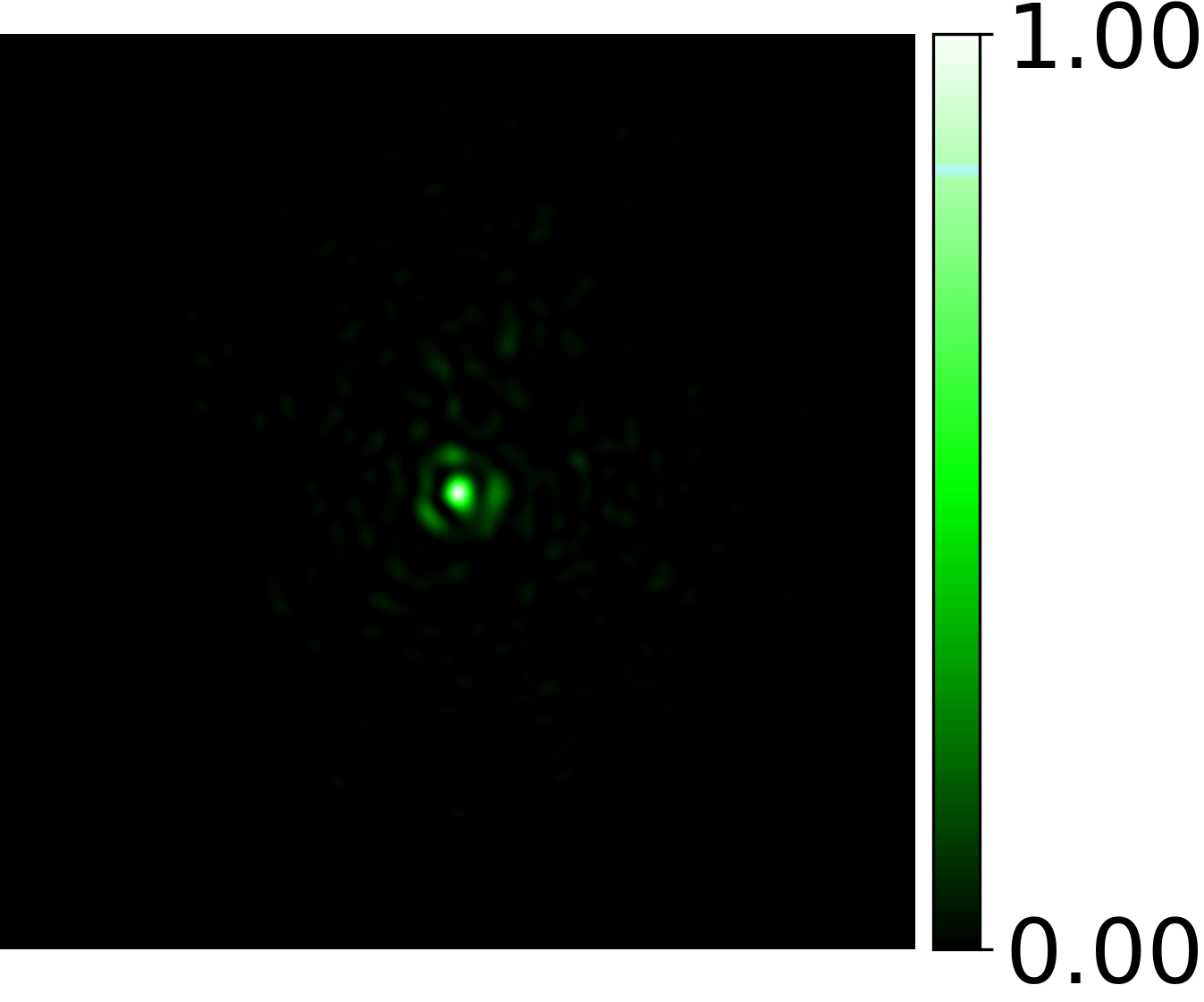}&			
			\includegraphics[width= 0.14\textwidth]{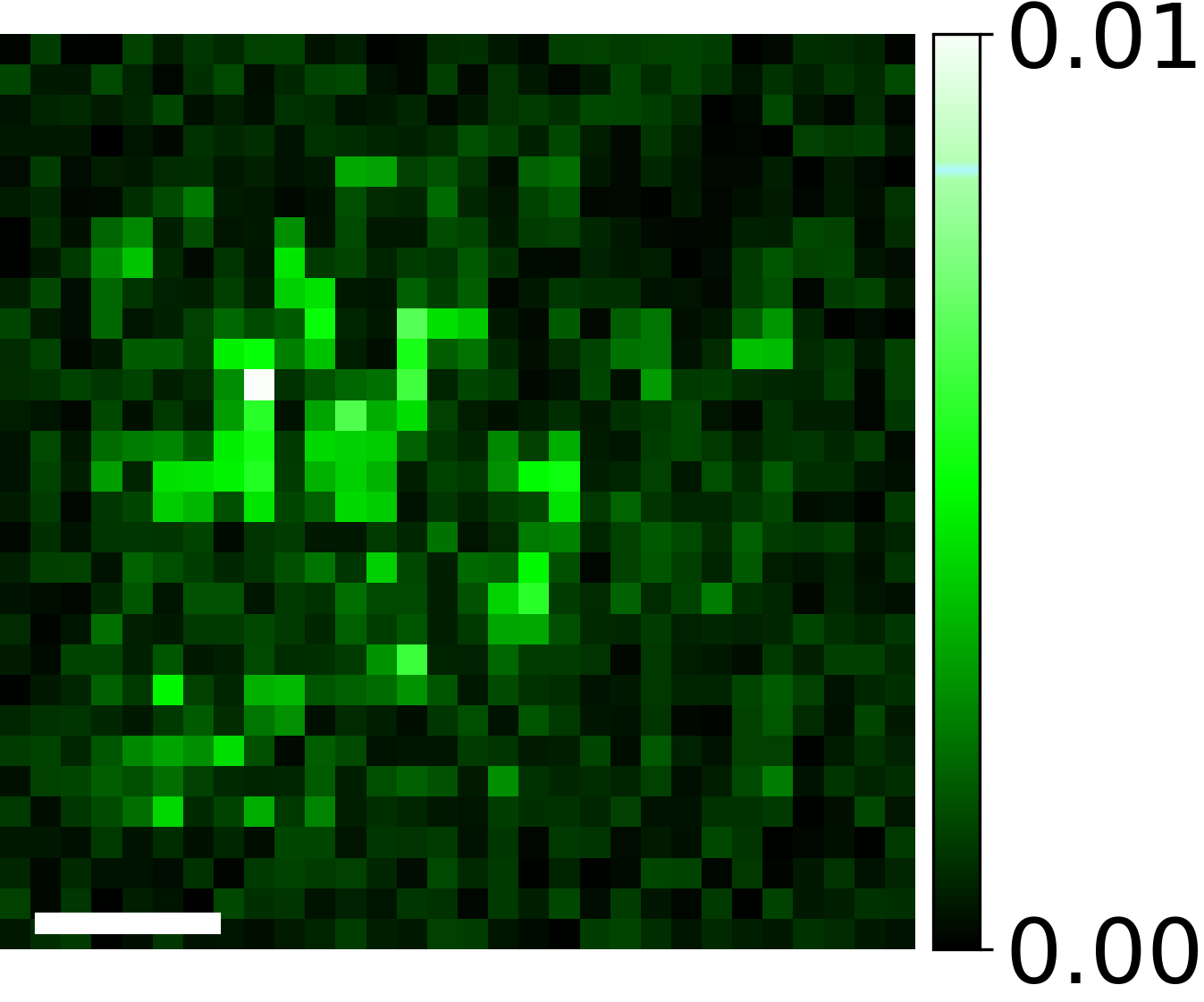}&
			\includegraphics[width= 0.14\textwidth]{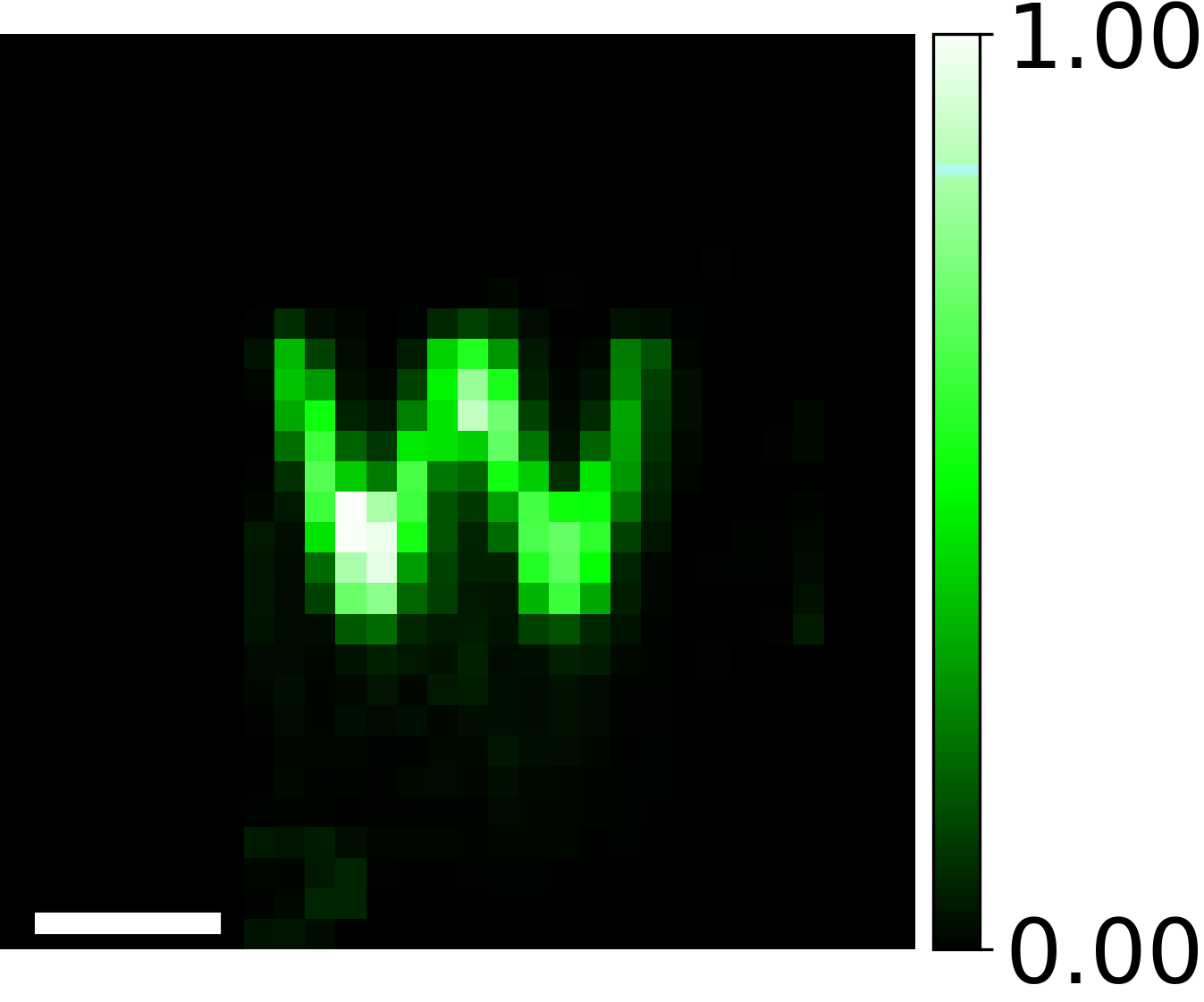}&
			\includegraphics[width= 0.14\textwidth]{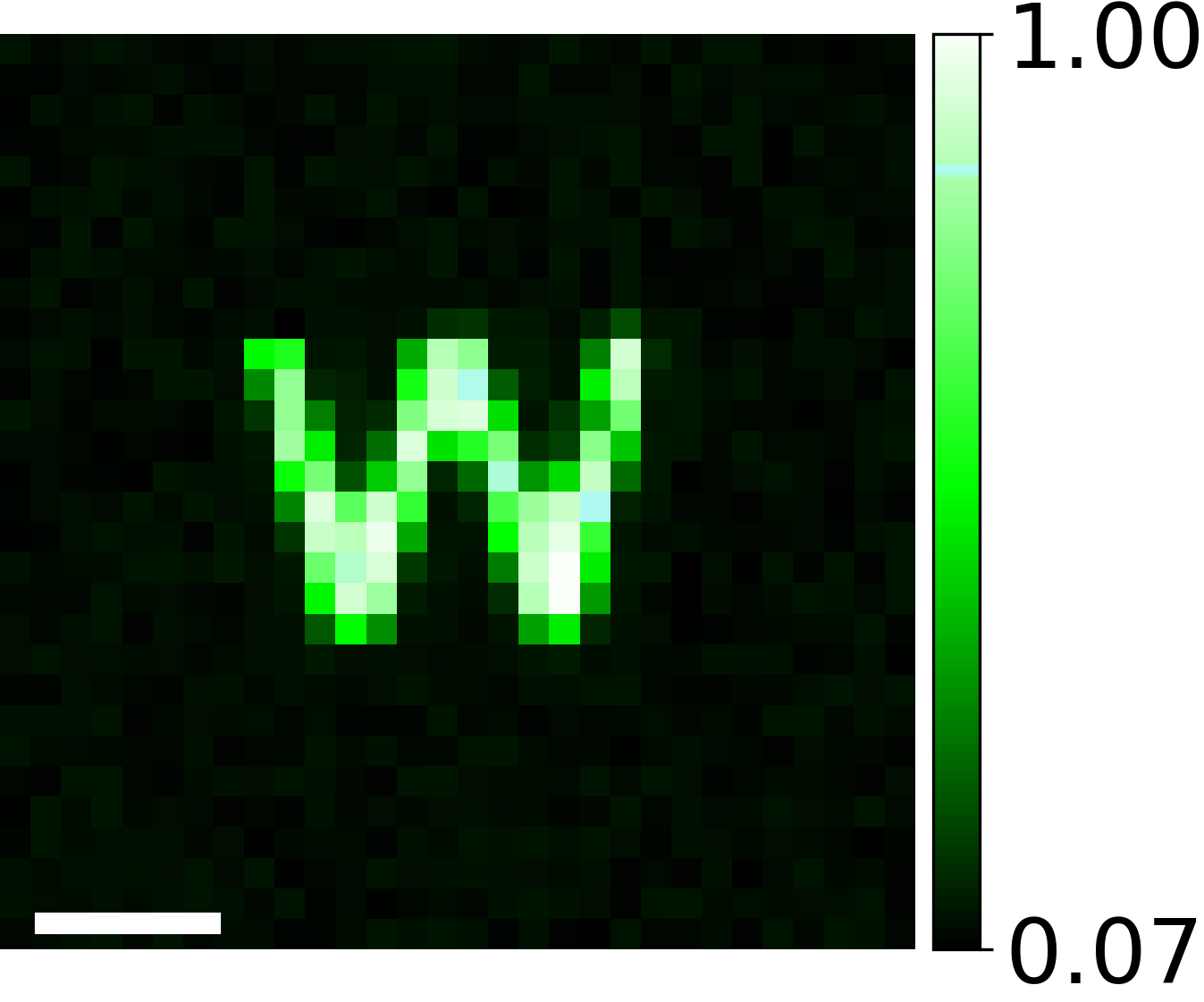}\\
			\includegraphics[width= 0.14\textwidth]{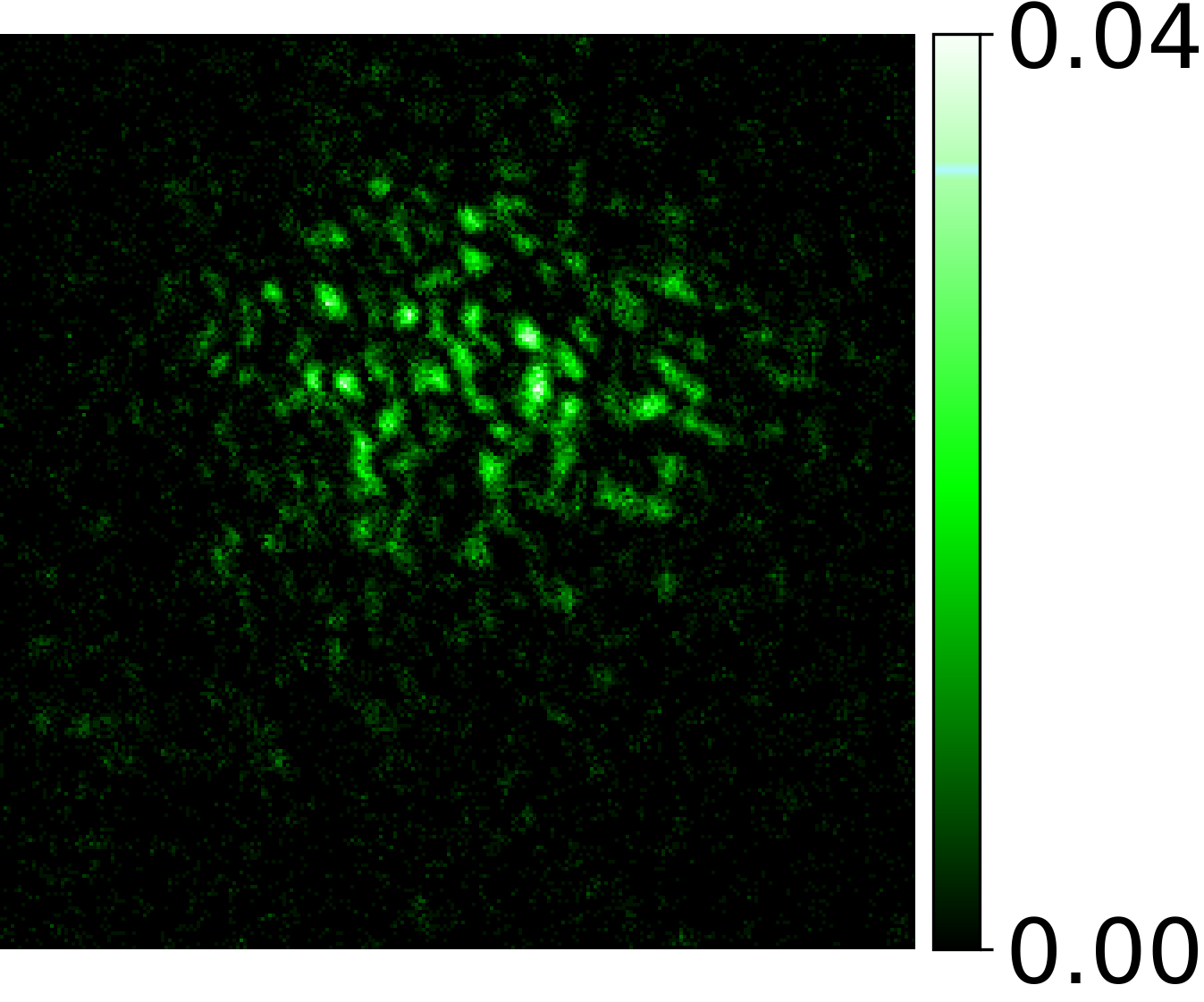}&
			\includegraphics[width= 0.14\textwidth]{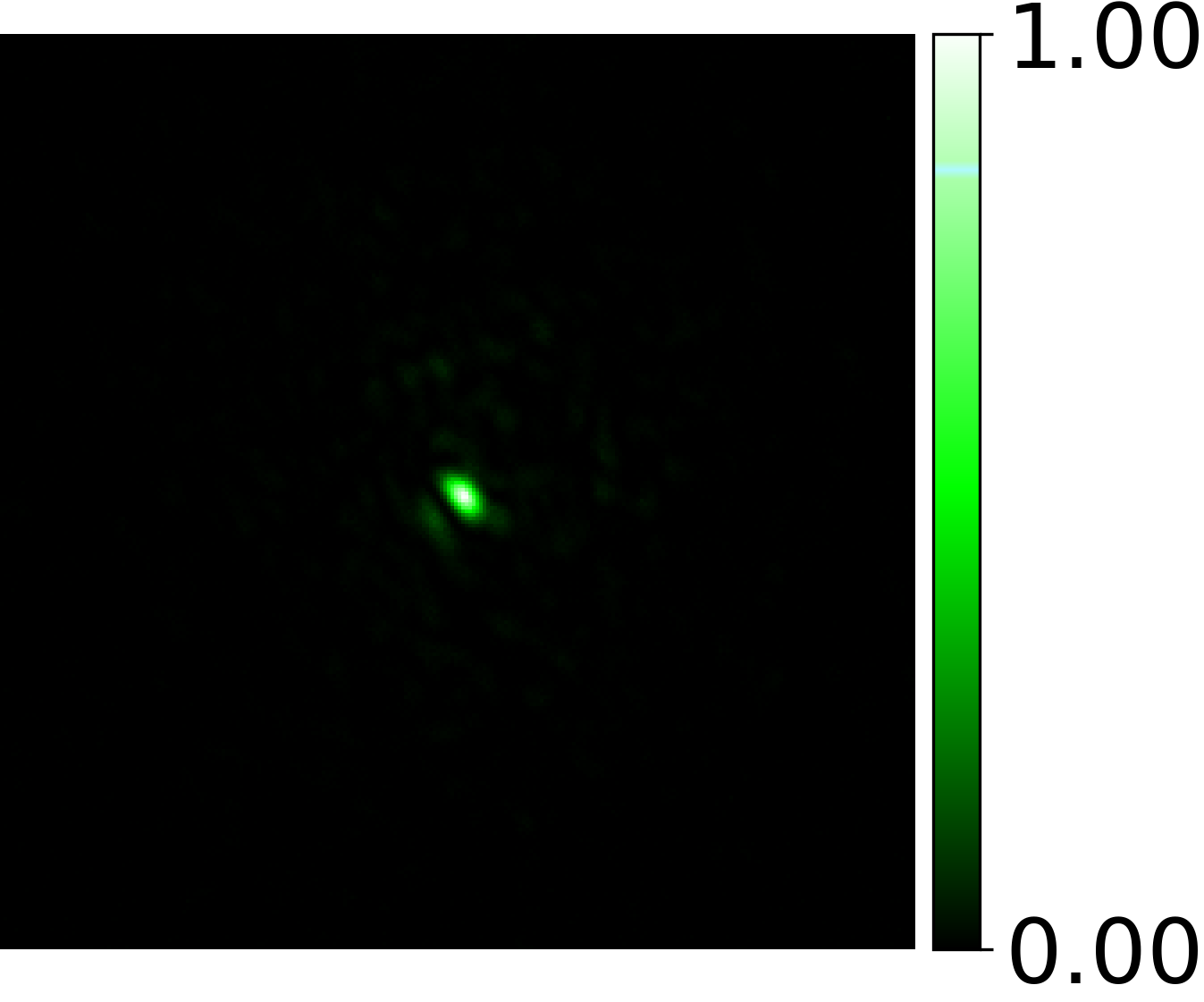}&
			\includegraphics[width= 0.14\textwidth]{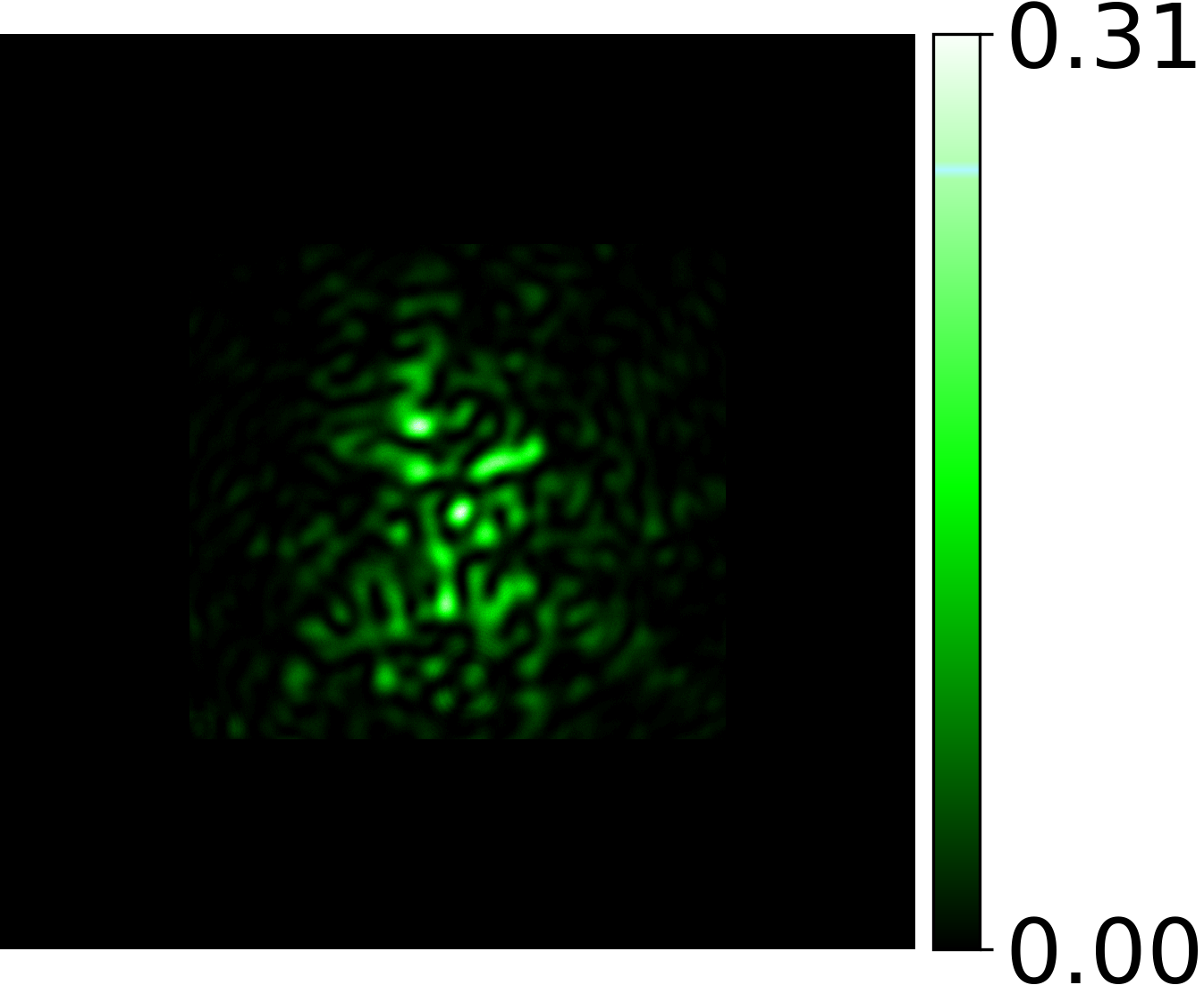}&
			\includegraphics[width= 0.14\textwidth]{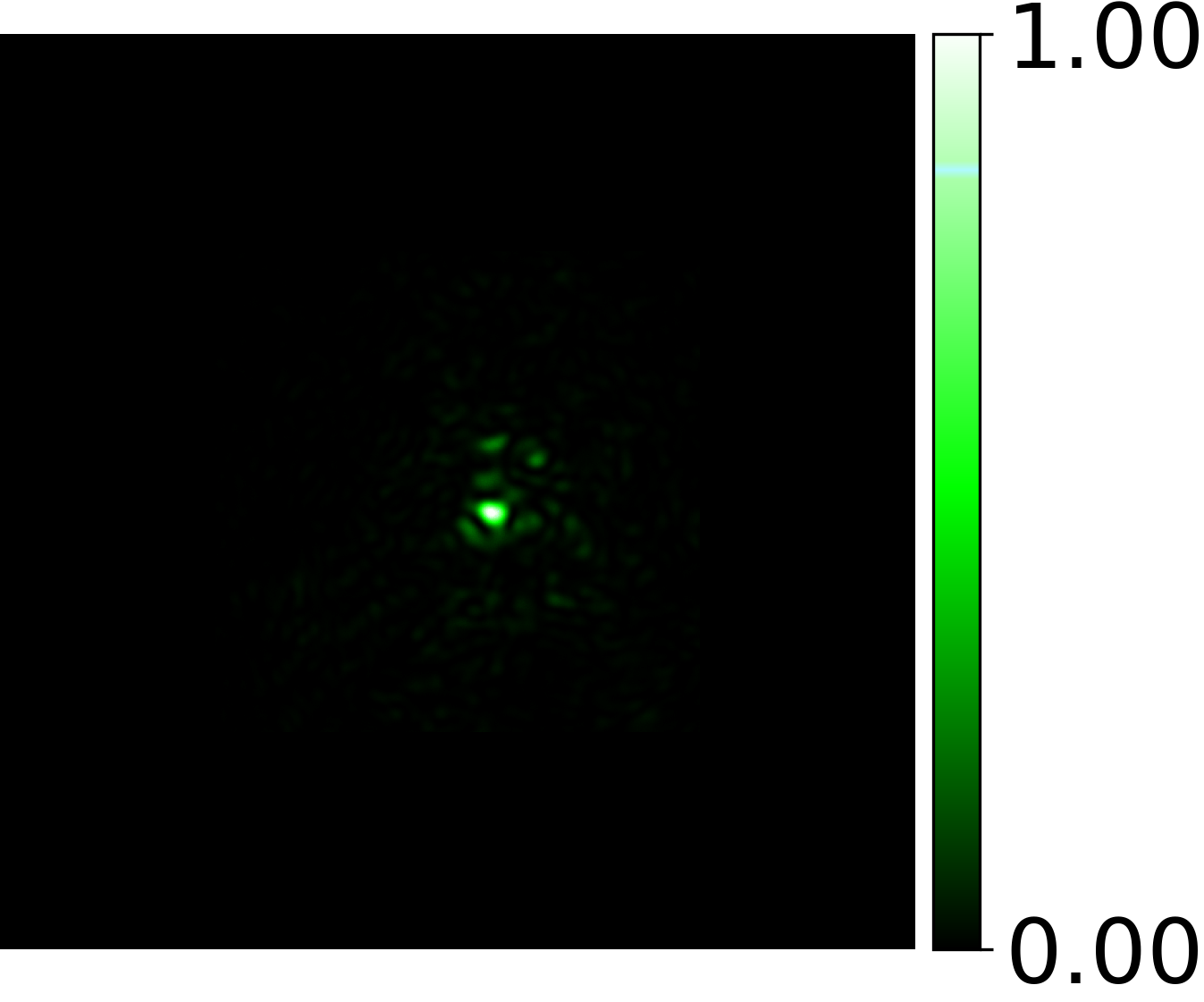}&
			\includegraphics[width= 0.14\textwidth]{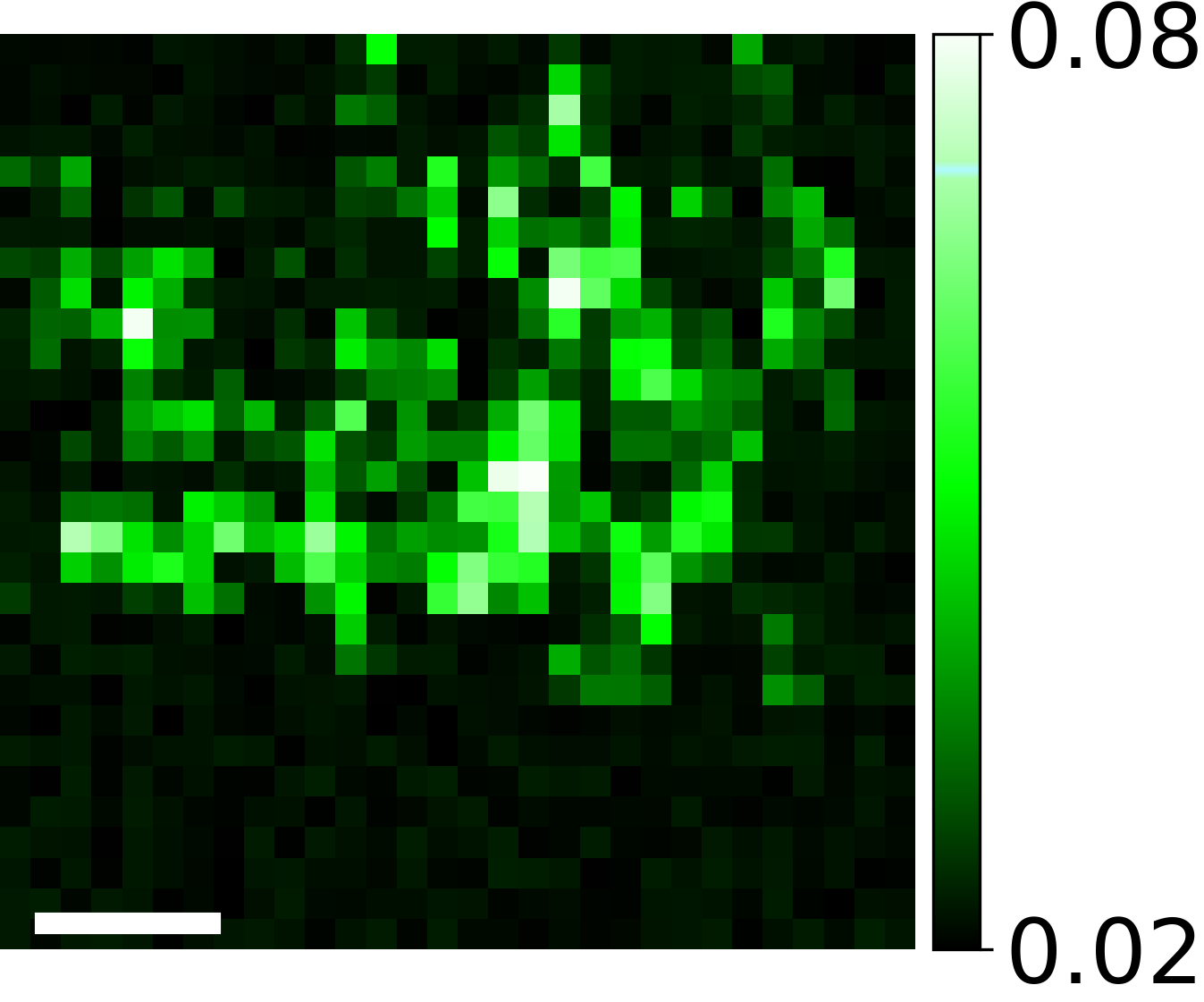}&
			\includegraphics[width= 0.14\textwidth]{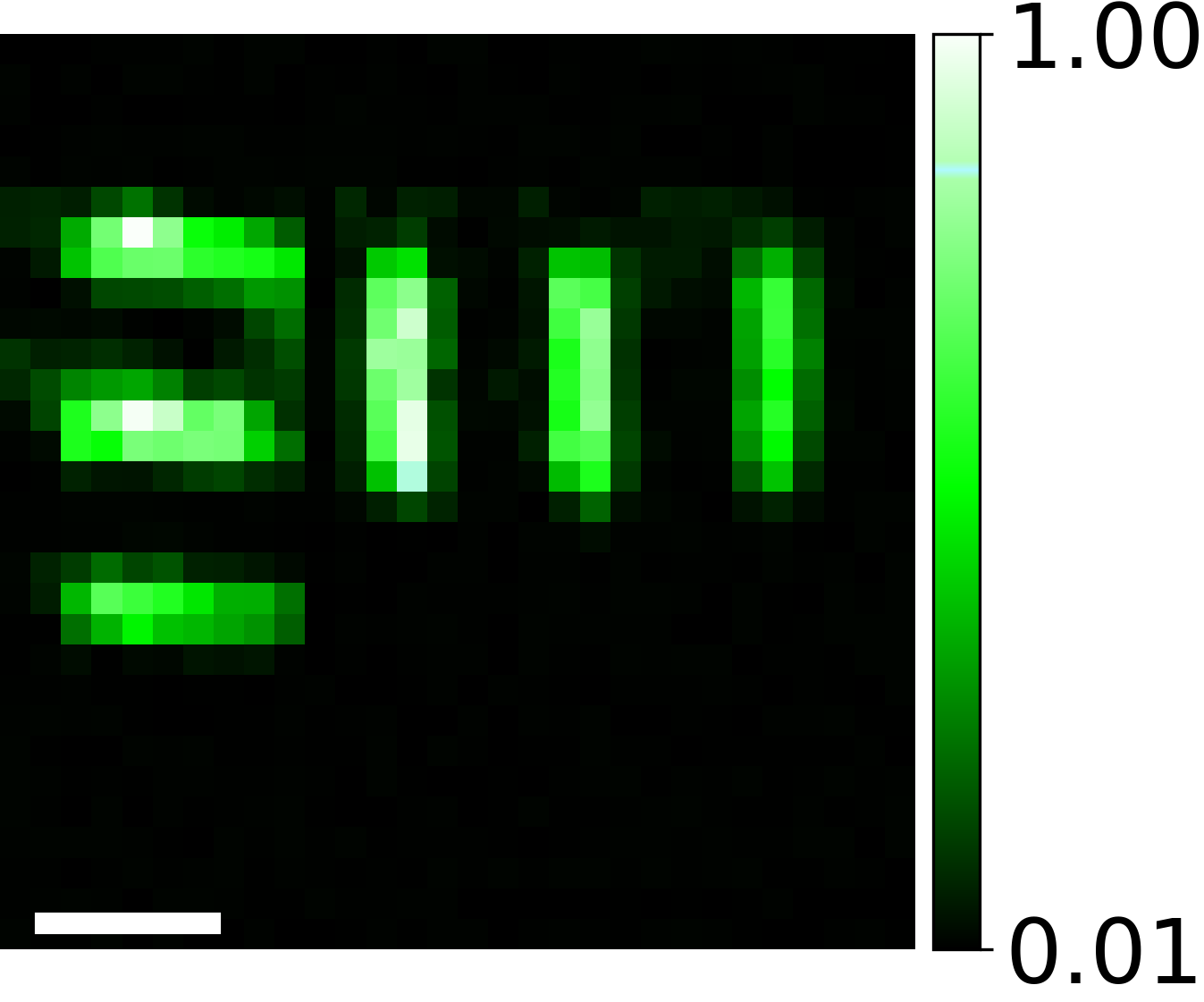}&
			\includegraphics[width= 0.14\textwidth]{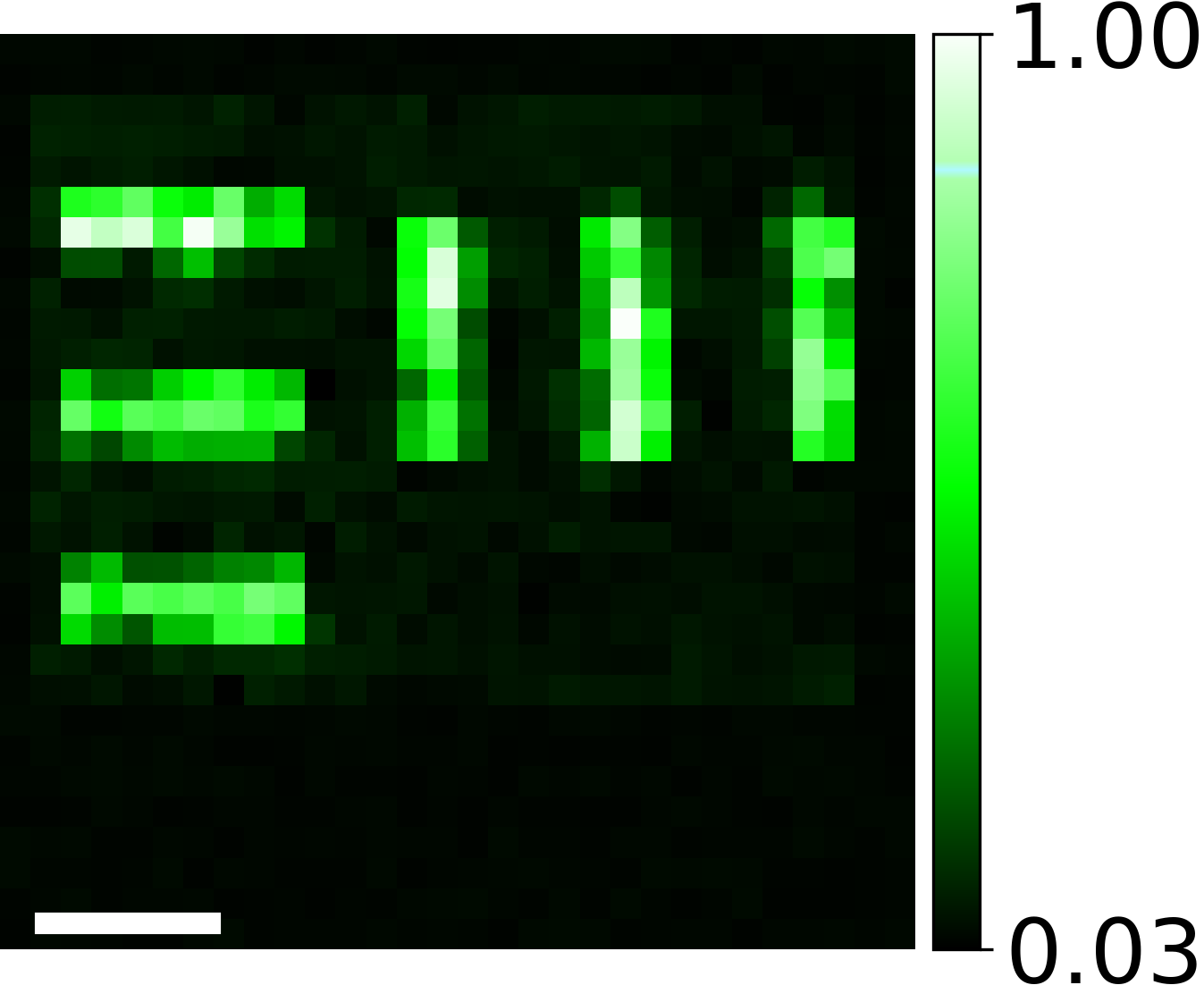} \\ 
		\end{tabular}
		\caption{\textbf{Confocal imaging:} of patterns printed on chrome mask, through a scattering layer.  Our algorithm successfully focuses light on the desired plane. Each row shows a different target. Columns 1-2: Main camera images of a single focal spot before and after optimization. Columns 3-4: Corresponding validation camera images. Columns 5-7: Confocal scanning results: Without aberration correction, with aberration correction, and reference scan of the target before applying scattering material, respectively. In the first three targets (top  rows) the scattering material is chicken breast of thickness of $130\um$, $150\um$ and $200\um$, respectively. In the fourth target the scattering material is two layers of parafilm. Scale bar on confocal images is $4\um$.}
	\label{fig:res-mask}
	\end{center}
\end{figure*}

%% file: fig_beads_3D_1.tex
\begin{figure*}[t!]
	\begin{center}		
		\begin{tabular}{@{}c@{~~~~~~~~~~~~~~~~~~}c@{~}c@{~}c@{~}c@{~}}			
			\multicolumn{1}{c}{\hspace{-0.6cm}\large Target }&&
			\multicolumn{3}{c}{\hspace{-0.6cm}\large Cross section}\\
			\multicolumn{1}{c}{\hspace{-0.6cm}\large illustration }&&
			\multicolumn{2}{c}{\hspace{-0.6cm} Main cam.} &
			\multicolumn{1}{c}{\hspace{-0.6cm} Valid. cam.} \\
			&&
			\multicolumn{1}{c}{\hspace{-0.6cm} \scriptsize w/o}&	
			\multicolumn{1}{c}{\hspace{-0.6cm} \scriptsize w/ }&	
			\multicolumn{1}{c}{\hspace{-0.6cm} \scriptsize Incoherent}\vspace{-0cm}\\
			&&
			\multicolumn{1}{c}{\hspace{-0.6cm} \scriptsize modulation}&	
			\multicolumn{1}{c}{\hspace{-0.6cm} \scriptsize modulation}&	
			\multicolumn{1}{c}{\hspace{-0.6cm} \scriptsize illumination}\\
			
			\multirow{7}{8em}{\hspace{-1cm}\vspace{-1cm} \includegraphics[width= 0.3\textwidth]{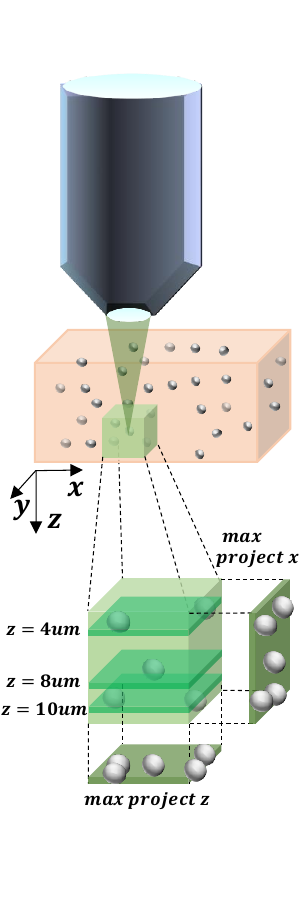}}&
			{\raisebox{0.7cm}	{\rotatebox[origin=c]{90}{~ {\tiny $z=4\um$} }}}&
			\includegraphics[width= 0.15\textwidth]{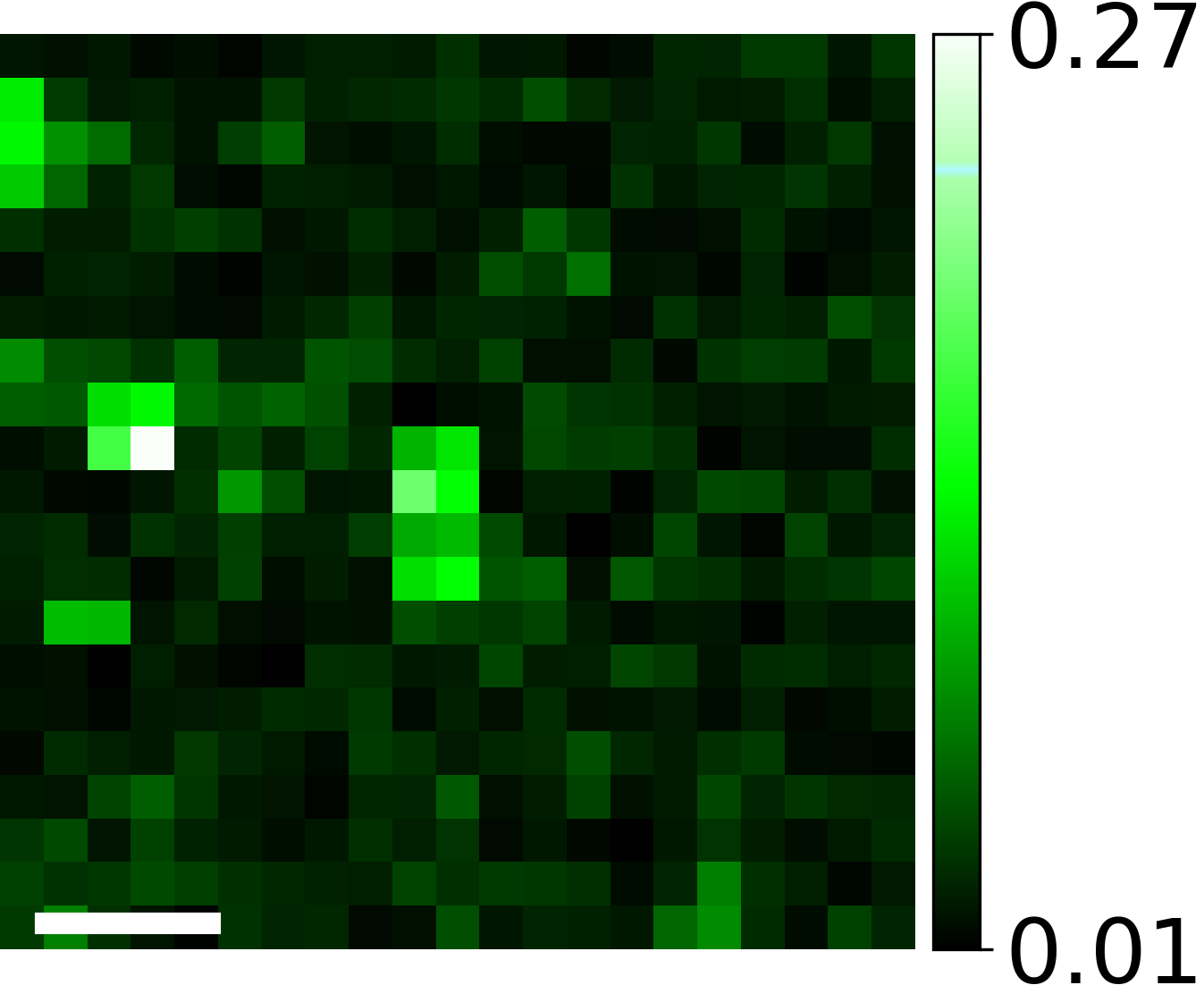}&
			\includegraphics[width= 0.15\textwidth]{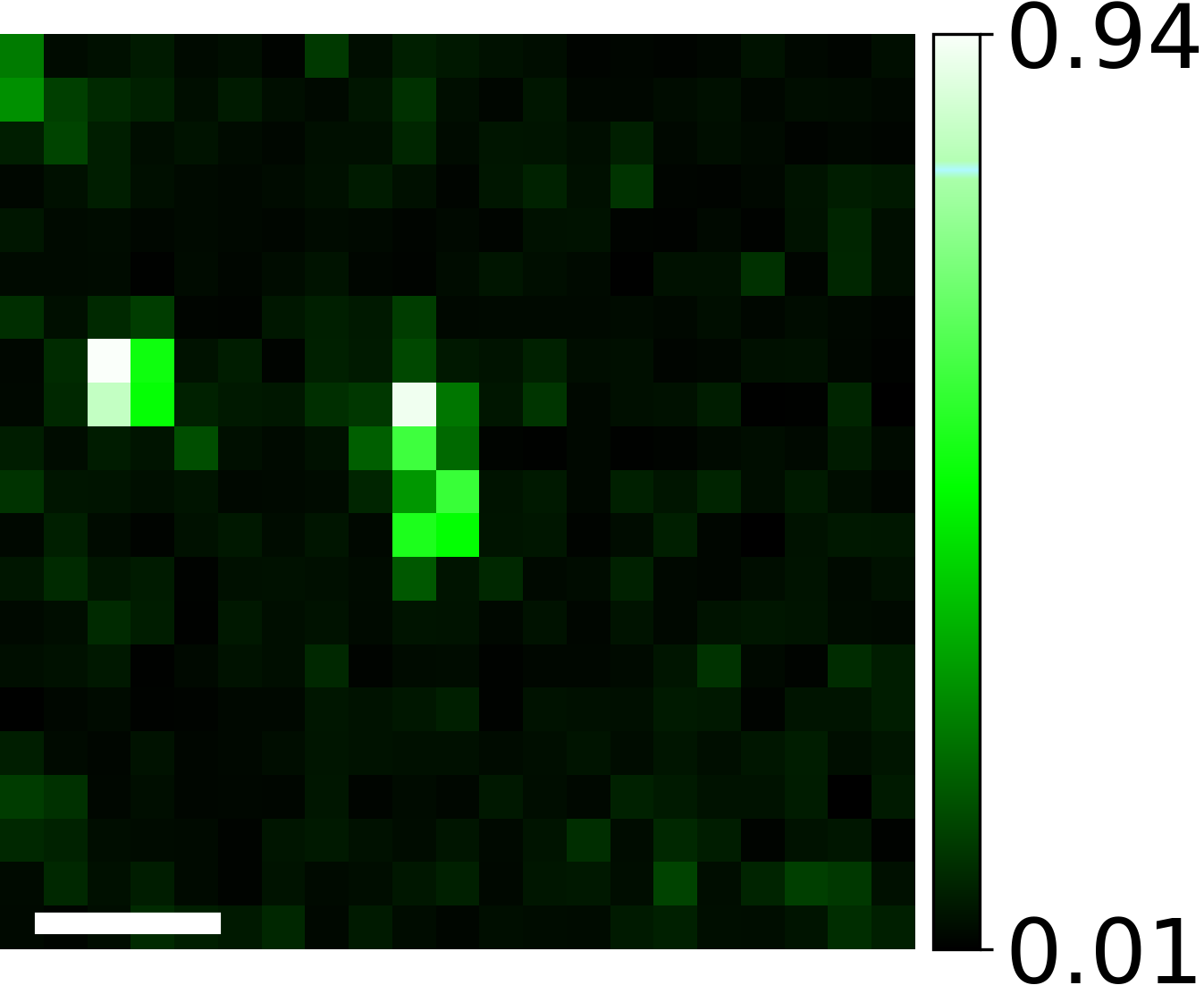}&
			\includegraphics[width= 0.15\textwidth]{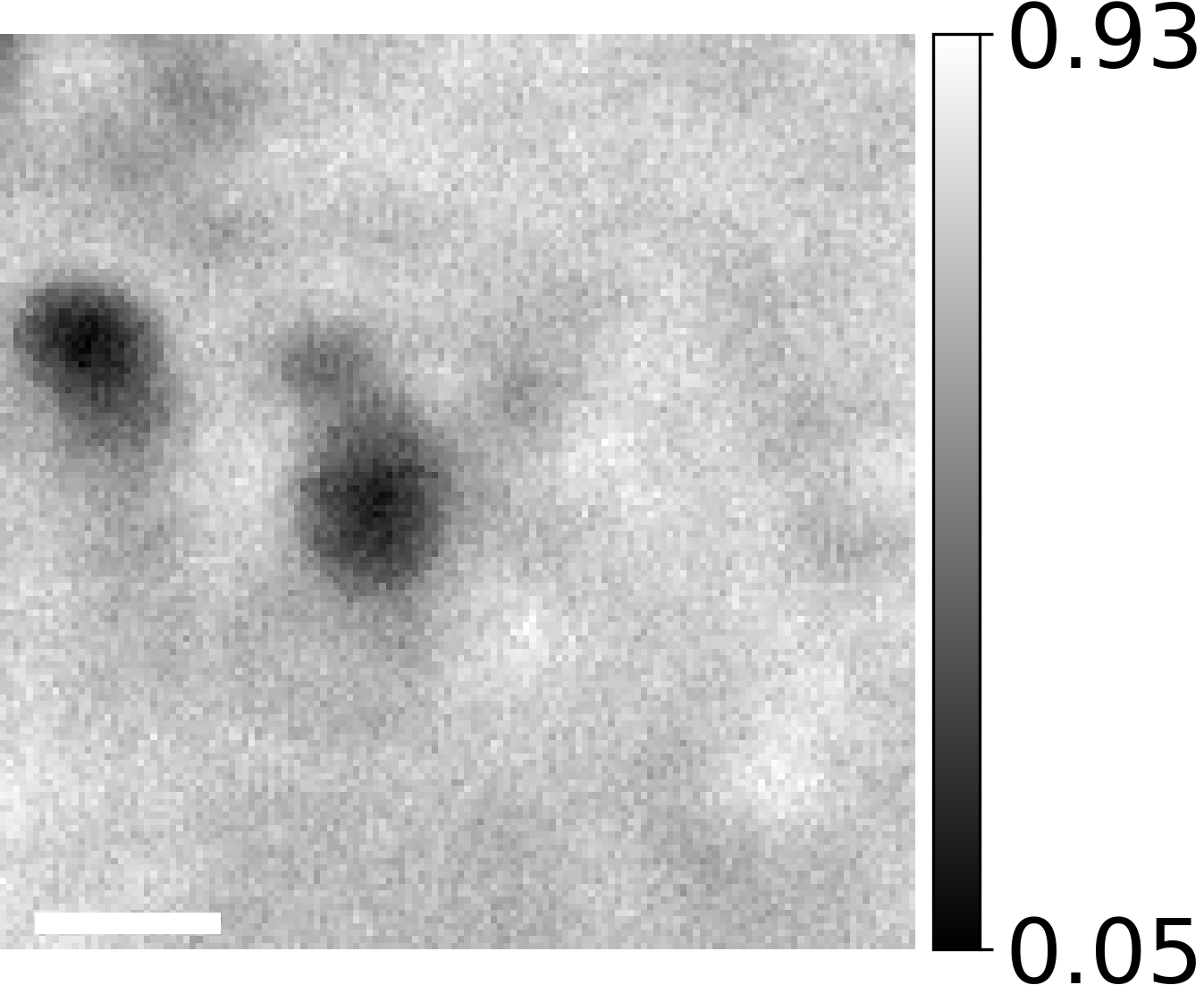}\\
			&{\raisebox{0.7cm}	{\rotatebox[origin=c]{90}{~ {\tiny $z=8\um$} }}}&
			\includegraphics[width= 0.15\textwidth]{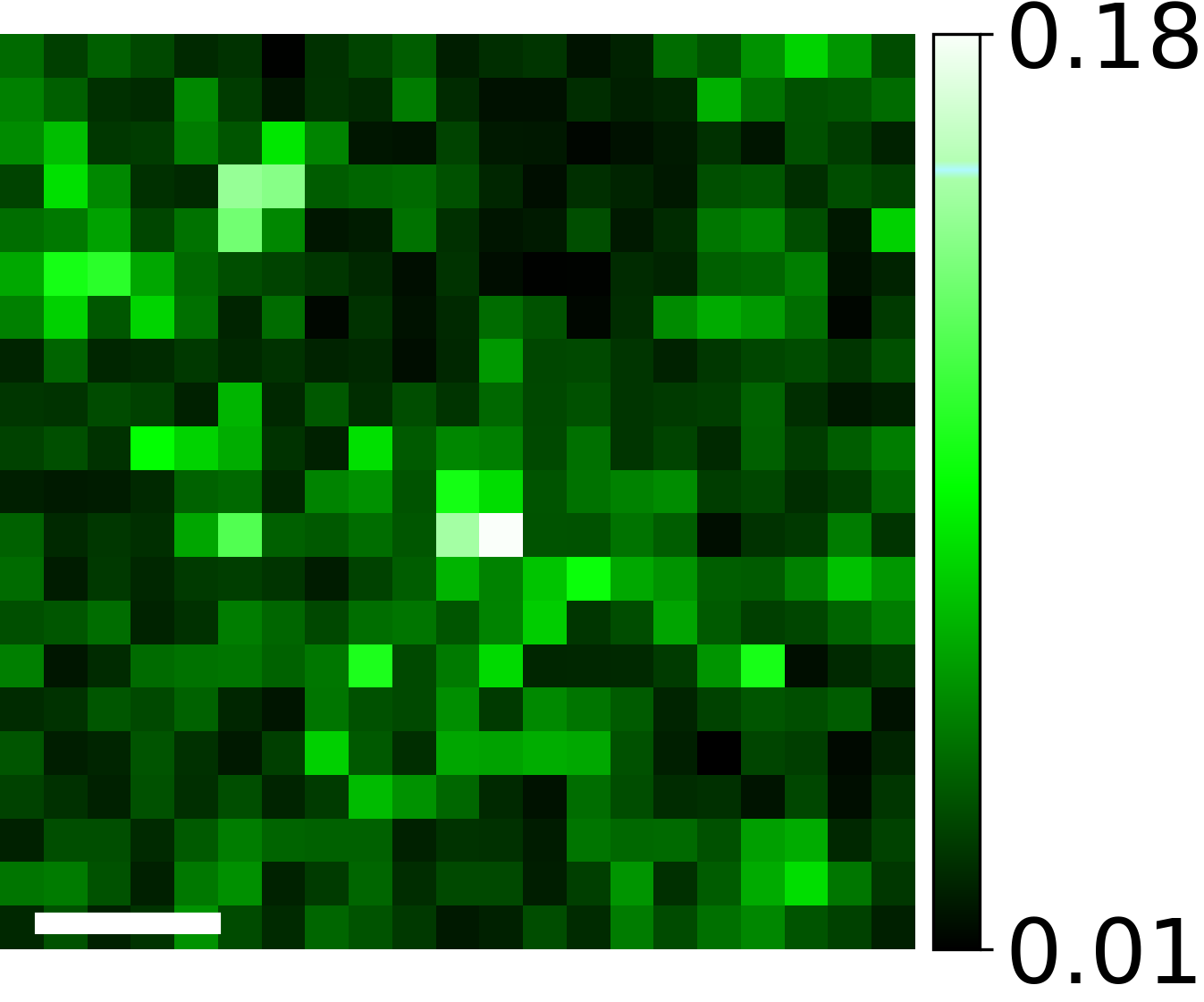}&
			\includegraphics[width= 0.15\textwidth]{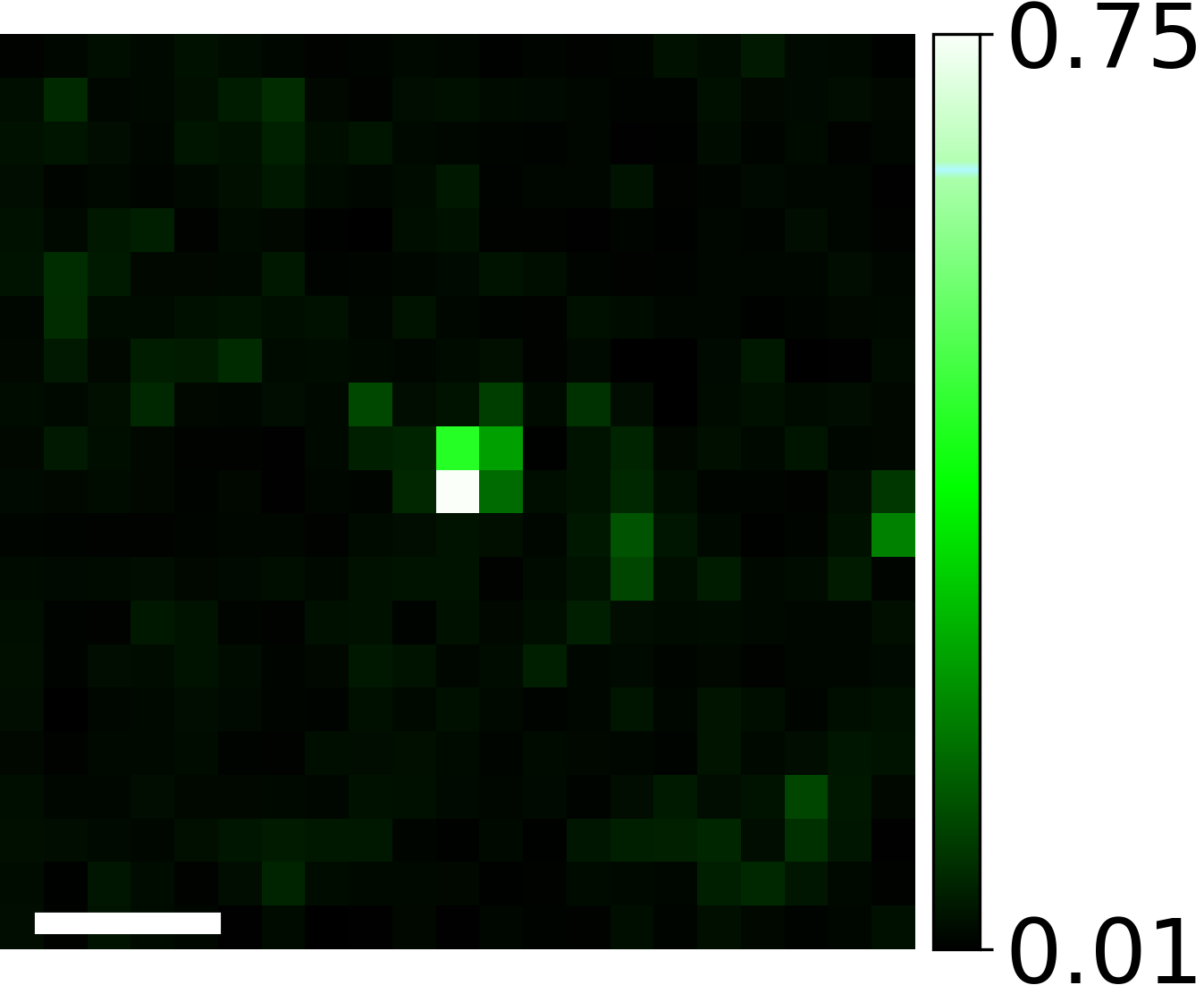}&
			\includegraphics[width= 0.15\textwidth]{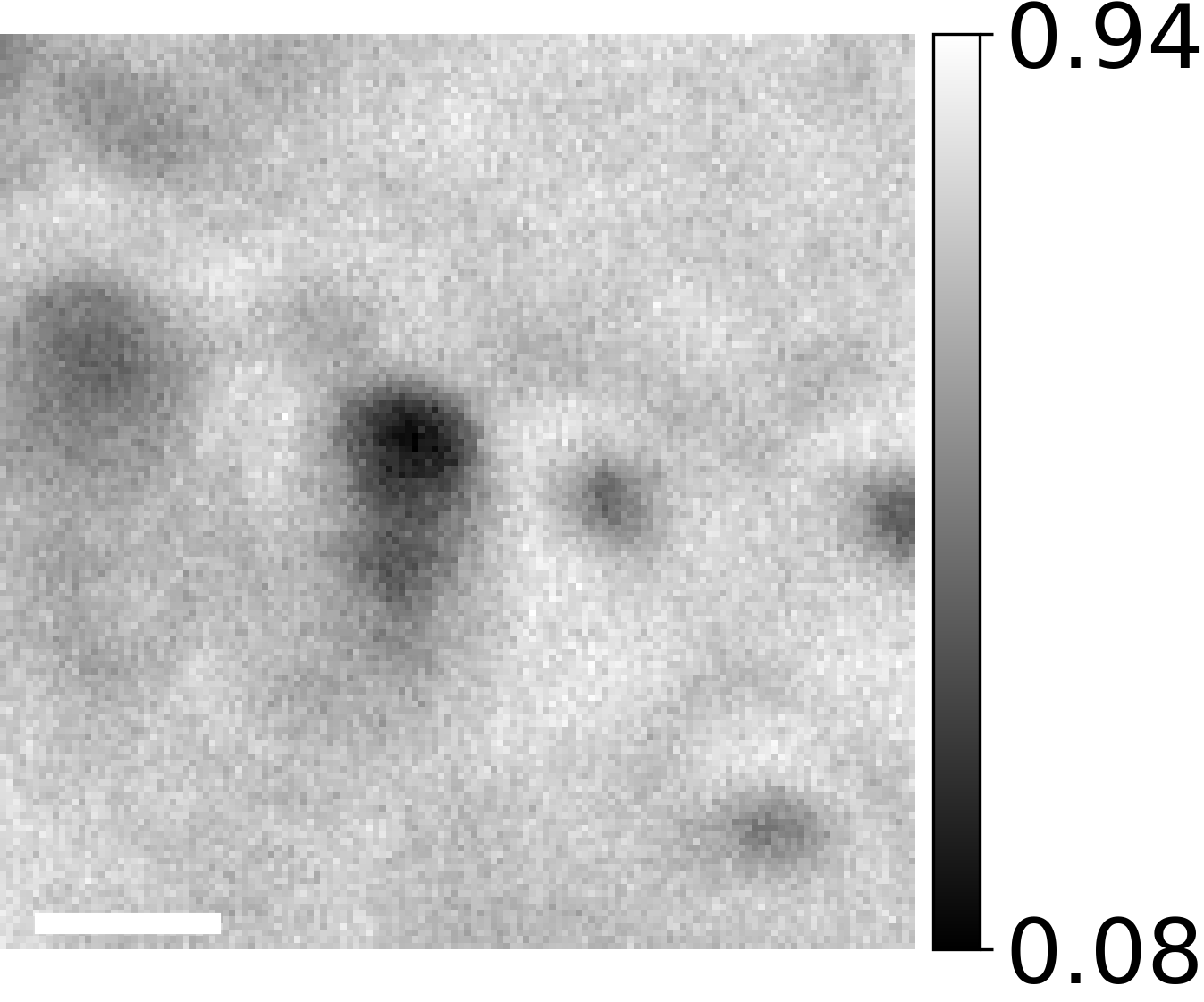}\\
			&{\raisebox{0.7cm}	{\rotatebox[origin=c]{90}{~ {\tiny $z=10\um$} }}}&
			\includegraphics[width= 0.15\textwidth]{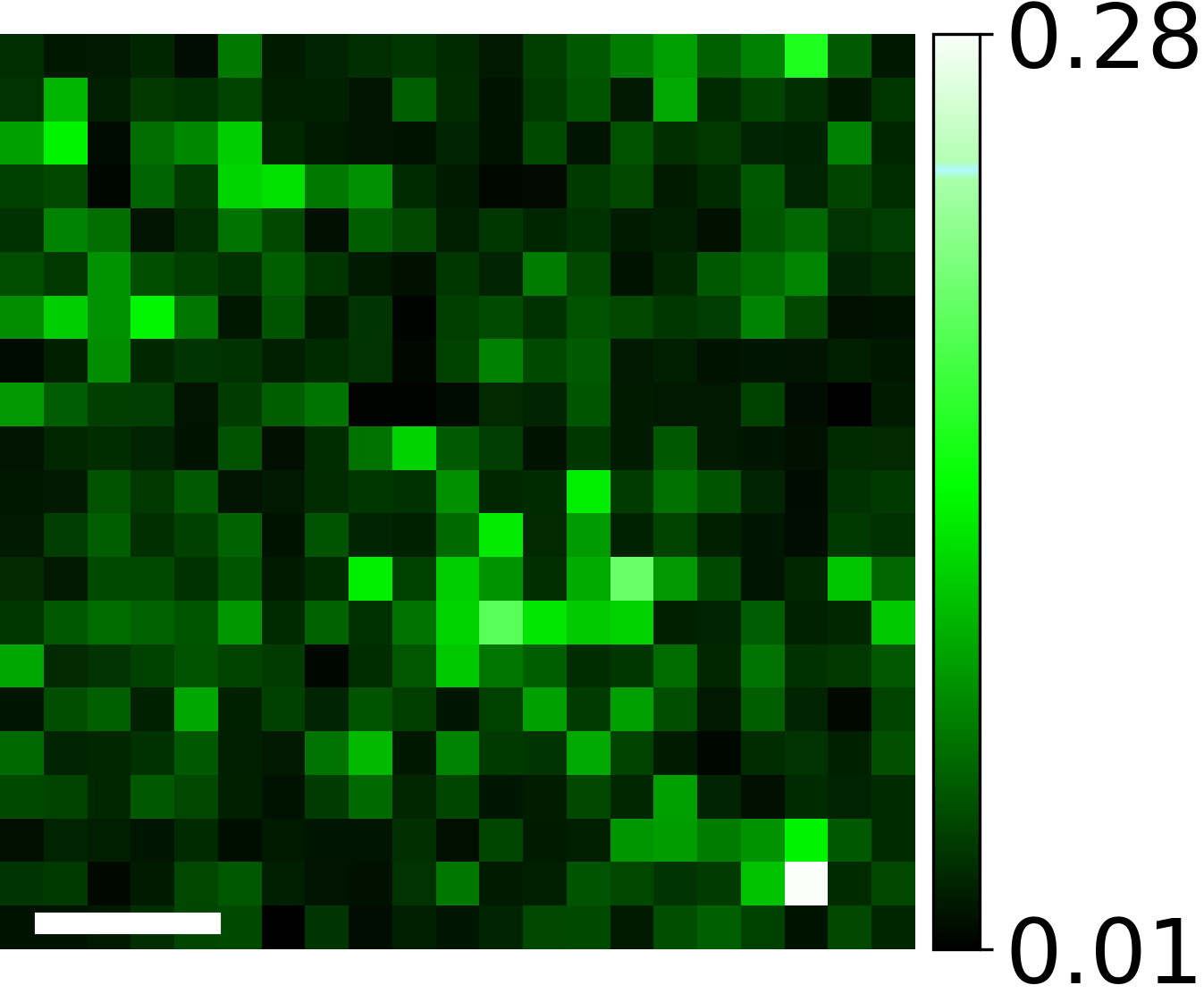}&
			\includegraphics[width= 0.15\textwidth]{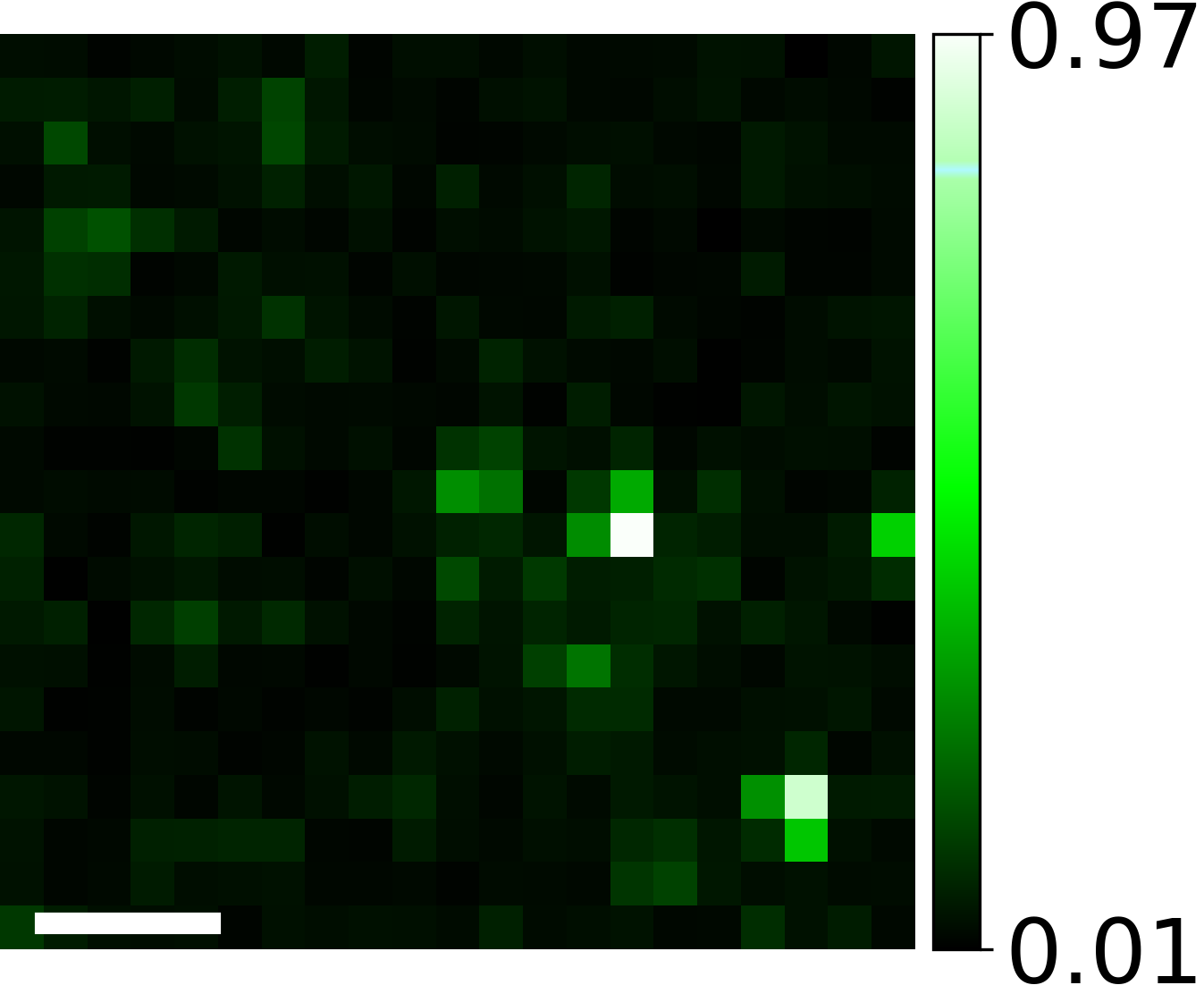}&
			\includegraphics[width= 0.15\textwidth]{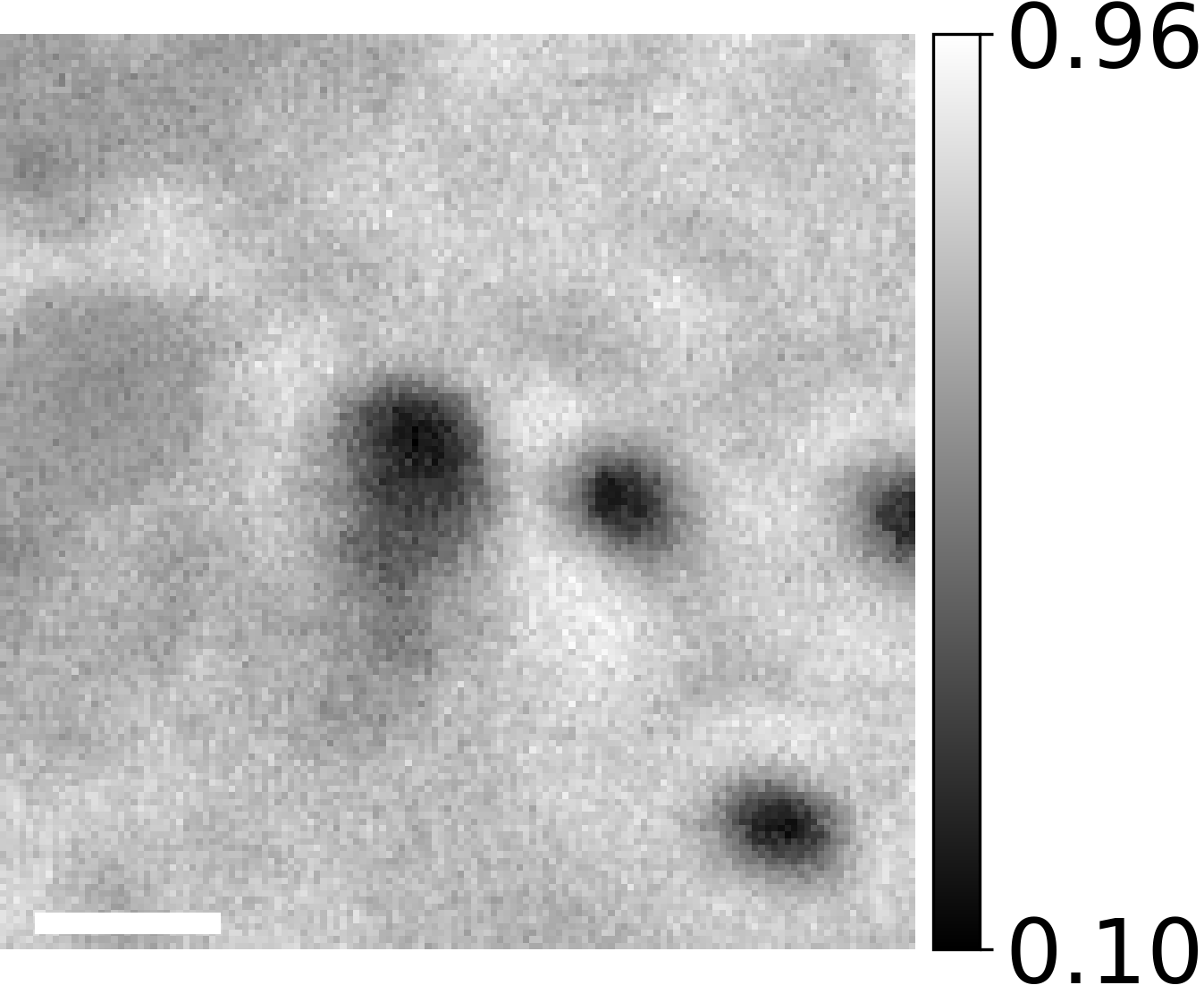}\vspace{-0.0cm}\\
								
			&&\multicolumn{3}{c}{\hspace{-0.6cm}\large Max projection}\vspace{-0.0cm}\\
			&{\raisebox{0.65cm}	{\rotatebox[origin=c]{90}{~ {\tiny Max projection $z$} }}}&
			\includegraphics[width= 0.15\textwidth]{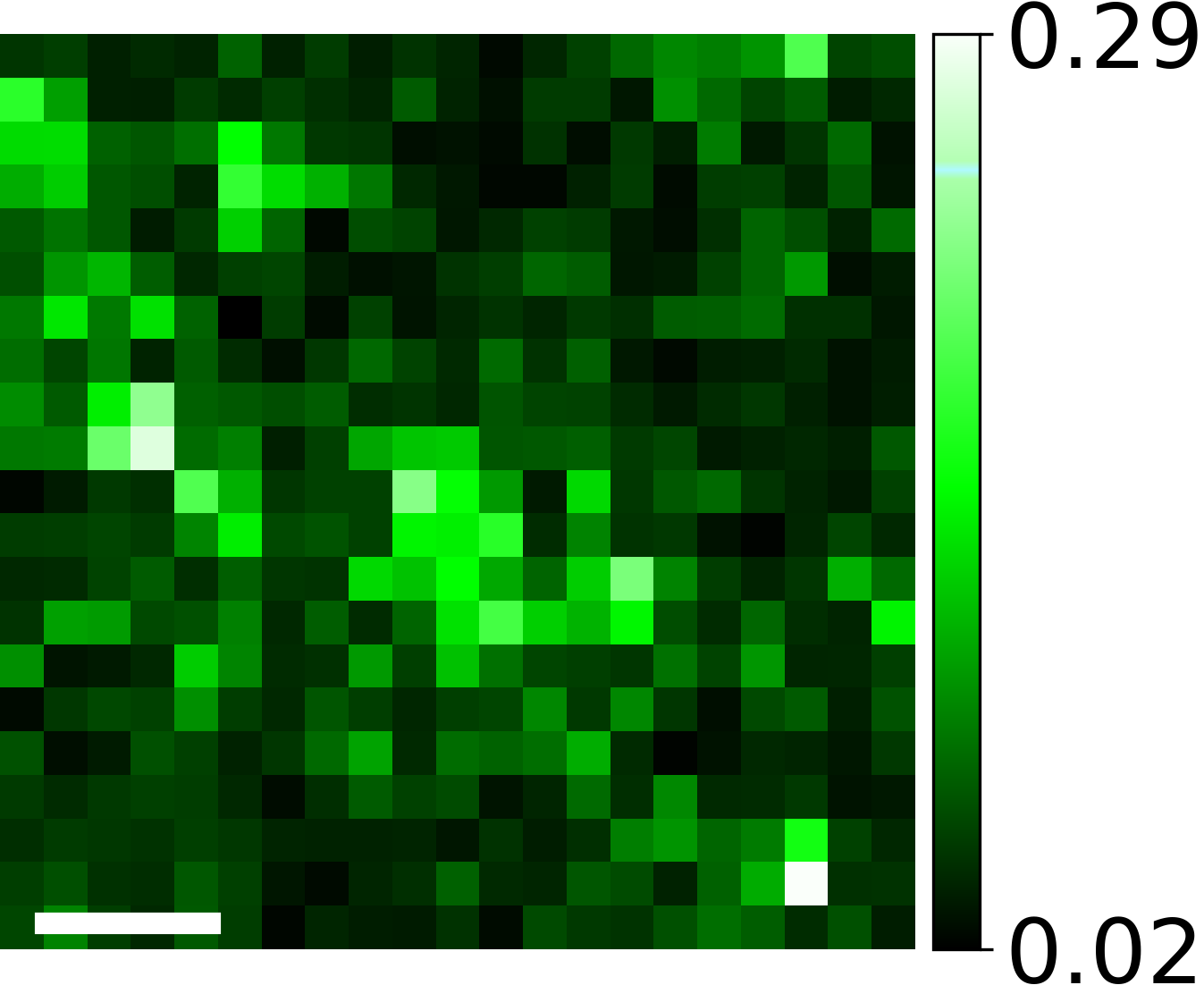}&
			\includegraphics[width= 0.15\textwidth]{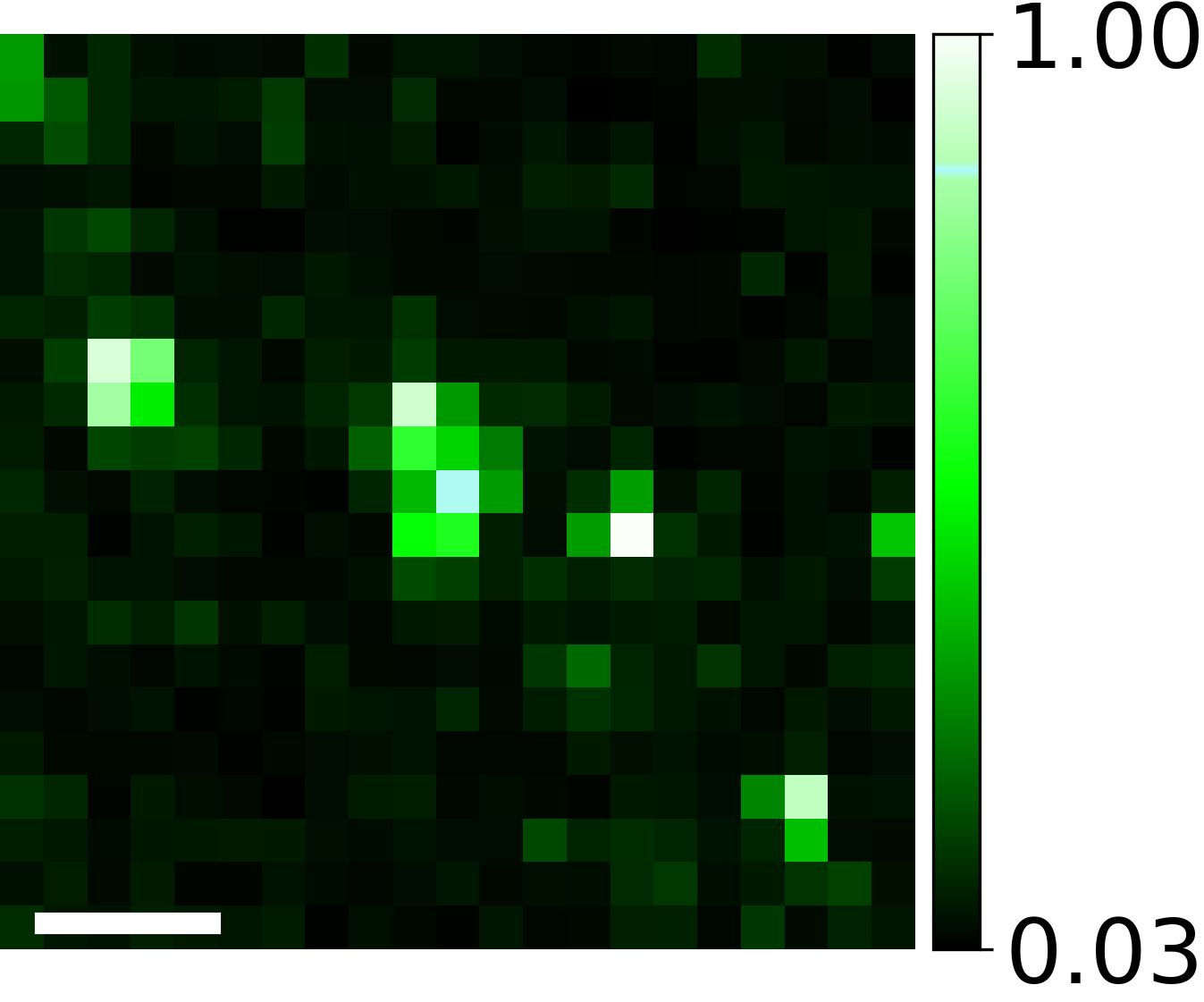}&
			\includegraphics[width= 0.15\textwidth]{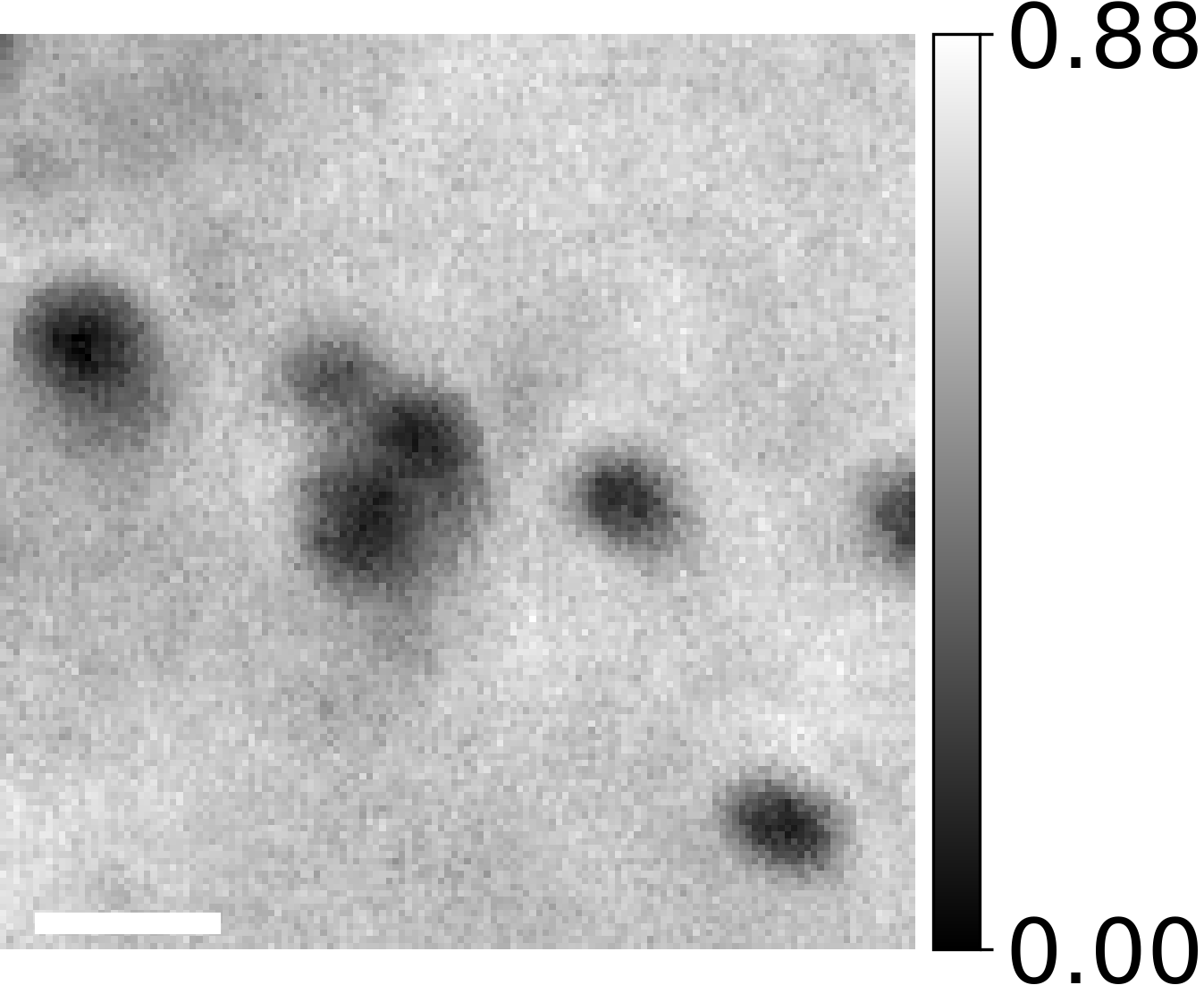}\\
			&{\raisebox{0.9cm}	{\rotatebox[origin=c]{90}{~ {\tiny Max projection $y$} }}}&
			\includegraphics[width= 0.15\textwidth]{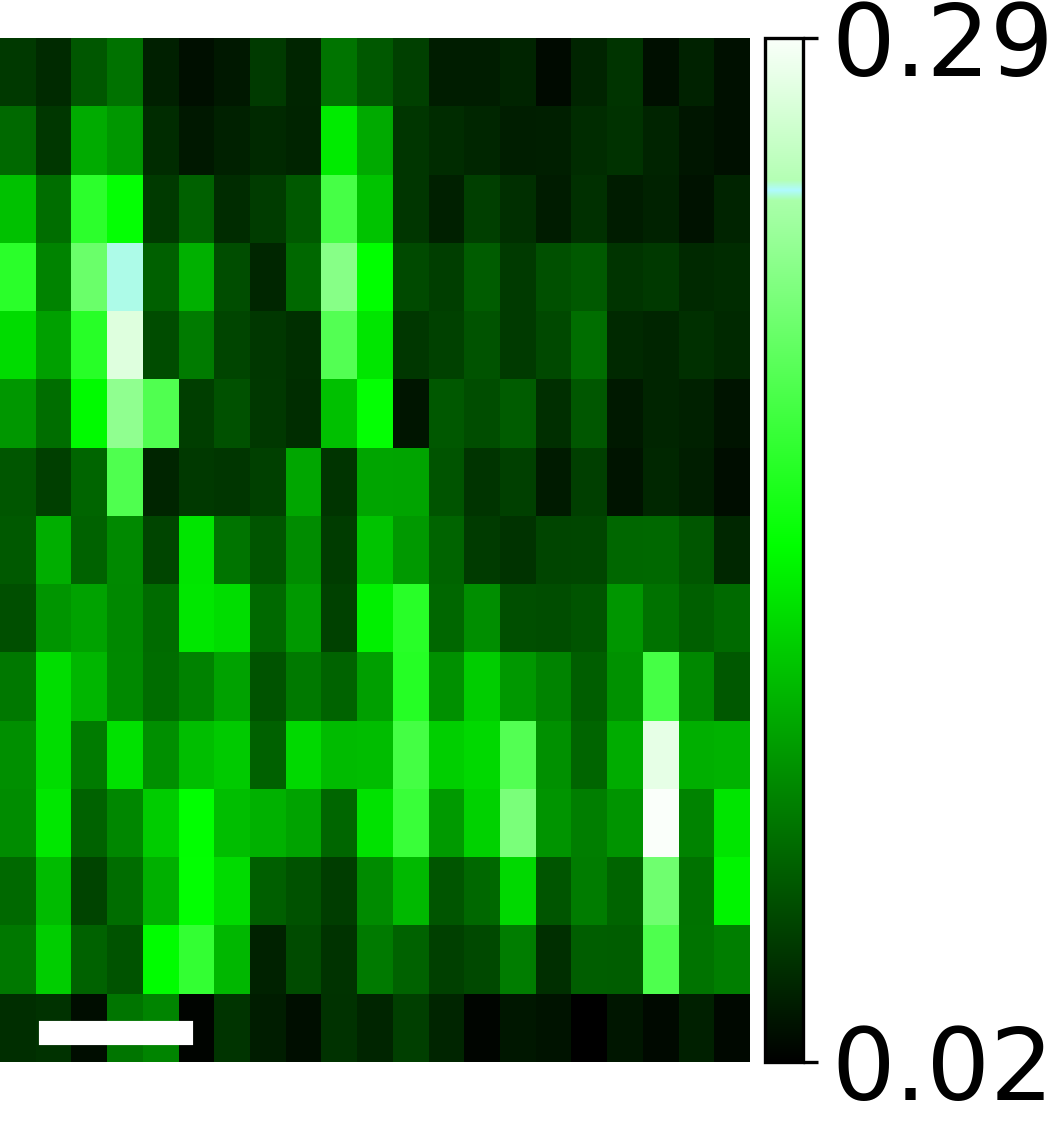}&
			\includegraphics[width= 0.15\textwidth]{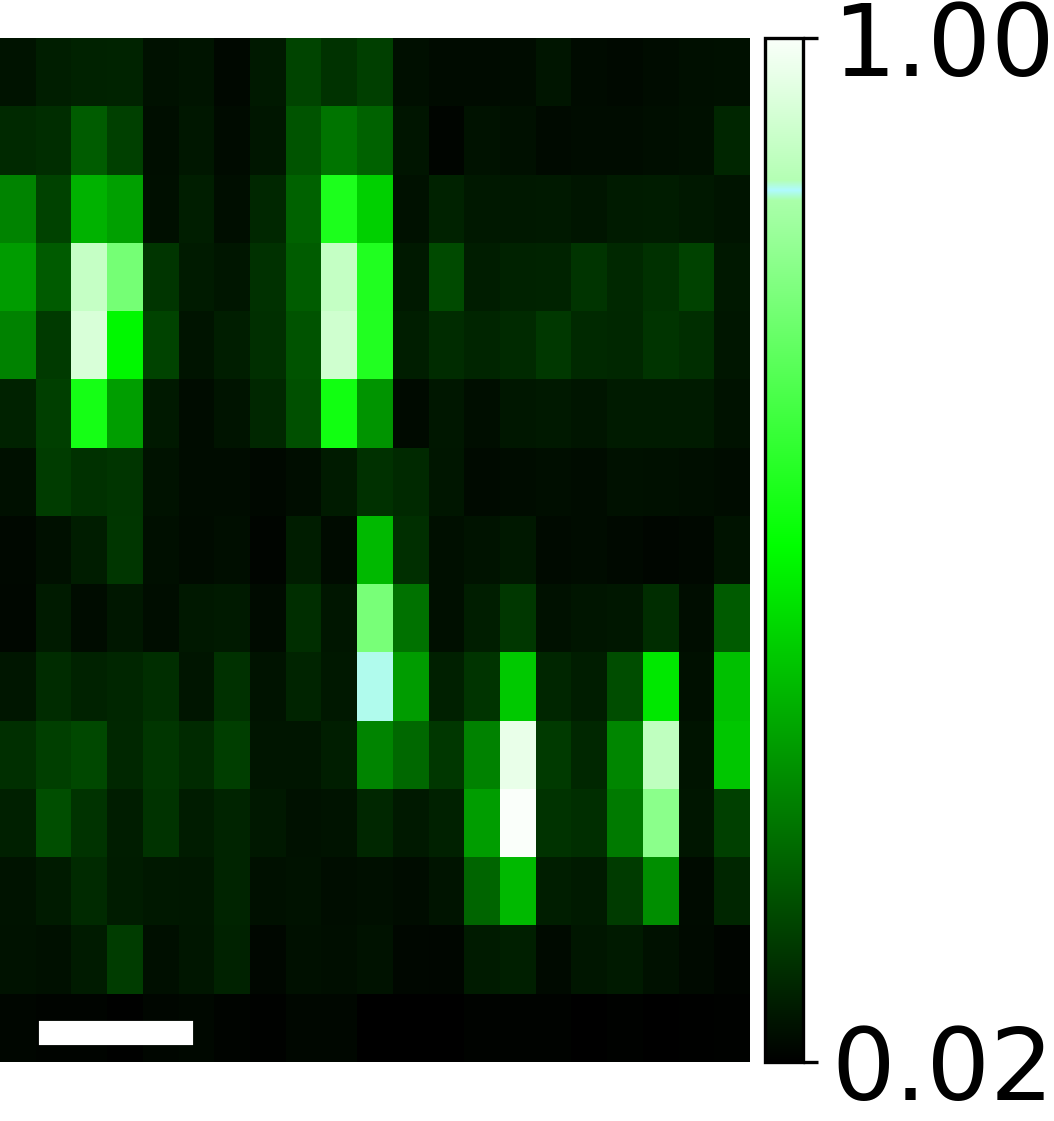}&
			\includegraphics[width= 0.15\textwidth]{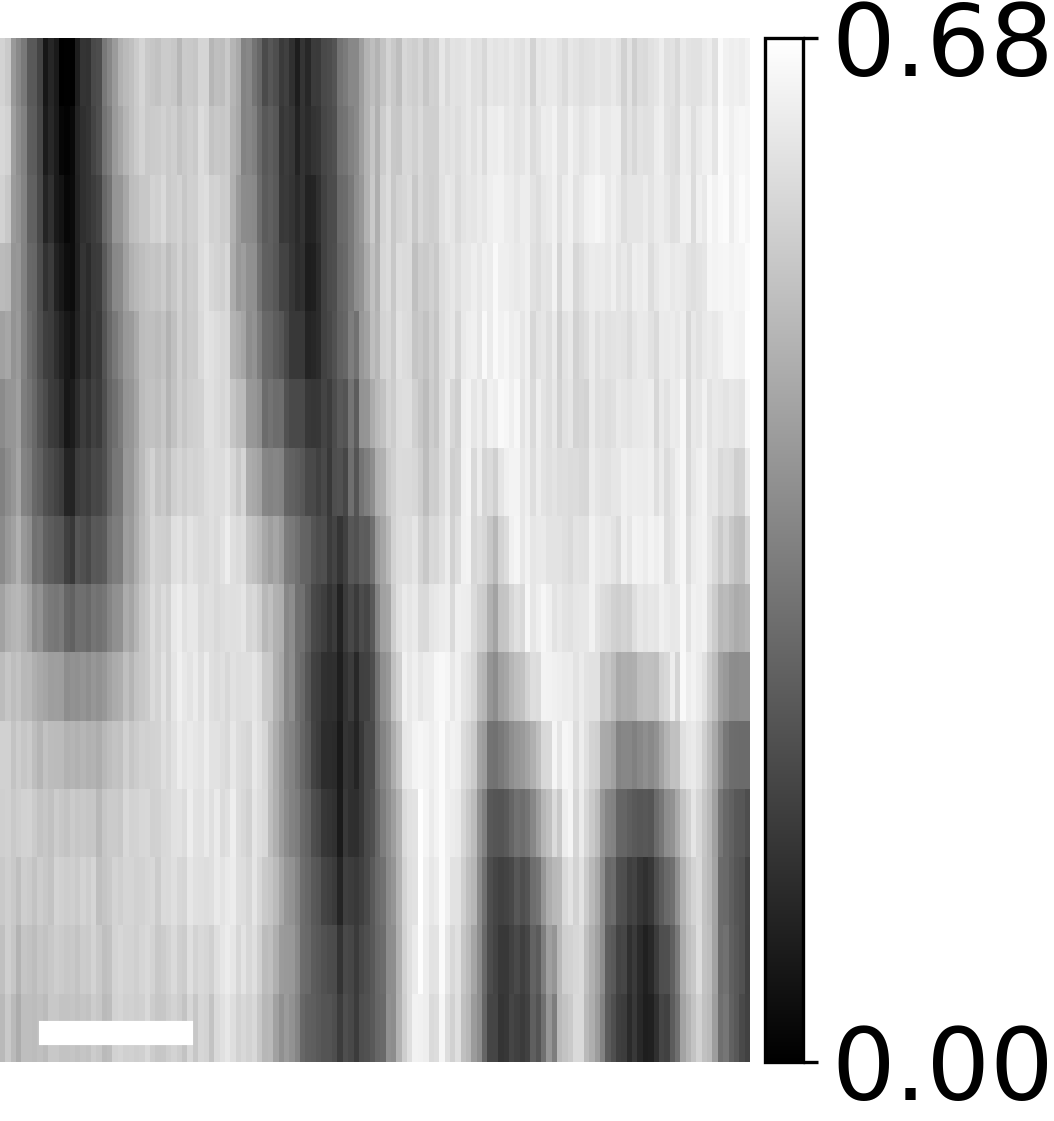}\\
			&{\raisebox{0.9cm}	{\rotatebox[origin=c]{90}{~ {\tiny Max projection $x$} }}}&
			\includegraphics[width= 0.15\textwidth]{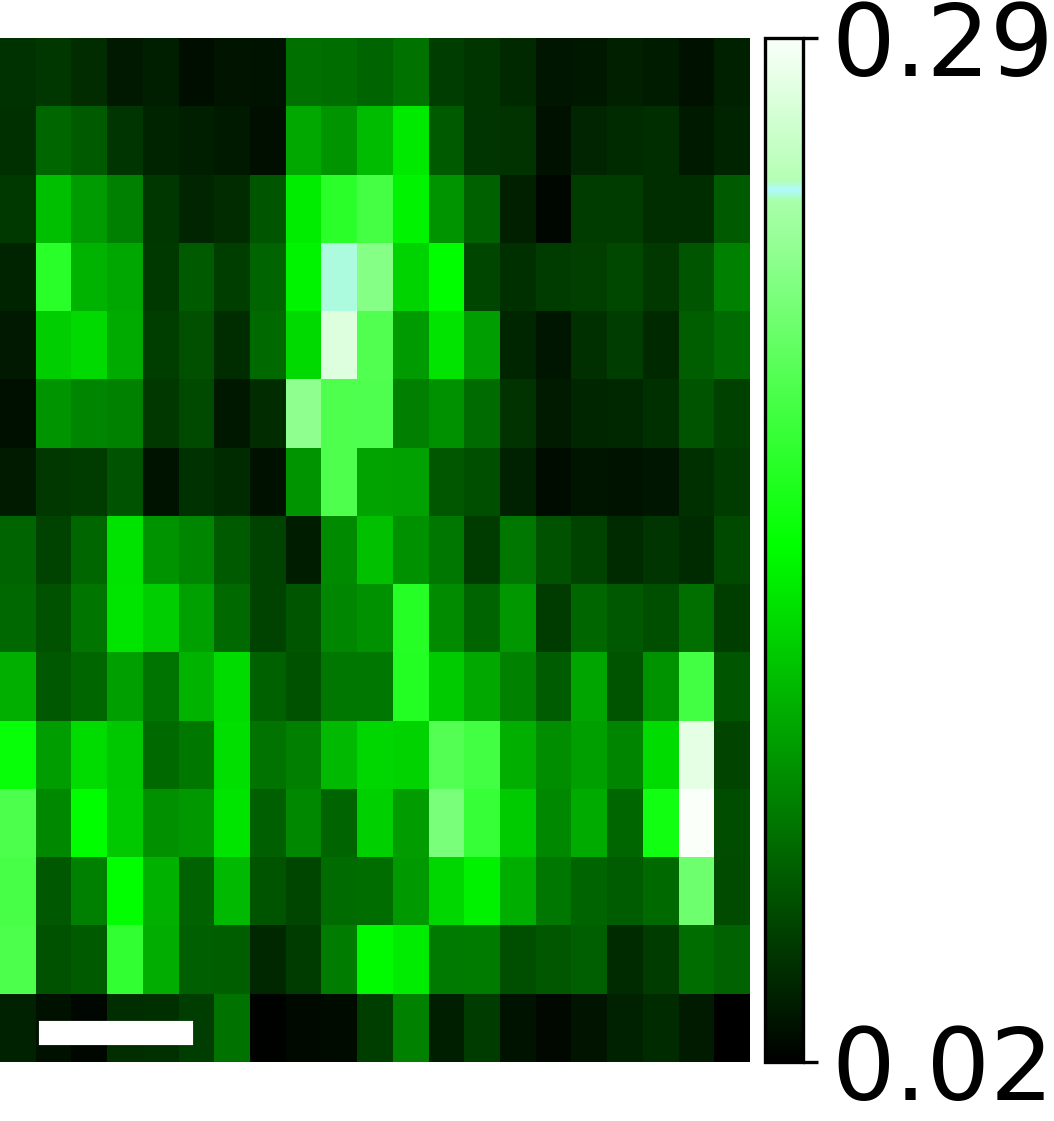}&
			\includegraphics[width= 0.15\textwidth]{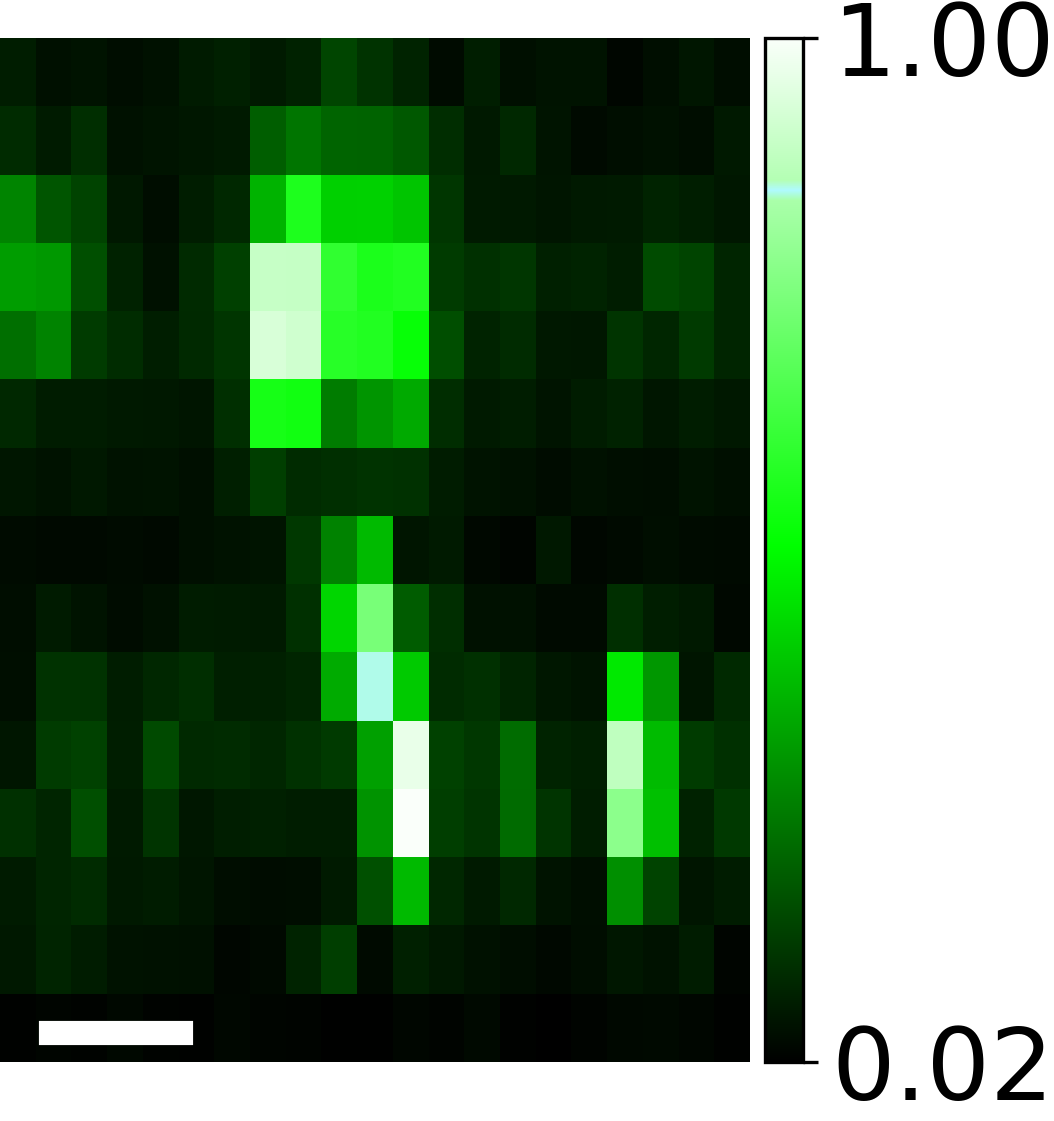}&
			\includegraphics[width= 0.15\textwidth]{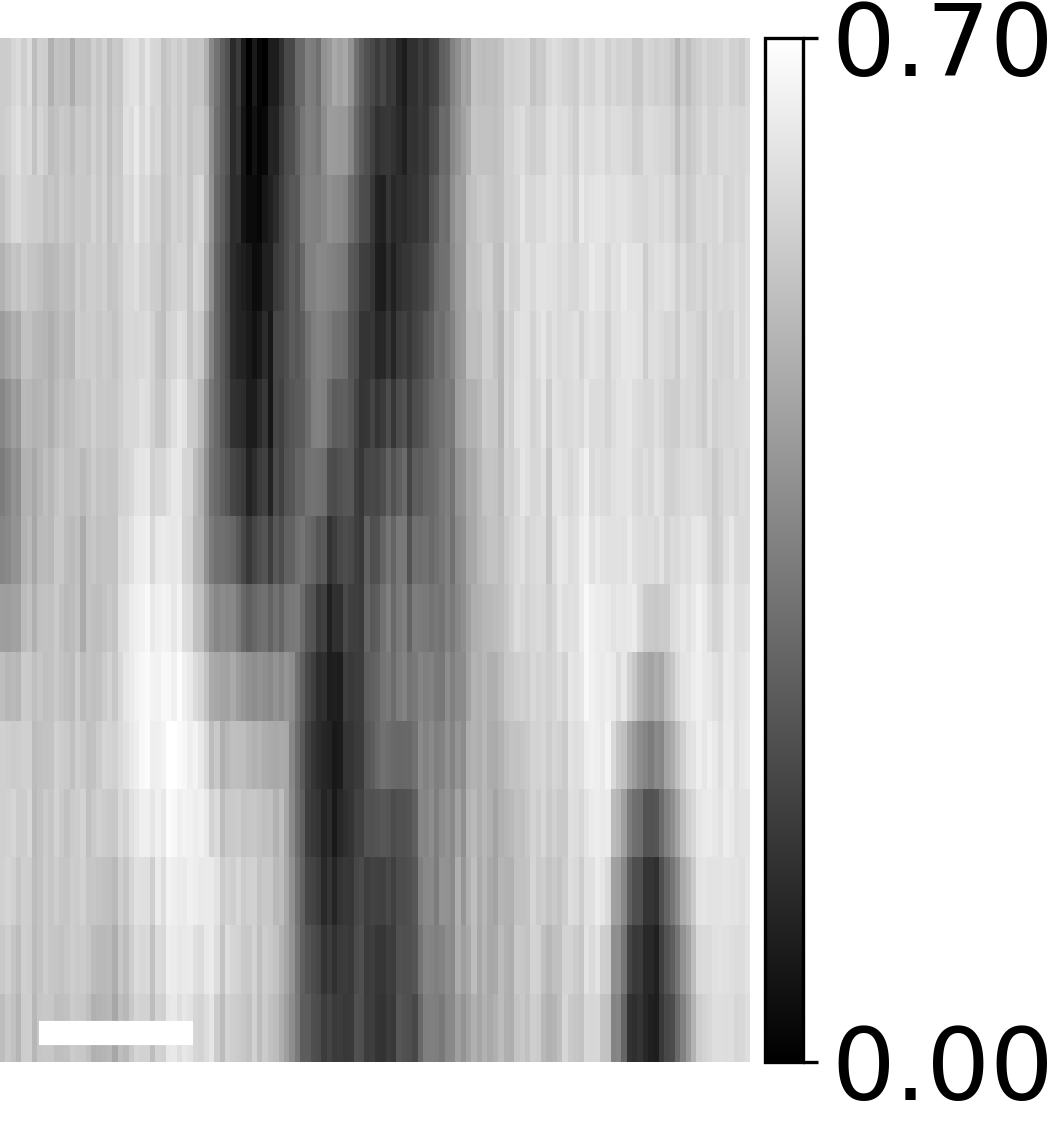}\vspace{-0.0cm}\\
			
		\end{tabular}
		\caption{\blue{\textbf{Confcal of 3D volume:} Our algorithm used to image a scattering dispersion of $0.5\um$ diameter polystyrene beads in agarose gel. The first column depicts a schematic view of the target, where we image a small volume of size $10.4\um \times 10.4\um \times 14\um$. The green bars inside the volume depcit different cross sections we presnt to the right. In addition, we present a maximum projection to each of the three axes. Columns 2-3: Confocal scanning results: with and without aberration correction. Column 4: A reference image of the target captured from the validation camera behind the target, under wide-filed incoherent illumination. Rows 1-3 show three $xy$ cross sections in different depths $4,8,10 \um$. Rows 4-6 present maximum projection onto each of the axes. Scale bar on confocal images is $2\um$.}\vspace{-0.1cm} }
		\label{fig:beads3d}
	\end{center}
\end{figure*}

%% file: fig_onion.tex
\begin{figure*}[t!]
	\begin{center}		
		\begin{tabular}{@{}c@{~}c@{~}c@{~}c@{~}c@{~}c@{~}}		
			\multicolumn{1}{c}{\hspace{-0.6cm}\large Target }& 	
			\multicolumn{2}{c}{\hspace{-0.6cm}\large One point }& 
			\multicolumn{2}{c}{\hspace{-0.6cm}\large Confocal scan}&
			\multicolumn{1}{c}{\hspace{-0.2cm}\large Wide field}\\
			\multicolumn{1}{c}{\hspace{-0.6cm}\large illustration }& 	
			\multicolumn{2}{c}{\hspace{-0.6cm} Main cam.} & \multicolumn{2}{c}{\hspace{-0.6cm} Main cam.} &
			\multicolumn{1}{c}{\hspace{-0.2cm} Valid. cam.} \\&
			\multicolumn{1}{c}{\hspace{-0.6cm} \scriptsize w/o}&	
			\multicolumn{1}{c}{\hspace{-0.6cm} \scriptsize w/ }&	
			\multicolumn{1}{c}{\hspace{-0.6cm} \scriptsize w/o}&	
			\multicolumn{1}{c}{\hspace{-0.6cm} \scriptsize w/ }&	
			\multicolumn{1}{c}{\hspace{-0.2cm} \scriptsize Incoherent}\vspace{-0.0cm}\\&
			\multicolumn{1}{c}{\hspace{-0.6cm} \scriptsize modulation}&	
			\multicolumn{1}{c}{\hspace{-0.6cm} \scriptsize modulation}&	
			\multicolumn{1}{c}{\hspace{-0.6cm} \scriptsize modulation}&	
			\multicolumn{1}{c}{\hspace{-0.6cm} \scriptsize modulation}&	
			\multicolumn{1}{c}{\hspace{-0.2cm} \scriptsize illumination}\\
			\raisebox{0cm}{\hspace{-0.9cm}\includegraphics[width= 0.18\textwidth]{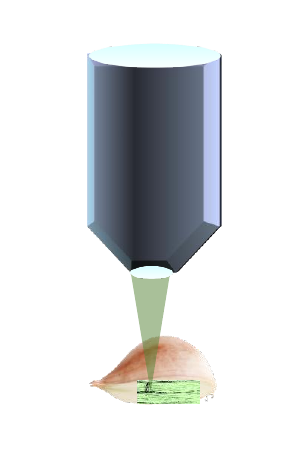}}&
			\raisebox{0.90cm}{\includegraphics[width= 0.18\textwidth]{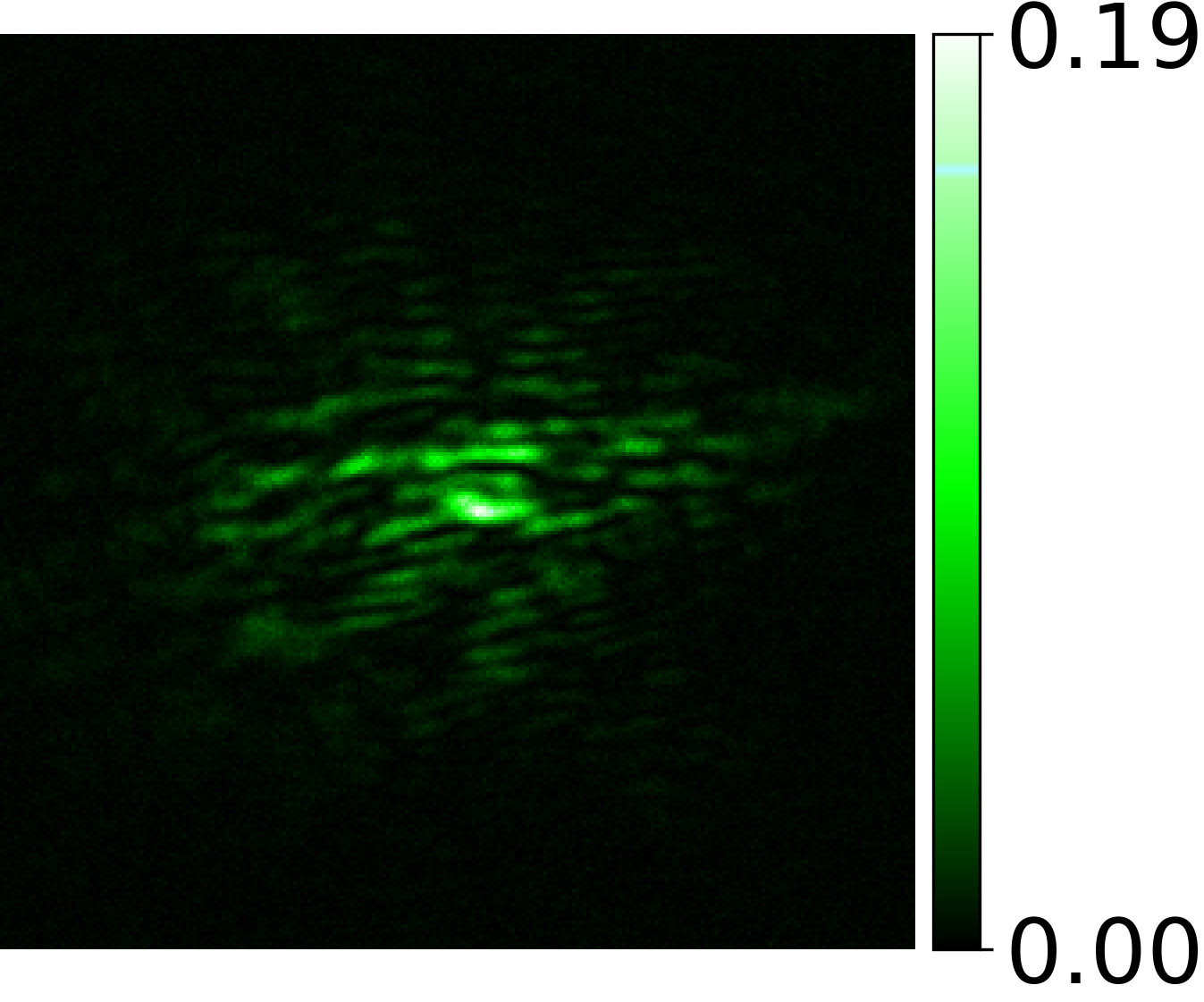}}&
			\raisebox{0.90cm}{\includegraphics[width= 0.18\textwidth]{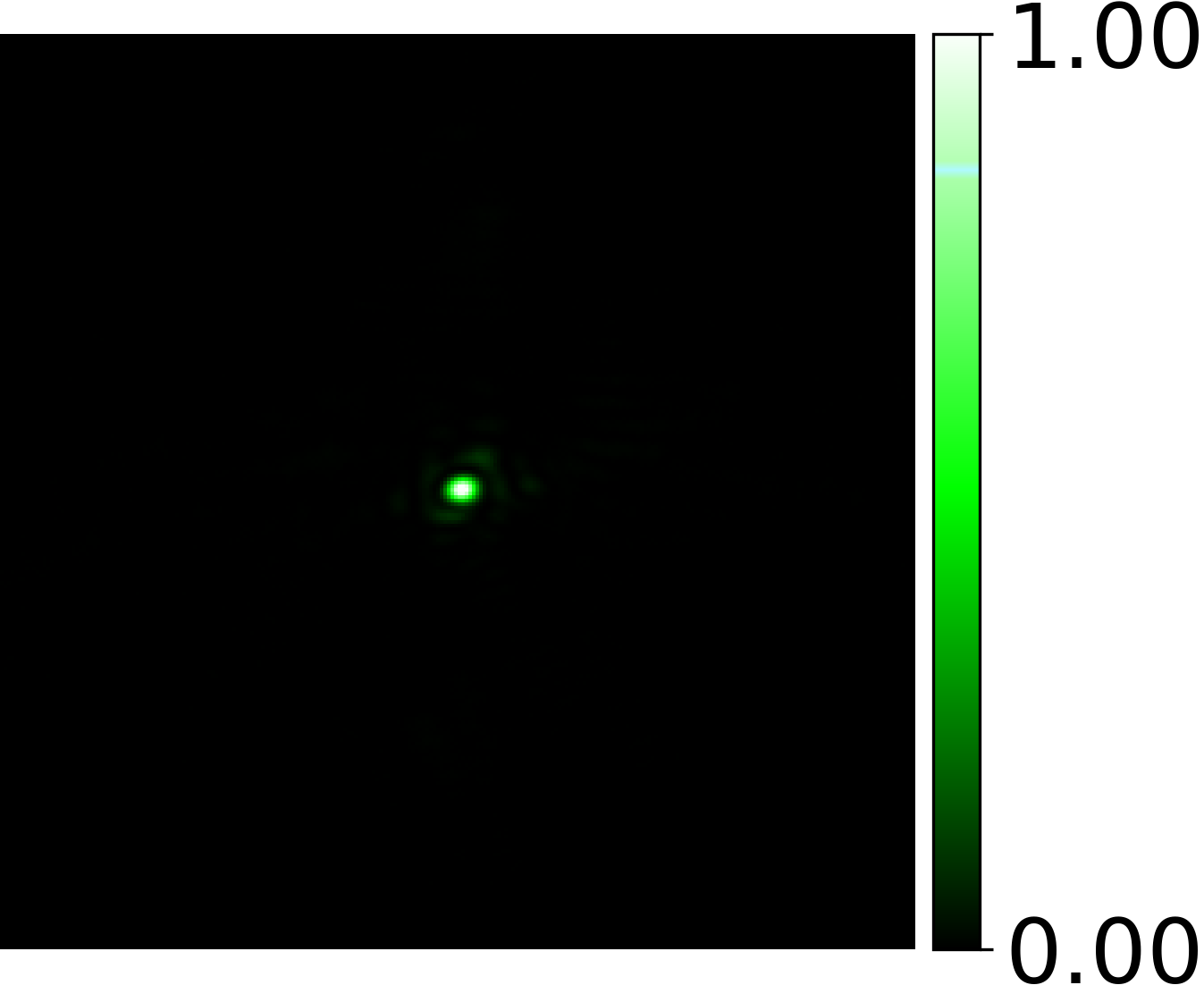}}&
			\includegraphics[width= 0.18\textwidth]{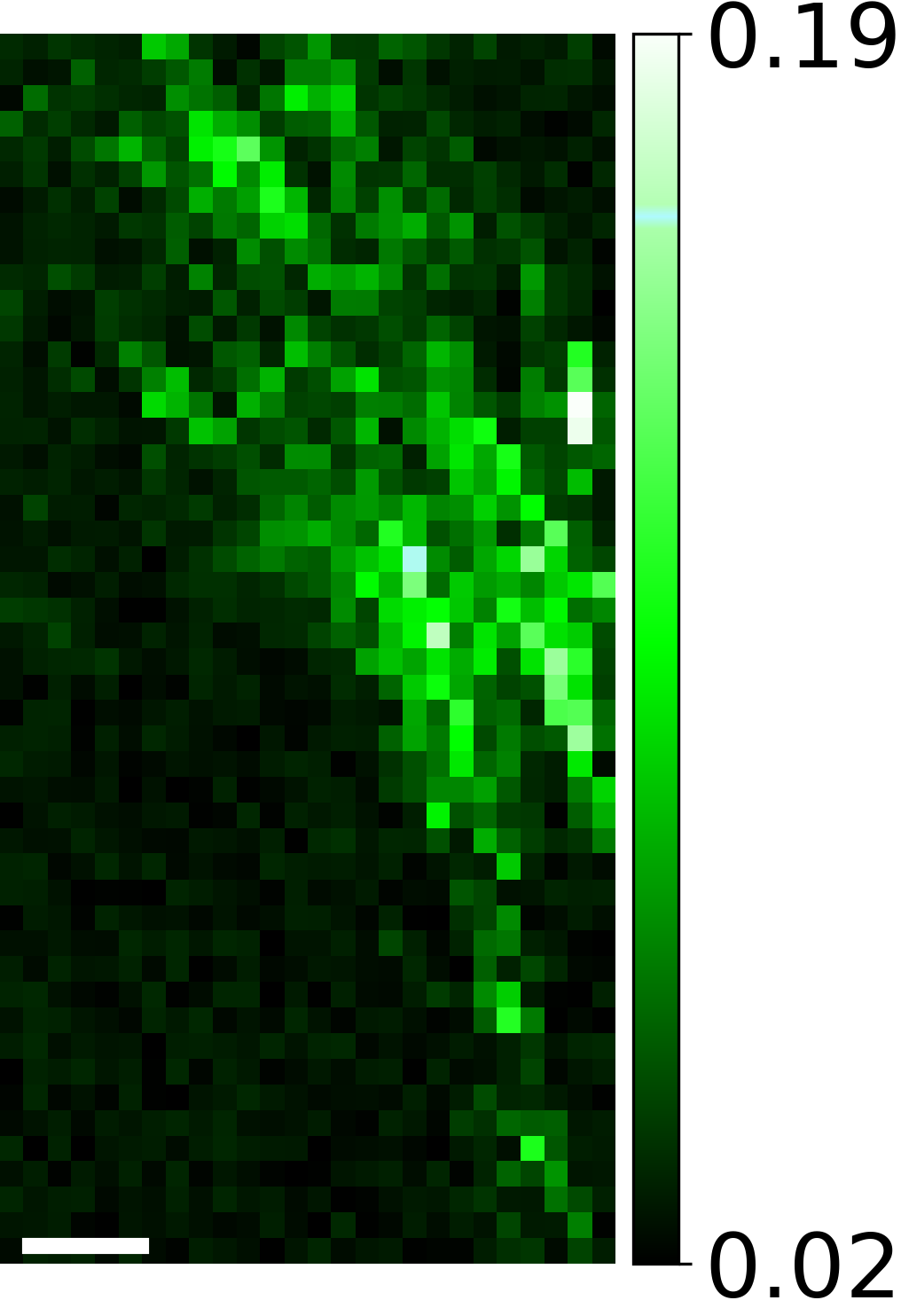}&
			\includegraphics[width= 0.18\textwidth]{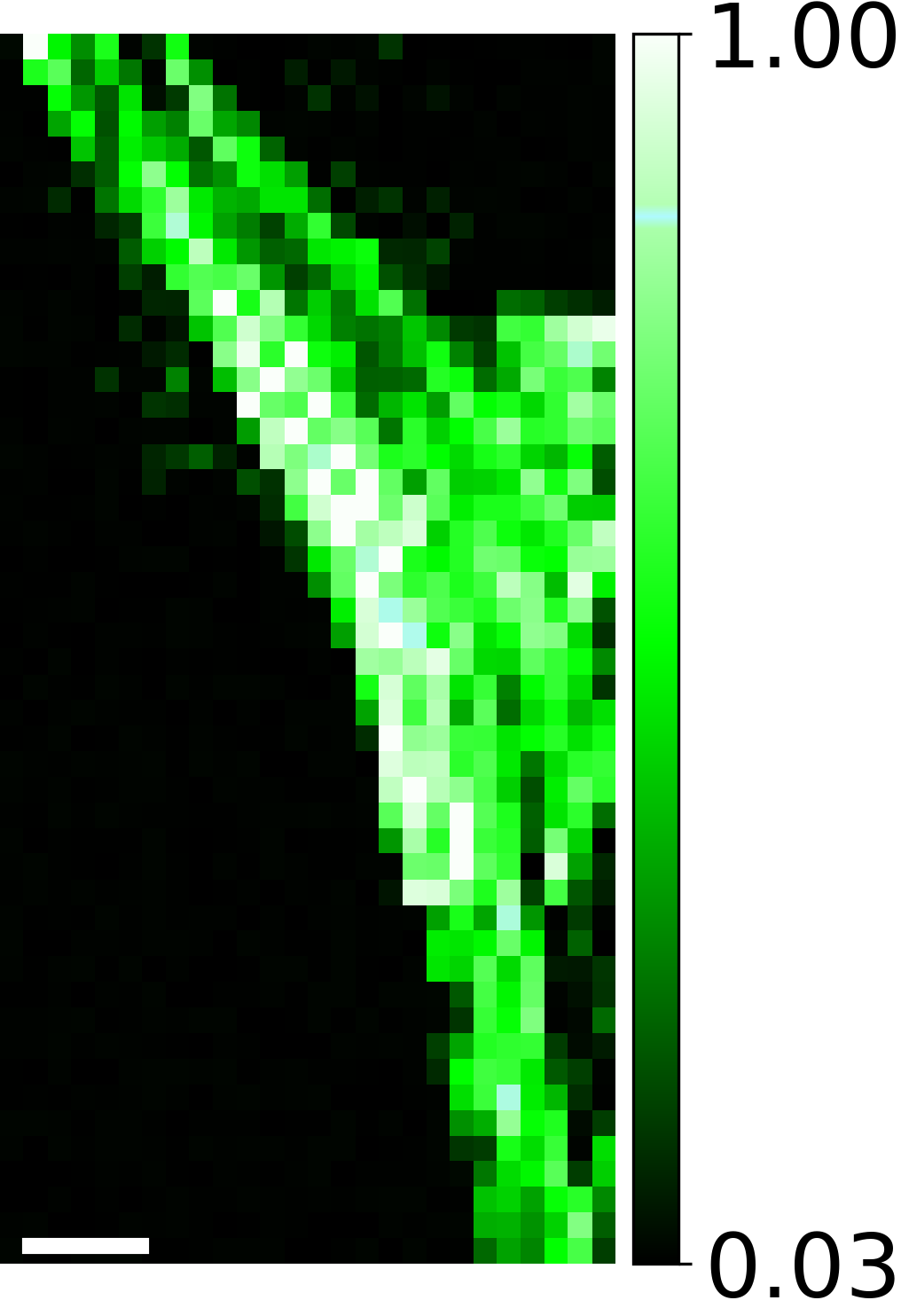}&
			\includegraphics[width= 0.13\textwidth]{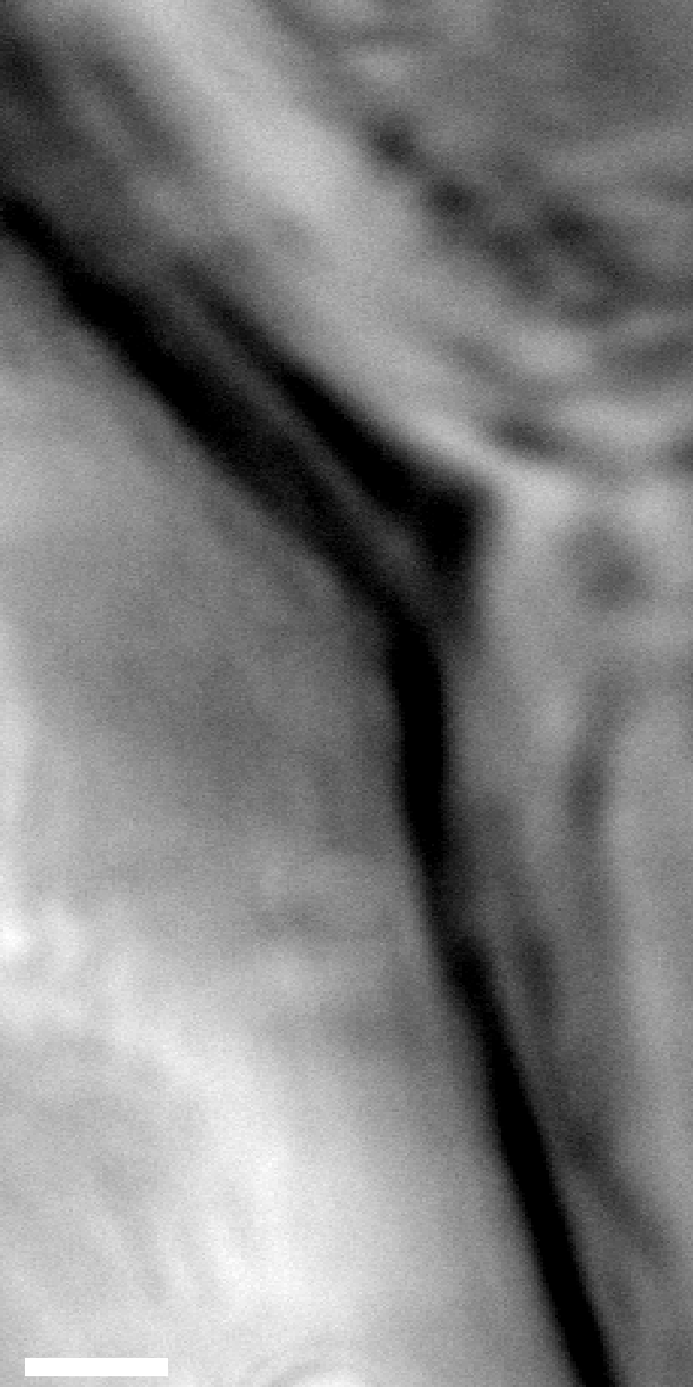}\\

		\end{tabular}
		\caption{\textbf{Confcal scan of onion cells:} \blue{Our algorithm used to image onion cells inside a $140 \um$ thick onion slice. First column:  an illustration of the target. Columns 2-3: Main camera images of a single focal spot before and after correction. Columns 4-5: Confocal scanning results: with and without aberration correction. Column 6: A reference image of the target captured from a validation camera behind the target, under wide-filed  incoherent illumination. Scale bar is $4\um$.} }
		\label{fig:onion}
	\end{center}
\end{figure*}

%% file: fig_onion_3D_1.tex
\begin{figure*}[t!]
	\begin{center}		
		\begin{tabular}{@{}c@{~~~~~~~~~~~~~~}c@{~}c@{~}c@{~}}			
			\multicolumn{1}{c}{\hspace{-0.6cm}\large Target }&&
			\multicolumn{2}{c}{\hspace{-0.6cm}\large Cross section X-Y}\\
			\multicolumn{1}{c}{\hspace{-0.6cm}\large illustration }&&
			\multicolumn{2}{c}{\hspace{-0.6cm} Main cam.} \\
			\multirow{5}{8em}{\hspace{0cm}\vspace{0cm}\includegraphics[width= 0.3\textwidth]{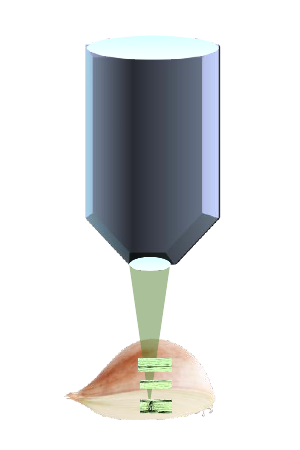}}&&
			\multicolumn{1}{c}{\hspace{-0.6cm} \scriptsize w/o}&	
			\multicolumn{1}{c}{\hspace{-0.6cm} \scriptsize w/ }\\
			&&
			\multicolumn{1}{c}{\hspace{-0.6cm} \scriptsize modulation}&	
			\multicolumn{1}{c}{\hspace{-0.6cm} \scriptsize modulation}\\

			&{\raisebox{0.7cm}	{\rotatebox[origin=c]{90}{~ {\tiny $z=80\um$} }}}&
			\includegraphics[width= 0.15\textwidth]{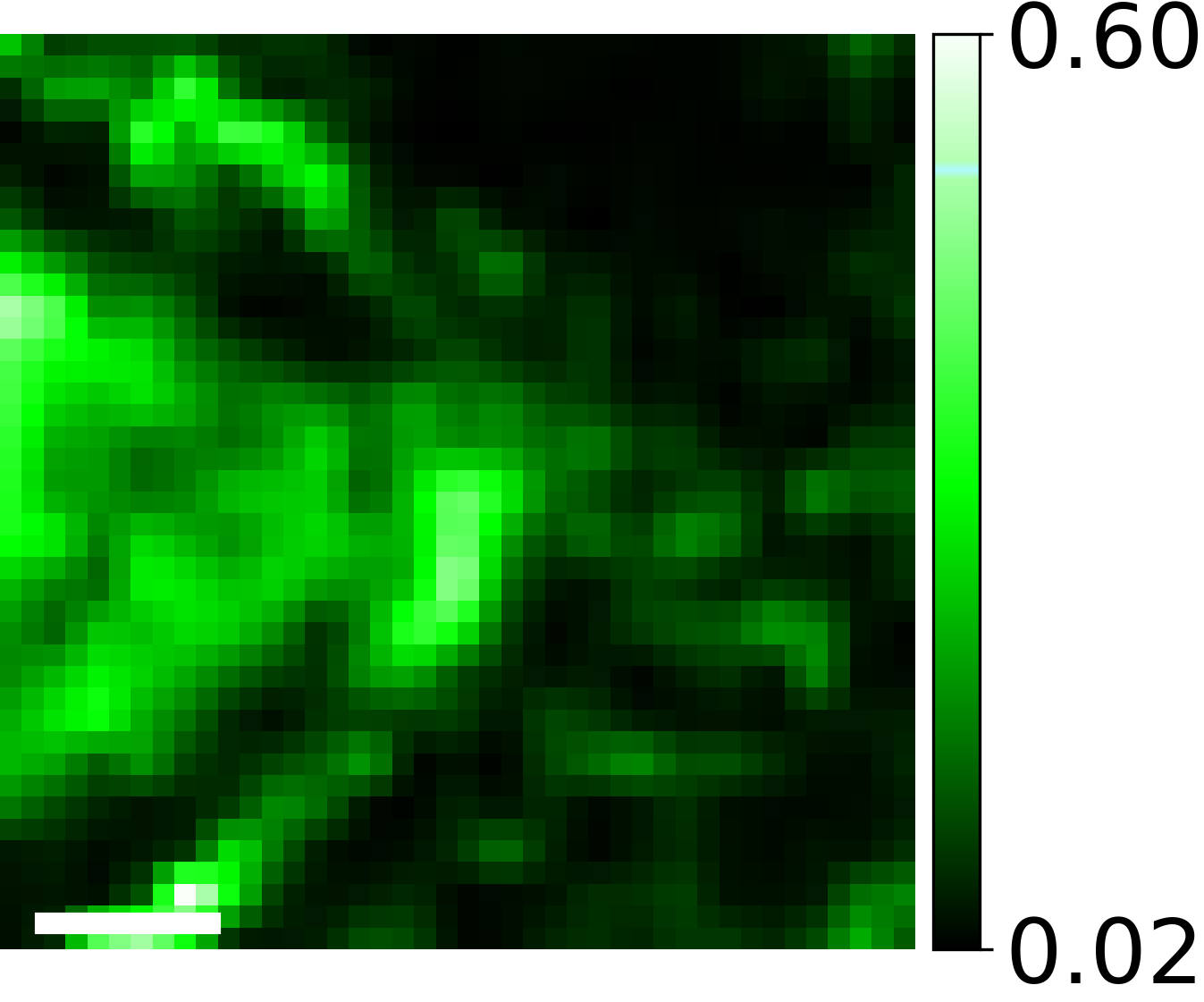}&
			\includegraphics[width= 0.15\textwidth]{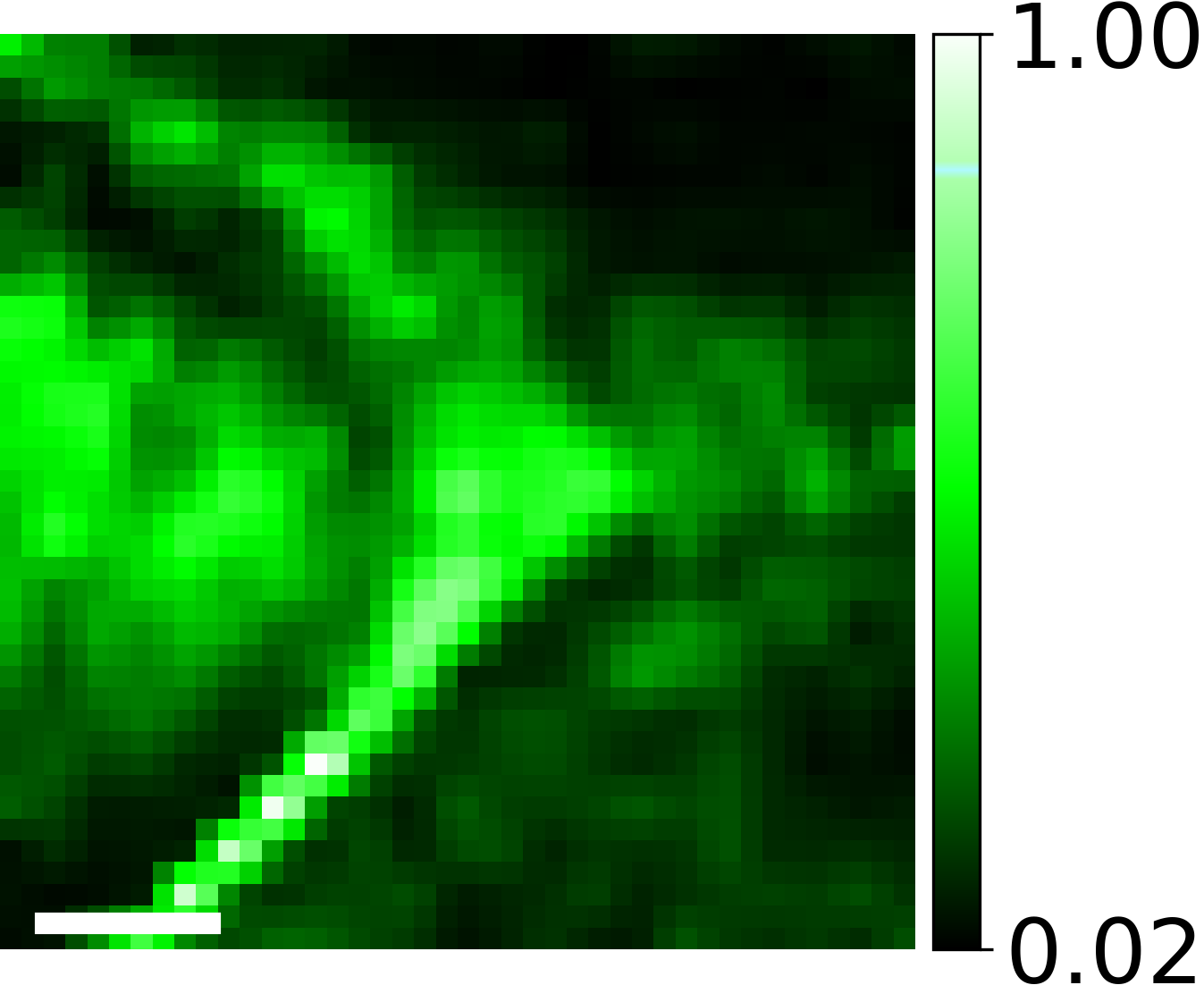}\\
			&{\raisebox{0.7cm}	{\rotatebox[origin=c]{90}{~ {\tiny $z=130\um$} }}}&
			\includegraphics[width= 0.15\textwidth]{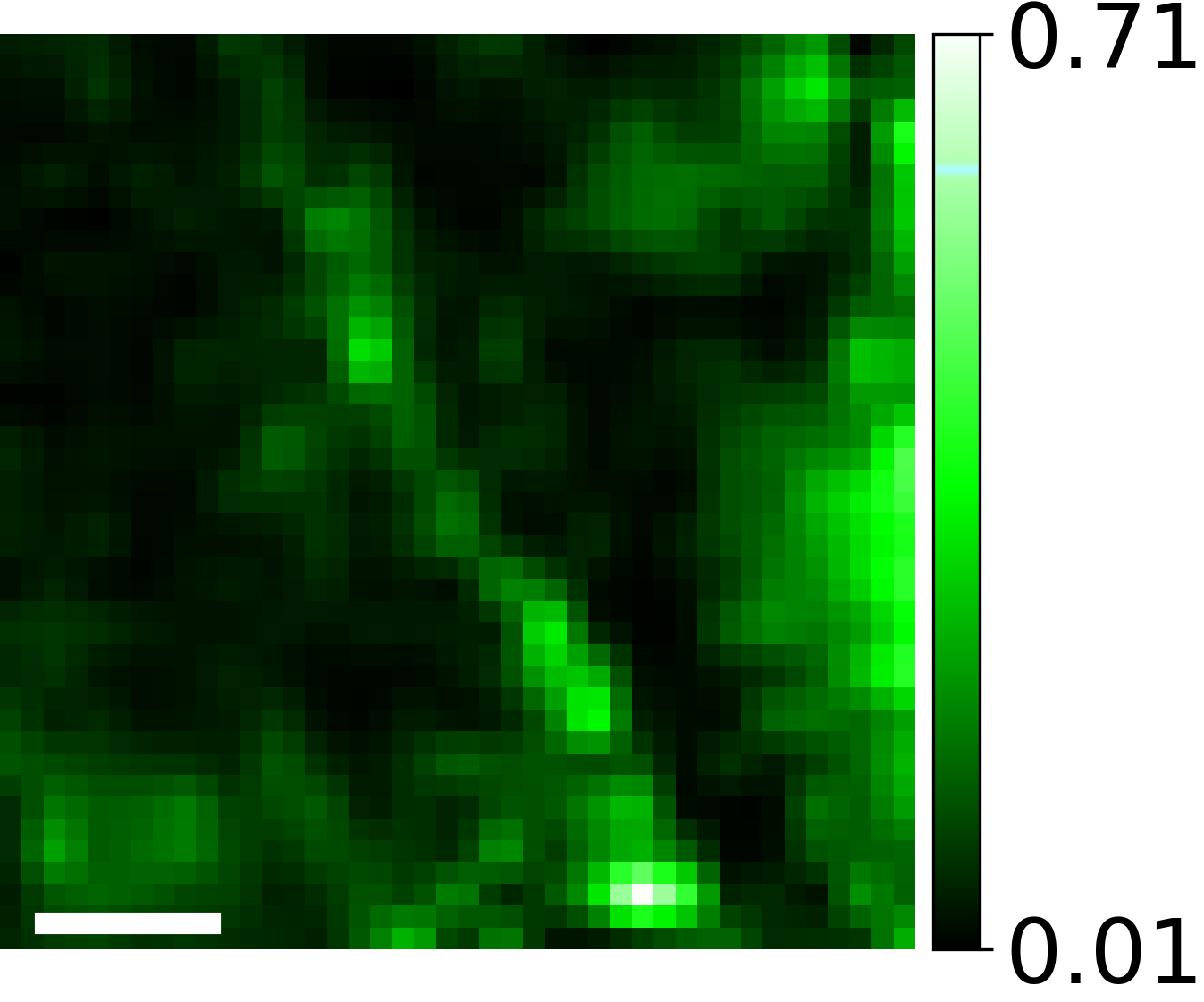}&
			\includegraphics[width= 0.15\textwidth]{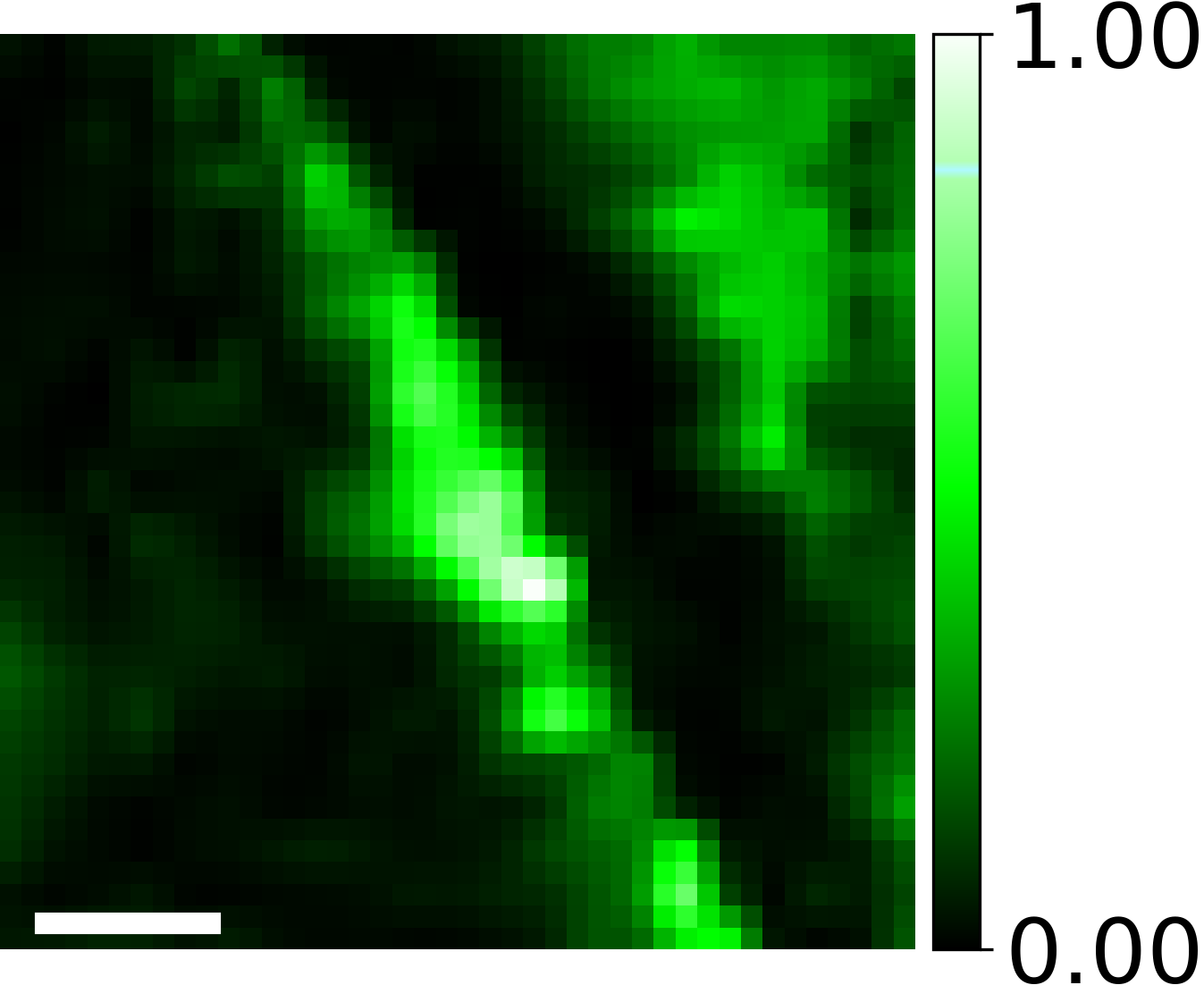}\\
			&{\raisebox{0.7cm}	{\rotatebox[origin=c]{90}{~ {\tiny $z=190\um$} }}}&
			\includegraphics[width= 0.15\textwidth]{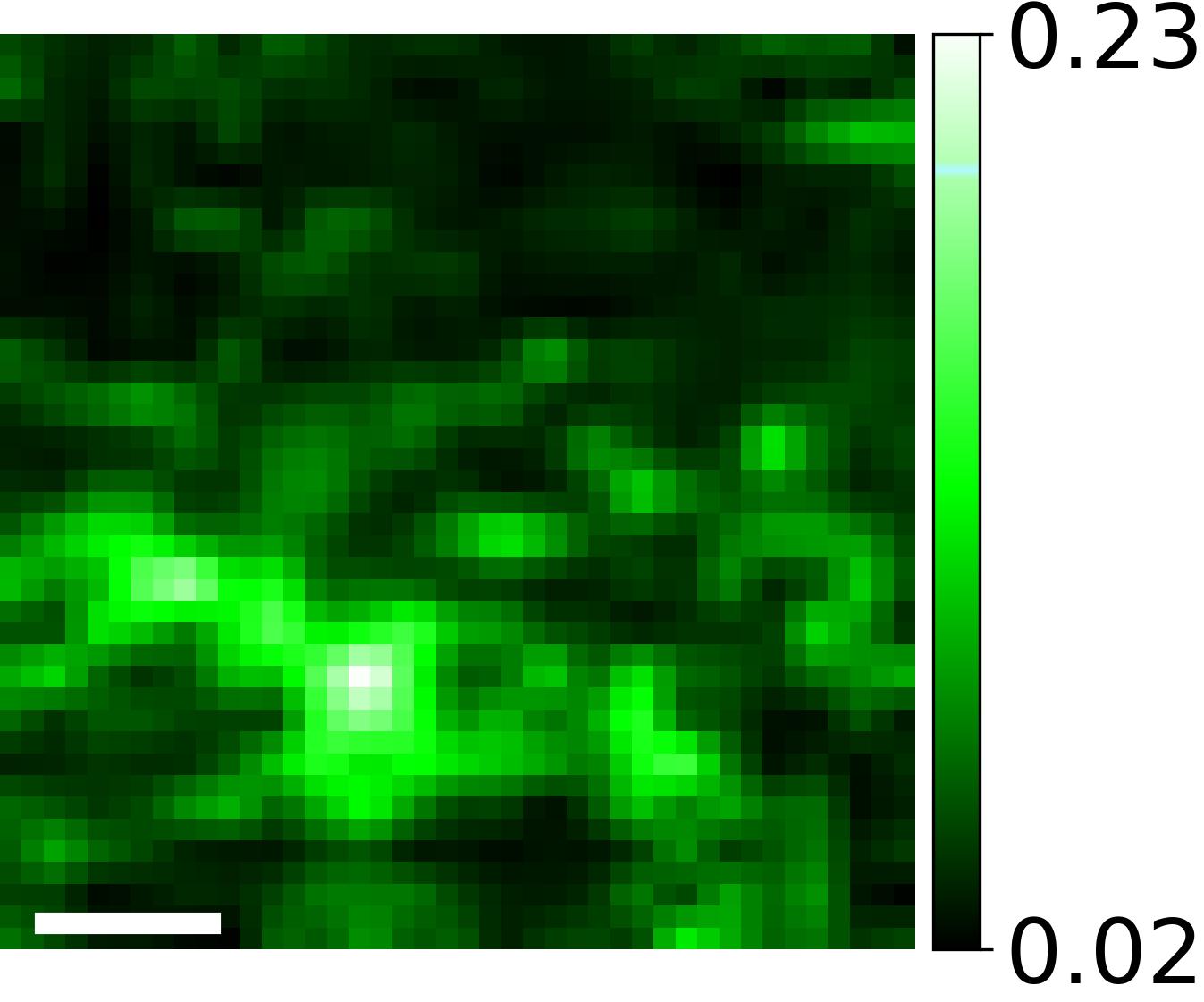}&
			\includegraphics[width= 0.15\textwidth]{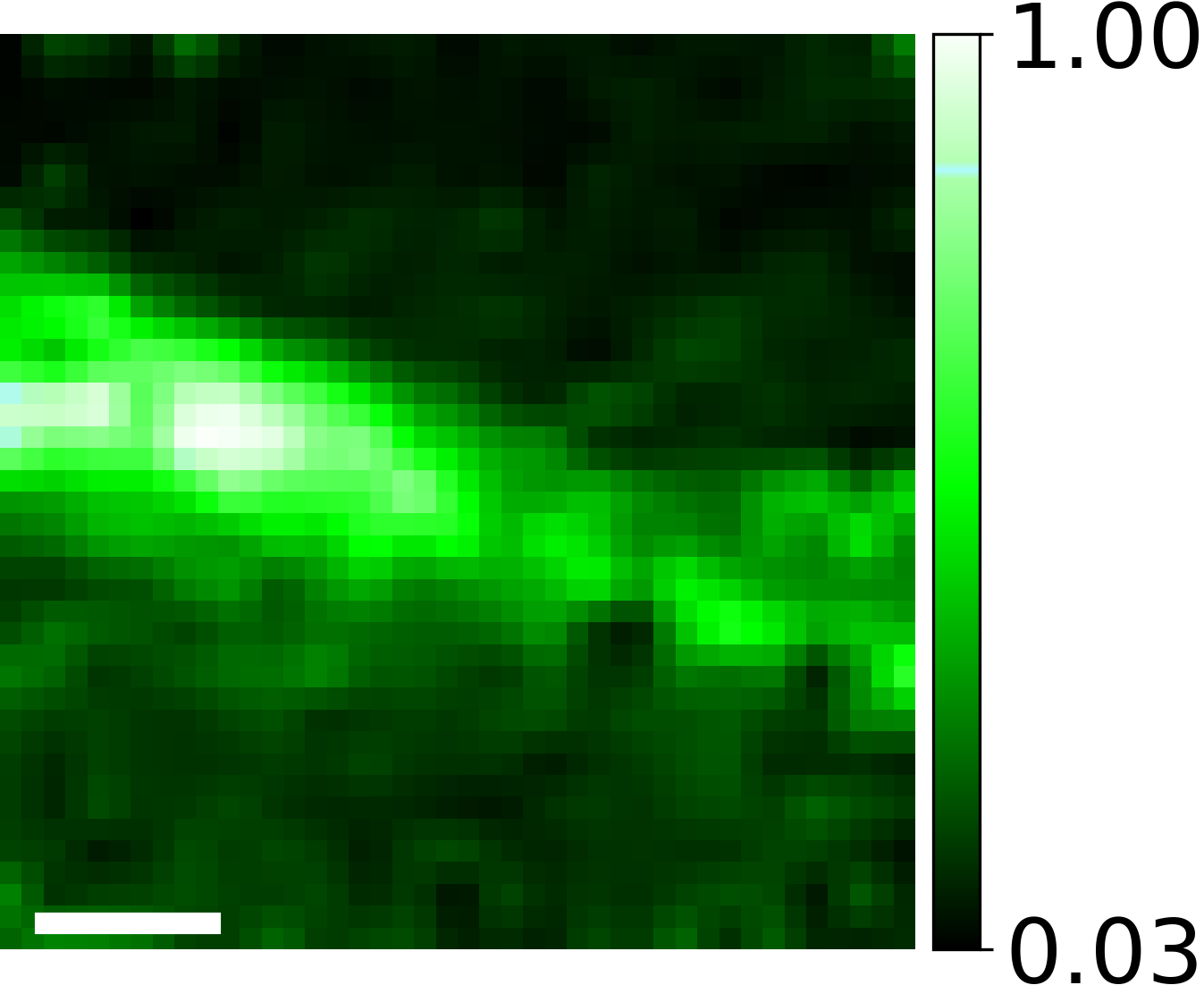}\\
			
		\end{tabular}
		\caption{\bblue{\textbf{3D aberration correction of an onion slice:} Our algorithm used to image onion cells at different depths of $80,130,190\um$. The first column depicts a schematic view of the target. For each layer we present x-y cross sections. Scale bar on confocal images is $5\um$.}\vspace{0.1cm} }
		\label{fig:onion3d}
	\end{center}
\end{figure*}

%% file: fig_cmp_resolution.tex
\begin{figure*}[t!]
	\begin{center}\begin{tabular}{{@{}c@{~}c@{~}c@{~}c@{~}c@{~}}}
			\multicolumn{1}{c}{}&
			\multicolumn{1}{c}{\hspace{-0.7cm} \small Init.}&
			\multicolumn{1}{c}{\hspace{-0.7cm} \small Res. $12{\times}12$}&
			\multicolumn{1}{c}{\hspace{-0.7cm} \small Res. $36{\times}36$}\\ 
			
			{\raisebox{0.9cm}	{\rotatebox[origin=c]{90}{~ {\small Main cam.} }}}&
			\includegraphics[width= 0.2\textwidth]{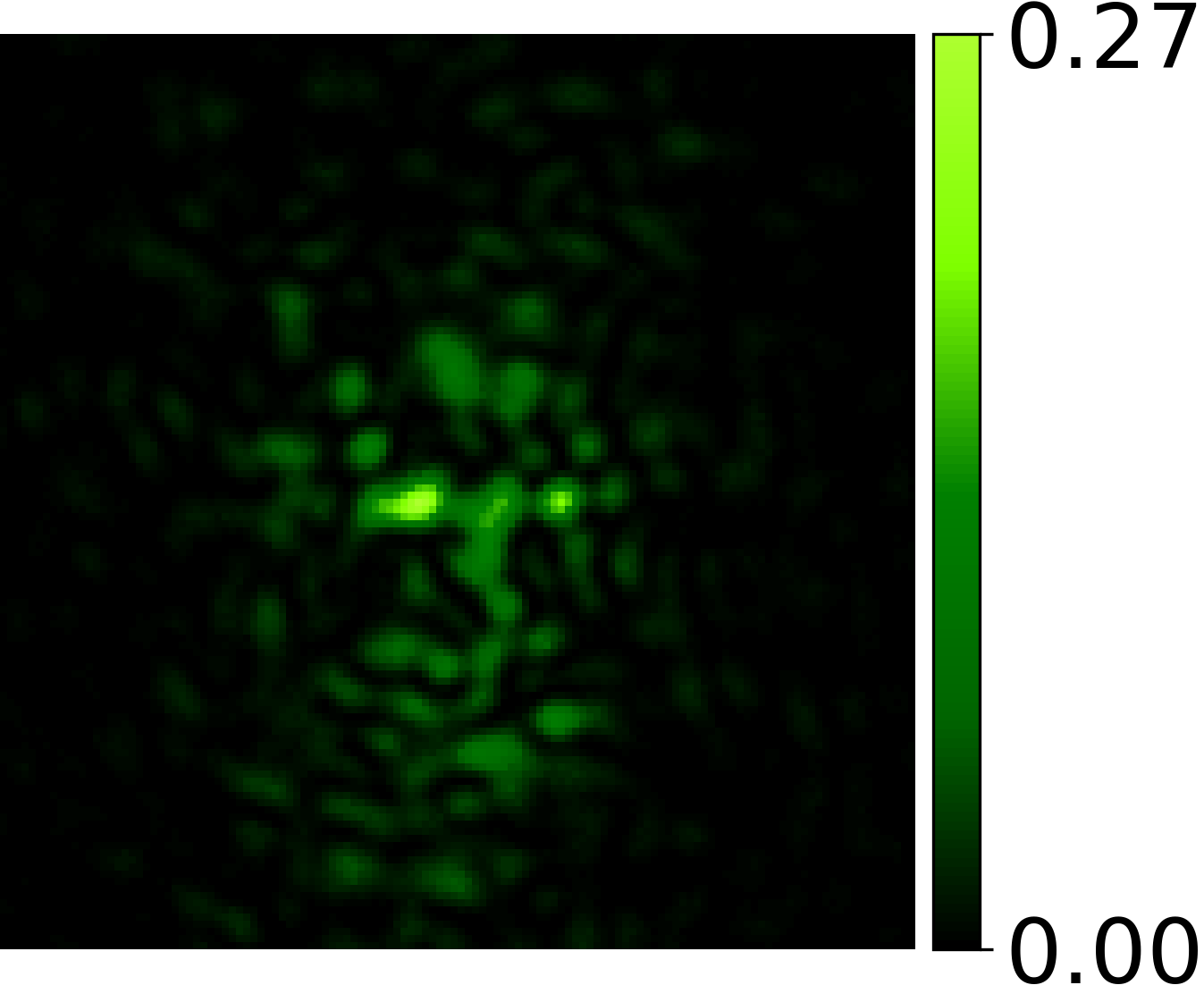}&
			\includegraphics[width= 0.2\textwidth]{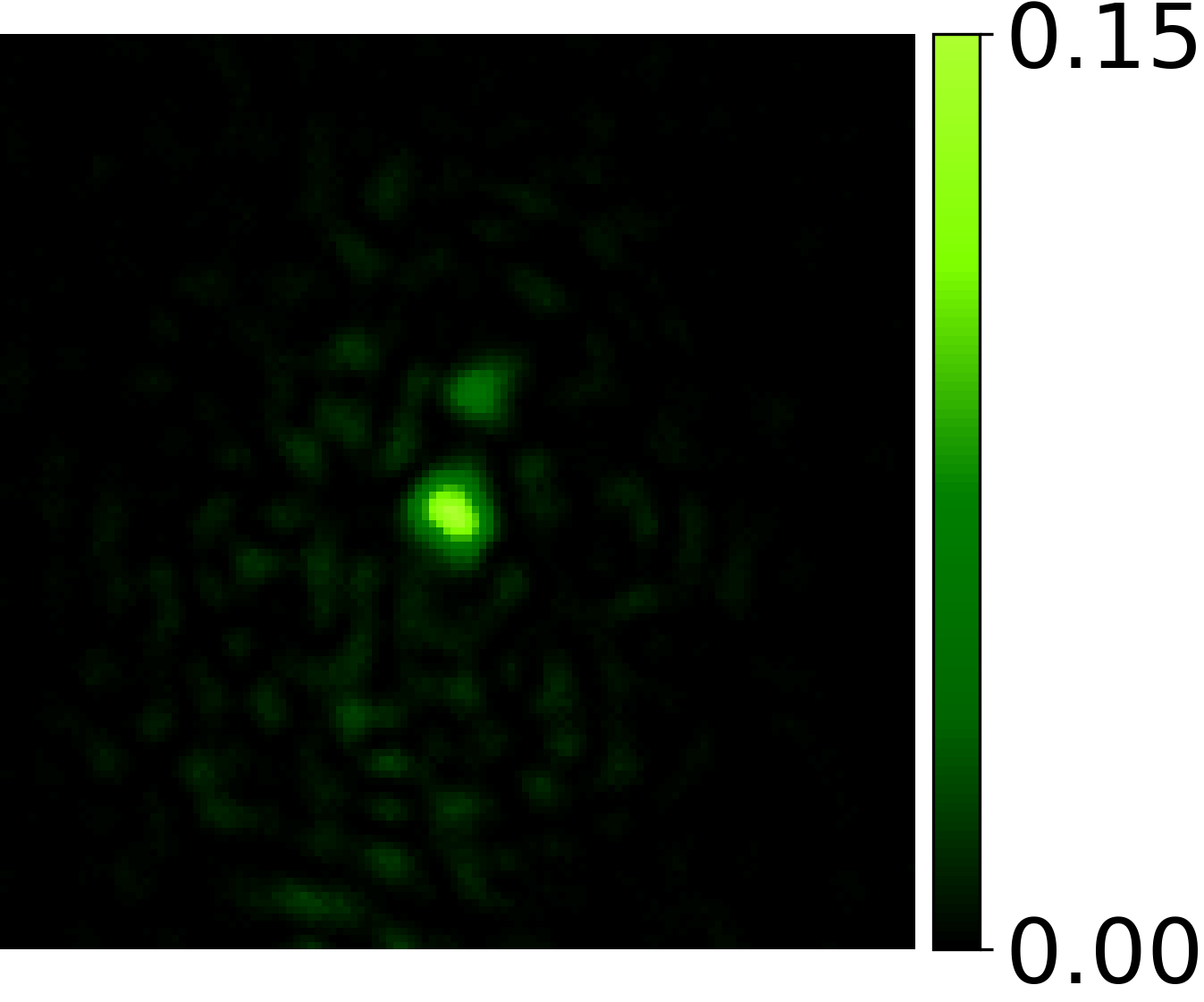}&
			\includegraphics[width= 0.2\textwidth]{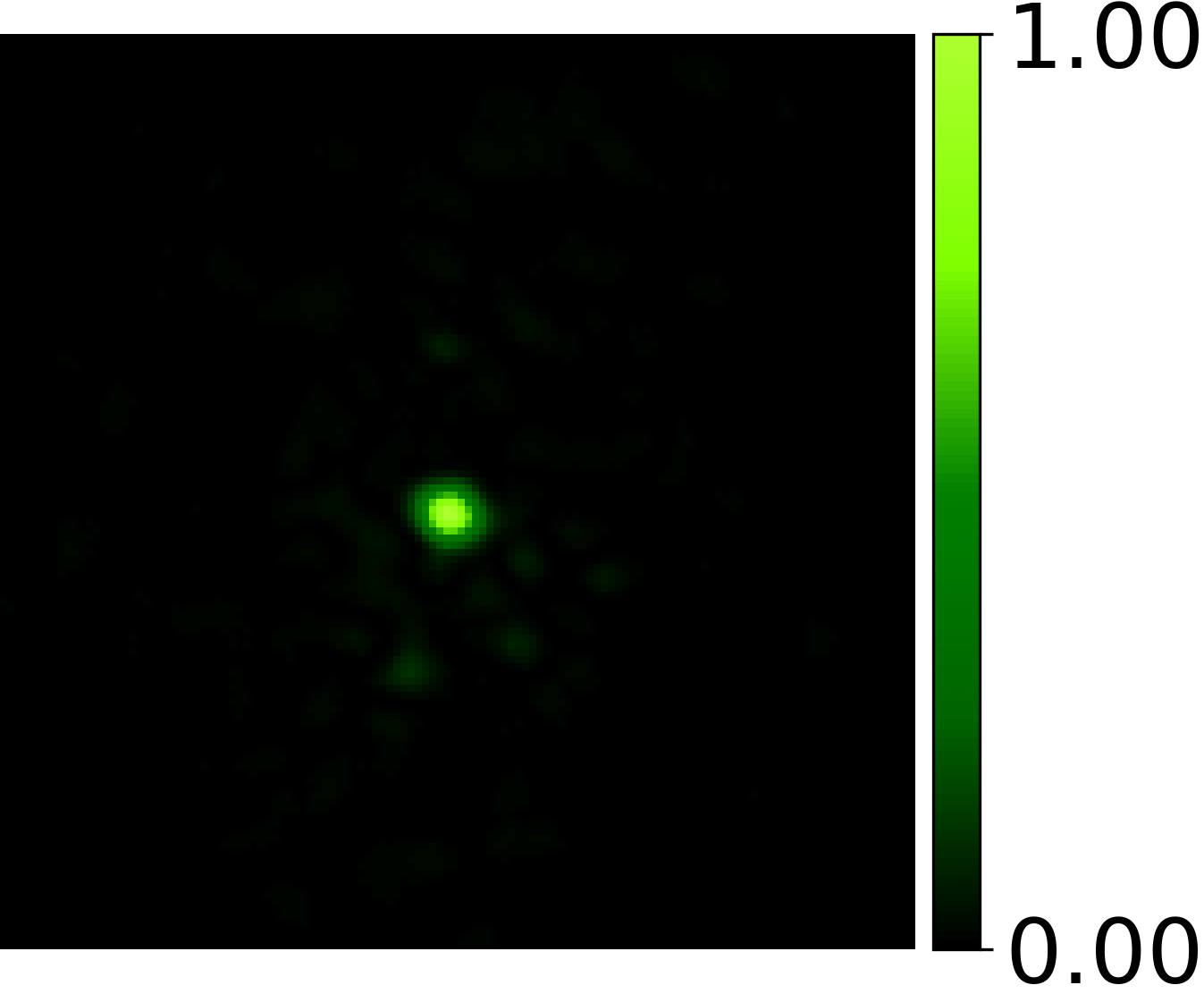}\\
		
			{\raisebox{0.9cm}	{\rotatebox[origin=c]{90}{~ {\small Valid. cam.} }}}&
			\includegraphics[width= 0.2\textwidth]{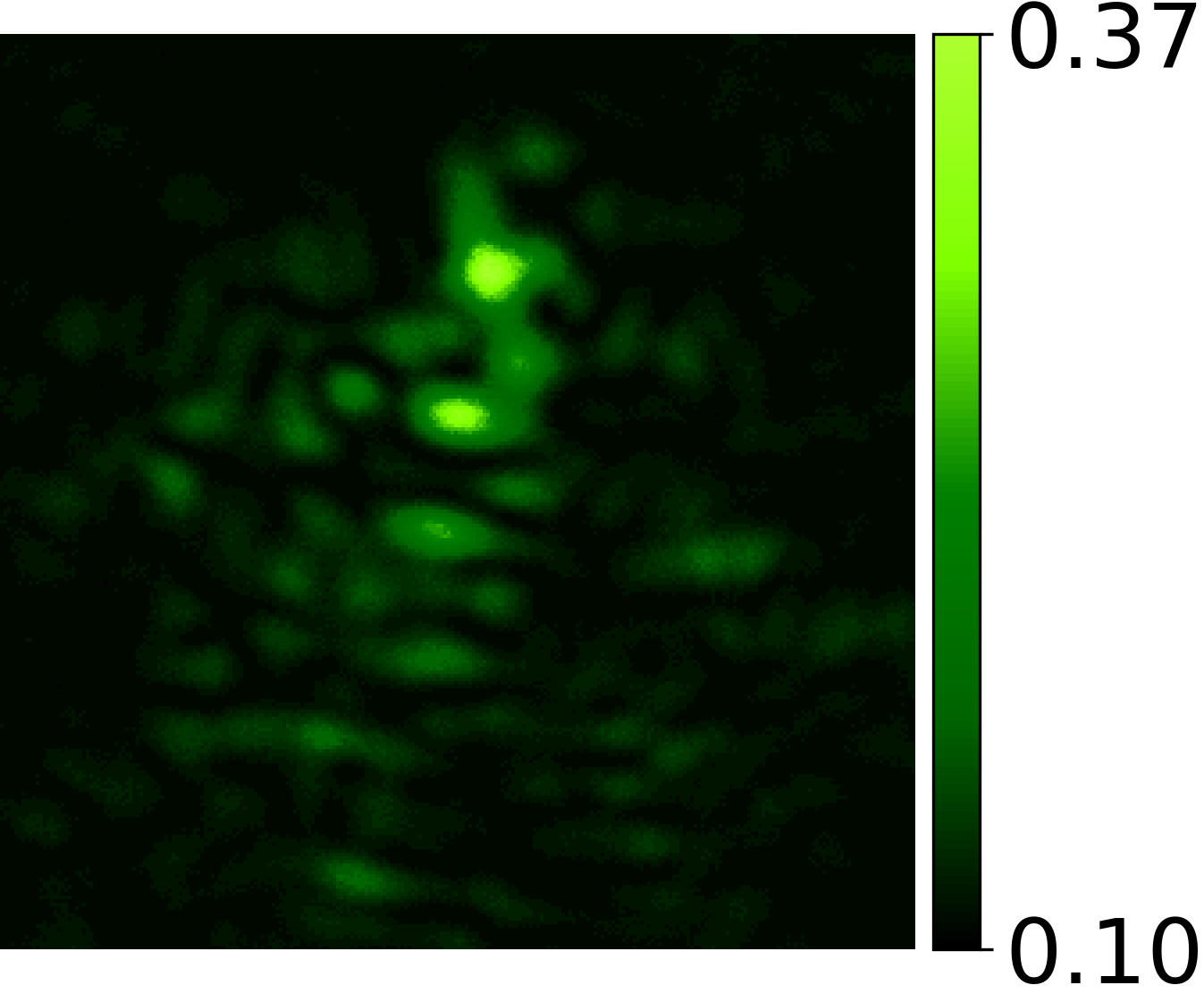}&
			\includegraphics[width= 0.2\textwidth]{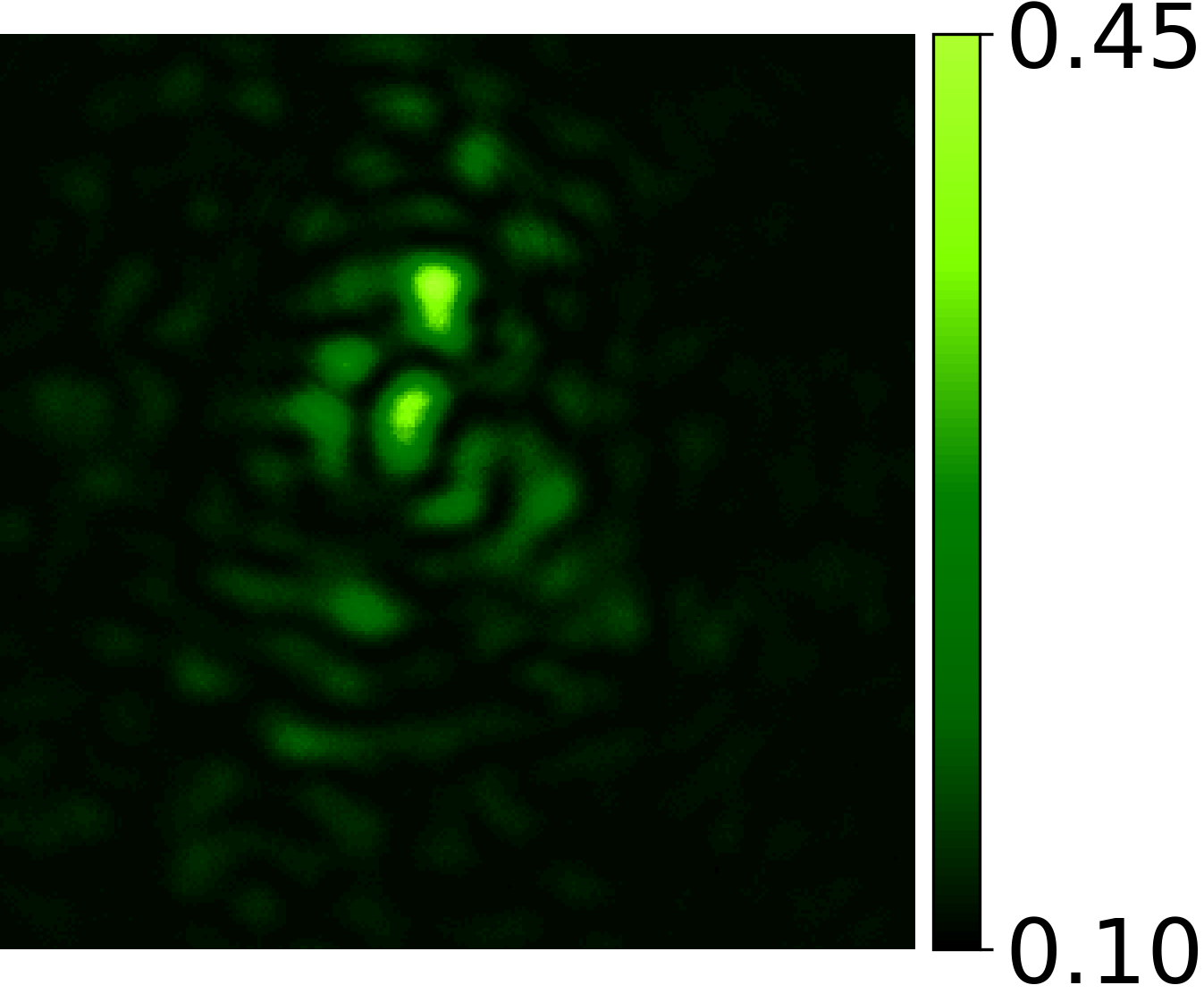}&
			\includegraphics[width= 0.2\textwidth]{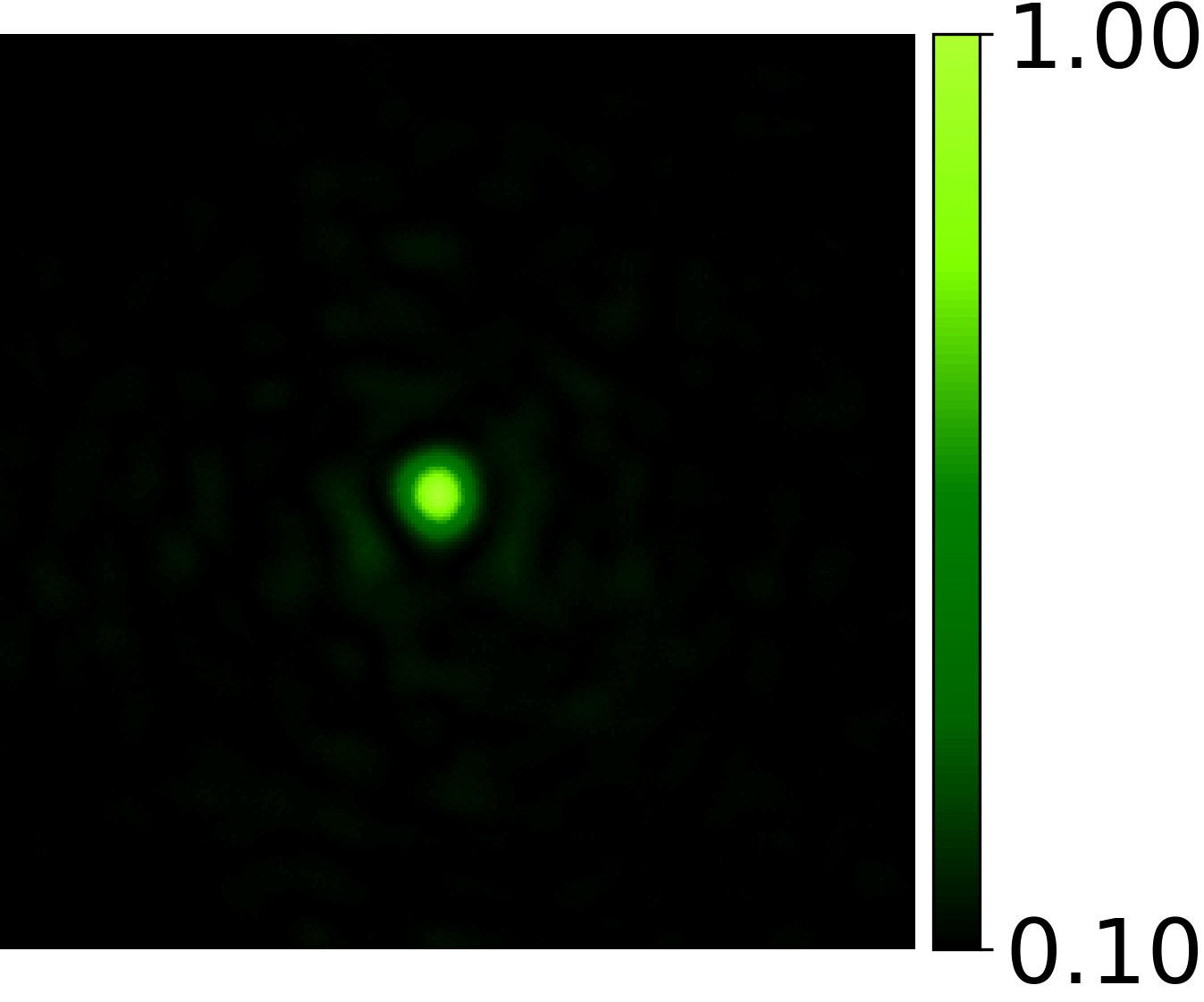}\\
			
			{\raisebox{0.9cm}	{\rotatebox[origin=c]{90}{~ {\small Phase mask} }}}&
			\hspace{-0.04\textwidth}\includegraphics[width= 0.15\textwidth]{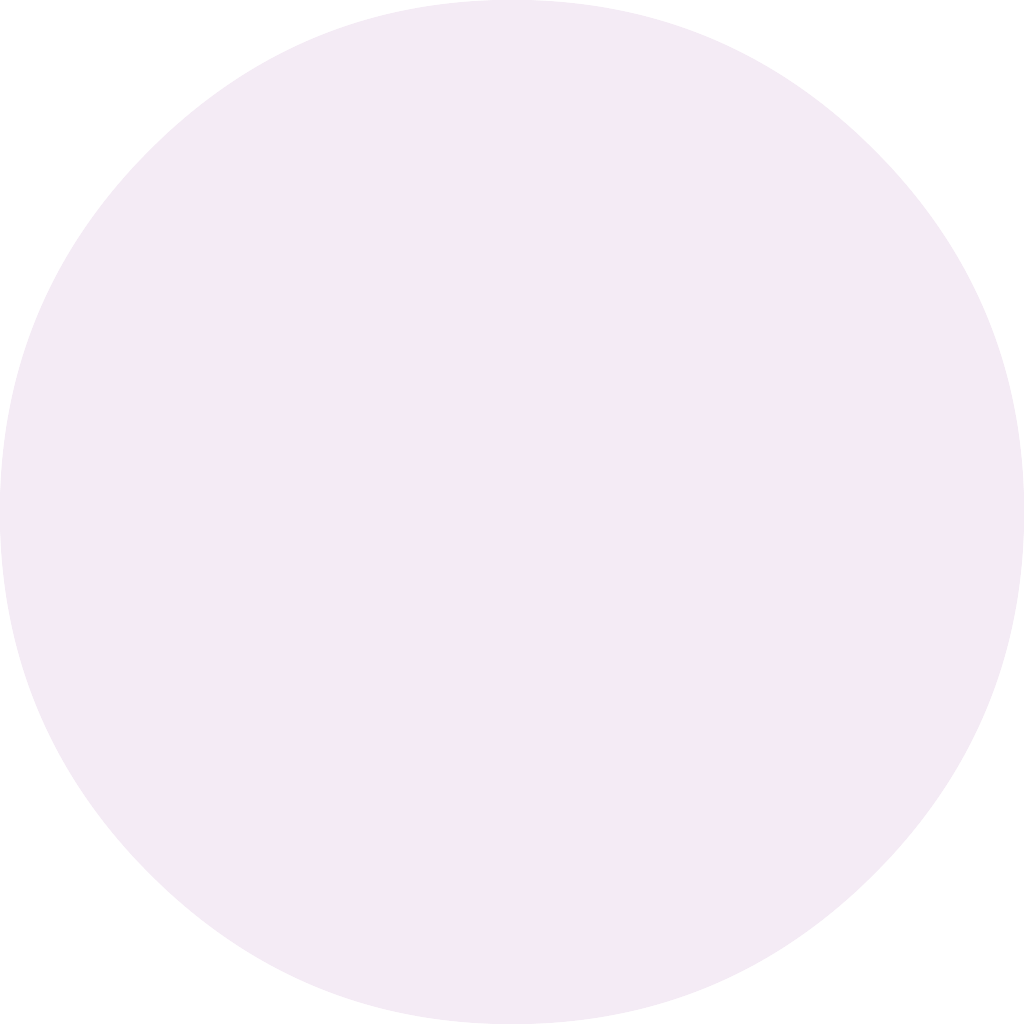}&
			\hspace{-0.04\textwidth}\includegraphics[width= 0.15\textwidth]{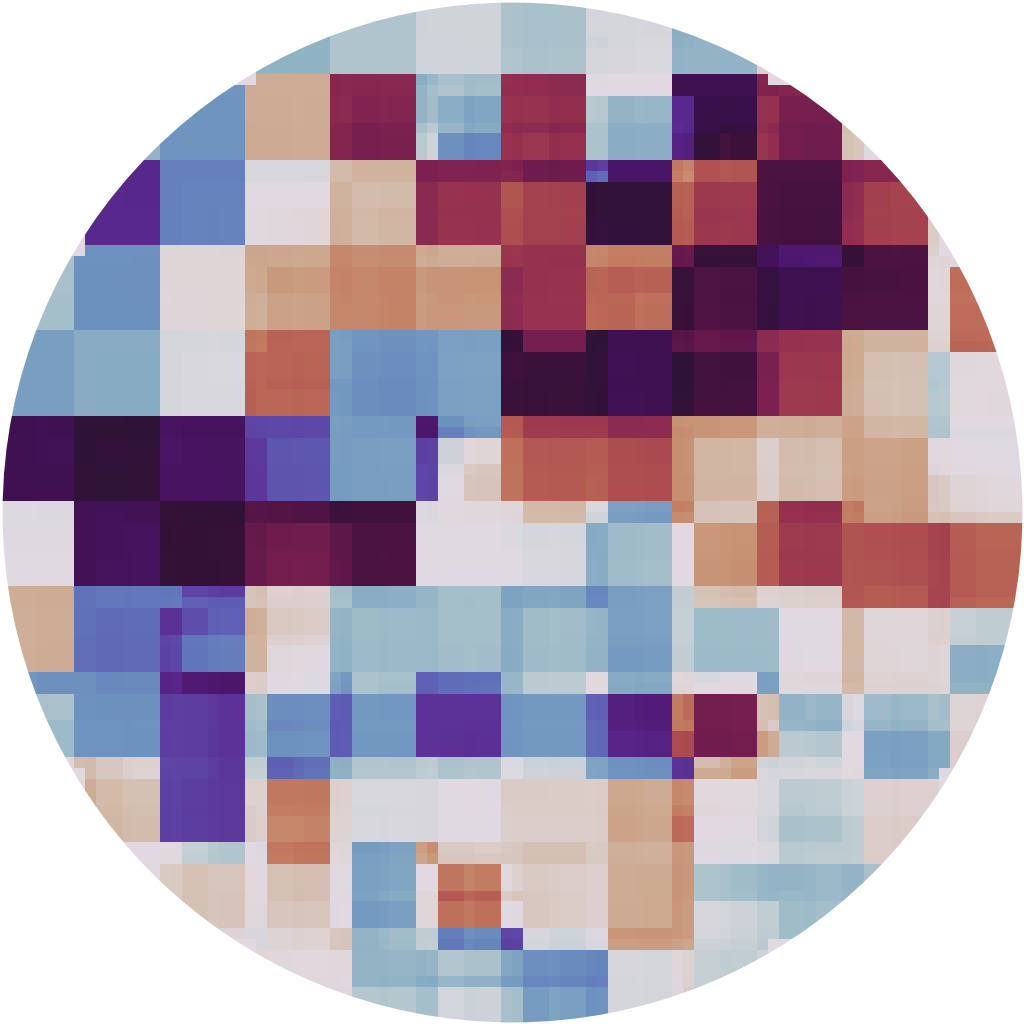}&
			\hspace{-0.04\textwidth}\includegraphics[width= 0.15\textwidth]{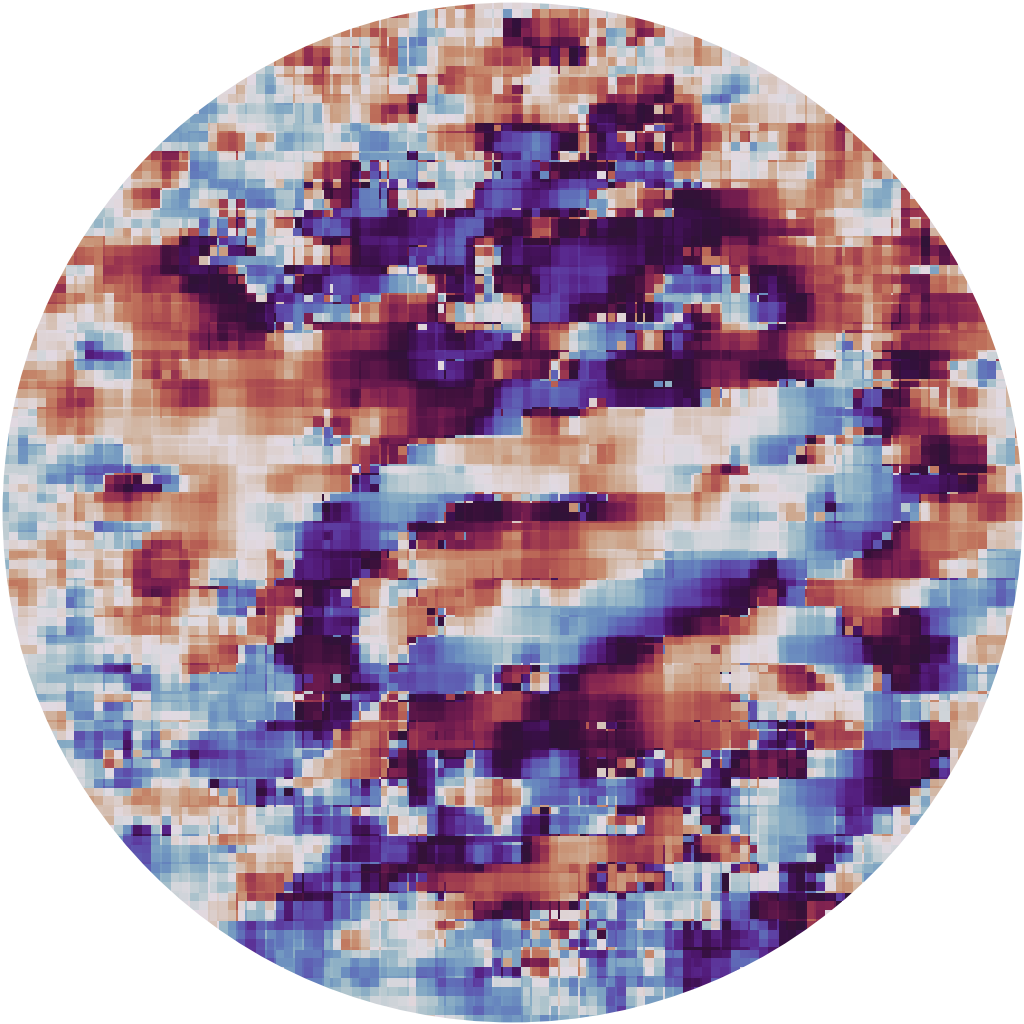}\\

		\end{tabular}
		\caption{\textbf{Impact of phase mask resolution:} 	We evaluate the effectiveness of gradient descent (GD) solutions using phase masks of varying resolutions. We place the chrome mask behind two layers of parafilm, creating a significant aberration. Results demonstrate that higher-resolution phase masks yield superior solutions, enabling convergence in both the main and validation cameras.}
	\label{fig:cmp_resultion}
\end{center}
\end{figure*}

%% file: fig_hadamard.tex
\begin{figure*}[t!]
	\begin{center}\begin{tabular}{{@{}c@{~}c@{~}c@{~}c@{~}c@{~}c@{~}c@{~}}}
			\multicolumn{1}{c}{}&
			\multicolumn{1}{c}{\hspace{-0.65cm} \small Init.}&
			\multicolumn{1}{c}{\hspace{-0.65cm} \small GD}&
			\multicolumn{1}{c}{\hspace{-0.65cm}\small CD}&
			\multicolumn{1}{c}{\hspace{-0.65cm}\small CD - OS}&
			\multicolumn{1}{c}{\small Score function}\\
			
			{\raisebox{0.67cm}	{\rotatebox[origin=c]{90}{~ {\small Main cam.} }}}&
			\includegraphics[width= 0.17\textwidth]{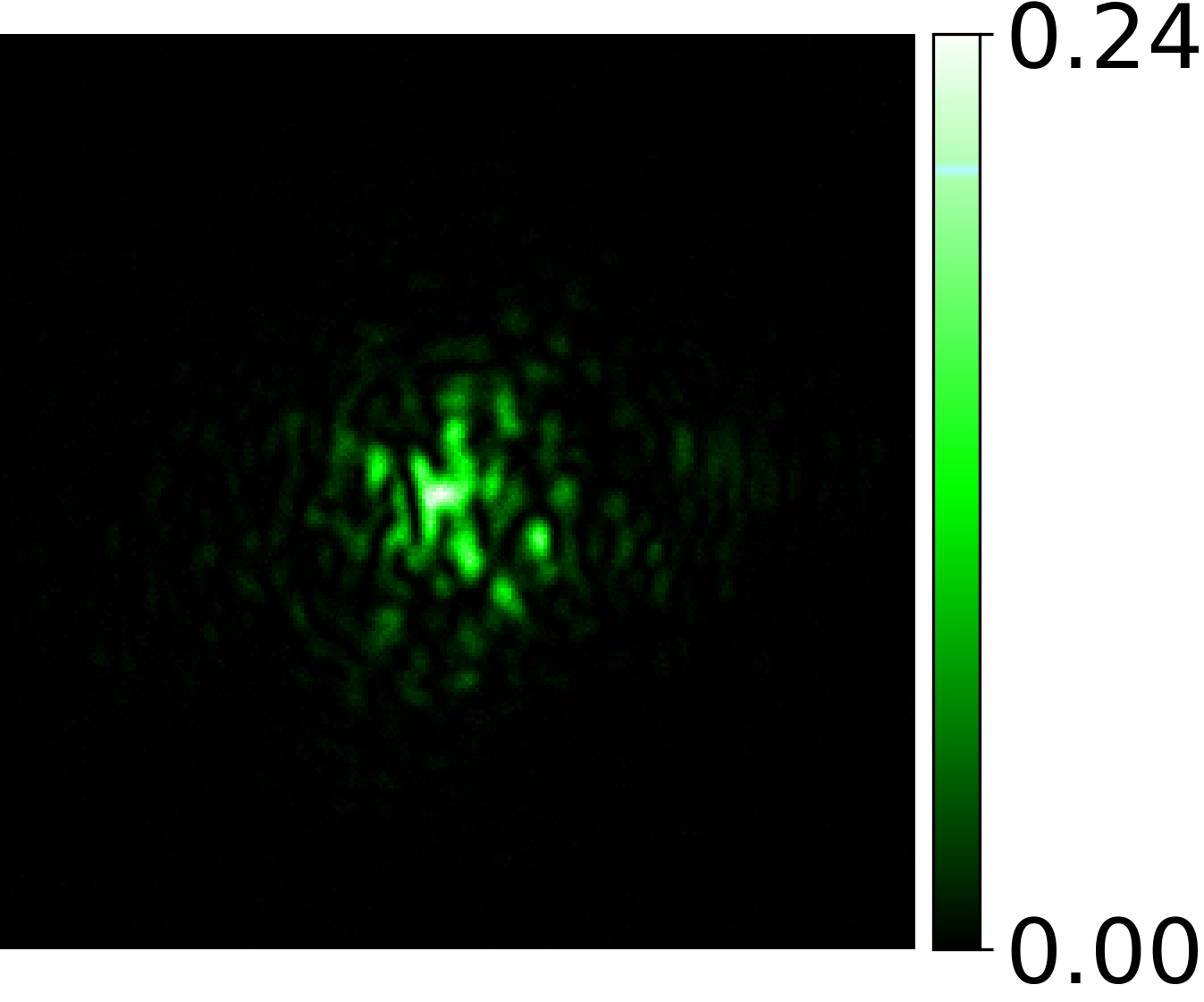}&
			\includegraphics[width= 0.17\textwidth]{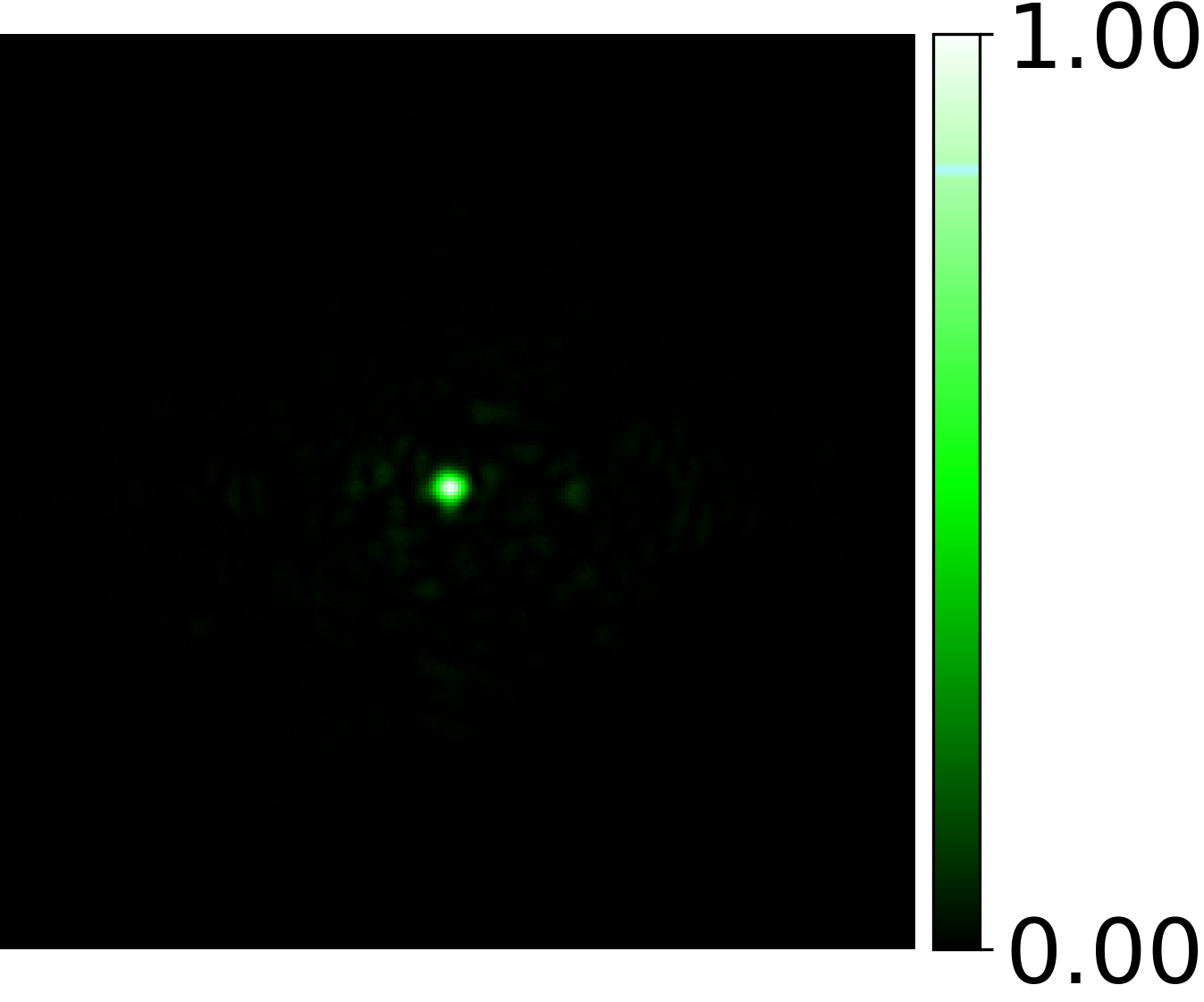}&
			\includegraphics[width= 0.17\textwidth]{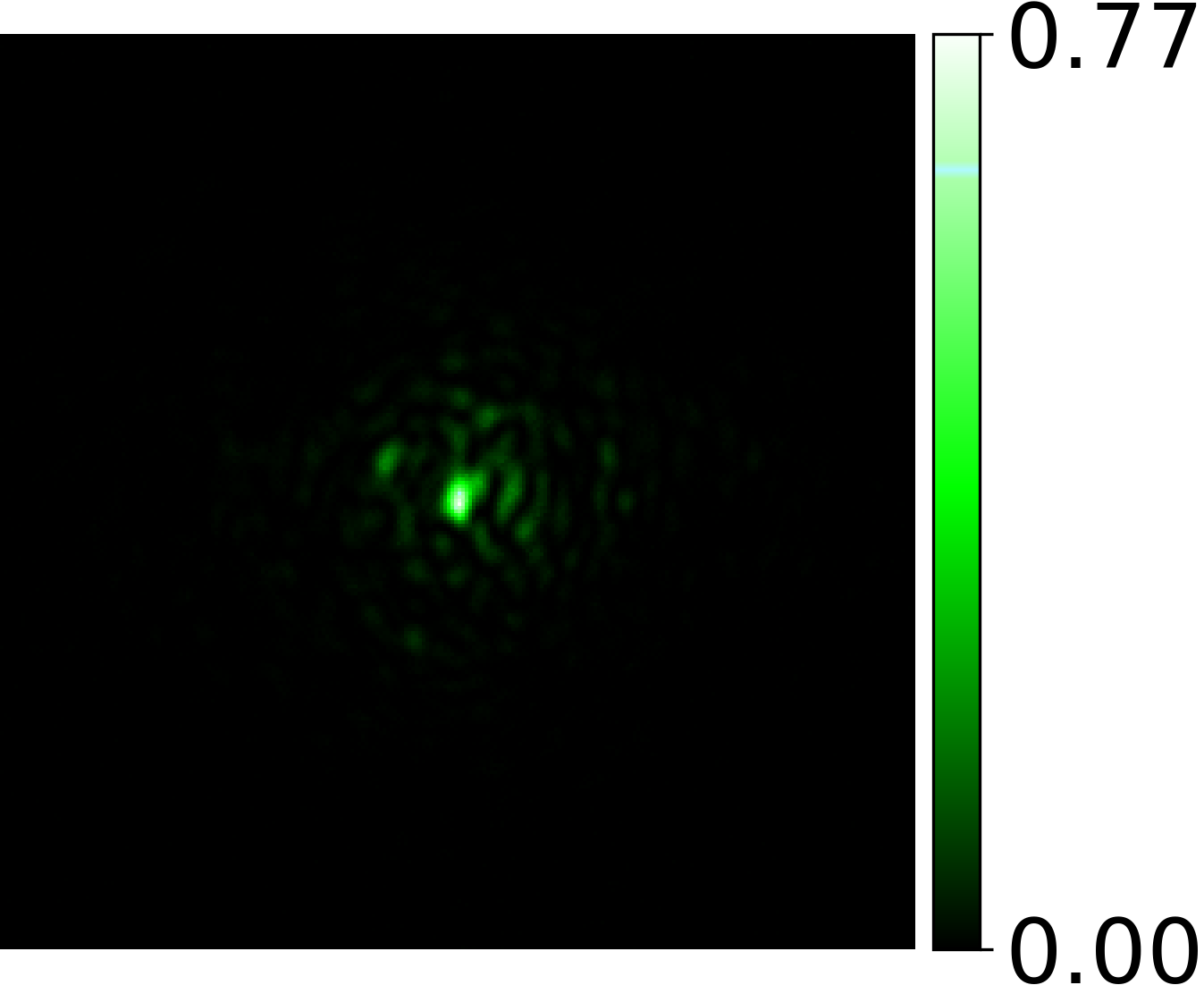}&
			\includegraphics[width= 0.17\textwidth]{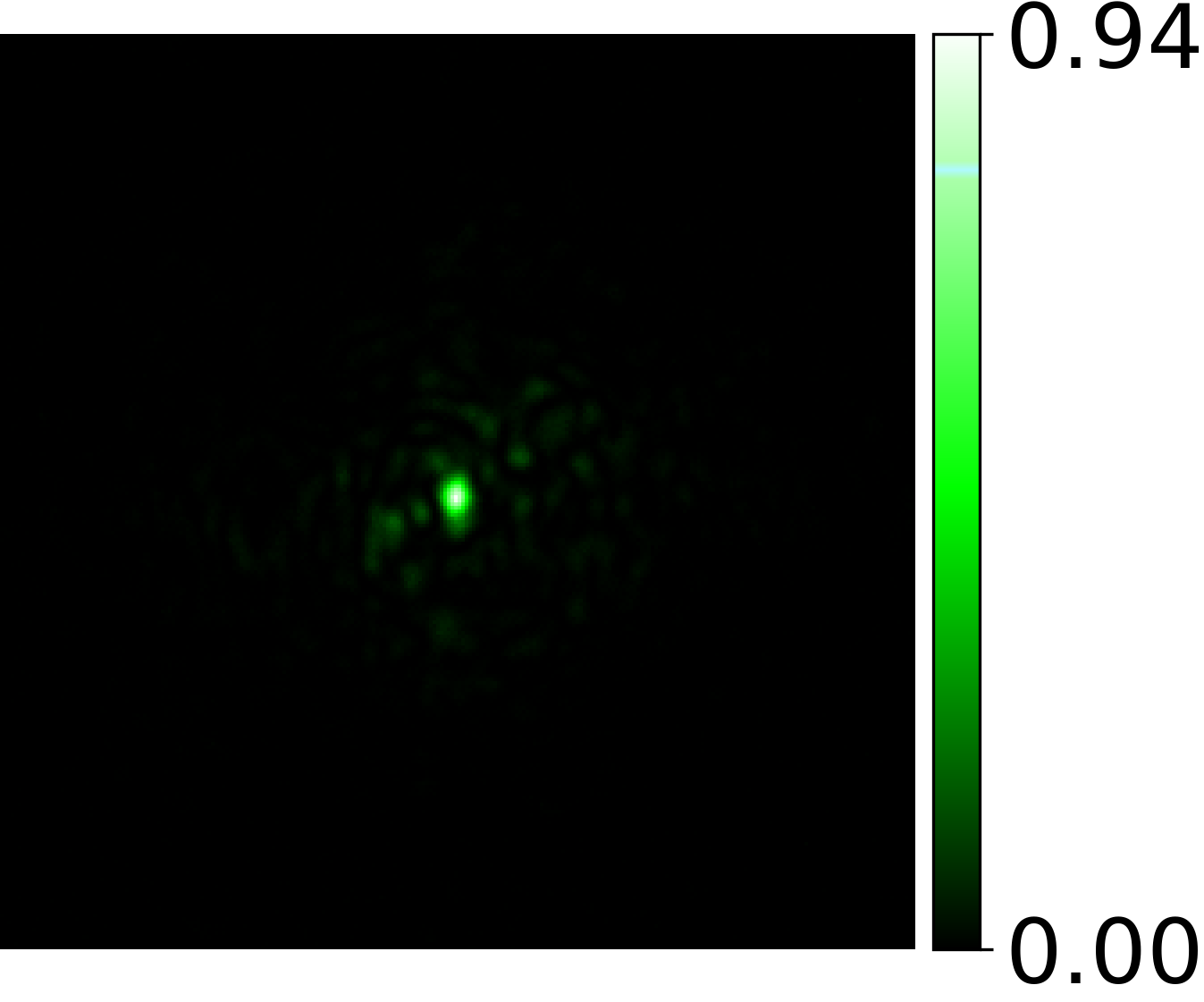}&
			\raisebox{-0.2cm}{\includegraphics[width= 0.25\textwidth]{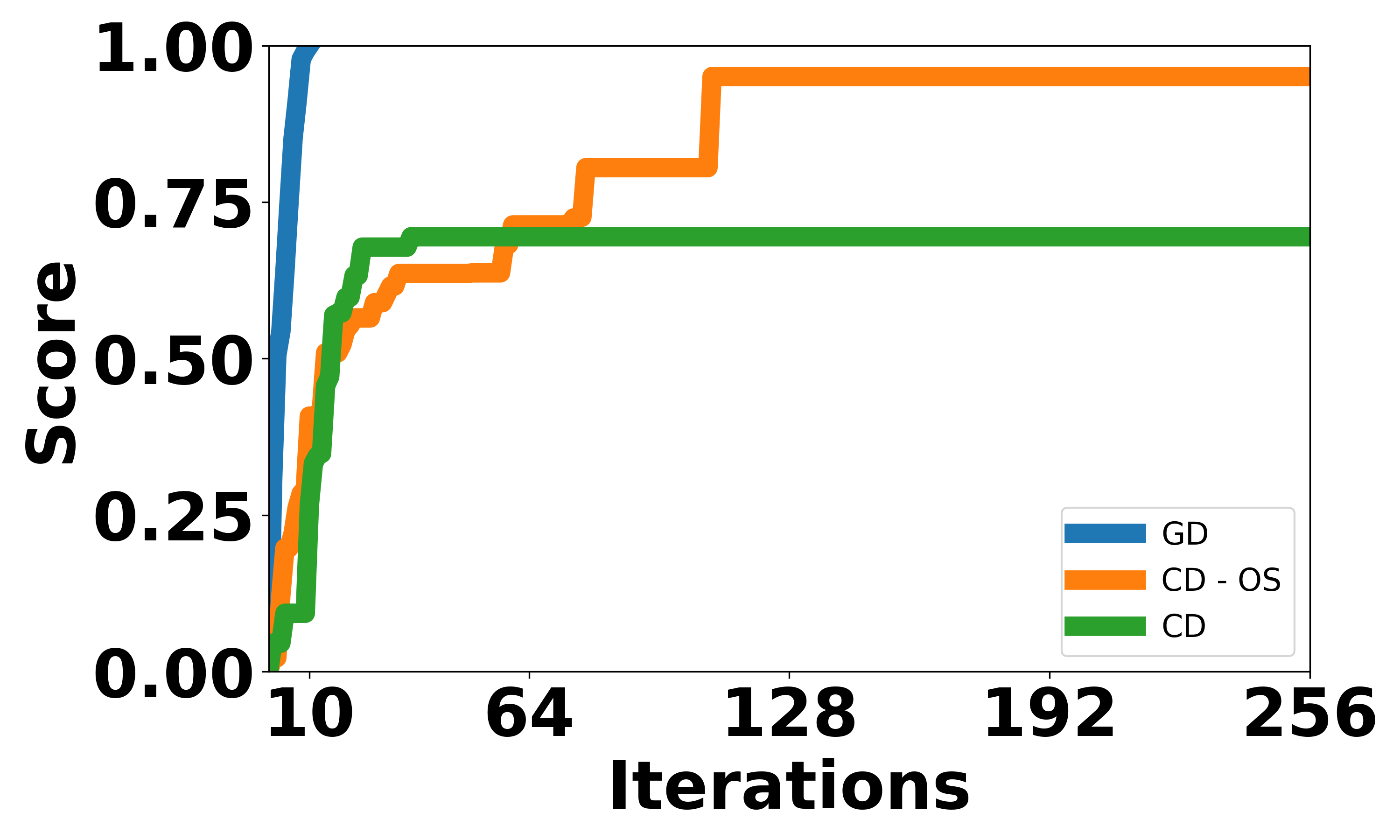}}\\
			
			{\raisebox{0.67cm}	{\rotatebox[origin=c]{90}{~ {\small Valid. cam.} }}}&
			\includegraphics[width= 0.17\textwidth]{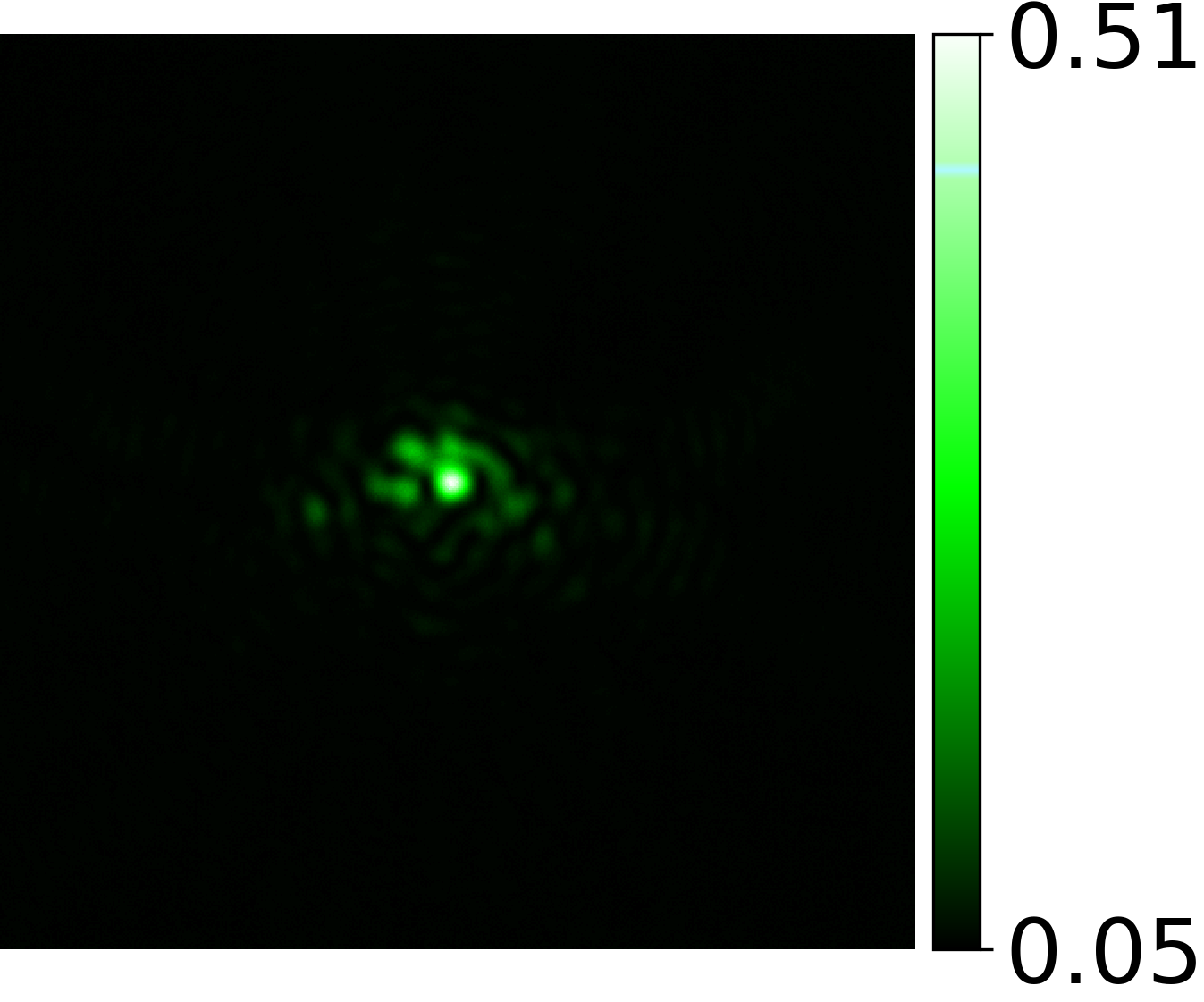}&
			\includegraphics[width= 0.17\textwidth]{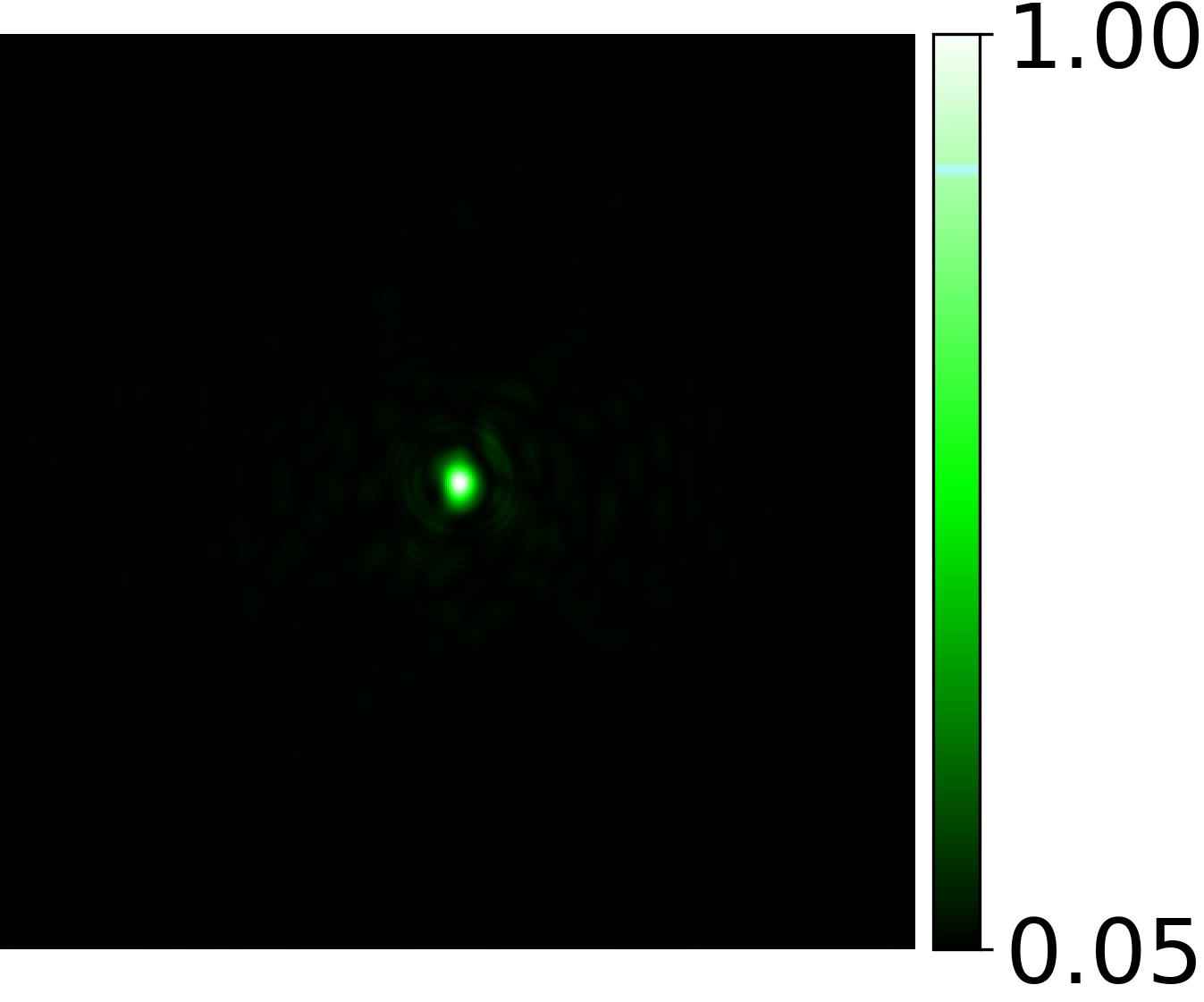}&
			\includegraphics[width= 0.17\textwidth]{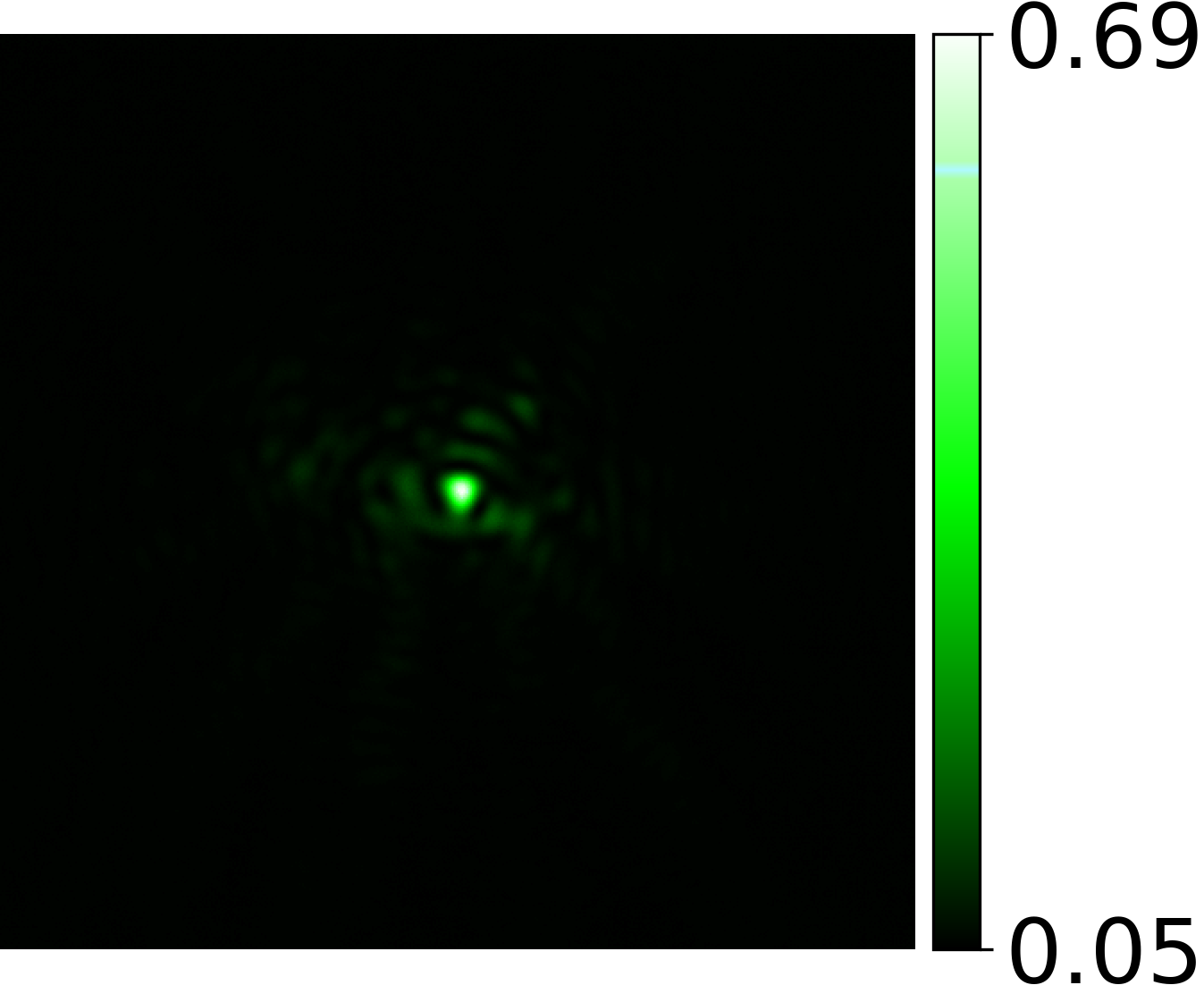}&
			\includegraphics[width= 0.17\textwidth]{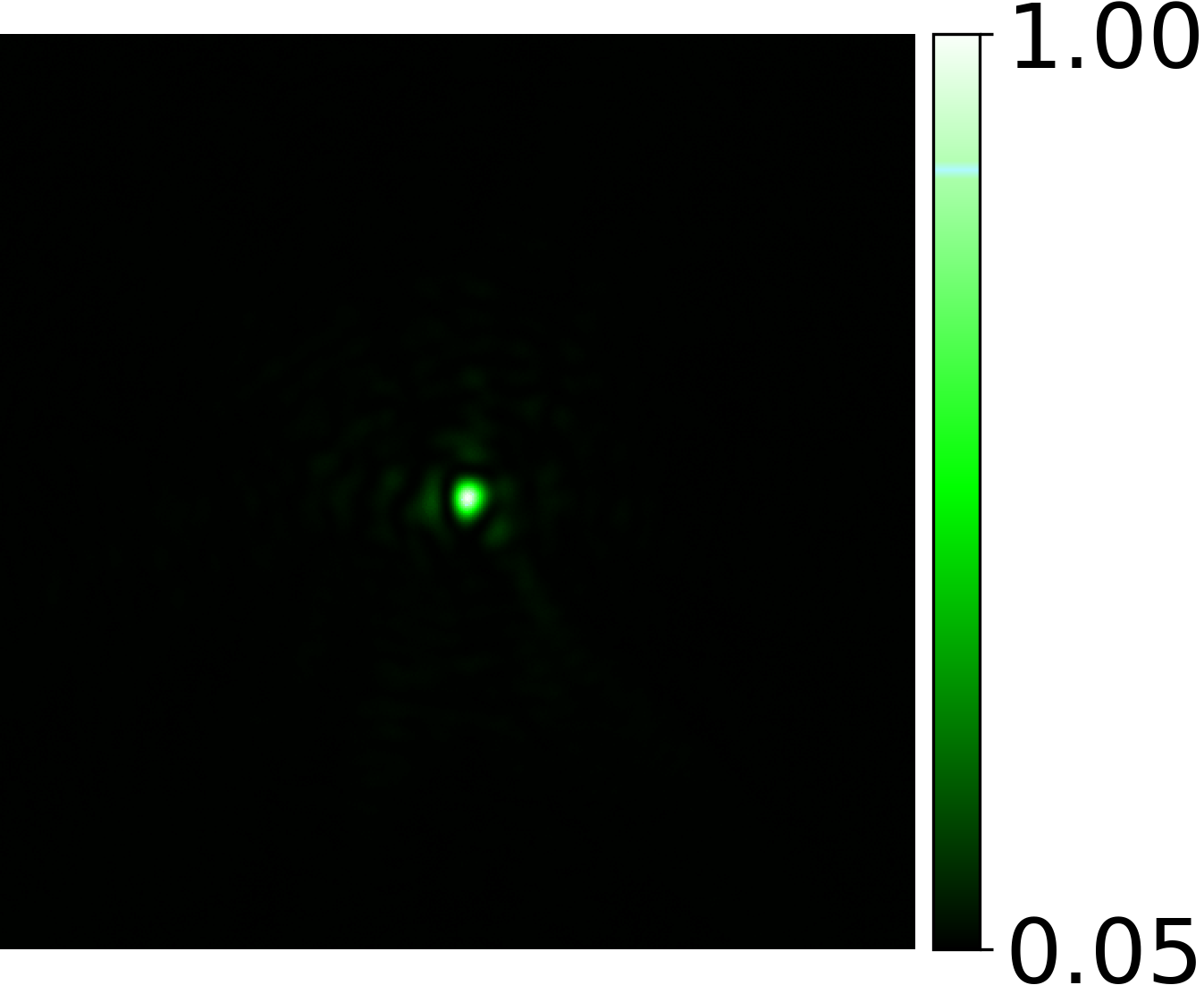}&\\
			
			{\raisebox{0.67cm}	{\rotatebox[origin=c]{90}{~ {\small Phase mask} }}}&
			\hspace{-0.03\textwidth}\includegraphics[width= 0.13\textwidth]{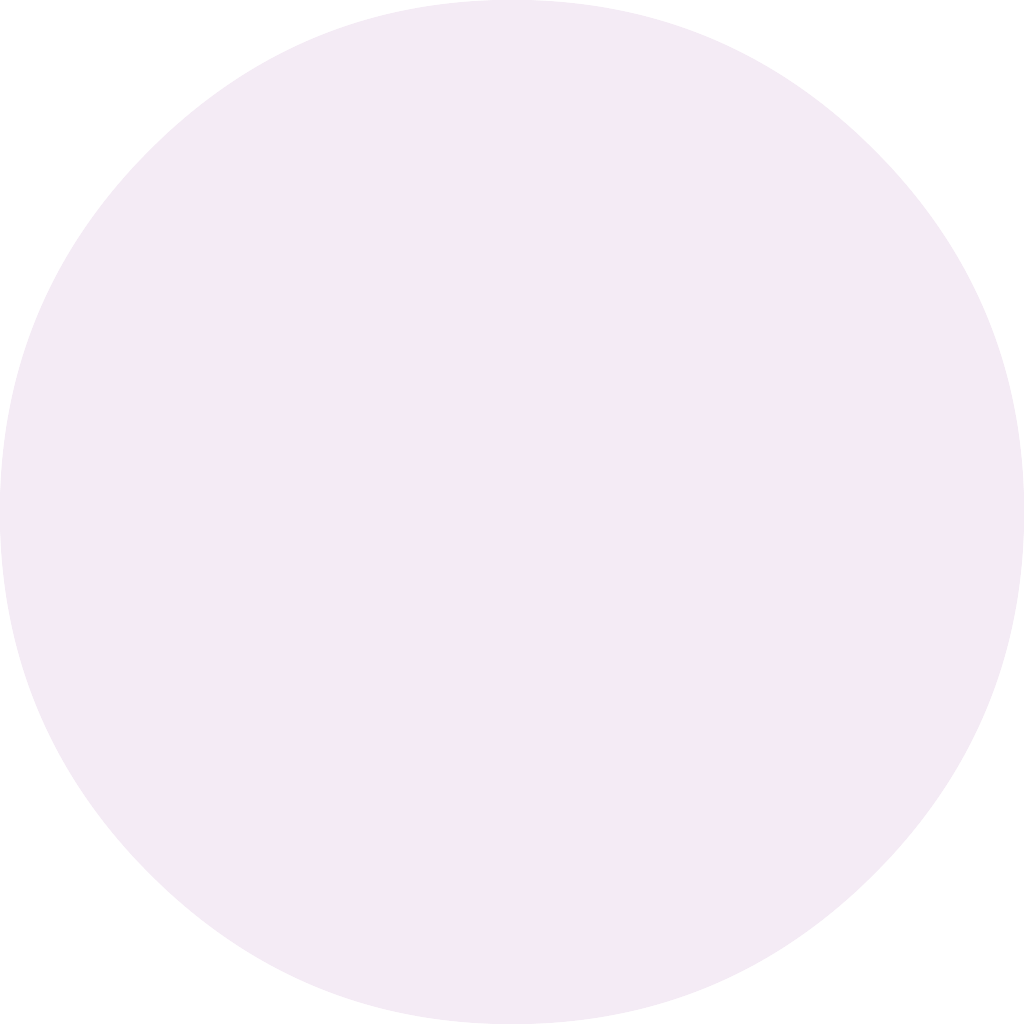}&
			\hspace{-0.03\textwidth}\includegraphics[width= 0.13\textwidth]{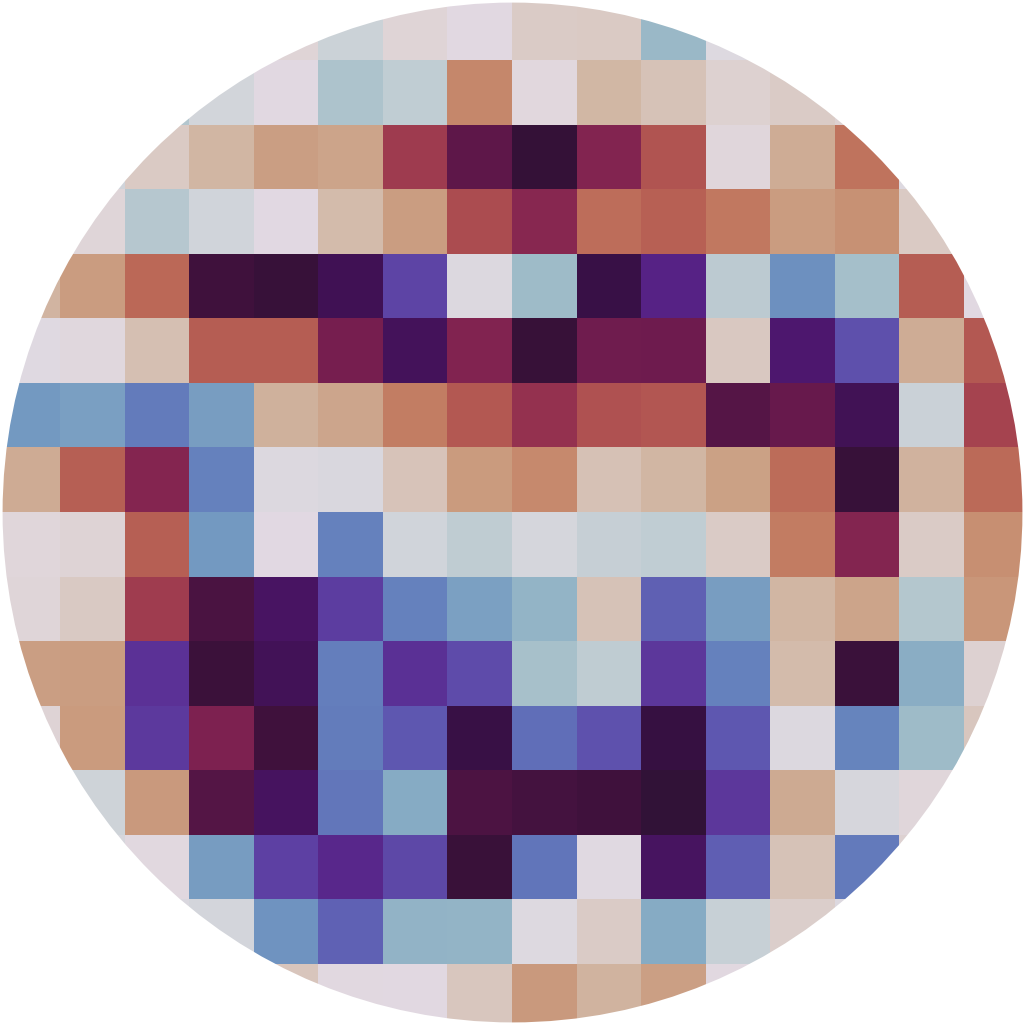}&
			\hspace{-0.03\textwidth}\includegraphics[width= 0.13\textwidth]{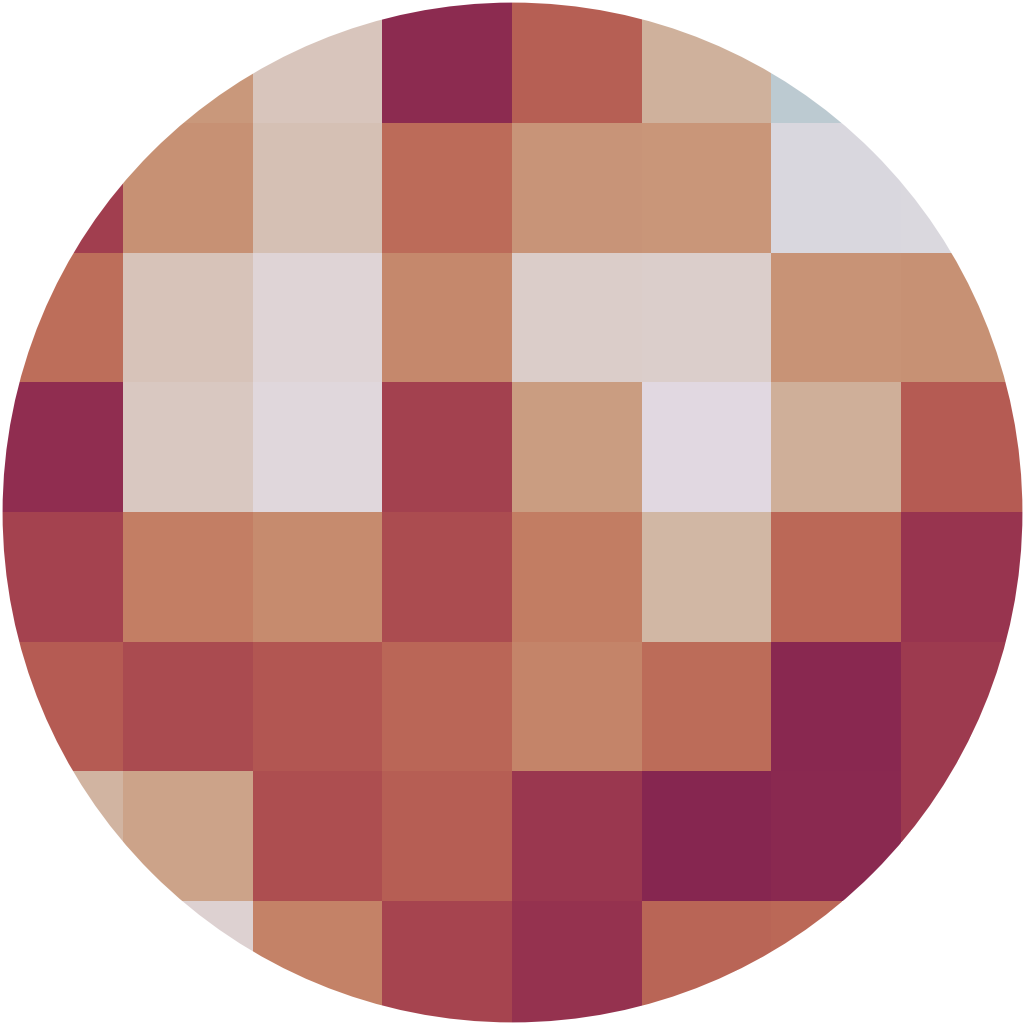}&
			\hspace{-0.03\textwidth}\includegraphics[width= 0.13\textwidth]{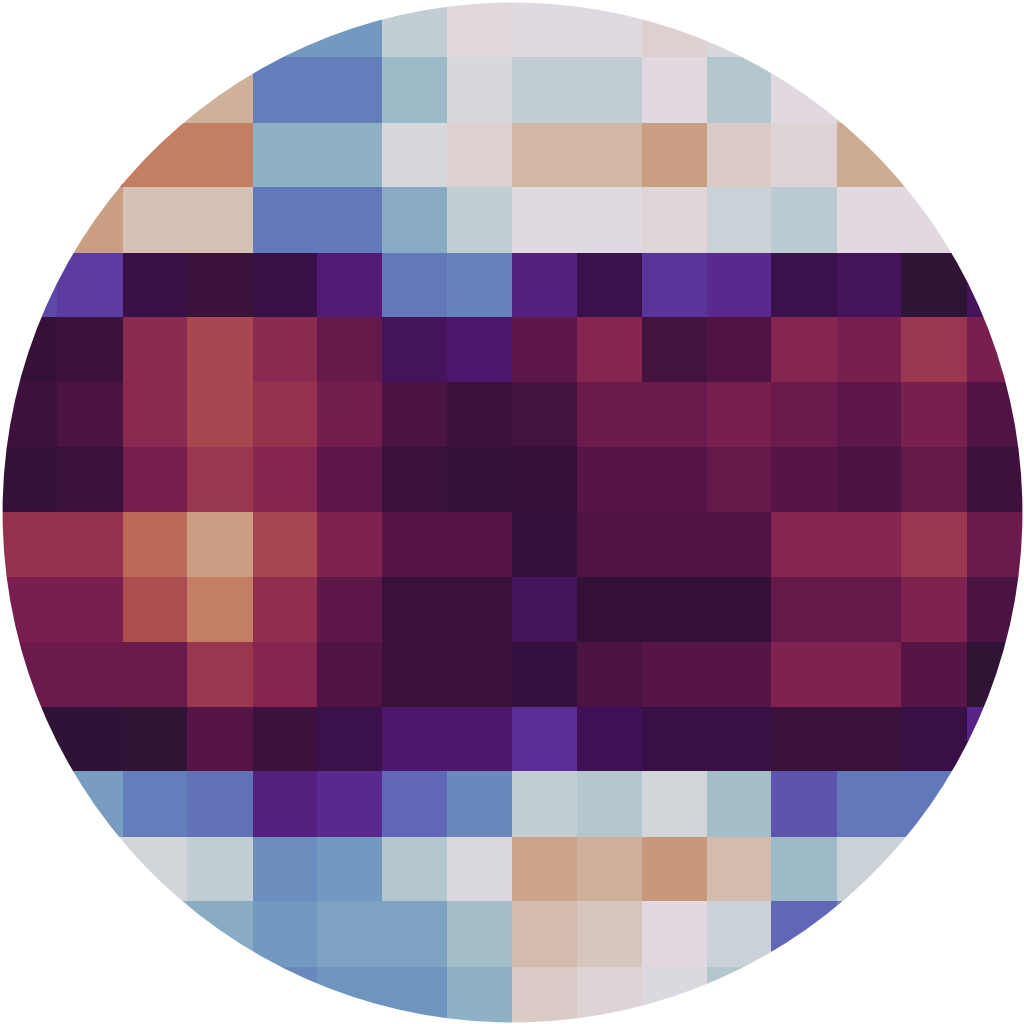}&
			
		\end{tabular}
		\caption{\textbf{Comparison of gradient descent and coordinate descent schemes:} We compare our gradient descent (GD) optimization to coordinate descent (CD) using the Hadamard basis. Due to CD's time-intensive nature, we optimize for a low-resolution mask and used a weak aberration volume (a single layer of parafilm tape). To address CD's susceptibility to noisy measurements, we test two configurations: Standard CD where we used 5 shots to estimate the phase of each basis element.   This approach could not improve the score beyond a certain point, because the improvement in the score is lower than the imaging noise.  We next used an over-sampled CD (CD-OS) capturing twenty-four images per basis element. This over-sampled version is less susceptible to noise and reached a higher  confocal score, at the cost of an even longer acquisition.   Overall, our results demonstrate that GD converges significantly faster than both CD variants and exhibits greater robustness to noise.}
		\label{fig:hadamarad}
	\end{center}
\end{figure*}

%% file: discussion.tex
\section{Discussion}
The challenge in computing a wavefront shaping modulation lies in its dependence on the unknown tissue structure. Consequently, most previous optimization strategies relied on slow coordinate descent strategies, which sequentially scan and query the modulation parameters one at a time.
Our contribution lies in deriving an analytic method to compute the gradient of the wavefront shaping score directly from the scattered wavefront. Crucially, we can compute the gradient in closed-form, despite the unknown tissue structure. This enables a transition from coordinate descent to gradient descent optimization, allowing simultaneous update to all SLM parameters. The resulting optimization strategy is orders of magnitude faster than previous coordinate descent approaches.  Moreover, our approach facilitates the estimation of high-resolution modulations,  leading to superior corrections when stronger scattering is present. 

We demonstrated the applicability of our gradient acquisition framework using a simple reflection-mode confocal-microscope setup. \bblue{This confocal system is inherently limited due to weak back-reflection of biological samples and the mixing of reflections from different depth layers, resulting in poor depth sectioning.  To enhance the utility of wavefront shaping corrections in real tissue imaging, our framework can be integrated into optical coherence tomography (OCT) or fluorescence microscope systems.
OCT enables imaging deep within tissue volumes by isolating content at different depths through path-length filtering.  However, in highly scattering tissues, OCT filtering is significantly affected by aberrations, which can be effectively corrected using a wavefront shaping system. Since OCT relies on coherent light, the optimization framework presented in this paper is directly applicable to OCT and can facilitate the development of a fast, aberration-corrected OCT system. }
\blue{Wavefront shaping is also highly effective in correcting the aberrations of fluorescence microscopes. We believe that a modification of our gradient formulation can be adapted to incoherent light sources. However, there are two main challenges for this approach. First, we need to retrieve the emitted complex wavefront from incoherent light sources. This could be tackled by an incoherent interferometer \cite{Chen:22, Chen_2024_CVPR,Horstmeyer15, YU2015632}. Second, the low photon count emitted by fluorescent targets makes it challenging to measure the gradient with sufficient SNR.}

Our rapid gradient-descent optimization algorithm resembles fast time-reversal algorithms~\cite{Dror22,doi:10.1121/1.424648,doi:10.1121/1.412285,Yang2014,Meng2012,Papadopoulos16} for wavefront shaping, which use the power iteration method to estimate the eigenvectors of the reflection matrix. The relationship with these algorithms is explored in detail in the supplementary file. We demonstrate that when the target area includes a single point, our derived gradient is equivalent to a power iteration on the reflection matrix.
However, as demonstrated in \figref{fig:cmp_size}, for coherent imaging, the largest eigenvector of the reflection matrix does not always  yield an optimal modulation, and adjustments to the power iteration method are not straightforward. In contrast, defining a target score and its gradient provides a principled method for guiding the modulation search towards desired outcomes. 

%% file: methods.tex
\section{Methods}

\boldstart{System design:}
Our setup is fully depicted in Fig. 1 of the supplementary material.
A collimated laser beam (CPS532 Thorlabs) illuminates a phase SLM (GAEA 2 Holoeye). The light is then focused with an objective lens (N20X-PF Nikon) on the target plane through the scattering material. The reflected light from the target is collected through the same objective lens and is modulated by a second SLM (LETO Holoeye). Finally, the light is imaged with a camera (Atlas-314S Lucid vision labs). The second SLM (LETO Holoeye) has two functions: create a phase pattern used for estimating the outgoing wavefront via a phase-retrieval algorithm. The second use is to place the conjugate of the retrieved phase in order to show confocal scanned images and calculate the score function. In our setup, the SLMs are placed in the Fourier plane of the target. Additional details on the setup, calibration and alignment can be found in the supplementary material. In Fig. 2 of the supplementary material we show how to calibrate our system with back illumination. 

\boldstart{Experimental targets:}
For our targets, we use high-reflecting chrome-coated mask (Nanofilm) and use in-house lithography process to create patterns with a $2\um$ resolution. For scattering material, we used chicken breast tissue with thickness of $130\um-240\um$ or a number of layers of parafilm~\cite{Boniface2020}, as stated in the text.
For all results of chrome-mask target except for the results in \figref{fig:cmp_size}, the target area for the algorithm was $\Area = 15.6\um \times 15.6\um$ and we scanned points inside this area at twice the Nyquist pitch, with a sampling interval of $1.3\um$. After the algorithm convergence, the final confocal scan was done with a diffraction limited sampling interval of $0.65\um$. For the target on the fourth row of \figref{fig:res-mask} the target spans an area larger than  $\Area = 15.6\um \times 15.6\um$. Hence, we ran the algorithm four times in partially overlapping areas and stitched the results together to form the final image. 


\blue{For the bead targets, we added quadratic phase in order to focus in different depths, similar to~\cite{baek2025three}. In \figref{fig:beads3d}, the target volume is $\Area = 10.4 \um \times 10.4\um \times 14\um$. The sampling interval in lateral axis was $0.5\um$ and sampling interval in the axial direction was $2\um$, which is half the diffraction limit.}

\bblue{For onion cell imaging the target area was $\Area = 26\um \times 26\um$ with sampling interval of $1.3\um$. To capture the full image in \figref{fig:onion} we ran the algorithm four times with partially overlapping areas and stitched the results together to form the final image.}

In our current setup, the runtime of the algorithm is mainly limited by two components. The first is the liquid-crystal SLM, which works at a rate of approximately $17Hz$ for a phase pattern to fully transform. The second is performing gradient acquisition using phase diversity optimization. Each iteration of our algorithm requires acquiring five images for phase diversity optimization for each sampling direction, solving a phase diversity optimization problem for each sampling direction, and performing a gradient step using backtracking line-search (usually a single measurement is needed). For the results in \figref{fig:hadamarad} this means a runtime of 14 minutes. This is in comparison to coordinate descent optimization, which took 900 minutes in the same setup. By adding a fast SLM (which currently run at $1.4kH$) and using off-axis interferogram~\cite{Kwon2023,haim2023imageguidedcomputationalholographicwavefront,Balondrade_2024,Najar2024,Choi2023AngWavelength}, runtime could be substantially decreased to mere seconds. In the supplementary material, we further suggest an even faster setup using point-interferometry, and in Fig. 4 in the supplementary material we present the suggested setup. We believe that with such a point diffraction interferometry implementation, we can estimate a modulation in less than one second.


%% file: supp_setup.tex
\section{Optical system}

\boldstart{System:} 
A drawing of our system is shown in \figref{fig:supp_setup}. A collimated laser passes through a polarizer to align the laser polarization with the SLM main-axis and is then expanded with a beam-expander (L1-L2). The wavefront is then modulated by the illumination SLM and reflected towards the sample. The modulated light passes through an objective lens and illuminates the sample. The forward scattered light continues to the validation camera, which is used for validating the focus of our algorithm on the image plane. The reflected light from the sample returns through the same objective lens and is reflected at a beam-splitter towards the imaging-SLM, which again modulates the wavefront. Finally, the light is collected by the main-camera sensor. The imaging-SLM has two functions: The first is to present the optimized pattern $\bu^\ell(\PrmVect)$ when measuring the score function and when confocal scanning a target. The second function is to present different defocus phase functions for capturing patterns for phase diversity optimization.

\input{fig_supp_setup}

A full component list of our system: 
$532nm$ laser (CPS532 Thorlabs), P - linear polarizer (LPNIRB100 Thorlabs), L1 - $100mm$ achromatic lens, L2 - $400mm$ achromatic lens, L3 - $150mm$ achromatic lens, L4 - $100mm$ achromatic lens, L5 - $200mm$ achromatic lens, TL - tube lens (TTL200-A Thorlabs), BS - beamsplitter (62-882 Edmund), OL1 - objective lens NA=0.5, MAG=$\times20$ (N20X-PF Nikon), OL2 - objective lens NA=0.7, MAG=$\times100$ (MY100X-806 Mitutoyo). Lens translation stage is a single-axis stage (PT1 Thorlabs) with a motorized actuator (Z925B Thorlabs). The sample holder was custom created with three translation stages (PT1 Thorlabs) allowing to adjust the target in all three axes. For the main camera we use A314S Atlas (Lucid Vision) and for the validation camera we use Grasshopper3 USB3 (Teledyne Flir).

\input{fig_supp_back_calib}

\boldstart{System alignment:} 
We start by explaining how we align our system. To modulate the Fourier transform of the wave, the illumination SLM needs to be at the focal plane of the lens following it (L3 in \figref{fig:supp_setup}), and the imaging-SLM at the focal plane of the lens before it (the same L3). We perform this alignment by using a third camera focused at infinity (we place an appropriate laser line filter ($532nm$) and focus the camera on a far building). Then we place two polarizers: before the SLM at a $45^o$ angle and after the SLM at a $135^o$ angle. Putting the polarizers in these angles converts the phase modulation to intensity variation (a dark pixel of the SLM pixel if it does not modulate the light, and a white pixel if it modulates by $\pi$). We then focus on the SLM through the relevant lens, forming a relay system. We display a checkerboard pattern on the SLM and adjust the distance between the SLM and the lens until we get a sharp image. We also calibrate the distance between the main/validation cameras and the lenses before them (L4/L5) to ensure they are focused at infinity.

\boldstart{Finding the active SLM area:} 
After the system is aligned, we identify the relevant illuminated areas on both SLMs that pass through the objective lens aperture. The size of these relevant areas can be calculated based on the aperture size and pixel size. We describe the calibration procedure for the illumination SLM; a similar procedure is followed for the imaging-SLM. To find the center $(x_c,y_c)$, we use the following procedure: We start by placing a mirror on the sample holder and moving the sample holder until we achieve focus at the main camera. To get an initial estimate of the centers we present a sinusoidal pattern of the same size as the aperture on the SLM and move it on $x-y$ axes of the SLM and look when we get a maximum intensity at the target pixel (if we do not modulate the right area on the SLM less light is directed to that target pixel), this gives us an initial estimate of  $(x_e,y_e)$.  We then place a defocus pattern (quadratic phase $((x-(x_c-x_e))^2+(y-(y_c-y_e))^2)$) on the illumination SLM and move the sample holder axially until again we achieve focus on the main camera. Ideally, if $(x_c,y_c) = (x_e,y_e)$ the location of the laser spot will be at the same $x-y$ location no matter what quadratic phase we introduce. If the location of the laser spot is different, we modify our estimation. We repeat this until the focus does not change. This is then repeated for the imaging-SLM while the illumination SLM presents a blank phase. 

\boldstart{Mapping of SLM to camera:} 
Next, we calibrate the mapping between the SLMs and the main camera. Using the focal length of the lens before the camera (L5), the SLM pitch, and the wavelength of the emitted light, we can map between SLM frequencies to the camera pixel using simple geometry. However, this mapping does not account for misalignment in the rotation of the SLMs. To overcome this, we fine-tune the calibration by displaying sinusoidal patterns and recording the shift presented on the main camera. Using these measurements, we estimate the rotation of the SLM.

\boldstart{Mapping between SLMs:} 
Finally, we need to accurately map every pixel on the illumination SLM to the corresponding pixel on the camera SLM. To do this, we follow a similar procedure as in~\cite{DrorNatureComm24,Dror22}. However, instead of using a fluorescent bead, we illuminate the target from the back, bypassing the illumination SLM. That is, we place mirrors so the laser is directed through O-L2, and after the sample the light continues to the main camera, as shown in \figref{fig:supp_back_setup}. We then recover the modulation pattern by applying phase-diversity optimization. The Helmholtz reciprocity (phase conjugation) principle~\cite{MUGNIER20061} dictates that if we place the conjugate of the recovered wavefront on the illumination SLM, it will focus into a point behind the scattering tissue. Hence, we redirect the laser to the forward path \figref{fig:supp_setup} (i.e. through the illumination SLM and O-L1) and place the modulation pattern on illumination SLM. We then expect to achieve a focused point on the validation camera. However, if the mapping between the SLMs is incorrect, we do not get a focused point on the validation camera. We can then find the mapping between the SLMs by shifting the pattern on the illumination SLM until the energy of the focused point on the validation camera is maximal.
 
\boldstart{Determining target area size $\Area$:} \blue{As demonstrated in Fig. 4 in the paper, to focus light inside the tissue, we need to optimize our score over a target area $\Area$ (also known as isoplanatic area). However, there is a compromise between the intensity of focus inside the tissue and the intensity of the focused spot on the main camera. In our work, to decide on the area size, we tried a few isoplanatic area sizes for each sample and kept the ones providing the best results. Once we found a good area for one isoplanatic window we could scan many windows in the same sample using the same $\Area$ parameter and stitch them together. An alternative approach would be to use a multi-scale $\Area$ size as presented in~\cite{Najar2024}, where we start with a large area and decrease its size as we increase the iterations.}

\boldstart{Tilt-shift parameter calibration:} 
As our score maximizes the confocal intensity over an area, we rely on the tilt-shift memory effect. To apply the scan, we need to recover the parameter determining the ratio between the tilt and shift. To determine this parameter, we again follow a similar calibration method as in~\cite{DrorNatureComm24,Dror22}. However, since they rely on fluorescent beads, we instead illuminate the target from the back, bypassing the illumination SLM. We then recover the modulation pattern by applying phase-diversity optimization. We then redirect the laser to the forward path (i.e., through the illumination SLM and O-L1) and place the modulation pattern on the illumination SLM and we use the validation camera to view the focused spot. We then adjust the ratio between tilt and shift of the modulation pattern so that we can move the focused spot in the validation camera while preserving maximal intensity.

\boldstart{Gradient step:} 
When performing gradient descent optimization, we need to choose a step-size at each iteration. In our system, we use backtracking line search~\cite{Armijo1966}. For each iteration, we start with a predefined step size and after performing a step, we measure the score function. If the score function increases, we perform the step; however, if the score decreases, we do not perform the step. Instead, we divide the step size by two and again measure the score. We repeat this until the score increases or until the step size is smaller than some threshold. If the score does not increase, we perform a large step (with the initial step size) as the optimization might be stuck in a local maximum. At the end of optimization, we choose the phase modulation pattern that scored the highest.

%% file: fig_supp_setup.tex
\begin{figure*}[t!]
	\begin{center}
		\includegraphics[width= 0.95\textwidth]{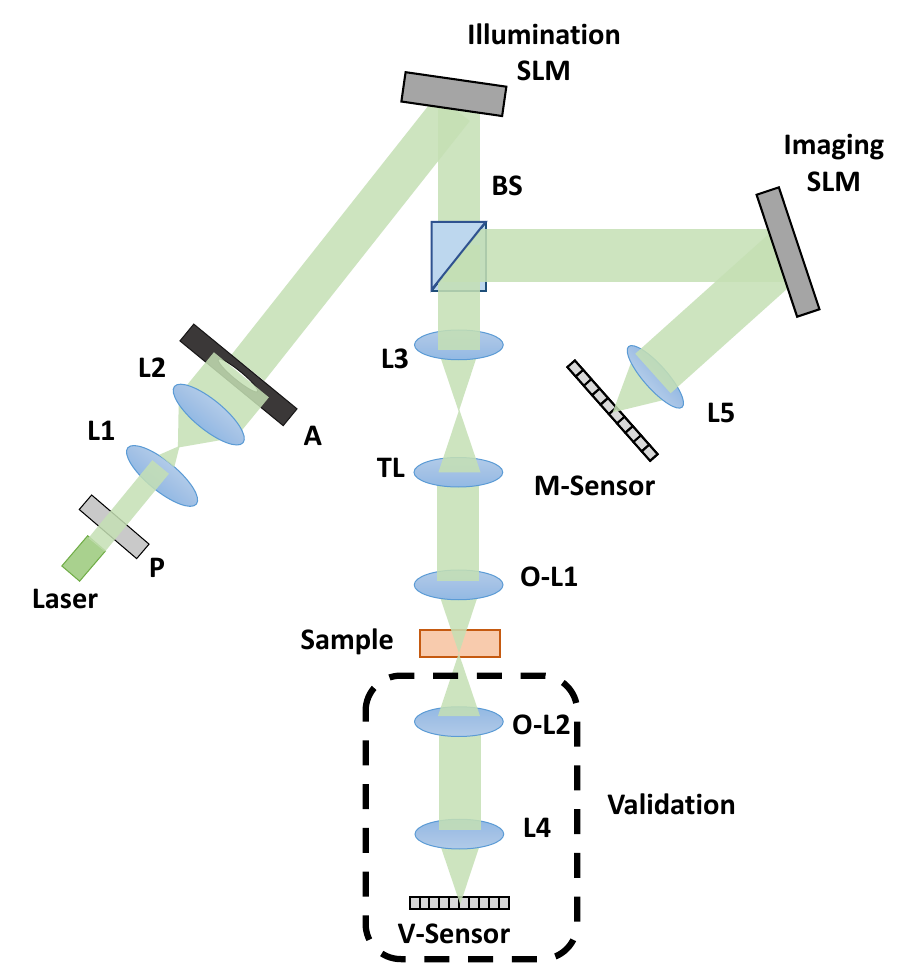}
	\end{center}
	\caption{{\bf{Setup:}} we present the full setup of optical system. Light emitted by the laser is reflected from the illumination-SLM towards the sample. The reflected light from the sample returns through the same path where it is reflected from the imaging-SLM and is then collected by the main-sensor. The transmitted light from the sample continues to the validation sensor which is only used to validate our results. P:polarizer, BS:beam-splitter, L-lens, O-L:objective lens, V-sensor:Validation sensor, M-Sensor:Main sensor.} 
	\label{fig:supp_setup}
\end{figure*}

%% file: fig_supp_back_calib.tex
\begin{figure*}[t!]
	\begin{center}
		\includegraphics[width= 0.95\textwidth]{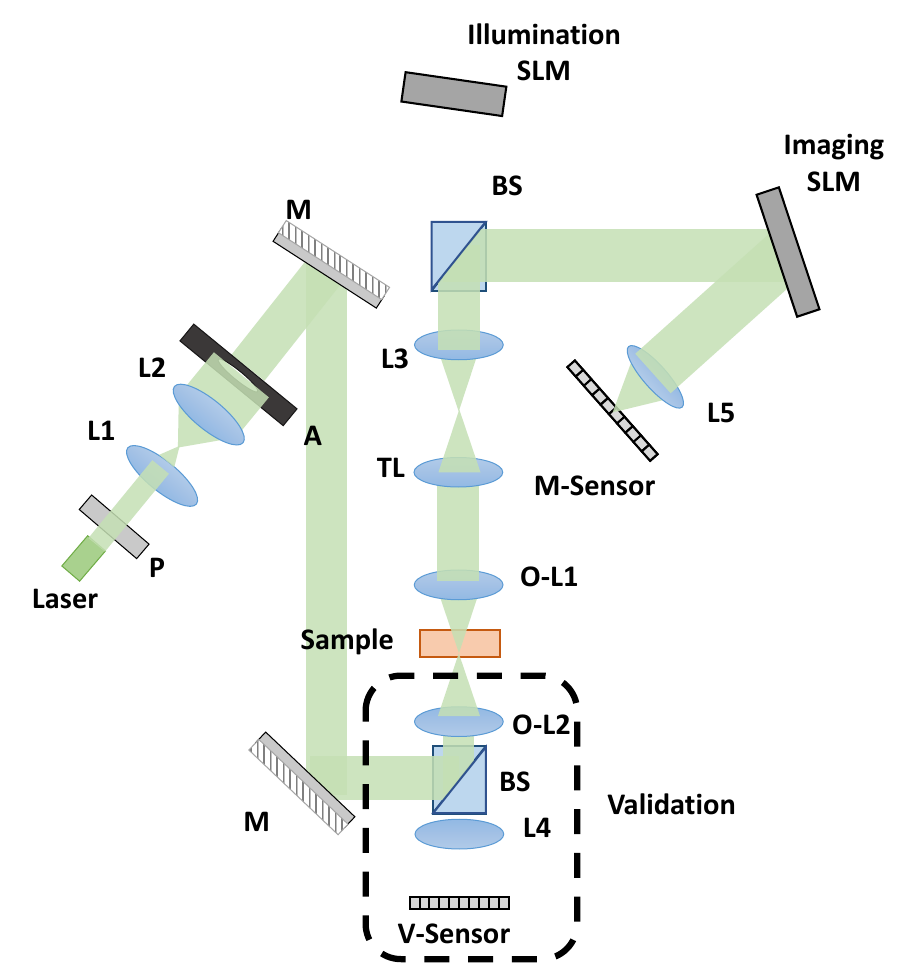}
	\end{center}
	\caption{{\bf{Back calibration:}} We use a back-path laser to calibrate our system, here we show that laser path using the back-path. }
	\label{fig:supp_back_setup}
\end{figure*}

%% file: score_sup.tex
\section{Justifying the confocal area score}
\blue{
To find a good wavefront shaping modulation, our algorithm relies on a score function that measures the confocal intensity over a small area $\Area$. To do this, we scan the area by tilting and shifting the modulation toward a point $\ell\in\Area$ and collect the averaged confocal intensity:
\BE\label{eq:score-confocal-area}
\Mtric(\PrmVect)\equiv \sum_{\ell\in \Area} |\bu^\ell(\PrmVect)^T\RM \bu^\ell(\PrmVect)|^2.
\EE
This score was used by several recent approaches for digital aberration correction~\cite{Balondrade_2024,zhang2024deepimaginginsidescattering,Choi2023AngWavelength}.
They optimize over finite isoplanatic patches to maximize
the diagonal elements of the reflection matrix, and the sum of these diagonal elements is equivalent to our confocal area score. 
To better explain why this score favors focusing wavefront-shaping modulations, we provide a derivation below.}
\blue{
For that, let us denote by $\TM$ the transmission matrix describing the propagation of light from the SLM plane to a mirror attached at the back of the scattering volume.  It has been shown that the reflection matrix can be expressed as a double pass through the transmission matrix~\cite{Balondrade_2024,zhang2024deepimaginginsidescattering,ChoiLyers2023}
\BE
\RM=\TM^T\cdot\TM,
\EE where $\TM^T$ is the transpose of the transmission matrix (this is just transpose, not a conjugate transpose).}

\blue{
We now consider an incoming modulation $\bu^\ell(\PrmVect)$ and denote the complex wavefront it generates at the back of the tissue as
\BE
\bou^\ell=\TM \bu^\ell(\PrmVect).
\EE 
}

\blue{We will express the confocal area score using the wavefront at the back layer and show that under idealized memory effect correlation, this score is equivalent to the nonlinear score of two-photon fluorescent excitation~\cite{Katz:14}, or the single photon confocal fluorescent feedback~\cite{DrorNatureComm24}. It has already been shown that such a score is maximized by focusing modulations, namely, when $\bou$ is a one-hot vector that brings all the energy into a single point and has zero intensity over the rest of its entries.}

\blue{\begin{claim}
	Let $\bou^o=\TM \bu^o(\PrmVect)$ denote the wavefront at the back of the tissue where the incoming illumination is directed to a point $\ell_o$ at the center of the patch $\Area$. If memory effect correlation holds over the area $\Area$, the confocal area score of \equref{eq:score-confocal-area} reduces to:
	\BE\Mtric(\PrmVect)=\sum_x |\bou^o(x)|^4, \EE
	where $x$ is a 2D position vector, running over the coordinates of $\bou^o$. 
\end{claim}}

\blue{
\proof	
The confocal intensity resulting from one modulation tilted toward point $\ell$ can be expressed as:
\BE
I^\ell(0)=\left|\sum_x (\bou^\ell(x))^2 \right|^2,
\EE
where $x$ runs over points on the back plane of the tissue, and at each such point we square the complex field while maintaining its phase (this is not an absolute value).
The confocal area score can then be expressed as: 
\BE\label{eq:conf-area-v}
\Mtric(\PrmVect)=\sum_\ell I^\ell(0)=\sum_\ell\left|\sum_x (\bou^\ell(x))^2 \right|^2
\EE
We now use the tilt-shift memory effect~\cite{osnabrugge2017generalized,SeeThroughSubmission} that states that if we tilt the incoming modulation towards two points $\ell$ and $\ell_o$ (where $\ell_o$ denotes the center of the patch, and $\ell$ is another point in $\Area$ ) separated by a 2D displacement vector $\Delta_\ell=\ell-\ell_o$, then the fields at the back of the tissue satisfy a tilt-shift correlation and
\BE
\bou^{\ell}(x)\approx\bou^{o}(x+\Delta_\ell)e^{\frac{2\pi i}{\lambda L}(\Delta_\ell\cdot x)},
\EE
where $L$ is $2/3$ of the tissue thickness. Assuming the memory effect correlation is strong enough, we can use the wavefront $\bou^0$ generated when directing light to the center of the patch $\Area$, and express all other wavefronts $\bou^\ell$, generated when directing light to the rest of the patch. With this, we can express the confocal area score of \equref{eq:conf-area-v} as\BE
\label{eq:conf-score-back-v-me}\Mtric(\PrmVect)=\sum_\ell \left|\sum_x (\bou^o(x))^2 e^{\frac{2\pi i}{\lambda L}(\Delta_\ell\cdot x)}  \right|^2.
\EE
Expanding \equref{eq:conf-score-back-v-me} we can write
\BE\label{eq:conf-score-back-v-me-x1x2}\Mtric(\PrmVect)=\sum_{x_1,x_2} (\bou^o(x_1))^2\cdot ({\bou^o(x_2)}^*)^{2}  \sum_\ell e^{\frac{2\pi i}{\lambda L}(\Delta_\ell\cdot (x_1-x_2))}
\EE
We now note that if $x_1\neq x_2$ and the range of $\Delta_\ell$ values is large enough terms of the form $\sum_\ell e^{\frac{2\pi i}{\lambda L}(\Delta_\ell\cdot (x_1-x_2))}$ are equivalent to the mean of a sinusoidal, which is $0$. Therefore, \equref{eq:conf-score-back-v-me-x1x2} reduces to: 
\BE\label{eq:conf-score-back-v-me-x1x2}\Mtric(\PrmVect)=\sum_{x} |\bou^o(x))|^4, 
\EE as desired.\eop}

\blue{
\begin{claim}
	Assuming sufficient memory effect correlation, the confocal area score is maximized when the incoming modulation $\PrmVect$ makes the wavefront $\bou$ at the back of the tissue a sparse one-hot vector.
	\end{claim}}
\blue{\proof
Using the previous claim, the confocal area score is equivalent to $\Mtric(\PrmVect)=\sum_{x} |\bou^o(x)|^4$. This is equivalent to the confocal score measured using incoherent fluorescent wavefront shaping~\cite{DrorNatureComm24} with single photon excitation, or to the nonlinear emission using two photon excitation~\cite{Katz:14}. It has already been shown that this score favors focusing  modulations. To see this, note that the total laser power is bounded and by using a modulation, we can spread the light in different ways, but we cannot increase its power. Therefore, for any modulation  $\PrmVect$ the norm of the wavefront $\bou^o$ behind the tissue is bounded.
\BE
\sum_x |\bou^o(x)|^2\leq C.
\EE
It is easy to see that to maximize $|\bou^o(x)|^4$ under the bounded norm constraint, it is best if $\bou^o$ is a one-hot vector that has all its power in one entry and zero energy at all other entries.
\eop}

\boldstart{Tilt-shift and time reversal:}
Our algorithm optimizes the wavefront to ensure that input and output fields are similar and exhibit the memory effect and ensure that focusing on adjacent points results in tilt-shifted versions of each other. To illustrate this, we present simulation results of our algorithm in \figref{fig:tilt_shift_sup}. Assuming the SLM is conjugate to the output plane, we demonstrate the focusing achieved for two spots inside the tissue before and after applying our algorithm.

We simulate a scenario without applying wavefront correction and present the intensity on the imaging plane. In this case, light is aberrated and spread across the target plane. This results in an uncorrelated reflected wavefront returning to the SLM plane $u^{w/o}_{out}$, when compared with the  conjugate phase presented on the SLM $u^{w/o}_{in}$, with correlation score: $\bm{C}\left(u^{w/o}_{in},u^{w/o}_{out} \right) = 0.51$. Additionally, since light reflects from a large area inside the tissue, wavefronts reflected from two different points $u^{w/o}_{out_c}, u^{w/o}_{out_l}$ are also uncorrelated, with  $\bm{C}\left(u^{w/o}_{out_c},u^{w/o}_{out_l} \right) = 0.47$.

Conversely, after applying our algorithm to optimize a wavefront that incorporates both memory effect and time-reversal principle, we achieve focus on the target plane. This improvement is exhibited by increased correlation between incoming SLM modulation and output wavefronts $\bm{C}\left(u^{w/}_{in},u^{w/}_{out} \right) = 0.87$, as well as enhanced correlation between wavefronts reflected by neighboring points on the target plane $\bm{C}\left(u^{w}_{out_c},u^{w/}_{out_l} \right) = 0.77.$

\input{fig_tilt_shift_sup}

%% file: fig_tilt_shift_sup.tex
\begin{figure}[t!]     
	\begin{center}
				{\includegraphics[width=1\textwidth]{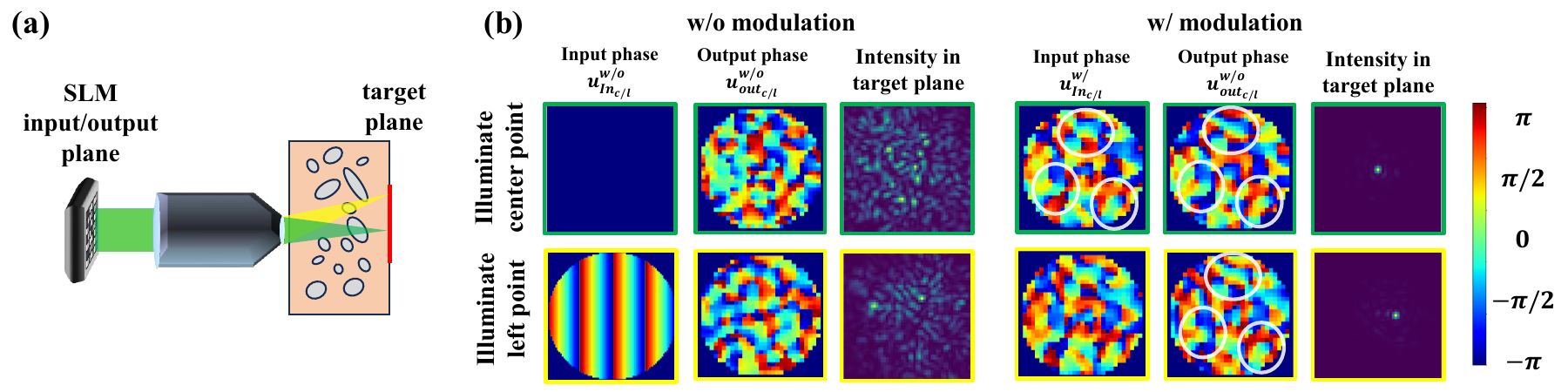}}
	\end{center}
	\caption{\textbf{Tilt-shift and time reversal effects:} (a) We show a simplified schematic of our system with two input wavefronts,  green focusing into the center of the imaging plane and yellow focusing to a neighboring point to the left. (b) Simulation results showing the incoming phase on the SLM, the conjugate of the output phase after tissue reflection (when reaching the SLM plane), and the intensity at the target plane. With light modulation, strong correlations are observed between the incoming SLM phase and the output phase, as well as between output phases for different points (white circles indicate areas of strong correlation). Without aberration correction, correlation decreases rapidly.}
	\label{fig:tilt_shift_sup}
\end{figure}

%% file: pd_interferometry.tex
\section{Gradient acquisition }
\input{fig_pd_interferometry}

\subsection{Score and gradient with definitions}
\blue{We seek to maximize the confocal intensity averaged over a target area $\Area$, where $\PrmVect$ is a complex wavefront displayed on the SLM:
\BE\label{eq:score-confocal-area}
\Mtric(\PrmVect)\equiv \sum_{\ell\in \Area} |\bu^\ell(\PrmVect)^T\RM \bu^\ell(\PrmVect)|^2.
\EE
Differentiating this score with respect to $\PrmVect$ provides:
\BE\label{eq:deriv-vect}
\frac{\partial \Mtric(\PrmVect)}{\partial \PrmVect}=2 \sum_\ell  \underbrace{ \left(\bu^\ell(\PrmVect)^T\RM \bu^\ell(\PrmVect) \right)^*}_{(1)}\cdot\underbrace{\left(\RM \bu^\ell(\PrmVect) + (\bu^\ell(\PrmVect)^T \RM)^T \right)}_{(2)}\hadprod \underbrace{\frac{\partial \bu^\ell(\PrmVect) }{\partial \PrmVect}}_{(3)}.
\EE}

\blue{In most SLMs we can only adjust the phase of the wavefront while its amplitude is fixed, so $\PrmVect$ is a wavefront of the form $\PrmVect(x)=e^{2\pi \PrmPhaseVect(x)}$. To further differentiate the score with respect to $\PrmPhaseVect$ one can use the chain rule once more, leading to
\BE\label{eq:deriv-vect-phase}
\frac{\partial \Mtric(\PrmVect)}{\partial \PrmPhaseVect}=
2\pi {Im}\left(\PrmVect \hadprod \frac{\partial \Mtric(\PrmVect)}{\partial \PrmVect}\right)
\EE
Where ${Im}$ denotes the imaginary part of the complex wavefront. }

\subsection{Gradient acquisition using phase diversity optimization}

To acquire the gradient of our wavefront shaping score, we need to capture the complex wavefront arriving at the sensor $\bv^\ell = \RM \bu^\ell(\PrmVect)$, for every point $\ell$ in the scanned area $\Area$. In our current implementation, we do this using a phase diversity optimization scheme~\cite{Dean2003,Robert1982} for each of the target points. That is, we sequentially place the modulation wavefronts $\bu^\ell(\PrmVect)$ on the illumination SLM and capture the resulting speckle intensity images at the sensor. We place on the camera SLM five known defocus wavefronts $\bu^j_{\text{defocus}}$ and capture five speckle intensity images $I^1_\ell,\ldots I^5_\ell$. We use these images to solve an optimization problem to to retrieve the phase of the complex wavefront from the measured intensity images~\cite{candes2015} resulting in the optimization problem:
\BE\label{eq:phase-diverse}
\hat{\bv}^\ell = \arg \max_{\bv^\ell} \sum_{j=1}^5  \norm{I^j_\ell -  |\Fr(\bv^\ell \hadprod \bu^j_{\text{defocus}})|^2 }^2.
\EE

We scan multiple points $\ell\in \Area$ and repeat the same phase optimization. These wavefronts are averaged to compute one instance of the gradient.

\blue{Note that as long as the magnification of the imaging system is set such that the speckle grain of the captured image $I^j$ is larger than a pixel, there is no inherent limitation on the resolution of the recovered wavefront $\bv^\ell$ .}

\subsection{Fast gradient acquisition using point diffraction interferometry}
We describe another future approach for gradient acquisition which would allow us to acquire it much faster, using a small number of only three shots. 
This possible setup would utilize a variant of a point diffraction interferometry system~\cite{Smartt_1975,Akondi2014PDI} that can capture the complex wavefront with no optimization, and can also average over all  $\ell\in \Area$ within one shot.

\boldstart{Setup for point diffraction interferometry:} 
\figref{fig:setup_interf} illustrates a setup for gradient acquisition with point diffraction interferometry. A laser beam illuminates
a tissue sample, and an SLM can modulate its shape. The SLM is placed such that it is conjugate to a
plane inside the tissue, where the aberration is actually happening. The illumination wavefront propagates
through the scattering tissue, reflects from its internal structure, and back-scatters to the camera. The  returning light passes through the
same SLM on its way to the camera.
The returning modulated light splits at a beam splitter toward two sensors. Before the sensors, the light passes through another relay system and a second SLM. This second SLM is placed such that it is conjugate to the target we wish to image, rather than conjugate to the aberration plane.
We refer to the first SLM (closer to the sample) as ``the modulation SLM'' and to the second one as ``the sensor SLM''.
After the sensor SLM, the light splits through another beam splitter toward two sensors.  The first sensor (which can also be a single point detector) is placed  such that it is conjugate to the target points we wish to image and to the sensor SLM, and would measure the confocal intensity score at the center of the frame. The second 2D sensor is placed such that it is conjugate to the modulation SLM and would measure the gradient directly. We refer to these two sensors as the
 {\em  score sensor} and the {\em gradient sensor}.

The setup also includes a galvo mirror that can tilt the incoming and outgoing light. While we can implement such a tilt with the SLM itself, a galvo mirror can be tilted faster than the refresh rate of the SLM.
By tilting the incoming beam, we can direct it toward different target points $\ell \in \Area$. Tilting the returning beam implies that the light scattered from target point $\ell$ is directed to the central pixel of the sensor, regardless of the actual spatial position of $\ell$.
In this setup, to illuminate the tissue with the modulation wavefront $\bu^\ell(\PrmVect)$, we place $\PrmVect$ on the SLM and use the galvo to tilt it toward the target point $\ell$. So, effectively we get:
\BE\label{eq:tilt-slm-modulation}
\bu^\ell(\PrmVect)(x)=\bzeta_x\hadprod e^{\frac{2\pi i}{\lambda}(\tau^\ell\cdot x)}\hadprod \PrmVect(x),
\EE
where $\tau^\ell$ represents the tilt toward point $\ell$.
$\bzeta_x$ represents here the shape of the illumination wavefront reaching the modulation SLM. Since our SLM is not placed at the Fourier plane of the system, it is illuminated by a spherical wavefront and not by a plane wave. Note that $\bzeta_x$ was defined in the main paper as the (known) propagation of the wavefront from the modulation SLM to the center of the sensor. Since we assume our system is fully symmetric the same $\bzeta_x$ applies both on the incoming and on the outgoing paths. 

\boldstart{Interferometric measurements:}
We start with a mathematical expression for the wavefront we will measure by the gradient sensor. We recall that this sensor is focused at the modulation SLM and directly images the wavefront at this plane.    If we use a flat phase on the sensor SLM and display the modulation $\bu^\ell(\PrmVect)$ (by displaying $\PrmVect$ on the modulation SLM and combining it with the proper galvo tilt), it will measure the 2D wavefront:
 \BE\label{eq:v-focus-on-mod-slm}
 \bv^\ell= \left(\bu^\ell(\PrmVect)\right)\hadprod \left(\RM \bu^\ell(\PrmVect)\right),
 \EE
 where $\RM \bu^\ell(\PrmVect)$ is the propagation of the incoming modulated illumination via the tissue and back to the SLM plane. On the SLM plane, each coordinate is multiplied again by $\bu^\ell(\PrmVect)$. We use the notation $\hadprod$ to denote an element-wise multiplication between two 2D fields. Since we image the modulation SLM plane, we measure this as a 2D wavefront and not as a scalar.
 
Next, denote by $\mu^\ell$ the complex scalar reaching the center of the score sensor when there is a blank phase on the sensor SLM. As this sensor is focused on the target plane (and on the sensor SLM, which is conjugate to the target plane):  
 \BE
 \mu^\ell=(\bu^\ell(\PrmVect))^T\RM \bu^\ell(\PrmVect),
 \EE
 where $\bu^\ell(\PrmVect)$ is defined in \equref{eq:tilt-slm-modulation} to include the SLM modulation, the tilt, and the propagation from the modulation plane to the score sensor.

 If we could place an aperture on pixel $x$ of the 2nd sensor SLM (so we block the parts of the wavefront that do not pass via point $x$),  the 2D wavefront we measure at the gradient sensor (which is focused on the modulation SLM) is governed by the scalar $\mu^\ell$ and corresponds to: 
 \BE\label{eq:v-focus-on-mod-slm-via-apt}
 \bv^\ell_1=\mu^\ell\cdot{\bzeta_x}^*,
 \EE
 where $\bzeta_x$, defined in Eq. (2) of the main paper, encodes the (known) propagation from the modulation SLM plane to the center of the score sensor.

 We denote the difference between the wavefronts in \equpref{eq:v-focus-on-mod-slm}{eq:v-focus-on-mod-slm-via-apt} by:  
 \BE
 \bv^\ell_2=\bv^\ell-\bv^\ell_1.
 \EE
 
 Moreover, rather than using a pinhole on the sensor SLM, we suggest varying the phase of pixel $x$ of the sensor SLM to $\phi_j$ while keeping all other pixels at phase $0$. This implies that the gradient sensor would see the wavefront $\bv^\ell_2+e^{i\phi_j}\bv^\ell_1$, where $\bv^\ell_2$ is the part of the wavefront that passes when blocking pixel $x$, and $e^{i\phi_j}\bv^\ell_1$ is the part that passed via pixel $x$ after its phase is adjusted. We can use standard phase shifting interferometry and measure $J=3$ intensity images while varying the phase $\phi_j$ of the central pixel equally between $0$ and $2\pi$.
 We measure
 \BE
 I_j^\ell=|\bv^\ell_2+e^{i\phi_j}\bv^\ell_1|^2=e^{-i\phi_j}{\bv^\ell_1}^*\bv^\ell_2+ e^{i\phi}{\bv^\ell_1}{\bv^\ell_2}^*+|\bv^\ell_1|^2+|\bv^\ell_2|^2.
 \EE
 By summing the intensity images with the corresponding phase, we can isolate the interference term
 \BE\label{eq:interf-v1v2}
 {\bv^\ell_1}^*\bv^\ell_2=\sum_j e^{i\phi} I_j^\ell.
 \EE
 
We can also measure $|\bv^\ell_1|^2$ and add it to \equref{eq:interf-v1v2} to receive
\BE
g^\ell={\bv^\ell_1}^*{\bv^\ell}.
\EE

We can now substitute Eqs. (\ref{eq:tilt-slm-modulation}), (\ref{eq:v-focus-on-mod-slm}) and (\ref{eq:v-focus-on-mod-slm-via-apt}),  and check the content of this interference term:
\BEA\label{eq:grad-interf}
g^\ell&=&  {\mu^\ell}^* \cdot \left(\bzeta_x\hadprod \bu^\ell(\PrmVect)\right) \hadprod \left(\RM \bu^\ell(\PrmVect)\right)\\
&=& \underbrace{{\mu^\ell}^*}_{(1)}  \cdot \underbrace{\bzeta_x\hadprod \PrmVect}_{(2)}\hadprod \underbrace{\bzeta_x\hadprod e^{\frac{2\pi\i}{\lambda}(\tau^\ell\cdot x)}}_{(3)} \hadprod \underbrace{\left(\RM \bu^\ell(\PrmVect)\right)}_{(4)}.
 \EEA
 Comparing this formula to the gradient derived in \equref{eq:deriv-vect},  we see that it corresponds to the desired gradient up to known multiplicative terms. Term (1) corresponds to term (1) of the gradient, term (2) is a known multiplicative factor, term (3) is effectively the derivative of the tilted modulation $\bu^\ell(\PrmVect)$ with respect to the SLM parameters
 $\frac{\partial \bu^\ell(\PrmVect)}{\partial \PrmVect},$
 which is term (3) in \equref{eq:deriv-vect},  see definition in \equref{eq:tilt-slm-modulation}. Term (4) is equivalent to term (2) in \equref{eq:deriv-vect}. 
 
 Moreover, note that to average over target points $\ell \in \Area$, all we need to do is to scan the galvo toward a different direction, without changing the pattern $\PrmVect$ on the SLM. This effectively implies that we can tilt the galvo {\em within exposure}, and capture three speckle images
 \BE
 I_j=\sum_\ell I_j^\ell.
 \EE
 From these three images we can compute the gradient directly. 
 
 The above derivation suggests a possible future implementation of our wavefront shaping algorithm which will allow us to image the gradient of the confocal score directly, using as little as three shots; where the scanning over the target area is done within exposure. 
Furthermore, since we use an interferometer, no phase diversity optimization is required, and the complex wavefront is measured directly.

\section{Equivalence to power iterations} The gradient derived above is similar to rapid
time reversal based approach for wavefront shaping~\cite{Dror22,doi:10.1121/1.424648,doi:10.1121/1.412285,Yang2014,Meng2012,Papadopoulos16}. Similar ideas were also used by digital correction algorithms~\cite{Kang2017, Balondrade_2024,Najar2024}.
These approaches start from the assumption that the desired modulation is the largest eigenvector of the reflection (or transmission) matrix of the tissue. This eigenvector can be estimated very efficiently using power iterations. In each iteration, one displays a modulation $\bu$ on the SLM, measures $\RM\bu$, and uses it as the next guess for the modulation.

We argue that the gradient we derived above collapses to such a power iteration 
if we restrict our score function to an area $\Area$ consisting of a single target point $\ell$. To see this, note that if we attempt to maximize the score

 \BE\label{eq:score-sup}
\Mtric(\PrmVect)= \left|{\bu^\ell(\PrmVect)}^T \RM \bu^\ell(\PrmVect)\right|^2,
\EE
under the constraint that $\PrmVect$ is a unit norm vector (since the SLM can only redistribute the laser energy but it cannot regenerate energy), then classical results in linear algebra imply it is maximized by the largest eigenvector of $\RM$. Moreover, up to known constants, the gradient derived in \equref{eq:grad-interf} is basically  a multiplication of the previous guess by the reflection matrix.
 However, as demonstrated in Fig. 4 of the main paper, given a coherent target, averaging over an area is crucial for getting a good wavefront correction. If one attempts to optimize the confocal intensity at a single point, interference can bring all light into a strong point at the sensor, without actually focusing the light into a single point inside the volume.

Attempting to adjust power algorithms to produce different modulations may require ad-hoc additions. 
However, defining a target score and deriving its gradient provides a more principled framework to impose desired properties on the modulation.

In particular~\cite{Dror22} used a variant of power iterations with incoherent fluorescent sources. The convergence of this incoherent version was a bit hard to analyze, and to justify it the supplementary file of~\cite{Dror22} uses a model assuming the phase of the wavefront is captured precisely with a point diffraction interferometer. Their incoherent phase estimation is equivalent to the point-diffraction interferometry setup  derived above.

%% file: fig_pd_interferometry.tex
\begin{figure*}[t!]
	\begin{center}
		\includegraphics[width= 0.95\textwidth]{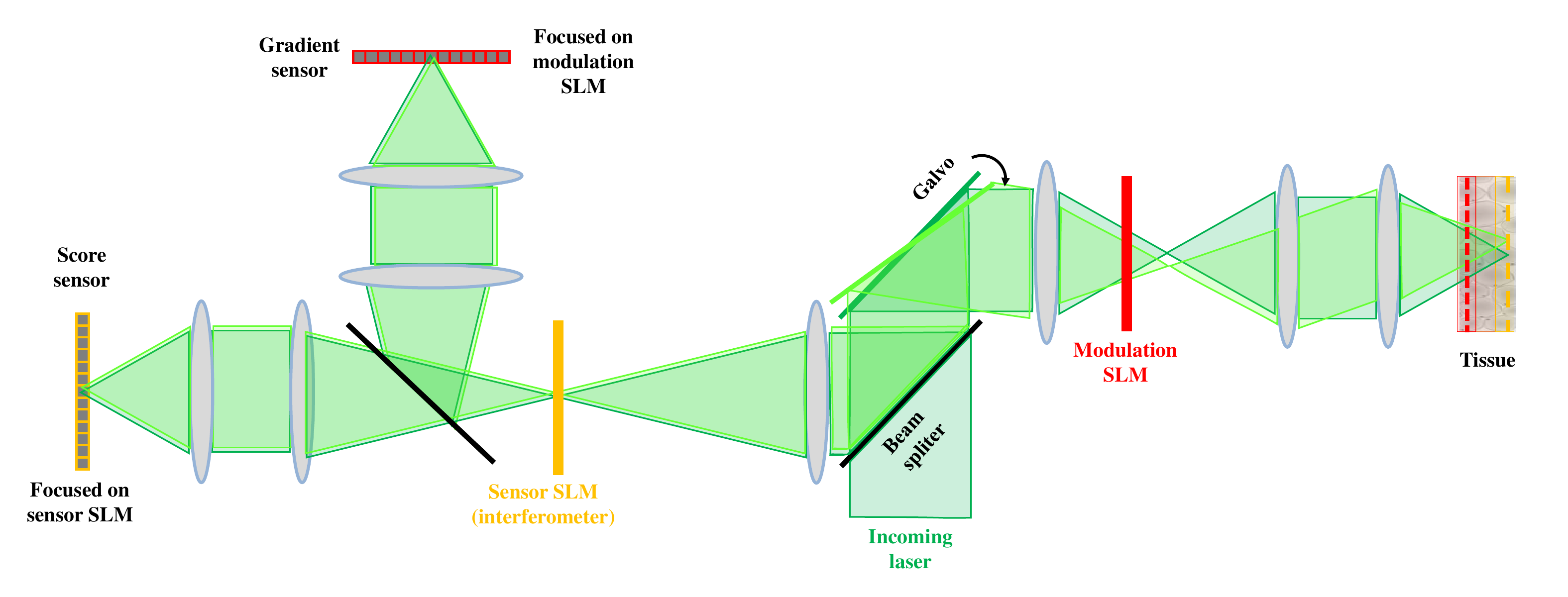}
	\end{center}
	\caption{{\bf{Setup for interferometric gradient acquisition:}} An incoming laser beam is directed to focus on a point inside the tissue, the focal point can be shifted with a galvo. The figure illustrates two such tilted paths in two shades of green. On its way to the sample the beam passes through a relay system,  containing an SLM for aberration correction. The SLM is placed such that it is conjugate to a  plane inside the tissue volume. Another SLM is used as an interferometer to modulate the reflected light captured by the sensor. Two cameras are used in this setup - the first is focused on the interferometer SLM and termed score sensor. The second sensor is focused on the modulation SLM and termed gradient sensor.} \label{fig:setup_interf}
\end{figure*}

%% file: sup_res.tex
\newpage
\section{Additional results }

\boldstart{Additional result of beads:}

\blue{We use our system to image polystyrene beads dispersed in agarose gel. We create two different targets, with bead diameters of $0.5\um$ and $3\um$. The slab thickness is $1.3mm$. The left part of \figref{fig:beads2} illustrates the schematic of this target, where we used our algorithm to focus inside this volume and image planar sub-regions at multiple depths. The optical depth (OD) is 3, where OD was estimated by measuring the attenuation of ballistic light in the validation camera.}
	
	\input{fig_beads2}

\blue{\figref{fig:beads2} visualizes confocal scanning of $x-y$ slices, and \figref{fig:beads_xy} visualizes a slice along the $x-z$ axes.  While a standard confocal scan of such beads is very noisy, with our estimated modulation, we can largely increase its contrast and achieve a clear image of spots corresponding to the beads position. To obtain a reference image of the bead positions, we image beads at the farther depth of the slab closer to the validation camera, so we are able to image a clear reference from the validation camera. For this reference image, we used wide-field incoherent illumination. }

\blue{For the bead targets in \figref{fig:beads2}, since the beads are sparser, we vary the target area, so we image multiple beads. The target area for the $3\um$ beads in the first two rows was $\Area = 16 \um \times 16\um$ and $ \Area = 32 \um \times 32\um$, respectively. The sampling interval was $0.65\um$ for both algorithm and final confocal scan. For the $0.5\um$ beads (rows 3-5) the area was $\Area = 13 \um \times 13\um$ with sampling interval of $0.5\um$ for both the algorithm and the final scan. In \figref{fig:beads_xy} the area was $\Area = 5.2\um \times 16\um$, the sampling interval in lateral axis was $0.5\um$ and sampling interval in axial was $2\um$. We note that even if the beads are larger than the diffraction limit the expected non-aberrated image of the bead in a reflection confocal scanning is a diffraction-limited spot, since the beads act as a mini-lens~\cite{Weise96}. Consequently, $3\um$ beads appear as narrow isolated spots after our correction is applied.}

\input{fig_beads_xy}

\pagebreak

\boldstart{Confocal scan captured images:}

In \figref{fig:stitch_sup}, we display a super-position of images captured by the main and validation cameras. We compose in one frame a sparse subset of the target points $\ell$ in the area we scanned. 
We show the results before and after optimization for the result presented in the second row of Fig. 5 of the main text. Before optimization, the wavefront reaching the main camera is significantly aberrated. After optimization, clear focal points reflected from the chrome are observed. In the validation camera, we observe the wavefront reaching the target plane. Without modulation, the wavefront is highly aberrated, whereas after optimization, a well-defined focus is visible on the target plane. Areas covered by chrome are not observable in the validation camera because the chrome attenuates light. For visualization purposes, all images were normalized by their maximum value. However, in reality, without modulation, the light reaching both cameras is scattered and spread across the sensor, and these images are much darker than the images captured using our modulation. 

\blue{The resolution target  in Fig. 5 in the main paper is constructed from four scanned isoplanatic patches. In \figref{fig:res-patches}, we show the different confocal scans from the different isoplanatic patches and their corresponding phase masks.}

\input{fig_stitch_sup}
\input{fig_res_patches}

\newpage

\boldstart{Onion cells:}
To better visualize the onion cell target we image in Fig. 7 of the main paper, we present in ~\figref{fig:onion_full_view_sup} a larger view through the validation camera.
The final result presented in the main paper in Fig. 7 is too big to be corrected by a single modulation and we have achieved the result by combining multiple isoplanatic patches, stitched to form the large field-of-view image. In~\figref{fig:onion-multi}, we show the individual isoplanatic patches. The patches partially overlap, and to find the overlap between patches we used template matching. The image in Fig. 7 in the paper was rotated for convenience of view.
\bblue{To understand the typical 3D structure of the onion, in~\figref{fig:ref-xy} we show confocal scanning at the front (shallow) parts of the onion where the light is not aberrated.}

\input{fig_onion_full_view}
\input{fig_onion_multi_separated}
\newpage
\input{fig_onion_ref_xy_slices}

\newpage
\bblue{In~\figref{fig:res-xy} we show more of the 3D structure results of Fig. 8 of the main paper. For that we include  x-y cross sections at multiple depths.  We scanned in intervals of $2\um$ spanning depth of $\pm 8$ from the focused depth. Layer 1 is at $80\um$, layer 2 is at $130\um$, and layer 3 is at $190\um$.}
\input{fig_onion_result_xy_slices}

\newpage
\pagebreak

\boldstart{Comparison with digital correction approaches:}\bblue{
	In \figref{fig:CLASS} we further compare our method with the CLASS algorithm~\cite{Choi2015}, a representative digital correction technique~\cite{Kwon2023,ChoiLyers2023,haim2023imageguidedcomputationalholographicwavefront,Balondrade_2024}. 	Digital approaches such as CLASS measure the scattered wavefront and numerically fit it with a parametric model describing the underlying aberration and hidden target. 
	The score function targeted by the CLASS algorithm is similar to the one used in this paper, it attempts to maximize energy along the diagonal of the reflection matrix, which is equivalent to maximizing the confocal energy, as we do in this paper.
	Because the measured scattered fields are typically very noisy, the reconstruction quality is ultimately limited by the signal-to-noise ratio (SNR) of the acquired data. 
	In contrast, performing the correction \textit{optically}—by applying the estimated modulation directly to the SLM—enhances the effective SNR of subsequent measurements, leading to cleaner reconstructions and improved correction fidelity. 
	In \figref{fig:CLASS}, we visualize CLASS applied to the wavefronts measured at the beginning of our algorithm when the SLM is blank. 
	We then apply CLASS to wavefronts acquired at the fifth and tenth  iterations, after placing the correction from the previous iteration on the SLM. With this optical pre-correction, the SNR of the measured data improves substantially, resulting in a markedly better reconstruction. }
	
	\bblue{	The CLASS reconstructions shown in \figref{fig:CLASS} should be regarded as illustrative rather than quantitative benchmarks of CLASS performance, due to two main differences. 
	First, the CLASS system measures the wavefront interferometrically, which can provide more accurate phase recovery than our phase-diversity estimate. 
	Second, the CLASS algorithm typically measures many more columns of the reflection matrix than our system acquires. Reflection-matrix imaging typically scans a large area that covers the speckle support around each corrected point, while our method scans only the corrected  region of interest.
	While this broader sampling can yield more complete information, it also requires scanning a much larger spatial region to cover the speckle support around each corrected point. 
	In contrast, our method scans only the local region of interest, making each acquisition significantly faster,  although multiple iterations are then required for convergence. }
	
	\bblue{An interesting future direction would be to combine these two approaches—using the scattered wavefronts measured in the first iteration to fit an aberration model through multiple gradient-descent updates, then placing this model on the SLM and re-measuring data under reduced scattering. 	Such a hybrid strategy could  reduce the number of capture iterations required, although given that our algorithm typically converges within about ten iterations, the expected improvement would likely be limited. }

 \input{fig_class_compare}

%% file: fig_beads2.tex
\begin{figure*}[h!]
	\begin{center}		
		\begin{tabular}{@{}c@{~}c@{~}c@{~}c@{~}c@{~}c@{~}}			
			\multicolumn{1}{c}{\hspace{-0.6cm}\large Target }& 
			\multicolumn{2}{c}{\hspace{-0.6cm}\large One point }& 
			\multicolumn{2}{c}{\hspace{-0.6cm}\large Confocal scan}&
			\multicolumn{1}{c}{\hspace{-0.6cm}\large Wide field}\\
			\multicolumn{1}{c}{\hspace{-0.6cm}\large illustration }& 
			\multicolumn{2}{c}{\hspace{-0.6cm} Main cam.} & \multicolumn{2}{c}{\hspace{-0.6cm} Main cam.} &
			\multicolumn{1}{c}{\hspace{-0.6cm} Valid. cam.} \\
			\multicolumn{1}{c}{\hspace{-0.6cm}}& 
			\multicolumn{1}{c}{\hspace{-0.6cm} \scriptsize w/o}&	
			\multicolumn{1}{c}{\hspace{-0.6cm} \scriptsize w/ }&	
			\multicolumn{1}{c}{\hspace{-0.6cm} \scriptsize w/o}&	
			\multicolumn{1}{c}{\hspace{-0.6cm} \scriptsize w/ }&	
			\multicolumn{1}{c}{\hspace{-0.6cm} \scriptsize Incoherent}\vspace{-0cm}\\
			\multicolumn{1}{c}{\hspace{-0.6cm}}& 
			\multicolumn{1}{c}{\hspace{-0.6cm} \scriptsize modulation}&	
			\multicolumn{1}{c}{\hspace{-0.6cm} \scriptsize modulation}&	
			\multicolumn{1}{c}{\hspace{-0.6cm} \scriptsize modulation}&	
			\multicolumn{1}{c}{\hspace{-0.6cm} \scriptsize modulation}&	
			\multicolumn{1}{c}{\hspace{-0.6cm} \scriptsize illumination}\\
			\multirow{5}{8em}{\includegraphics[width= 0.18\textwidth]{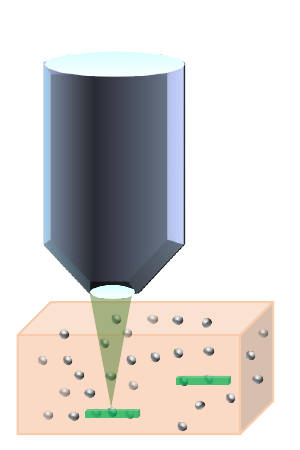}}&
			\includegraphics[width= 0.16\textwidth]{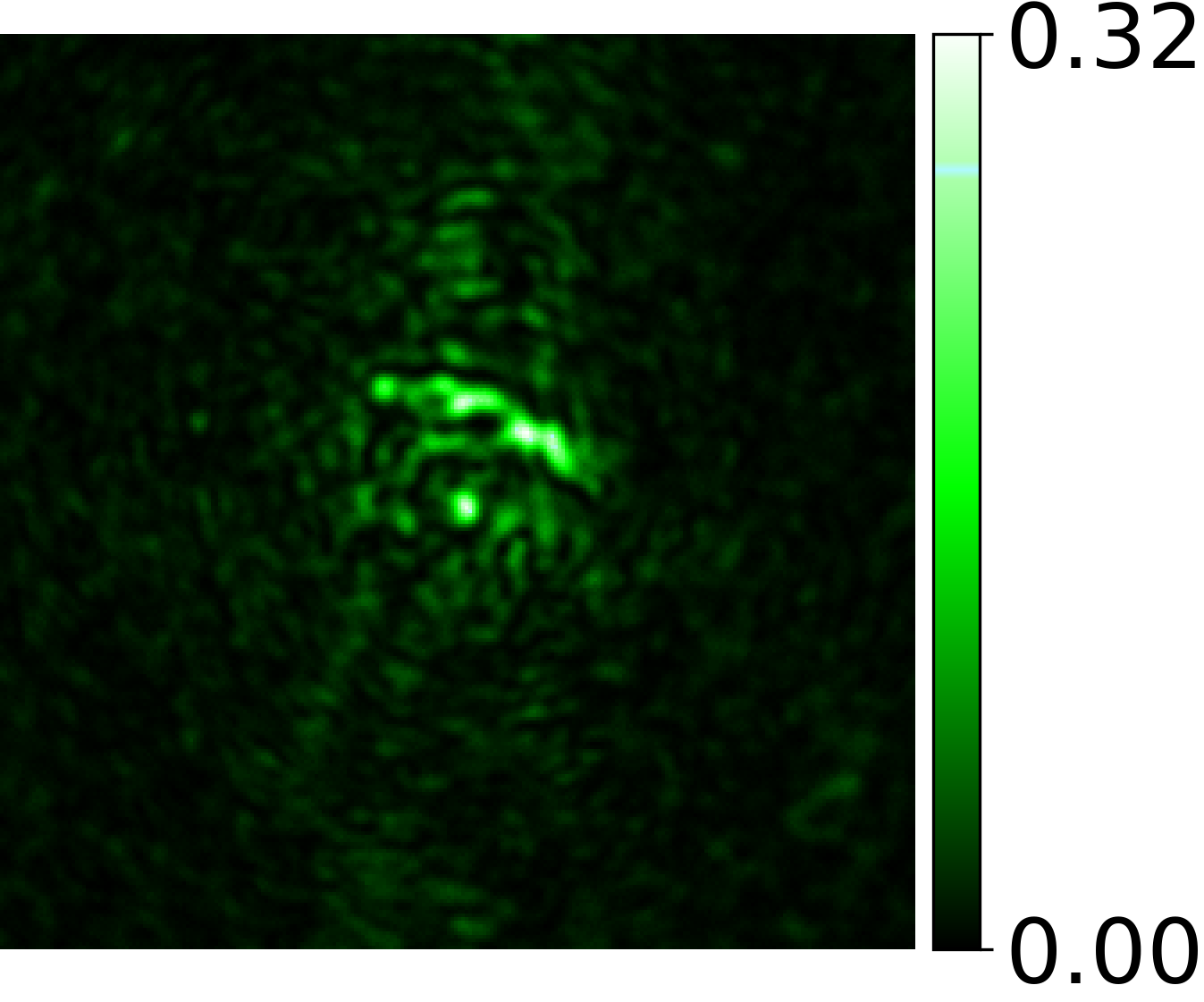}&
			\includegraphics[width= 0.16\textwidth]{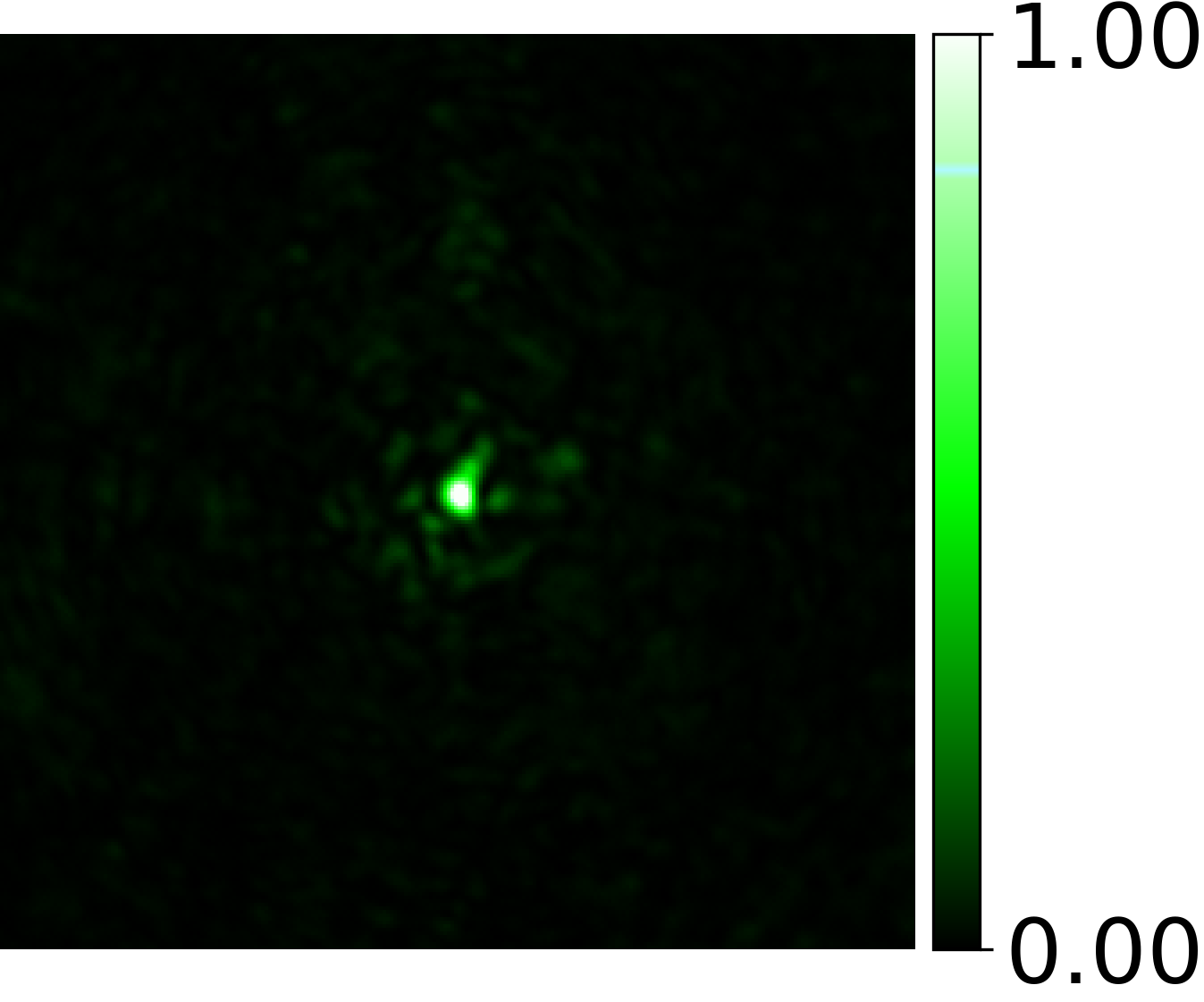}&
			\includegraphics[width= 0.16\textwidth]{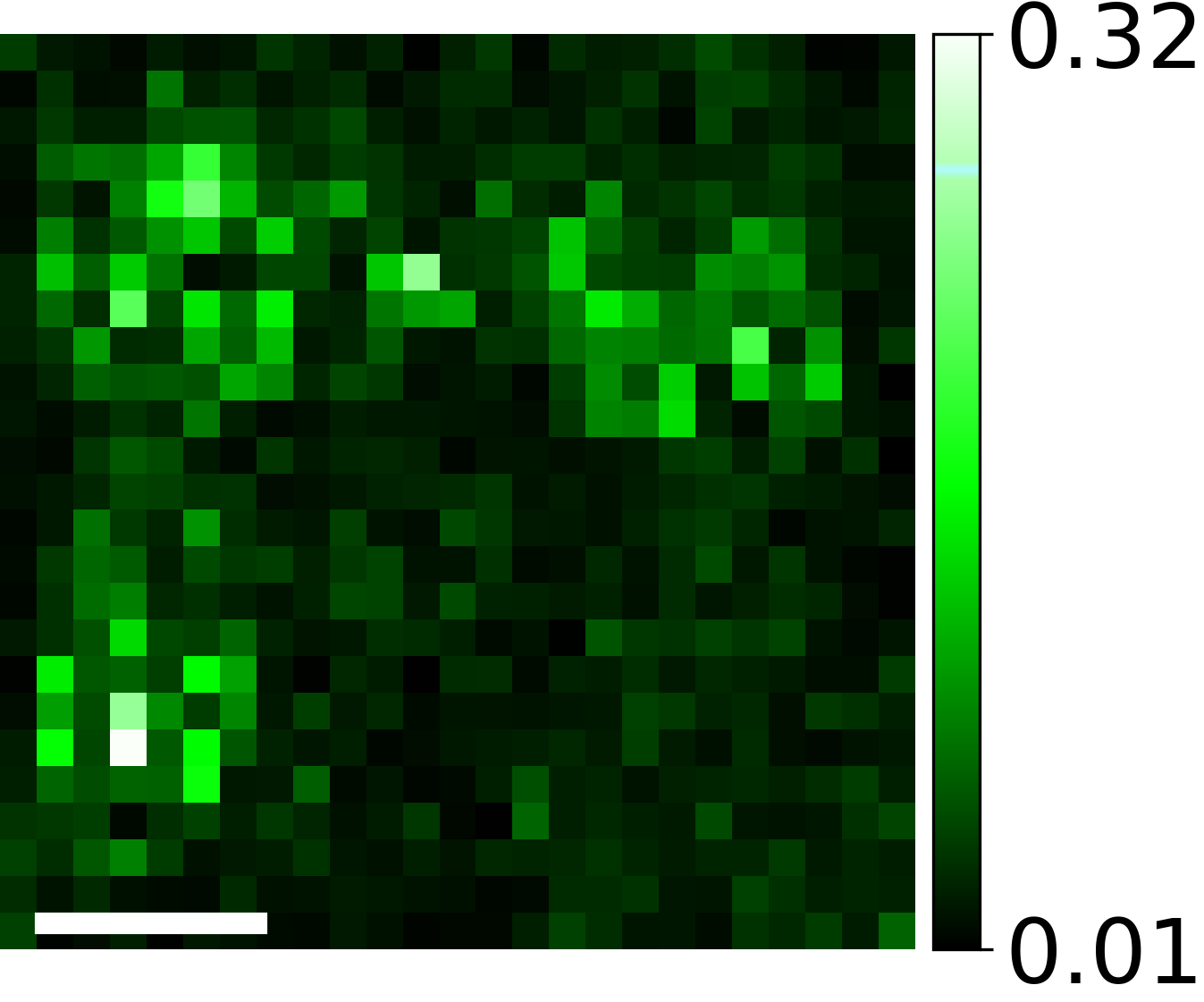}&
			\includegraphics[width= 0.16\textwidth]{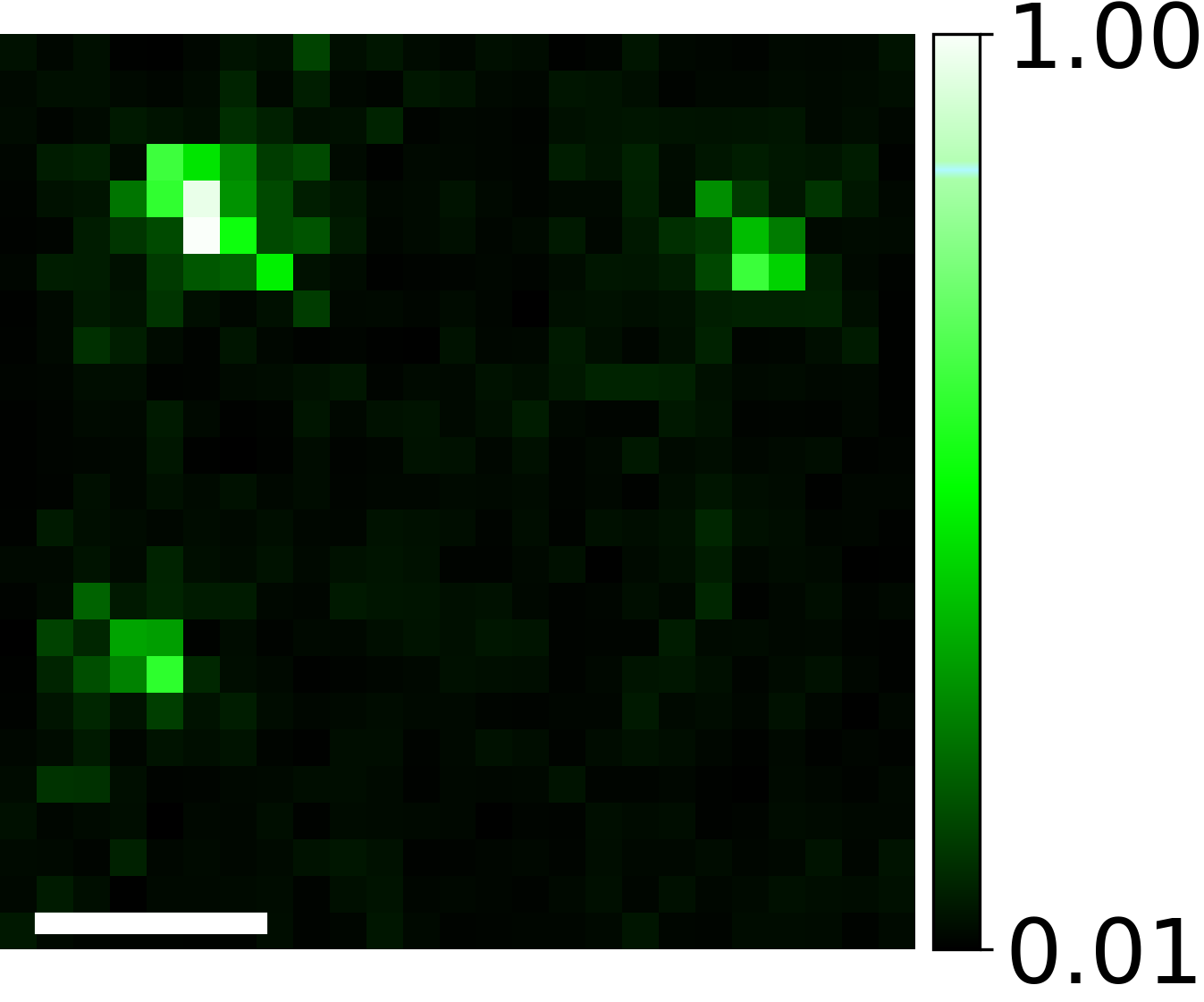}&
			\includegraphics[width= 0.16\textwidth]{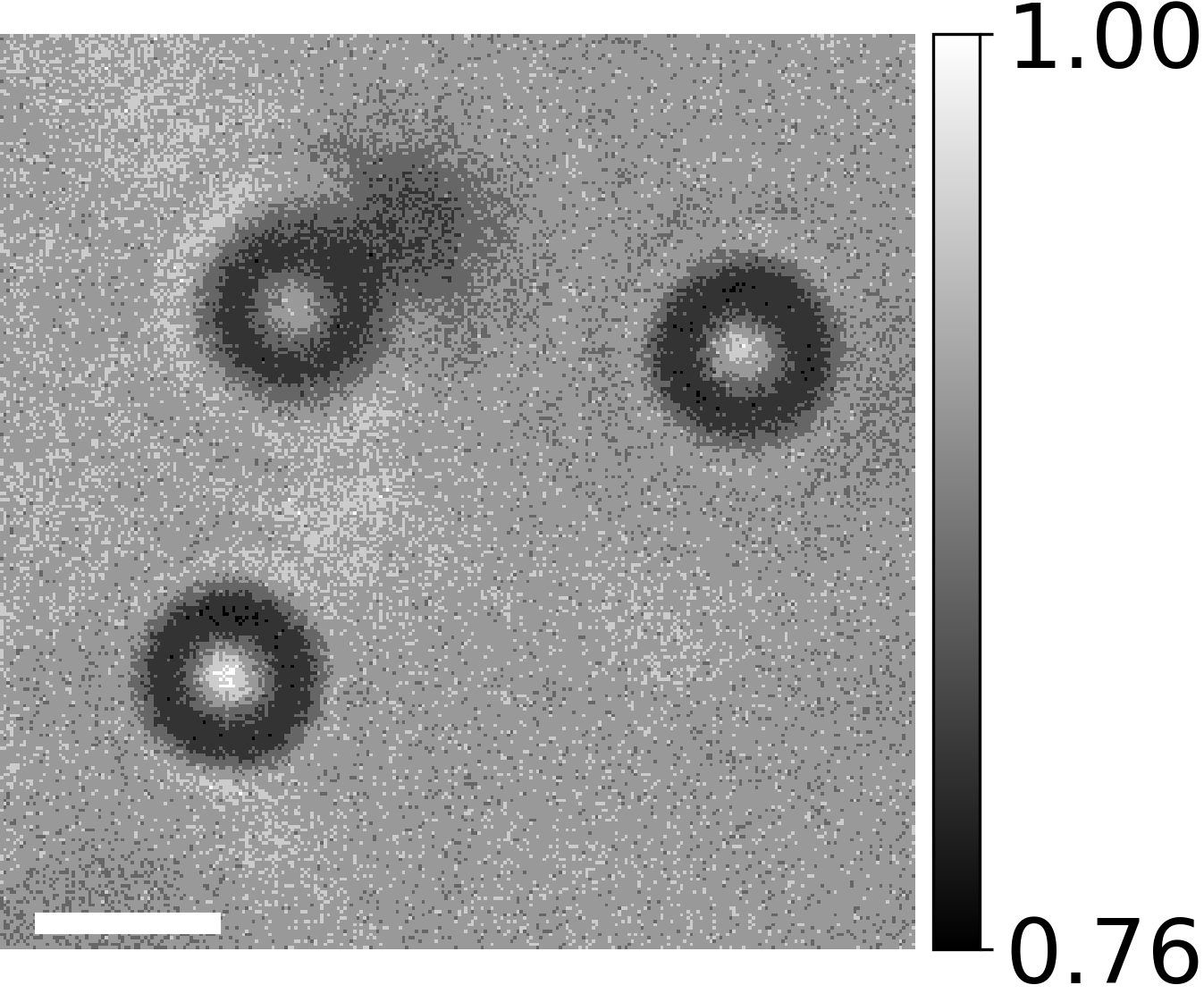}\\&
			\includegraphics[width= 0.16\textwidth]{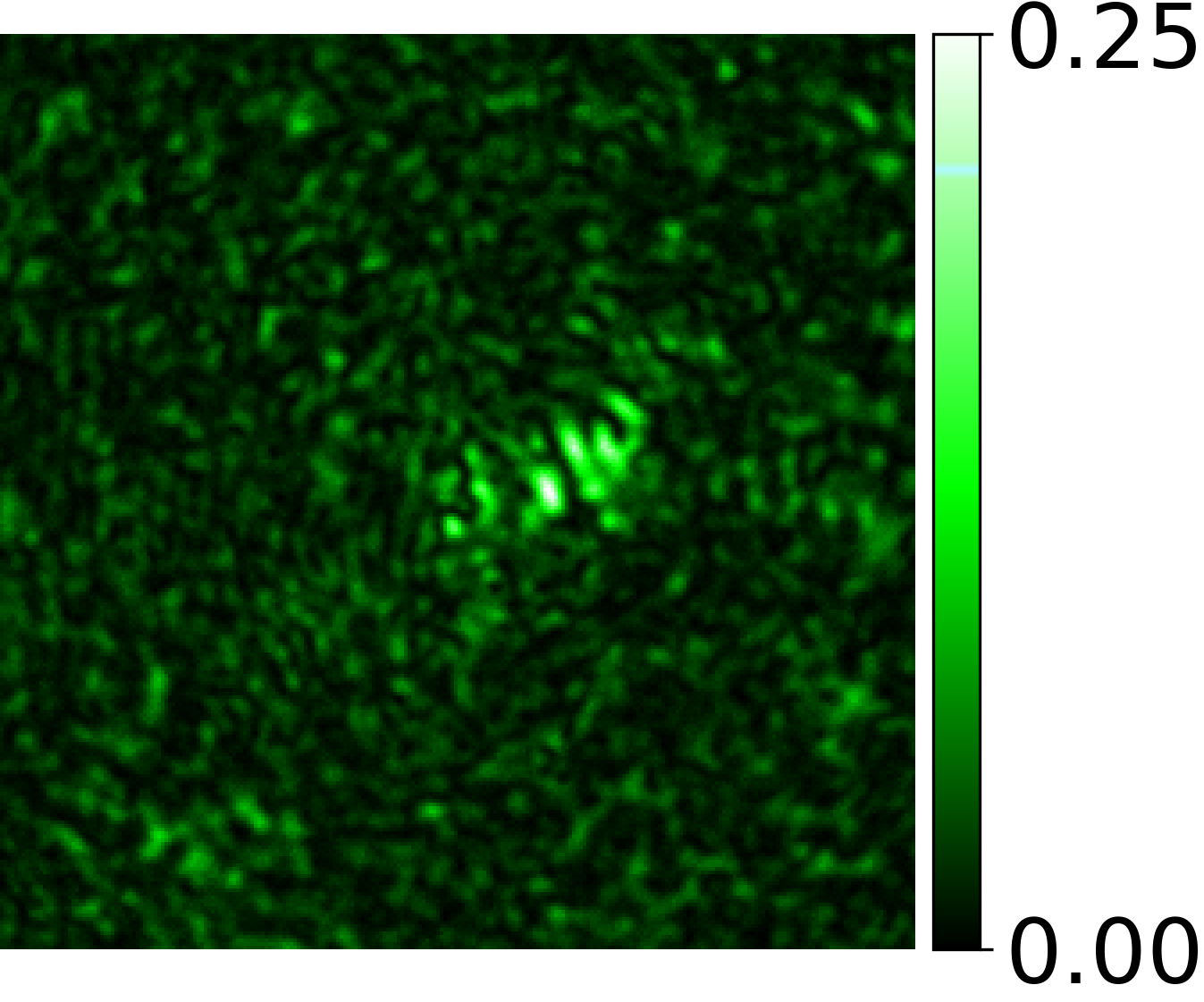}&
			\includegraphics[width= 0.16\textwidth]{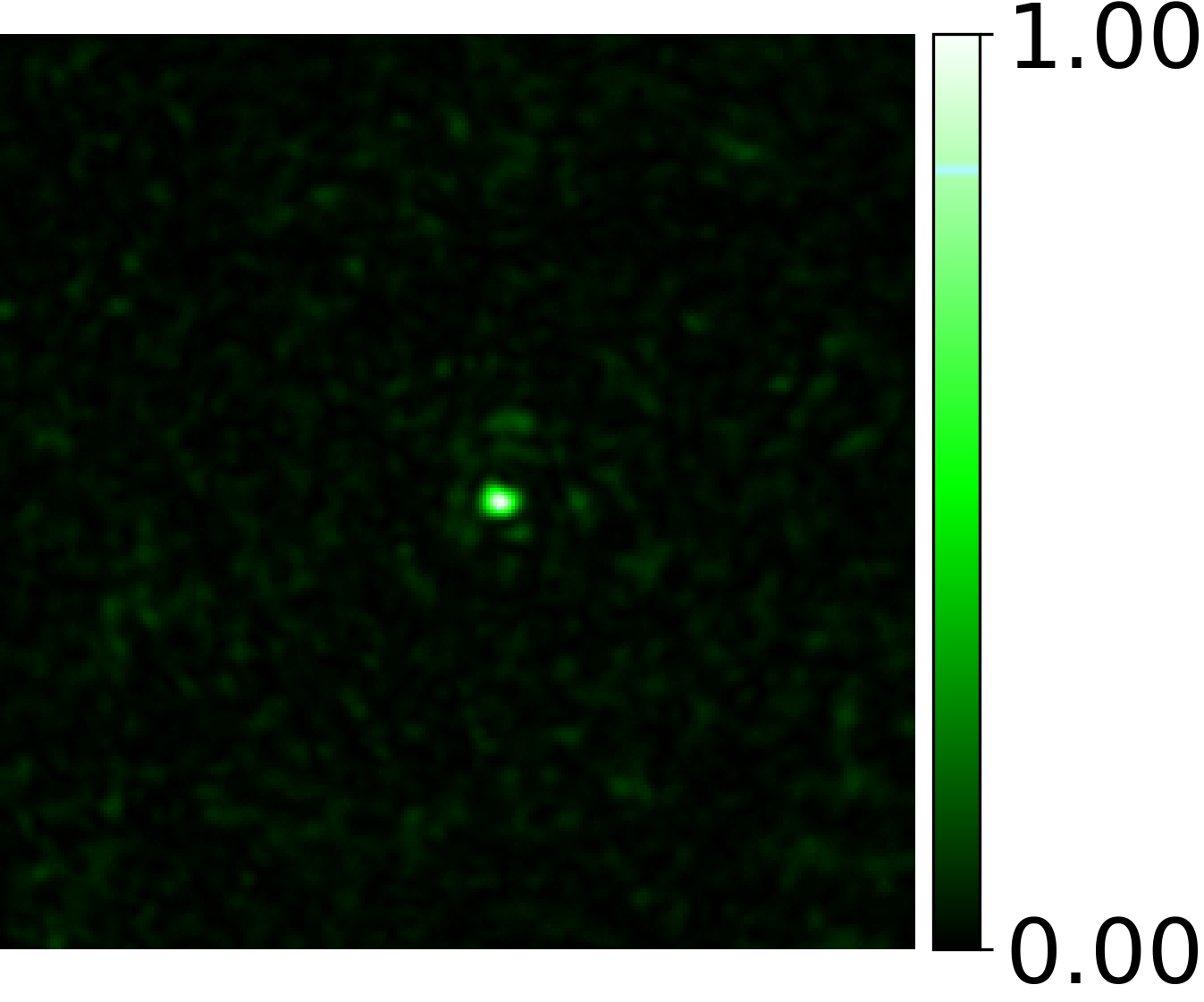}&
			\includegraphics[width= 0.16\textwidth]{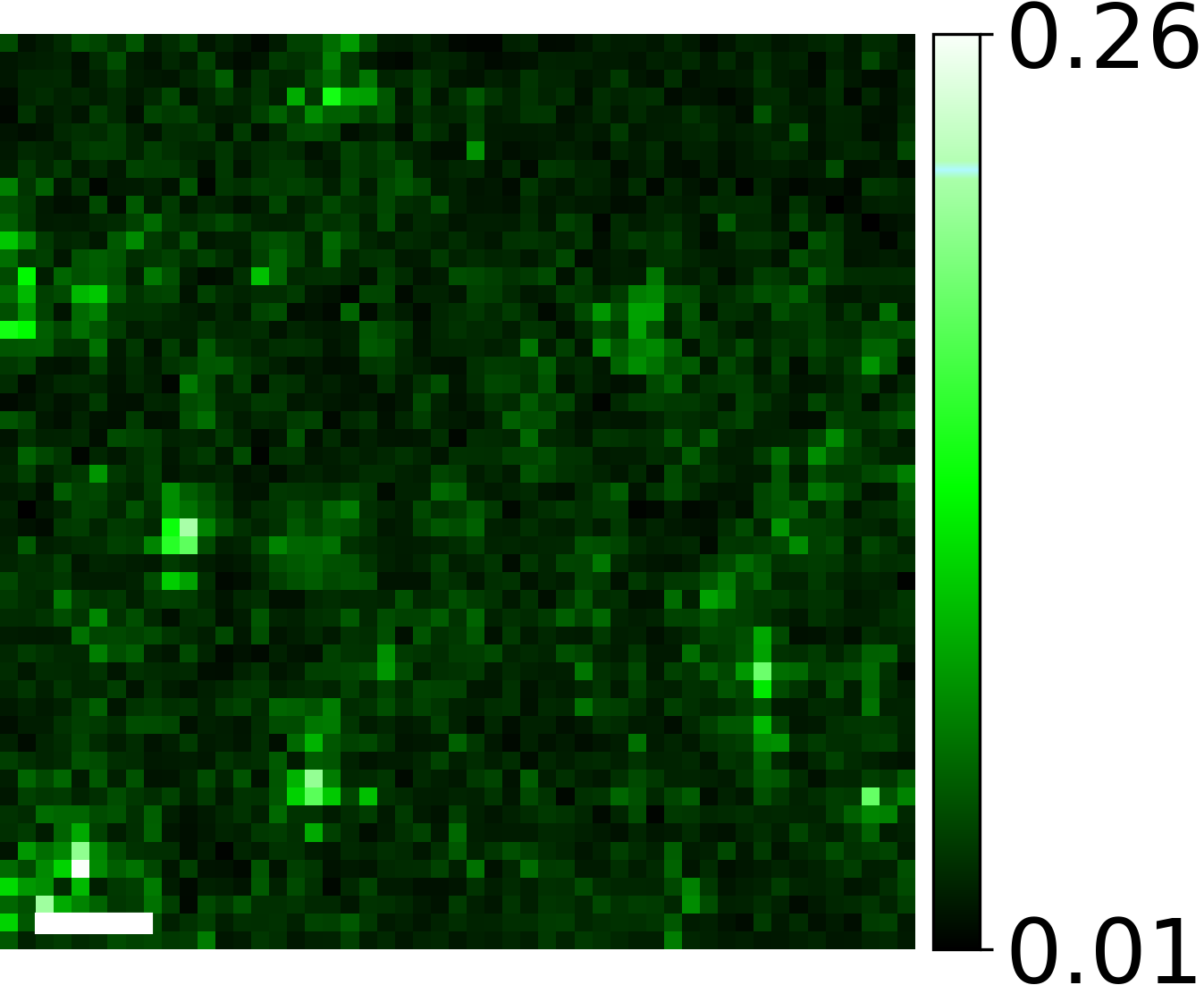}&
			\includegraphics[width= 0.16\textwidth]{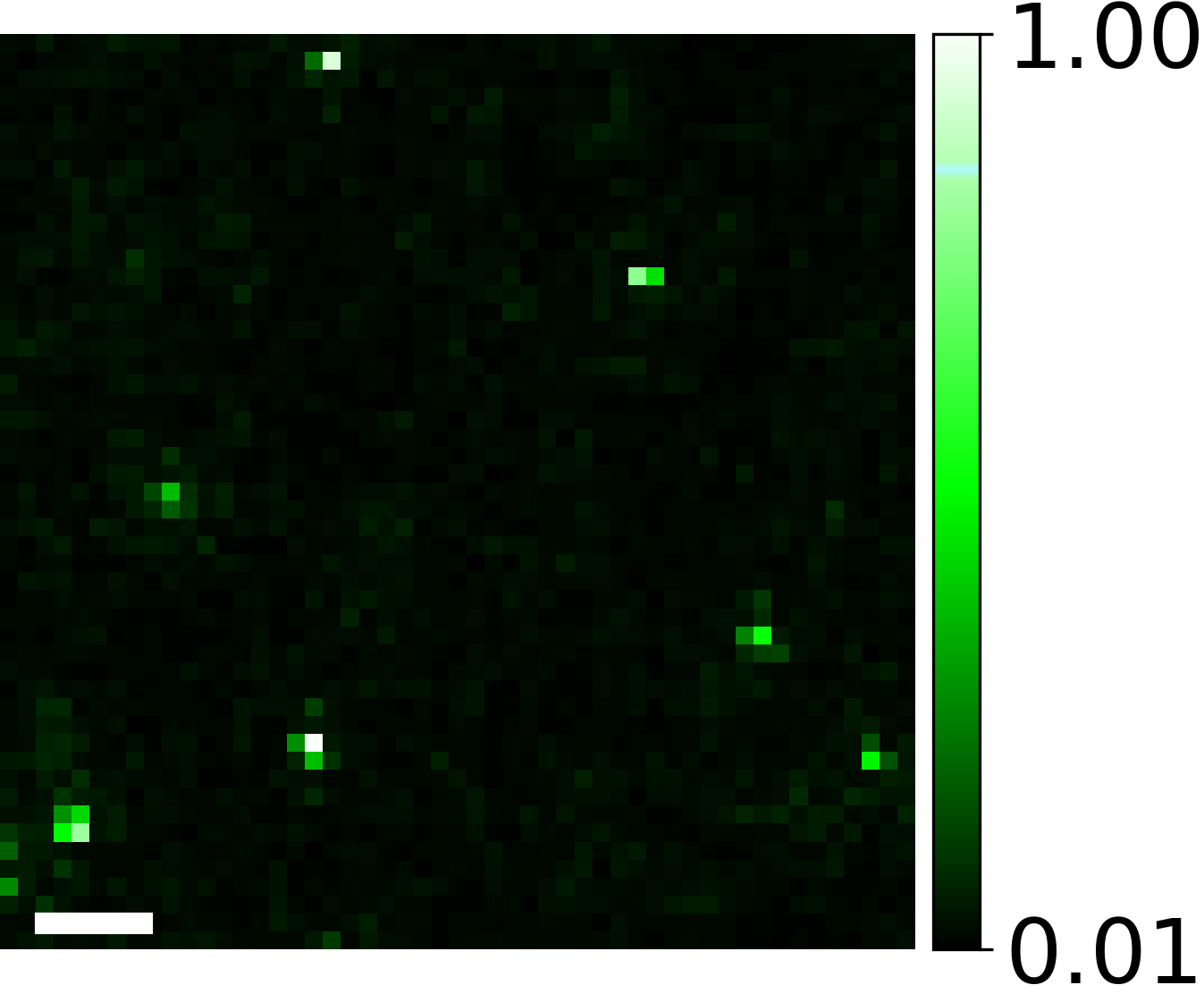}&
			\includegraphics[width= 0.16\textwidth]{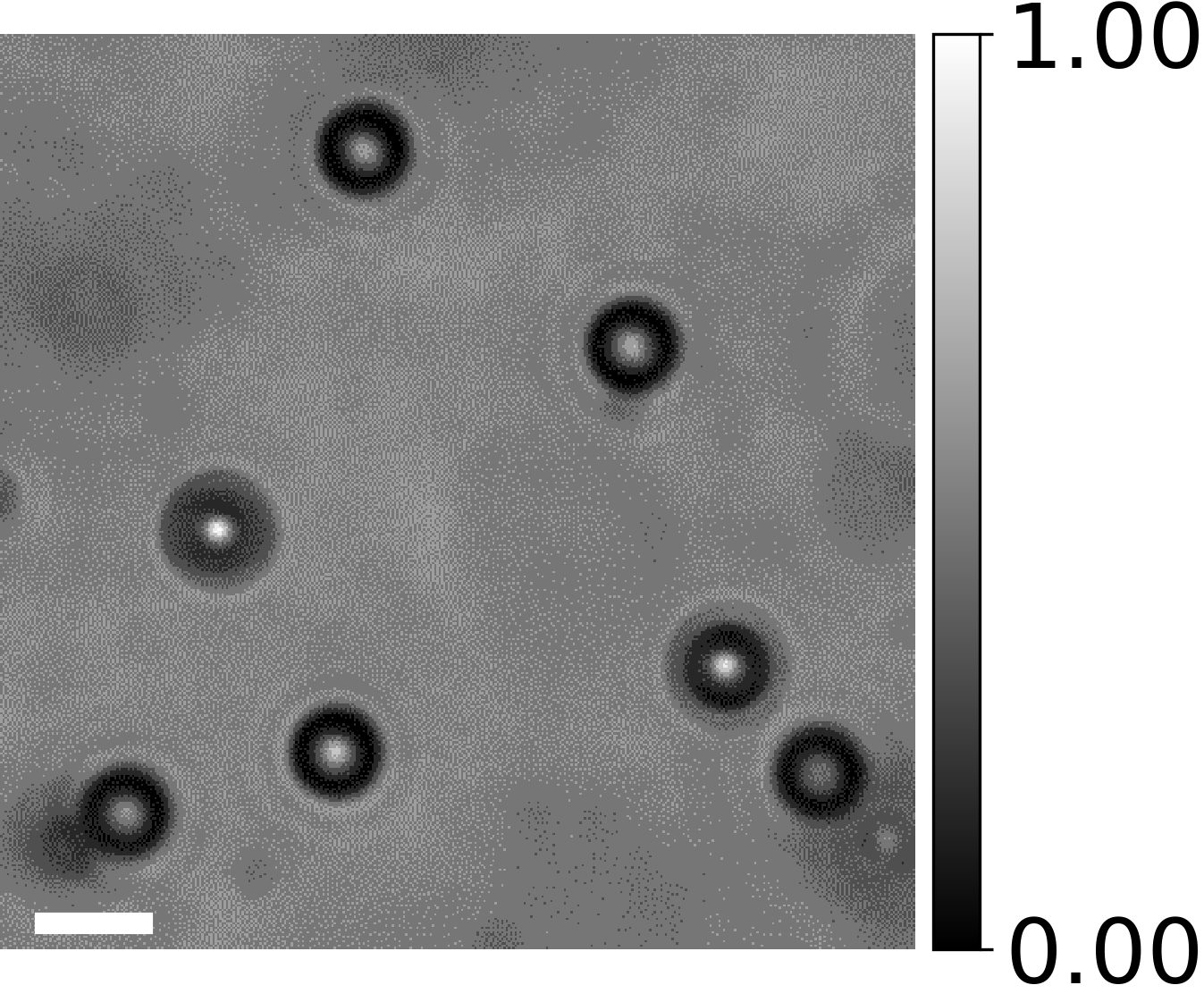}\\&
			\includegraphics[width= 0.16\textwidth]{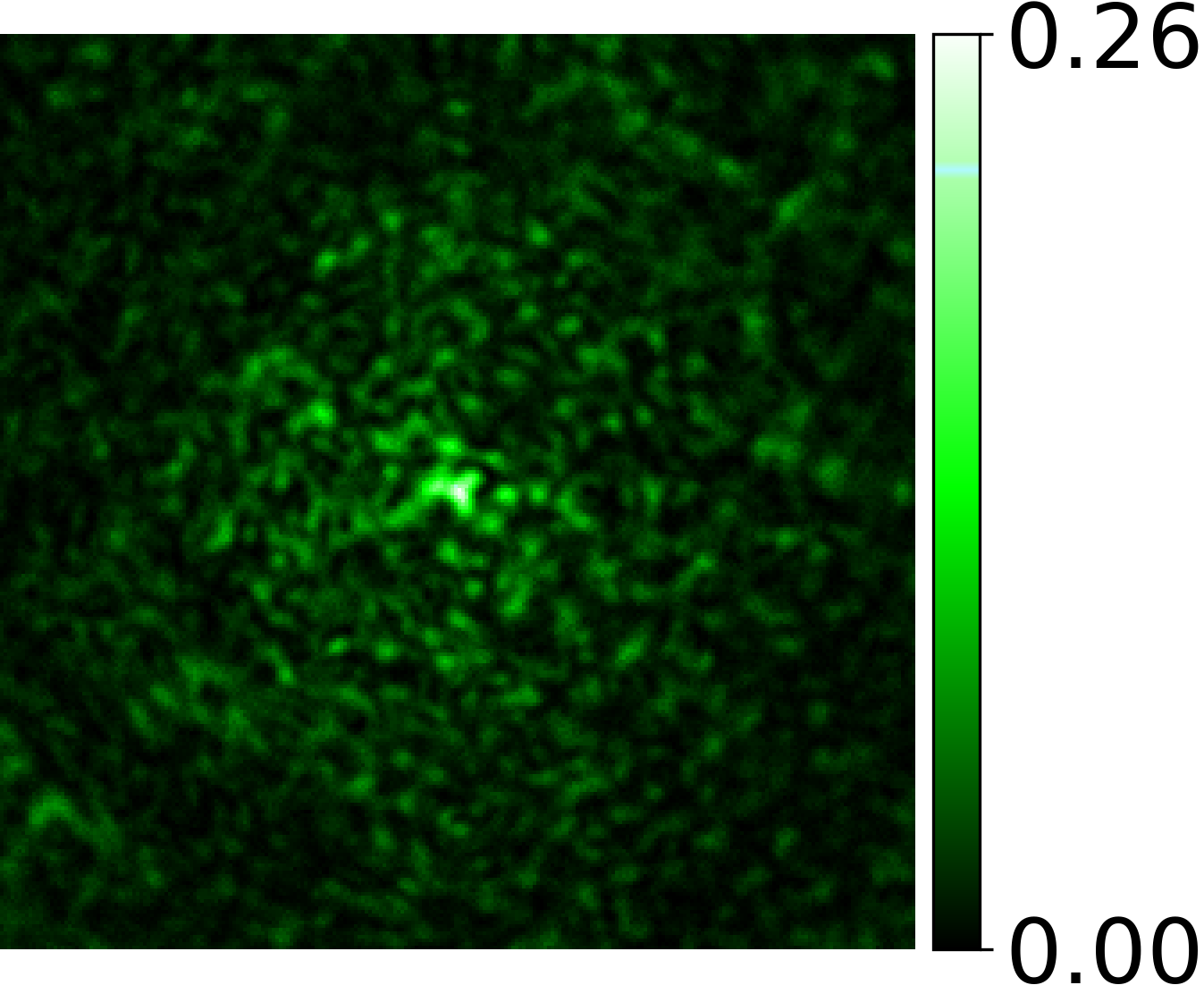}&
			\includegraphics[width= 0.16\textwidth]{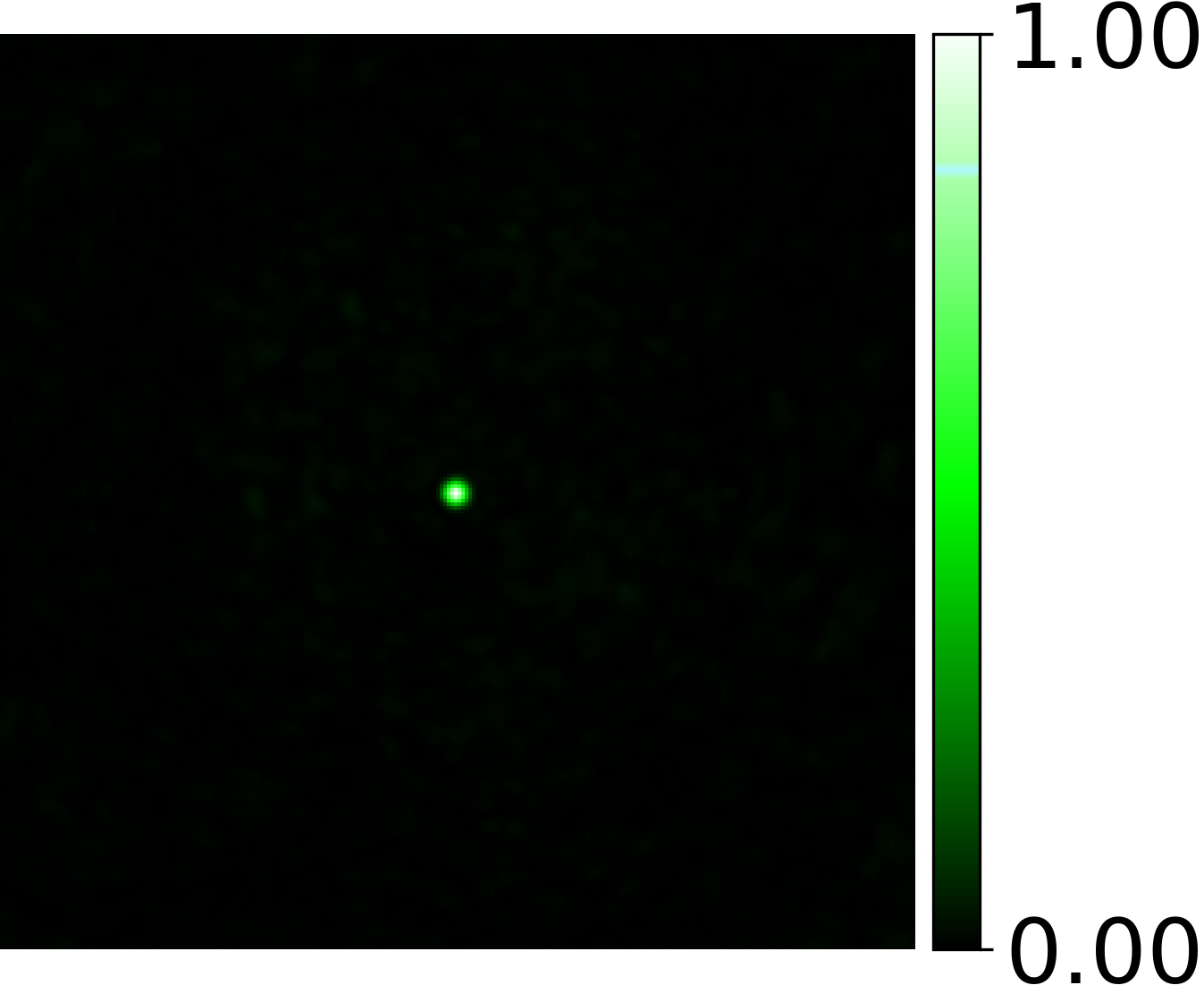}&
			\includegraphics[width= 0.16\textwidth]{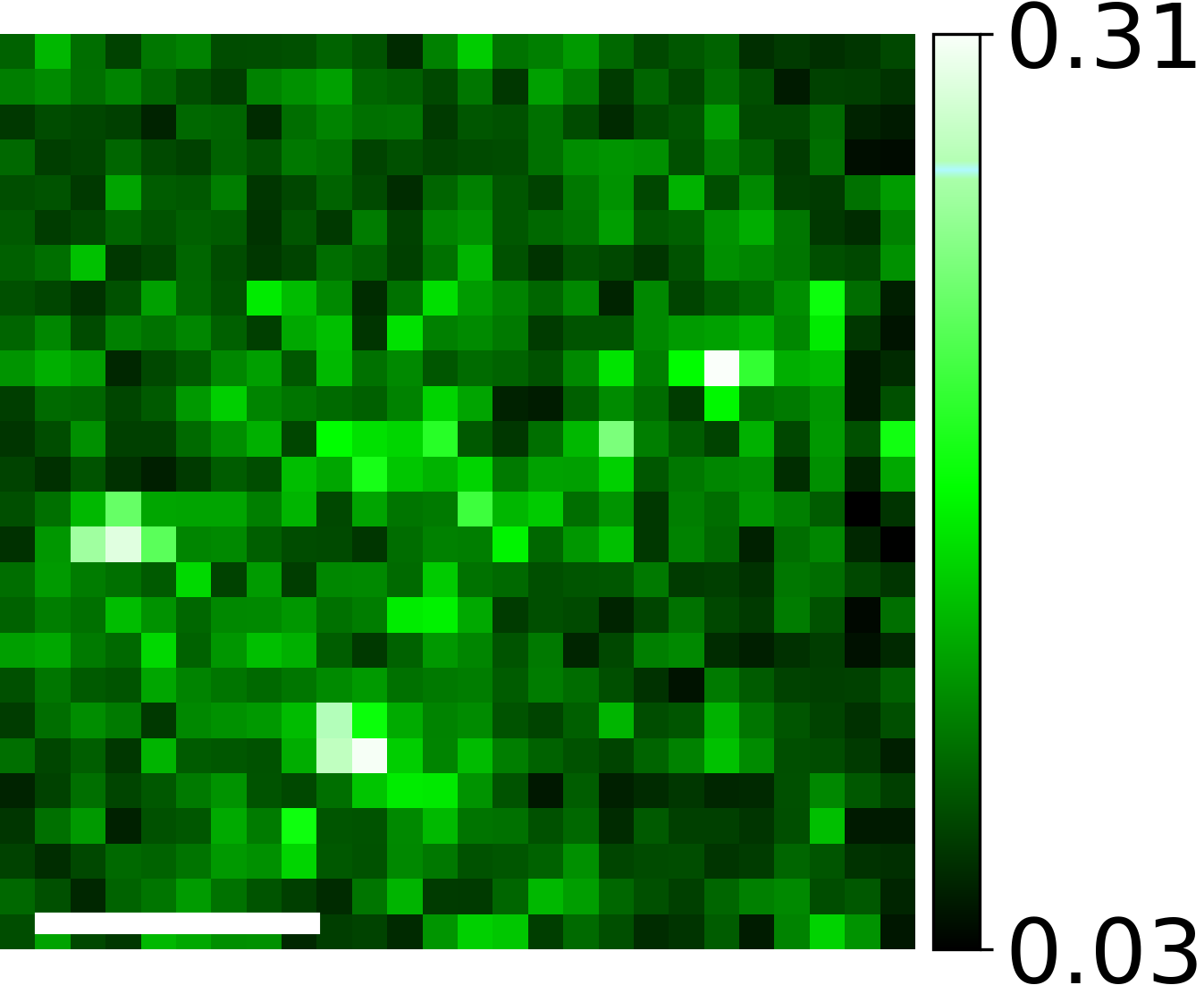}&
			\includegraphics[width= 0.16\textwidth]{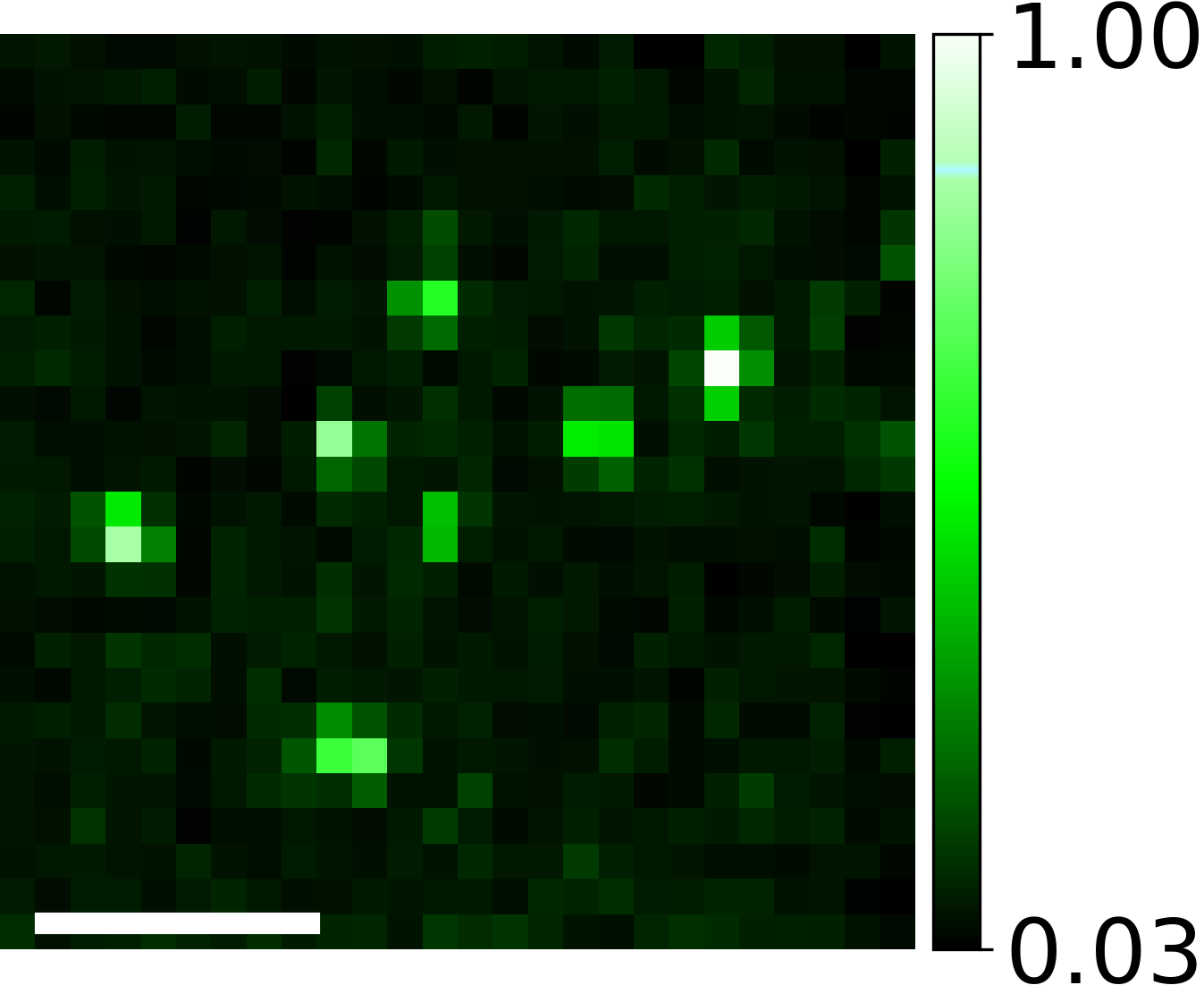}&
			\includegraphics[width= 0.16\textwidth]{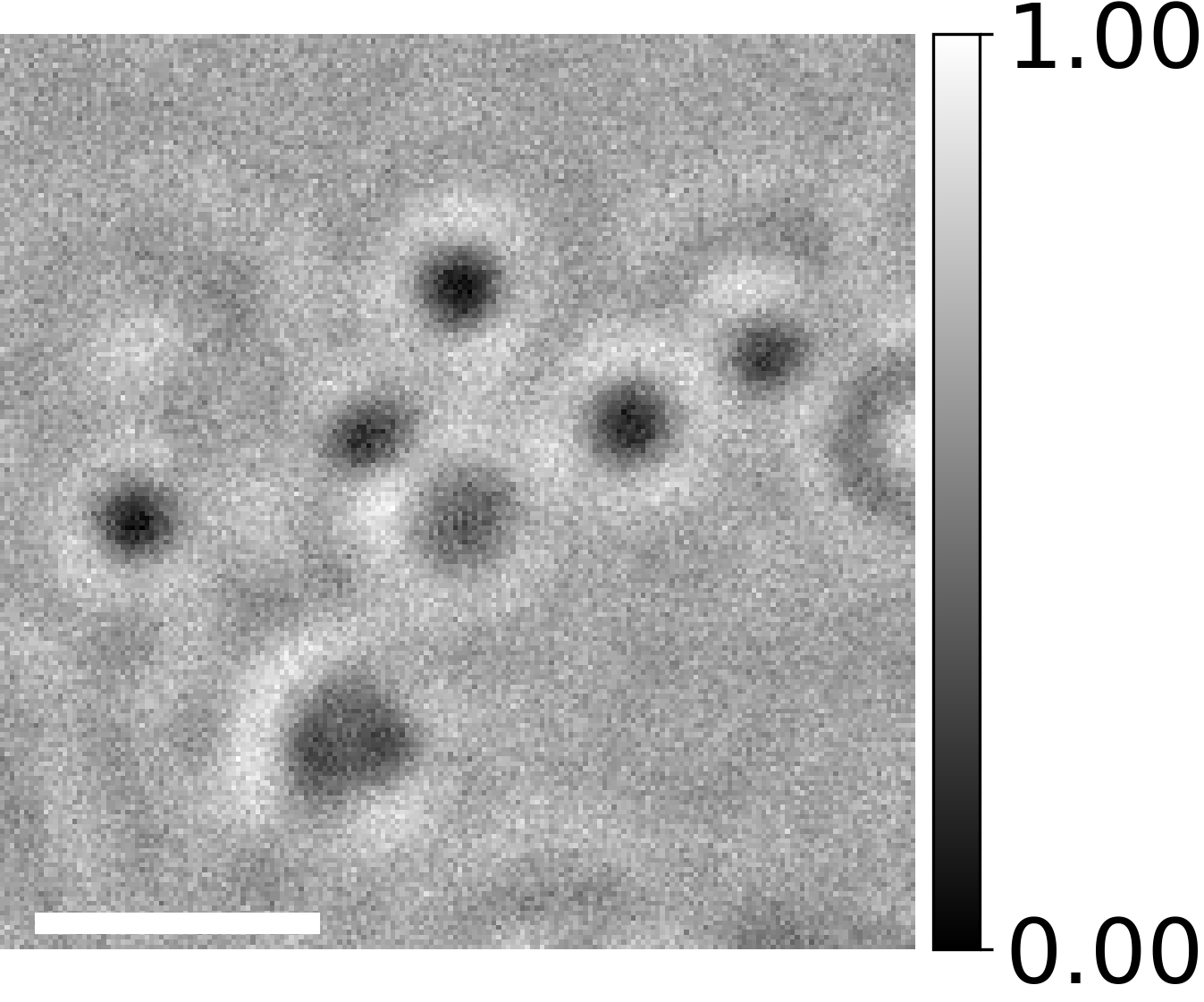}\\&
			\includegraphics[width= 0.16\textwidth]{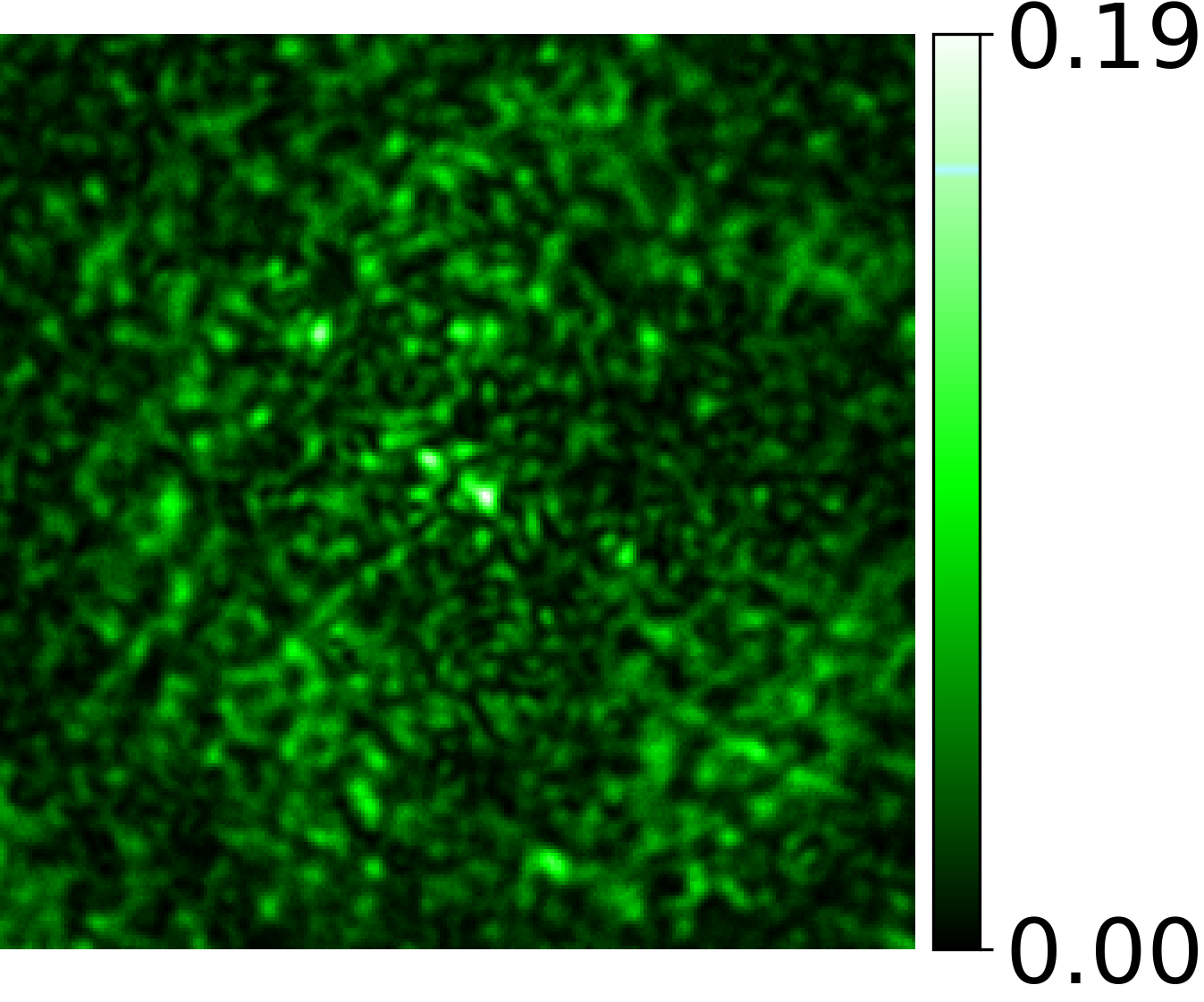}&
			\includegraphics[width= 0.16\textwidth]{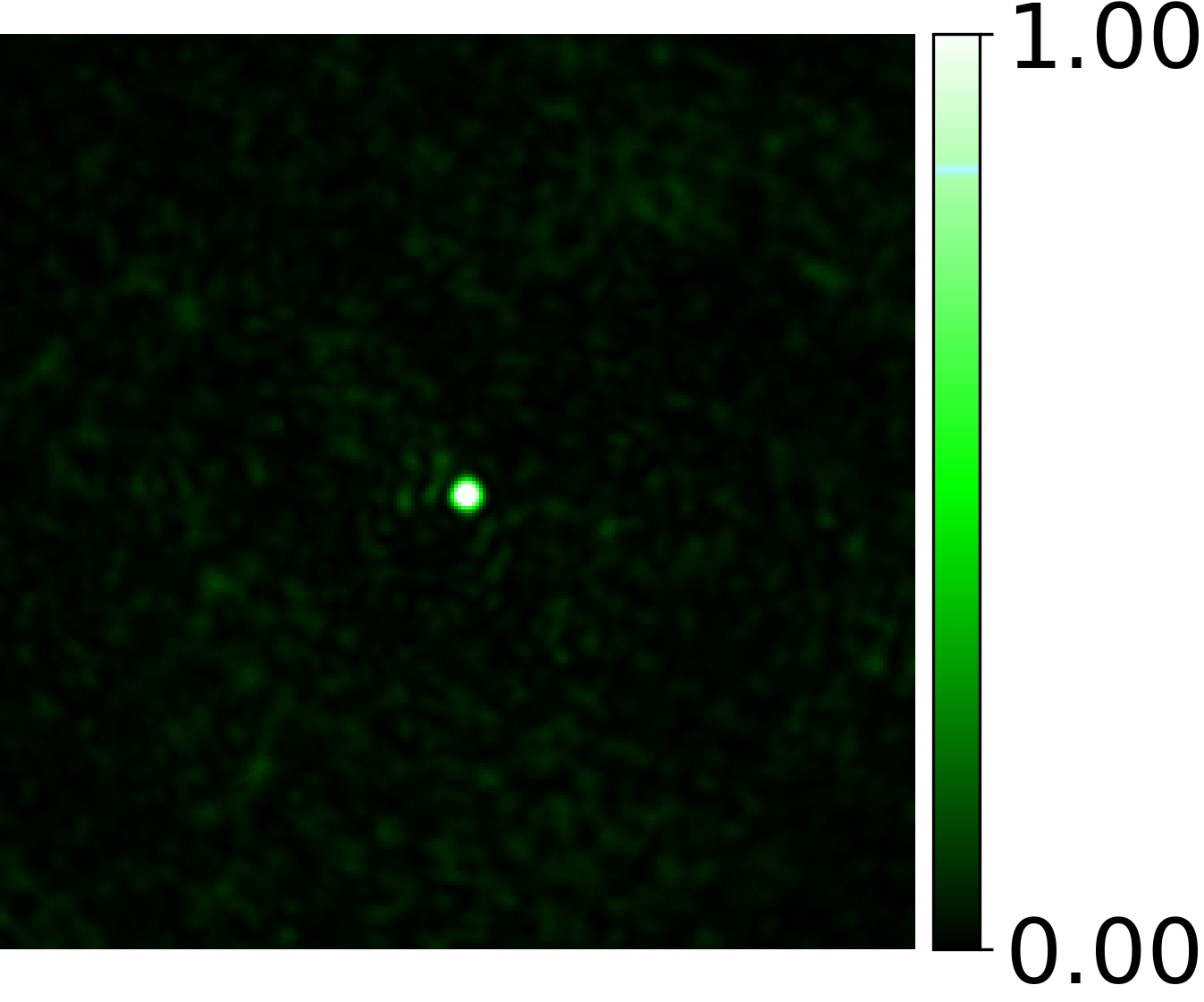}&
			\includegraphics[width= 0.16\textwidth]{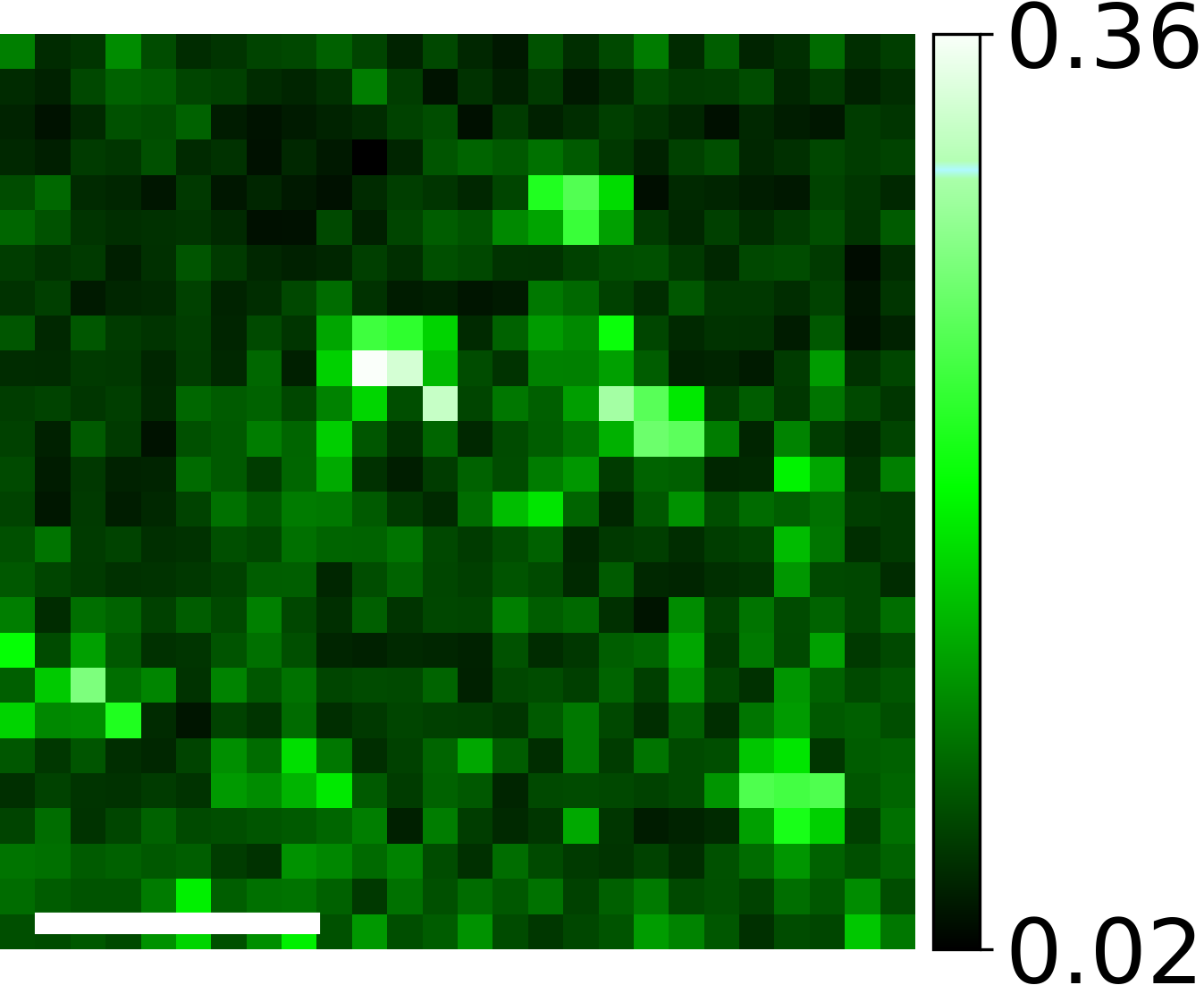}&
			\includegraphics[width= 0.16\textwidth]{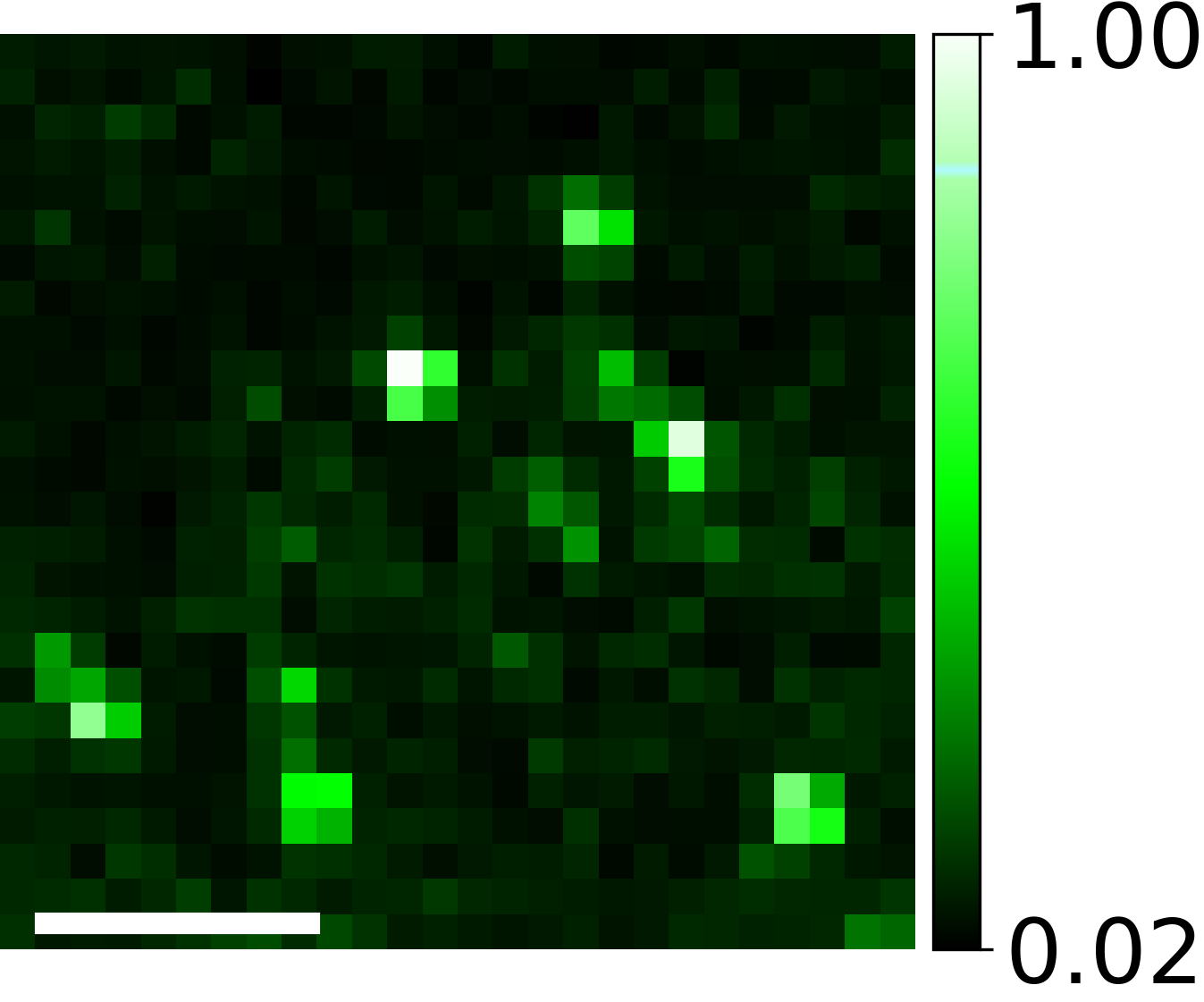}&
			\includegraphics[width= 0.16\textwidth]{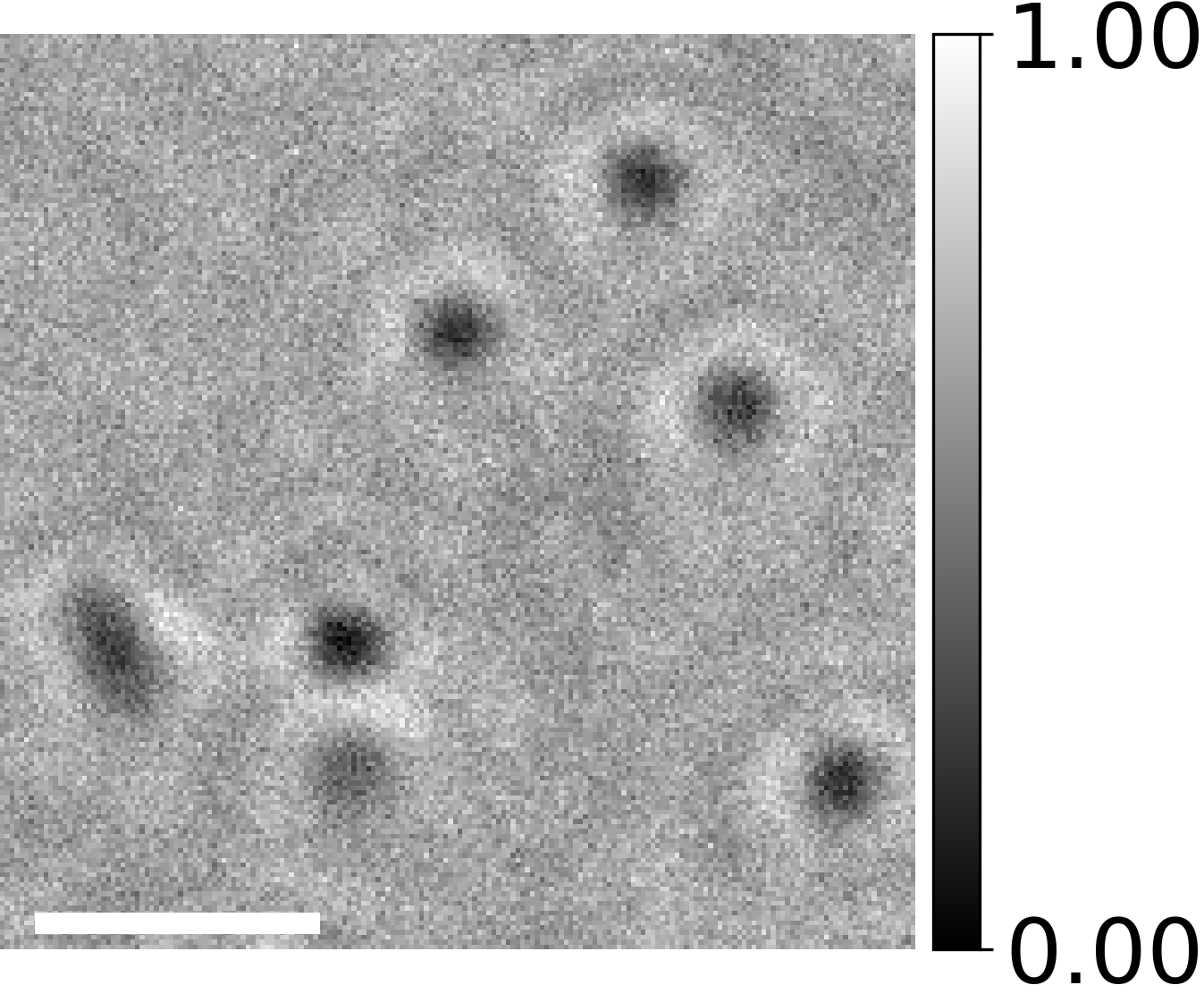}\\&
			\includegraphics[width= 0.16\textwidth]{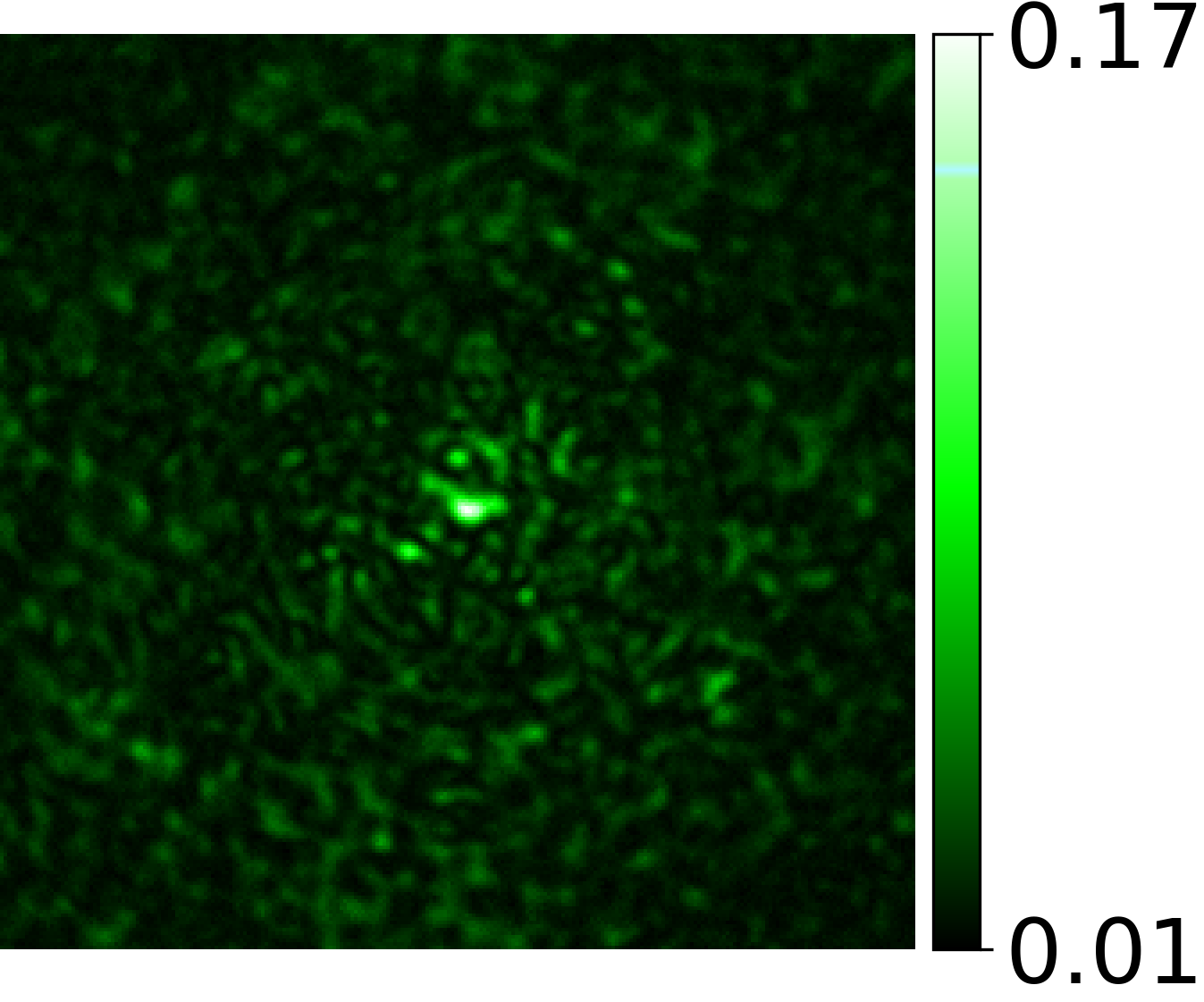}&
			\includegraphics[width= 0.16\textwidth]{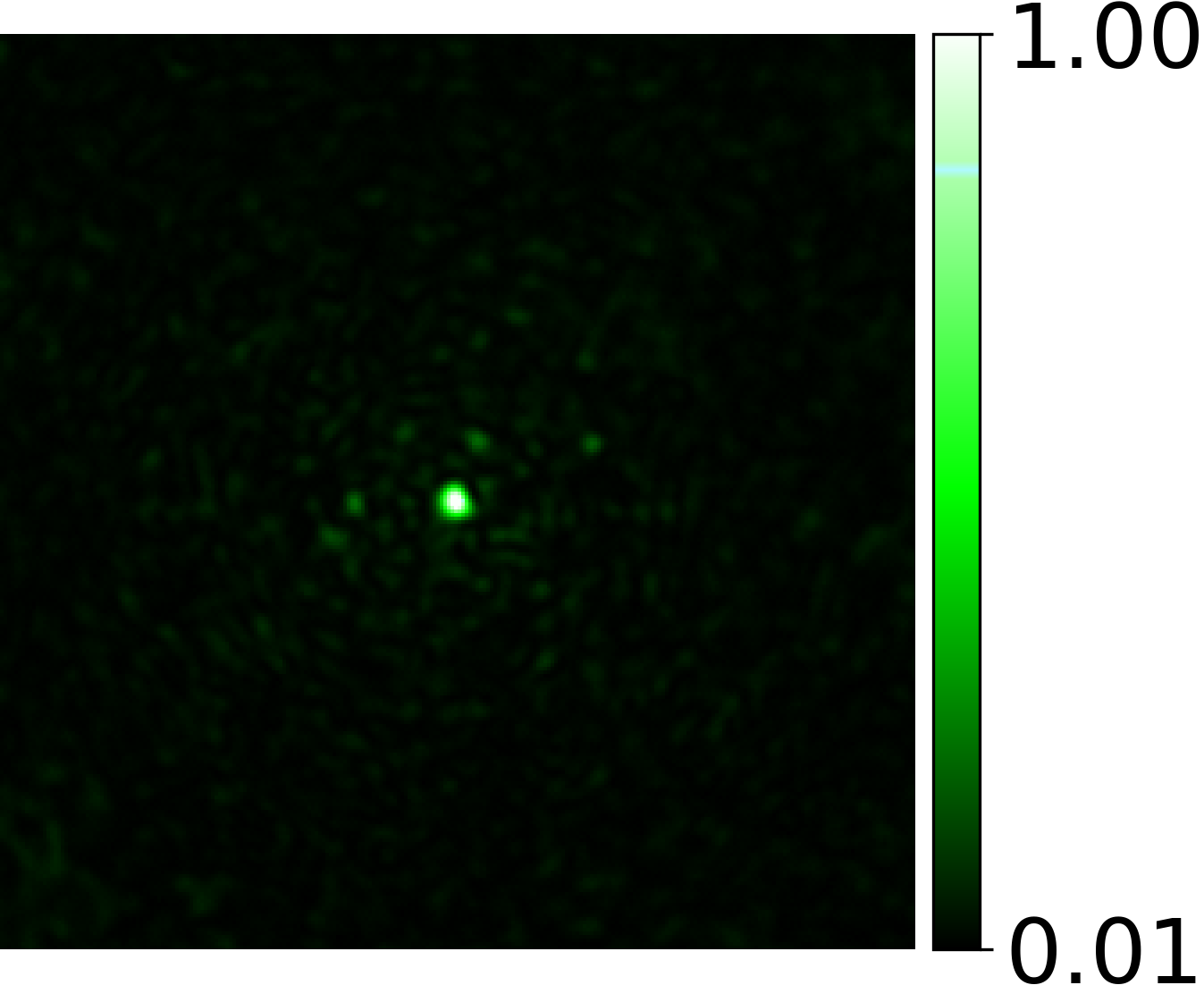}&
			\includegraphics[width= 0.16\textwidth]{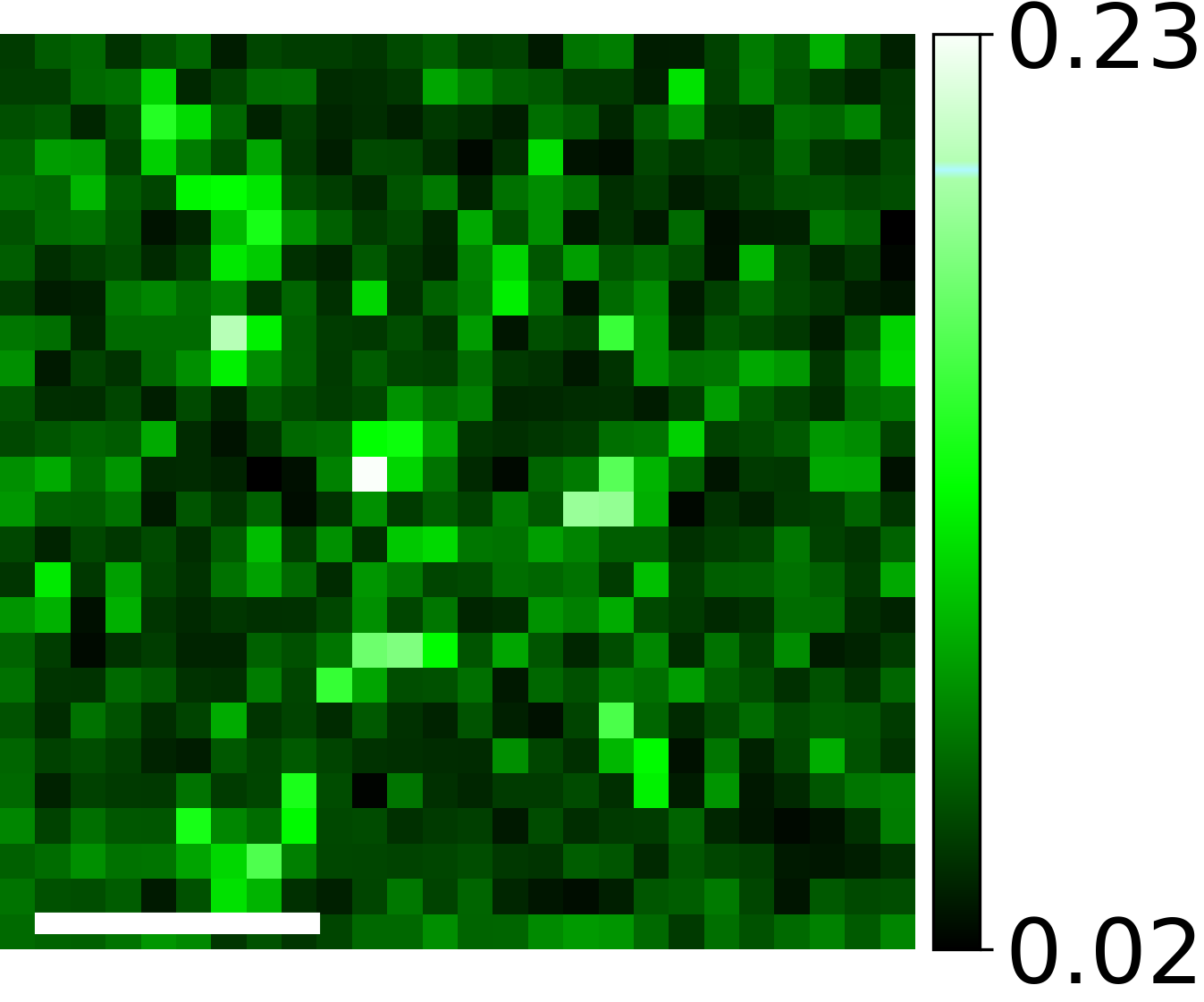}&
			\includegraphics[width= 0.16\textwidth]{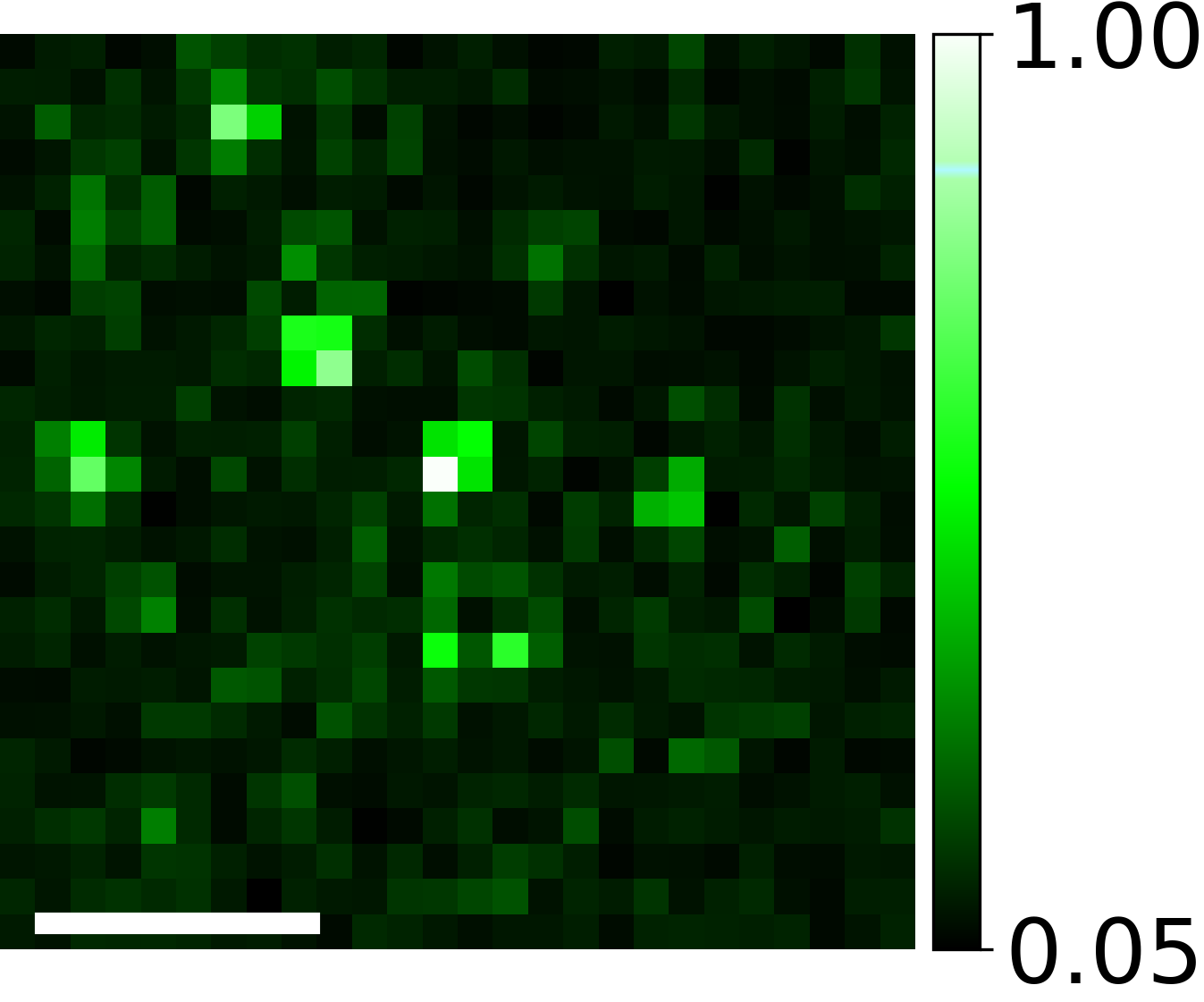}&
			\includegraphics[width= 0.16\textwidth]{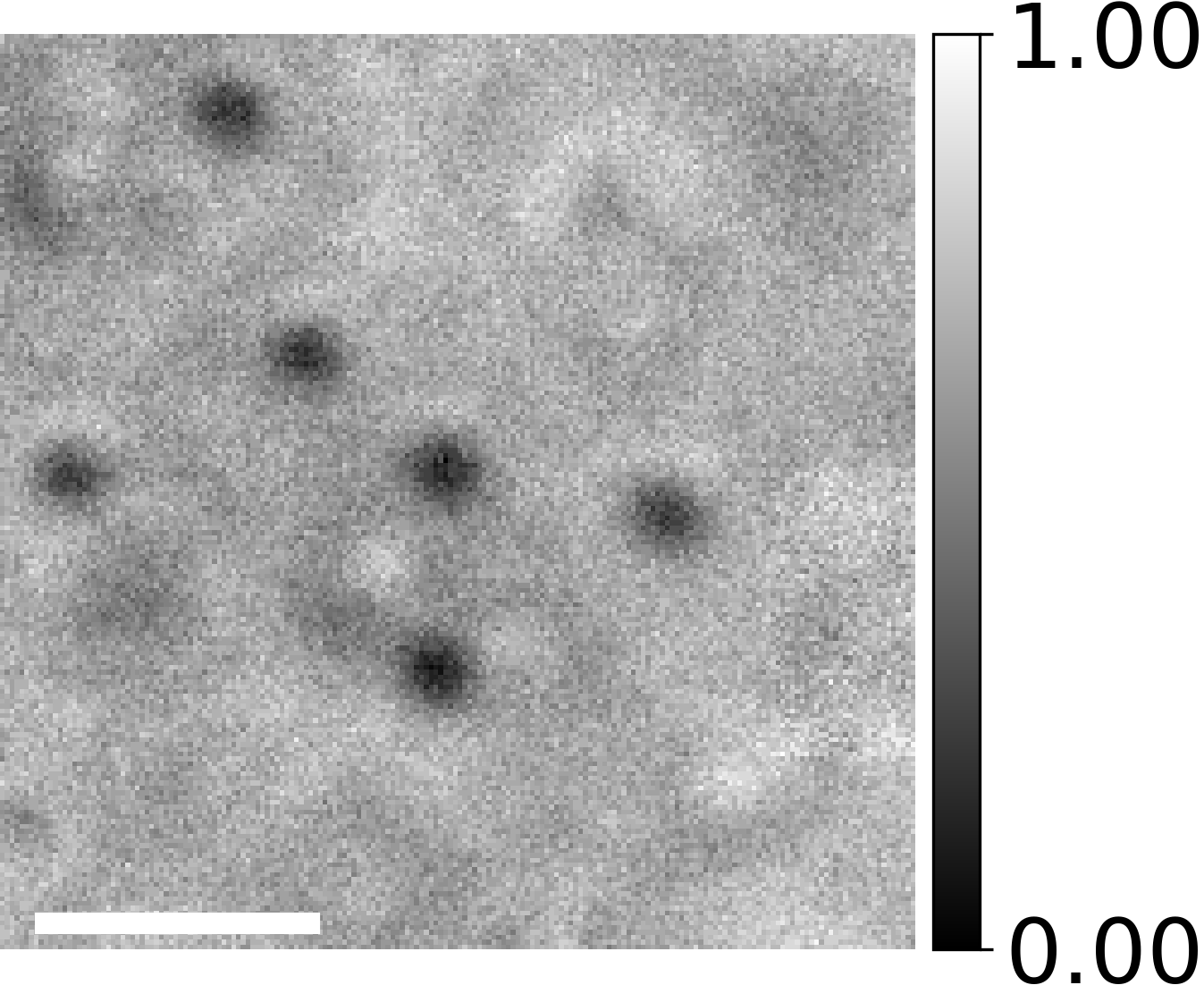}
		\end{tabular}
		\caption{\textbf{Confcal scan of beads:} Our algorithm used to image a scattering dispersion of  polystyrene beads in agarose gel. \blue{The first column depicts a schematic view of the target, where we image small regions in different spatial areas and varying depth of the gel. The green bars inside the target depict the different areas which were imaged.} Columns 2-3: Main camera images of a single focal spot before and after correction. Columns 4-5: Confocal scanning results: with and without aberration correction. Column 6: A reference image of the target captured from the validation camera behind the target, under wide-filed  incoherent illumination. Rows 1-2 use $3\um$ diameter beads and rows 3-5 use $0.5\um$ diameter beads. Scale bar on confocal images is $4\um$. Note that an ideal aberration free image of beads  in a reflection mode confocal  microscope is a diffraction limited spot~\cite{Weise96}, even if the beads are larger than the diffraction limit since they act as a mini-lens focusing light. }
		\label{fig:beads2}
	\end{center}
\end{figure*}

%% file: fig_beads_xy.tex
\begin{figure*}[h!]
	\begin{center}		
		\begin{tabular}{@{}c@{~}c@{~}c@{~}}			
			\multicolumn{2}{c}{\hspace{-1.5cm}\large Confocal scan}&
			\multicolumn{1}{c}{\hspace{-1.5cm}\large Wide field}\\
			\multicolumn{2}{c}{\hspace{-1.5cm} Main cam.} & 
			\multicolumn{1}{c}{\hspace{-1.5cm} Valid. cam.} \\
			\multicolumn{1}{c}{\hspace{-1.5cm} \scriptsize w/o}&	
			\multicolumn{1}{c}{\hspace{-1.5cm} \scriptsize w/ }&		
			\multicolumn{1}{c}{\hspace{-1.5cm} \scriptsize Incoherent}\vspace{-0cm}\\
			\multicolumn{1}{c}{\hspace{-1.5cm} \scriptsize modulation}&	
			\multicolumn{1}{c}{\hspace{-1.5cm} \scriptsize modulation}&	
			\multicolumn{1}{c}{\hspace{-1.5cm} \scriptsize illumination}\\
			\includegraphics[width= 0.15\textwidth]{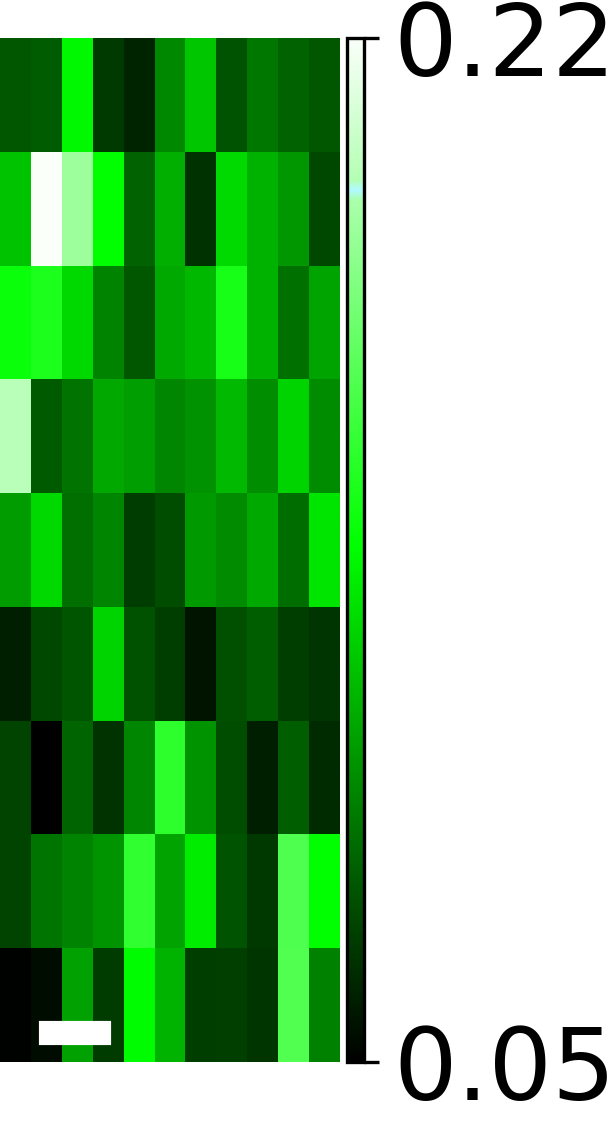}&
			\includegraphics[width= 0.15\textwidth]{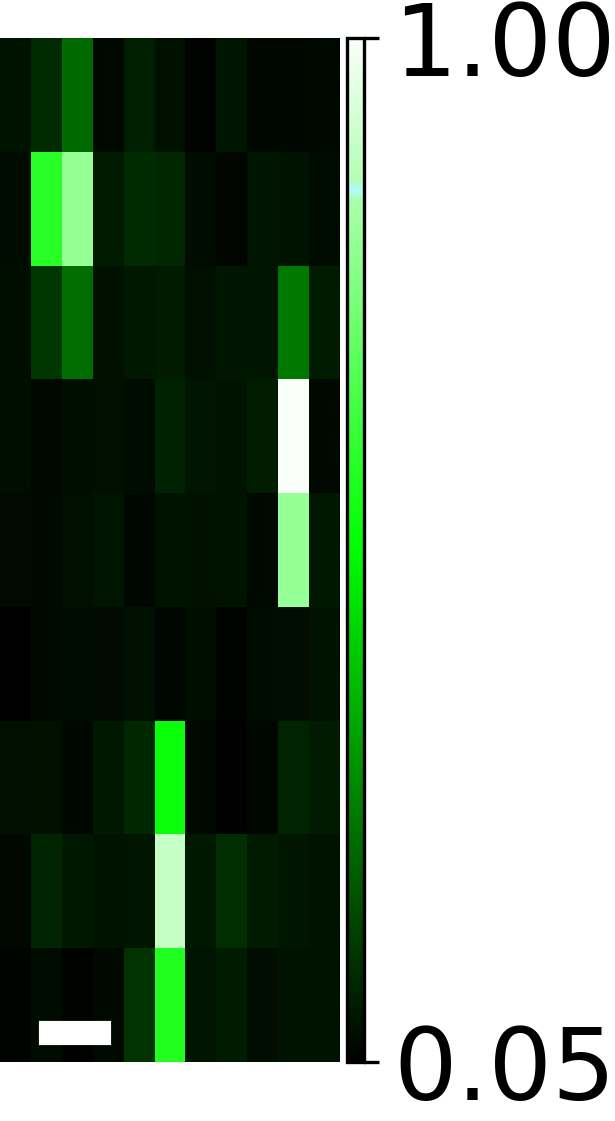}&
			\includegraphics[width= 0.15\textwidth]{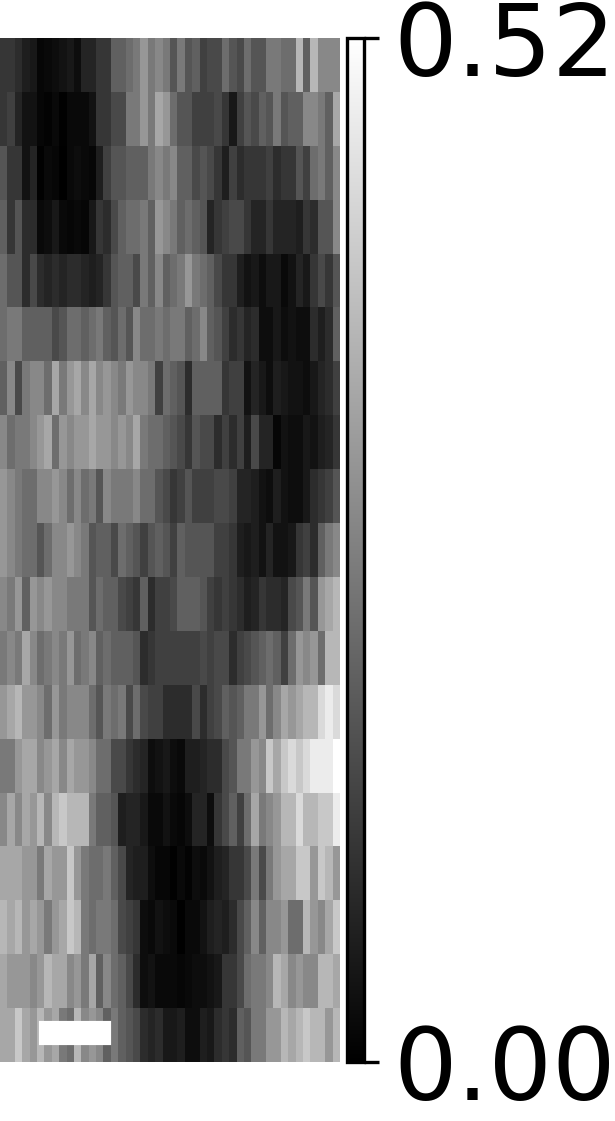}
		\end{tabular}
		\caption{\blue{\textbf{Confocal scan of beads, $x-z$ slice:} We apply our algorithm on a 2D $x-z$ slice, by utilizing the memory effect on the lateral axis and add quadratic phase to change focusing plane in the axial direction. We image an area $\Area = 5.2\um \times 16\um$. The scale bar is $1.2\um$.} }
		\label{fig:beads_xy}
	\end{center}
\end{figure*}

%% file: fig_stitch_sup.tex
\begin{figure*}[h!]
	\begin{center}		
		\begin{tabular}{@{}c@{~}c@{~}c@{~}c@{~}}			
			\multicolumn{2}{c}{\hspace{-0.6cm}\large Main cam. }& 
			\multicolumn{2}{c}{\hspace{-0.6cm}\large Valid. cam.}\\
			\multicolumn{1}{c}{\hspace{-0.6cm} w/o}&	
			\multicolumn{1}{c}{\hspace{-0.6cm} w/ }&	
			\multicolumn{1}{c}{\hspace{-0.6cm} w/o}&	
			\multicolumn{1}{c}{\hspace{-0.6cm} w/ }\\
			\multicolumn{1}{c}{\hspace{-0.6cm} modulation}&	
			\multicolumn{1}{c}{\hspace{-0.6cm} modulation}&	
			\multicolumn{1}{c}{\hspace{-0.6cm} modulation}&	
			\multicolumn{1}{c}{\hspace{-0.6cm} modulation}\\
			\includegraphics[width= 0.24\textwidth]{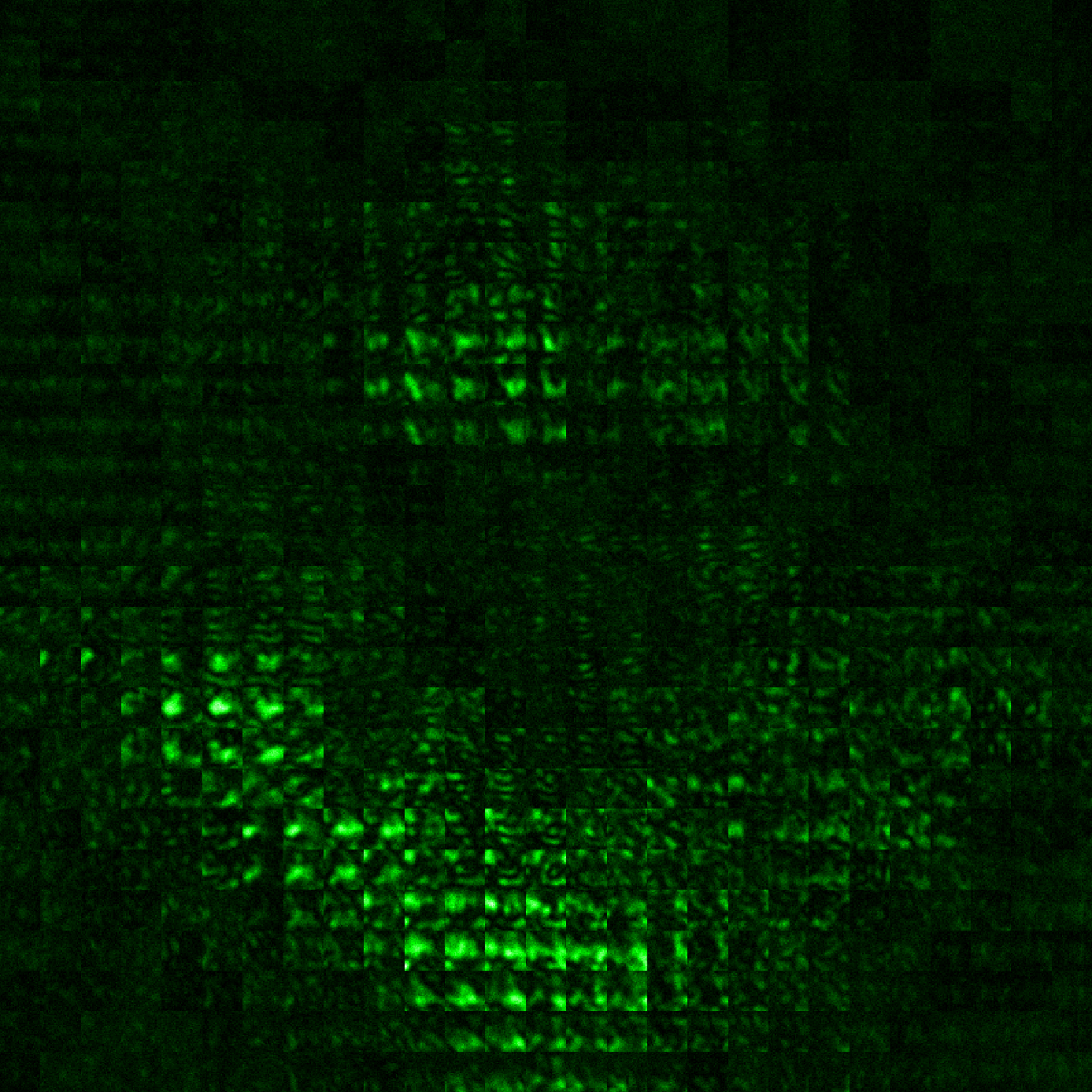}&
			\includegraphics[width= 0.24\textwidth]{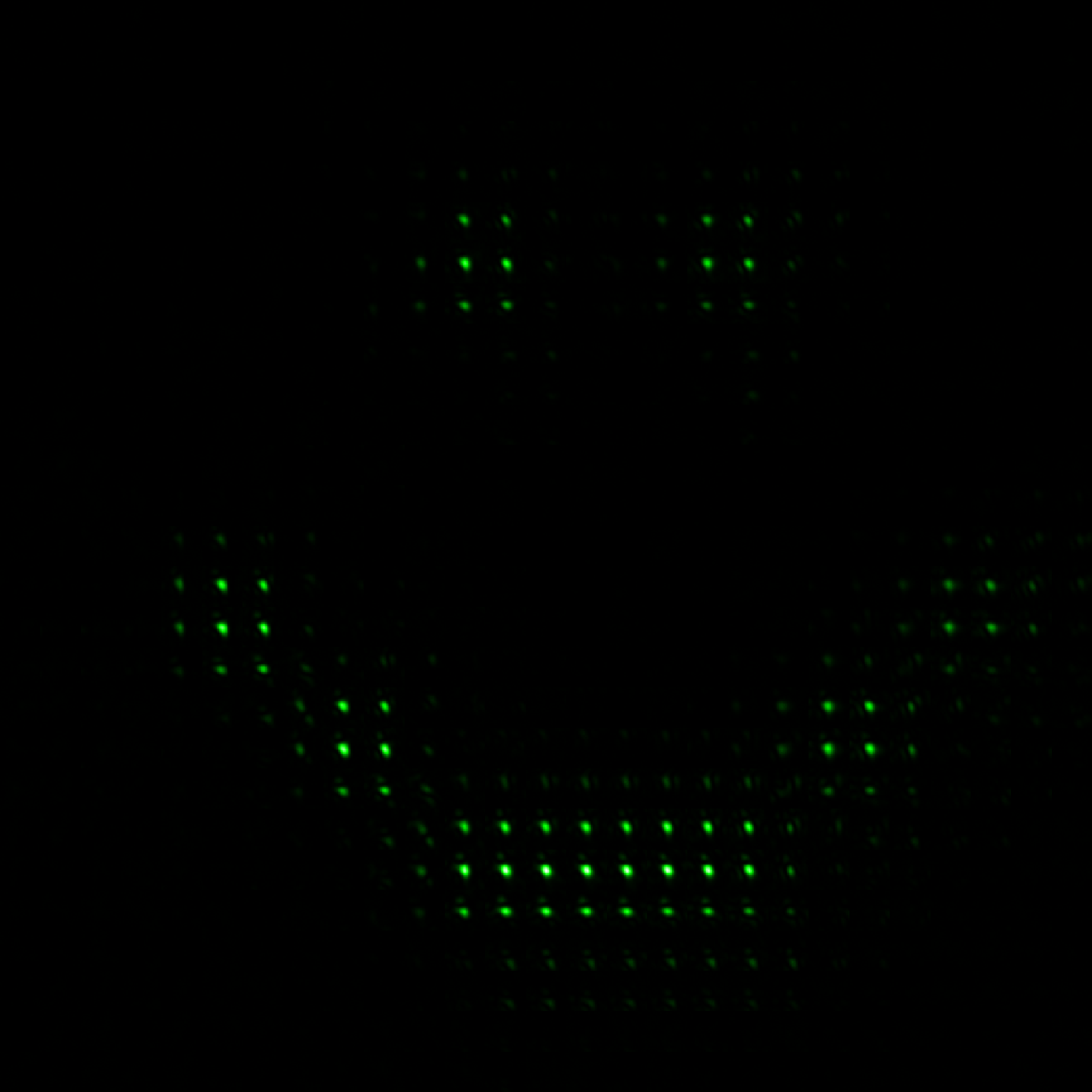}&
			\includegraphics[width= 0.24\textwidth]{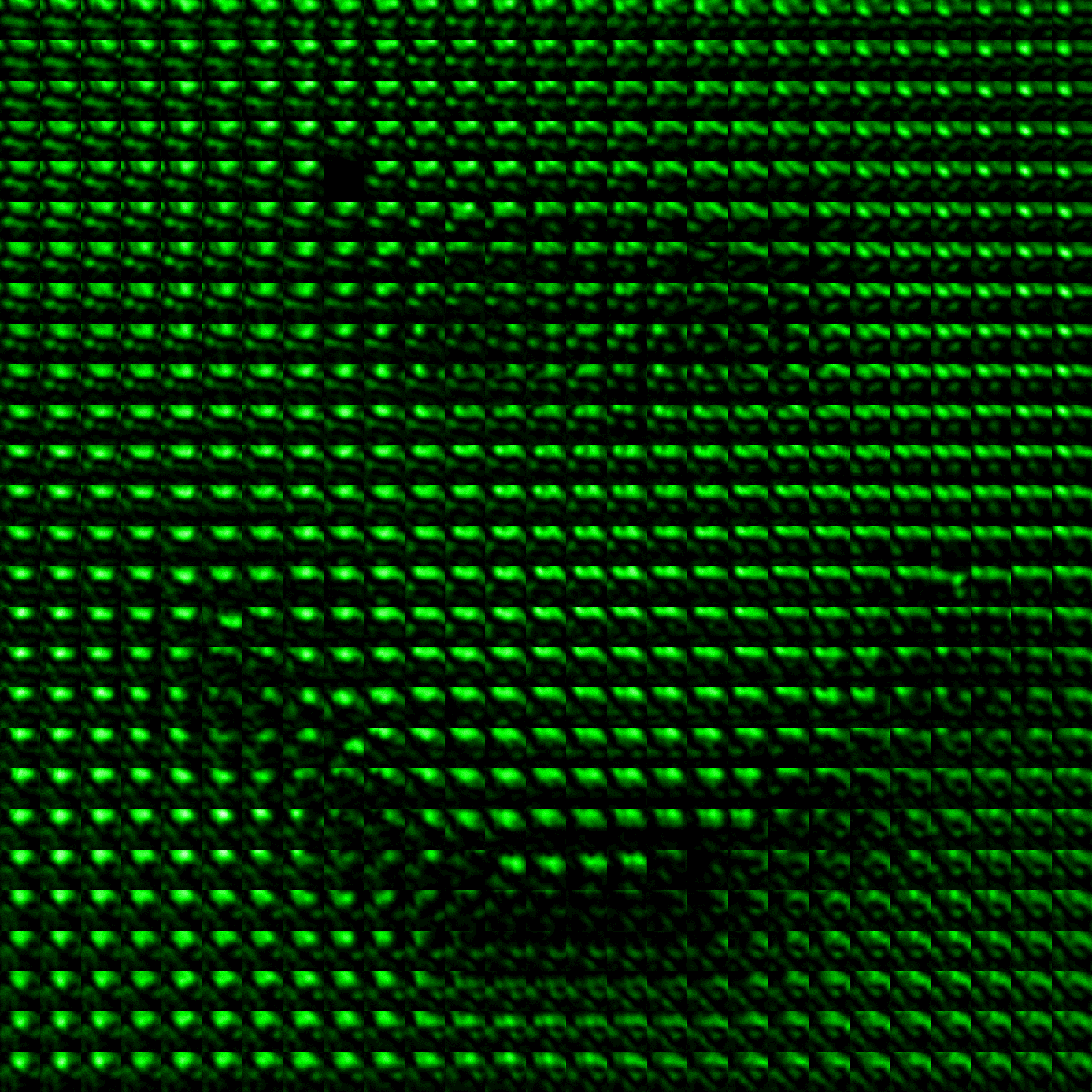}&
			\includegraphics[width= 0.24\textwidth]{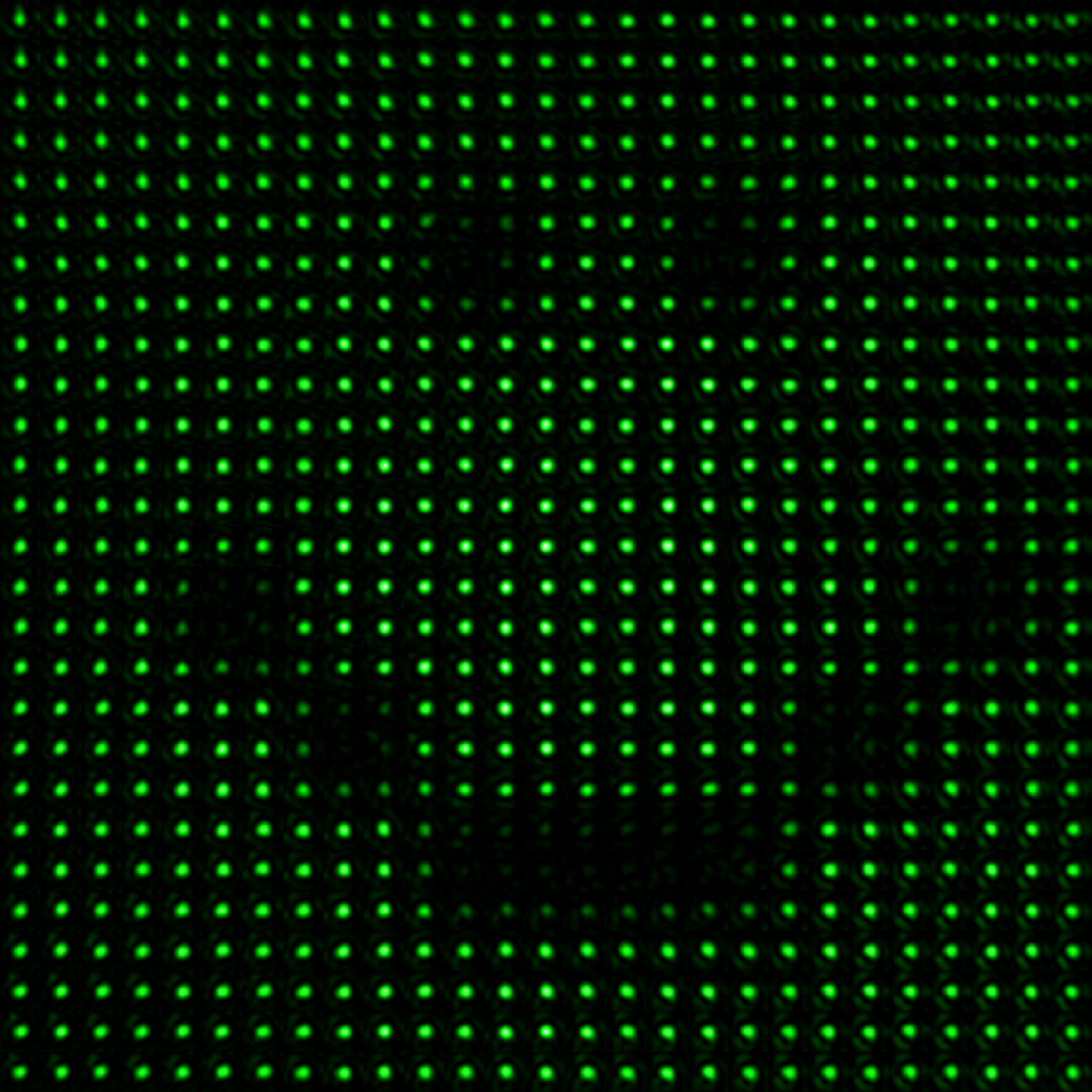}
		\end{tabular}
		\caption{\textbf{Individual scanning images:} We present the scanning results before and after optimization from the Main camera and Validation camera. Before optimization the light is scattered at both cameras. After optimization, we get good focus on both main and validation cameras. Areas with coated chrome reflect more light resulting in brighter spots on the main camera. The chrome significantly attenuates the forward scattered light, hence they appear as black spots in the validation camera. All images were normalized to the maximum value of the image for visualization purposes, in practice the images with no correction are much darker. }
		\label{fig:stitch_sup}
	\end{center}
\end{figure*}

%% file: fig_res_patches.tex
\begin{figure*}[h!]
	\begin{center}		
		\begin{tabular}{@{}c@{~}c@{~}c@{~}c@{~}c@{~}}			
			\multicolumn{1}{c}{\hspace{-1cm}\large Combined confocal scan }& 
			\multicolumn{2}{c}{\hspace{-0.1cm}\large Confocal scan }& 
			\multicolumn{2}{c}{\hspace{-0.3cm}\large Phase mask }\\
			\raisebox{1.2cm}{\multirow{2}{14em}{\includegraphics[width= 0.32\textwidth]{figs/res_mask/resolution/Ifinal_res.png}}}&
			\includegraphics[width= 0.12\textwidth]{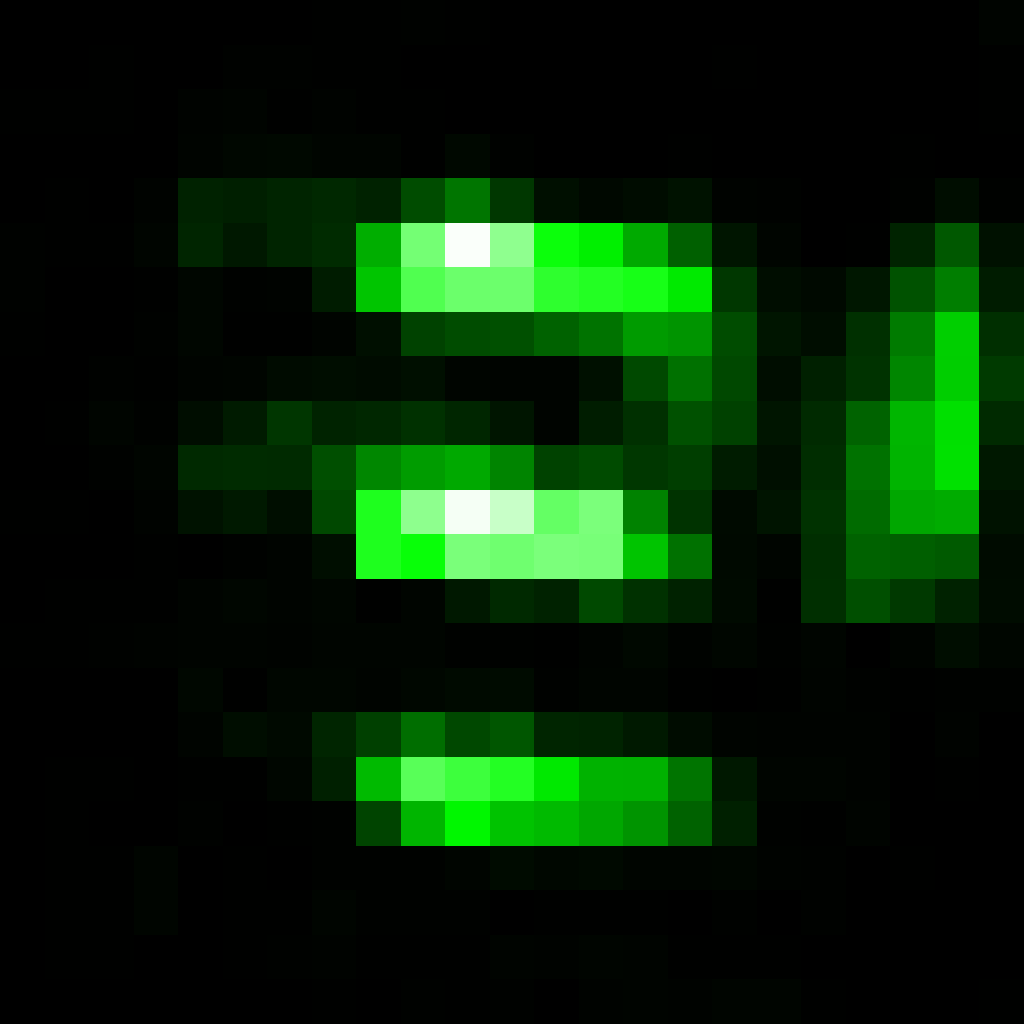}&
			\includegraphics[width= 0.12\textwidth]{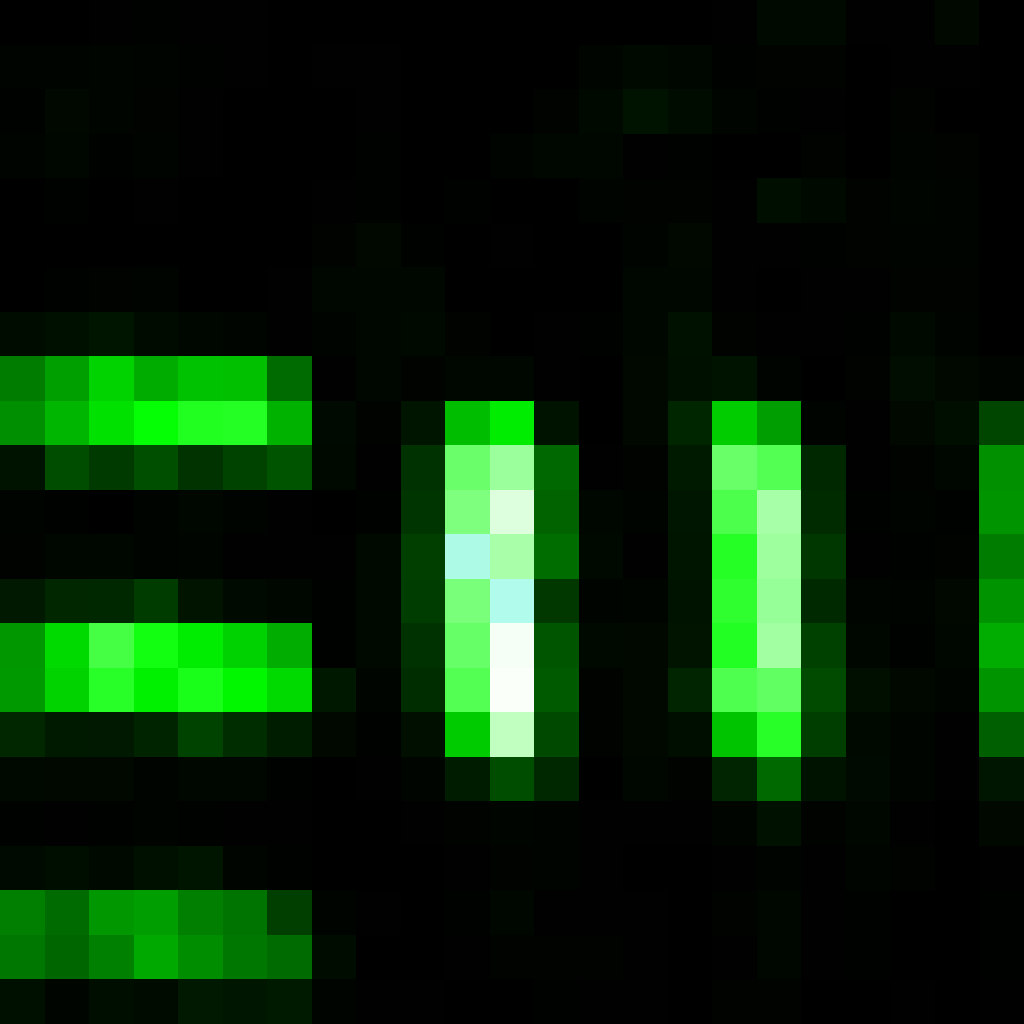}&
			\includegraphics[width= 0.12\textwidth]{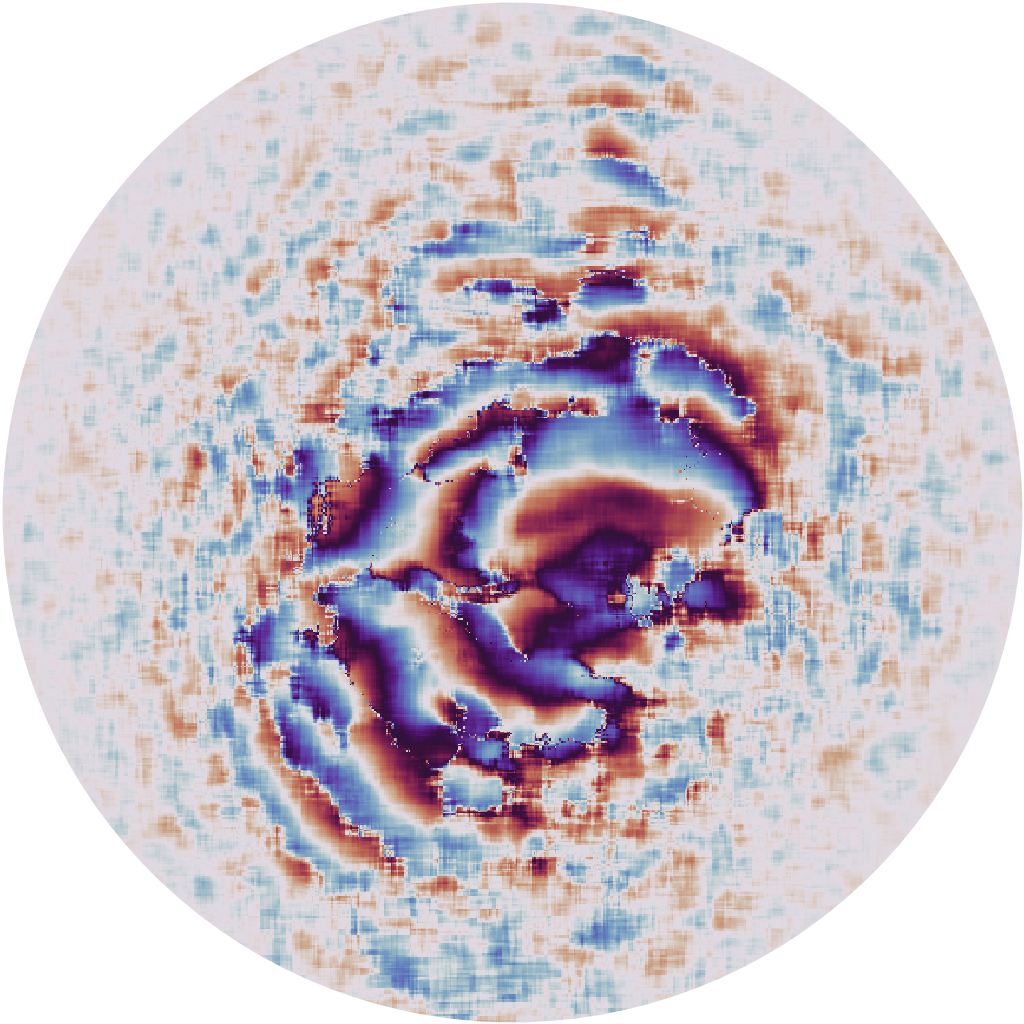}&
			\includegraphics[width= 0.12\textwidth]{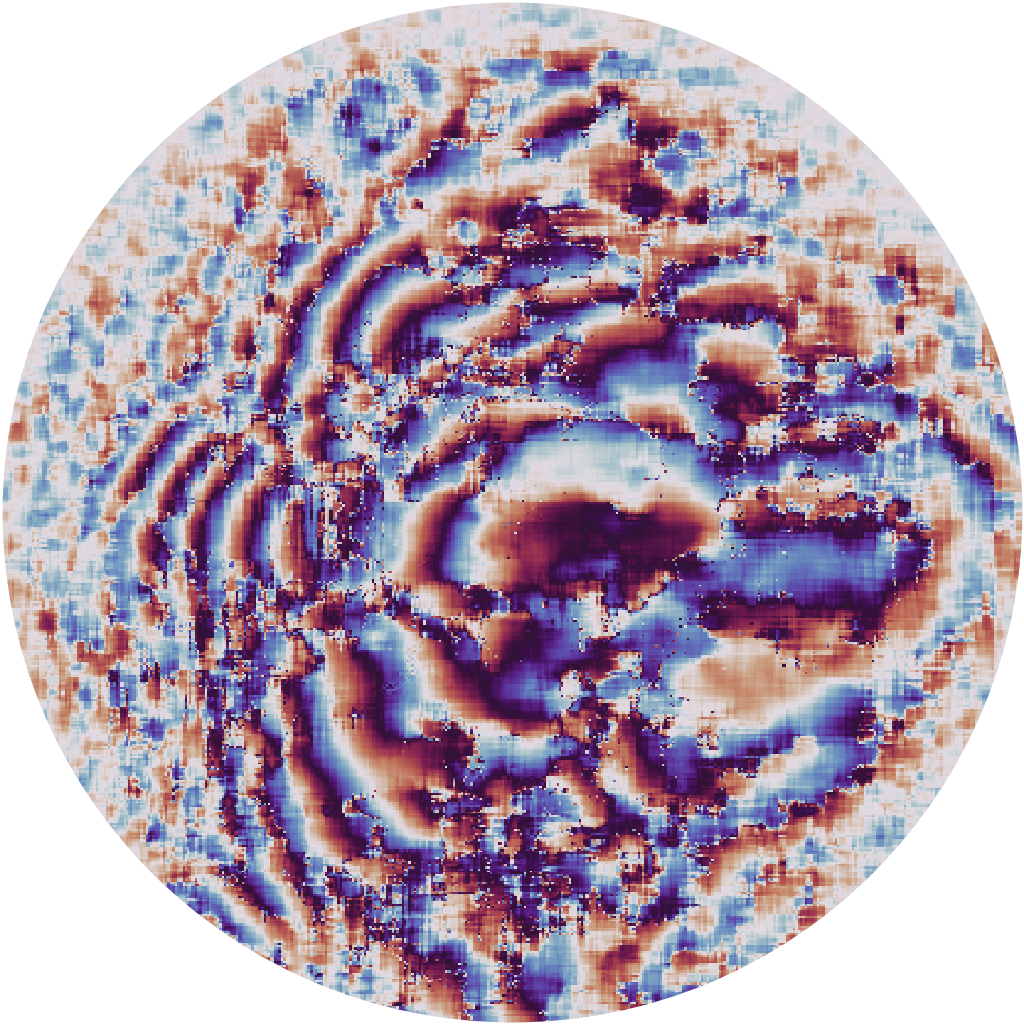}\\&			
			\includegraphics[width= 0.12\textwidth]{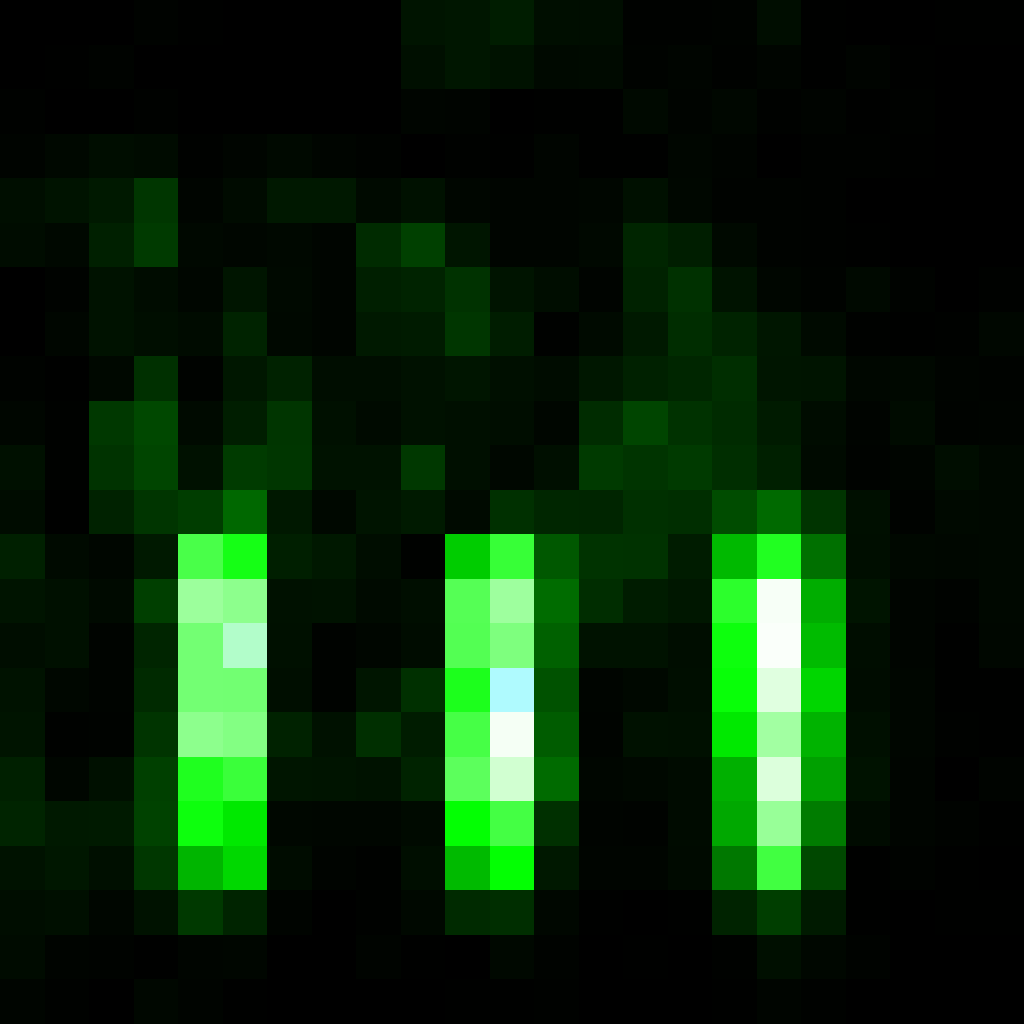}&
			\includegraphics[width= 0.12\textwidth]{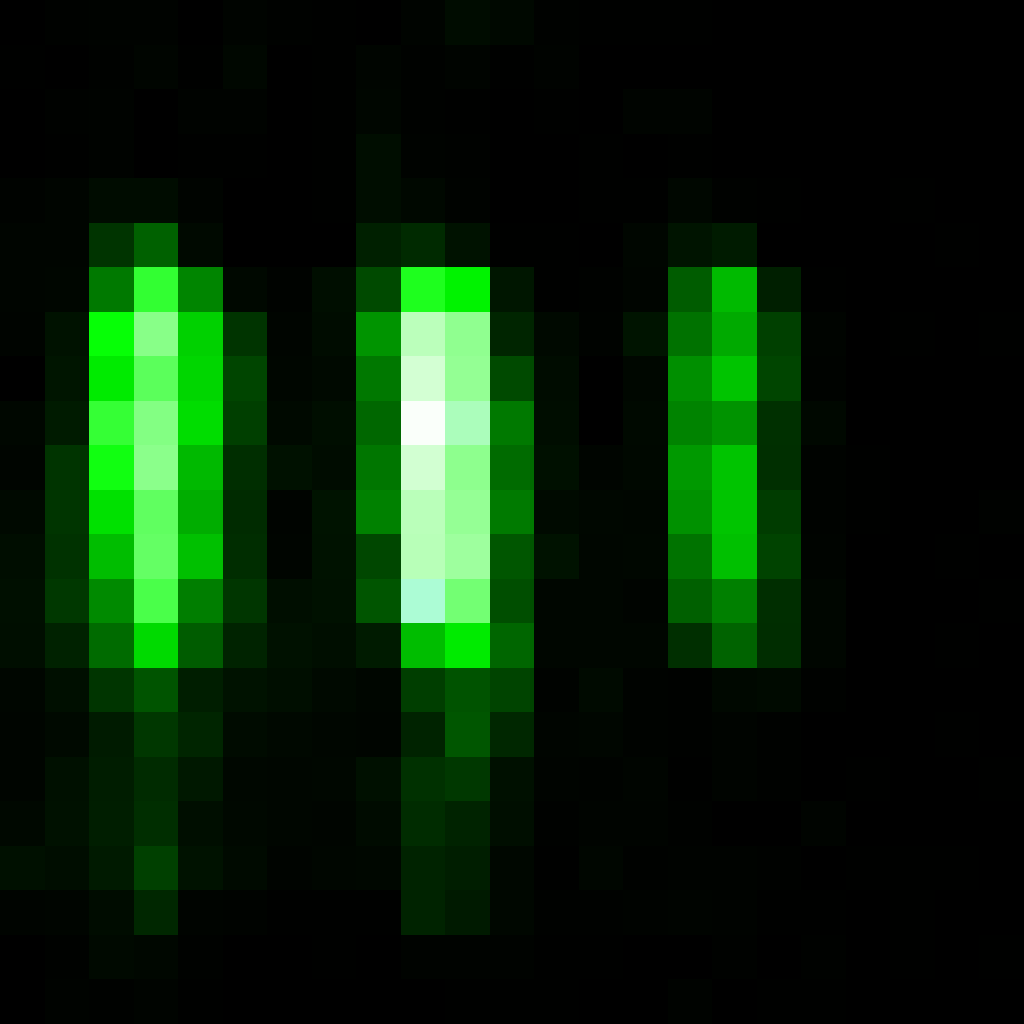}&
			\includegraphics[width= 0.12\textwidth]{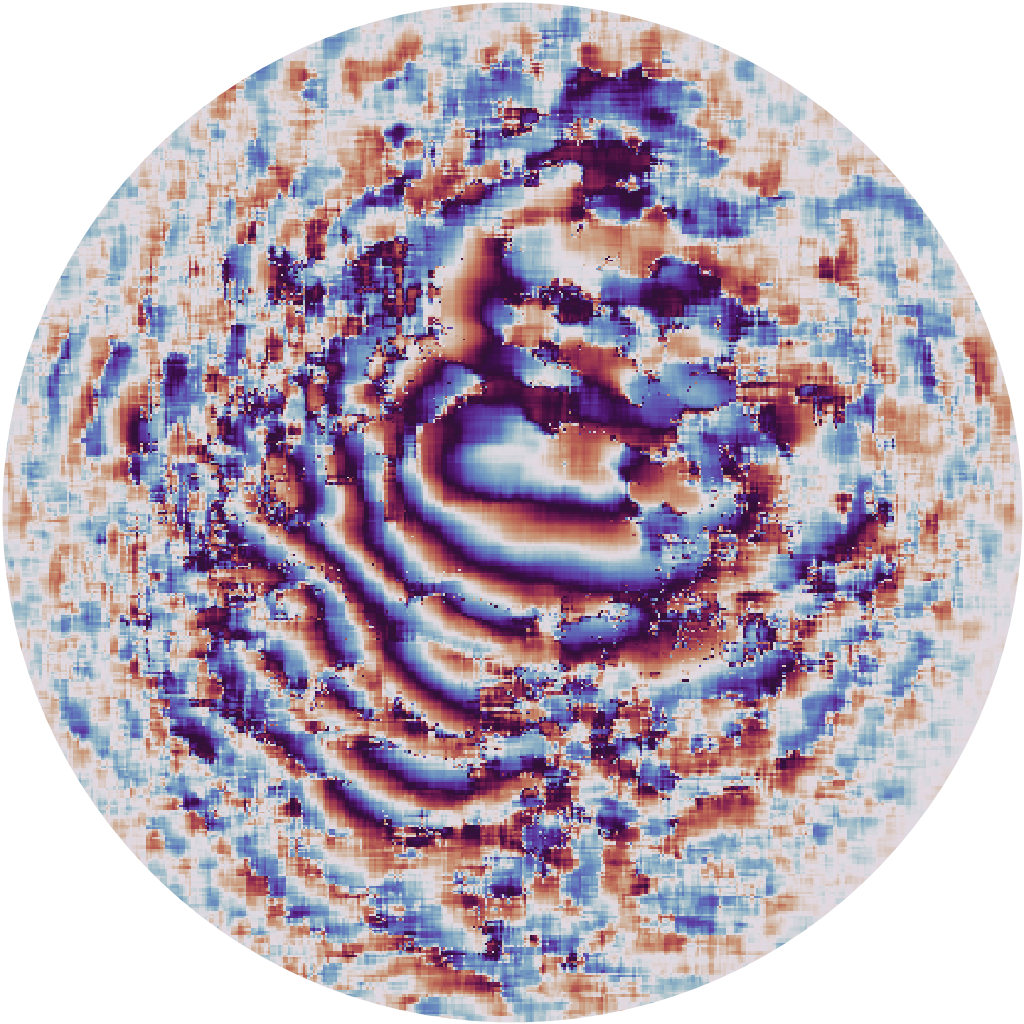}&			
			\includegraphics[width= 0.12\textwidth]{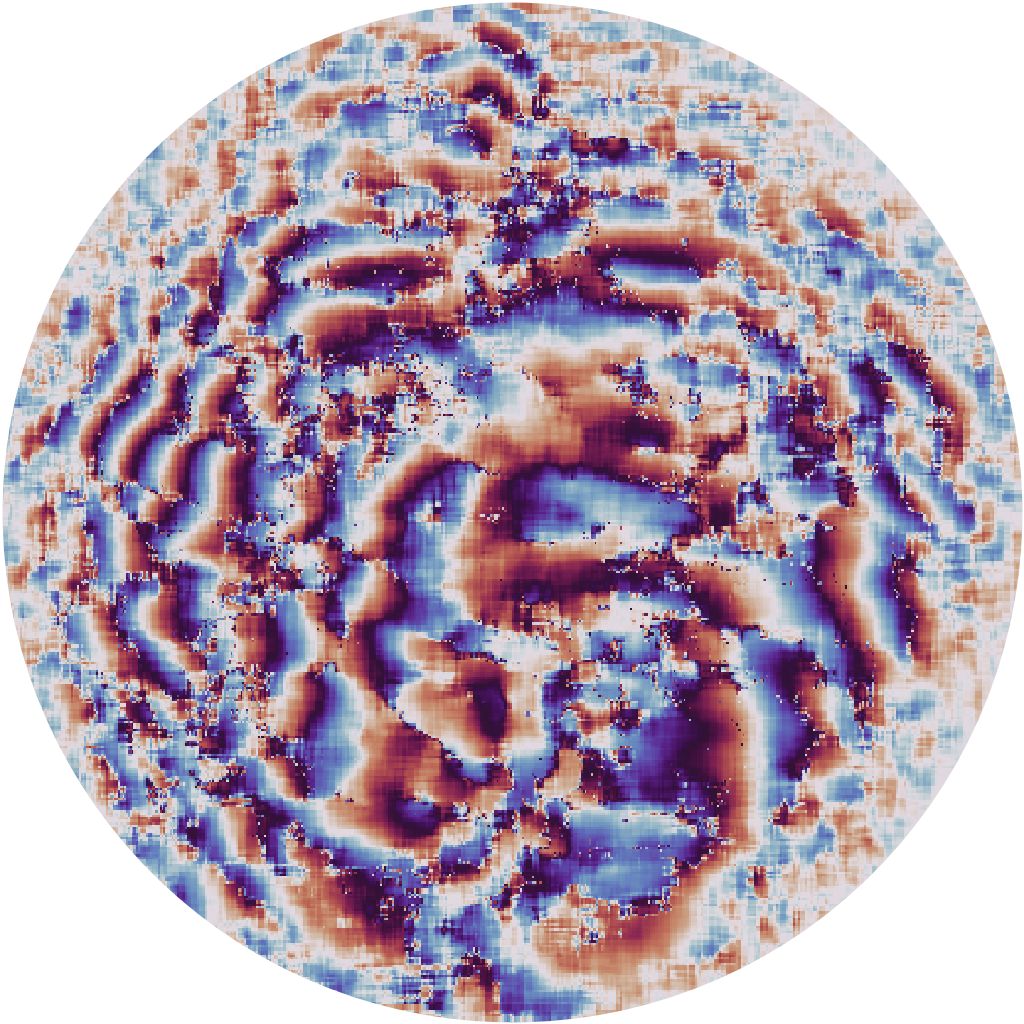}

		\end{tabular}
		\caption{\textbf{Iso-planatic patches  of resolution target:} \blue{We show the iso-planatic patches that compose the combined result of the resolution target presented in Fig. 5 in the main paper. The first column shows the result presented in the main paper. The second and third columns show the iso-planatic patches which were scanned by our algorithm. The fourth and fifth columns shows the phase mask used for each iso-planatic patch. The iso-planatic patches were combined using a template matching algorithm.}  }
		\label{fig:res-patches}
	\end{center}
\end{figure*}

%% file: fig_onion_full_view.tex
\begin{figure*}[h!]
	\begin{center}		
		\begin{tabular}{@{}c@{~}c@{~}}			
			\multicolumn{1}{c}{\hspace{0cm}\large Onion cells }& 
			\multicolumn{1}{c}{\hspace{0cm}\large Confocal scanned area}\\
			\includegraphics[width= 0.5\textwidth]{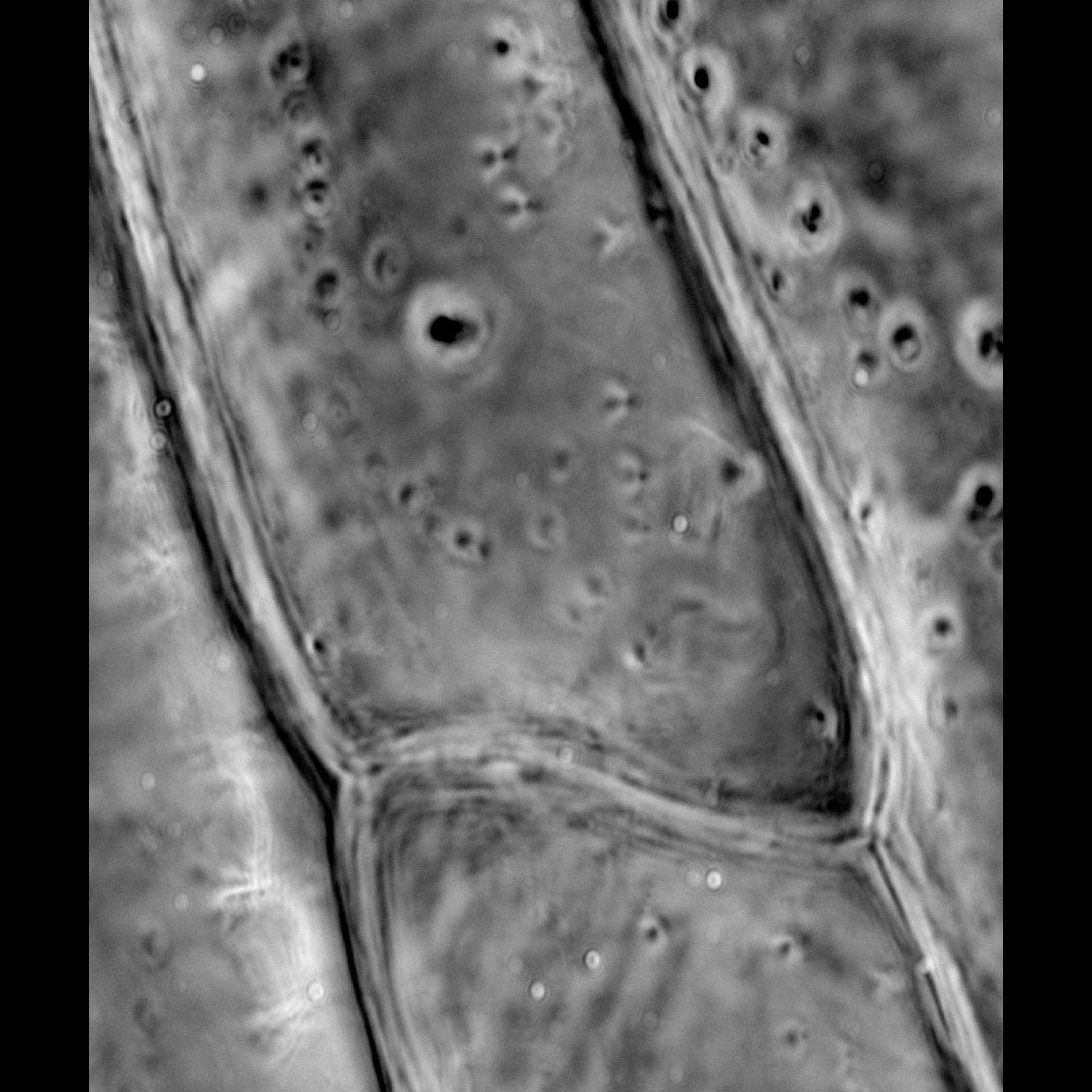}&
			\raisebox{1.5cm}{\includegraphics[width= 0.13\textwidth]{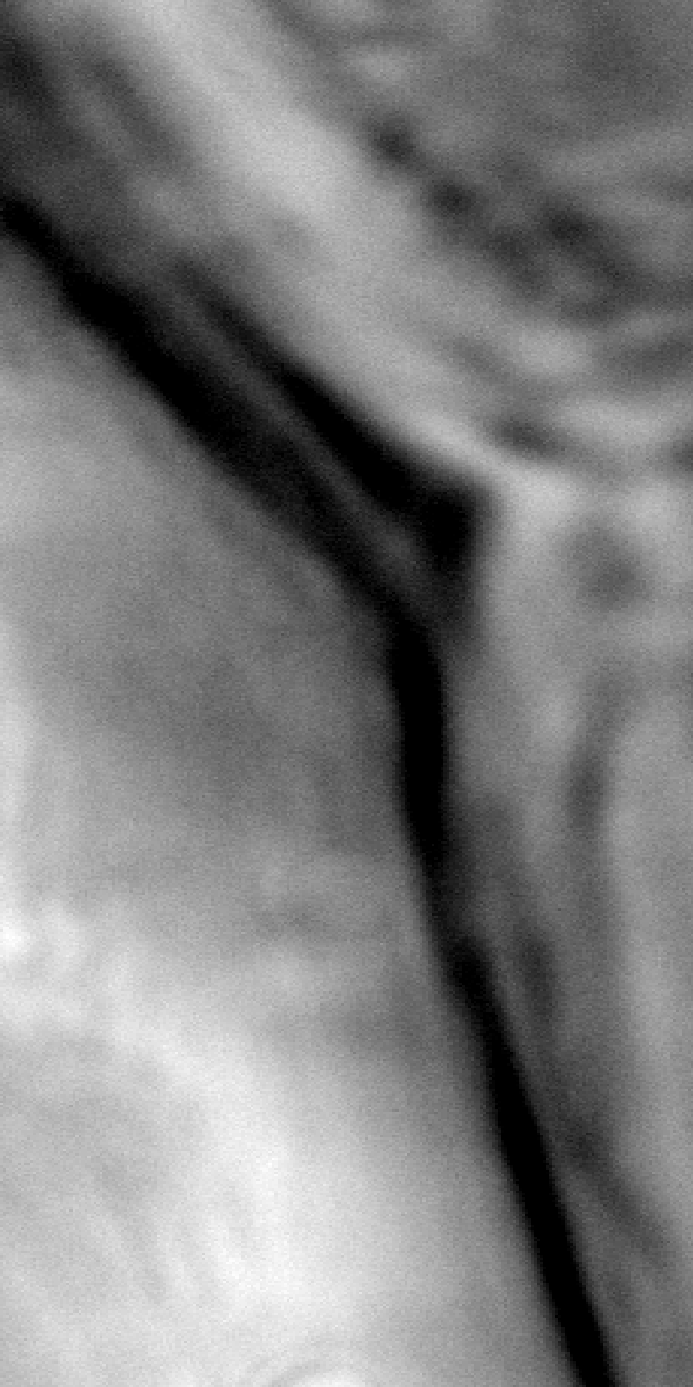}}\\

		\end{tabular}
		\caption{\textbf{Onion cells:} \blue{We show a larger view of the onion cells example from Fig. 7 of the main paper, captured by our validation camera. }}
		\label{fig:onion_full_view_sup}
	\end{center}
\end{figure*}

%% file: fig_onion_multi_separated.tex
\begin{figure*}[t!]
	\begin{center}		
		\begin{tabular}{@{}c@{}c@{~~~~~~~~~~~~~~~~~~}c@{~~~~~}}			
			\multicolumn{1}{c}{\hspace{-1.5cm}\large Combined confocal scan }& 
			\multicolumn{1}{c}{\hspace{-2cm}\large Confocal scan }& 
			\multicolumn{1}{c}{\hspace{-0.9cm}\large Phase mask }\\
			\raisebox{1.2cm}{\multirow{4}{14em}{\includegraphics[width= 0.355\textwidth]{figs/onion/I_fix.png}}}&
			\includegraphics[width= 0.12\textwidth]{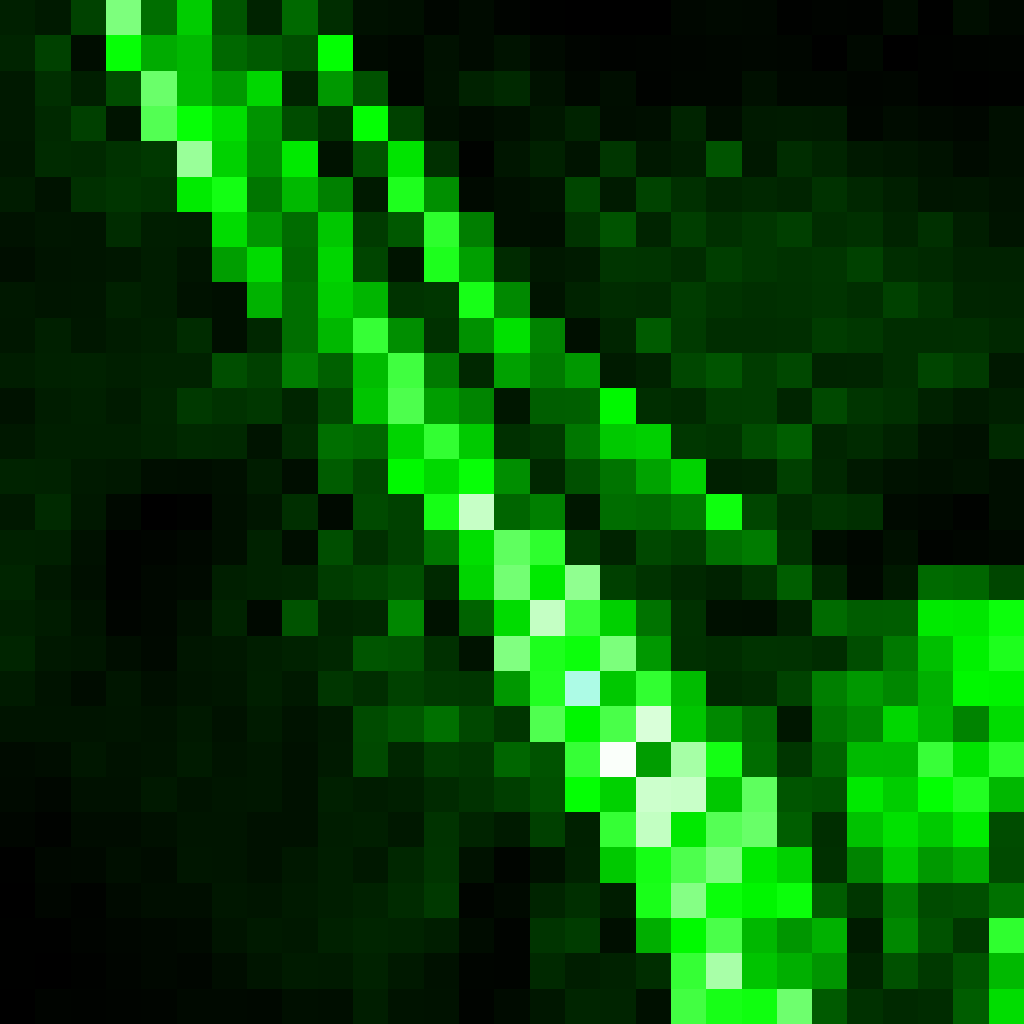}&
			\includegraphics[width= 0.12\textwidth]{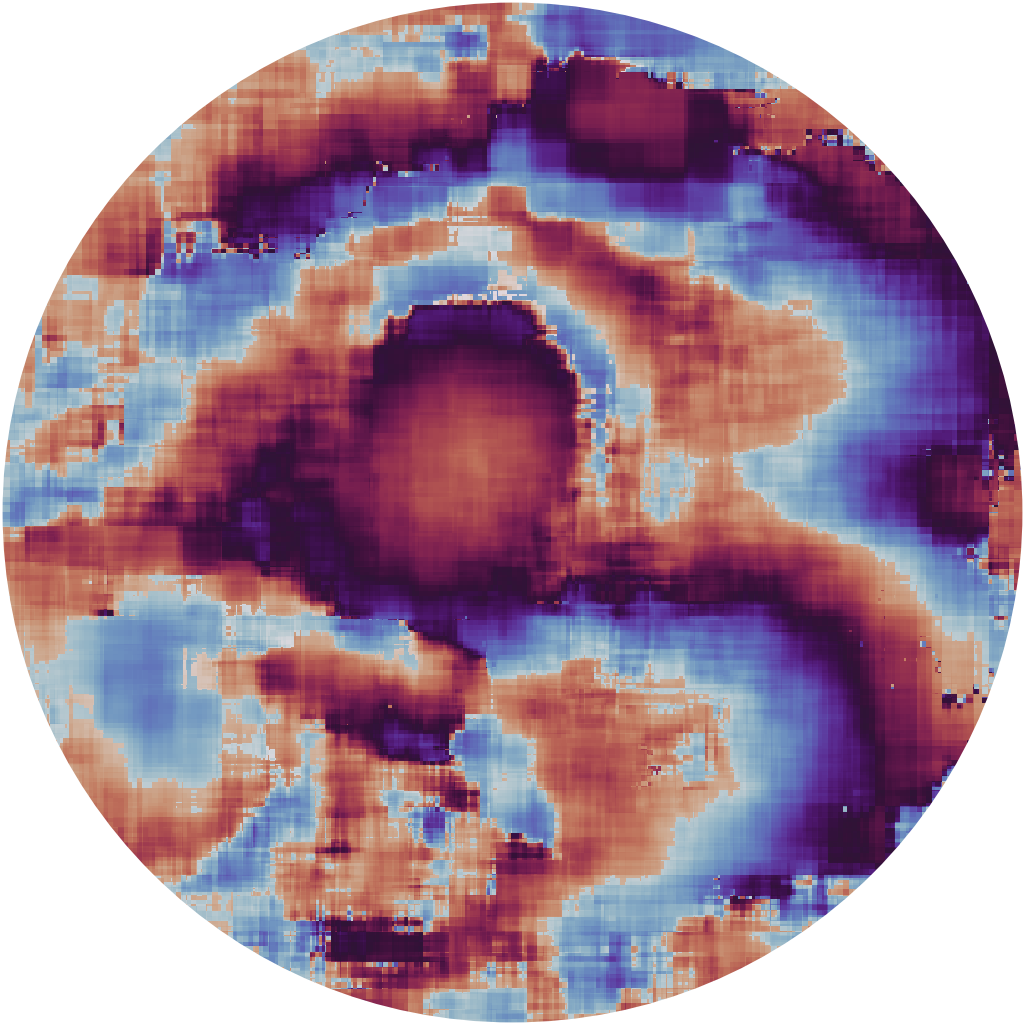}\\&
			\includegraphics[width= 0.12\textwidth]{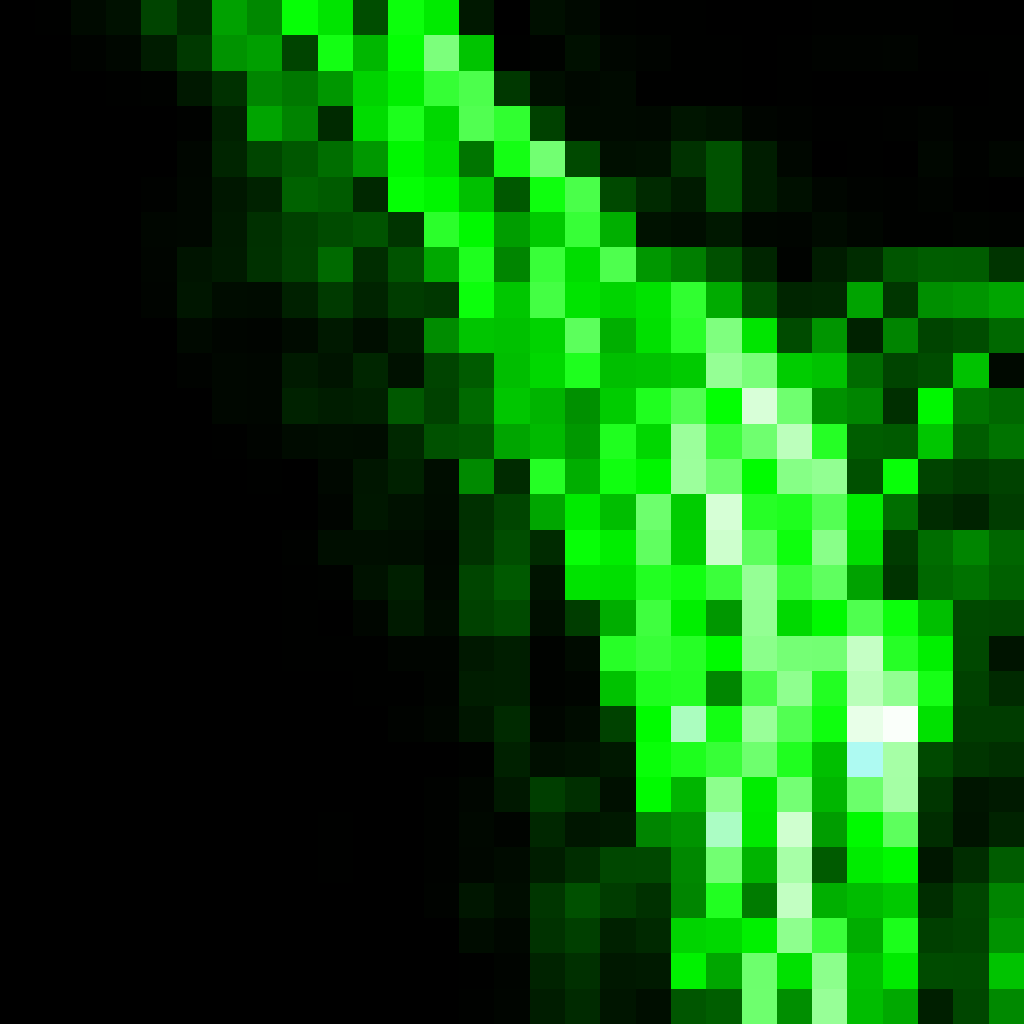}&
			\includegraphics[width= 0.12\textwidth]{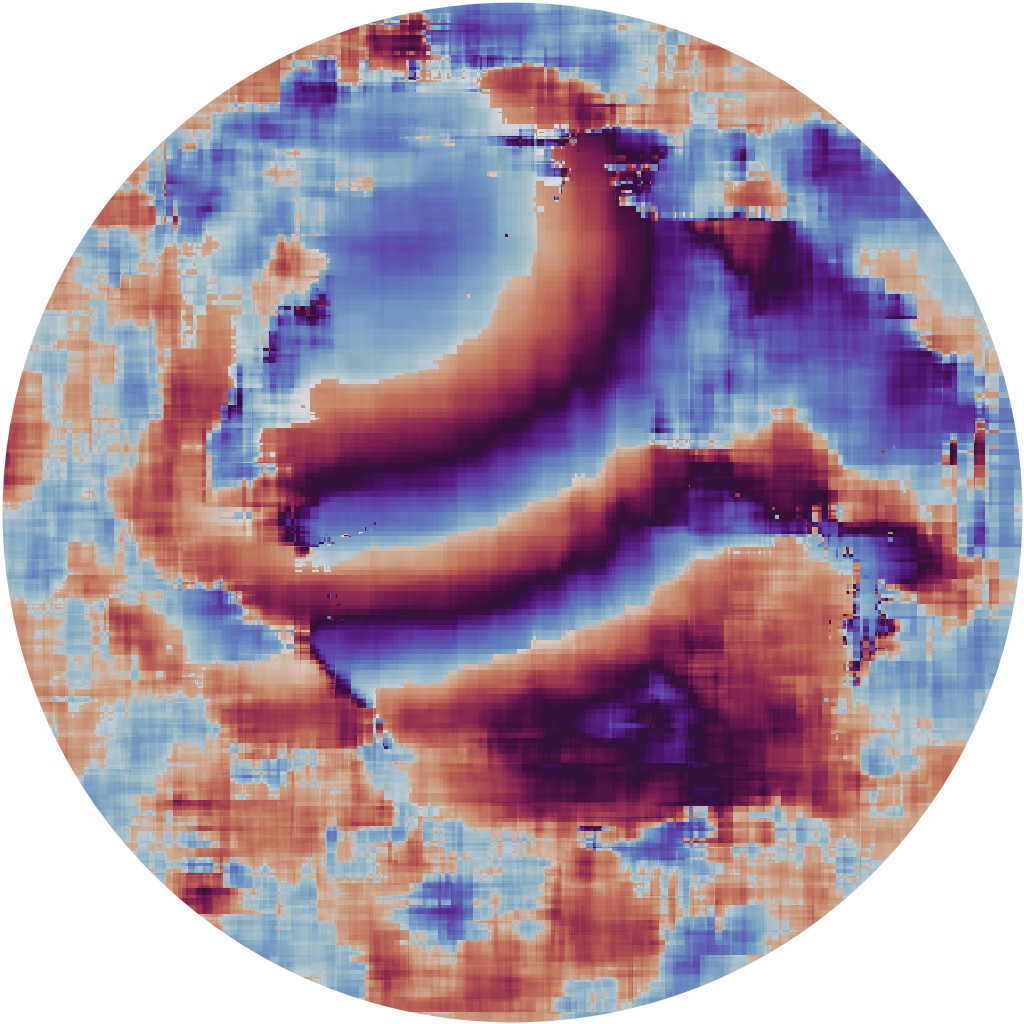}\\&
			\includegraphics[width= 0.12\textwidth]{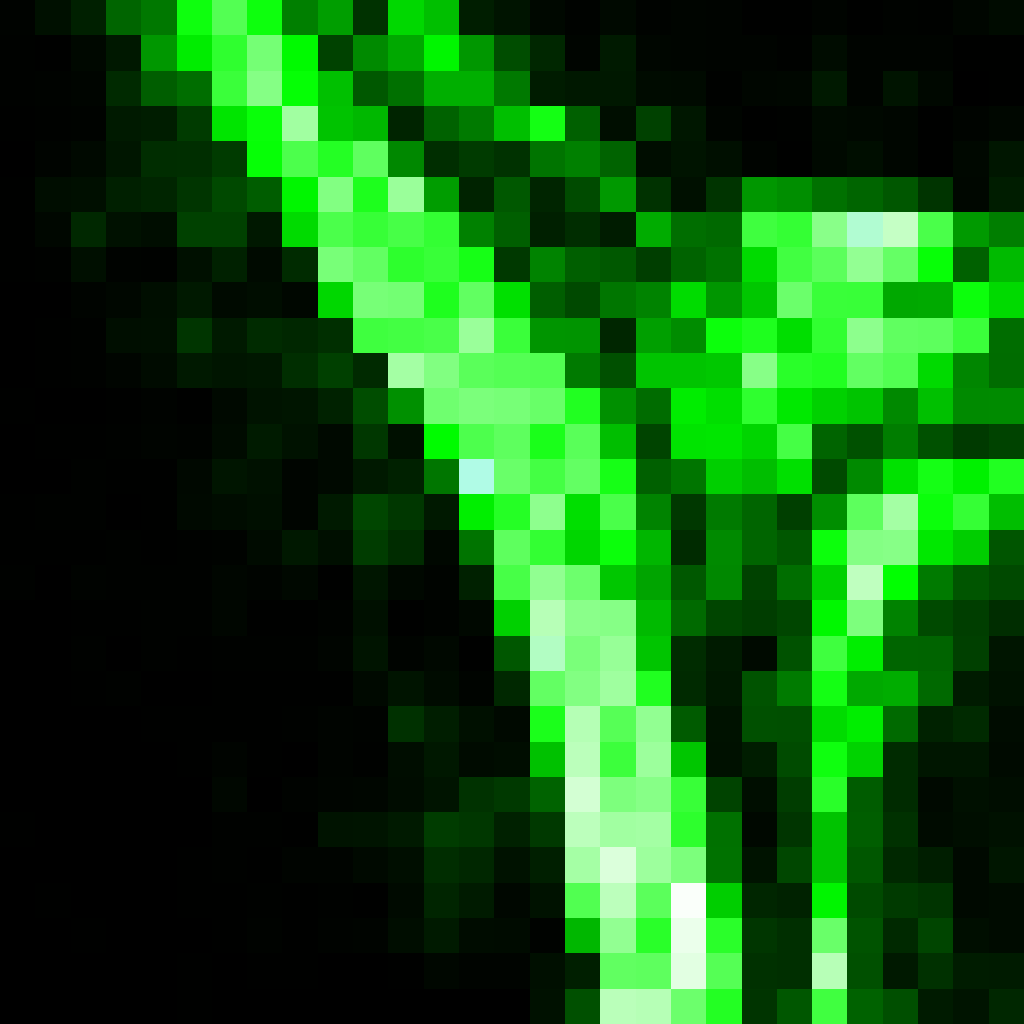}&
			\includegraphics[width= 0.12\textwidth]{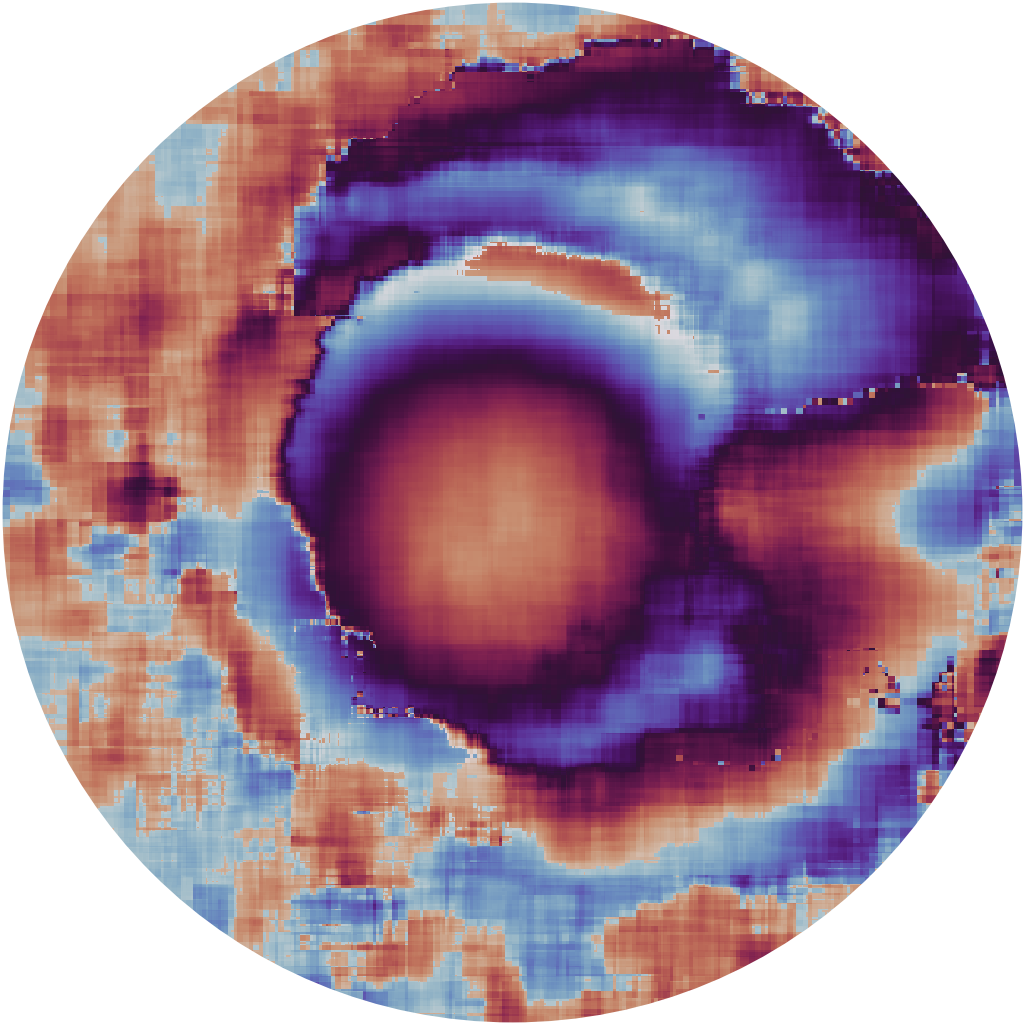}\\&
			\includegraphics[width= 0.12\textwidth]{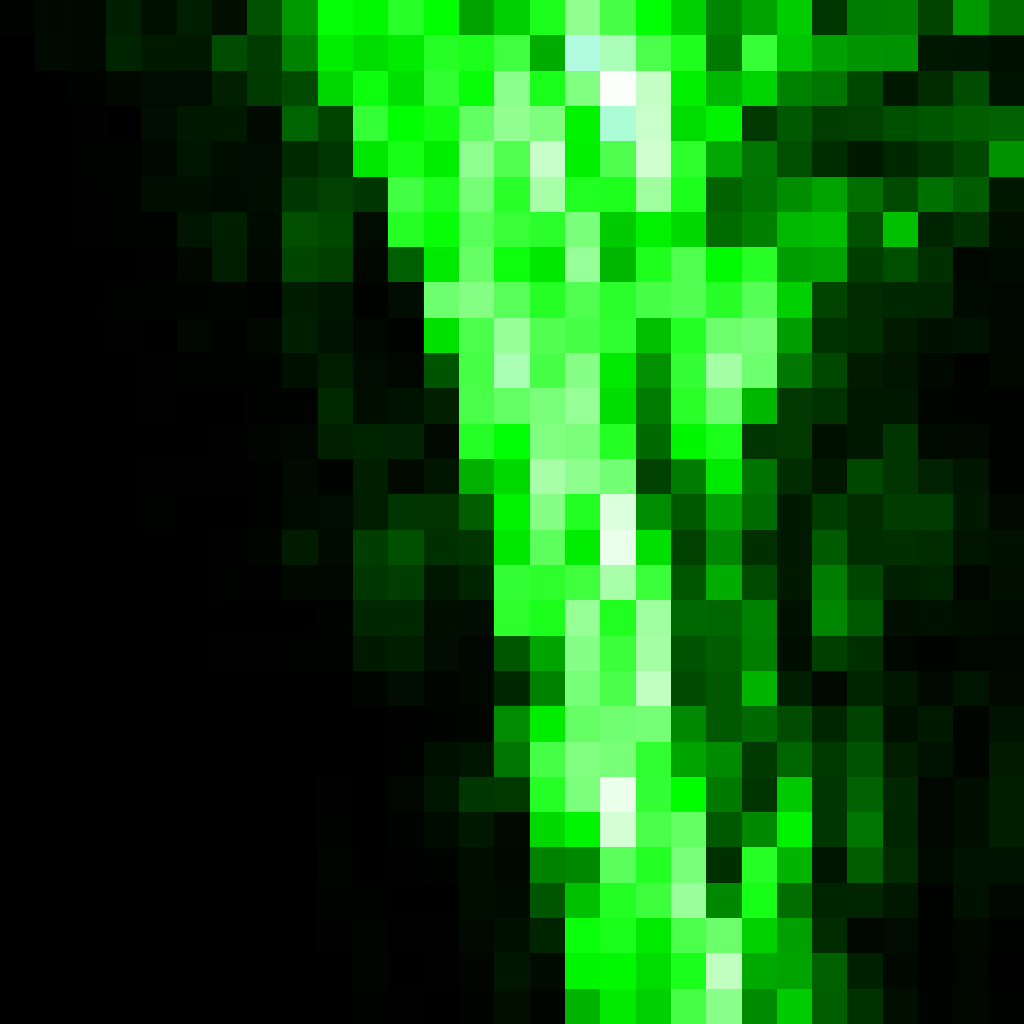}&
			\includegraphics[width= 0.12\textwidth]{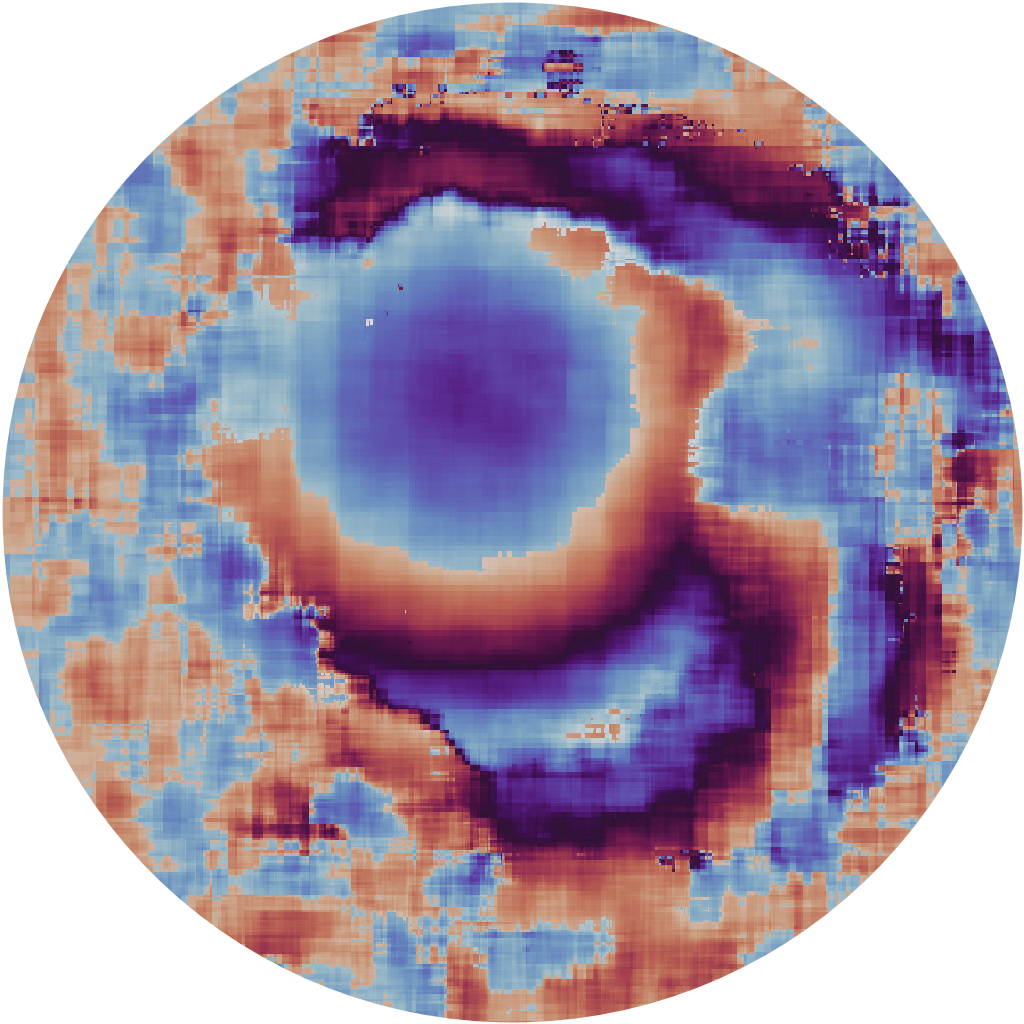}\\&

		\end{tabular}
		\caption{\textbf{Iso-planatic patches in onion result:} \blue{We show the iso-planatic patches that compose the combined result of the onion cells boundary presented in Fig. 7 in the main paper. The first column shows the result presented in the main paper. The second column shows the iso-planatic patches which were scanned by our algorithm. The third column shows the phase mask used for each iso-planatic patch. The iso-planatic patches were combined using a template matching algorithm.}  }
		\label{fig:onion-multi}
	\end{center}
\end{figure*}

%% file: fig_onion_ref_xy_slices.tex
\begin{figure*}[b!]
	\begin{center}		
		\begin{tabular}{@{}c@{}c@{}}			
			\multicolumn{2}{c}{\hspace{0cm}\large Confocal scan }\\
			\multicolumn{2}{c}{\hspace{0cm}\large No abberation }\\
			{\raisebox{0.7cm}	{\rotatebox[origin=c]{90}{~ {\scriptsize $z=-8\um$} }}}&
			\includegraphics[width= 0.12\textwidth]{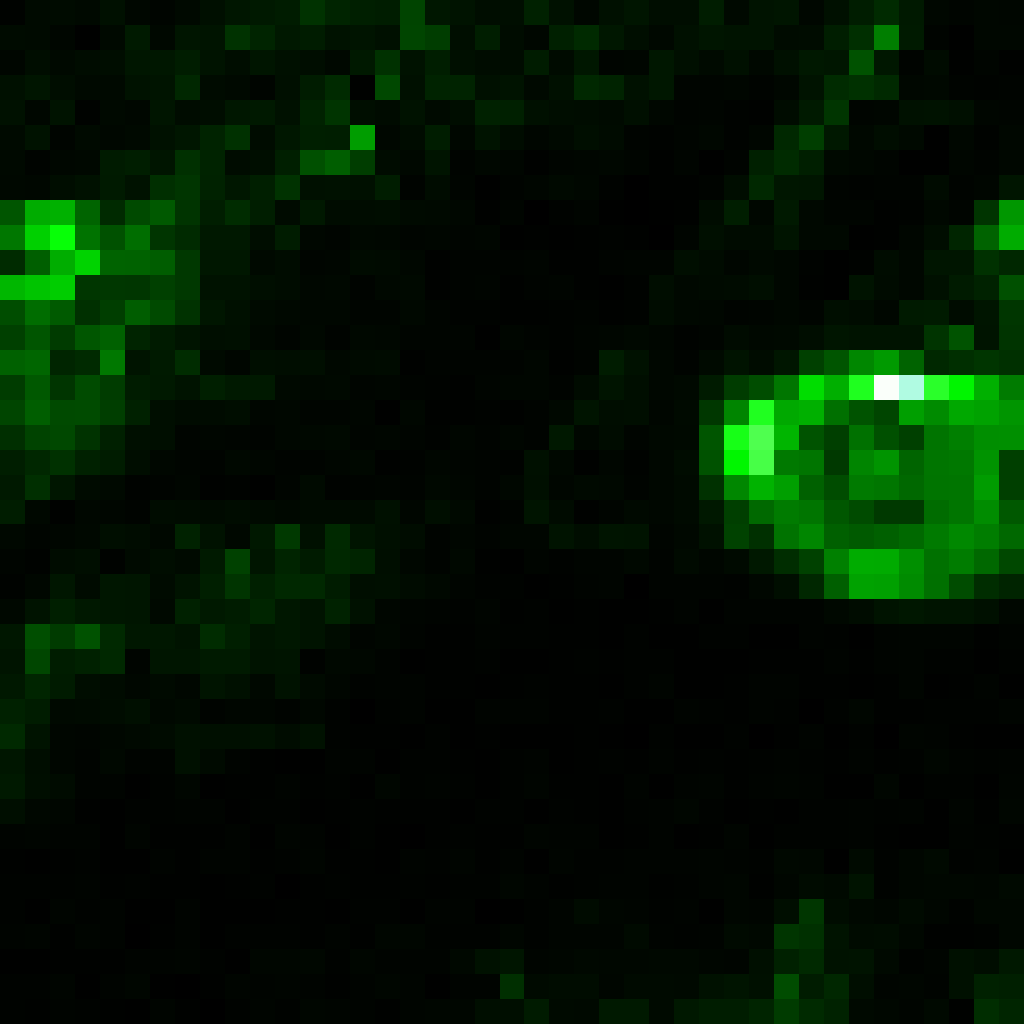}\\
			{\raisebox{0.7cm}	{\rotatebox[origin=c]{90}{~ {\scriptsize $z=-6\um$} }}}&
			\includegraphics[width= 0.12\textwidth]{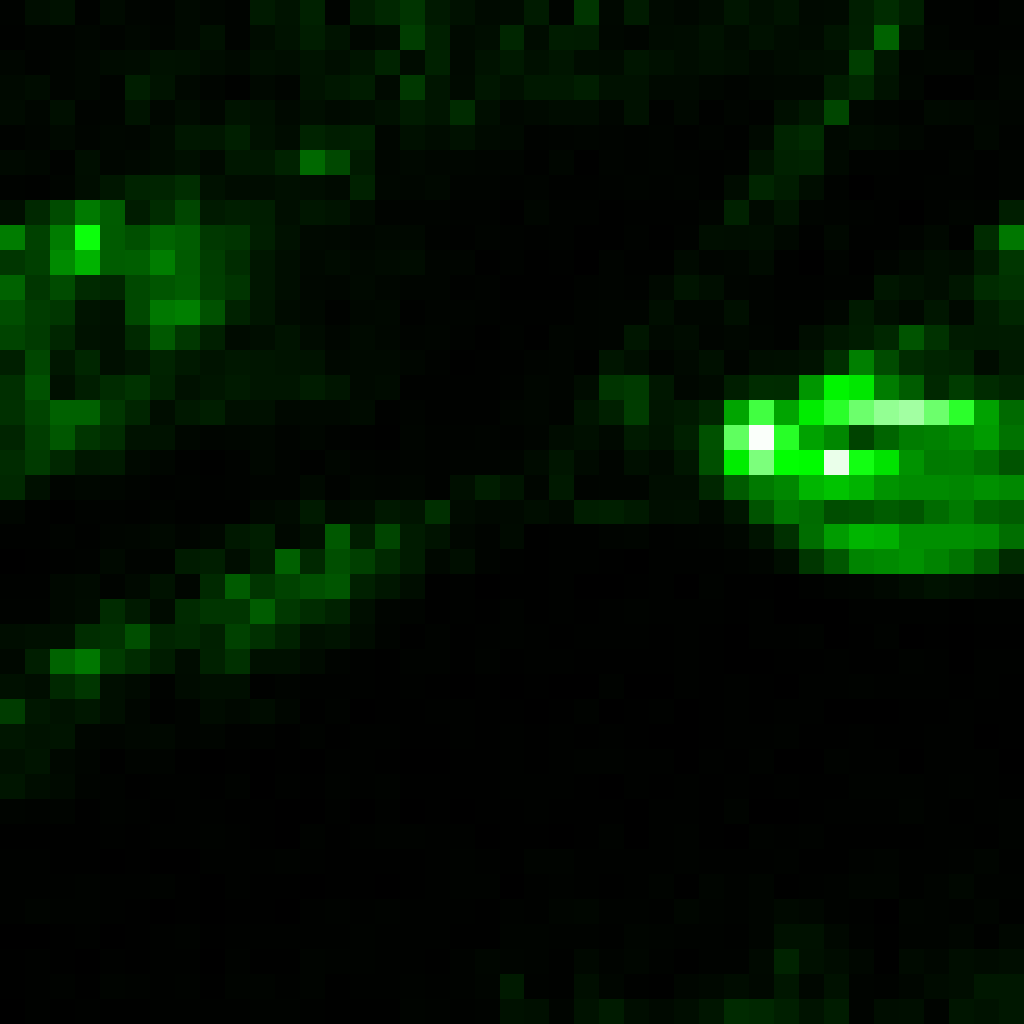}\\
			{\raisebox{0.7cm}	{\rotatebox[origin=c]{90}{~ {\scriptsize $z=-4\um$} }}}&
			\includegraphics[width= 0.12\textwidth]{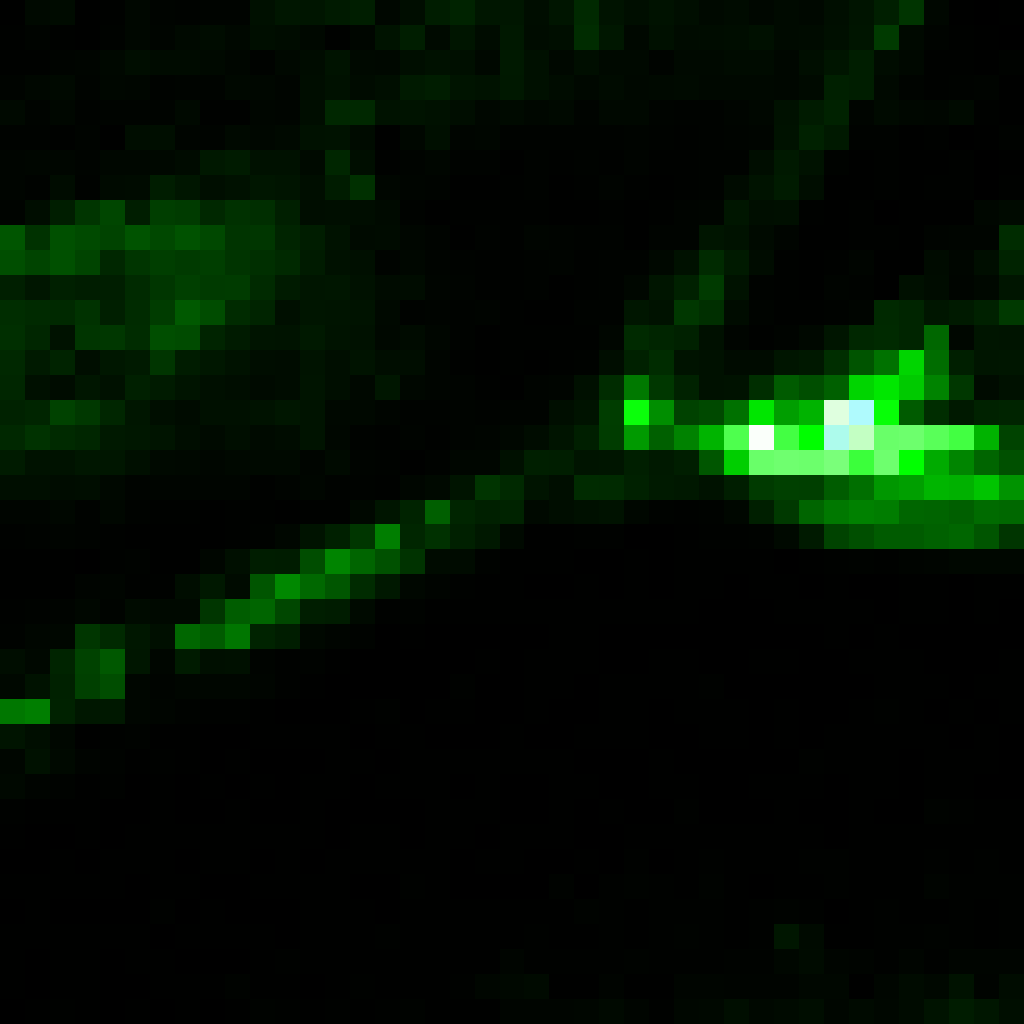}\\
			{\raisebox{0.7cm}	{\rotatebox[origin=c]{90}{~ {\scriptsize $z=-2\um$} }}}&
			\includegraphics[width= 0.12\textwidth]{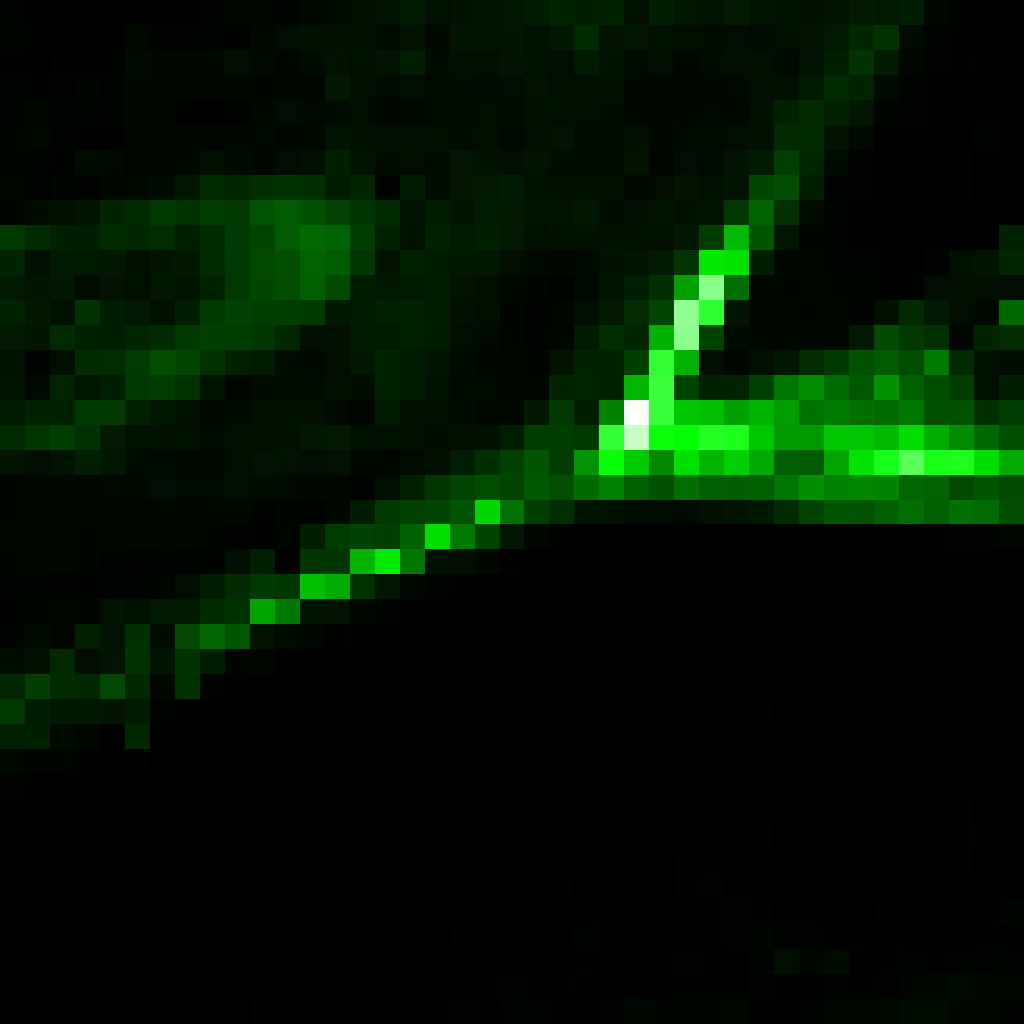}\\
			{\raisebox{0.7cm}	{\rotatebox[origin=c]{90}{~ {\scriptsize $z=0\um$} }}}&
			\includegraphics[width= 0.12\textwidth]{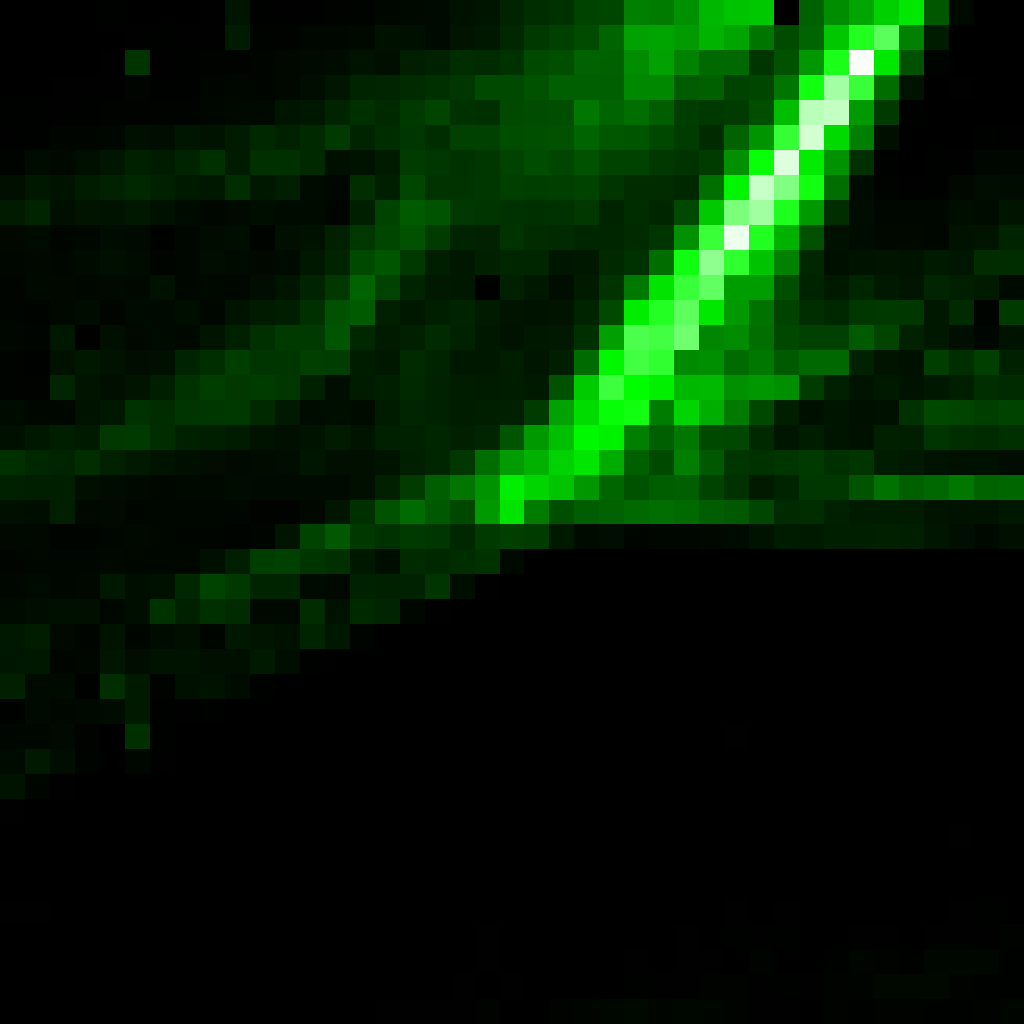}\\
			{\raisebox{0.7cm}	{\rotatebox[origin=c]{90}{~ {\scriptsize $z=2\um$} }}}&
			\includegraphics[width= 0.12\textwidth]{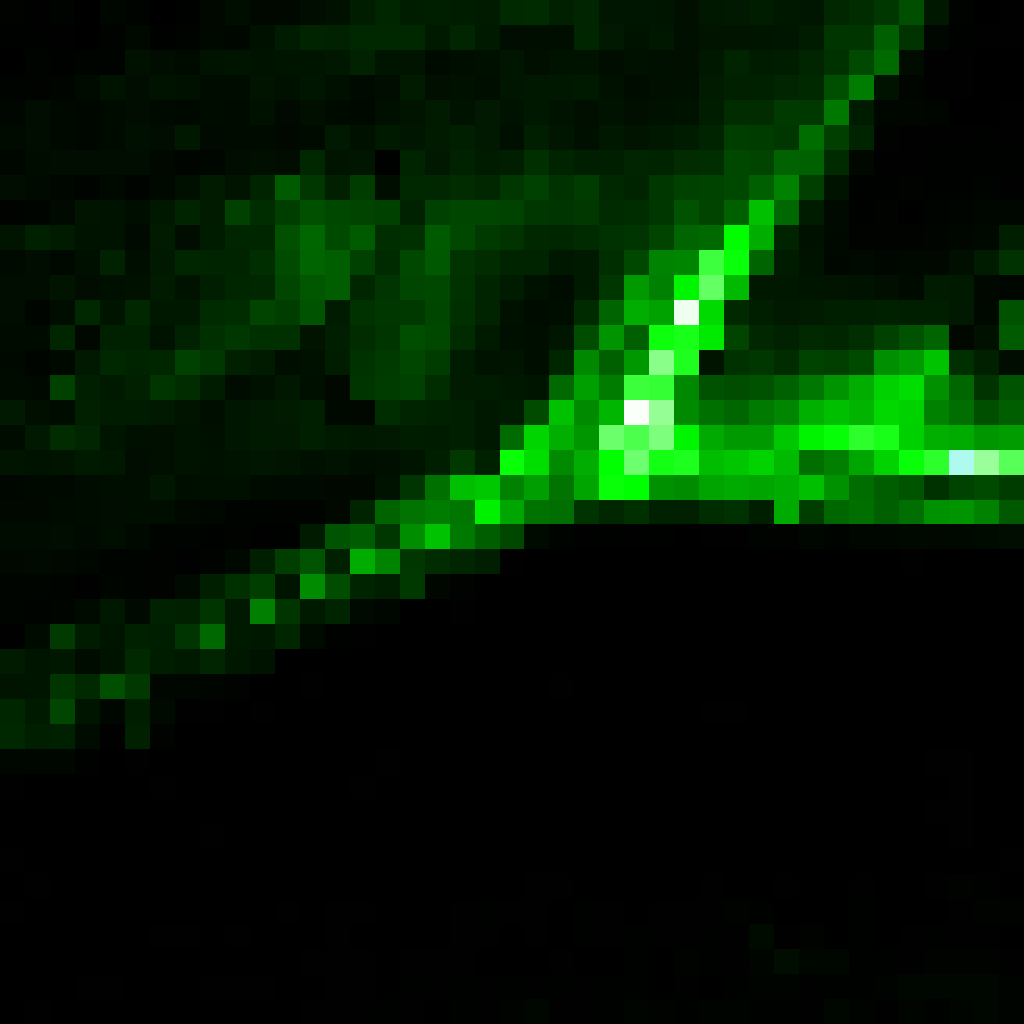}\\
			{\raisebox{0.7cm}	{\rotatebox[origin=c]{90}{~ {\scriptsize $z=4\um$} }}}&
			\includegraphics[width= 0.12\textwidth]{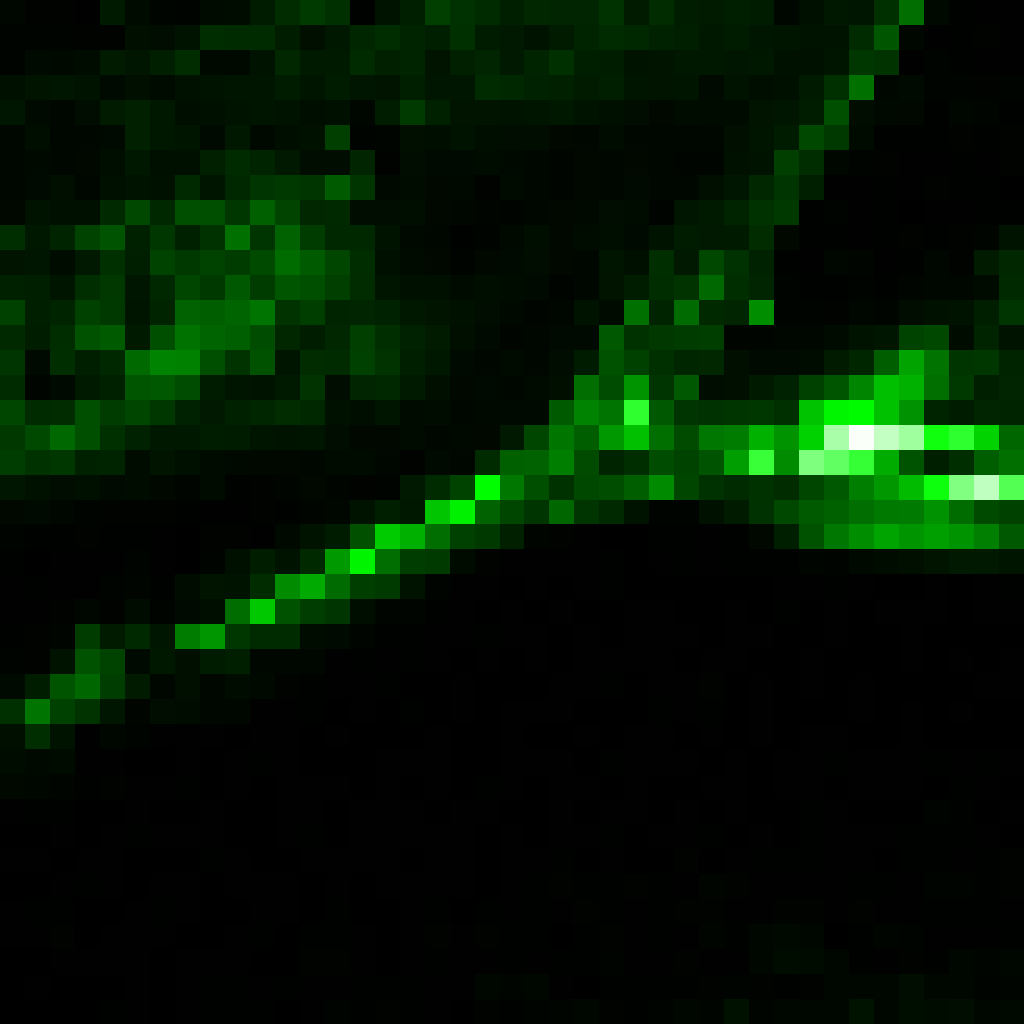}\\
			{\raisebox{0.7cm}	{\rotatebox[origin=c]{90}{~ {\scriptsize $z=6\um$} }}}&
			\includegraphics[width= 0.12\textwidth]{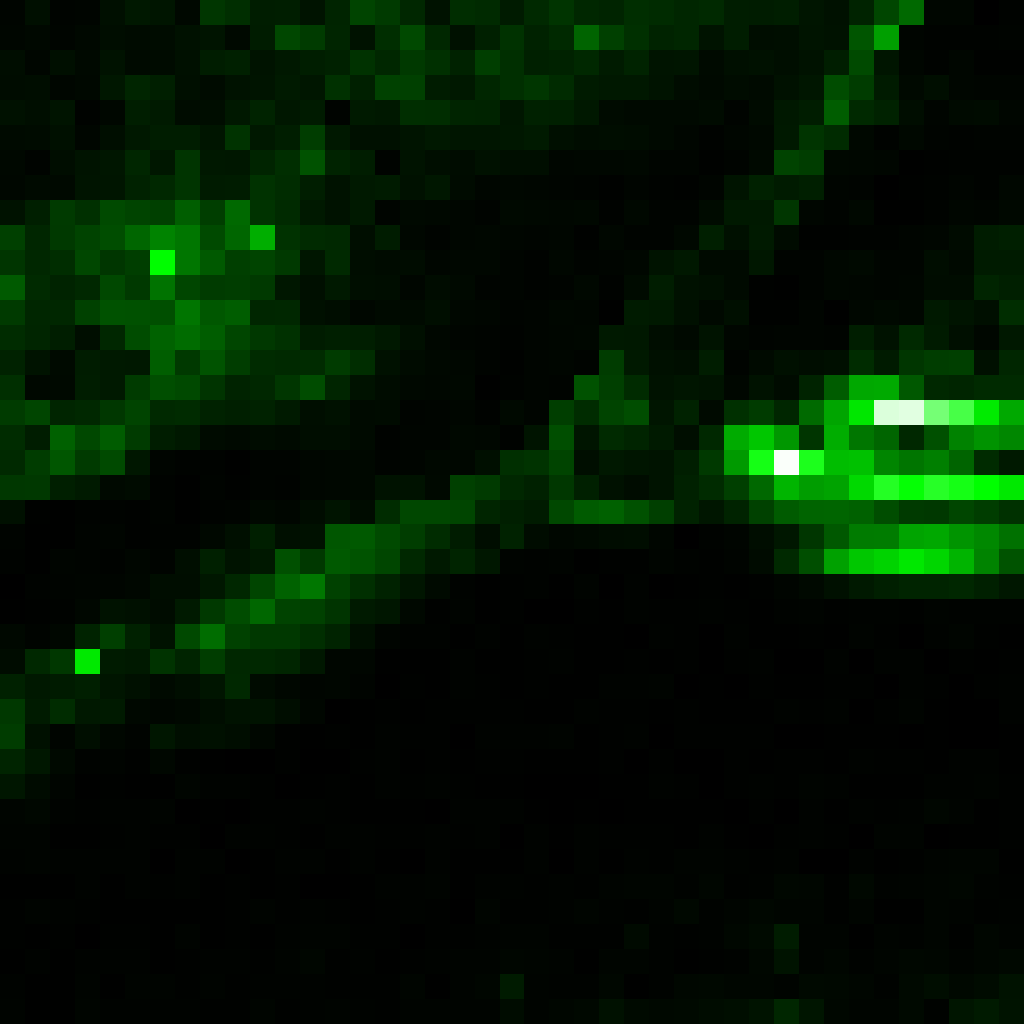}\\
			{\raisebox{0.7cm}	{\rotatebox[origin=c]{90}{~ {\scriptsize $z=8\um$} }}}&
			\includegraphics[width= 0.12\textwidth]{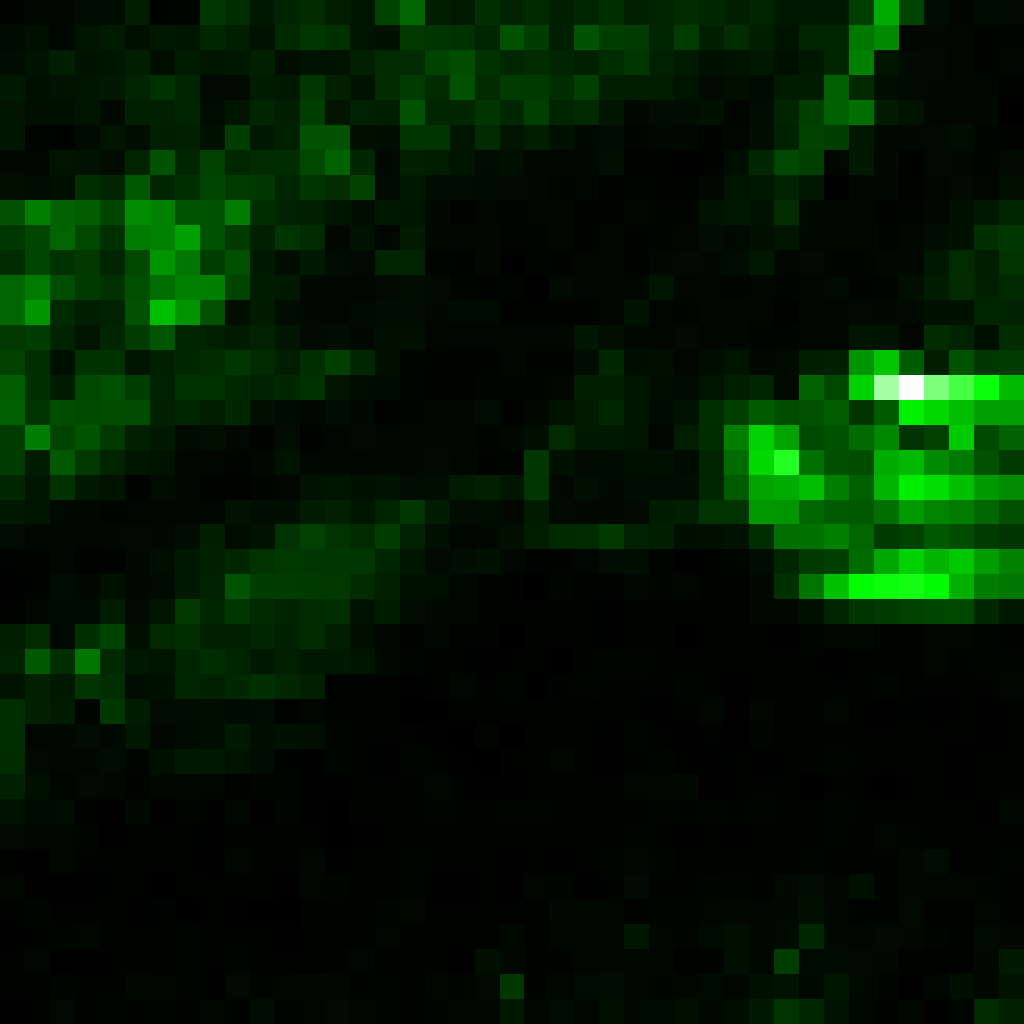}\\

		\end{tabular}
		\caption{\textbf{X-Y cross sections of reference target:} \bblue{To visualize the typical 3D structure of an onion, we show x-y cross sections at multiple depths. This example uses a shallow onion layer without aberration.}  }
		\label{fig:ref-xy}
	\end{center}
\end{figure*}

%% file: fig_onion_result_xy_slices.tex
\begin{figure*}[b!]
	\begin{center}		
		\begin{tabular}{@{}c@{~~~~}c@{~~~~}c@{~~~~}c@{~~~~}c@{~~~~}c@{~~~~}c@{~~~~}}			
			&
			\multicolumn{2}{c}{\hspace{-0.6cm}\scriptsize Layer 1 }&
			\multicolumn{2}{c}{\hspace{-0.6cm}\scriptsize Layer 2 }&
			\multicolumn{2}{c}{\hspace{-0.6cm}\scriptsize Layer 3 }\\
			
			&
			\multicolumn{1}{c}{\hspace{-0.6cm} \scriptsize w/o}&	
			\multicolumn{1}{c}{\hspace{-0.6cm} \scriptsize w/ }&
			\multicolumn{1}{c}{\hspace{-0.6cm} \scriptsize w/o}&	
			\multicolumn{1}{c}{\hspace{-0.6cm} \scriptsize w/ }&	
			\multicolumn{1}{c}{\hspace{-0.6cm} \scriptsize w/o}&	
			\multicolumn{1}{c}{\hspace{-0.6cm} \scriptsize w/ }\\
			&
			\multicolumn{1}{c}{\hspace{-0.6cm} \scriptsize modulation}&	
			\multicolumn{1}{c}{\hspace{-0.6cm} \scriptsize modulation}&	
			\multicolumn{1}{c}{\hspace{-0.6cm} \scriptsize modulation}&	
			\multicolumn{1}{c}{\hspace{-0.6cm} \scriptsize modulation}&	
			\multicolumn{1}{c}{\hspace{-0.6cm} \scriptsize modulation}&	
			\multicolumn{1}{c}{\hspace{-0.6cm} \scriptsize modulation}\\
			
			{\raisebox{0.7cm}	{\rotatebox[origin=c]{90}{~ {\scriptsize $z=-8\um$} }}}&
			\includegraphics[width= 0.12\textwidth]{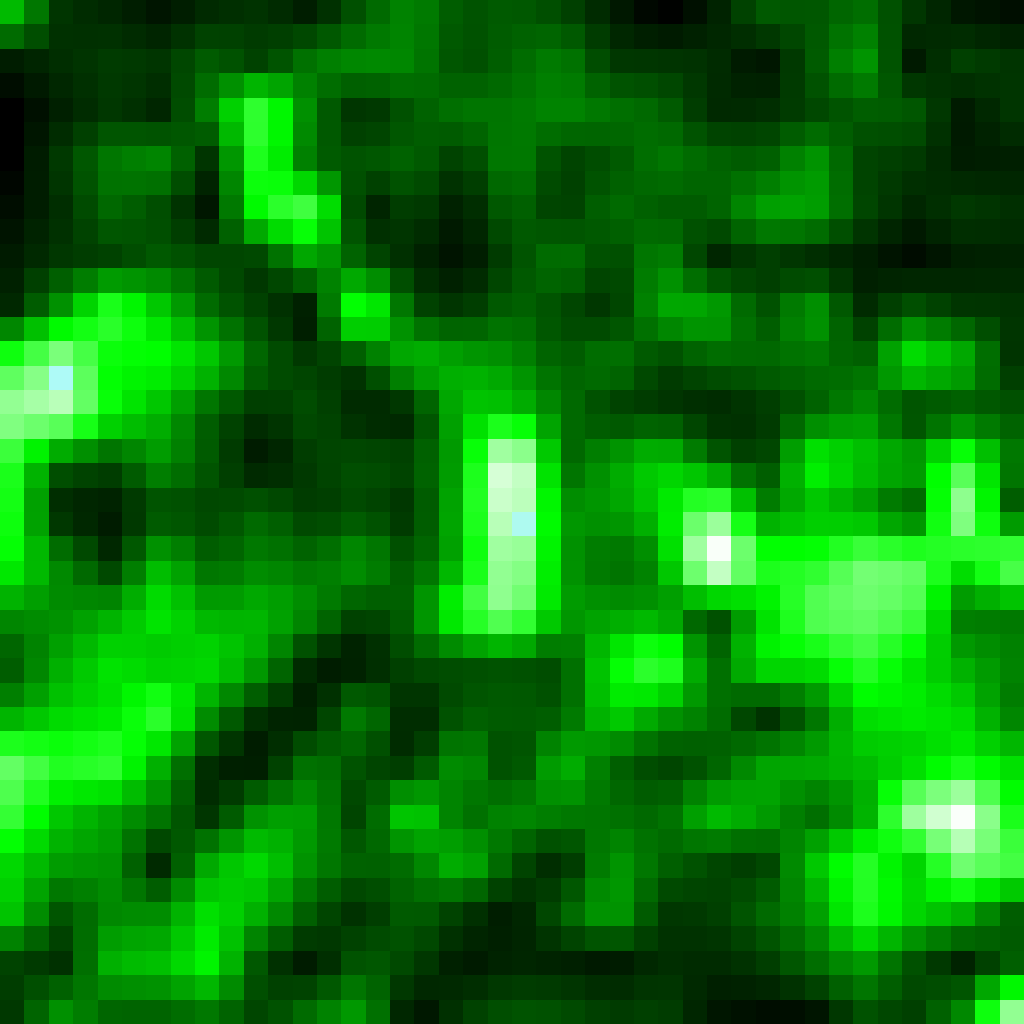}&
			\includegraphics[width= 0.12\textwidth]{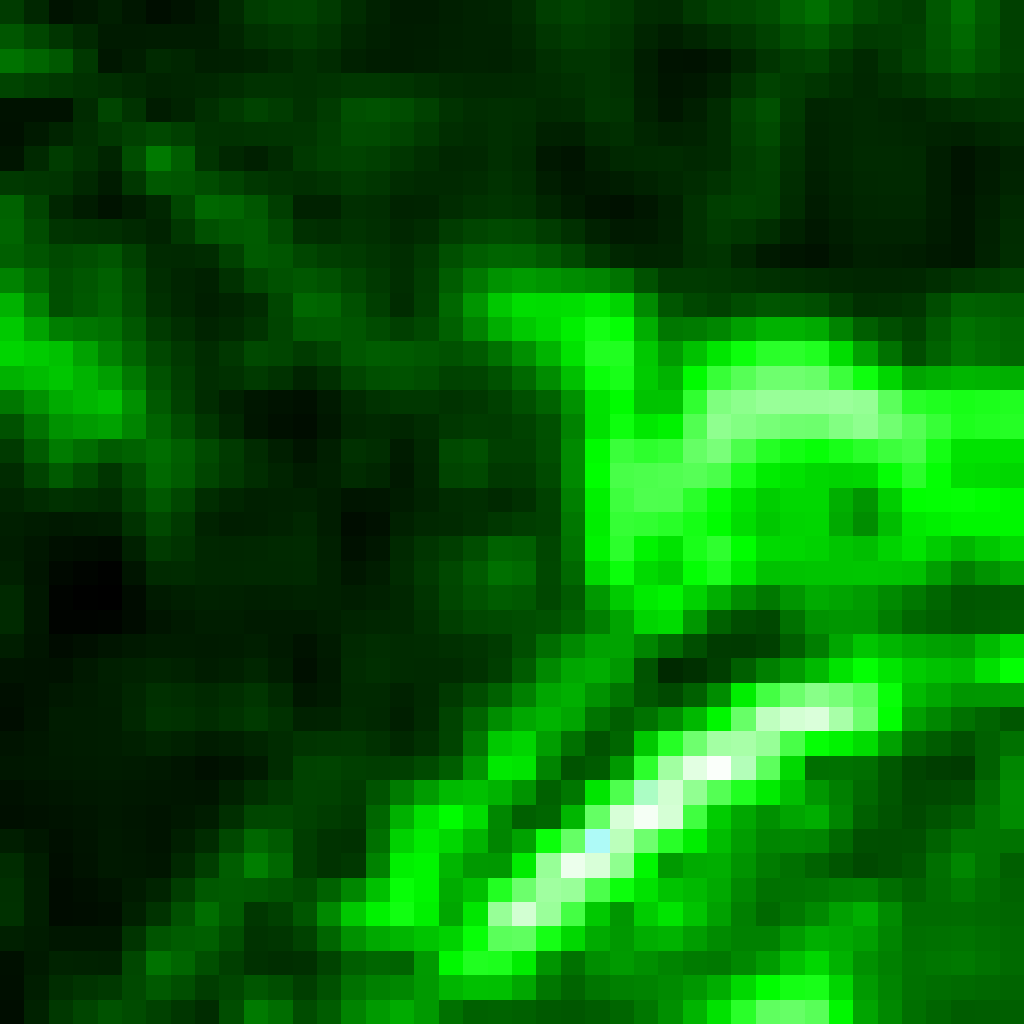}&
			\includegraphics[width= 0.12\textwidth]{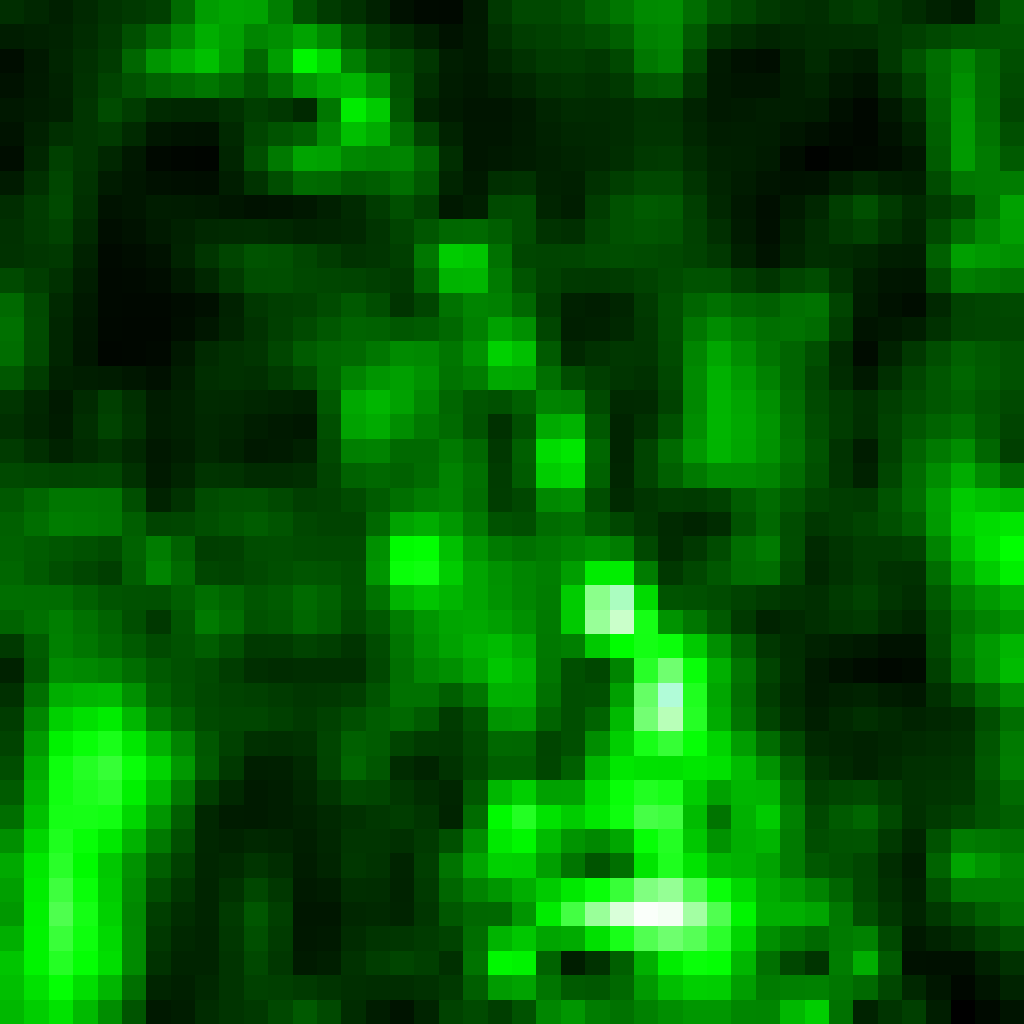}&
			\includegraphics[width= 0.12\textwidth]{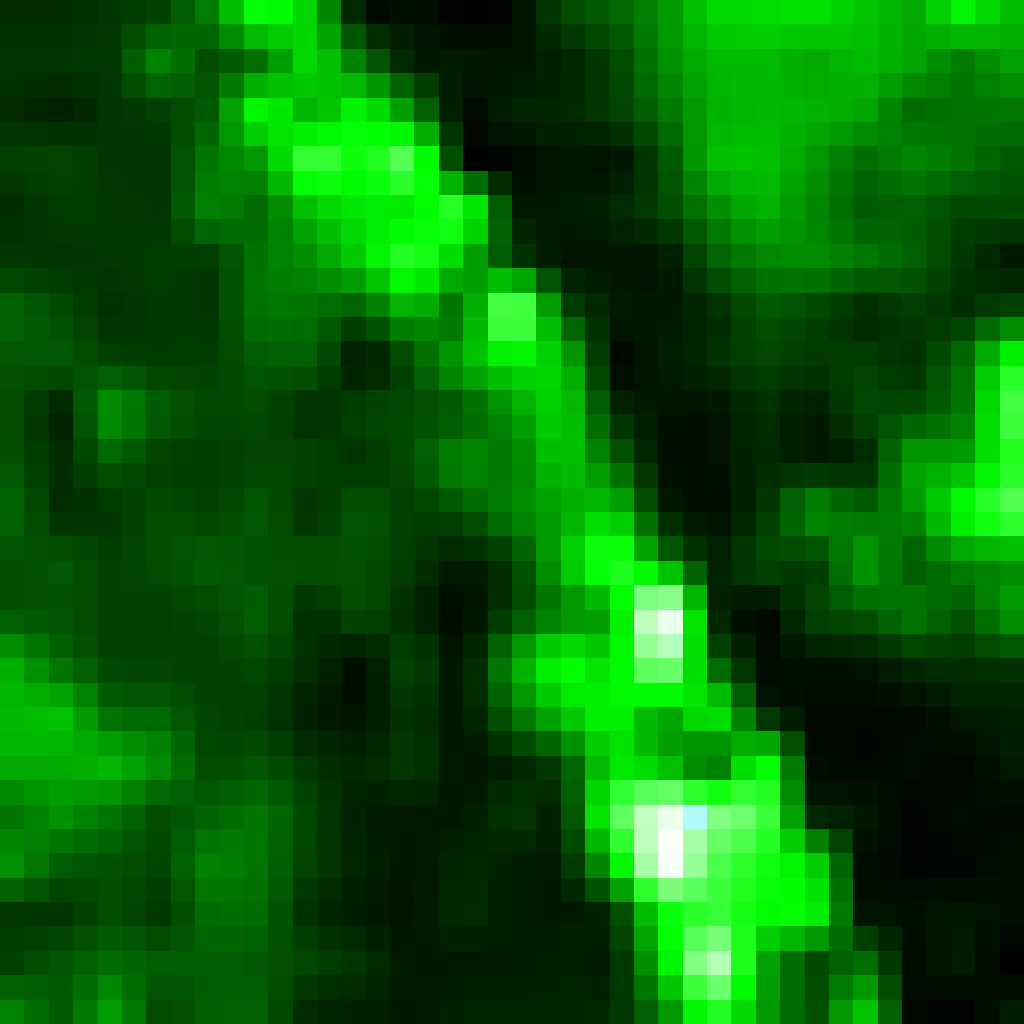}&
			\includegraphics[width= 0.12\textwidth]{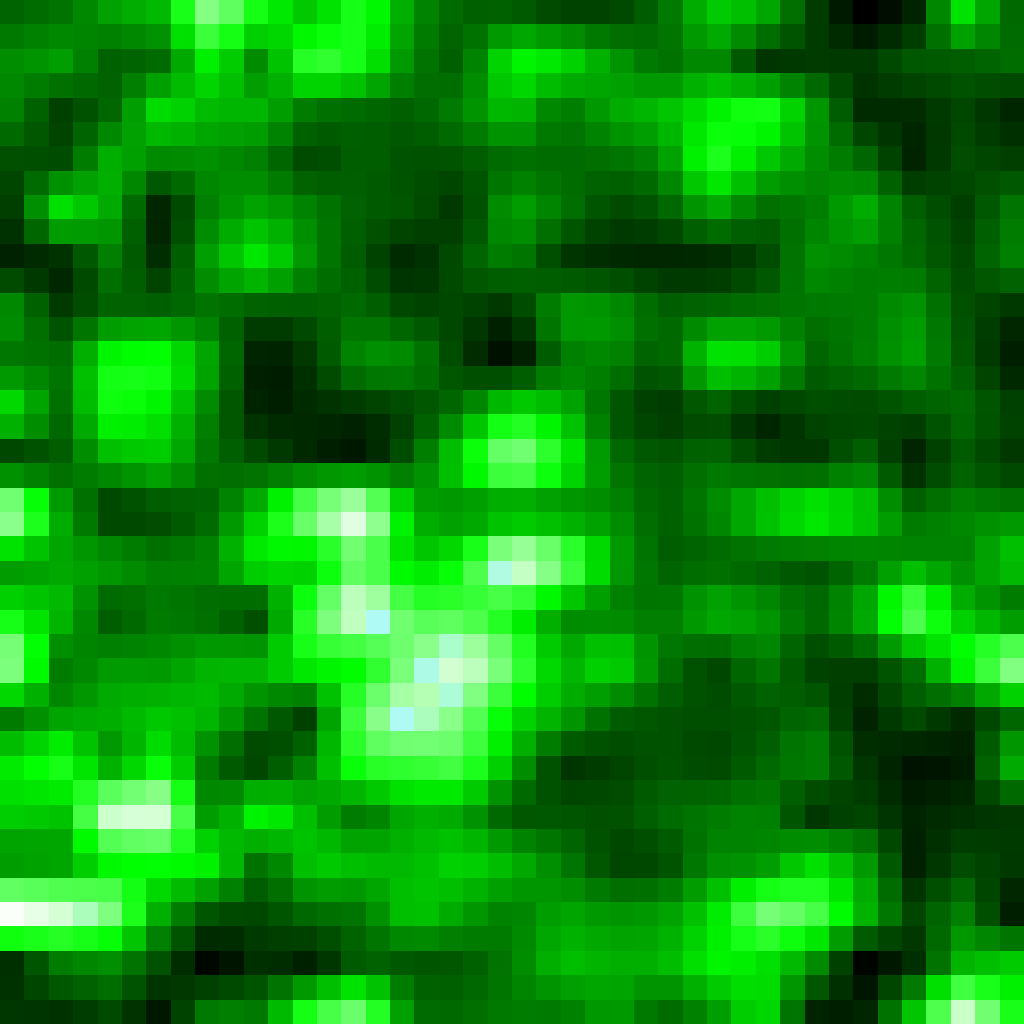}&
			\includegraphics[width= 0.12\textwidth]{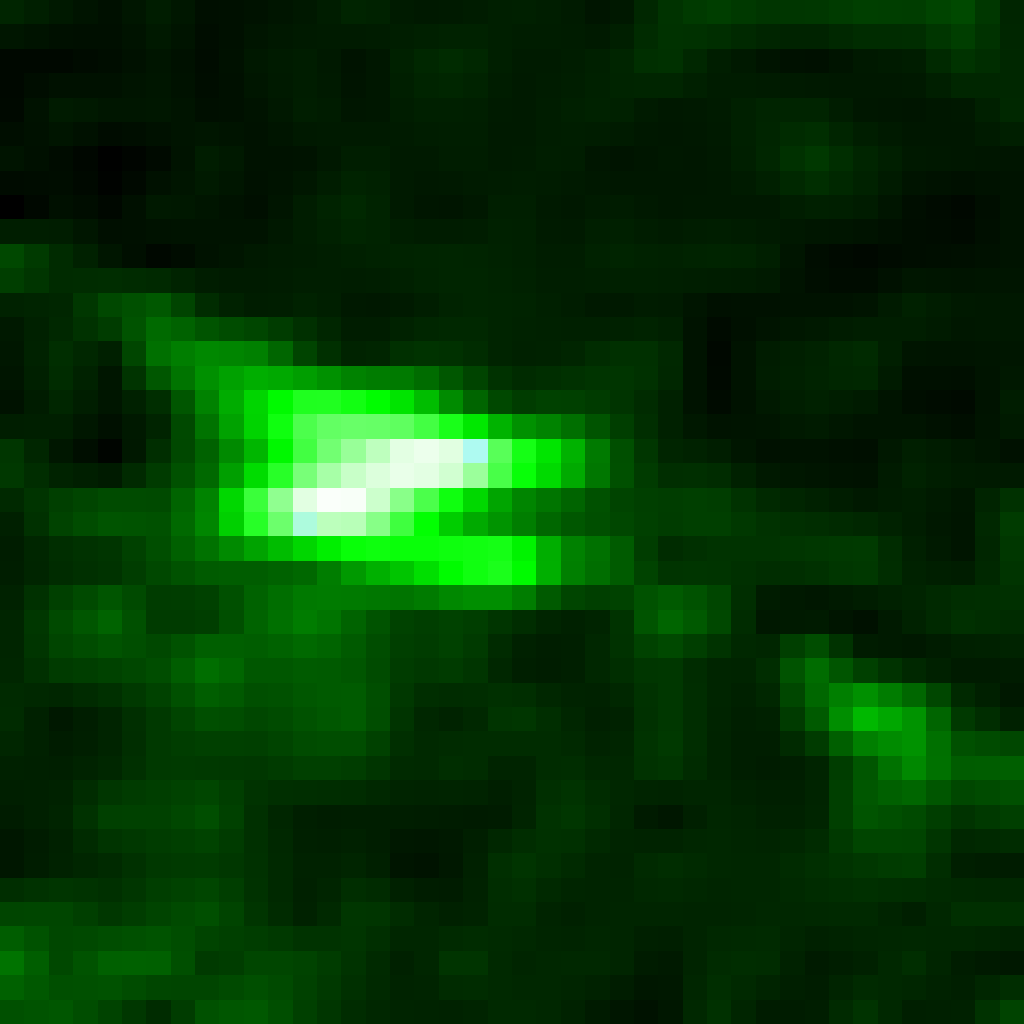}\\
			{\raisebox{0.7cm}	{\rotatebox[origin=c]{90}{~ {\scriptsize $z=-6\um$} }}}&
			\includegraphics[width= 0.12\textwidth]{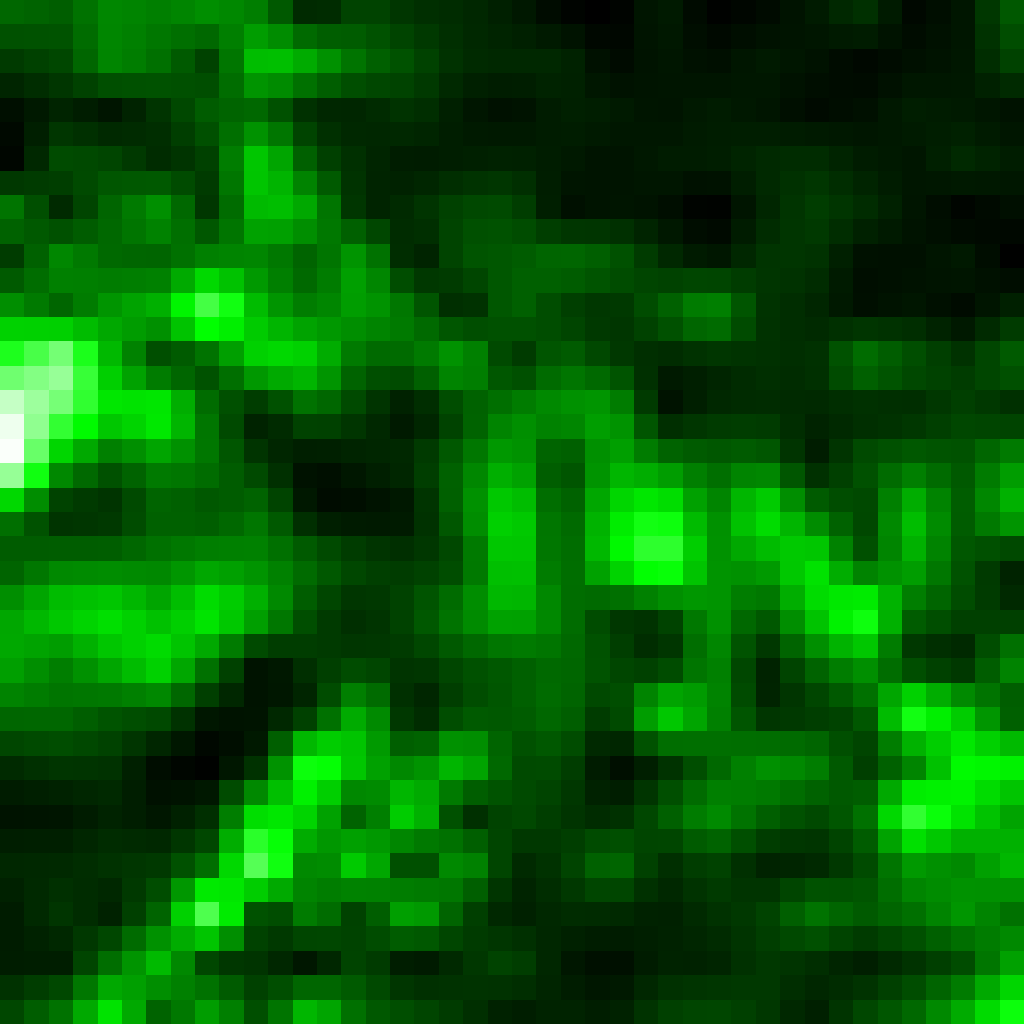}&
			\includegraphics[width= 0.12\textwidth]{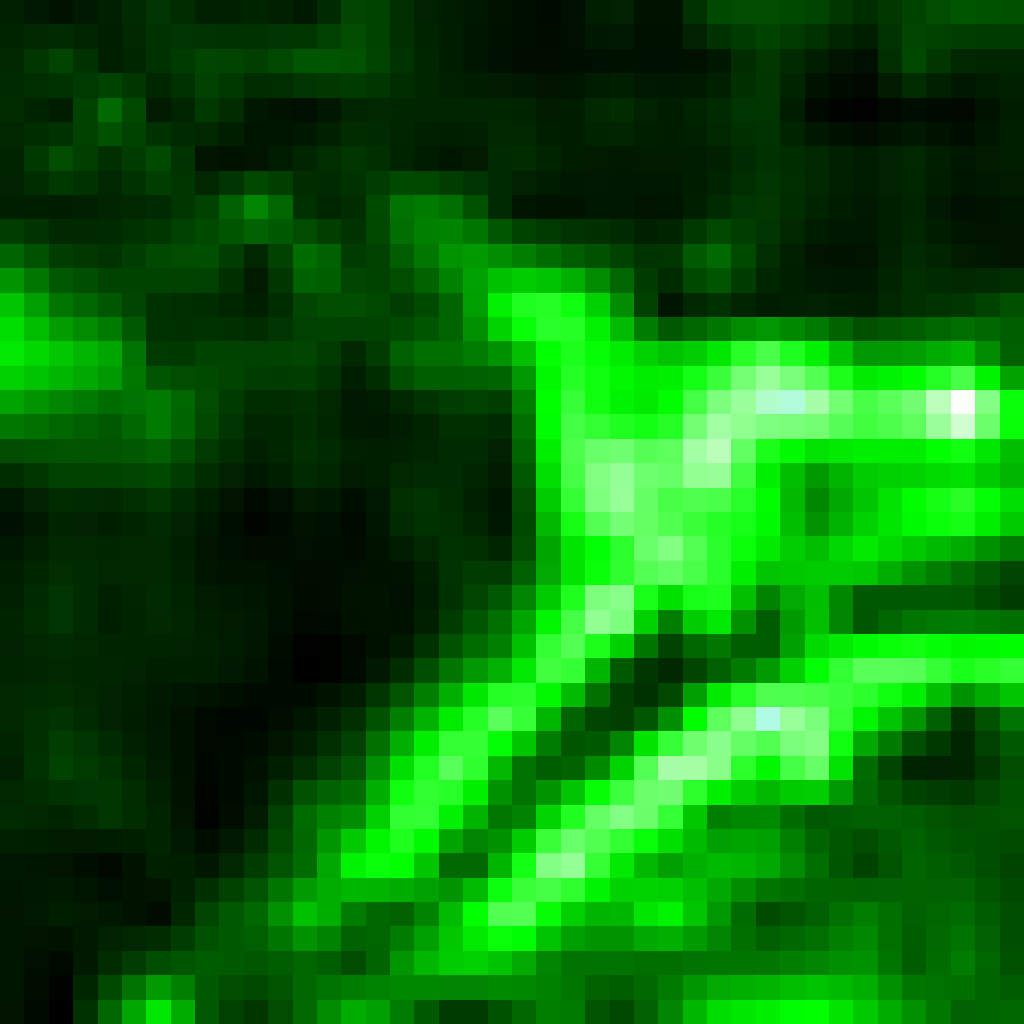}&
			\includegraphics[width= 0.12\textwidth]{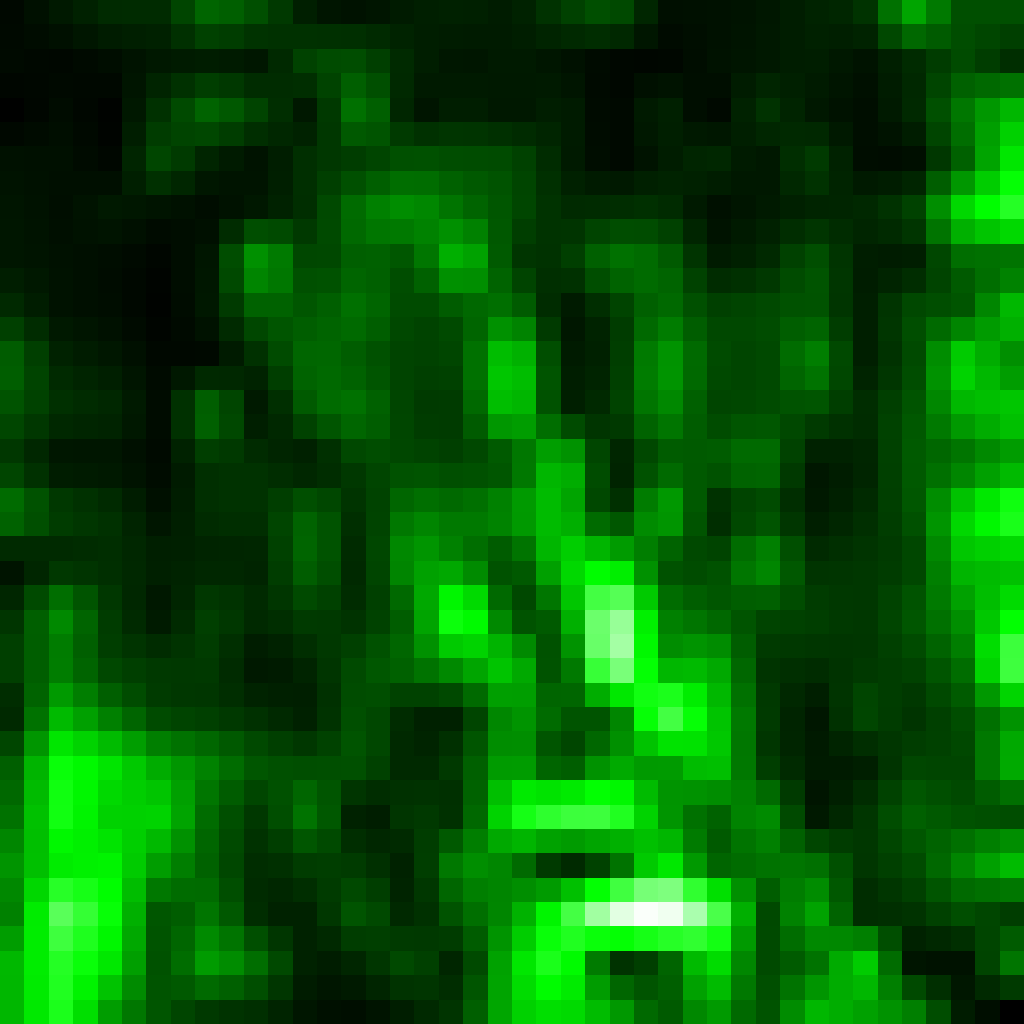}&
			\includegraphics[width= 0.12\textwidth]{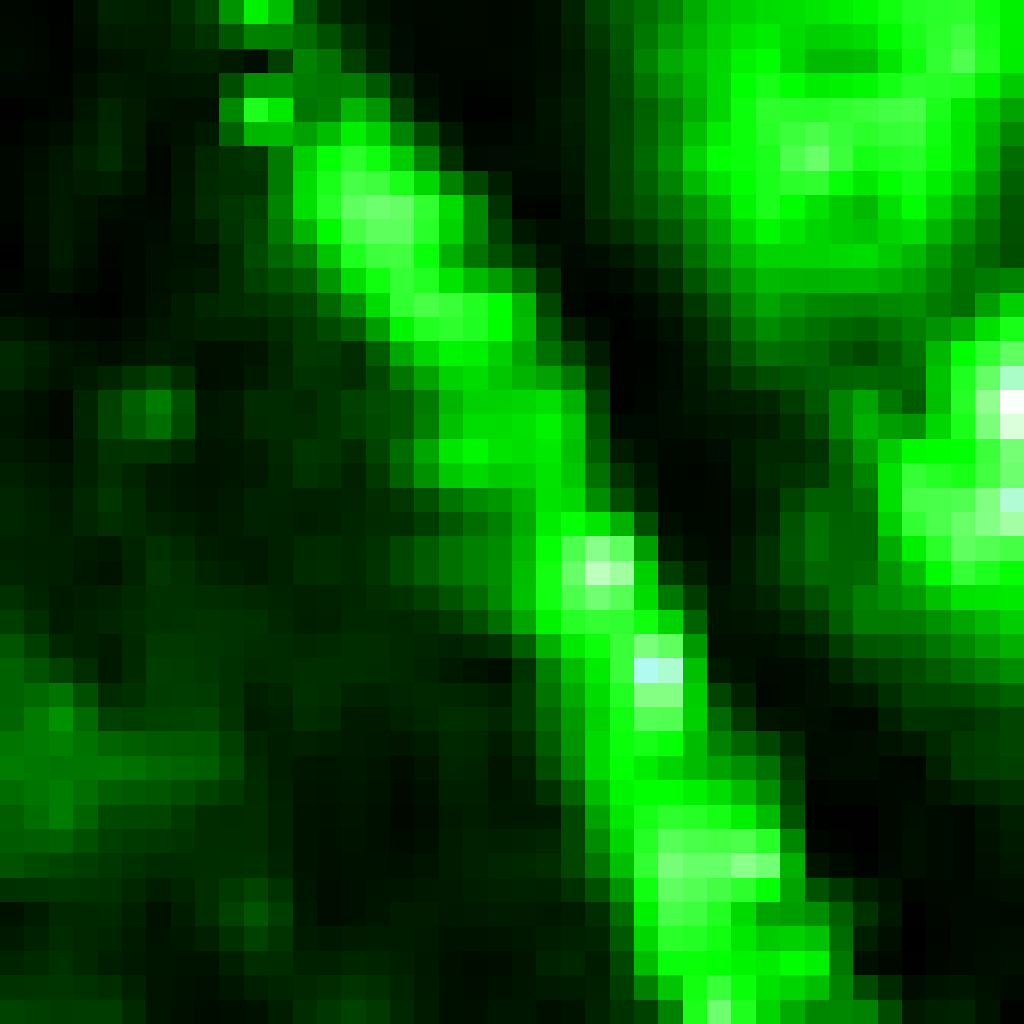}&
			\includegraphics[width= 0.12\textwidth]{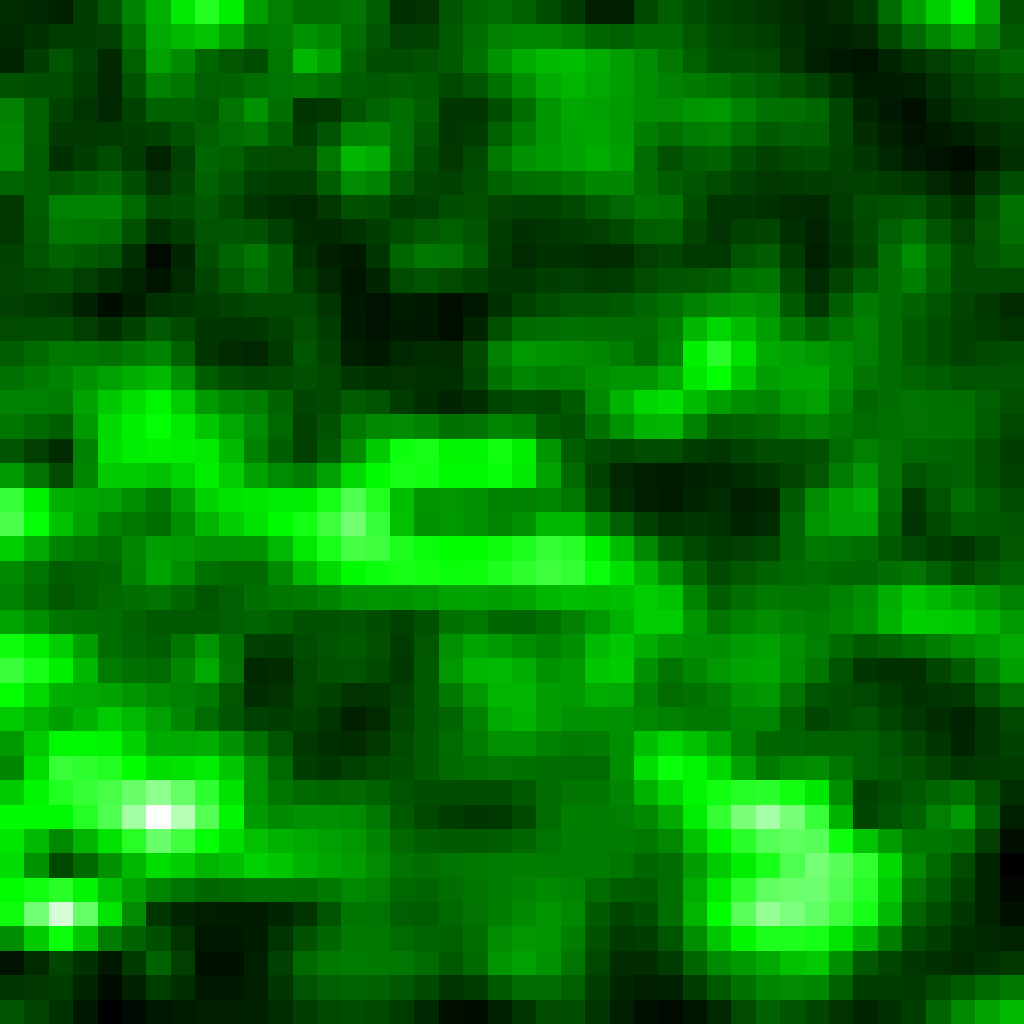}&
			\includegraphics[width= 0.12\textwidth]{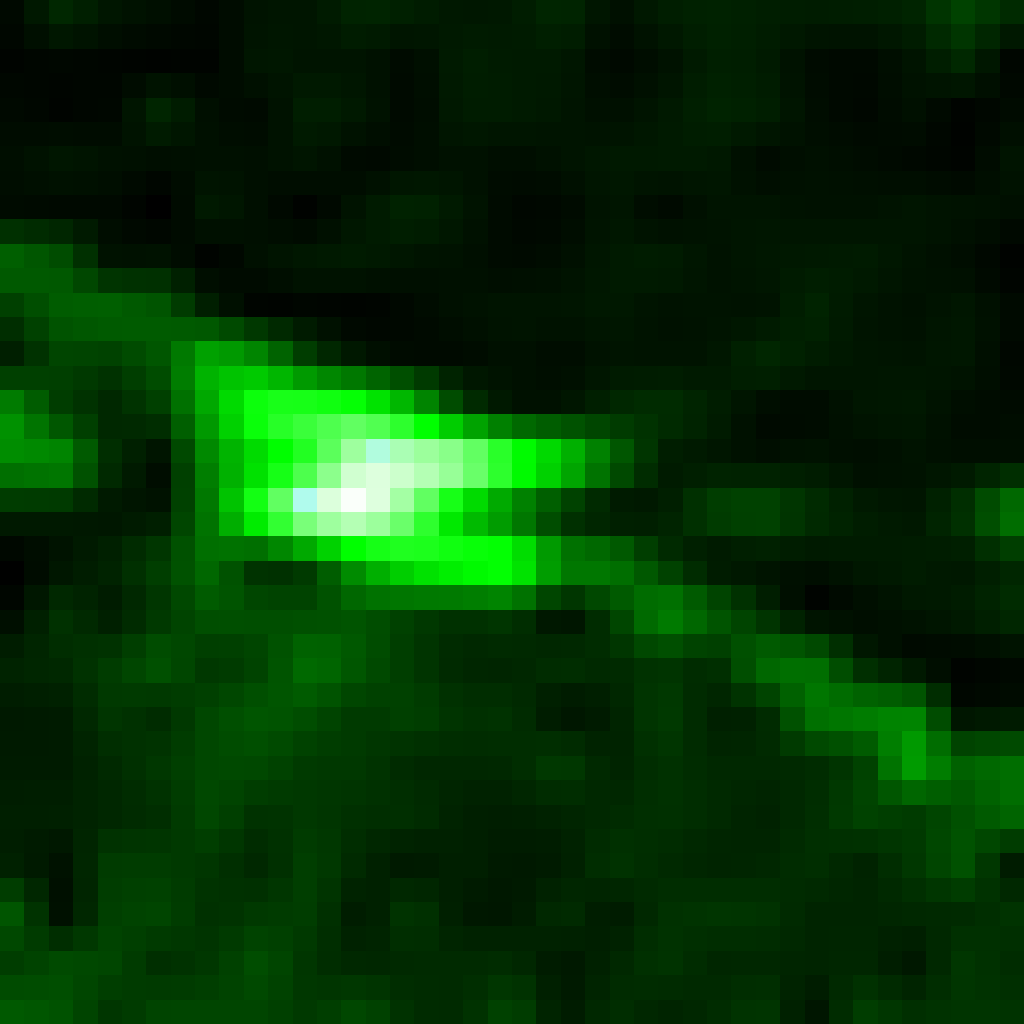}\\
			{\raisebox{0.7cm}	{\rotatebox[origin=c]{90}{~ {\scriptsize $z=-4\um$} }}}&
			\includegraphics[width= 0.12\textwidth]{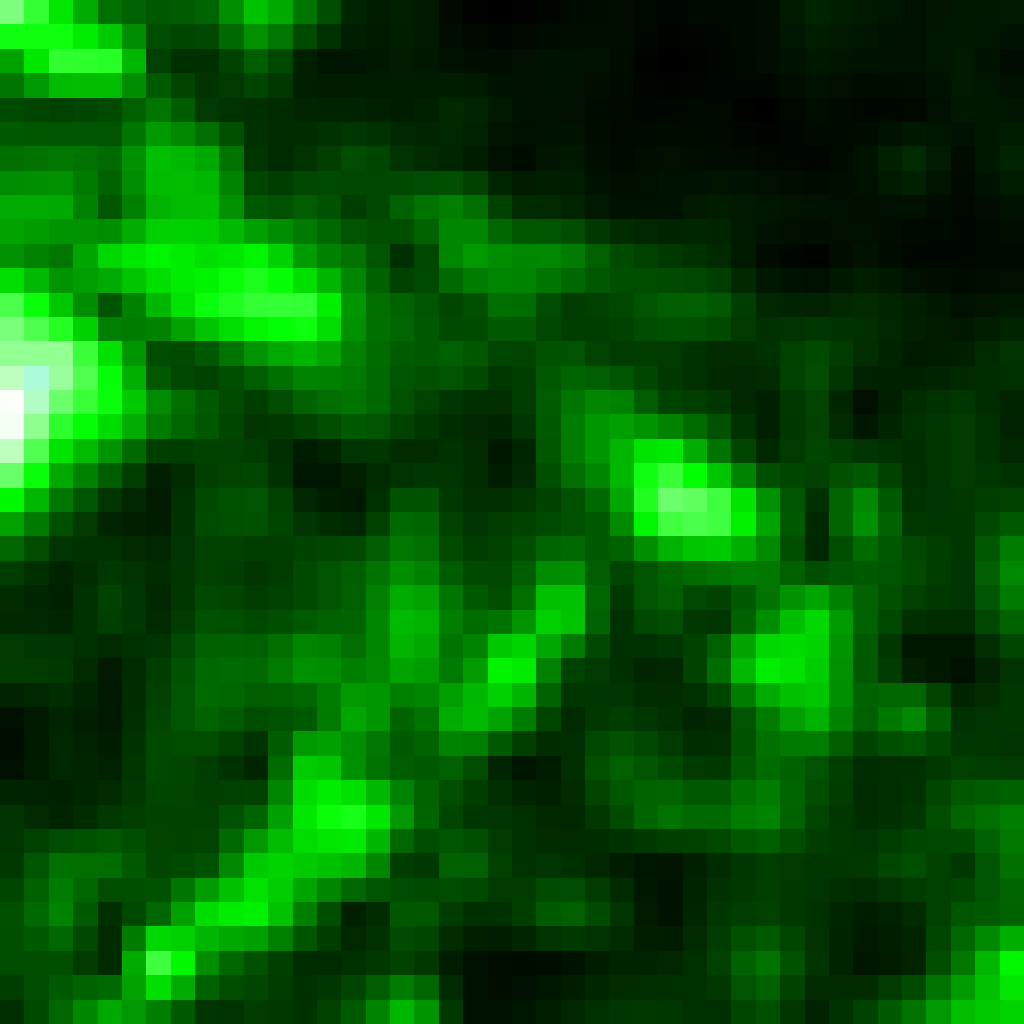}&
			\includegraphics[width= 0.12\textwidth]{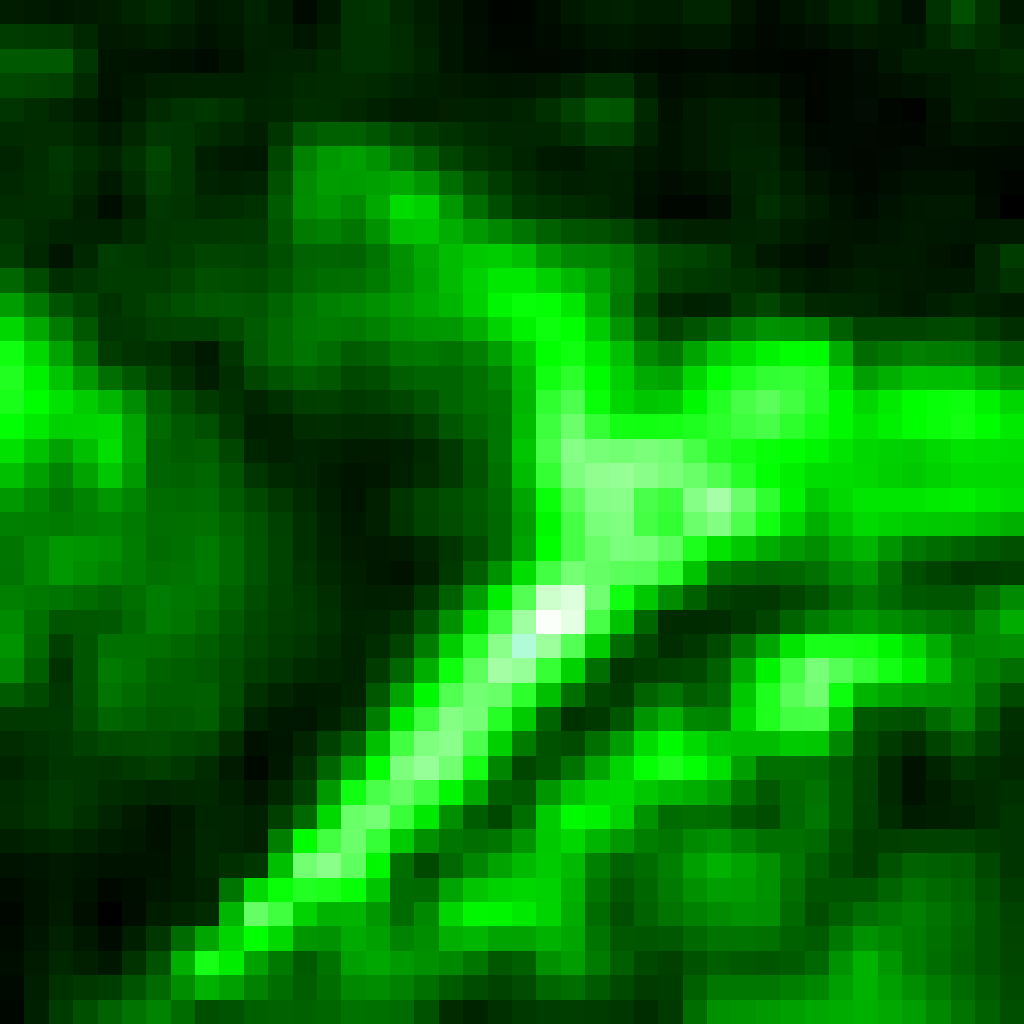}&
			\includegraphics[width= 0.12\textwidth]{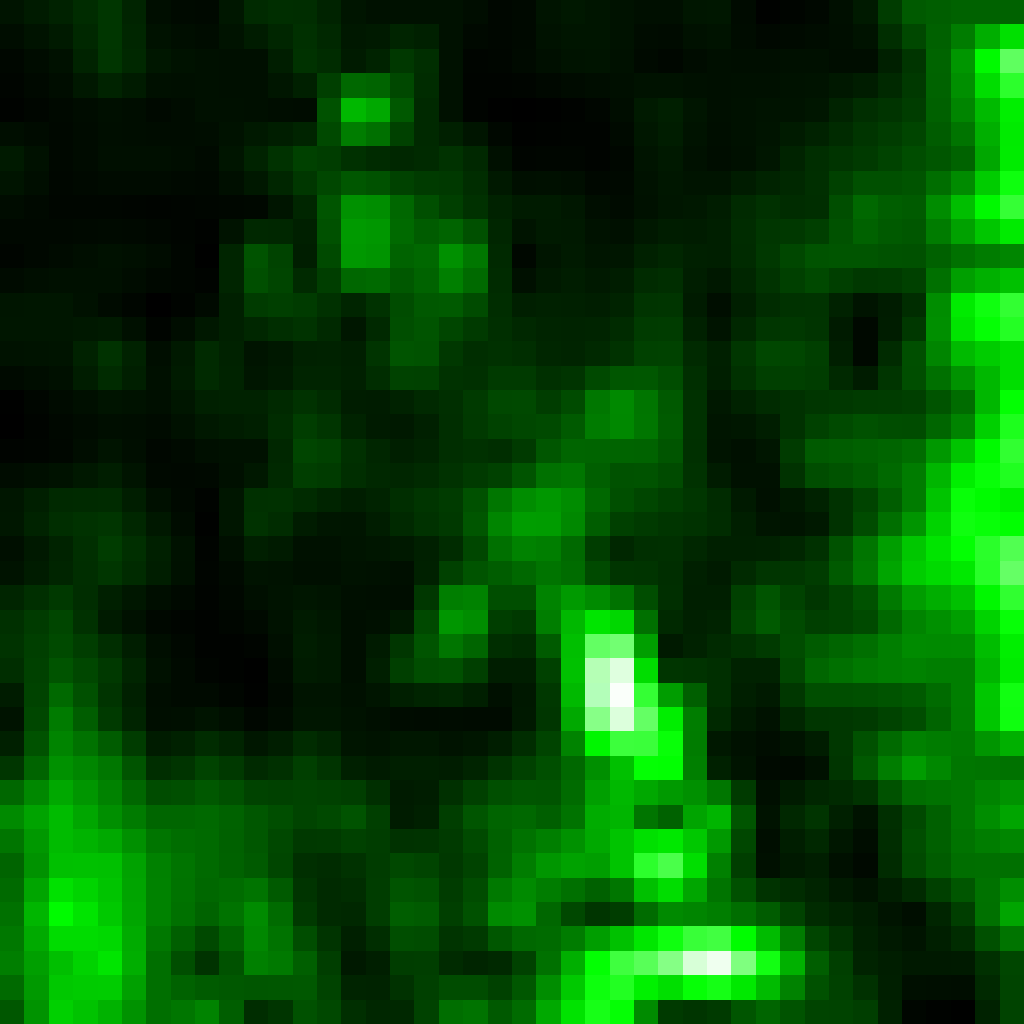}&
			\includegraphics[width= 0.12\textwidth]{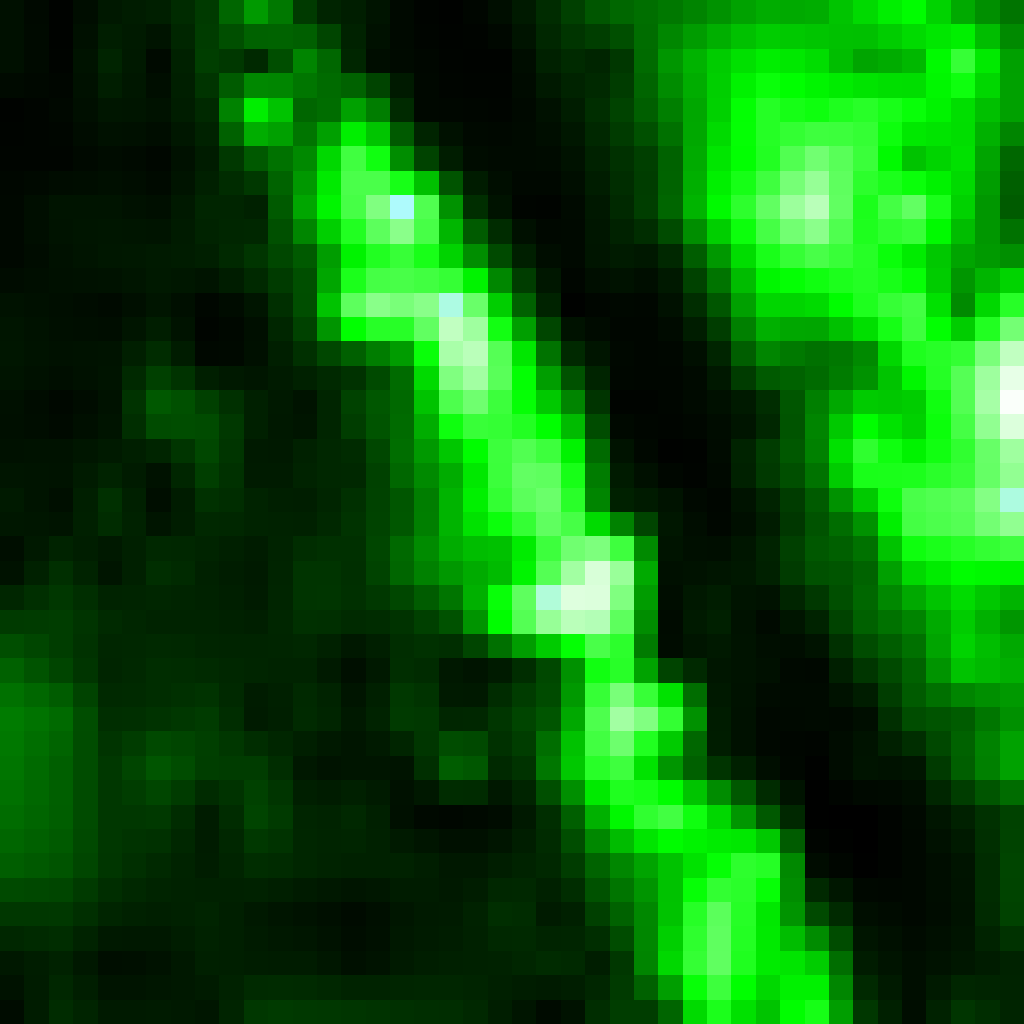}&
			\includegraphics[width= 0.12\textwidth]{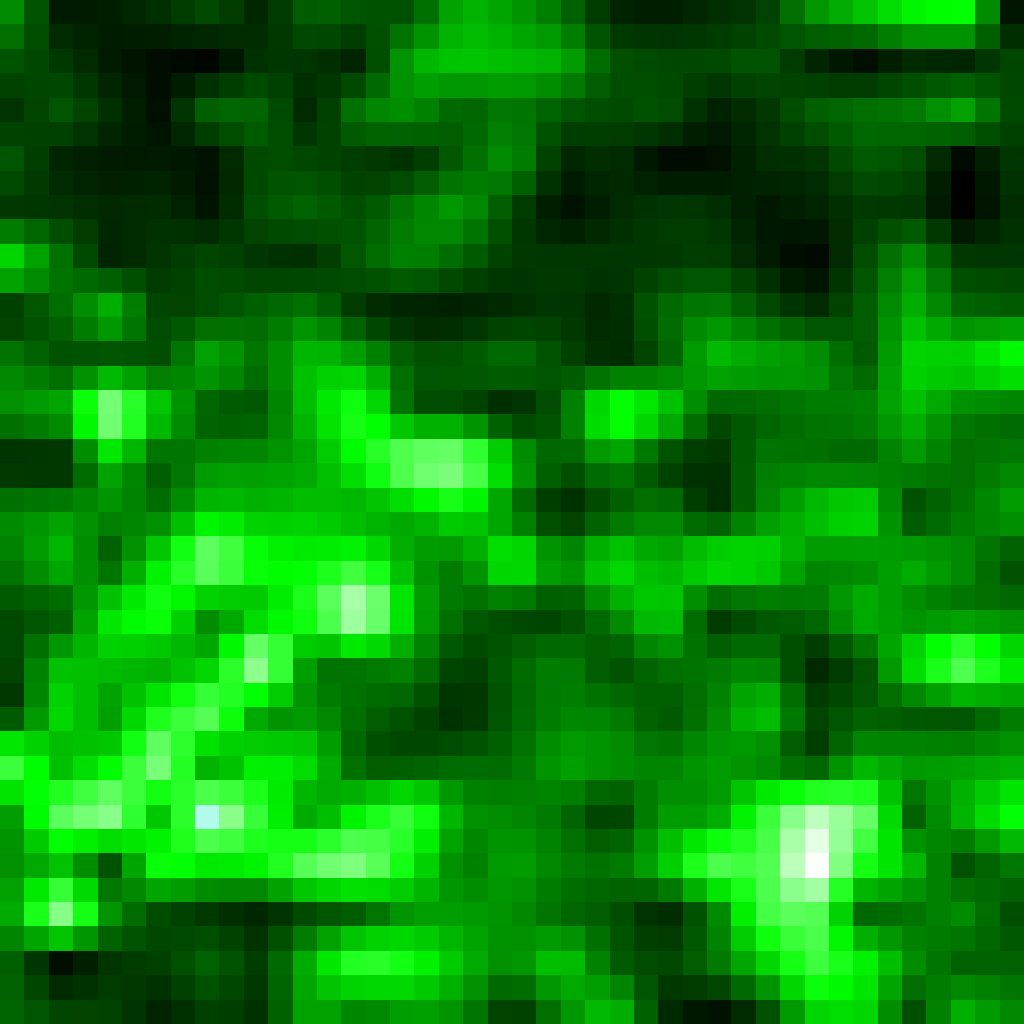}&
			\includegraphics[width= 0.12\textwidth]{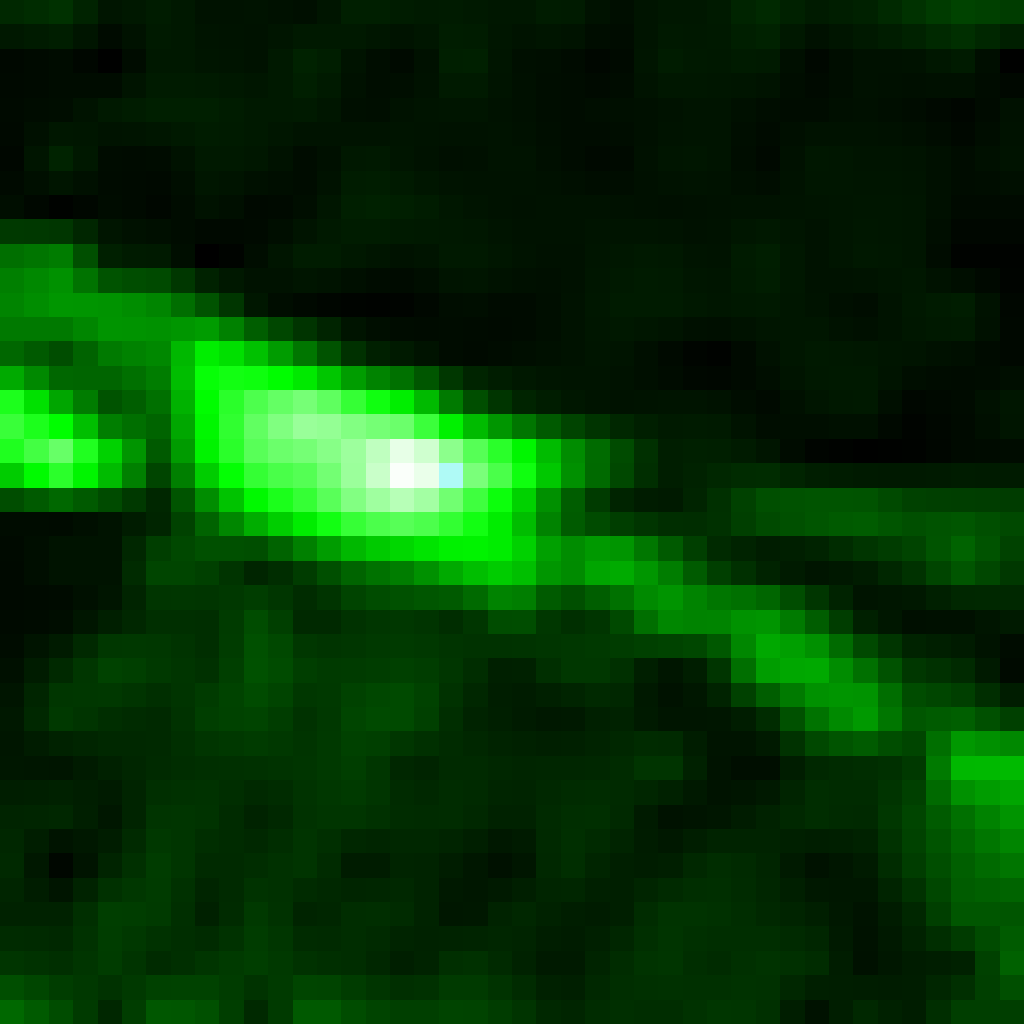}\\
			{\raisebox{0.7cm}	{\rotatebox[origin=c]{90}{~ {\scriptsize $z=-2\um$} }}}&
			\includegraphics[width= 0.12\textwidth]{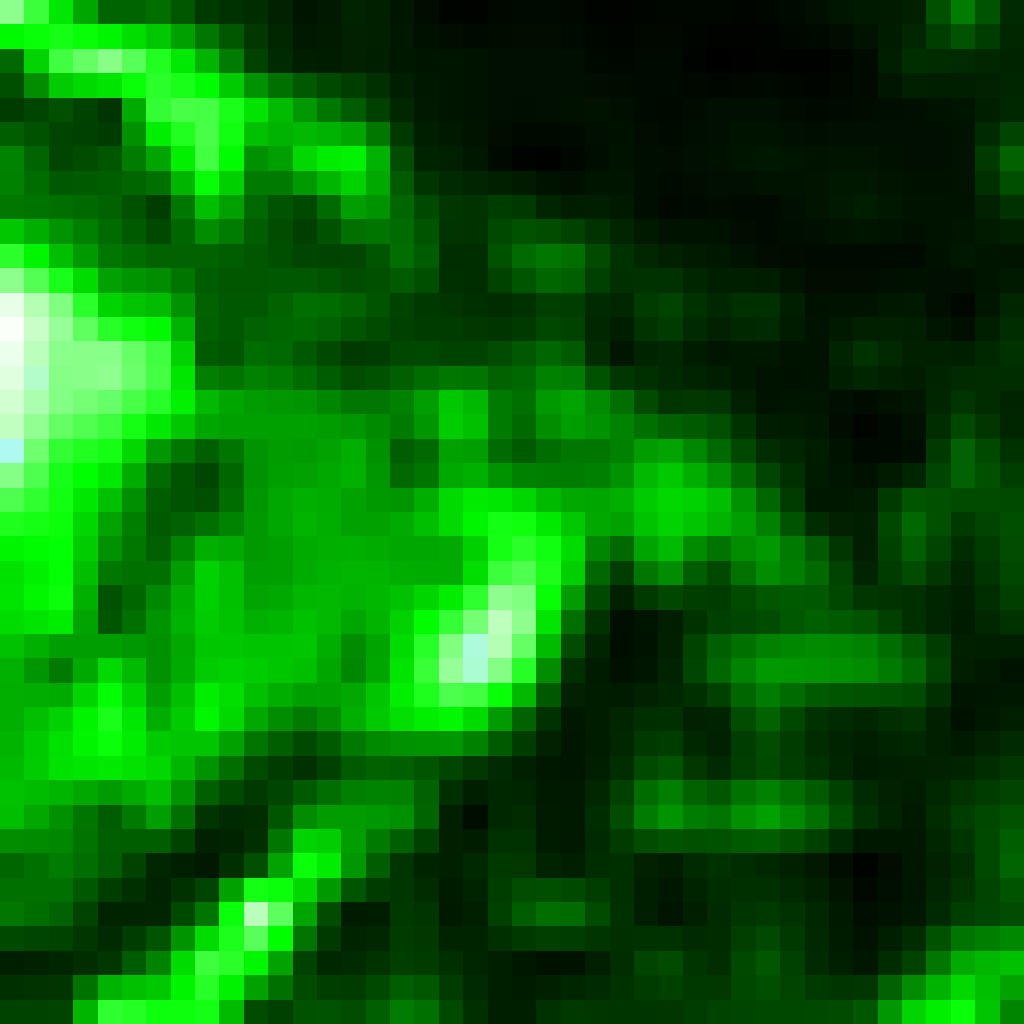}&
			\includegraphics[width= 0.12\textwidth]{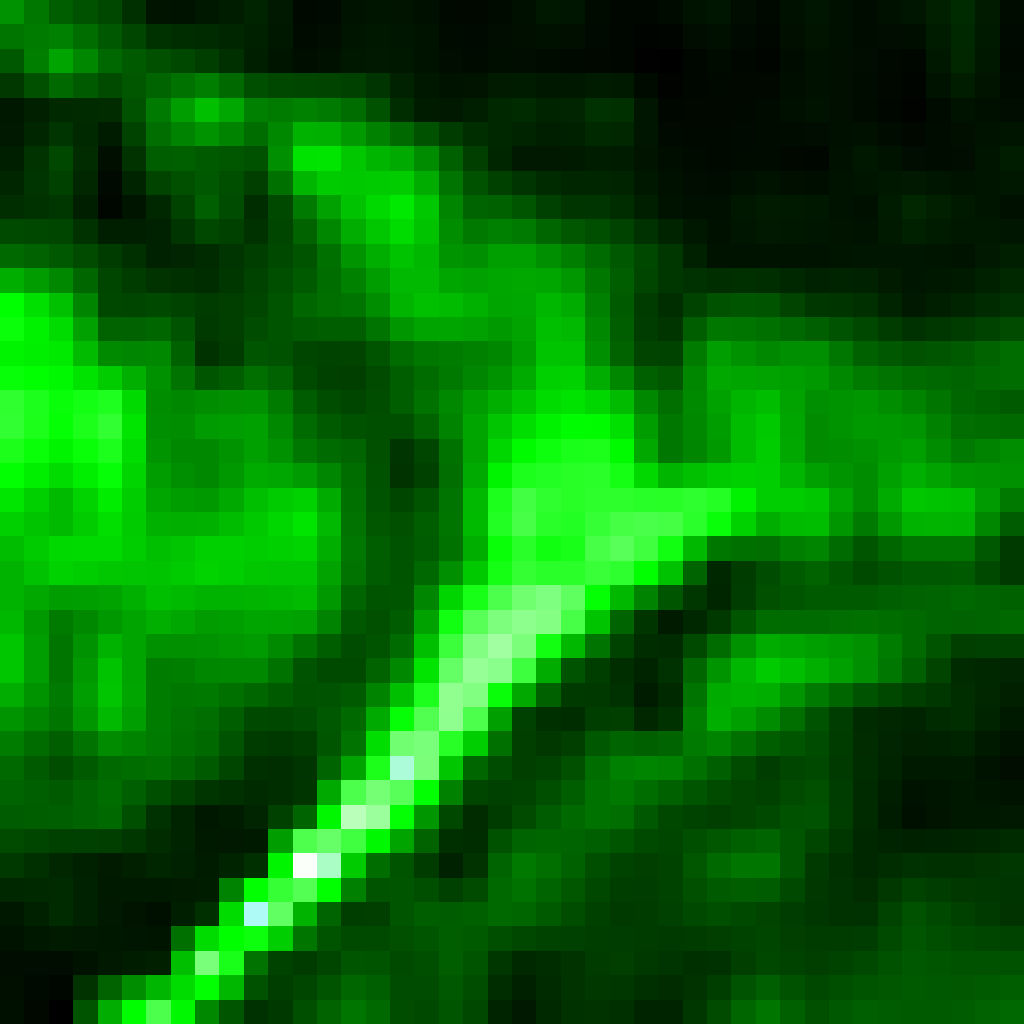}&
			\includegraphics[width= 0.12\textwidth]{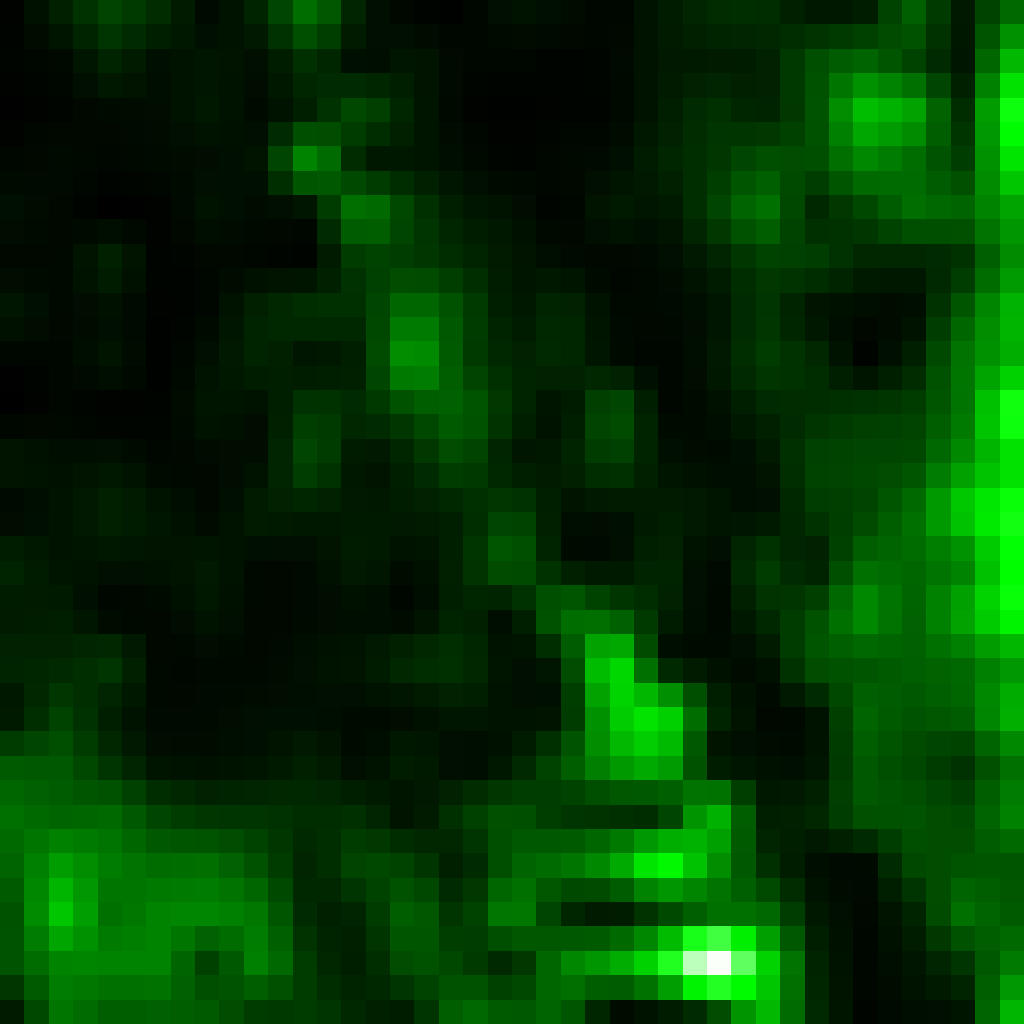}&
			\includegraphics[width= 0.12\textwidth]{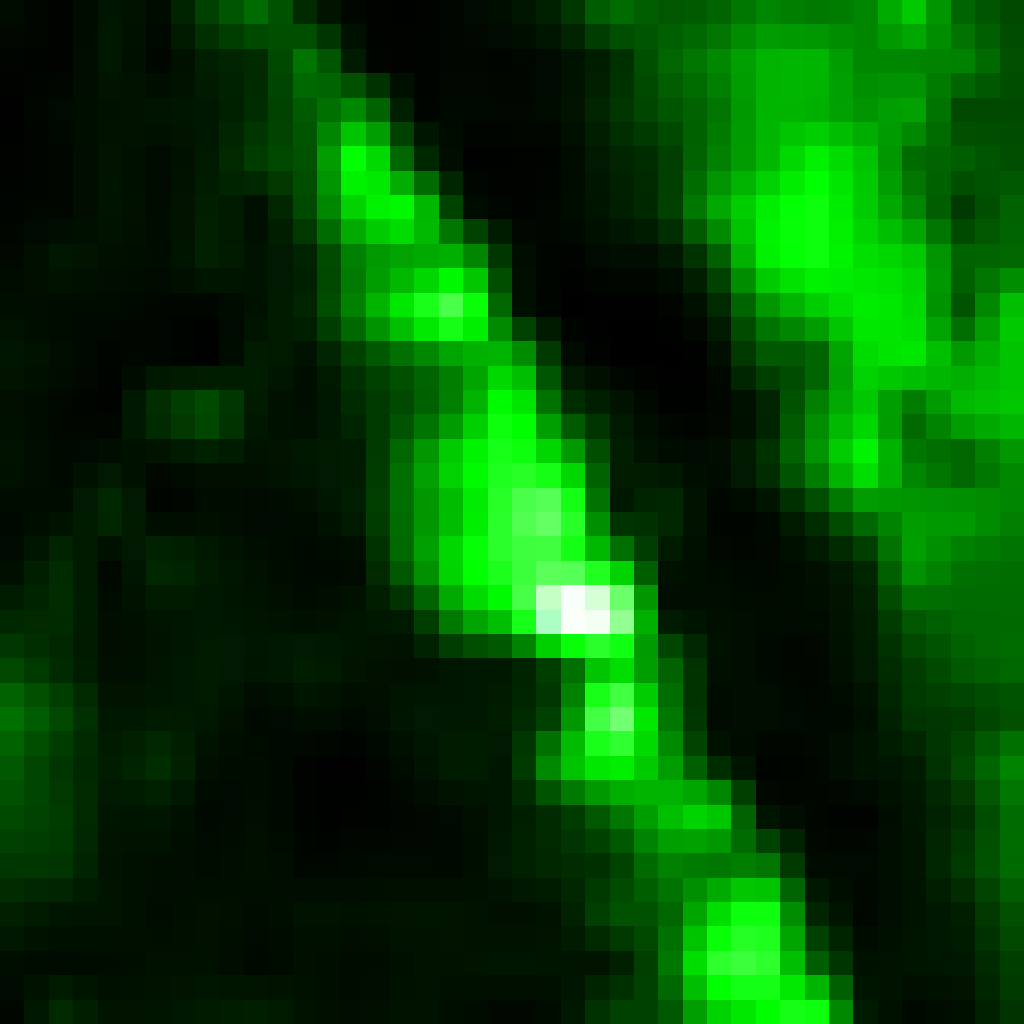}&
			\includegraphics[width= 0.12\textwidth]{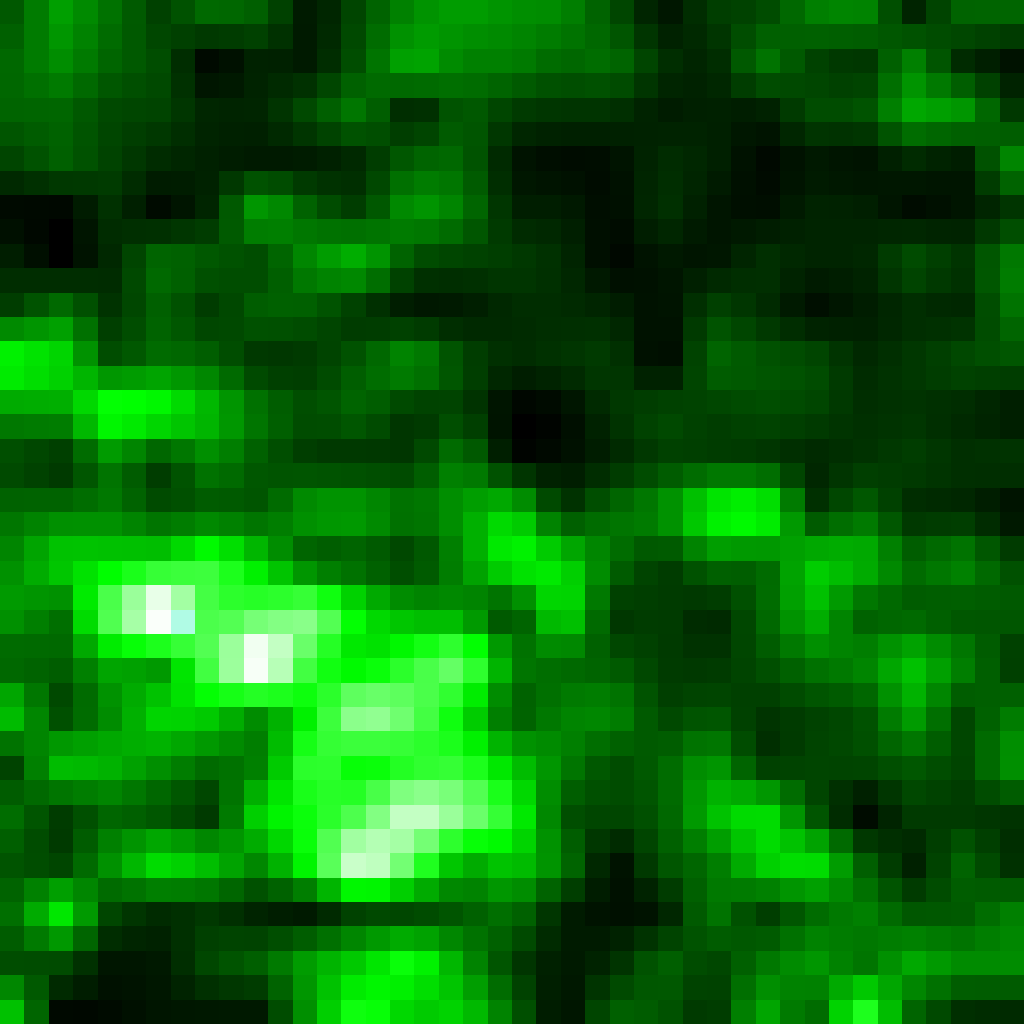}&
			\includegraphics[width= 0.12\textwidth]{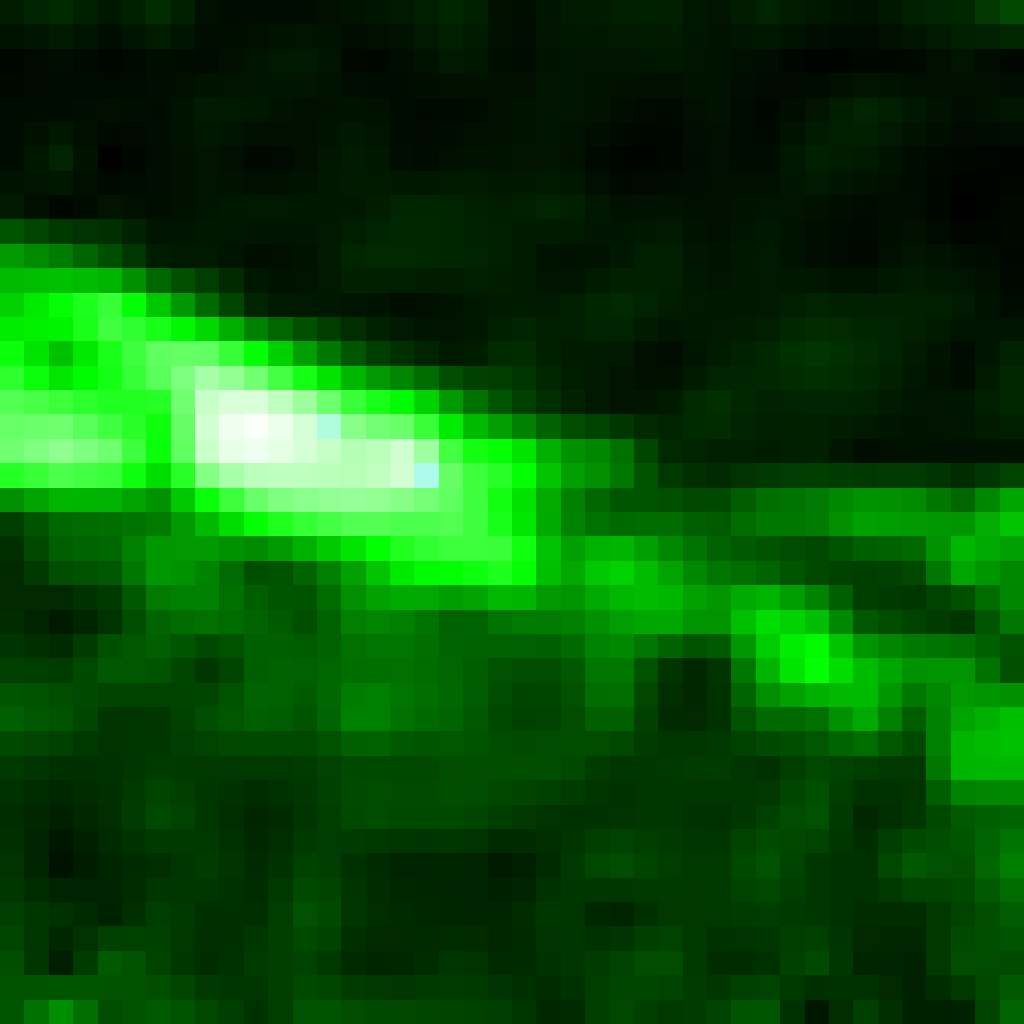}\\
			{\raisebox{0.7cm}	{\rotatebox[origin=c]{90}{~ {\scriptsize $z=0\um$} }}}&
			\includegraphics[width= 0.12\textwidth]{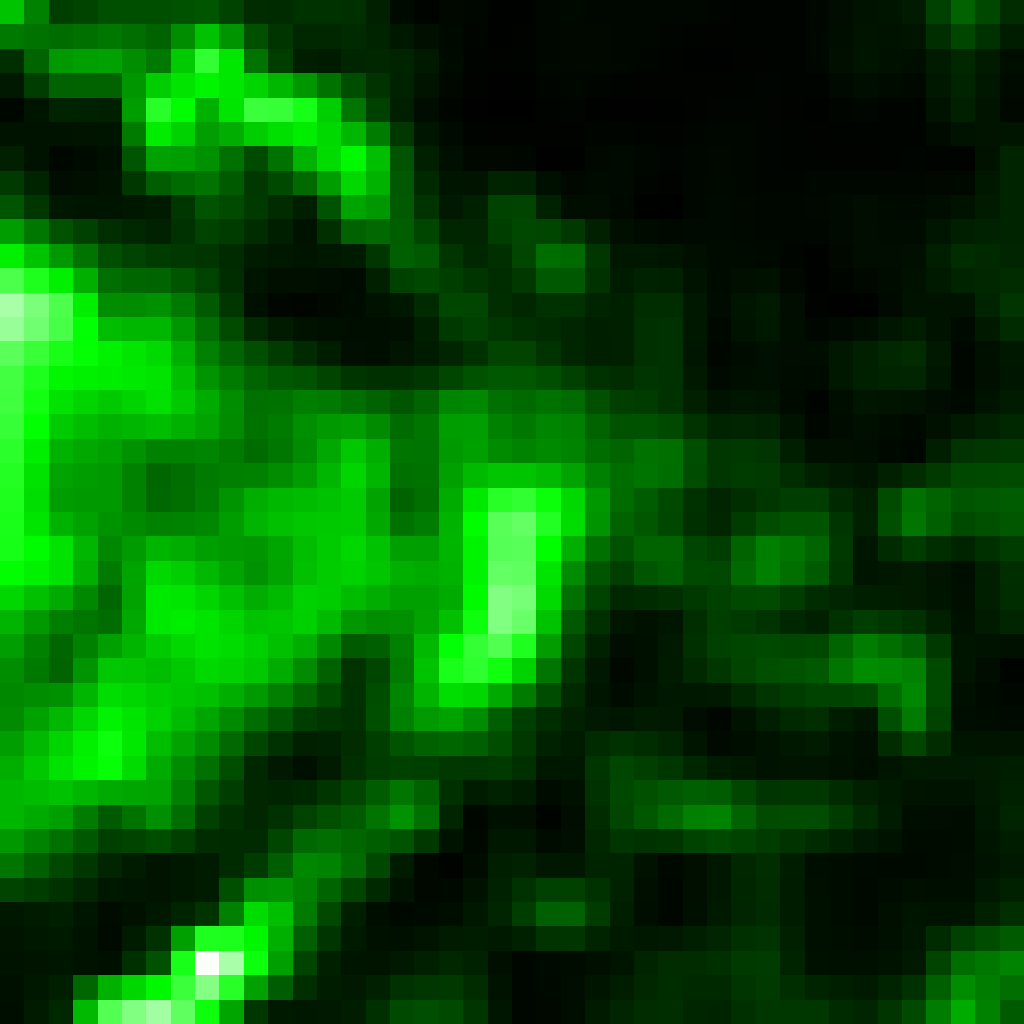}&
			\includegraphics[width= 0.12\textwidth]{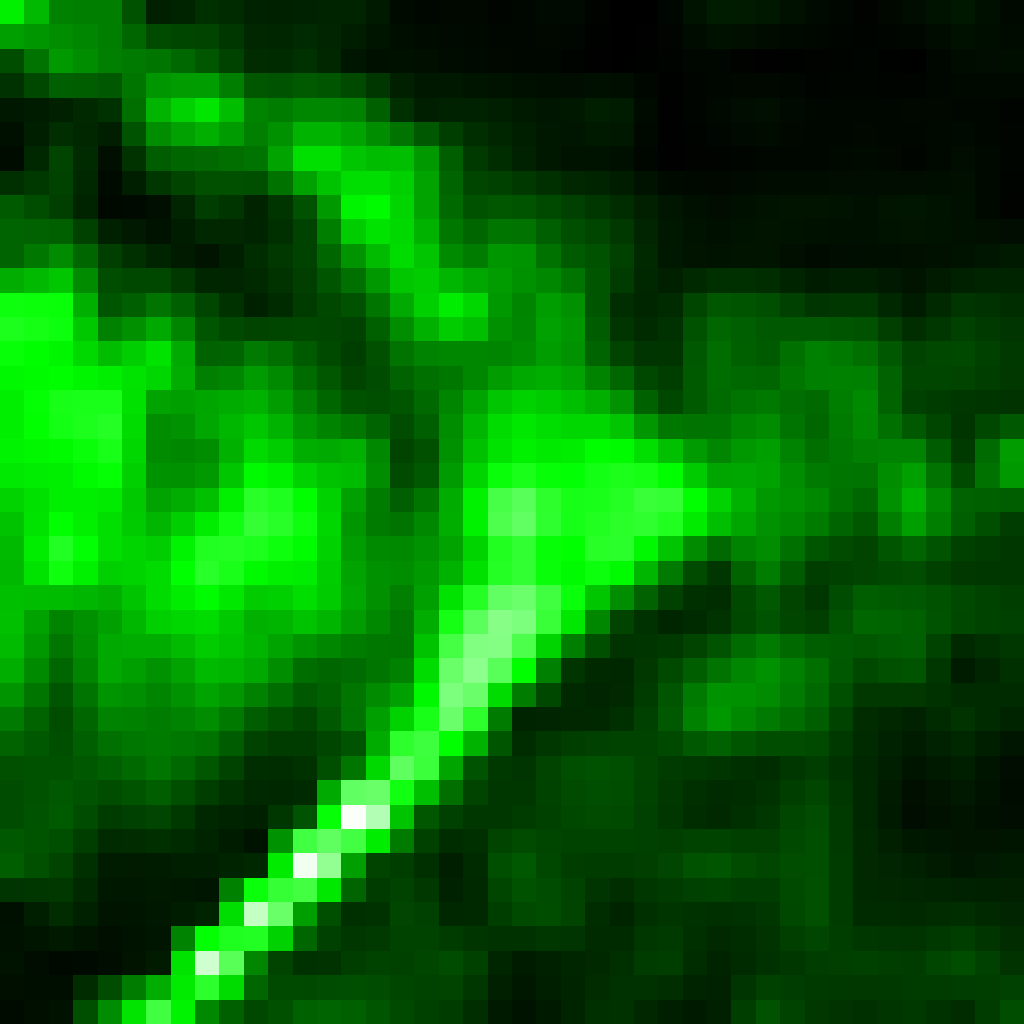}&
			\includegraphics[width= 0.12\textwidth]{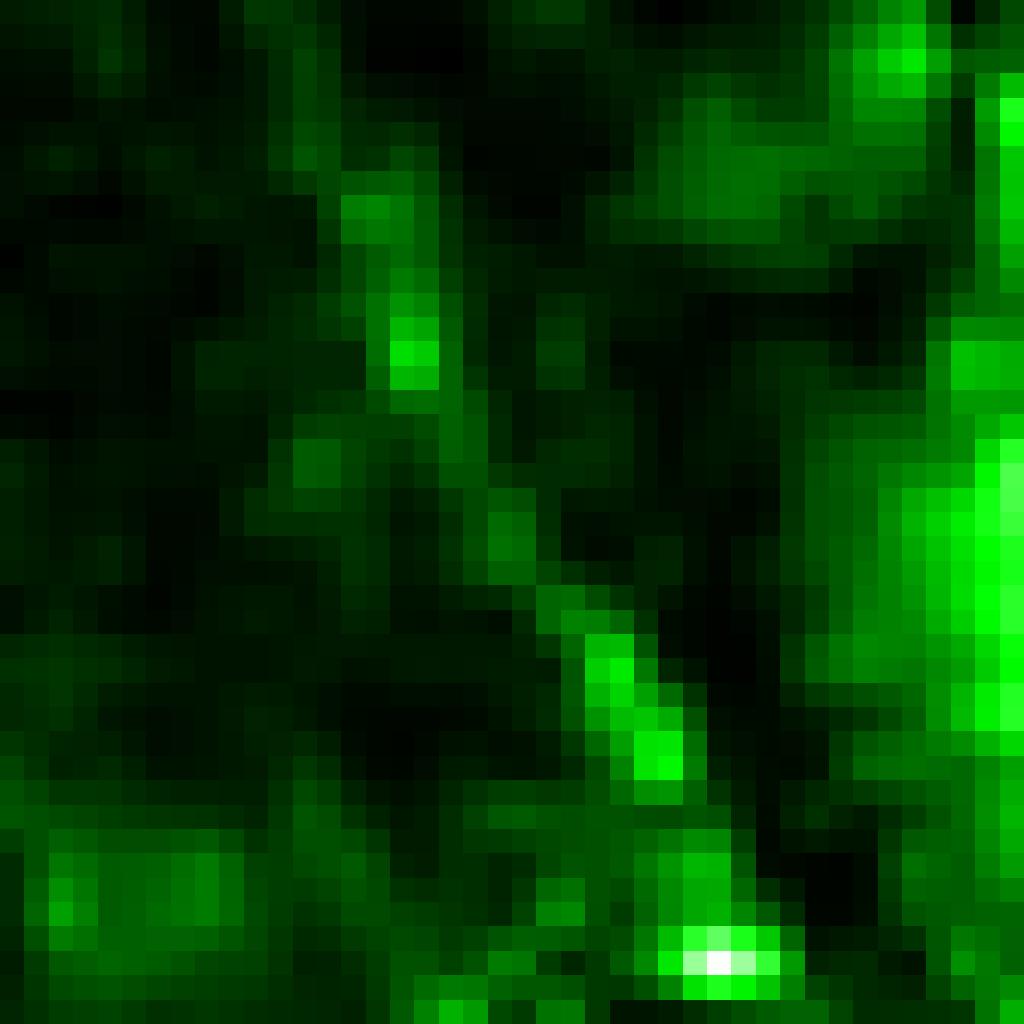}&
			\includegraphics[width= 0.12\textwidth]{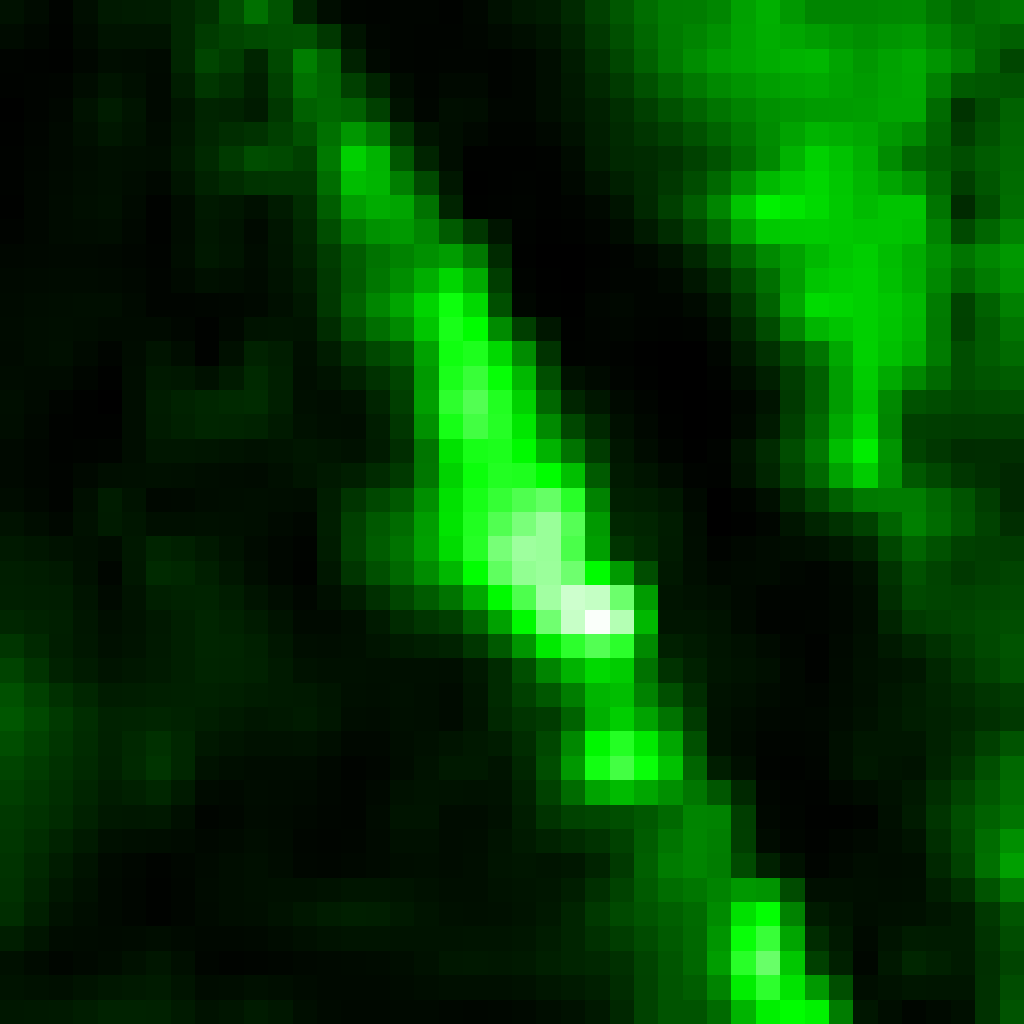}&
			\includegraphics[width= 0.12\textwidth]{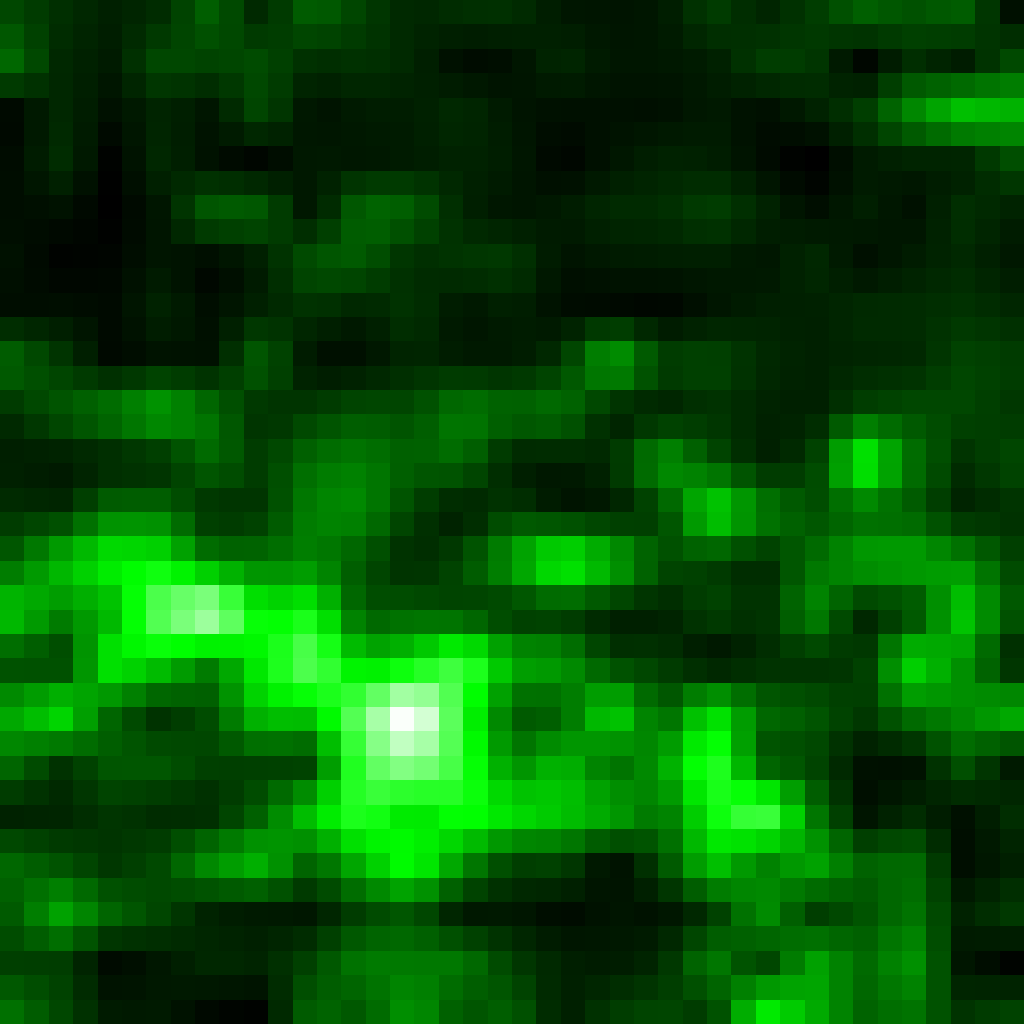}&
			\includegraphics[width= 0.12\textwidth]{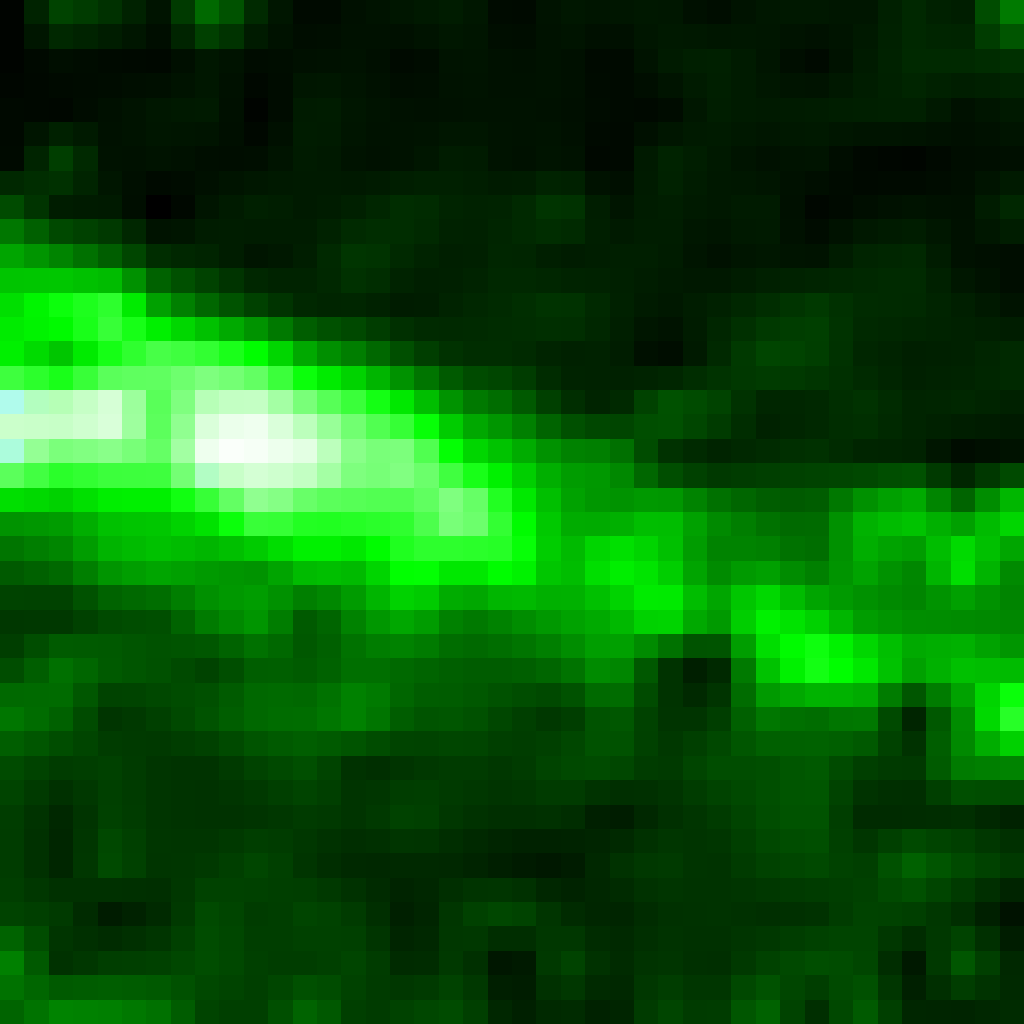}\\
			{\raisebox{0.7cm}	{\rotatebox[origin=c]{90}{~ {\scriptsize $z=2\um$} }}}&
			\includegraphics[width= 0.12\textwidth]{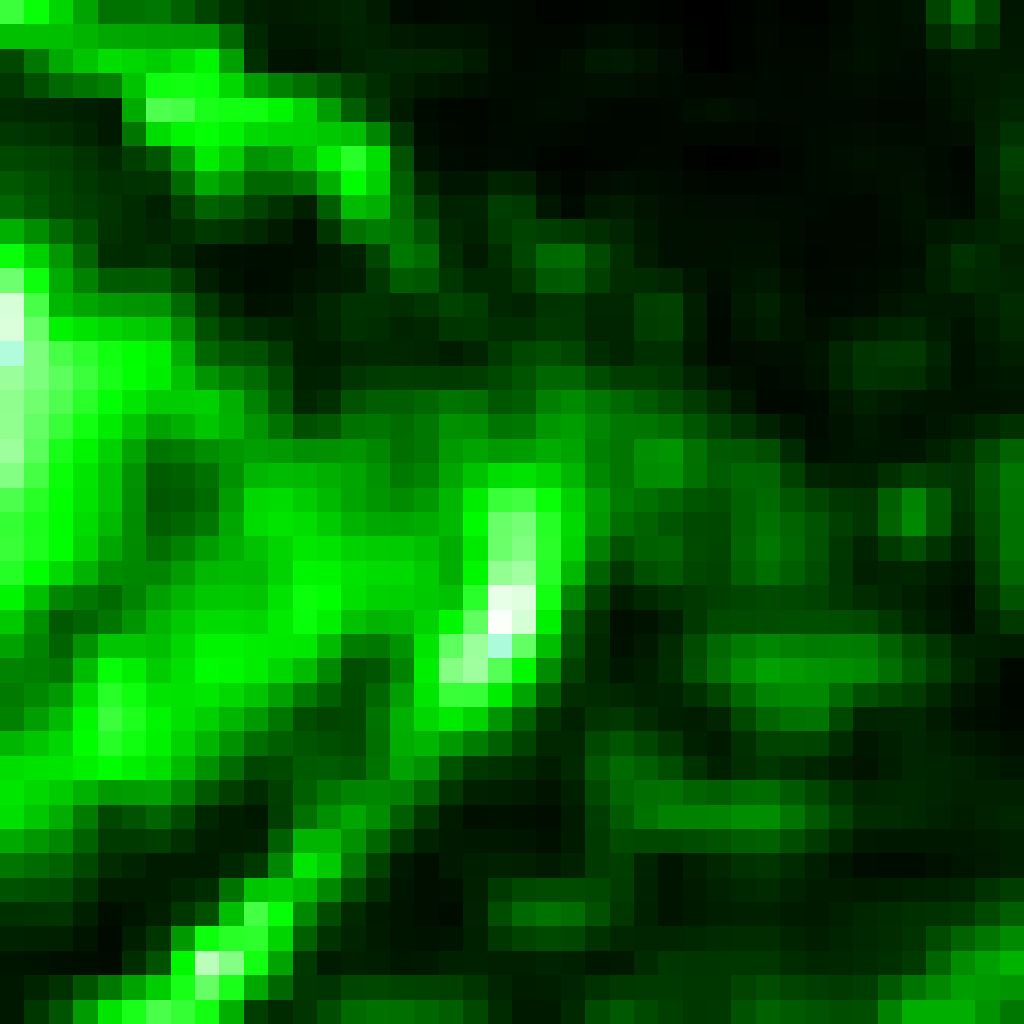}&
			\includegraphics[width= 0.12\textwidth]{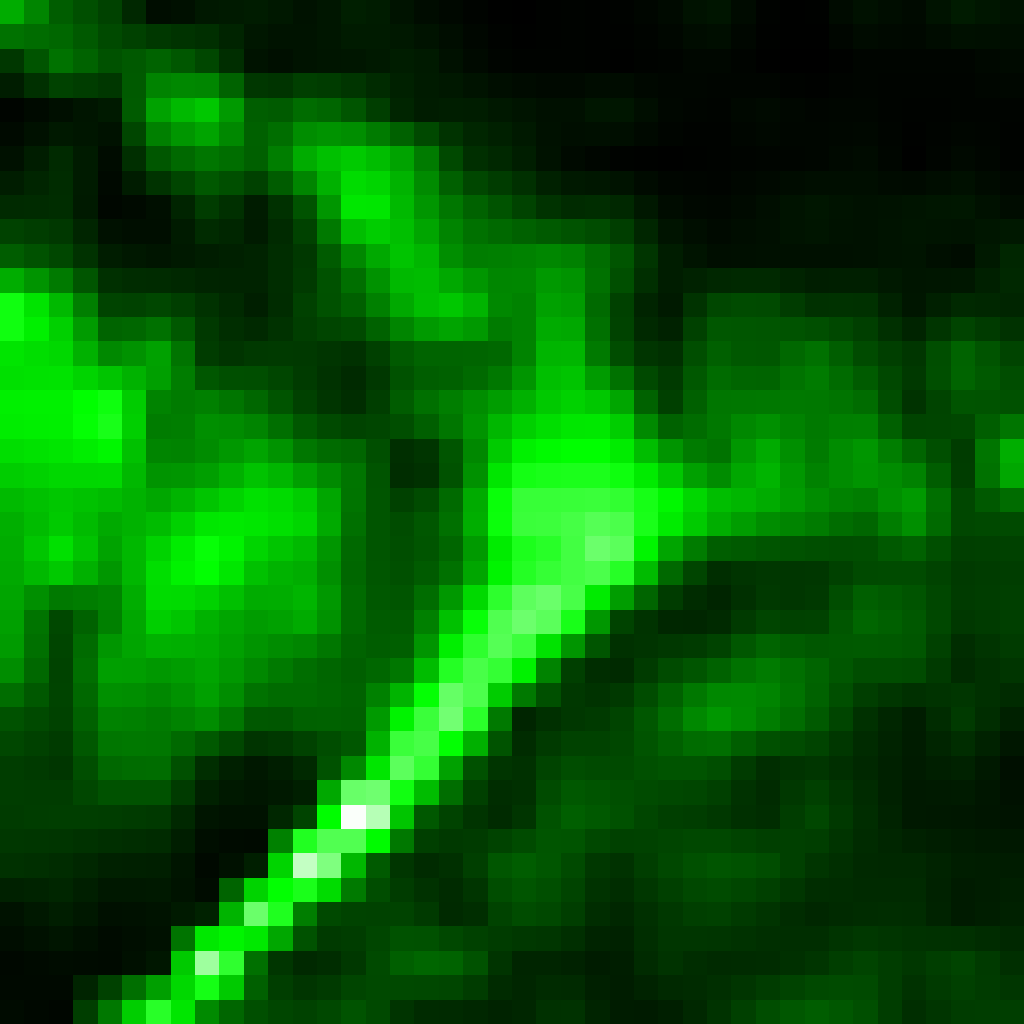}&
			\includegraphics[width= 0.12\textwidth]{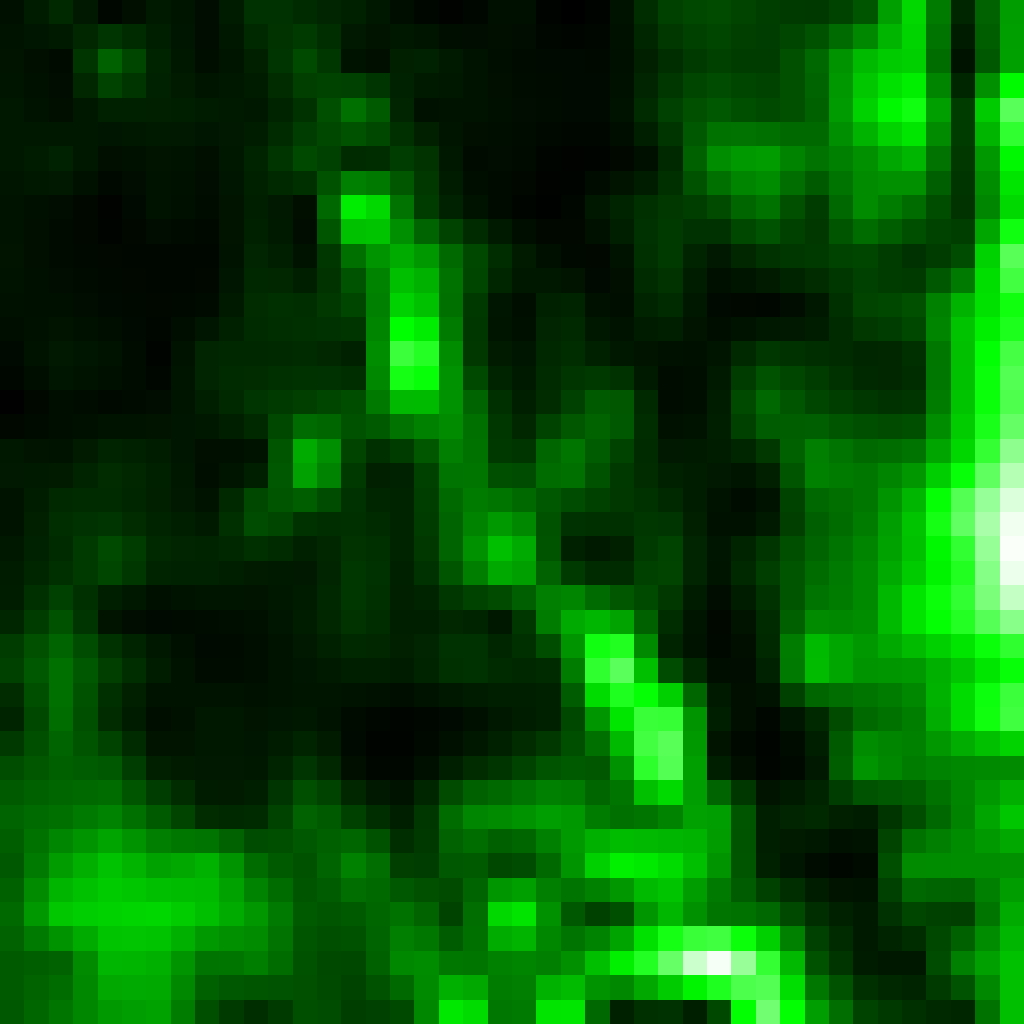}&
			\includegraphics[width= 0.12\textwidth]{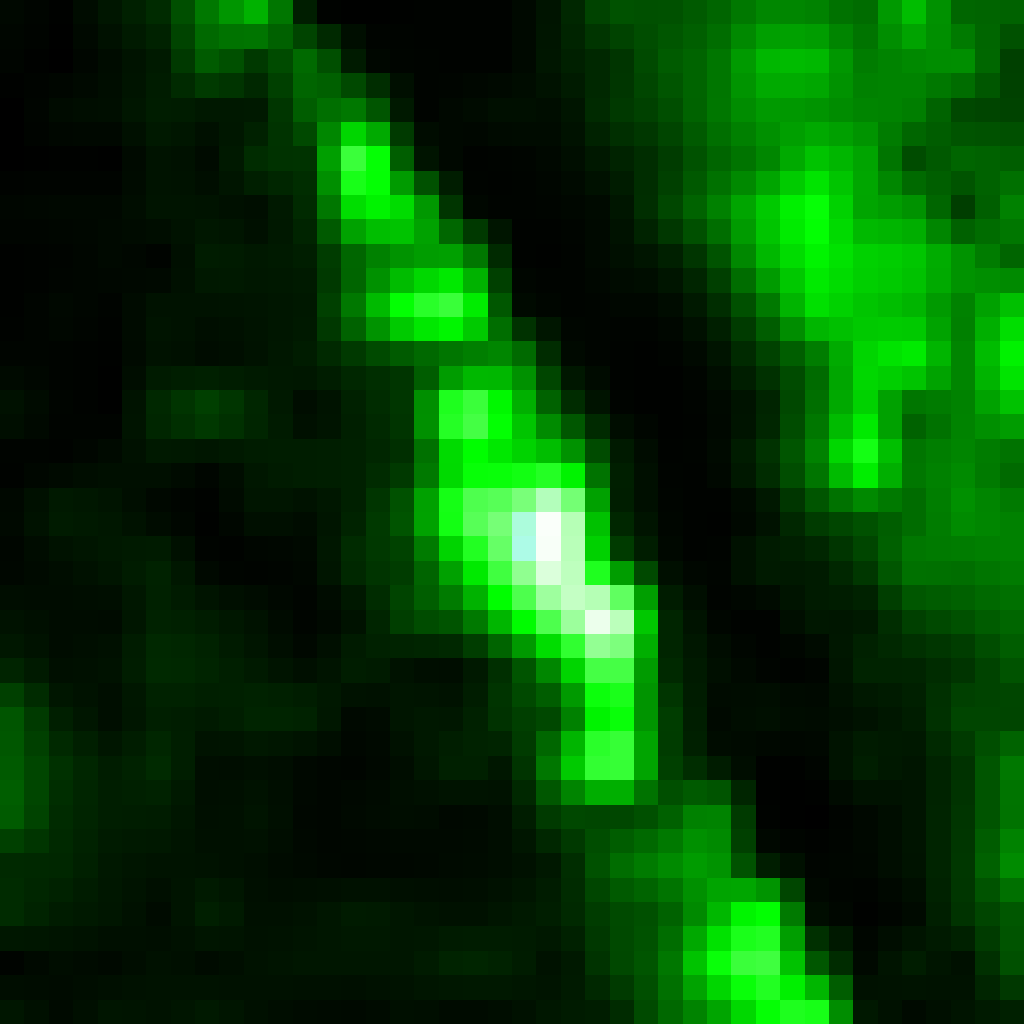}&
			\includegraphics[width= 0.12\textwidth]{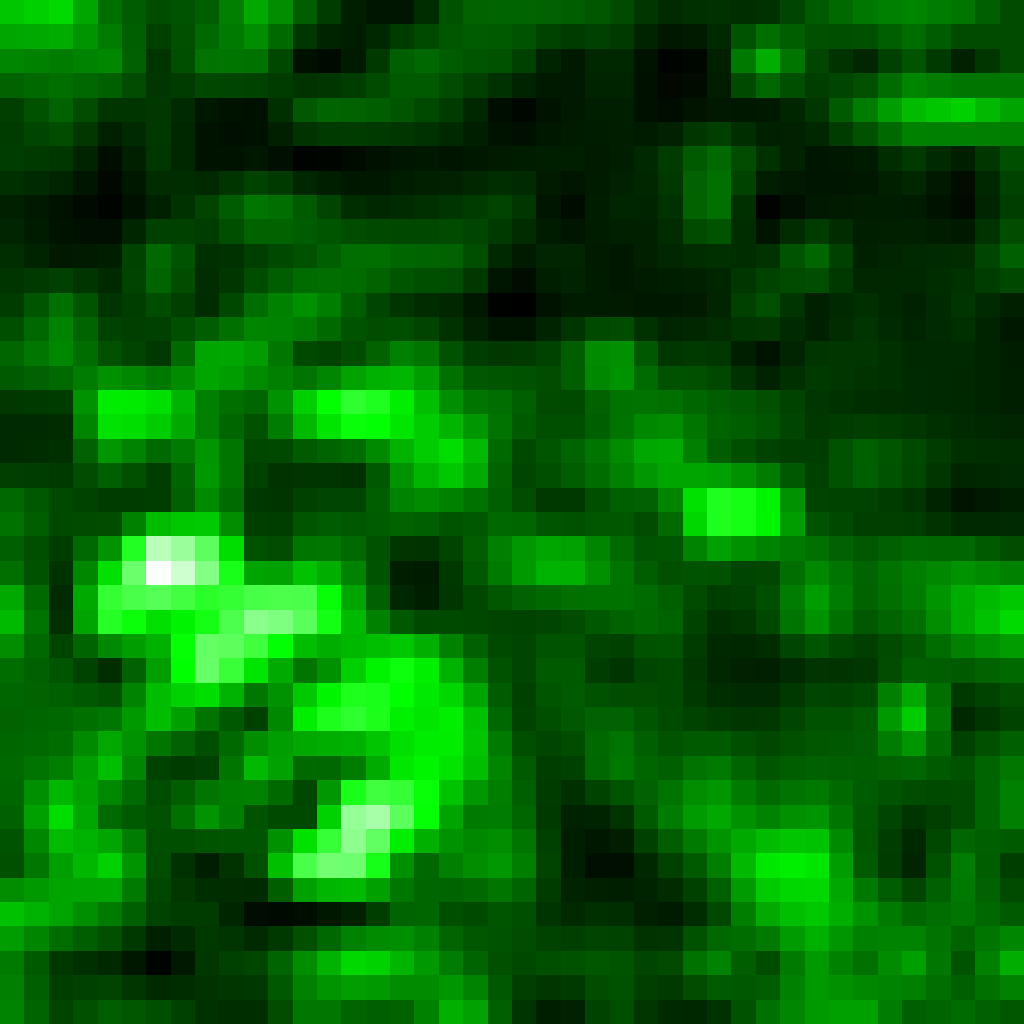}&
			\includegraphics[width= 0.12\textwidth]{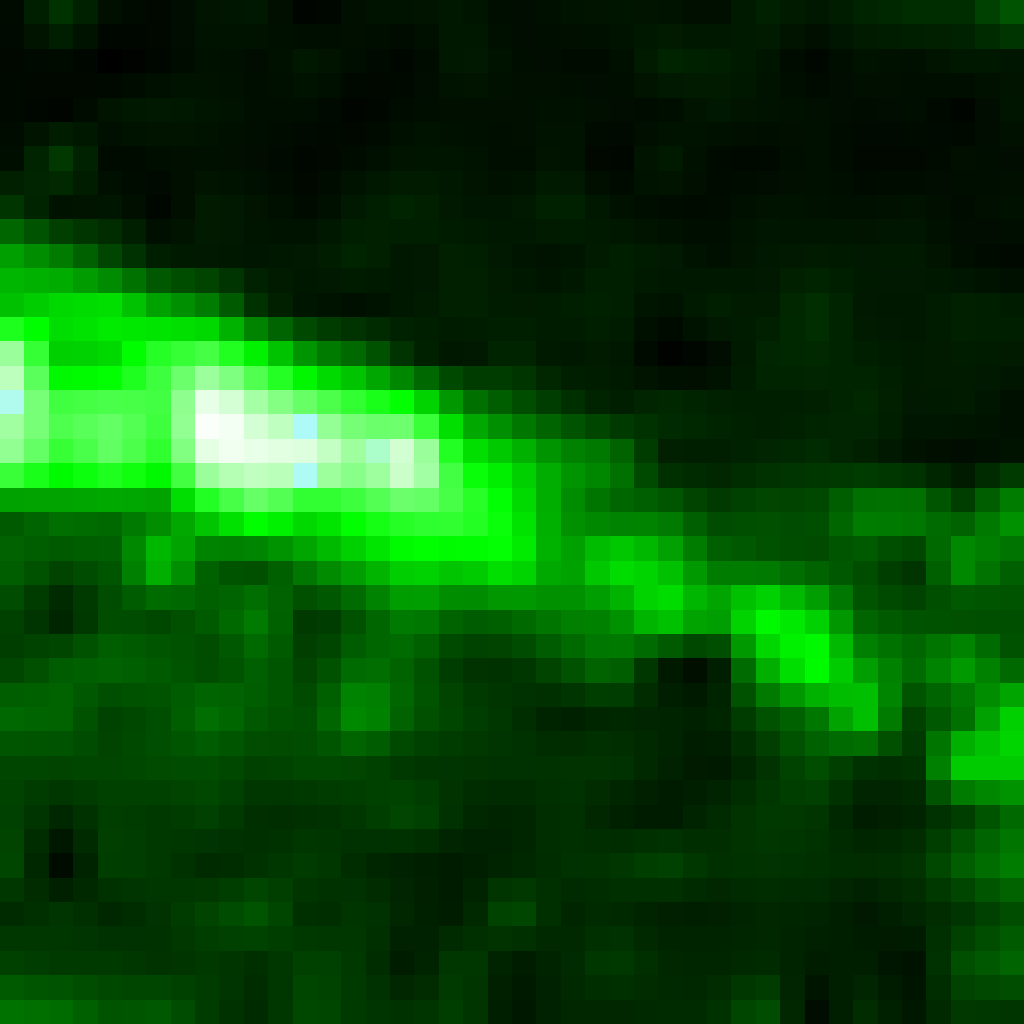}\\
			{\raisebox{0.7cm}	{\rotatebox[origin=c]{90}{~ {\scriptsize $z=4\um$} }}}&
			\includegraphics[width= 0.12\textwidth]{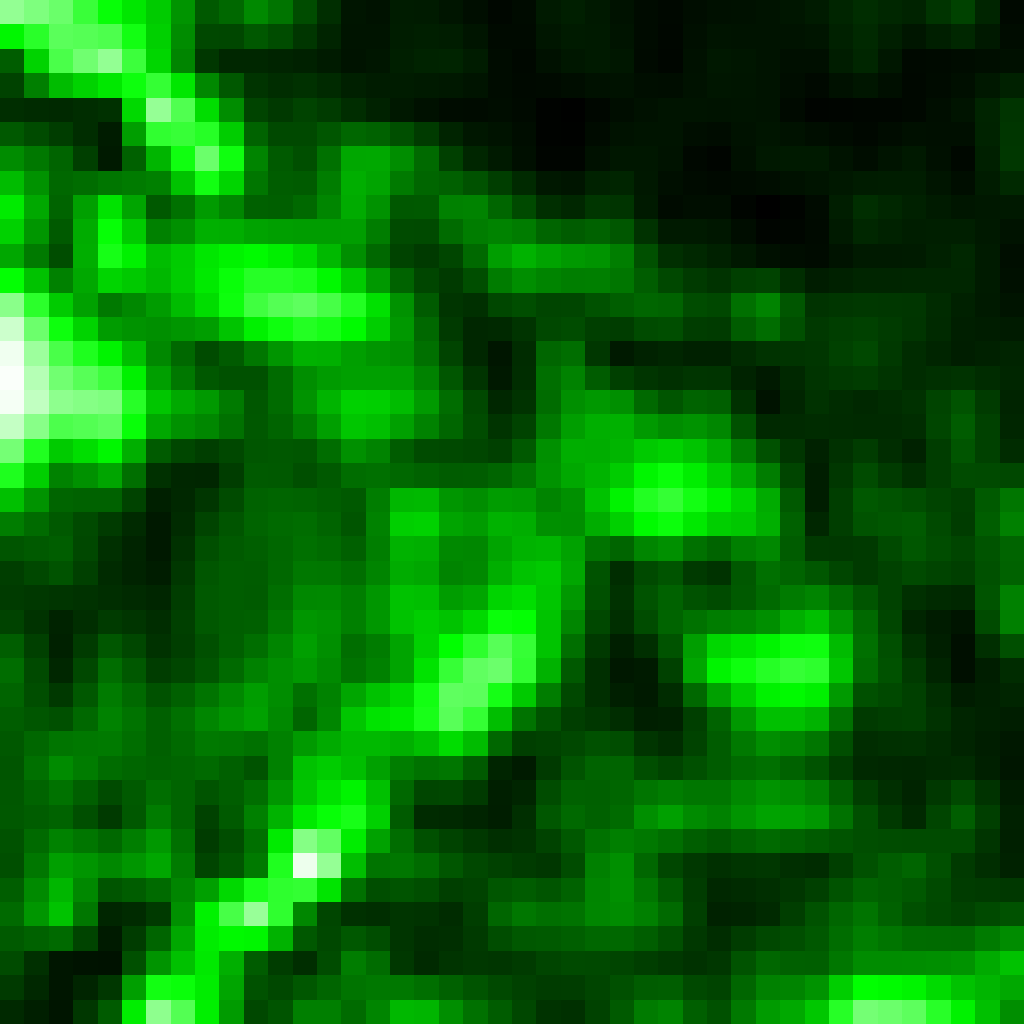}&
			\includegraphics[width= 0.12\textwidth]{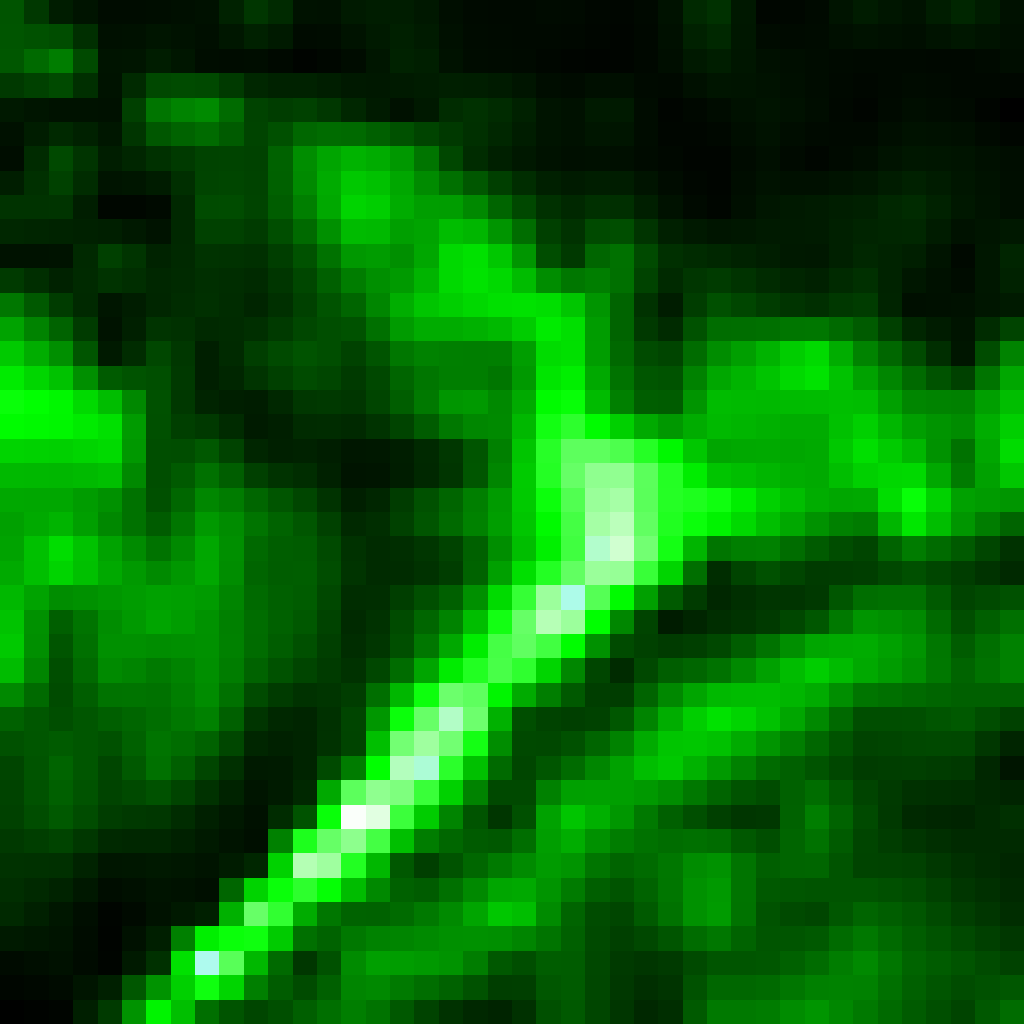}&
			\includegraphics[width= 0.12\textwidth]{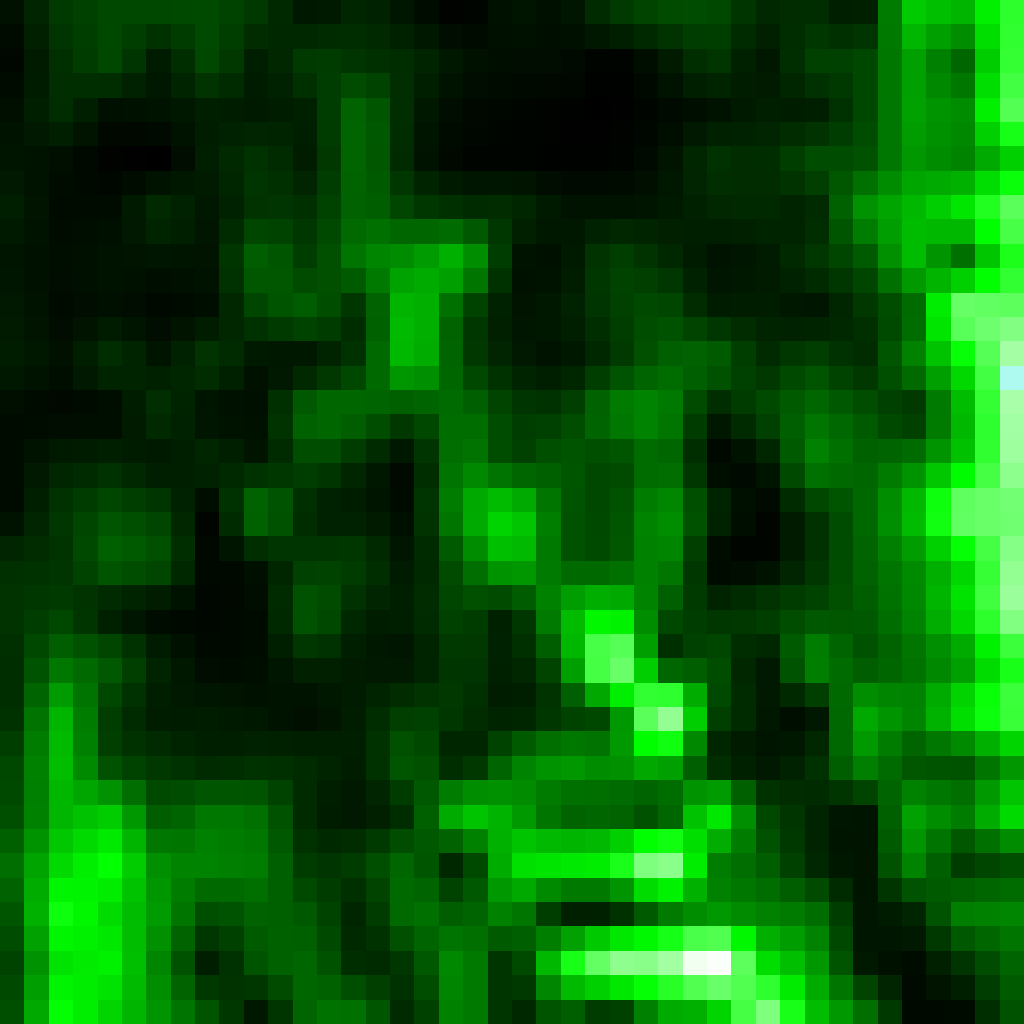}&
			\includegraphics[width= 0.12\textwidth]{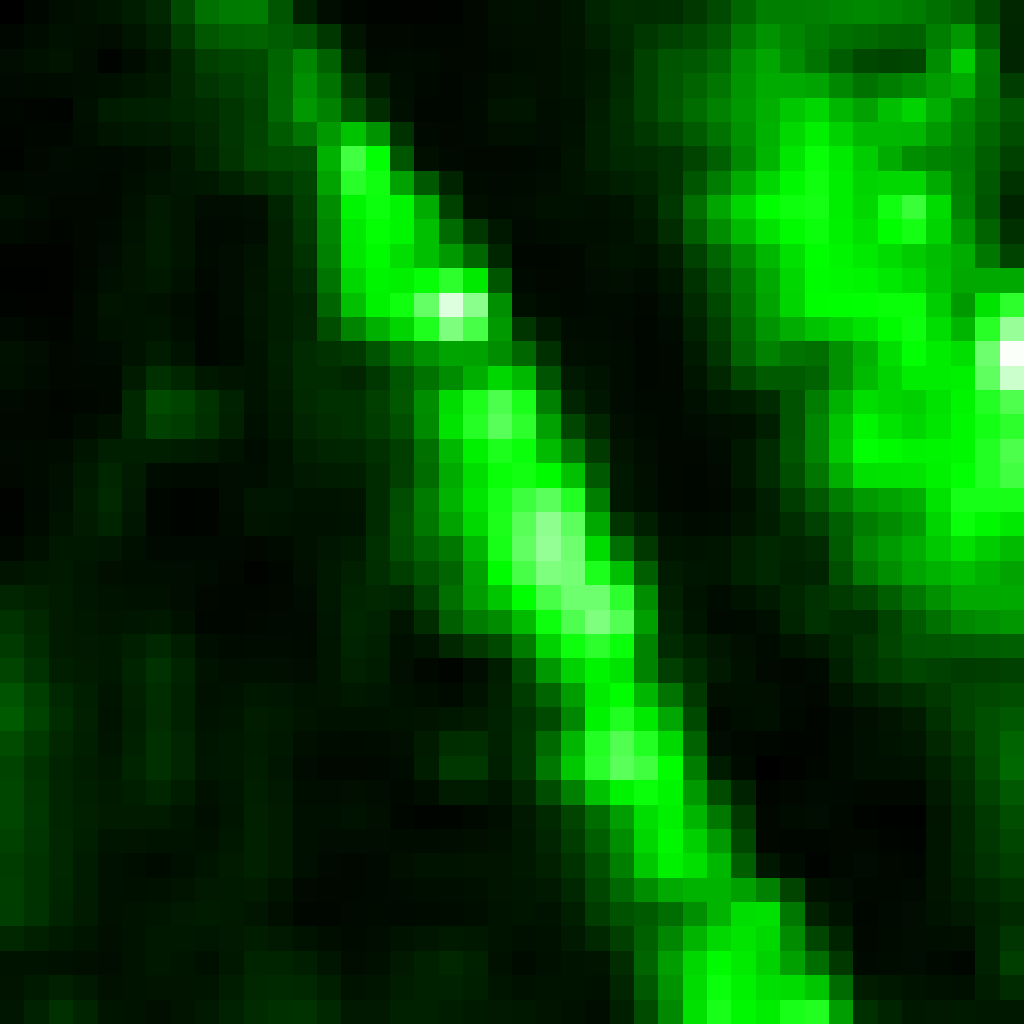}&
			\includegraphics[width= 0.12\textwidth]{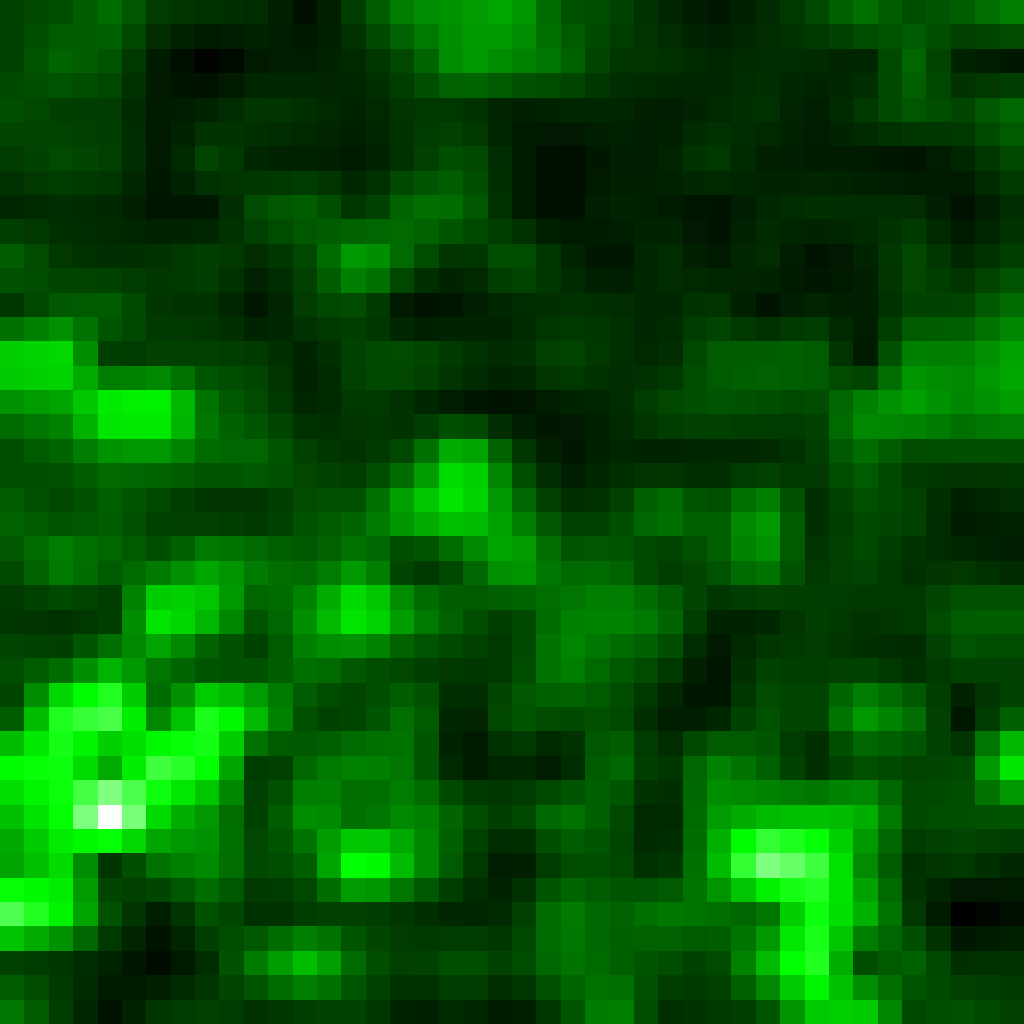}&
			\includegraphics[width= 0.12\textwidth]{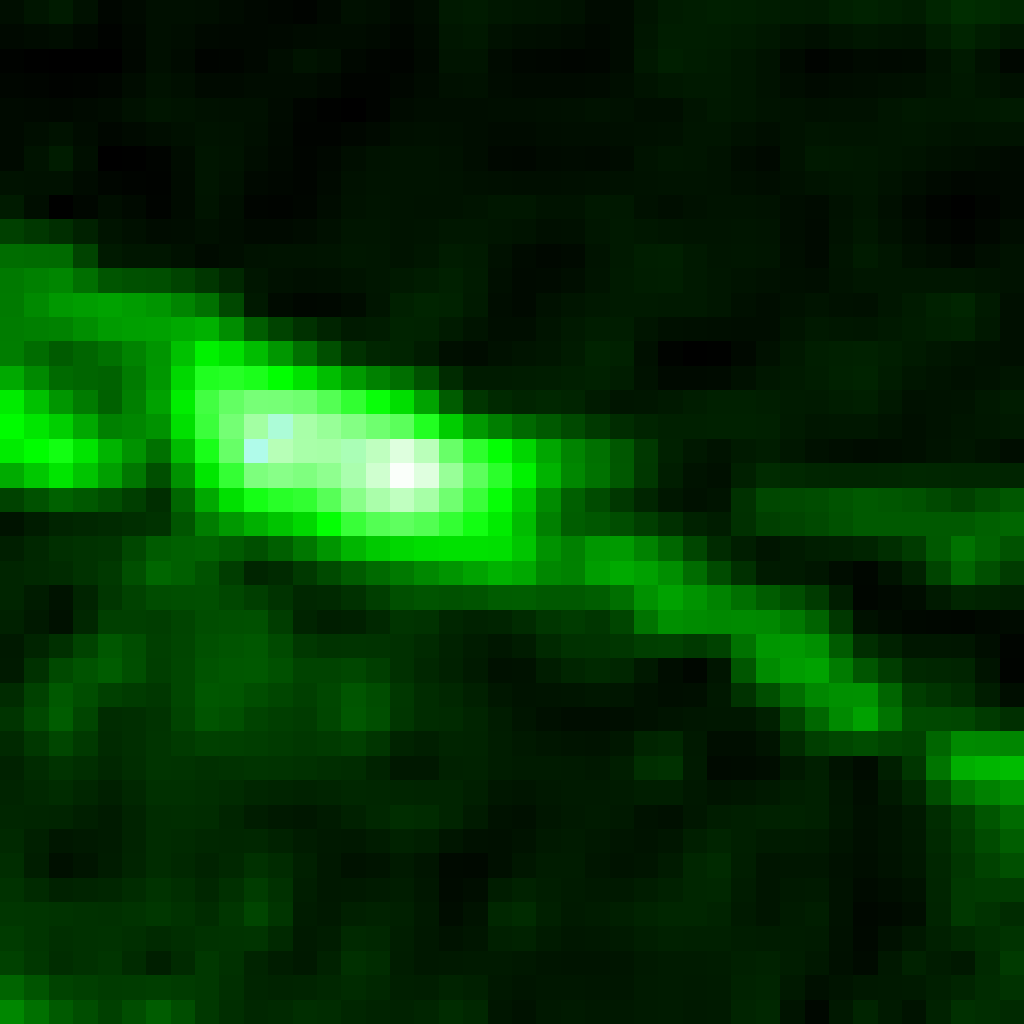}\\
			{\raisebox{0.7cm}	{\rotatebox[origin=c]{90}{~ {\scriptsize $z=6\um$} }}}&
			\includegraphics[width= 0.12\textwidth]{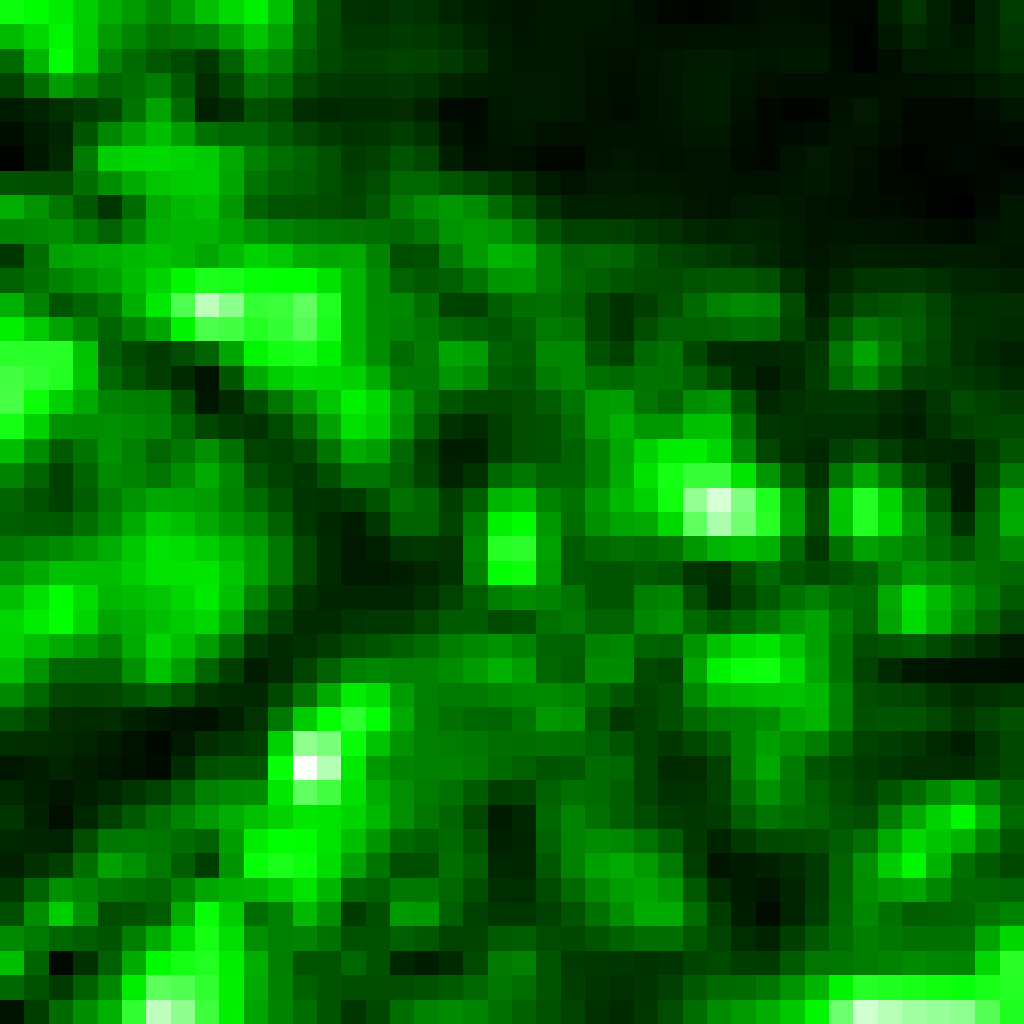}&
			\includegraphics[width= 0.12\textwidth]{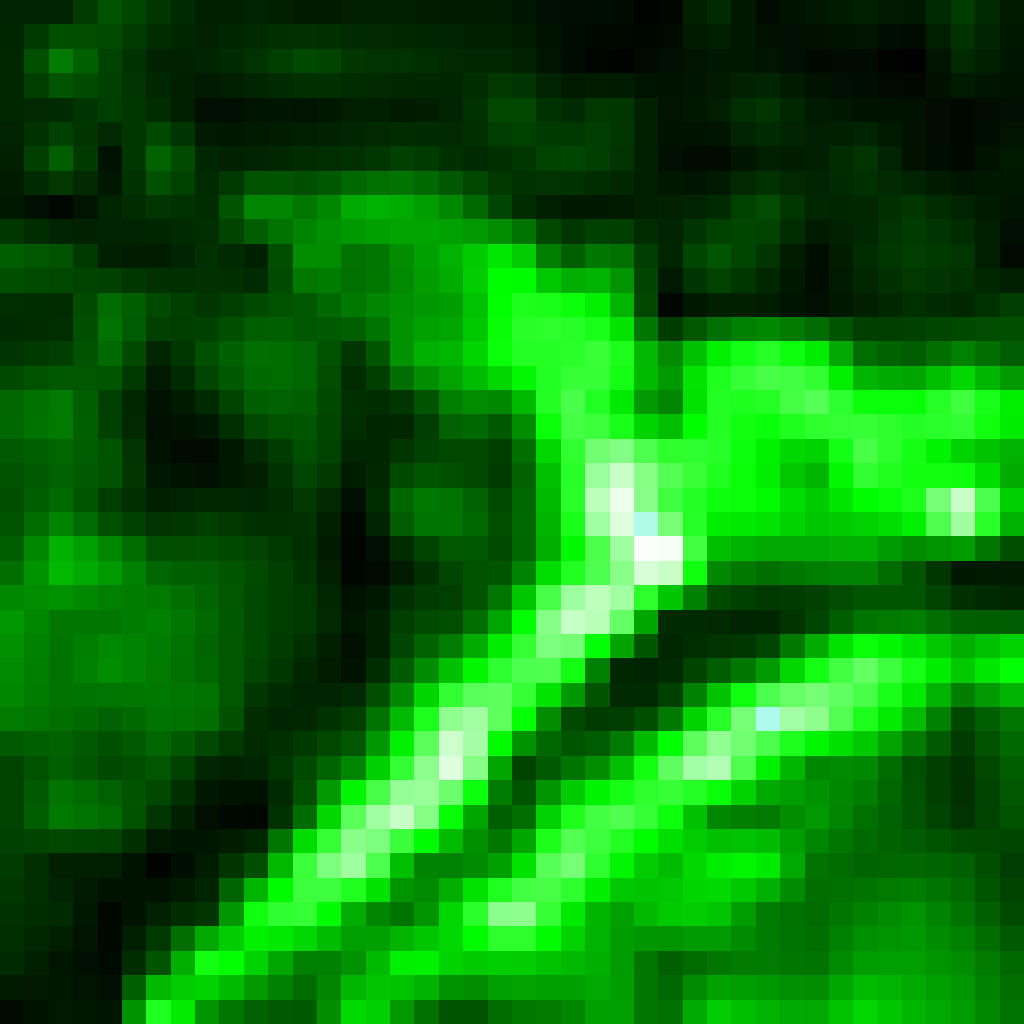}&
			\includegraphics[width= 0.12\textwidth]{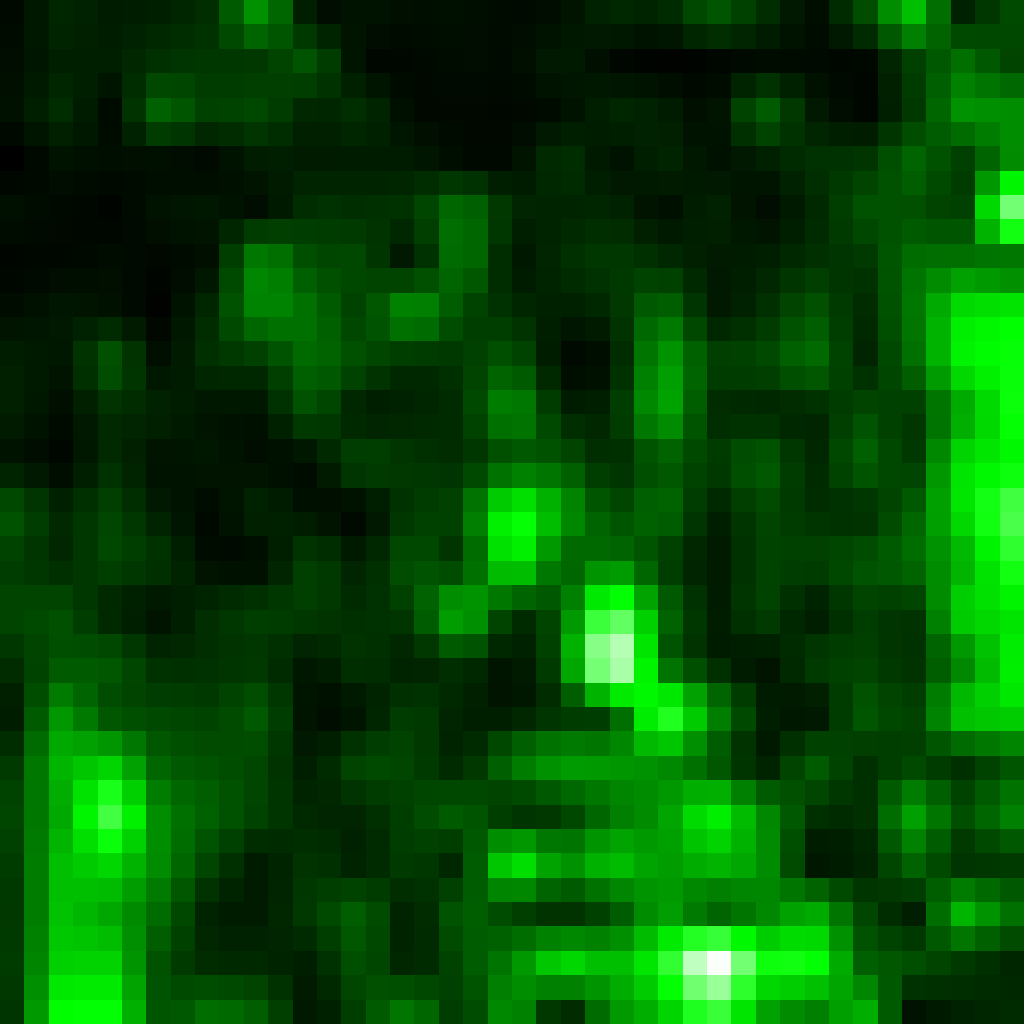}&
			\includegraphics[width= 0.12\textwidth]{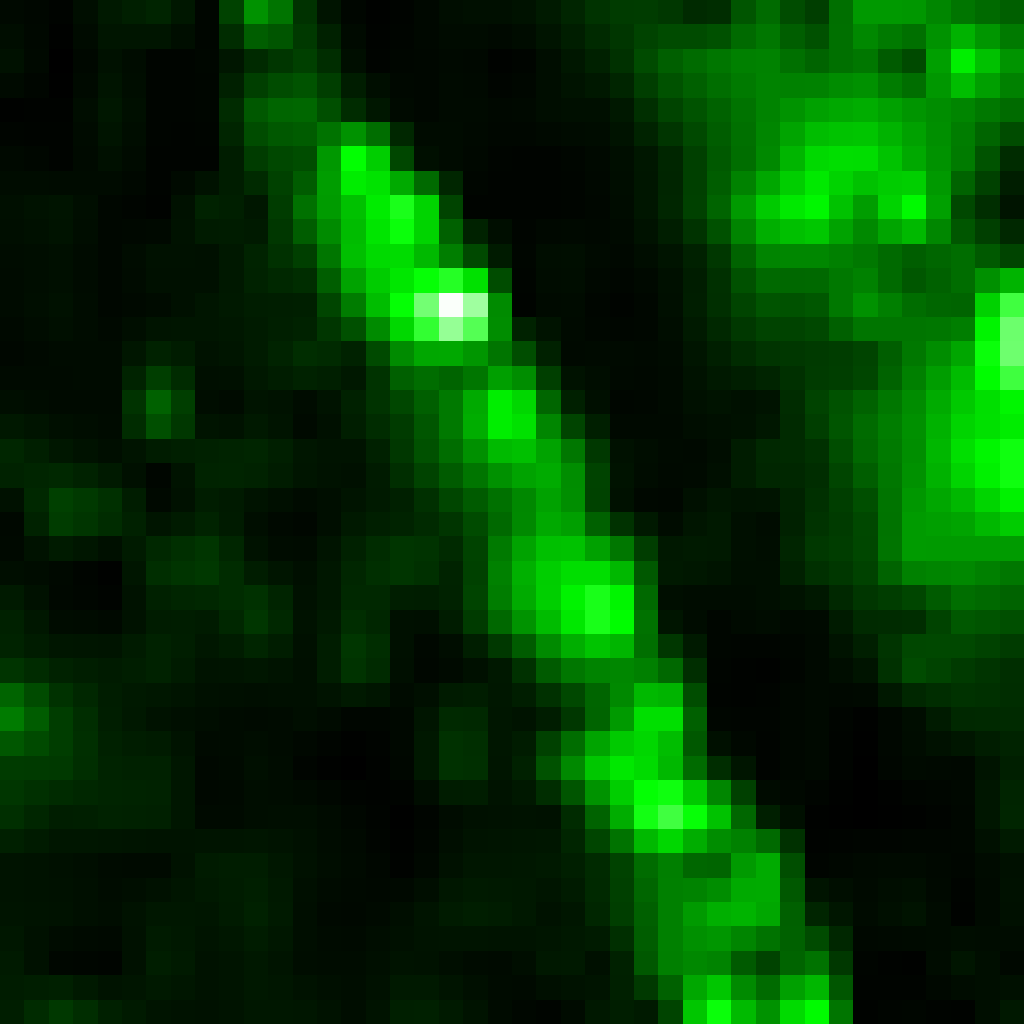}&
			\includegraphics[width= 0.12\textwidth]{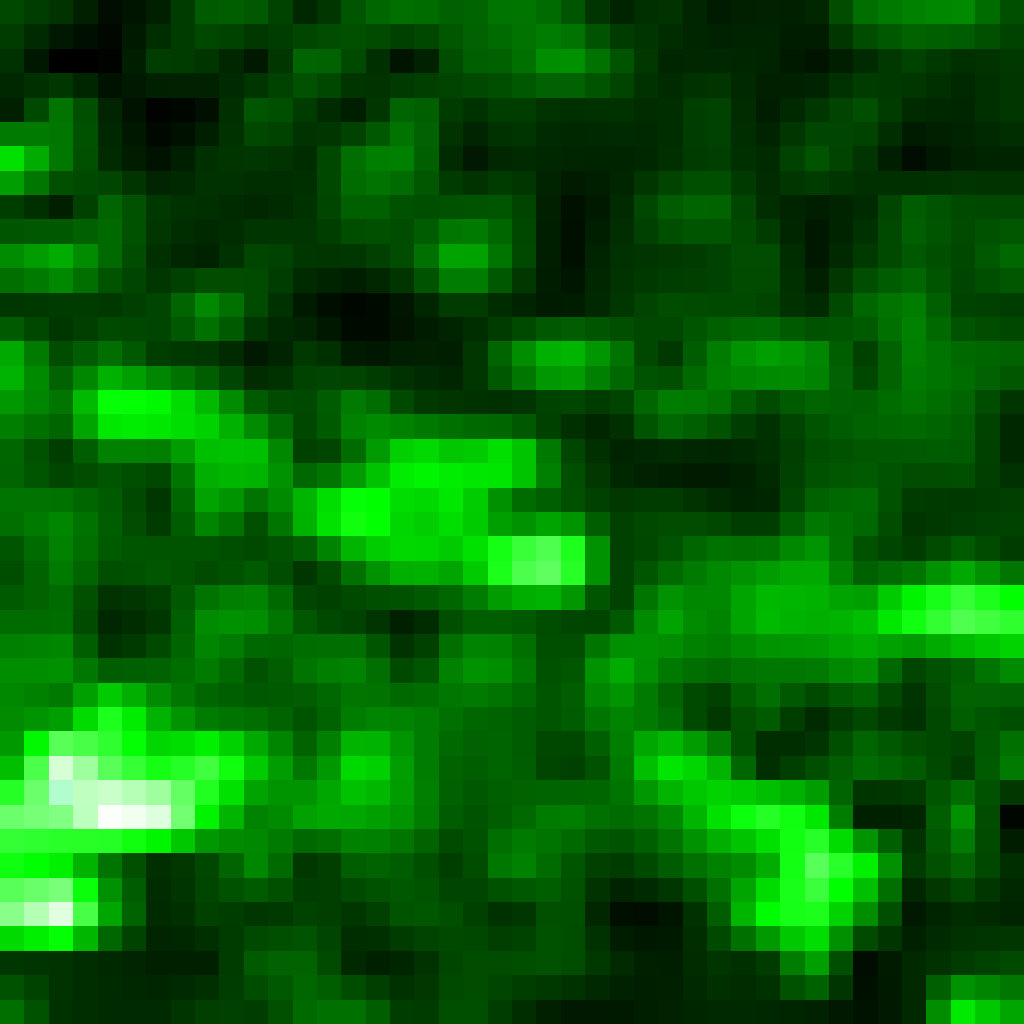}&
			\includegraphics[width= 0.12\textwidth]{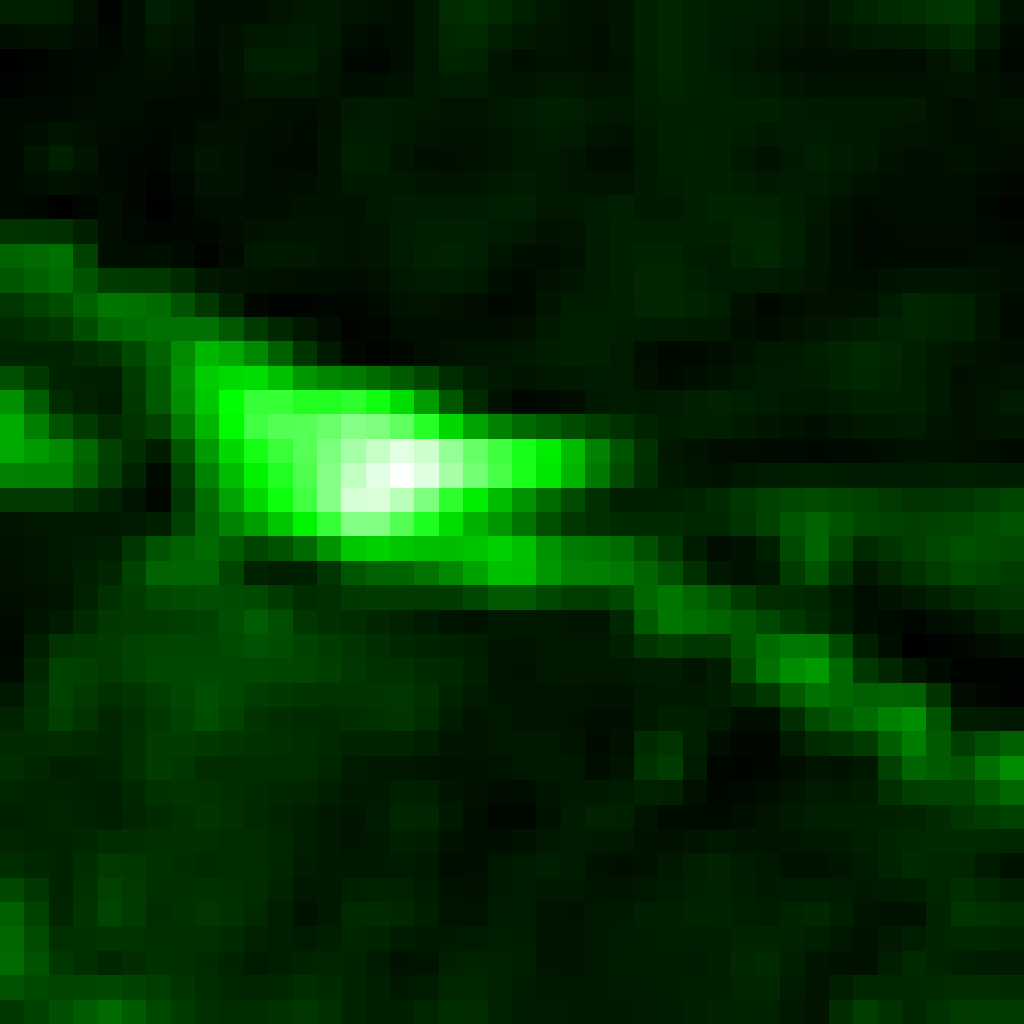}\\
			{\raisebox{0.7cm}	{\rotatebox[origin=c]{90}{~ {\scriptsize $z=8\um$} }}}&
			\includegraphics[width= 0.12\textwidth]{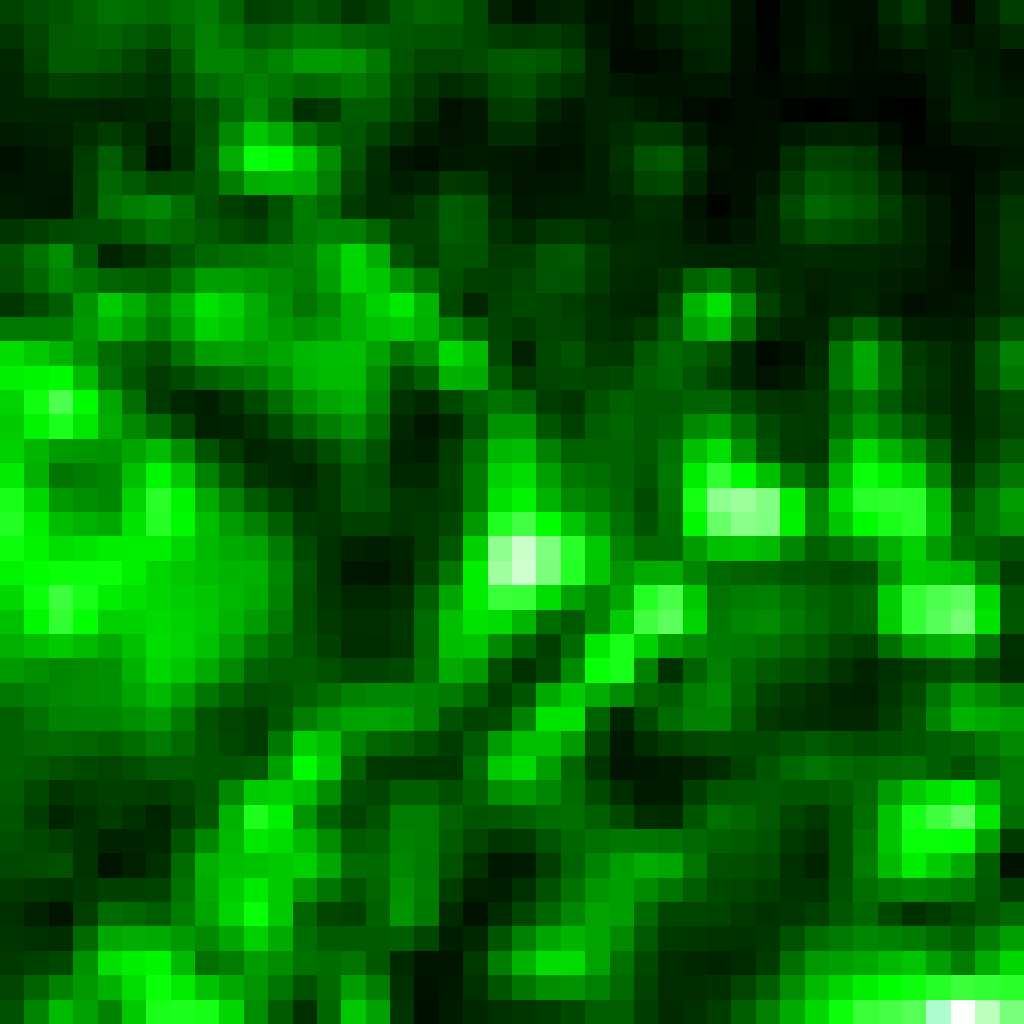}&
			\includegraphics[width= 0.12\textwidth]{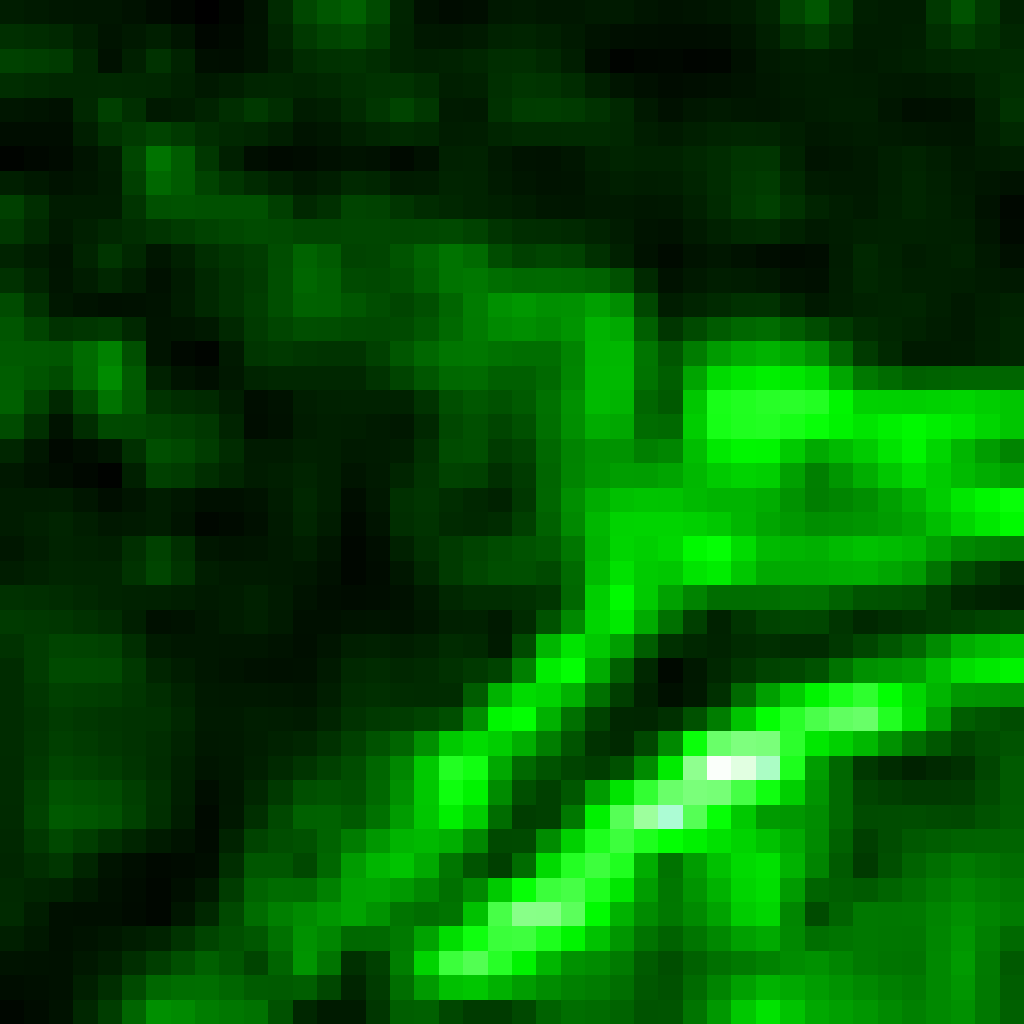}&
			\includegraphics[width= 0.12\textwidth]{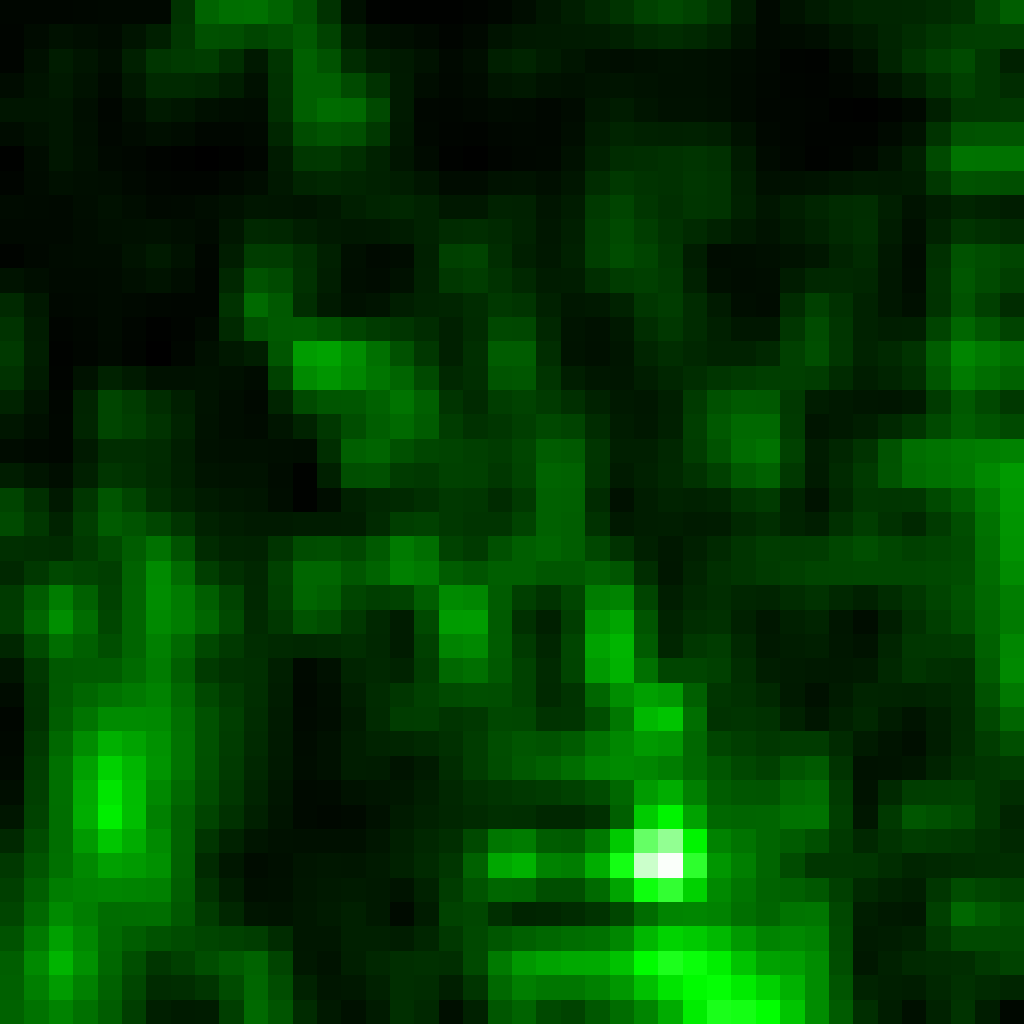}&
			\includegraphics[width= 0.12\textwidth]{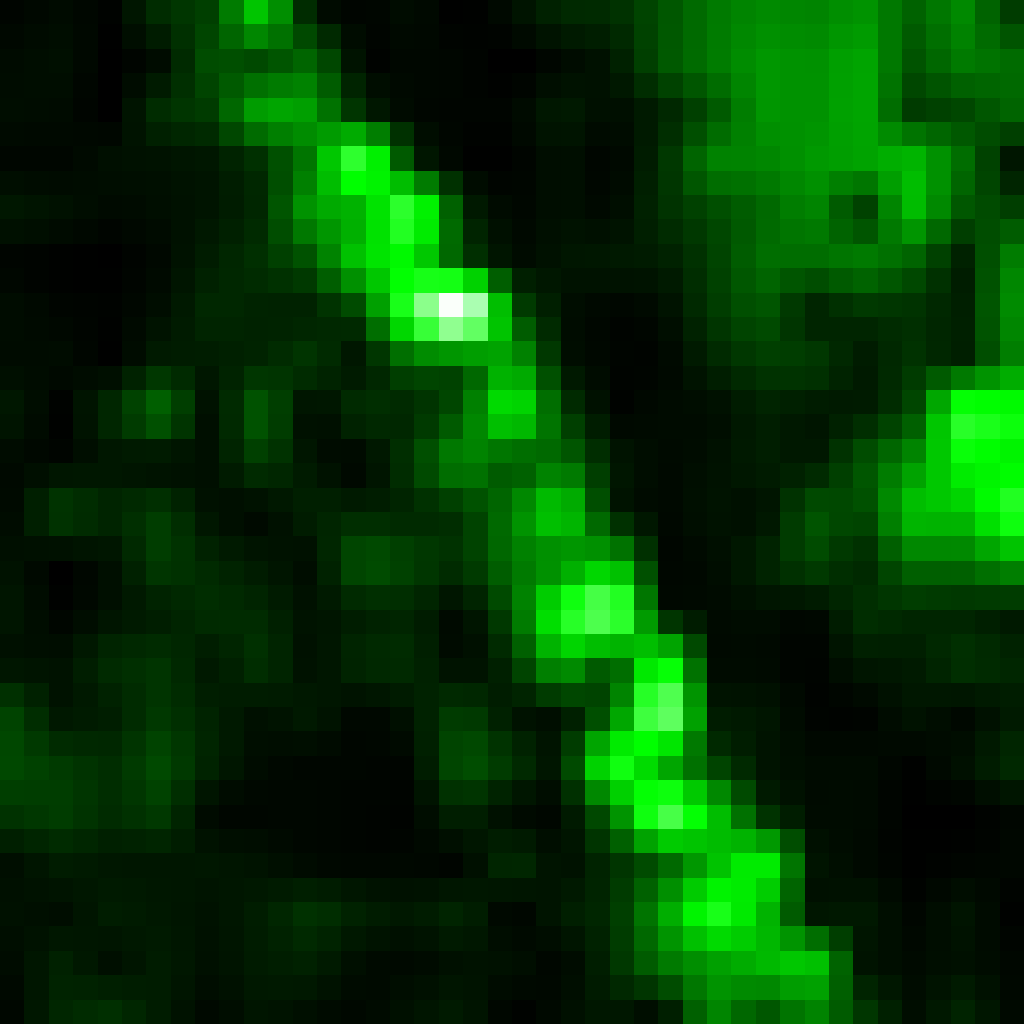}&
			\includegraphics[width= 0.12\textwidth]{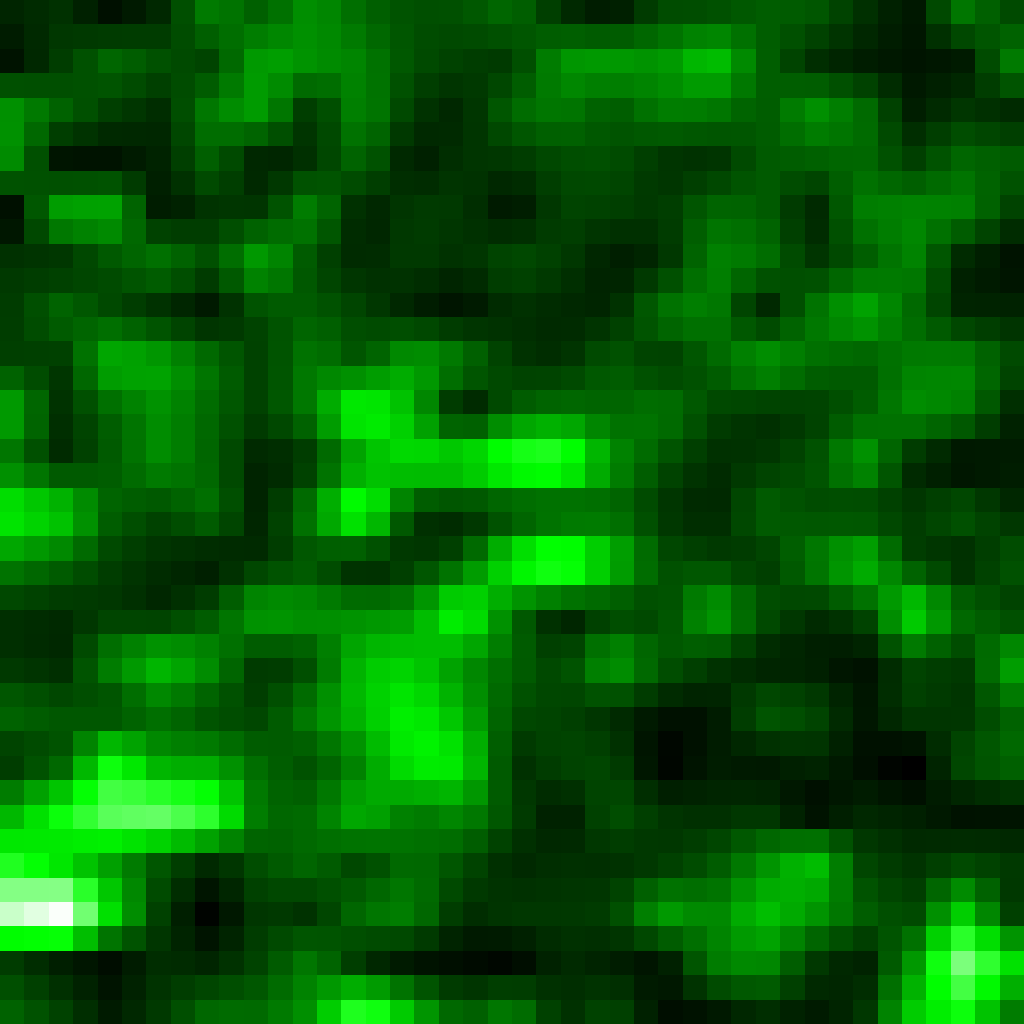}&
			\includegraphics[width= 0.12\textwidth]{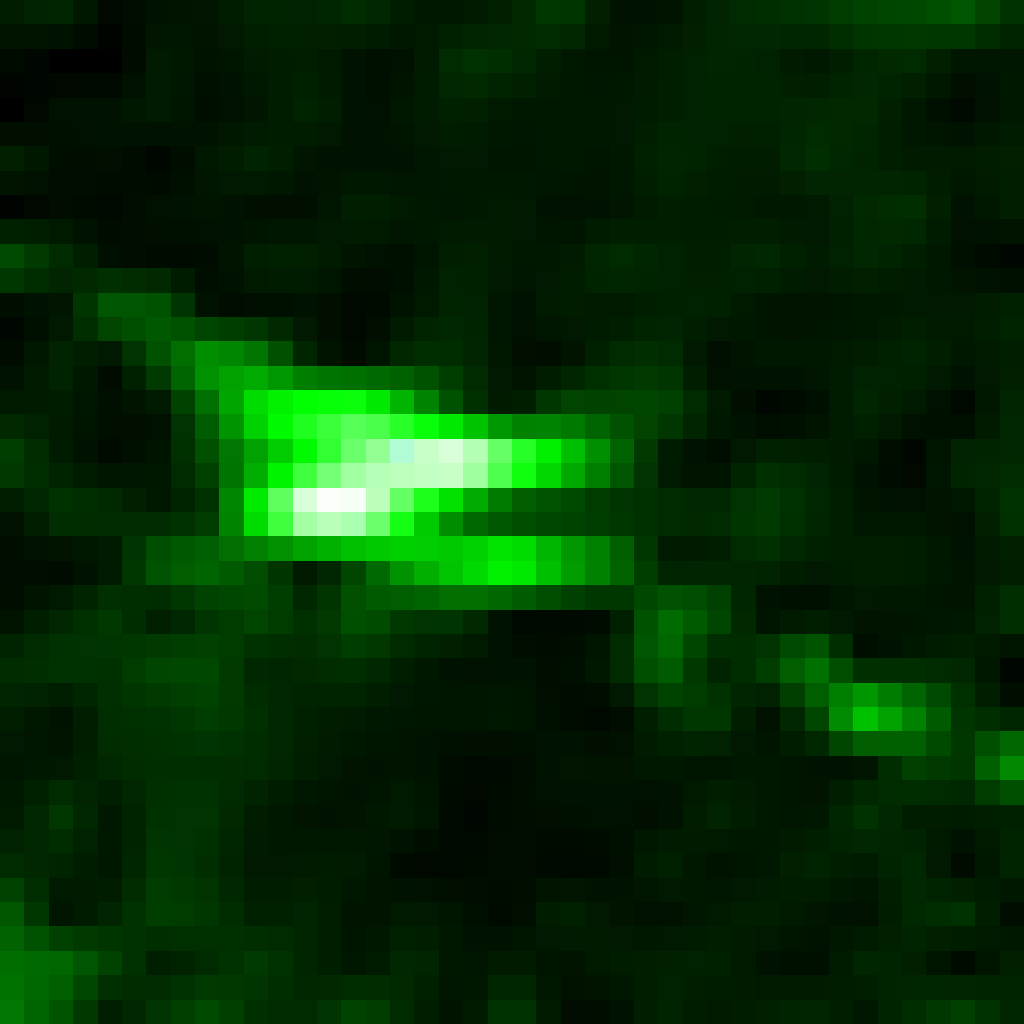}\\

		\end{tabular}
		\caption{\textbf{X-Y cross sections:} \bblue{To better visualize the 3D structure, we show additional x-y cross sections of our results of Fig. 8 from the main paper. We scanned in intervals of $2\um$ spanning depth of $\pm 8$ from the focused depth. Layer 1 is at $80\um$, layer 2 is at $130\um$, and layer 3 is at $190\um$.}  }
		\label{fig:res-xy}
	\end{center}
\end{figure*}

%% file: fig_class_compare.tex
\begin{figure*}[h!]
	\begin{center}		
		\begin{tabular}{@{}c@{~}c@{~}c@{~}c@{~}c@{~}}			
			\multicolumn{1}{c}{\hspace{-0.6cm} \small w/o optical }&	
			\multicolumn{1}{c}{\hspace{-0.6cm} \small  w/ optical}&	
			\multicolumn{1}{c}{\hspace{-0.6cm} \small CLASS }&	
			\multicolumn{1}{c}{\hspace{-0.6cm} \small CLASS  }&	
			\multicolumn{1}{c}{\hspace{-0.6cm} \small CLASS  }\\
			\multicolumn{1}{c}{\hspace{-0.6cm} \small modulation}&	
			\multicolumn{1}{c}{\hspace{-0.6cm} \small modulation}&	
			\multicolumn{1}{c}{\hspace{-0.6cm} \small first iteration}&	
			\multicolumn{1}{c}{\hspace{-0.6cm} \small fifth iteration}&	
			\multicolumn{1}{c}{\hspace{-0.6cm} \small tenth iteration}\\
			\includegraphics[width= 0.18\textwidth]{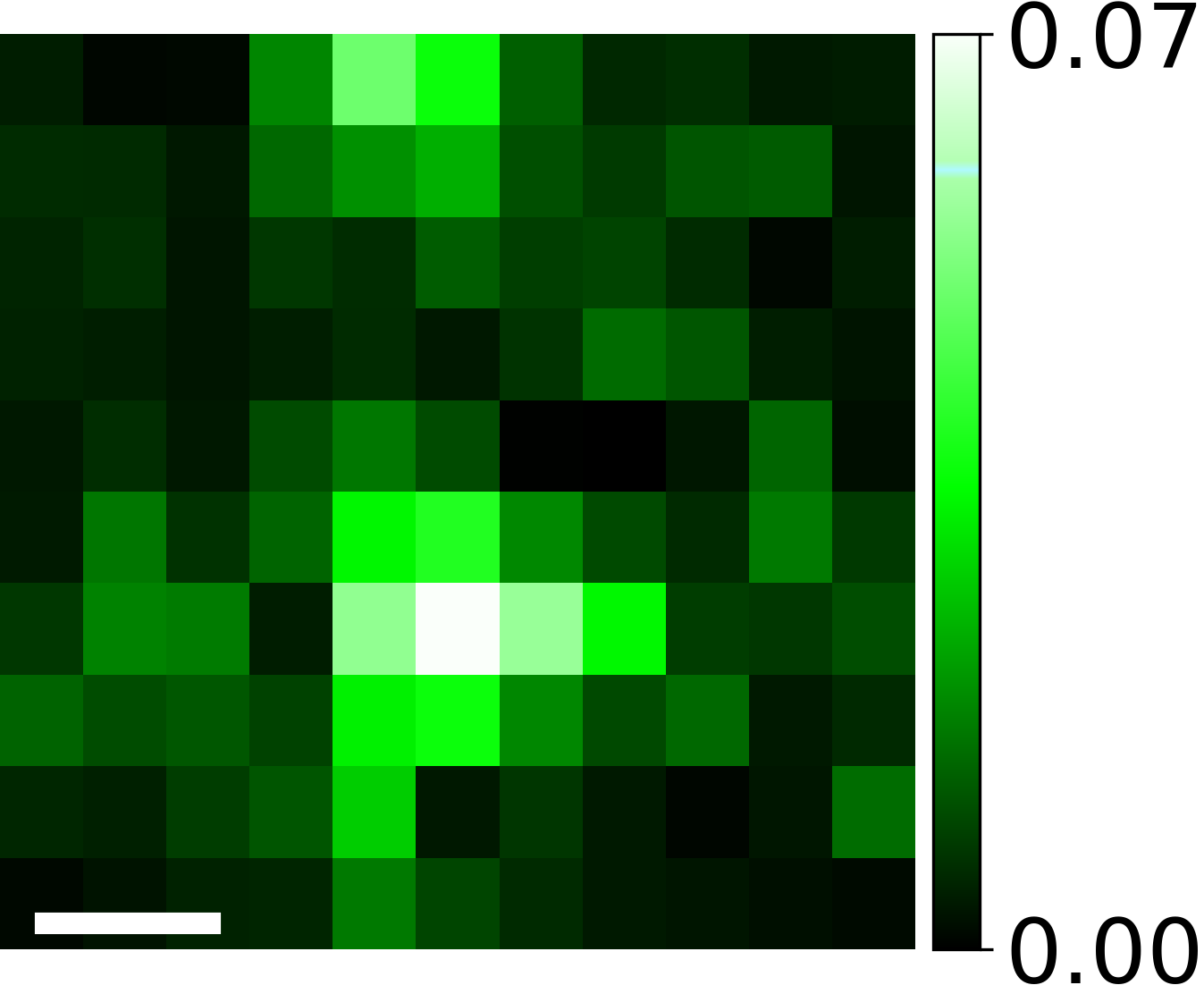}&
			\includegraphics[width= 0.18\textwidth]{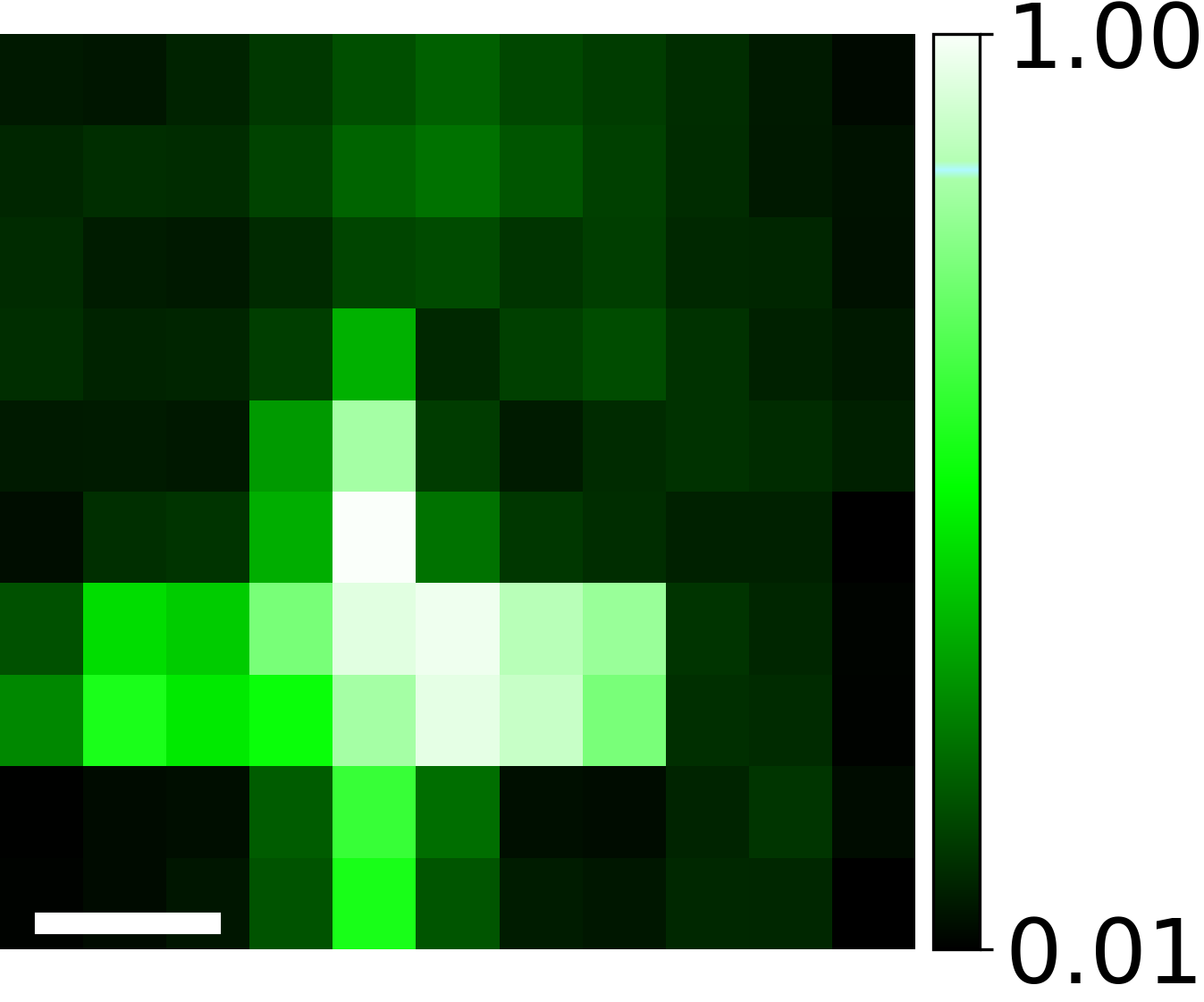}&
			\includegraphics[width= 0.18\textwidth]{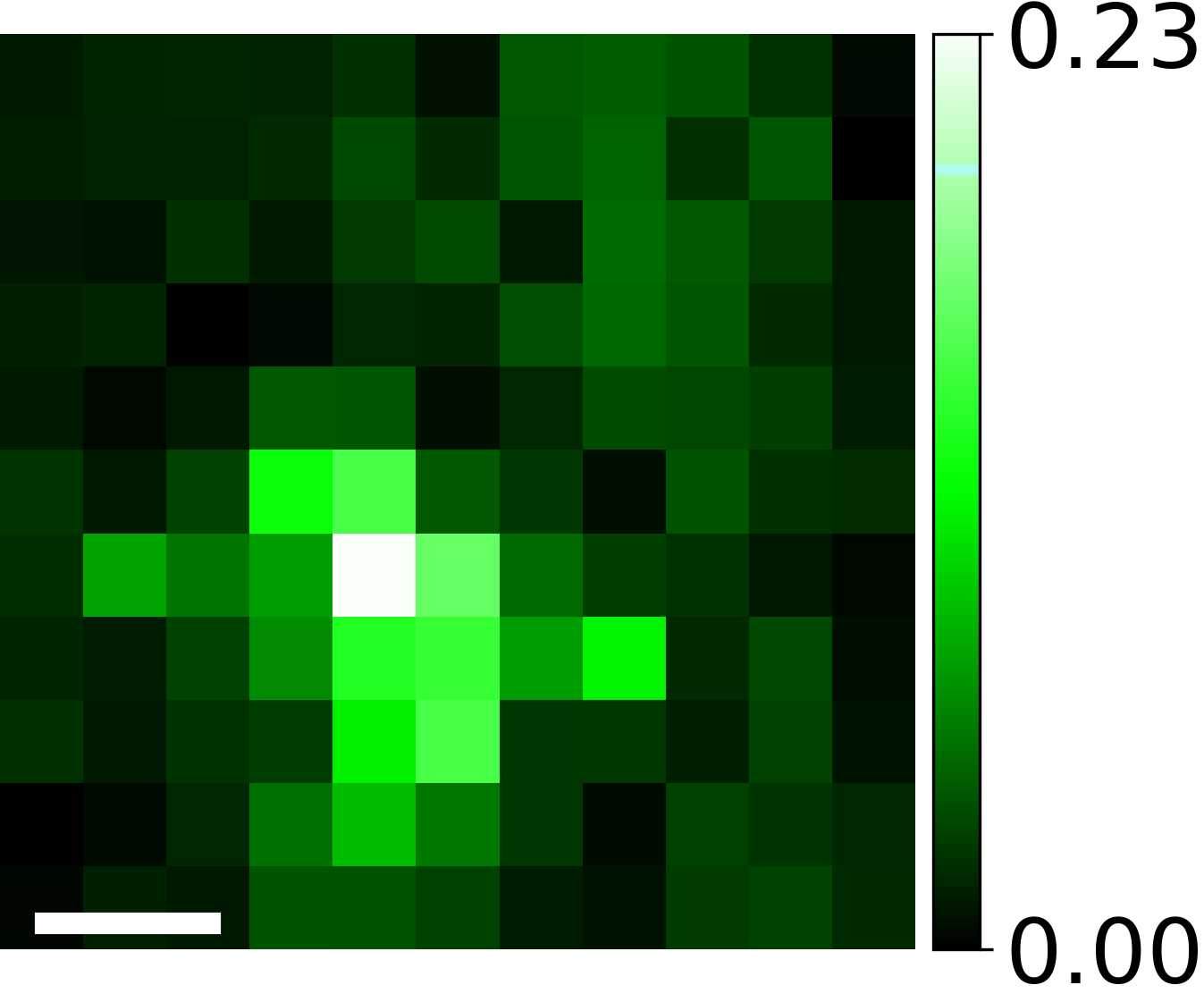}&
			\includegraphics[width= 0.18\textwidth]{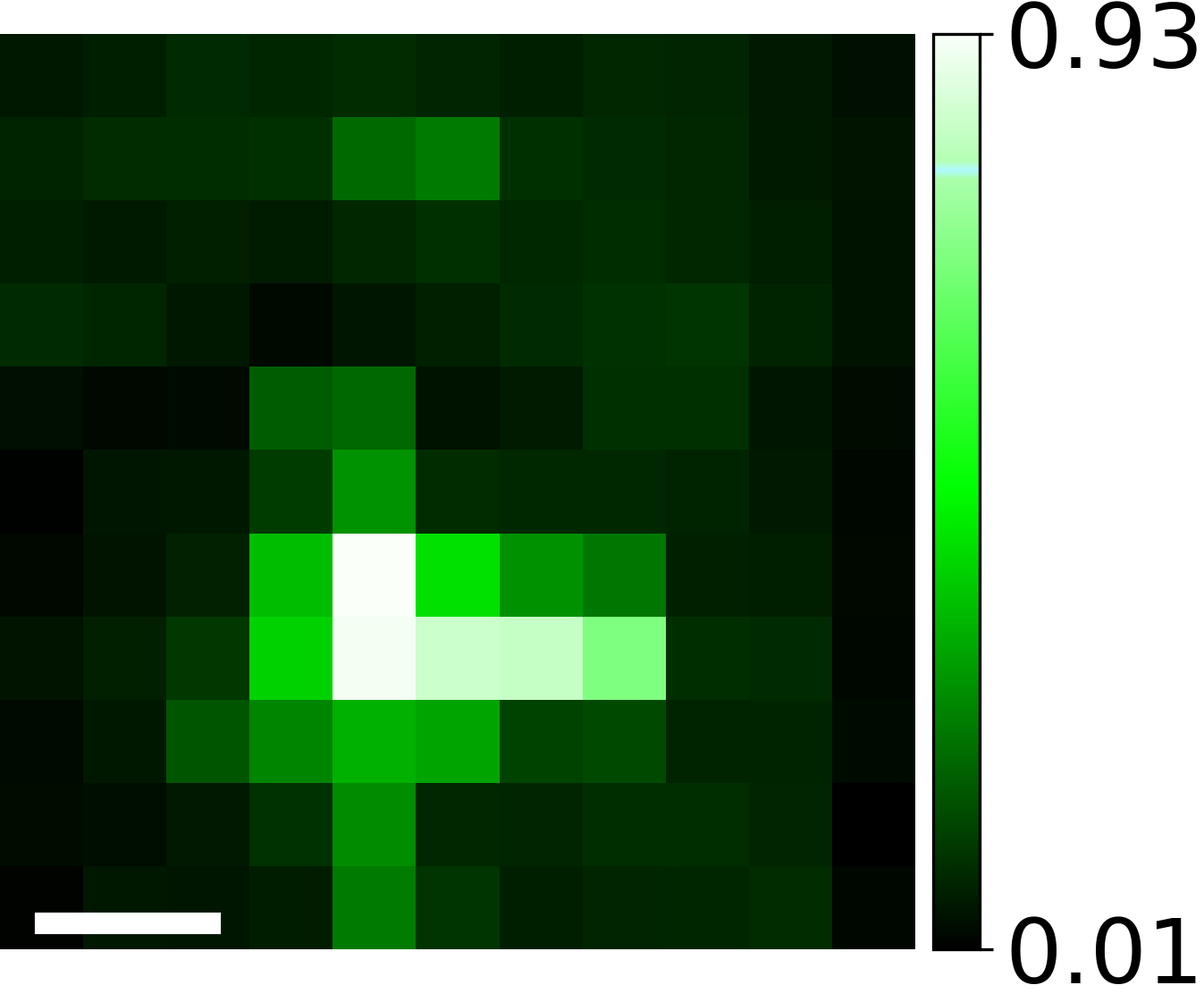}&
			\includegraphics[width= 0.18\textwidth]{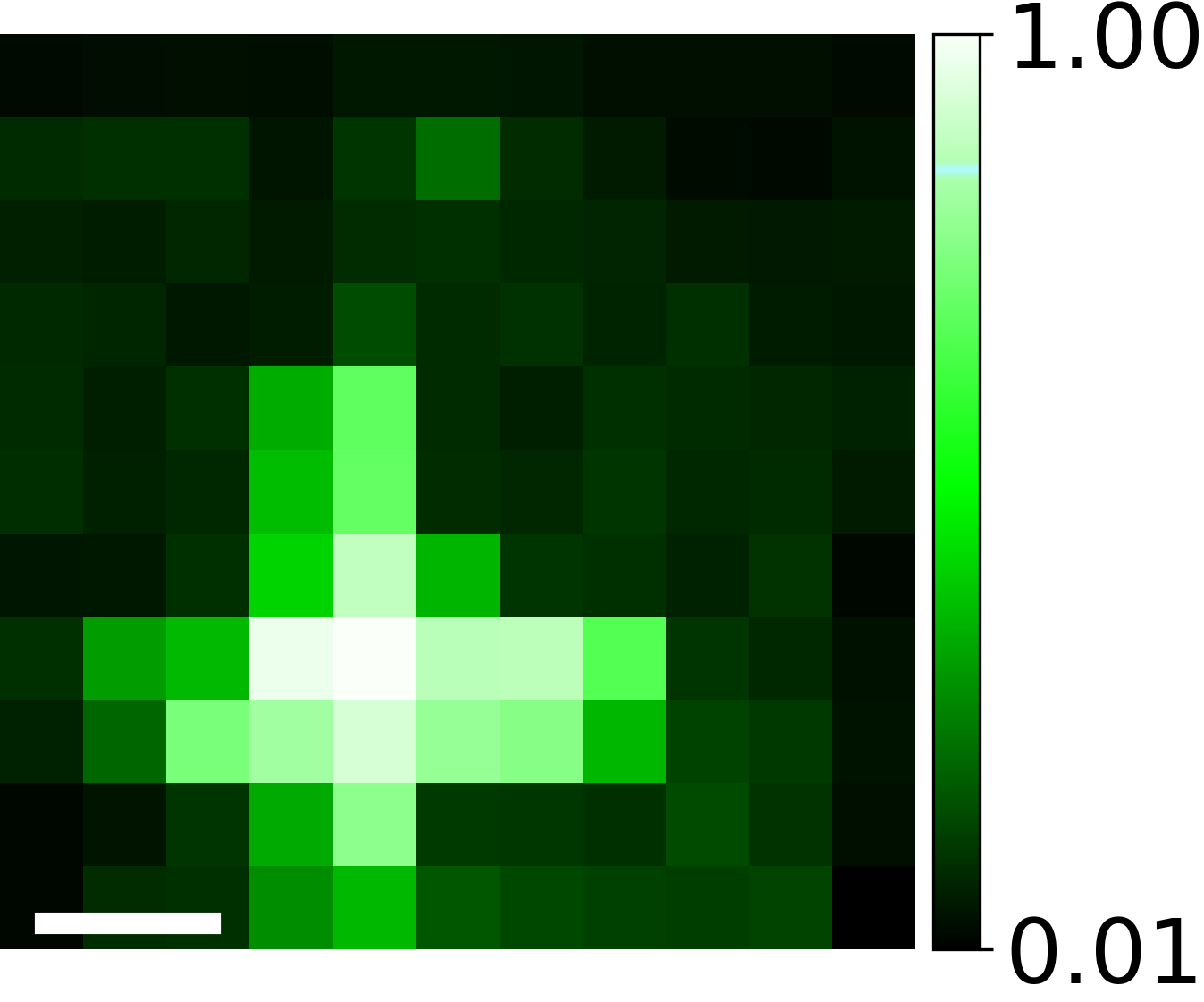}
		\end{tabular}
		\caption{\textbf{Comparison to CLASS algorithm:} \bblue{We compare our optical correction approach, with the digital correction CLASS algorithm. We visualize our captured result without applying optical modulation, with applying the optical modulation, and three results of the CLASS algorithm, applied to the wavefronts measured at the beginning of our algorithm when the SLM is blank, and  to the wavefronts acquired at the fifth and tenth iterations. Applying optical correction increases the SNR of the measured wavefronts and, as a result, allows the CLASS algorithm to achieve better  results. }}
		\label{fig:CLASS}
	\end{center}
\end{figure*}